\documentclass[a4paper,twoside,11pt]{book}

\usepackage{mathdots}
\usepackage{amsmath, amsthm, amssymb}
\usepackage{graphicx}
\usepackage{tabularx}
\usepackage{fancybox,fancyhdr}
\usepackage{epic,eepic,amsmath,amsthm,amssymb}
\usepackage{epsfig,cite,indentfirst,oldgerm}
\usepackage{multicol,boxedminipage}
\usepackage{float}
\usepackage{color}
\numberwithin{equation}{section}

\makeatletter
\def\cleardoublepage{\clearpage\if@twoside \ifodd\c@page\else
        \hbox{}
        \thispagestyle{empty}
        \newpage
        \if@twocolumn\hbox{} \newpage\fi\fi\fi}
\makeatother

\newtheorem{abs}{ }
\newtheorem{pref}{ }
\newtheorem{declaration}{ }
\def\notad{\mathbin{\triangleright{\mkern-10mu}{/}}}
\def\phis{\phi^{*}}
\def\nome{\textsf{q}}
\def\t{\tilde}

\def\nodd{n\in\tilde{\mathbb{N}}}

\def\no{\nonumber}
\def\ni{\noindent}
\def\proofend{\ensuremath{\square}}
\def\pr{'}
\def\beqa{\begin{eqnarray}}
\def\eeqa{\end{eqnarray}}
\def\ba{\begin{array}}
\def\ea{\end{array}}
\def\gl{\begin{swabfamily}gl\end{swabfamily}}
\def\psis{\psi^{*}}
\def\Psis{\Psi^{*}}
\def\union{\mathop{\bigcup}}
\def\vac{|\mbox{vac}\rangle}
\def\cav{\langle\mbox{vac}|}
\def\dprod{\mathop{\prod{\mkern-29.5mu}{\mathbf\longleftarrow}}}
\def\rprod{\mathop{\prod{\mkern-28.0mu}{\mathbf\longrightarrow}}}
\def\r{\rangle}
\def\l{\langle}
\def\a{\alpha}
\def\b{\beta}
\def\hb{\hat\beta}
\def\d{\delta}
\def\g{\gamma} 
\def\e{\epsilon}
\def\tg{\operatorname{tg}}
\def\ctg{\operatorname{ctg}}
 \def\sh{\operatorname{sh}}
 \def\ch{\operatorname{ch}}
\def\cth{\operatorname{cth}}
 \def\th{\operatorname{th}}
\def\eps{\varepsilon}
 \def\la{\lambda}
\def\tla{\tilde{\lambda}}
\def\Gh{\widehat{\Gamma}}
\def\tmu{\tilde{\mu}}
\def\s{\sigma}
\def\sul{\sum\limits}
\def\pl{\prod\limits}
\def\lt({\left(}
\def\rt){\right)}

\def\const{{\rm const}}
\def\argum{\{\mu_j\},\{\la_k\}} 
\def\umarg{\{\la_k\},\{\mu_j\}} 
\def\prodmu #1{\prod\limits_{j #1 k} \sinh(\mu_k-\mu_j)}
\def\prodla #1{\prod\limits_{j #1 k} \sinh(\lambda_k-\lambda_j)}
\def\tr{\operatorname{tr}}
\def\Res{\operatorname{Res}}
\def\det{\operatorname{det}}
\def\ivac{\langle\Omega|}
\def\fvac{|\Omega\rangle}
\def\fdirac{|0\rangle}
\def\idirac{\langle0|}
\def\psis{\psi^{*}}
\def\Psis{\Psi^{*}}
\def\lprod{\mathop{\prod{\mkern-29.5mu}{\mathbf\longleftarrow}}}
\def\rprod{\mathop{\prod{\mkern-28.0mu}{\mathbf\longrightarrow}}}
\def\complex{\mathbb{C}}
\def\integer{\mathbb{Z}}
\def\complex{\mathbb{C}}
\def\integer{\mathbb{Z}}
\def\cla{{\it Cl}_{\psi}}
\def\psis{\psi^*}
\def\Psis{\Psi^*}
\def\vac{|0\rangle}
\def\dvac{\langle 0|}

\newtheorem{definition}{Definition}
\newtheorem{example}{Example}

\newtheorem{lemma}{Lemma}

\newtheorem{remark}{Remark}

\newtheorem{theorem}{Theorem}

\def\no{\nonumber}
\def\ni{\noindent}
\def\proofend{\ensuremath{\square}}
\def\pr{'}
\def\beqa{\begin{eqnarray}}
\def\eeqa{\end{eqnarray}}
\def\ba{\begin{array}}
\def\ea{\end{array}}
\def\gl{\begin{swabfamily}gl\end{swabfamily}}
\def\psis{\psi^{*}}
\def\Psis{\Psi^{*}}
\def\union{\mathop{\bigcup}}
\def\dprod{\mathop{\prod{\mkern-29.5mu}{\mathbf\longleftarrow}}}
\def\rprod{\mathop{\prod{\mkern-28.0mu}{\mathbf\longrightarrow}}}
\def\r{\rangle}
\def\l{\langle}
\def\a{\alpha}
\def\b{\beta}
\def\hb{\hat\beta}
\def\d{\delta}
\def\g{\gamma}
\def\e{\epsilon}
\def\tg{\operatorname{tg}}
\def\ctg{\operatorname{ctg}}
 \def\sh{\operatorname{sh}}
 \def\ch{\operatorname{ch}}
\def\cth{\operatorname{cth}}
 \def\th{\operatorname{th}}
\def\eps{\varepsilon}
 \def\la{\lambda}
\def\tla{\tilde{\lambda}}
\def\Gh{\widehat{\Gamma}}
\def\tmu{\tilde{\mu}}
\def\s{\sigma}
\def\sul{\sum\limits}
\def\pl{\prod\limits}
\def\lt({\left(}
\def\rt){\right)}

\def\const{{\rm const}}
\def\argum{\{\mu_j\},\{\la_k\}} 
\def\umarg{\{\la_k\},\{\mu_j\}} 
\def\prodmu #1{\prod\limits_{j #1 k} \sinh(\mu_k-\mu_j)}
\def\prodla #1{\prod\limits_{j #1 k} \sinh(\lambda_k-\lambda_j)}

\def\tr{\operatorname{tr}}
\def\Res{\operatorname{Res}}
\def\det{\operatorname{det}}

\def\bd{b^{\dag}}

\def\phid{\phi^{\dagger}}
\def\varphid{\varphi^{\dagger}}

\begin{document}

\bibliographystyle{plain}


\thispagestyle{empty}

\begin{center}

\vspace*{\stretch{2}}

{\Huge Free fermions in classical and quantum integrable models}

\vspace*{\stretch{0.7}}

{\Large Michael Alan Wheeler}

\vspace*{\stretch{2}}

\large{
Submitted in total fulfilment of the requirements
of the degree of
\\  
Doctor of Philosophy 
\\ 
\bigskip\bigskip\smallskip 
December 2010 
}
         
\vspace*{\stretch{1}}
         
{\bf 
Department of Mathematics and Statistics 
\\ 
\smallskip 
The University of Melbourne
}

\vspace*{\stretch{0.5}}


\end{center}


\newpage

\thispagestyle{empty}

\phantom{nothing}


\setcounter{secnumdepth}{-4}
\chapter*{Abstract}

The aim of this thesis is to study the role of free fermions in certain classical and quantum integrable models. The classical models studied are the KP and BKP hierarchies of partial differential equations. We review the presence of free fermions in the theory of these hierarchies, a topic which was established in the series of papers \cite{dkm},\cite{djkm3},\cite{djkm4},\cite{djkm5},\cite{djkm6},\cite{djkm7},\cite{km},\cite{sat2},\cite{sat1} by the Kyoto school. The quantum models studied are descendents and relatives of the XYZ model, or its lattice equivalent, the eight-vertex model \cite{bax3},\cite{bax2},\cite{bax4},\cite{bax1}. We give a number of new results in the context of these models, revealing the presence of fermions in typical quantities such as Bethe eigenvectors, partition functions and scalar products. The appearance of classical fermions in these quantum mechanical objects has some powerful accompanying features.

\begin{abs}
{\rm Generally speaking, the fermions facilitate the calculation of the objects themselves. Often their presence causes complete factorization, and we study a number of partition functions and scalar products which fall under this category.}
\end{abs}

\begin{abs}
{\rm In special cases, the fermions allow us to prove that the object under consideration is a solution of a classical hierarchy. Since classical hierarchies and quantum models are both solved by a variant of inverse scattering, it is natural to expect that their solutions should be related. We hope that the results of this thesis indicate a deeper correspondence than that which is presently accepted in the literature.}
\end{abs}

The first two chapters of this thesis are introductory, and designed to fix the notations and concepts which appear later. Chapter 1 is based almost exclusively on \cite{jm1}, and reviews the polynomial solutions of the KP and BKP hierarchies via the fermionic approach. Chapter 2 describes the quantum inverse scattering method of the Leningrad school \cite{kbi}, in the setting of a generalized discrete model.

The last four chapters of the thesis contain new material. In chapter 3 we study the phase model of \cite{bik1}, \cite{bik2}, in parallel with the closely related $i$-boson model. We observe that the Bethe eigenvectors of these models admit a natural description in terms of charged and neutral fermions, respectively. This proves that the scalar products of these models are solutions of the KP and BKP hierarchies, respectively. We also derive generating functions for ordinary and strict plane partitions using the calculus of free fermions \cite{fw1},\cite{fwz3}. In chapter 4 we consider the $q$-boson model of \cite{bb}, which specializes to the models studied in chapter 3 in the respective limits $q \rightarrow \infty$ and $q\rightarrow i$. We provide a description of the Bethe eigenvectors using the more complicated algebra of $t$-deformed fermions \cite{jin1}, \cite{jin2}. This leads to a fermionic proof of a generating function for $t$-weighted plane partitions \cite{fw2},\cite{vul2}.

Chapter 5 studies the trigonometric limit of the XYZ model, the XXZ model. We review the determinant expressions discovered in \cite{ize} and \cite{sla} for the partition function and scalar product, respectively. By writing these objects as expectation values of charged fermions we prove that they are solutions of the KP hierarchy \cite{fwz4}, \cite{fwz5}. We also derive an explicit expression for the Bethe eigenvectors of the model. Chapter 6 considers the trigonometric Felderhof model \cite{da1},\cite{fel1},\cite{fel2},\cite{fel3}, which generalizes the free fermion point of the six-vertex model, and the elliptic Deguchi-Akutsu height model \cite{da2}, which generalizes the free fermion point of the eight-vertex SOS model \cite{bax4}. Motivated by the fermionic nature of these models, we give factorized expressions for their partition functions \cite{fwz1},\cite{fwz2}.


\newpage

\thispagestyle{empty}

\phantom{nothing}


\setcounter{secnumdepth}{-3}
\chapter*{Declaration}
 
This is to certify that

\begin{declaration}
{\rm
This thesis comprises only my original work towards
the PhD. 
}
\end{declaration}

\begin{declaration}
{\rm
Due acknowledgement has been made in the text to
all other material used.
}
\end{declaration}

\begin{declaration}
{\rm
This thesis is less than 100,000 words in length.
}
\end{declaration}

\bigskip
\bigskip
\bigskip

Michael Wheeler


\newpage

\thispagestyle{empty}

\phantom{nothing}


\setcounter{secnumdepth}{-2}
\chapter*{Preface}

This thesis contains material from the papers

\begin{pref}
{\rm A~Caradoc, O~Foda, M~Wheeler, M~Zuparic,
\textit{On the trigonometric Felderhof model with domain wall boundary conditions,}
J. Stat. Mech. 0703:P010 (2007)
}
\end{pref}
\begin{pref}
{\rm
O~Foda, M~Wheeler,
\textit{BKP plane partitions,}
JHEP01 (2007) 075
}
\end{pref}
\begin{pref}
{\rm
O~Foda, M~Wheeler,
\textit{Hall-Littlewood plane partitions and KP,}
Int. Math. Res. Notices (2009),
2597--2619
}
\end{pref}
\begin{pref}
{\rm
O~Foda, M~Wheeler, M~Zuparic,
\textit{Factorized domain wall partition functions in trigonometric vertex models,}
J. Stat. Mech. (2007) P10016
}
\end{pref}
\begin{pref}
{\rm
O~Foda, M~Wheeler, M~Zuparic,
\textit{Two elliptic height models with factorized domain wall partition functions,}
J. Stat. Mech. (2008) P02001
}
\end{pref}
\begin{pref}
{\rm
O~Foda, M~Wheeler, M~Zuparic,
\textit{On free fermions and plane partitions,}
Journal of Algebra {\bf 321} (2009),
3249--3273
}
\end{pref}
\begin{pref}
{\rm
O~Foda, M~Wheeler, M~Zuparic,
\textit{Domain wall partition functions and KP,}
J. Stat. Mech. (2009) P03017
}
\end{pref}
\begin{pref}
{\rm
O~Foda, M~Wheeler, M~Zuparic,
\textit{XXZ scalar products and KP,}
Nucl. Phys. B {\bf 820} (2009),
649--663
}
\end{pref}

\noindent which were written by the author (MW) and collaborators. The results and methods which have been selected from these papers are, in all instances, the original work of the author and his supervisor (OF).  


\newpage

\thispagestyle{empty}

\phantom{nothing}


\tableofcontents


\newpage

\thispagestyle{empty}

\phantom{nothing}


\listoffigures


\newpage

\thispagestyle{empty}

\phantom{nothing}

%
%
%
%
%
%

\setcounter{secnumdepth}{1}
\chapter{Free fermions in classical hierarchies}
\setcounter{secnumdepth}{2}
\setcounter{section}{-1}

\section{Introduction}

The study of classical integrable models, in modern times at least, dates back to the work of C~S~Gardner, J~M~Greene, M~D~Kruskal and R~M~Miura in \cite{ggkm}. In this seminal paper, the authors discovered the classical inverse scattering method for solving the non-linear Korteweg-deVries (KdV) equation. The central idea behind this method is the observation that the KdV equation can be written as the compatibility condition for two linear partial differential equations, the Lax pair \cite{lax}, effectively linearizing the problem. Soon afterwards, in \cite{miu} and \cite{mgk}, it was shown that solutions to the KdV equation give rise to infinitely many conserved quantities, proving that KdV is integrable in the sense of Liouville.

This chapter reviews the polynomial solutions of the KP hierarchy (which contains the KdV equation as a special case), and of the BKP hierarchy. The technique we employ stems from the collective works of M~Sato, Y~Sato and E~Date, M~Jimbo, M~Kashiwara, T~Miwa in the 1980s, which culminated in the beautiful paper \cite{jm1}. It involves embedding the infinite dimensional Lie algebras $A_{\infty}$ and $B_{\infty}$ in the algebra of free fermions, and constructing a highest weight representation for the latter. The solutions of the bilinear equations of the KP and BKP hierarchies are recovered from the orbit of the highest weight vector under the corresponding transformation groups, $GL_{\infty}$ and $O_{\infty}$, respectively.  

We have written this chapter in a manner which emphasizes the similarities between the theory of the KP and BKP hierarchies. The reader will find that every result which applies to one hierarchy has a parallel result in the context of the other. Except where otherwise indicated, the material of this chapter is taken from \cite{jm1}, and we have adopted most of the notations contained therein. Alternative references for the material on the KP hierarchy include chapter 9 of \cite{bbt}, \cite{kr} and the detailed book \cite{mjd}. Introductory material on the BKP hierarchy can also be found in \cite{djkm4}, \cite{hir}, \cite{orl} and \cite{you}. 

The synopsis of the first part of this chapter, on the KP hierarchy, is as follows. In section \ref{c-fermi} we introduce the algebra of charged fermions $Cl_{\psi}$ and construct a representation of this algebra on the Fock space $\mathcal{F}_{\psi}$. We write down the partition basis for $\mathcal{F}_{\psi}$ and define a bilinear form between this vector space and its dual. The Lie algebra $A_{\infty}$ is given as a subalgebra of $Cl_{\psi}$, and we define the KP evolution operator as an element of the corresponding transformation group $GL_{\infty}$. We conclude the section by calculating the form between the dual vacuum vector under time evolution, and a basis vector of $\mathcal{F}_{\psi}$. The result is a Schur polynomial, which collectively comprise a basis for the space of all polynomials.

In section \ref{c-KP} we start from the KP bilinear identity, which is an integral equation, and show that it gives rise to the infinitely many partial differential equations of the KP hierarchy. Using the basis of Schur polynomials, we write solutions of the KP hierarchy as forms between the dual vacuum under time evolution and special states $g_{\psi}|0\rangle$ in $\mathcal{F}_{\psi}$. We derive the necessary and sufficient condition on $g_{\psi}$ to ensure KP solubility, and call it the charged fermionic bilinear identity (CFBI).

In section \ref{c-sol} we construct solutions of the CFBI. We prove that $g_{\psi}$ satisfies the CFBI if and only if $g_{\psi} \in GL_{\infty}$. As an example, we write the partition basis vectors of $\mathcal{F}_{\psi}$ in this way, showing that every Schur polynomial is a solution of the KP hierarchy. There exists another perspective through which the CFBI can be solved. Expanding $g_{\psi}|0\rangle$ in the canonical basis of $\mathcal{F}_{\psi}$, we show that the CFBI is satisfied if and only if the expansion coefficients obey the KP Pl\"ucker relations. As an example, we give an explicit determinant solution of the KP Pl\"ucker relations. 

The second part of this chapter, on the BKP hierarchy, has an almost identical structure. In section \ref{n-fermi} we introduce the algebra of neutral fermions $Cl_{\phi}$ and construct a representation of this algebra on the Fock space $\mathcal{F}_{\phi}$. We give a basis for $\mathcal{F}_{\phi}$ and apply the previous bilinear form to this vector space and its dual. The Lie algebra $B_{\infty}$ is given as a subalgebra of $Cl_{\phi}$, and we define the BKP evolution operator as an element of the corresponding transformation group $O_{\infty}$. We conclude the section by calculating the form between the dual vacuum vector under time evolution, and a basis vector of $\mathcal{F}_{\phi}$. The result is a Schur $Q$-polynomial, which collectively comprise a basis for the space of polynomials in odd variables.

In section \ref{n-BKP} we start from the BKP bilinear identity, and show that it gives rise to the infinitely many partial differential equations of the BKP hierarchy. Using the basis of Schur $Q$-polynomials, we write solutions of the BKP hierarchy as forms between the dual vacuum under time evolution and special states $g_{\phi}|0\rangle$ in $\mathcal{F}_{\phi}$. We derive the necessary and sufficient condition on $g_{\phi}$ to ensure BKP solubility, and call it the neutral fermionic bilinear identity (NFBI).

In section \ref{n-sol} we construct solutions of the NFBI. We prove that $g_{\phi}$ satisfies the NFBI if and only if $g_{\phi} \in O_{\infty}$. As an example, we write the strict partition basis vectors of $\mathcal{F}_{\phi}$ in this way, showing that every Schur $Q$-polynomial is a solution of the BKP hierarchy. The NFBI can also be solved through another perspective. Expanding $g_{\phi}|0\rangle$ in the canonical basis of $\mathcal{F}_{\phi}$, we show that the NFBI is satisfied if and only if the expansion coefficients obey the BKP Pl\"ucker relations. As an example, we give an explicit Pfaffian solution of the BKP Pl\"ucker relations. 

\section{Charged fermions and related definitions}
\label{c-fermi}

\subsection{Charged fermions}
\label{cfermi}

Consider two infinite sets $\{\psi_m\}_{m \in \mathbb{Z}}$ and $\{\psis_m\}_{m \in \mathbb{Z}}$, where $m$ is assumed to run over all integers. The elements in these sets are called {\it charged fermions}. Each fermion $\psi_m$ is assigned positive charge $(+1)$, while each fermion $\psis_m$ is assigned negative charge $(-1)$. The charged fermions obey the anticommutation relations

\begin{align}
&
[\psi_m,\psi_n]_{+}=[\psis_m,\psis_n]_{+}=0 
\label{cfermi1} 
\\
&
[\psi_m,\psis_n]_{+}=\delta_{m,n} 
\nonumber
\end{align}

\noindent for all $m,n \in \mathbb{Z}$, where we have defined the anticommutator $[a,b]_{+} = ab+ba$. These equations contain as a special case $\psi_m^2={\psis_m}^2=0$ for all $m \in \mathbb{Z}$, which is a defining property of fermions.

\subsection{Clifford algebra $\cla$}
\label{clA}

The {\it Clifford algebra} $Cl_{\psi}$ is the associative algebra generated by $1$ and the charged fermions $\{\psi_m\}_{m \in \mathbb{Z}}$ and $\{\psis_m\}_{m \in \mathbb{Z}}$, modulo the anticommutation relations (\ref{cfermi1}). Considered as a vector space, $Cl_{\psi}$ has the basis 

\begin{align}
{\rm Basis}(Cl_{\psi})
=
\Big\{
1,
\psi_{m_1}\ldots \psi_{m_r},
\psis_{n_s} \ldots \psis_{n_1},
\psi_{m_1}\ldots \psi_{m_r} \psis_{n_s} \ldots \psis_{n_1}
\Big\}
\label{clA1}
\end{align}

\noindent where $\{m_1 > \cdots > m_r\}$ and $\{n_s < \cdots < n_1\}$ range over all integers, and the cardinalities of these sets take all values $r,s \geq 1$. The Clifford algebra $\cla$ splits into the following direct sum of subalgebras

\begin{align}
\cla = \bigoplus_{i \in \mathbb{Z}} \cla^{(i)}
\label{clA2}
\end{align}

\noindent where $\cla^{(i)}$ is the linear span of all monomials comprised of $r$ positive $(+1)$ fermions and $s$ negative $(-1)$ fermions, such that $r-s=i$. In this chapter we are mainly interested in the subalgebra $Cl_{\psi}^{(0)}$ which has the basis 

\begin{align}
{\rm Basis}\left(Cl_{\psi}^{(0)}\right)
=
\Big\{
1,
\psi_{m_1}\ldots\psi_{m_r}
\psis_{n_r}\ldots\psis_{n_1}
\Big\}
\label{clA3}
\end{align}

\noindent where $\{m_1>\cdots>m_r\}$ and $\{n_r <\cdots < n_1\}$ range over all integers, and the cardinality of these sets takes all values $r \geq 1$.
 
\subsection{Fock representations of $\cla$}
\label{repclA} 

We introduce a {\it vacuum vector} $\vac$ and {\it dual vacuum vector} $\dvac$, and define actions of $\cla$ on them by setting 
 
\begin{align}
\psi_m\vac=\psis_n\vac=0, 
\quad 
\dvac\psis_m=\dvac\psi_n=0
\label{repclA1} 
\end{align}

\noindent for all integers $m < 0, n \geq 0$. The {\it Fock space} $\mathcal{F}_{\psi}$ and {\it dual Fock space} $\mathcal{F}_{\psi}^{*}$ are the vector spaces generated linearly by the action of $Cl_{\psi}$ on $\vac$ and $\dvac$, respectively. By virtue of the annihilation relations (\ref{repclA1}), they have the bases

\begin{align}
{\rm Basis}(\mathcal{F}_{\psi})
=
\Big\{
\vac, 
\psi_{m_1}\ldots \psi_{m_r}\vac,
\psis_{n_s}\ldots \psis_{n_1}\vac,
\psi_{m_1}\ldots \psi_{m_r}
\psis_{n_s}\ldots \psis_{n_1}\vac
\Big\}
\label{repclA2}
\end{align}

\noindent where $\{m_1 > \cdots > m_r \geq 0\}$ and $\{n_s < \cdots < n_1 < 0 \}$ range over all non-negative and negative integers, respectively, and the cardinalities of these sets take all values $r,s\geq 1$, and

\begin{align}
{\rm Basis}(\mathcal{F}_{\psi}^{*})
=
\Big\{
\dvac,
\dvac \psi_{m_1} \ldots \psi_{m_r},
\dvac \psis_{n_s} \ldots \psis_{n_1},
\dvac \psi_{m_1} \ldots \psi_{m_r}
\psis_{n_s} \ldots \psis_{n_1}
\Big\}
\label{repclA3}
\end{align}

\noindent where $\{0 > m_1 > \cdots > m_r\}$ and $\{0 \leq n_s < \cdots < n_1 \}$ range over all negative and non-negative integers, respectively, and the cardinalities of these sets take all values $r,s \geq 1$. The representations of $Cl_{\psi}$ on the vector spaces (\ref{repclA2}) and (\ref{repclA3}) are called the {\it Fock representations.}

Using the definition of the Clifford subalgebras $(\ref{clA2})$, we decompose $\mathcal{F}_{\psi}$ and $\mathcal{F}_{\psi}^{*}$ into the following direct sums of subspaces

\begin{align}
\mathcal{F}_{\psi}=\bigoplus_{i\in\integer}\mathcal{F}_{\psi}^{(i)}, 
\quad
\mathcal{F}_{\psi}^*=\bigoplus_{i\in\integer}\mathcal{F}_{\psi}^{*(i)} 
\label{repclA4}
\end{align}

\noindent where $\mathcal{F}_{\psi}^{(i)}$ and $\mathcal{F}_{\psi}^{*(i)}$ are the subspaces generated linearly by the action of $\cla^{(i)}$ on $\vac$ and $\dvac$, respectively. The bases of $\mathcal{F}_{\psi}^{(0)}$ and $\mathcal{F}_{\psi}^{*(0)}$ are given by

\begin{align}
{\rm Basis}\left( \mathcal{F}_{\psi}^{(0)} \right)
=
\Big\{
\vac,
\psi_{m_1}\ldots\psi_{m_r}
\psis_{n_r}\ldots\psis_{n_1}\vac
\Big\}
\label{repclA5}
\end{align}

\noindent where $\{m_1 > \cdots > m_r \geq 0\}$ and $\{ n_r < \cdots < n_1 < 0 \}$ range over all non-negative and negative integers, respectively, and the cardinality of these sets takes all values $r\geq 1$, and

\begin{align}
\quad\quad\ \ 
{\rm Basis} \left( \mathcal{F}_{\psi}^{*(0)} \right)
=
\Big\{
\dvac,
\dvac \psi_{m_1}\ldots \psi_{m_r}
\psis_{n_r} \ldots \psis_{n_1}
\Big\}
\label{repclA6}
\end{align} 

\noindent where $\{0 > m_1 > \cdots > m_r \}$ and $\{0 \leq n_r < \cdots < n_1 \}$ range over all negative and non-negative integers, respectively, and the cardinality of these sets takes all values $r \geq 1$.

\begin{remark}
{\rm
In future sections we will occasionally exploit the notation

\begin{align}
\psi_{\{m\}} = \psi_{m_1}\ldots \psi_{m_r},
\quad\quad
\psis_{\{n\}} = \psis_{n_s}\ldots \psis_{n_1}
\end{align}



\noindent for all ordered sets $\{m\} = \{m_1 > \cdots > m_r\}$ and $\{n\} = \{n_s < \cdots < n_1\}$ with cardinality $r,s \geq 1$. We also define $\psi_{\{m\}} = \psis_{\{n\}} =1$ when $\{m\}$ and $\{n\}$ are empty. For example, using this notation the basis (\ref{repclA5}) can be written as

\begin{align}
{\rm Basis}\left( \mathcal{F}_{\psi}^{(0)} \right)
=
\Big\{ \psi_{\{m\}} \psis_{\{n\}} |0\rangle\ \Big|\ {\rm card}\{m\} = {\rm card}\{n\} \Big\}
\label{repclA7}
\end{align}

\noindent where $\{m\}$ and $\{n\}$ range over all ordered sets of non-negative and negative integers, with the same cardinality. Similarly the basis (\ref{repclA6}) can be written as

\begin{align}
{\rm Basis}\left(\mathcal{F}_{\psi}^{*(0)}\right)
=
\Big\{ \langle 0| \psi_{\{m\}} \psis_{\{n\}}\ \Big| \ {\rm card}\{m\} = {\rm card}\{n\} \Big\}
\label{repclA8}
\end{align}

\noindent where $\{m\}$ and $\{n\}$ range over all ordered sets of negative and non-negative integers, with the same cardinality. 
}
\end{remark}

\subsection{Partition basis}
\label{partitions}

For all integers $l \geq 1$ we define the {\it charged vacua}

\begin{align}
|l\rangle
=
\psi_{l-1} \ldots \psi_0 \vac,
\quad
|-l\rangle
=
\psis_{-l} \ldots \psis_{-1} \vac
\label{partitions1}
\end{align}

\noindent and the {\it dual charged vacua}

\begin{align}
\langle l|
=
\dvac \psis_0 \ldots \psis_{l-1},
\quad
\langle -l|
=
\dvac \psi_{-1} \ldots \psi_{-l}
\label{partitions2}
\end{align}

\noindent It is possible to express the elements of the bases (\ref{repclA5}) and (\ref{repclA6}) in terms of the charged vacua, as we see in the following lemma.

\begin{lemma}
{\rm 
The Fock subspaces $\mathcal{F}_{\psi}^{(0)}$ and $\mathcal{F}_{\psi}^{*(0)}$ have the equivalent bases

\begin{align}
\ \
{\rm Basis}\left( \mathcal{F}_{\psi}^{(0)} \right)
=
\Big\{
\vac,
\psi_{m_1} \ldots \psi_{m_l} |-l\rangle
\Big\}
\label{partitions3}
\end{align}

\noindent where $\{m_1 > \cdots > m_l > -l\}$ range over all integers greater than $-l$, and all values $l \geq 1$ are allowed, and

\begin{align}
{\rm Basis}\left( \mathcal{F}_{\psi}^{*(0)} \right)
=
\Big\{
\dvac,
\langle -l| \psis_{n_l} \ldots \psis_{n_1}
\Big\}
\label{partitions4}
\end{align}

\noindent where $\{-l < n_l < \cdots < n_1 \}$ range over all integers greater than $-l$, and all values $l \geq 1$ are allowed.
}
\end{lemma}

\begin{proof}
{\rm
 Firstly, we show that every element 
$
\psi_{m_1} \ldots \psi_{m_l} |-l\rangle
$ 
of (\ref{partitions3}) is an element of (\ref{repclA5}), up to a minus sign. When $l = 1$ we have $\psi_{m_1}|-1\rangle = \psi_{m_1} \psis_{-1} \vac$, which is clearly an element of (\ref{repclA5}). For $l > 1$, there exists some $r \geq 1$ such that

\begin{align}
\psi_{m_1} \ldots \psi_{m_l} |-l\rangle
=
\psi_{m_1} \ldots \psi_{m_r}
\psi_{m_{r+1}} \ldots \psi_{m_l}
\psis_{-l} \ldots \psis_{-1} \vac
\label{partitions5}
\end{align}

\noindent with 
$
\{m_1 > \cdots > m_r \geq 0 > m_{r+1} > \cdots > m_l\}
$. 
Let $\{n_r < \cdots < n_1\}$ be the set 
$\{-l < \cdots < -1\}$ with the integers 
$
\{m_{l} < \ldots < m_{r+1}\}
$ 
omitted. Using the anticommutation relations (\ref{cfermi1}) and the fact that the string of fermions $\psi_{m_{r+1}} \ldots \psi_{m_l}$ annihilates the vacuum $|0\rangle$, equation (\ref{partitions5}) becomes  

\begin{align}
\psi_{m_1} \ldots \psi_{m_l} |-l\rangle
=
(-)^{\sum_{i=r+1}^{l}(m_i+i)}
\psi_{m_1} \ldots \psi_{m_r}
\psis_{n_r} \ldots \psis_{n_1} \vac
\label{partitions6}
\end{align}

\noindent where the right hand side is an element of (\ref{repclA5}), up to an irrelevant minus sign. Secondly, we show that every element 
$
\psi_{m_1}\ldots \psi_{m_r} \psis_{n_r}\ldots \psis_{n_1}\vac
$ 
of (\ref{repclA5}) is an element of (\ref{partitions3}), up to a minus sign. Defining $-l = n_r$, we have 

\begin{align}
\psi_{m_1}\ldots \psi_{m_r}
\psis_{n_r} \ldots \psis_{n_1}
|0\rangle
=
\psi_{m_1}\ldots\psi_{m_r}
\psis_{n_r}\ldots\psis_{n_1}
\psi_{-1}\ldots\psi_{-l}
|-l\rangle
\label{partitions7}
\end{align}

\noindent and we define 
$
\{ m_{r+1} > \cdots > m_{l}\}
$ 
to be the set 
$\{-1 > \cdots > -l \}$ with the omission of $\{n_1 > \cdots > n_r\}$. Using the anticommutation relations (\ref{cfermi}) and the fact that the string of fermions $\psis_{n_r}\ldots\psis_{n_1}$ annihilates the charged vacuum $|-l\rangle$, equation (\ref{partitions7}) becomes

\begin{align}
\psi_{m_1}\ldots \psi_{m_r}
\psis_{n_r} \ldots \psis_{n_1}
|0\rangle
=
(-)^{\sum_{i=1}^{r}(n_i+i)}
\psi_{m_1}\ldots\psi_{m_r}
\psi_{m_{r+1}} \ldots \psi_{m_l}
|-l\rangle
\label{partitions8}
\end{align}

\noindent where the right hand side is an element of (\ref{partitions3}), up to an irrelevant minus sign. This concludes the proof that the elements of the bases (\ref{repclA5}) and (\ref{partitions3}) are in one-to-one correspondence. The proof that the elements of the bases (\ref{repclA6}) and (\ref{partitions4}) are in one-to-one correspondence proceeds along completely analogous lines.
}
\end{proof}

A {\it partition} $\mu = \{\mu_1 \geq \cdots \geq \mu_l > \mu_{l+1} = \cdots = 0\}$ is a set of weakly decreasing non-negative integers, of which finitely many are non-zero. We refer to each integer $\mu_i$ as a {\it part} of $\mu$, and to the sum of all parts 

\begin{align}
|\mu| = \sum_{i=1}^{l} \mu_i
\end{align}

\noindent as the {\it weight} of $\mu$. The number $l$ of non-zero parts is called the {\it length} of $\mu$, and denoted by $\ell(\mu)$. 

Every partition has a pictorial representation called a {\it Young diagram}. This is a collection of $l$ rows of boxes, such that the $i^{\rm th}$ row is $\mu_i$ boxes long. Throughout the thesis, we will use the notation $[n,m]$ to denote a rectangular Young diagram with $n$ rows and $m$ columns. A partition $\mu$ will be said to satisfy $\mu \subseteq [n,m]$ if its Young diagram fits inside the rectangle $[n,m]$. For more information on the theory of partitions, we refer the reader to section 1 in chapter I of \cite{mac}. 

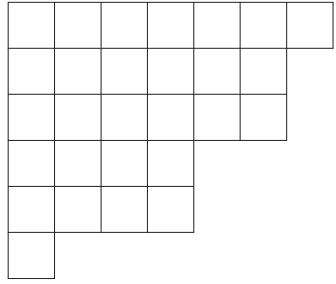
\begin{figure}[H]

\begin{center}
\begin{minipage}{4.3in}

\setlength{\unitlength}{0.0003cm}
\begin{picture}(20000,12000)(-10000,10000)

\path(0,22000)(14000,22000)
\path(0,20000)(14000,20000)
\path(0,18000)(12000,18000)
\path(0,16000)(12000,16000)
\path(0,14000)(8000,14000)
\path(0,12000)(8000,12000)
\path(0,10000)(2000,10000)

\path(0,10000)(0,22000)
\path(2000,10000)(2000,22000)
\path(4000,12000)(4000,22000)
\path(6000,12000)(6000,22000)
\path(8000,12000)(8000,22000)
\path(10000,16000)(10000,22000)
\path(12000,16000)(12000,22000)
\path(14000,20000)(14000,22000)

\end{picture}

\end{minipage}
\end{center}

\caption[Young diagram of the partition $\mu = \{7,6,6,4,4,1\}$]{Young diagram of the partition $\mu = \{7,6,6,4,4,1\}$. In this case $|\mu| = 7+6+6+4+4+1 = 28$ and $\ell(\mu) = 6$. This partition satisfies $\mu \subset [6,7]$.}
\end{figure}

\noindent It is possible to match the elements of the bases (\ref{partitions3}) and (\ref{partitions4}) with partitions. We define $|0\rangle = |\emptyset)$ and $\langle 0 | = ( \emptyset |$, and for all sets of integers $\{m_1 > \cdots > m_l >-l\}$ we write

\begin{align}
\psi_{m_1}\ldots \psi_{m_l} |-l\rangle
=
|\mu_1,\ldots,\mu_l )
=
|\mu),
\quad
\langle -l| \psis_{m_l} \ldots \psis_{m_1}
=
(\mu_1,\ldots,\mu_l|
=
(\mu|
\label{cfermi-part}
\end{align}

\noindent where $\mu_i = m_i +i$ for all $1\leq i \leq l$. Under this identification, we let $|\emptyset)$ and $(\emptyset|$ be copies of the {\it empty partition} $\emptyset$, and $|\mu)$ and $(\mu|$ be copies of the partition $\mu = \{\mu_1 \geq \cdots \geq \mu_l > 0\}$. This gives a one-to-one correspondence between the elements of the bases (\ref{partitions3}) and (\ref{partitions4}) and the elements of the set of all partitions. This will prove useful in later sections, when we encounter functions which are naturally indexed by partitions. 
     
\subsection{Charged fermion expectation values}
\label{I}
    
For arbitrary $g \in Cl_{\psi}$ we define its {\it vacuum expectation value} $\langle g \rangle \in \mathbb{C}$ by

\begin{align}
\langle g \rangle
=
\langle 0| g|0\rangle
\label{I1}
\end{align}

\noindent where it is assumed that $\langle 0| 1 |0\rangle =1$ and

\begin{align}
\langle 0| (g_1+g_2) |0\rangle
=
\langle 0| g_1 |0\rangle + \langle 0| g_2 |0\rangle,
\quad
\langle 0| \kappa g_1 |0\rangle
=
\kappa \langle 0| g_1 |0\rangle
\label{I2}
\end{align}

\noindent for all $g_1,g_2 \in Cl_{\psi}$ and $\kappa \in \mathbb{C}$. It is straightforward to show that the anticommutation relations (\ref{cfermi1}), the annihilation actions (\ref{repclA1}) and the conditions (\ref{I2}) define the value of $\langle g \rangle$ completely and unambiguously. 

\begin{lemma}
{\rm 
Let $\{m_1,\ldots,m_r\}$ and $\{n_1,\ldots,n_r\}$ be two arbitrary sets of integers. We claim that

\begin{align}
\langle 
\psi_{m_1} \ldots \psi_{m_r}
\psis_{n_r} \ldots \psis_{n_1}
\rangle
=
\det
\left(
\langle \psi_{m_i} \psis_{n_j} \rangle
\right)_{1 \leq i,j \leq r}
\label{KPpoly-1}
\end{align}

\noindent This is a special case of {\it Wick's theorem}, see chapter 4 of \cite{mjd}.
}
\end{lemma}

\begin{proof}
{\rm
The case $r=1$ is trivial, since  

\begin{align}
\langle \psi_{m_1} \psis_{n_1} \rangle 
= 
\det \Big( \langle \psi_{m_i} \psis_{n_j} \rangle \Big)_{i,j = 1}
\end{align}

\noindent Using this case as the basis for induction, we assume there exists $r\geq 2$ such that 

\begin{align}
\langle \psi_{m_1} \ldots \psi_{m_{r-1}}
\psis_{n_{r-1}} \ldots \psis_{n_1} \rangle
=
\det\Big( \langle \psi_{m_i} \psis_{n_j} \rangle \Big)_{1 \leq i,j \leq r-1}
\label{inductass}
\end{align}

\noindent for all sets of integers $\{m_1,\ldots,m_{r-1}\}$ and $\{n_1,\ldots,n_{r-1}\}$. We define 

\begin{align}
I_1
&=
\langle \psi_{m_1}\ldots \psi_{m_r}\psis_{n_r} \ldots \psis_{n_1} \rangle
\label{KPpoly5}
\\
I_2
&=
\sum_{k=1}^{r} (-)^{k+1} 
\langle \psi_{m_1} \psis_{n_k} \rangle
\langle \psi_{m_2}\ldots \psi_{m_r} 
\psis_{n_r} \ldots \widehat{\psis_{n_k}} \ldots \psis_{n_1}
\rangle
\nonumber
\end{align}

\noindent where $\{m_1,\ldots,m_r\}$ and $\{n_1,\ldots,n_r\}$ are arbitrary sets of integers, and $\widehat{\psis_{n_k}}$ means the omission of the indicated fermion. By the annihilation properties (\ref{repclA1}) of the fermions, $I_1=I_2=0$ unless both $m_1 < 0$ and $m_1 = \{n_{k_1},\ldots,n_{k_l}\}$ for some $1 \leq k_1 < \cdots < k_l \leq r$. When both these conditions are satisfied it is readily verified that

\begin{align}
I_1
=
I_2
=
\sum_{k \in \{k_1,\ldots,k_l\}}
(-)^{k+1}
\langle \psi_{m_1} \psis_{n_k} \rangle
\langle \psi_{m_2} \ldots \psi_{m_r} 
\psis_{n_r} \ldots \widehat{\psis_{n_k}} 
\ldots \psis_{n_1} \rangle
\end{align}

\noindent which proves that $I_1=I_2$ for all $m_1$. Therefore we obtain

\begin{align}
I_1
&=
\sum_{k=1}^{r}
(-)^{k+1}
\langle \psi_{m_1} \psis_{n_k}\rangle
\langle \psi_{m_2} \ldots \psi_{m_r}
\psis_{n_r} \ldots \widehat{\psis_{n_k}} \ldots \psis_{n_1} \rangle
\label{proofend}
\\
&=
\sum_{k=1}^{r}
(-)^{k+1}
\langle \psi_{m_1} \psis_{n_k} \rangle
\det \Big( 
\langle \psi_{m_i} \psis_{n_j} \rangle
\Big)_{2 \leq i \leq r, j\not= k}
=
\det
\Big(
\langle \psi_{m_i} \psis_{n_j} \rangle
\Big)_{1 \leq i,j \leq r}
\nonumber
\end{align}

\noindent where we used the assumption (\ref{inductass}) to produce the $(r-1) \times (r-1)$ determinant in the second line of (\ref{proofend}). This completes the proof by induction.
}
\end{proof}

We define a {\it bilinear form} $\langle , \rangle$ which maps $\mathcal{F}_{\psi}^{*} \times \mathcal{F}_{\psi} \rightarrow \mathbb{C}$. Its action on the arbitrary vectors $\langle 0|g_1 \in \mathcal{F}_{\psi}^{*}$ and $g_2|0\rangle \in \mathcal{F}_{\psi}$ is given by

\begin{align}
\Big\langle\langle 0|g_1, g_2|0\rangle \Big\rangle
=
\Big\langle g_1 g_2 \Big\rangle
\label{form}
\end{align}

\noindent Let $(\mu| = \langle -l|\psis_{m_l}\ldots\psis_{m_1}$ and $|\nu) = \psi_{n_1}\ldots\psi_{n_k}|-k\rangle$ be partition vectors. By a straightforward calculation, we obtain

\begin{align}
\Big\langle
(\mu|,|\nu)
\Big\rangle
&=
\langle -l|
\psis_{m_l}\ldots\psis_{m_1}
\psi_{n_1}\ldots \psi_{n_k}
|-k\rangle
=
\delta_{k,l}
\prod_{i=1}^{l}
\delta_{m_i,n_i}
=
\delta_{\mu,\nu}
\label{part-orth}
\end{align}

\noindent In other words, the bilinear form (\ref{form}) induces orthonormality between the partition elements of $\mathcal{F}_{\psi}^{*(0)}$ and $\mathcal{F}_{\psi}^{(0)}$, which confirms that these vector spaces are genuinely dual.

\subsection{Lie algebra $A_{\infty}$}
\label{ainfty}

Let $A_{\infty} \subset Cl_{\psi}^{(0)}$ be the vector space whose elements $X \in A_{\infty}$ are given by

\begin{align}
X
=
\sum_{i,j \in \mathbb{Z}}
a_{i,j} :\psi_i \psis_j:
+
\kappa
\label{ainfty1}
\end{align}


\noindent where we have defined the normal ordering $:\psi_i \psis_j: = \psi_i \psis_j - \langle 0| \psi_i \psis_j |0\rangle$, and where the coefficients satisfy $a_{i,j} = 0$ for $|i-j|$ sufficiently large, with $\kappa \in \mathbb{C}$.

\begin{lemma}
{\rm The vector space $A_{\infty}$ becomes a Lie algebra when it is equipped with the commutator as Lie bracket.}
\end{lemma}

\begin{proof}
{\rm We prove the closure of $A_{\infty}$ under commutation. Let $X$ be given by (\ref{ainfty1}) and define 

\begin{align}
X_1
=
\sum_{i,j \in \mathbb{Z}}
a^{(1)}_{i,j} : \psi_i \psis_j :
+
\kappa_1
\label{ainfty2}
\end{align}


\noindent where $a_{i,j}^{(1)}= 0$ for $|i-j|$ sufficiently large. By a simple calculation, we obtain 

\begin{align}
[X,X_1]
=
\sum_{i,j \in \mathbb{Z}}
a^{(2)}_{i,j} :\psi_i \psis_j:
+
\kappa_2
\label{ainfty3}
\end{align}


\noindent where for all $i,j \in \mathbb{Z}$ we have defined

\begin{align}
a^{(2)}_{i,j} 
= 
\sum_{k \in \mathbb{Z}} 
\Big( a_{i,k} a^{(1)}_{k,j} 
- 
a^{(1)}_{i,k}a_{k,j} \Big)
\label{ainfty4}
\end{align}

\noindent and where

\begin{align}
\kappa_2 
= 
\left( \sum_{i < 0, j \geq 0} - \sum_{i \geq 0, j<0} \right) a_{i,j}a^{(1)}_{j,i}
\label{ainfty5}
\end{align}

\noindent Clearly the right hand side of equation $(\ref{ainfty3})$ is also an element of $A_{\infty}$, proving the closure of the set under commutation. 
}
\end{proof}

\subsection{$A_{\infty}$ Heisenberg subalgebra}
\label{aheisen}

We define operators $H_m \in A_\infty$ which are given by
 
\begin{align}
H_m 
= 
\sum_{i \in \mathbb{Z}}:\psi_{i}\psis_{i+m}:
\label{aheisen1}
\end{align}

\noindent for all $m \in \mathbb{Z}$. Together with the central element $1$, the operators $\{H_m\}_{m \in \mathbb{Z}}$ generate a Heisenberg subalgebra. The closure of this set follows from the commutation relation  

\begin{align}
\left[ H_m, H_n \right] 
= 
m \delta_{m + n, 0}
\label{aheisen2}
\end{align}

\noindent for all $m,n \in \mathbb{Z}$.\footnote{For a detailed proof of (\ref{aheisen2}), see the exercise 5.1 in chapter 4 of \cite{mjd}. Alternatively, one can notice that $H_m,H_n$ are obtained from $X,X_1$ by setting $a_{i,j} = \delta_{i+m,j},\ a^{(1)}_{i,j} = \delta_{i+n,j}$ and $\kappa=\kappa_1=0$. Substituting these values into (\ref{ainfty4}) and (\ref{ainfty5}), we obtain the desired commutator (\ref{aheisen2}) by virtue of (\ref{ainfty3}).} The Heisenberg generators (\ref{aheisen1}) also have simple commutation relations with the charged fermions (\ref{cfermi1}), given by 

\begin{align}
\left[H_m, \psi_n \right] =  \psi_{n-m}, 
\quad 
\left[H_m,\psis_n \right] = -\psis_{m+n}
\label{aheisen3}
\end{align}

\noindent for all $m,n \in \mathbb{Z}$.

\subsection{KP evolution operators}
\label{KPev}

We introduce the Hamiltonian

\begin{align}
H\{t\}
=
\sum_{n=1}^{\infty}
t_n H_n
\end{align}

\noindent where $\{t\}=\{t_1,t_2,t_3,\ldots\}$ is an infinite set of free variables, and define the generating functions

\begin{align}
\Psi(k) = \sum_{i \in \mathbb{Z}} \psi_i k^i,
\quad
\Psis(k) = \sum_{i \in \mathbb{Z}} \psis_i k^{-i}
\label{KPexpI9}
\end{align}

\noindent where $k$ is an indeterminate. Using these definitions and the equations (\ref{aheisen3}), we obtain the commutation relations

\begin{align}
[H\{t\},\Psi(k)]
&
=
\sum_{n=1}^{\infty}
\sum_{i \in \mathbb{Z}}
[H_n,\psi_i]
t_n k^i
\label{KPexpI10}
=
\left(
\sum_{n=1}^{\infty}
t_n k^n
\right)
\Psi(k)
\\
[H\{t\},\Psis(k)]
&
=
\sum_{n=1}^{\infty}
\sum_{i \in \mathbb{Z}}
[H_n,\psis_i]
t_n k^{-i}
\label{KPexpI11}
=
-\left(
\sum_{n=1}^{\infty}
t_n k^{n}
\right)
\Psis(k)
\end{align}

\noindent which, in turn, imply that

\begin{align}
e^{H\{t\}}\Psi(k) 
&
= 
\exp\left(\sum_{n=1}^{\infty} t_n k^n \right)
\Psi(k) e^{H\{t\}}
\label{KPexpI12}
\\
e^{H\{t\}}\Psis(k) 
&
= 
\exp\left(-\sum_{n=1}^{\infty} t_n k^{n} \right)
\Psis(k) e^{H\{t\}}
\label{KPexpI13} 
\end{align}

\noindent Following the terminology of \cite{jm1}, the operator $e^{H\{t\}}$ in this pair of equations is called a {\it KP evolution operator}.  As we shall see, the KP evolution operator plays an essential role in constructing solutions of the KP hierarchy of partial differential equations.

\subsection{Schur polynomials}
\label{KPpoly}

For all $m \geq 0$, the {\it one-row Schur polynomial} $\chi_m\{t\}$ in the infinite set of variables $\{t\}$ is defined by

\begin{align}
\chi_{m}\{t\}
=
{\rm Coeff}_{k^m}
\left[
\exp\left( \sum_{n=1}^{\infty} t_n k^n \right)
\right]
\label{KPbil-1}
\end{align}

\noindent where ${\rm Coeff}_{k^m} [f(k)]$ denotes the coefficient of $k^m$ in the Taylor series expansion of $f(k)$. We also define $\chi_m\{t\} = 0$ when $m < 0$. From this definition, the {\it Schur polynomial} \footnote{The polynomials $\chi_{\mu}$ are sometimes called {\it character polynomials,} in reference to their connection with the linear representations of the symmetric groups, see section 7 in chapter I of \cite{mac}.} $\chi_{\mu}\{t\}$ associated to the arbitrary partition $\mu = \{\mu_1 \geq \cdots \geq \mu_l \geq 0\}$ is given by

\begin{align}
\chi_{\mu}\{t\}
=
\det\Big(
\chi_{(\mu_i -i+j)}\{t\}
\Big)_{1 \leq i,j \leq l}
\label{surprising}
\end{align}

\noindent where $\chi_{(\mu_i-i+j)}\{t\}$ represents a one-row Schur polynomial. The following result maps the partition elements of $\mathcal{F}_{\psi}^{(0)}$ to their corresponding Schur polynomials.

\begin{lemma}
{\rm
 Let $|\mu) = |\mu_1,\ldots, \mu_l)=\psi_{m_1} \ldots \psi_{m_l} |-l\rangle$ be an element of the partition basis of $\mathcal{F}_{\psi}^{(0)}$, where $\mu_i = m_i + i$ for all  $1 \leq i \leq l$. We claim that

\begin{align}
\chi_{\mu}\{t\}
=
(\emptyset| e^{H\{t\}} |\mu)
\label{fermi-char}
\end{align}
}
\end{lemma}

\begin{proof}
{\rm 
Using the definition of the one-row Schur polynomials (\ref{KPbil-1}) in the commutation relation (\ref{KPexpI12}) and then extracting the coefficient of $k^m$ from the resulting equation, we obtain

\begin{align}
e^{H\{t\}}
\psi_m
&
=
\left(
\sum_{i=0}^{\infty}
\psi_{(m-i)} \chi_{i}\{t\}
\right)
e^{H\{t\}}
\label{KPpoly8}
\end{align}


\noindent By definition, we have

\begin{align}
(\emptyset|e^{H\{t\}}|\mu)
=
\langle 0|  e^{H\{t\}} \psi_{m_1} \ldots \psi_{m_l} 
|-l \rangle
\label{KPpoly7}
\end{align}

\noindent and using the commutation relation(\ref{KPpoly8}) we move the evolution operator $e^{H\{t\}}$ in (\ref{KPpoly7}) towards the right, obtaining

\begin{align}
(\emptyset| e^{H\{t\}} |\mu)
=
\sum_{i_1,\ldots, i_l = 0}^{\infty}
\Big\langle  
\psi_{(m_1-i_1)} 
\ldots
\psi_{(m_l-i_l)} 
\psis_{-l}\ldots \psis_{-1}
\Big\rangle
\chi_{i_1}\{t\}
\ldots
\chi_{i_l} \{t\}
\end{align}

\noindent where we have used the fact that $e^{H\{t\}} |-l\rangle = |-l\rangle$. Applying lemma 2 to the previous vacuum expectation value, we find

\begin{align}
(\emptyset| e^{H\{t\}} |\mu)
=
\sum_{i_1,\ldots,i_l = 0}^{\infty}
\det\Big(\langle
\psi_{(m_p-i_p)} \psis_{-q}
\rangle \Big)_{1\leq p,q \leq l}
\chi_{i_1}\{t\} \ldots \chi_{i_l} \{t\}
\end{align}

\noindent Collecting the one-row Schur polynomial $\chi_{i_p}\{t\}$ as 
a factor multiplying the $p^{\rm th}$ row of the determinant, we obtain

\begin{align}
(\emptyset| e^{H\{t\}} |\mu)
=
\det\left(
\sum_{i_p=0}^{\infty}
\langle 
\psi_{(m_p-i_p)} \psis_{-q}
\rangle
\chi_{i_p}\{t\}
\right)_{1\leq p,q \leq l}
\label{KPpoly10}
\end{align}

\noindent Using the anticommutation relations (\ref{cfermi1}), the annihilation properties (\ref{repclA1}), and the fact that $\chi_m\{t\}=0$ for all $m < 0$, equation (\ref{KPpoly10}) becomes

\begin{align}
(\emptyset|e^{H\{t\}}|\mu)
=
\det\left(
\sum_{i_p \in \mathbb{Z}}
\delta_{i_p,(m_p+q)}
\chi_{i_p}\{t\}
\right)_{1\leq p,q \leq l}
\end{align}

\noindent Finally, using the Kronecker delta to truncate the sum occurring in the previous determinant, we obtain

\begin{align}
(\emptyset|e^{H\{t\}}|\mu)
=
\det\Big(
\chi_{(m_p+q)}\{t\}
\Big)_{1\leq p,q \leq l}
=
\det\Big(
\chi_{(\mu_p -p+q)}\{t\}
\Big)_{1 \leq p,q \leq l}
\end{align}

\noindent and the final determinant is the Schur polynomial $\chi_{\mu}\{t\}$.
}
\end{proof}

\begin{lemma}
{\rm
Let $(\mu \pm 1) = \Big\{(\mu_1 \pm 1) \geq \cdots \geq (\mu_l \pm 1) \geq 0\Big\}$ be a pair of partitions, and set $m_i = \mu_i -i$ for all $1 \leq i \leq l$. We claim that

\begin{align}
\chi_{(\mu+1)}\{t\}
&=
\det\Big(
\chi_{(\mu_i - i+j+1)}\{t\}
\Big)_{1 \leq i,j \leq l}
=
\langle -1|
e^{H\{t\}}
\psi_{m_1}
\ldots
\psi_{m_l}
|-l-1\rangle
\label{KPpoly-2}
\\
\chi_{(\mu -1)}\{t\}
&=
\det\Big(
\chi_{(\mu_i -i+j-1)}\{t\}
\Big)_{1 \leq i,j \leq l}
=
\langle 1|
e^{H\{t\}}
\psi_{m_1}
\ldots
\psi_{m_l}
|-l+1\rangle
\label{KPpoly-3}
\end{align}
}
\end{lemma}

\begin{proof}
{\rm
Using the commutation relation (\ref{KPpoly8}) we move the evolution operator $e^{H\{t\}}$ in (\ref{KPpoly-2}) and (\ref{KPpoly-3}) towards the right, obtaining 

\begin{align}
\chi_{(\mu+1)}\{t\}
&=
\Big\langle
\psi_{-1}
\psi_{(m_1-i_1)} 
\ldots
\psi_{(m_l-i_l)} 
\psis_{(-l-1)}
\ldots
\psis_{-1}
\Big\rangle
\chi_{i_1}\{t\}
\ldots
\chi_{i_l}\{t\}
\label{KPpoly-4}
\\ 
\chi_{(\mu-1)}\{t\}
&=
\Big\langle
\psis_0
\psi_{(m_1-i_1)} 
\ldots
\psi_{(m_l-i_l)} 
\psis_{(-l+1)}
\ldots
\psis_{-1}
\Big\rangle
\chi_{i_1}\{t\}
\ldots
\chi_{i_l}\{t\}
\label{KPpoly-5}
\end{align}

\noindent where in both cases summation over all $\{0 \leq i_1,\ldots,i_l <\infty\}$ is implied. Using the anticommutation relations (\ref{cfermi1}) to move $\psi_{-1}$ and $\psis_{0}$ towards the right in their respective equations (\ref{KPpoly-4}) and (\ref{KPpoly-5}), and then applying lemma 2, we find


%

\begin{align}
\chi_{(\mu+1)}\{t\}
&=
\det\Big(
\sum_{i_p \in \mathbb{Z}}
\langle
\psi_{(m_p-i_p)} \psis_{(-q-1)}
\rangle
\chi_{i_p}\{t\}
\Big)_{1 \leq p,q \leq l}
\\
\chi_{(\mu-1)}\{t\}
&=
\sum_{j=1}^{l}
(-)^{j+1}
\chi_{m_j}\{t\}
\det\Big(
\sum_{i_p \in \mathbb{Z}}
\langle \psi_{(m_p-i_p)} \psis_{(-q+1)} \rangle
\chi_{i_p}\{t\}
\Big)_{p\not=j,2 \leq q \leq l}
\end{align}

\noindent with $\chi_{m}\{t\} = 0$ for all $m<0$. Finally, evaluating the expectation values within these determinants explicitly, we recover

\begin{align}
\chi_{(\mu+1)}\{t\}
&=
\det\Big(
\chi_{(m_p+q+1)}\{t\}
\Big)_{1 \leq p,q \leq l}
=
\det\Big(
\chi_{(\mu_p - p +q +1)}\{t\}
\Big)_{1 \leq p,q \leq l}
\\
\chi_{(\mu-1)}\{t\}
&=
\sum_{j=1}^{l}
(-)^{j+1}
\chi_{m_j}\{t\}
\det\Big( \chi_{(m_p+q-1)}\{t\} \Big)_{p\not=j,2\leq q\leq l}
\label{KPpoly-6}
\\
&=
\det\Big(
\chi_{(m_p+q-1)}\{t\}
\Big)_{1 \leq p,q \leq l}
=
\det\Big(
\chi_{(\mu_p - p +q -1)}\{t\}
\Big)_{1 \leq p,q \leq l}
\nonumber
\end{align}

\noindent where the second line of (\ref{KPpoly-6}) follows from the expansion of an $l\times l$ determinant down its first column.
}
\end{proof}

\subsection{Schur functions}
\label{schur}

Following section 2 in chapter I of \cite{mac}, the {\it complete symmetric function} $h_m\{x\}$ in the infinite set of variables $\{x\}=\{x_1,x_2,x_3,\ldots\}$ is defined as

\begin{align}
h_m\{x\}
=
{\rm Coeff}_{k^m}
\left[
\prod_{i=1}^{\infty}
\frac{1}{1-x_i k}
\right]
\label{complete}
\end{align}

\noindent From this definition, the {\it Schur function}\footnote{Notice that we use the words {\it polynomial} and {\it function} to distinguish between $\chi_{\mu}\{t\}$ and $s_{\mu}\{x\}$. Unlike $\chi_{\mu}\{t\}$, which always depends on finitely many of the variables $\{t\}$, the function $s_{\mu}\{x\}$ can depend on infinitely many variables $\{x\}$, and it would be improper to call it a polynomial.} $s_{\mu}\{x\}$ associated to the partition $\mu = \{\mu_1 \geq \cdots \geq \mu_l \geq 0\}$ is given by

\begin{align}
s_{\mu}\{x\}
=
\det\Big(
h_{(\mu_i - i+j)}\{x\}
\Big)_{1 \leq i,j \leq l}
\label{schurfunct}
\end{align}

\begin{lemma}
{\rm
The Schur polynomial $\chi_{\mu}\{t\}$ and the Schur function $s_{\mu}\{x\}$ are equal under the change of variables $t_n = \frac{1}{n} \sum_{i=1}^{\infty} x_i^n$ for all $n \geq 1$.  
}
\end{lemma}

\begin{proof}
{\rm
Fixing $t_n = \frac{1}{n} \sum_{i=1}^{\infty} x_i^n$ for all $n \geq 1$, we obtain

{\small
\begin{align}
{\rm Coeff}_{k^m}
\left[
\exp\left(
\sum_{n=1}^{\infty}
t_n k^n
\right)
\right]
=
{\rm Coeff}_{k^m}
\left[
\exp\left(
\sum_{i=1}^{\infty}
\sum_{n=1}^{\infty}
\frac{1}{n} (x_i k)^n
\right)
\right]
=
{\rm Coeff}_{k^m}
\left[
\prod_{i=1}^{\infty}
\frac{1}{1-x_i k}
\right]
\end{align}
}

\noindent implying that $\chi_{m} \{t\} = h_m\{x\}$ for all $m \geq 0$. Therefore under the prescribed change of variables we have

\begin{align}
\det\Big(
\chi_{(\mu_i -i+j)}\{t\}
\Big)_{1 \leq i,j \leq l}
=
\det\Big(
h_{(\mu_i-i+j)}\{x\}
\Big)_{1 \leq i,j \leq l}
\end{align}

\noindent which completes the proof.
}
\end{proof}

It is well known that the Schur functions $s_{\mu}\{x\}$ comprise a basis for the ring of symmetric functions in $\{x\}$.\footnote{See section 3 in chapter I of \cite{mac}.} This fact, together with the equality of $\chi_{\mu}\{t\}$ and $s_{\mu}\{x\}$ under the previous change of variables, proves that the Schur polynomials $\chi_{\mu}\{t\}$ are a basis for the set of all polynomials in $\{t\}$.

Throughout the rest of the thesis, we will usually consider Schur functions $s_{\mu}\{x\}$ in a finite number of variables $\{x\} = \{x_1,\ldots,x_N\}$. These are obtained from the formulae (\ref{complete}), (\ref{schurfunct}) by setting $x_n= 0$ for all $n > N$. In these cases, $s_{\mu}\{x\} = 0$ if $\ell(\mu) > N$, which is explained in section 3, chapter I of \cite{mac}.  







\section{KP hierarchy}
\label{c-KP}

\subsection{KP hierarchy in bilinear form}
\label{KPbil}

The {\it KP hierarchy} is an infinite set of partial differential equations in infinitely many independent variables $\{t\} = \{t_1,t_2,t_3,\ldots\}$. This hierarchy of differential equations actually derives from a single integral equation, called the {\it KP bilinear identity}. A function $\tau\{t\}$ which satisfies every differential equation in the hierarchy, or equivalently, satisfies the KP bilinear identity, is called a {\it KP $\tau$-function.}

Define the shifted sets of variables

\begin{align}
\{t \pm \epsilon_k\} 
= 
\left\{
t_1 \pm k^{-1},
t_2 \pm \frac{1}{2} k^{-2},
t_3 \pm \frac{1}{3} k^{-3},
\ldots
\right\}
\label{KPbil0}
\end{align}

\noindent where $k$ is a free parameter. Using this notation, $\tau\{t\pm\epsilon_k\}$ should be understood as a function with $n^{\rm th}$ argument $t_n\pm\frac{1}{n}k^{-n}$. The KP bilinear identity is the equation 

\begin{align}
\oint
\exp{\left( \sum_{n=1}^{\infty} (t_n-s_n)k^n \right)}
\tau\{t-\epsilon_k\}
\tau\{s+\epsilon_k\}
\frac{dk}{2\pi i}
=
0
\label{KPbil1}
\end{align}

\noindent where $\{t\}=\{t_1,t_2,t_3,\ldots\}$ and $\{s\}=\{s_1,s_2,s_3,\ldots\}$ are two infinite sets of variables, and the integration in the $k$-plane is taken around a small contour at $k = \infty$. Equivalently, this equation says that the sum of residues of the integrand in the $k$-plane is equal to zero. 

In this thesis we will always assume that $\tau\{t\}$ is a polynomial in its variables. In this special case, $\tau\{t-\epsilon_k\}$ and 
$\tau\{s+\epsilon_k\}$ have singularities in $k$ only at $k=0$. This means
that the integral $(\ref{KPbil1})$ is equal to the coefficient of $k^{-1}$ in the Laurent series of the integrand. Therefore, for polynomial $\tau$-functions, the KP bilinear identity becomes

\begin{align}
{\rm Coeff}_{k^{-1}}
\left[
\exp\left( \sum_{n=1}^{\infty} (t_n-s_n)k^n \right)
\tau\{t-\epsilon_k\}
\tau\{s+\epsilon_k\}
\right]
=
0
\label{KPbil2}
\end{align}

\noindent where ${\rm Coeff}_{k^{-1}}\left[f(k)\right]$ denotes the coefficient of $k^{-1}$ in the Laurent series of $f(k)$. We expose the infinitely many differential equations which underly the equation (\ref{KPbil2}) by making the substitutions $\{t\} \rightarrow \{t-s\}$ and $\{s\} \rightarrow \{t+s\}$, giving
 
\begin{align}
{\rm Coeff}_{k^{-1}}
\left[
\exp\left(
-2 \sum_{n=1}^{\infty} s_n k^n
\right)
\tau\{t-s-\epsilon_k\}
\tau\{t+s+\epsilon_k\}
\right]
=
0
\label{KPbil5}
\end{align}

%

\noindent Using the definition (\ref{KPbil-1}), the exponential term in 
$(\ref{KPbil5})$ can be replaced with a sum over one-row Schur polynomials, producing the equation

\begin{align}
{\rm Coeff}_{k^{-1}}
\left[ 
\sum_{m=0}^{\infty} \chi_m\{-2s\} k^m
\tau\{t-s-\epsilon_k\}
\tau\{t+s+\epsilon_k\}
\right]
\label{KPbil6}
=
0
\end{align}

\noindent or equivalently,

\begin{align}
\sum_{m=0}^{\infty}
\chi_m\{-2s\}
{\rm Coeff}_{k^{-m-1}}
\Big[
\tau\{t-s-\epsilon_k\}
\tau\{t+s+\epsilon_k\}
\Big]
=
0
\label{KPbil-2}
\end{align}

\noindent In order to progress from this last equation to an infinite set of differential equations, we need to introduce the notion of bilinear differential operators.

\begin{definition}
{\rm 
The {\it bilinear differential operators}\footnote{The definition of these operators is due to R~Hirota, see \cite{hir}.} $\{D\} = \{D_1,D_2,D_3,\ldots\}$ act on {\it ordered pairs} of functions $f\{t\} \cdot g\{t\}$. Letting $P\{D\}$ denote an arbitrary polynomial combination of these operators, we define 

\begin{align}
\label{KPbil4}
P\{D\} f\{t\} \cdot g\{t\}
=
P\{\partial_{z}\} 
\Big(
f\{t+z\} g\{t-z\}
\Big)_{\{z\}\rightarrow \{0\}}
\end{align}

\noindent where $\{\partial_z\}= \{\partial_{z_1},\partial_{z_2},\partial_{z_3},\ldots\}$ and $\{t\pm z\}= \{t_1\pm z_1,t_2\pm z_2,t_3\pm z_3,\ldots\}$, and where we have set $\{z\} \rightarrow \{0\}$ after differentiation.
}
\end{definition}

\begin{lemma}
{\rm
\begin{align}
\tau\{t-s-\epsilon_k\}
\tau\{t+s+\epsilon_k\}
=
\exp
\left(
\sum_{n=1}^{\infty}
(s_n+\frac{1}{n}k^{-n})
D_n
\right)
\tau\{t\} \cdot \tau\{t\}
\label{KPbil7}
\end{align}
}
\end{lemma}

\begin{proof}
{\rm
For an arbitrary function $f(z)$ and $\kappa \in \mathbb{C}$, we have the Taylor series identity

\begin{align}
f(\kappa)
=
\sum_{n=0}^{\infty}
\frac{f^n(0)}{n!} \kappa^n
=
\Big(
e^{\kappa \partial_z} f(z)
\Big)_{z\rightarrow 0}
\label{KPbil8}
\end{align}

\noindent Extending this identity to infinitely many variables, we find

\begin{align}
\tau\{t+s+\epsilon_k\}
\tau\{t-s-\epsilon_k\}
=
&
\exp\left(
\sum_{n=1}^{\infty}
\kappa_n
\partial_{z_n}
\right)
\Big(
\tau\{t+z\} \tau\{t-z\}
\Big)_{\{z\}\rightarrow \{0\}}
\nonumber
\\
= 
& 
\exp
\left(
\sum_{n=1}^{\infty}
\kappa_n D_n
\right)
\tau\{t\} \cdot \tau\{t\}
\label{KPbil9}
\end{align}

\noindent where we have defined $\kappa_n =(s_n+\frac{1}{n}k^{-n})$ for all $n \geq 1$. The final line (\ref{KPbil9}) follows from the definition (\ref{KPbil4}) of the action of the operators $\{D\}$, and achieves the proof of (\ref{KPbil7}).
}
\end{proof}

Returning to equation (\ref{KPbil-2}), we employ the result (\ref{KPbil7}) to obtain

\begin{align}
\sum_{m=0}^{\infty}
\chi_{m}\{-2s\}
{\rm Coeff}_{k^{-m-1}}
\left[
\exp\left(
\sum_{n=1}^{\infty}
(s_n+\frac{1}{n}k^{-n})D_n
\right)
\right]
\tau\{t\}\cdot\tau\{t\}
=
0
\label{KPbil-3}
\end{align}

\noindent Recalling the definition (\ref{KPbil-1}) of the one-row Schur polynomials once again, equation (\ref{KPbil-3}) becomes

\begin{align}
\sum_{m=0}^{\infty}
\chi_{m}\{-2s\}
\chi_{(m+1)}\{\mathbb{D}\}
\exp\left(
\sum_{n=1}^{\infty}
s_n D_n
\right)
\tau\{t\} \cdot \tau\{t\}
=
0
\label{KPbil10}
\end{align}

\noindent where we have defined the set of operators $\{\mathbb{D}\} = \{D_1,\frac{1}{2}D_2,\frac{1}{3}D_3,\ldots\}$. For all $\{1\leq m_1 < \cdots < m_l\}$ and $\{n_1,\ldots,n_l \geq 1\}$
the coefficient of the monomial $s_{m_1}^{n_1}\ldots s_{m_l}^{n_l}$ on the left hand side of $(\ref{KPbil10})$ must vanish, giving rise to infinitely many consistency equations, which are the differential equations of the KP hierarchy.

\begin{example}
{\rm
Up to an irrelevant factor the coefficient of $s_1^3$ on the left hand side 
of $(\ref{KPbil10})$ is equal to 
$
\left( D_1^4 - 4D_1 D_3 +3D_2^2 \right) \tau\{t\} \cdot \tau\{t\}
$,
implying that

\begin{align}
\left( D_1^4 - 4D_1 D_3 +3D_2^2 \right) 
\tau\{t\} \cdot \tau\{t\}
=
0
\label{KPbil11}
\end{align}

\noindent which is the KP equation in bilinear form. Higher equations in the hierarchy are obtained from the coefficients of different monomials.
}
\end{example} 

\subsection{Charged fermion bilinear identity} 
\label{KPexpI}

We turn to constructing solutions of the KP bilinear identity (\ref{KPbil1}) using the calculus of the charged fermions $\{\psi_m\}_{m \in \mathbb{Z}}$ and $\{\psis_m\}_{m \in \mathbb{Z}}$. The following result states that certain special vacuum expectation values are KP $\tau$-functions.  


\begin{theorem}
{\rm 
Let $g_{\psi}$ be a finite element of $Cl_{\psi}^{(0)}$ and define 

\begin{align}
\tau\{t\}
=
\langle 0|e^{H\{t\}} g_{\psi}|0\rangle
=
\Big\langle
e^{H\{t\}} g_{\psi}
\Big\rangle
\label{KPexpI1}
\end{align}

\noindent The polynomial $\tau\{t\}$ satisfies the KP bilinear identity (\ref{KPbil1}) if and only if $g_{\psi}$ satisfies the {\it charged fermion bilinear identity} (CFBI)

\begin{align}
\sum_{i \in \mathbb{Z}}
\psi_i g_{\psi}|0\rangle
\otimes
\psis_i g_{\psi}|0\rangle
=
0
\label{KPhelp}
\end{align}
}
\end{theorem}






\begin{proof} 
{\rm
We split the proof into three steps. In the first two steps we prove that if (\ref{KPhelp}) holds, then $\tau\{t\}$ as given by (\ref{KPexpI1}) satisfies the KP bilinear identity (\ref{KPbil1}). In the third step we prove the converse statement.


\medskip
\noindent
{\bf Step 1.} Acting upon the left hand side of $(\ref{KPhelp})$ with the tensored dual states
$\langle 1|e^{H\{t\}} 
\otimes
\langle -1|e^{H\{s\}}
$ we have the result

\begin{align}
\sum_{i\in \mathbb{Z}}
\langle 1|
e^{H\{t\}}
\psi_i
g_{\psi}|0\rangle
\langle -1|
e^{H\{s\}}
\psis_i 
g_{\psi}|0\rangle
=
0
\label{KPexpI8}
\end{align}

\noindent We convert the sum on the left hand side of $(\ref{KPexpI8})$ into a contour integral, using the generating functions (\ref{KPexpI9}) to write

%

\begin{align}
\oint
\langle 1|
e^{H\{t\}}
\Psi(k) 
g_{\psi}|0\rangle
\langle -1|
e^{H\{s\}}
\Psis(k) 
g_{\psi}|0\rangle
\frac{dk}{2\pi i k}
=
0
\label{KPexpI-2}
\end{align}

\noindent where the contour of integration surrounds the pole at $k=0$. By virtue of the commutation relations (\ref{KPexpI12}) and (\ref{KPexpI13}), it is possible to switch the order of $e^{H\{t\}}$ and $\Psi(k)$, and likewise $e^{H\{s\}}$ and $\Psis(k)$ in (\ref{KPexpI-2}), giving

\begin{align}
\oint
\exp\left(\sum_{n=1}^{\infty}(t_n-s_n)k^n\right)
\langle 1|
\Psi(k) 
e^{H\{t\}} 
g_{\psi}|0\rangle
\langle -1|
\Psis(k) 
e^{H\{s\}} 
g_{\psi}|0\rangle
\frac{dk}{2\pi i k}
=
0
\label{KPexpI14}
\end{align}

\medskip
\noindent
{\bf Step 2. (Lemma 8.)} 
We propose the pair of identities

\begin{align}
\langle 1|\Psi(k)
&
=
\langle 0|
\exp\Big(-\sum_{n=1}^{\infty}\frac{1}{n}k^{-n}H_n\Big)
\label{KPexpI17}
\\
\langle -1|\Psis(k)
&
=
k 
\langle 0|
\exp\Big(\sum_{n=1}^{\infty}\frac{1}{n}k^{-n}H_n\Big)
\label{KPexpI18}
\end{align}

\begin{proof}
{\it (Lemma 8.)}\  
{\rm 
From the definitions of the dual charged vacua (\ref{partitions2}) and the generating functions (\ref{KPexpI9}), we obtain

\begin{align}
\langle 1|\Psi(k) 
&
= 
\langle 0|\psis_0 \sum_{i \in \mathbb{Z}} \psi_i k^i 
=
\langle 0|\psis_0 \sum_{i=0}^{\infty}  \psi_{-i} k^{-i} 
\label{KPexpI19}
\\
\langle -1|\Psis(k) 
&
= 
\langle 0|\psi_{-1} \sum_{i \in \mathbb{Z}} \psis_i k^{-i} 
=
k \langle 0|\psi_{-1}  \sum_{i=0}^{\infty}  \psis_{i-1} k^{-i} 
\label{KPexpI20}
\end{align}

\noindent where the annihilation properties (\ref{repclA1}) have been used to truncate the sums. Rearranging the right hand sides of these equations, we find

\begin{align}
\langle 1| \Psi(k)
&=
\langle 0|
+
\sum_{l=1}^{\infty}
(-k)^{-l}
\langle -l| \psis_{-l+1} \ldots \psis_{0}
\label{KPexpI-4}
\\
\langle -1| \Psis(k)
&=
k
\left(
\langle 0|
+
\sum_{i=1}^{\infty}
k^{-i}
\langle -1| \psis_{i-1}
\right)
\label{KPexpI-5}
\end{align}


\noindent Now consider the expression (\ref{KPbil-1}) for the one-row Schur polynomial. When the variables $\{t\}$ are set to $t_n = \pm k^{-n}/n$ for all $n \geq 1$, this expression simplifies greatly. We obtain

\begin{align}
\chi_m\{t\}
\Big|_{t_n = -k^{-n}/n}
=
\left\{
\begin{array}{ll}
(-k)^{-m}, & m \leq 1
\\
0, & {m \geq 2}
\end{array}
\right.
,
\quad
%
%
\chi_m\{t\}
\Big|_{t_n = k^{-n}/n}
=
k^{-m}
\end{align}

\noindent and substitute these formulae into the expression (\ref{surprising}) for the Schur polynomial associated to $\mu = \{\mu_1 \geq \cdots \geq \mu_l > 0\}$, giving

\begin{align}
\chi_{\mu}\{t\}
\Big|_{t_n = -k^{-n}/n}
=
(-k)^l
\prod_{i=1}^{l}
\delta_{\mu_i,1},
%
\quad
\chi_{\mu}\{t\}
\Big|_{t_n = k^{-n}/n}
=
k^{-\mu_1}
\delta_{l,1}
\end{align}

\noindent Using the result of lemma 4, these equations become

\begin{align}
\langle 0|
\exp\Big(
-\sum_{n=1}^{\infty}
\frac{1}{n} k^{-n} H_n
\Big)
\psi_{m_1}\ldots\psi_{m_l}
|-l\rangle
&=
(-k)^{-l} \prod_{i=1}^{l} \delta_{m_i+i,1}
\\
\langle 0|
\exp\Big(
\sum_{n=1}^{\infty}
\frac{1}{n} k^{-n} H_n
\Big)
\psi_{m_1}\ldots\psi_{m_l}
|-l\rangle
&=
k^{-m_1-1}
\delta_{l,1}
\end{align}

\noindent where we have defined $|\mu)=\psi_{m_1}\ldots\psi_{m_l}|-l\rangle$ as usual. Finally, due to the orthonormality of partition vectors (\ref{part-orth}), we obtain 

\begin{align}
\langle 0|
\exp\Big(
-\sum_{n=1}^{\infty}
\frac{1}{n} k^{-n} H_n
\Big)
&=
\langle 0|
+
\sum_{l=1}^{\infty}
(-k)^{-l}
\langle -l|
\psis_{-l+1}\ldots \psis_0
\\
\langle 0|
\exp\Big(
\sum_{n=1}^{\infty}
\frac{1}{n} k^{-n} H_n
\Big)
&=
\langle 0|
+
\sum_{i=1}^{\infty}
k^{-i}
\langle -1| \psis_{i-1}
\end{align}

\noindent Comparing these equations with (\ref{KPexpI-4}) and (\ref{KPexpI-5}), we complete the proof of (\ref{KPexpI17}) and (\ref{KPexpI18}).
}
\end{proof} 

Applying (\ref{KPexpI17}) and (\ref{KPexpI18}) to equation (\ref{KPexpI14}), we obtain

\begin{align}
\oint
\exp\left(\sum_{n=1}^{\infty}(t_n-s_n)k^n \right)
\langle 0| 
e^{H\{t-\epsilon_k\}}
g_{\psi}|0\rangle
\langle 0| 
e^{H\{s+\epsilon_k\}}
g_{\psi}|0\rangle
\frac{dk}{2\pi i}
=
0
\label{KPexpI-3}
\end{align}

\noindent Note the disappearance of the factor of $\frac{1}{k}$ from the integrand of 
$(\ref{KPexpI14})$, which is due to the cancelling factor of $k$ in 
$(\ref{KPexpI18})$. Equation (\ref{KPexpI-3}) proves that if (\ref{KPhelp}) holds, functions given by (\ref{KPexpI1}) satisfy the KP bilinear identity (\ref{KPbil1}).

\medskip
\noindent
{\bf Step 3.} For any finite $g_{\psi} \in Cl_{\psi}^{(0)}$ there exists an integer $l \geq 1$ and coefficients $\kappa_{\{m\},\{n\}}$ such that

\begin{align}
\sum_{i\in\mathbb{Z}}
\psi_i g_{\psi} |0\rangle
\otimes
\psis_i g_{\psi} |0\rangle
=
\sum_{\substack{{\rm card}\{m\}=l \\ {\rm card}\{n\}=l}}
\kappa_{\{m\},\{n\}}
\psi_{{\{m\}}} |-l+1\rangle
\otimes
\psi_{{\{n\}}} |-l-1\rangle
\label{pluck}
\end{align}

\noindent where the sum is over all sets of integers $\{m\} = \{m_1 > \cdots > m_{l} \geq -l+1\}$ and $\{n\} = \{n_1 > \cdots > n_{l} \geq -l-1\}$, of fixed cardinality $l$. Acting upon both sides of this equation with the tensored dual states $\langle 1|e^{H\{t\}} \otimes \langle -1|e^{H\{s\}}$ we find

{\small
\begin{align}
\oint
\exp\left(\sum_{n=1}^{\infty}(t_n-s_n)k^n \right)
\langle 0| 
e^{H\{t-\epsilon_k\}}
g_{\psi}|0\rangle
\langle 0| 
e^{H\{s+\epsilon_k\}}
&
g_{\psi}
|0\rangle
\frac{dk}{2\pi i}
=
\\
&
\sum_{\substack{{\rm card}\{m\}=l \\ {\rm card}\{n\}=l}}
\kappa_{\{m\},\{n\}}
\chi_{(\mu-1)}\{t\}
\chi_{(\nu+1)}\{s\}
\nonumber
\end{align}
}

\noindent where the left hand side has already been derived in steps 1 and 2, and the right hand side follows from lemma 5, with $\mu_i = m_i+i$ and $\nu_i = n_i+i$. Assuming that $\tau\{t\}$ as given by (\ref{KPexpI1}) satisfies the KP bilinear identity, we thus obtain

\begin{align}
\sum_{\substack{{\rm card}\{m\}=l \\ {\rm card}\{n\}=l}}
\kappa_{\{m\},\{n\}}
\chi_{(\mu-1)}\{t\}
\chi_{(\nu+1)}\{s\}
=
0
\end{align}

\noindent which can only be true if all of the coefficients $\kappa_{\{m\},\{n\}}=0$, since the Schur polynomials are linearly independent. Substituting this trivial value for the coefficients into (\ref{pluck}), we recover the CFBI (\ref{KPhelp}). This completes the proof of the converse statement.
}
\end{proof}

\section{Solutions of the CFBI}
\label{c-sol}

\subsection{Orbit of $GL_{\infty}$}

\begin{theorem}
{\rm Suppose $g_{\psi}$ is a finite element of $Cl_{\psi}^{(0)}$. Then $g_{\psi}$ solves the CFBI (\ref{KPhelp}) if and only if

\begin{align}
g_{\psi}|0\rangle
=
e^{X_1}\ldots e^{X_l}
|0\rangle
\label{KPfermI-2}
\end{align}

\noindent for some $\{X_1,\ldots,X_l\} \in A_{\infty}$. In other words, the solution space of (\ref{KPhelp}) is generated by the orbit of the Lie group

\begin{align}
GL_{\infty}
=
\Big\{
e^{X_1}\ldots e^{X_l}\ 
\Big|\ 
X_i \in A_{\infty}\
{\rm for\ all}\
1\leq i \leq l
\Big\}
\end{align}

}
\end{theorem}

\medskip
\noindent 
{\it Proof.}\ We split the proof into two steps. In the first step we prove the forward statement, in the second step we prove its converse. 

\medskip
\noindent
{\bf Step 1. (Lemma 9.)} Let $|u\rangle$ and $|v\rangle$ be arbitrary state vectors in $\mathcal{F}_{\psi}$, and let $g_{\psi} = e^{X_1}\ldots e^{X_l}$ with each $X_i \in A_{\infty}$. We have

\begin{align}
\sum_{i\in \mathbb{Z}} 
\psi_i g_{\psi} |u\rangle 
\otimes
\psis_i g_{\psi} |v\rangle
=
\sum_{i \in \mathbb{Z}}
g_{\psi} \psi_i |u\rangle
\otimes
g_{\psi} \psis_i |v\rangle
\label{KPexpI2}
\end{align}

\begin{proof}
{\it (Lemma 9.)}\ 
{\rm 
For $m \geq 0$ and arbitrary $X \in A_{\infty}$, let $\mathcal{P}_m$ denote the proposition

\begin{align}
(\psi_i \otimes \psis_i)
\sum_{n=0}^{m}
\binom{m}{n}
X^n \otimes  X^{m-n}
=
\sum_{n=0}^{m}
\binom{m}{n}
X^n  \otimes X^{m-n} 
(\psi_i \otimes \psis_i)
\label{KPexpI3}
\end{align}

\noindent where summation over all integers $i$ is implied. The proposition $\mathcal{P}_0$ is trivial. Furthermore, letting $X \in A_{\infty}$ be given by (\ref{ainfty1}), by direct calculation we obtain the commutation relations  

%
 
 \begin{align}
 [\psi_i,X] = -\sum_{j \in \mathbb{Z}} a_{j,i} \psi_j, 
 \quad
 [\psis_i,X] = \sum_{j \in \mathbb{Z}} a_{i,j} \psis_j
 \end{align}

\noindent Using these commutators in the left hand side of $\mathcal{P}_1$, we obtain

\begin{align}
(\psi_i \otimes \psis_i)
(1\otimes X + X\otimes 1)
=
&
(1 \otimes X + X \otimes 1)
(\psi_i \otimes \psis_i)
+
a_{i,j}
\psi_i \otimes \psis_j 
-
a_{j,i}
\psi_j  \otimes \psis_i
\nonumber
\\
=
&
(1 \otimes X + X \otimes 1)
(\psi_i \otimes \psis_i)
\end{align}

\noindent where summation over all integers $i,j$ is implied. This proves $\mathcal{P}_1$ is true. Now suppose $\mathcal{P}_m$ is true for some $m \geq 1$. Using the identity $\binom{m}{n} + \binom{m}{n-1} = \binom{m+1}{n}$ for all $1 \leq n \leq m$,  we write

\begin{align}
\label{KPexpI4}
&
(\psi_i \otimes \psis_i)
\sum_{n=0}^{m+1}
\binom{m+1}{n}
 X^n \otimes X^{m+1-n}
 =
\\
\nonumber
&
(\psi_i \otimes \psis_i)
\left(
\sum_{n=0}^{m}
\binom{m}{n} 
X^n \otimes X^{m+1-n}
+
\sum_{n=1}^{m+1}
\binom{m}{n-1}
X^n \otimes X^{m+1-n}
\right)
\end{align}

\noindent with summation implied over all integers $i$. Shifting the summation index of the second sum on the right hand side of (\ref{KPexpI4}), we obtain

{\small
\begin{align}
\label{KPexpI5}
&
(\psi_i \otimes \psis_i)
\sum_{n=0}^{m+1}
\binom{m+1}{n}
 X^n \otimes X^{m+1-n}
 =
(\psi_i \otimes \psis_i)
\sum_{n=0}^{m}
\binom{m}{n} 
X^n \otimes X^{m-n} 
(1 \otimes X+X \otimes 1)
\end{align}
}

\noindent Now it is possible to use $\mathcal{P}_m$ and $\mathcal{P}_1$ in the right hand side of $(\ref{KPexpI5})$, to obtain

{\small
\begin{align}
\nonumber
(\psi_i \otimes \psis_i)
\sum_{n=0}^{m+1}
\binom{m+1}{n}
 X^n \otimes X^{m+1-n}
 &=
\sum_{n=0}^{m}
\binom{m}{n} 
X^n \otimes X^{m-n}
(\psi_i \otimes \psis_i)
(1 \otimes X + X \otimes 1)
\\
\label{KPexpI6}
&=
\sum_{n=0}^{m}
\binom{m}{n}
X^n \otimes X^{m-n}
(1 \otimes X + X \otimes 1)
(\psi_i \otimes \psis_i)
\end{align}
}

\noindent Finally, we use the reverse of (\ref{KPexpI5}) in the right hand side of (\ref{KPexpI6}), giving

\begin{align}
&
(\psi_i \otimes \psis_i)
\sum_{n=0}^{m+1}
\binom{m+1}{n} 
X^n \otimes X^{m+1-n}
=
\sum_{n=0}^{m+1}
\binom{m+1}{n}
X^n \otimes X^{m+1-n}
(\psi_i \otimes \psis_i)
\end{align}

\noindent Therefore $\mathcal{P}_m$ true $\implies$ $\mathcal{P}_{m+1}$ true, and the proposition $(\ref{KPexpI3})$ holds for all $m \geq 0$ by induction. By virtue of the proposition (\ref{KPexpI3}), for any $X \in A_{\infty}$ we have

\begin{align}
(\psi_i \otimes \psis_i)
(e^X \otimes e^X)
|u\rangle \otimes |v\rangle
=
&
(\psi_i \otimes \psis_i)
\sum_{m=0}^{\infty}
\frac{1}{m!}
\sum_{n=0}^{m}
\binom{m}{n}
X^{n}
\otimes
X^{m-n}
|u\rangle \otimes |v\rangle
\nonumber
\\
=
&
\sum_{m=0}^{\infty}
\frac{1}{m!}
\sum_{n=0}^{m}
\binom{m}{n}
X^{n}
\otimes
X^{m-n}
(\psi_i \otimes \psis_i)
|u\rangle \otimes |v\rangle
\nonumber
\\
=
&
(e^X \otimes e^X)(\psi_i \otimes \psis_i)|u\rangle \otimes |v\rangle 
\end{align}

\noindent Therefore we have proved that 

\begin{align}
\sum_{i \in \mathbb{Z}}
\psi_i e^{X} |u\rangle
\otimes
\psis_i e^{X} |v\rangle
=
\sum_{i \in \mathbb{Z}}
e^{X} \psi_i |u\rangle 
\otimes
e^{X} \psis_i |v\rangle
\label{KPfermI-3}
\end{align}

\noindent for arbitrary $X \in A_{\infty}$. Using (\ref{KPfermI-3}) $l$ times successively, once for each $e^{X_i}$ in $g_{\psi}$, we obtain equation (\ref{KPexpI2}).
}
\end{proof}

\medskip
\noindent
{\bf Corollary.} Having established the validity of equation (\ref{KPexpI2}) we employ a particular case of it, namely when $|u\rangle = |v\rangle = |0\rangle$, which gives

\begin{align}
\sum_{i\in \mathbb{Z}} 
\psi_i g_{\psi} |0\rangle 
\otimes
\psis_i g_{\psi} |0\rangle
=
\sum_{i \in \mathbb{Z}}
g_{\psi} \psi_i |0\rangle
\otimes
g_{\psi} \psis_i |0\rangle
=
0
\label{KPexpI7}
\end{align}

\noindent where the final equality is due to the fact that for all $i \in \mathbb{Z}$ either $\psi_i|0\rangle = 0$ or $\psis_i|0\rangle = 0$. Equation (\ref{KPexpI7}) completes the proof that when $g_{\psi}|0\rangle$ is of the form (\ref{KPfermI-2}), $g_{\psi}$ satisfies the CFBI (\ref{KPhelp}).

\medskip
\noindent
{\bf Step 2. (Lemma 10.)}\ Let $g_{\psi} \in Cl_{\psi}^{(0)}$ satisfy the CFBI (\ref{KPhelp}). Then for suitable $\{X_1,\ldots,X_l\} \in A_{\infty}$ we can write $g_{\psi}|0\rangle = e^{X_1} \ldots e^{X_l} |0\rangle$.

\begin{proof}
{\it (Lemma 10.)}\ 
{\rm We use the method of proof given in chapter 5 of \cite{mjd}. Since $g_{\psi}|0\rangle \in \mathcal{F}_{\psi}^{(0)}$ we can expand it in terms of the basis (\ref{repclA5}) to give

\begin{align}
g_{\psi}|0\rangle
=
c_{\emptyset} |0\rangle
+
\sum_{m \geq 0, n < 0}
c_{m,n}
\psi_{m} \psis_{n}
|0\rangle
+
g_{\psi}^{(1)}|0\rangle
\end{align}

\noindent for some suitable coefficients $c_{\emptyset}$ and $c_{m,n}$, and where all monomials within $g_{\psi}^{(1)} \in Cl_{\psi}^{(0)}$  consist of at least two positive $(+1)$ fermions and two negative $(-1)$ fermions.\footnote{Throughout the rest of the proof, we will always use $g_{\psi}^{(i)}$ to denote an element of $Cl_{\psi}^{(0)}$ with precisely this property.} From here, we need to consider the cases $c_{\emptyset}\not= 0$ and $c_{\emptyset} = 0$ separately.

\medskip
\noindent
{\bf Case 1. ($c_{\emptyset} \not= 0$)} We define elements $X_1,X_2$ of $A_{\infty}$ as follows

\begin{align}
X_1= \log c_{\emptyset},
\quad
X_2 
= 
\sum_{m \geq 0, n < 0}
c^{(1)}_{m,n} \psi_m \psis_n
\end{align}

\noindent where $c^{(1)}_{m,n} = c_{m,n} / c_{\emptyset}$ for all $m \geq 0, n < 0$. We trivially obtain

\begin{align}
e^{-X_1} g_{\psi}|0\rangle
=
|0\rangle
+
\sum_{m \geq 0,n<0} c^{(1)}_{m,n} \psi_m\psis_n |0\rangle
+
g_{\psi}^{(2)} |0\rangle
\label{KPblah}
\end{align}

\noindent where we have defined $g_{\psi}^{(2)} = g_{\psi}^{(1)} / c_{\emptyset}$. Next, we act on equation (\ref{KPblah}) with the operator $e^{-X_2}$. Term by term we have

\begin{align}
&
e^{-X_2} |0\rangle 
= 
|0\rangle 
-
\sum_{m \geq 0, n<0}
c^{(1)}_{m,n}
\psi_m \psis_n |0\rangle
+
g_{\psi}^{(3)} |0\rangle
\\
&
e^{-X_2} 
\sum_{m \geq 0,n<0} 
c^{(1)}_{m,n} \psi_m \psis_n |0\rangle
=
\sum_{m \geq 0,n<0}
c^{(1)}_{m,n} \psi_m \psis_n |0\rangle
+
g_{\psi}^{(4)} |0\rangle
\\
&
e^{-X_2} g_{\psi}^{(2)} |0\rangle
=
g_{\psi}^{(5)} |0\rangle
\end{align}

\noindent for some suitable $g_{\psi}^{(3)},g_{\psi}^{(4)},g_{\psi}^{(5)} \in Cl_{\psi}^{(0)}$. Combining these three results, we obtain

\begin{align} 
e^{-X_2} e^{-X_1} g_{\psi}|0\rangle
=
|0\rangle
+
g_{\psi}^{(6)} |0\rangle
\label{KPblah2}
\end{align}

\noindent where we have defined $g_{\psi}^{(6)} = g_{\psi}^{(3)}+g_{\psi}^{(4)}+g_{\psi}^{(5)}$. By virtue of equation (\ref{KPexpI2}) and the fact that $g_{\psi}$ obeys the CFBI (\ref{KPhelp}), we have

\begin{align}
0
&
=
\sum_{i \in \mathbb{Z}}
e^{-X_2} e^{-X_1} \psi_i g_{\psi} |0\rangle
\otimes
e^{-X_2} e^{-X_1} \psis_i g_{\psi} |0\rangle
\label{KPblah3}
\\
&
=
\sum_{i \in \mathbb{Z}}
\psi_i e^{-X_2} e^{-X_1} g_{\psi} |0\rangle
\otimes
\psis_i e^{-X_2} e^{-X_1} g_{\psi} |0\rangle
\nonumber
\end{align}

\noindent Substituting equation (\ref{KPblah2}) for $e^{-X_2} e^{-X_1} g_{\psi} |0\rangle$ into the second line of (\ref{KPblah3}) and using the annihilation properties (\ref{repclA1}), we find 

\begin{align}
\sum_{i \geq 0}
\psi_i |0\rangle
\otimes
\psis_i g_{\psi}^{(6)} |0\rangle
+
\sum_{i < 0}
\psi_i g_{\psi}^{(6)} |0\rangle
\otimes
\psis_i |0\rangle
+
\sum_{i \in \mathbb{Z}}
\psi_i g_{\psi}^{(6)} |0\rangle
\otimes
\psis_i g_{\psi}^{(6)} |0\rangle
=
0
\label{KPblah4}
\end{align}

\noindent We recall that all monomials within $g_{\psi}^{(6)} \in Cl_{\psi}^{(0)}$ consist of at least two positive $(+1)$ fermions and two negative $(-1)$ fermions. Therefore the left hand side of (\ref{KPblah4}) vanishes if and only if

\begin{align}
\psi_m g_{\psi}^{(6)} |0\rangle
=
\psis_n g_{\psi}^{(6)} |0\rangle
=
0
\end{align}

\noindent for all $m < 0, n \geq 0$. The only possible resolution of this equation is that $g_{\psi}^{(6)}|0\rangle = 0$. Substituting this value of $g_{\psi}^{(6)}|0\rangle$ into (\ref{KPblah2}) we see that $e^{-X_2} e^{-X_1} g_{\psi} |0\rangle = |0\rangle$, or equivalently, $g_{\psi}|0\rangle = e^{X_1} e^{X_2} |0\rangle$. This completes the proof in the case $c_{\emptyset} \not= 0$. 

\medskip
\noindent
{\bf Case 2. ($c_{\emptyset} = 0$)} We begin by stating an identity which we use in the proof. Fix two integers $p \geq 0, q < 0$ and two sets $\{m\}=\{m_1 > \cdots > m_r \geq 0\}$ and $\{n\}=\{n_1 < \cdots < n_r < 0\}$. The identity reads

\begin{align}
e^{-\psi_p \psis_q}
e^{\psi_q \psis_p}
\psi_{\{m\}} \psis_{\{n\}}
|0\rangle
=
\left\{
\begin{array}{ll}
(-)^{i+j+r+1}
\psi_{\{m\backslash m_i\}} \psis_{\{n \backslash n_j\}}
|0\rangle,
&
p = m_i
\\
&
q = n_j
\\
\\
\psi_{\{m\}} \psis_{\{n\}}|0\rangle
-
\psi_p \psi_{\{m\}} \psis_{\{n\}} \psis_q |0\rangle,
&
p \not\in \{m\}
\\
&
q \not\in \{n\}
\\
\\
\psi_{\{m\}} \psis_{\{n\}}
|0\rangle,
&
{\rm otherwise}
\end{array}
\right.
\label{KPexpII5}
\end{align}

\noindent where we have used the notation $\{m\backslash m_i\}, \{n \backslash n_j\}$ to denote the omission of the $i^{\rm th}$ and $j^{\rm th}$ elements from the sets $\{m\}$ and $\{n\}$, respectively. Returning to the proof, we observe that since $c_{\emptyset} = 0$ we can write

\begin{align}
g_{\psi}|0\rangle
=
\sum_{\substack{{\rm card}\{m\}=r \\ {\rm card}\{n\}=r}}
c_{\{m\},\{n\}}
\psi_{\{m\}} \psis_{\{n\}}
|0\rangle
+
g_{\psi}^{(7)}|0\rangle
\label{KPblah5}
\end{align}

\noindent where the sum is taken over all ordered sets of integers $\{m_1 > \cdots > m_r \geq 0\}$ and $\{n_1 < \cdots < n_r < 0\}$ of some fixed cardinality $r \geq 1$, and all monomials within $g_{\psi}^{(7)} \in Cl_{\psi}^{(0)}$ consist of at least $r+1$ positive $(+1)$ fermions and $r+1$ negative $(-1)$ fermions. Let $c_{\{p\},\{q\}}$ be a particular non-zero coefficient in the sum (\ref{KPblah5}), corresponding to the sets $\{p\} = \{p_1 > \cdots > p_r \geq 0\}$ and $\{q\} = \{q_1 < \cdots < q_r < 0\}$, and define

\begin{align}
X_{2i-1} = -\psi_{q_i} \psis_{p_i},
\quad
X_{2i} = \psi_{p_i} \psis_{q_i}
\end{align}

\noindent for all $1\leq i \leq r$. Successively applying the identity $(\ref{KPexpII5})$ to $g_{\psi}|0\rangle$, we obtain

\begin{align}
e^{-X_{2r}}
e^{-X_{2r-1}}
\ldots
e^{-X_2}
e^{-X_1}
g_{\psi}|0\rangle
=
c^{(2)}_{\emptyset} |0\rangle
+
\sum_{m \geq 0, n< 0}
c^{(2)}_{m,n} \psi_m \psis_n |0\rangle
+
g_{\psi}^{(8)} |0\rangle
\label{KPexpII6}
\end{align}

\noindent with $c^{(2)}_{\emptyset}=(-)^{r(r-1)/2}c_{\{p\},\{q\}}$ and the remaining coefficients $c^{(2)}_{m,n}$ suitably chosen, and where all monomials within $g_{\psi}^{(8)} \in Cl_{\psi}^{(0)}$ consist of at least two positive $(+1)$ fermions and two negative $(-1)$ fermions. Since $c^{(2)}_{\emptyset} \not= 0$, we can apply the procedure of case 1 to (\ref{KPexpII6}), ultimately obtaining

\begin{align}
g_{\psi}|0\rangle
=
e^{X_1} e^{X_2}
\ldots
e^{X_{2r+1}} e^{X_{2r+2}}
|0\rangle
\label{KPblah6}
\end{align}

\noindent where we have defined 

\begin{align}
X_{2r+1}
=
\log c^{(2)}_{\emptyset},
\quad
X_{2r+2}
=
\sum_{m \geq 0,n<0}
c^{(2)}_{m,n} / c^{(2)}_{\emptyset}
\psi_m \psis_n 
\end{align}

\noindent Since all $\{X_1,\ldots,X_{2r+2}\} \in A_{\infty}$, equation (\ref{KPblah6}) completes the proof in the $c_{\emptyset} = 0$ case. We have therefore proved lemma 10 which, in turn, finishes the proof of theorem 2. 
}
\end{proof}

\subsection{Schur polynomials and the orbit of $GL_{\infty}$}

\begin{example}
{\rm As a particular case of theorem 2, we show that every Schur polynomial (\ref{surprising}) is a KP $\tau$-function. Let $|\mu)=|\mu_1,\ldots,\mu_l)$ be an arbitrary partition equal to the Fock space vector $\psi_{m_1}\ldots\psi_{m_l}|-l\rangle$, where $m_i=\mu_i-i$ for all $1\leq i \leq l$. Using equation (\ref{partitions8}) from the proof of lemma 1, we obtain

\begin{align}
\chi_{\mu}\{t\}
&=
\langle 0|e^{H\{t\}} \psi_{m_1}\ldots\psi_{m_l}|-l\rangle
\label{charac-exp}
=
(-)^{\sum_{i=1}^{r}(n_i+i)}
\Big\langle
e^{H\{t\}}
\psi_{m_1}\ldots\psi_{m_r}
\psis_{n_r}\ldots\psis_{n_1}
\Big\rangle
\end{align}

\noindent where $m_1 > \cdots > m_r \geq 0 > m_{r+1} > \cdots > m_l > -l$, and $\{n_r < \cdots < n_1\}$ is the set $\{-l < \cdots < -1\}$ with $\{m_l < \cdots < m_{r+1}\}$ omitted. Defining 

\begin{align}
X_{2i-1}
=
-\psi_{n_i} \psis_{m_i},
\quad
X_{2i}
=
\psi_{m_i} \psis_{n_i}
\end{align}

\noindent for all $1\leq i \leq r$ and using the identity (\ref{KPexpII5}) from the last subsection, we obtain

\begin{align}
e^{-X_{2r}} e^{-X_{2r-1}}
\ldots
e^{-X_2} e^{-X_1}
\psi_{m_1} \ldots \psi_{m_r}
\psis_{n_r} \ldots \psis_{n_1} |0\rangle
=
(-)^{r(r-1)/2}
|0\rangle
\end{align}

\noindent or equivalently,

\begin{align}
\psi_{m_1}\ldots \psi_{m_r}
\psis_{n_r} \ldots \psis_{n_1} |0\rangle
=
(-)^{r(r-1)/2}
e^{X_1}e^{X_2}\ldots e^{X_{2r-1}} e^{X_{2r}} |0\rangle
\label{charac-const}
\end{align}

\noindent Substituting (\ref{charac-const}) into (\ref{charac-exp}), we find

\begin{align}
\chi_{\mu}\{t\}
=
(-)^{r+\sum_{i=1}^{r}n_i}
\Big\langle
e^{H\{t\}}
e^{X_1}\ldots e^{X_{2r}}
\Big\rangle
\end{align} 

\noindent Hence any Schur polynomial can be written as an expectation value of the form (\ref{KPexpI1}), with $g_{\psi} \in GL_{\infty}$. By theorem 2, the Schur polynomials are therefore $\tau$-functions of the KP hierarchy.
}
\end{example} 

\subsection{KP Pl\"ucker relations}
\label{KPpluck}

In this subsection we solve the CFBI (\ref{KPhelp}) from another, more direct perspective. Let $g_{\psi}|0\rangle$ be a finite element of $\mathcal{F}_{\psi}^{(0)}$ expanded in terms of the basis $(\ref{partitions3})$. Since $g_{\psi}|0\rangle$ is finite, there exists some $l \geq 0$ and coefficients $c_{\{m\}}$ such that

\begin{align}
g_{\psi}|0\rangle
=
\sum_{{\rm card} \{m\}=l}
c_{\{m\}}
\psi_{\{m\}}|-l\rangle
\label{pluck-exp}
\end{align}

\noindent where the sum is over all sets of integers $\{m\} = \{m_1 > \cdots > m_l \geq -l\}$. Using this expansion of $g_{\psi}|0\rangle$, we obtain

\begin{align}
\sum_{i \in \mathbb{Z}}
\psi_i g_{\psi}|0\rangle
\otimes
\psis_i g_{\psi}|0\rangle
&
=
\sum_{i \in \mathbb{Z}}
\sum_{\substack{{\rm card}\{m\}= l \\ {\rm card}\{n\}= l}}
c_{\{m\}} c_{\{n\}}
\psi_i \psi_{\{m\}} |-l\rangle
\otimes
\psis_i \psi_{\{n\}} |-l\rangle
\label{pluck-2}
\end{align}

\noindent where the second sum is over all sets of integers $\{m_1 > \cdots > m_l \geq -l\}$ and $\{n_1 > \cdots > n_l \geq -l\}$ of fixed cardinality $l$. Switching the order of the sums in (\ref{pluck-2}) and using the annihilation properties (\ref{repclA1}) of the fermions, we find

\begin{align}
&
\sum_{i \in \mathbb{Z}}
\psi_i g_{\psi}|0\rangle
\otimes
\psis_i g_{\psi}|0\rangle
\label{pluck-3}
=
\sum_{\substack{{\rm card}\{m\}= l \\ {\rm card}\{n\}= l}}
\sum_{i=1}^{l}
(-)^{i-1}
c_{\{m\}} c_{\{n\}}
\psi_{\{n_i,m\}} |-l\rangle
\otimes
\psi_{\{n\backslash n_i\}} |-l\rangle
\end{align}

\noindent where $\psi_{\{n_i,m\}} = 0$ if $n_i \in \{m\}$. Changing the indexing sets of the first sum in (\ref{pluck-3}), we obtain the equivalent expression

\begin{align}
&
\sum_{i \in \mathbb{Z}}
\psi_i g_{\psi}|0\rangle
\otimes
\psis_i g_{\psi}|0\rangle
\label{pluck-4}
=
\sum_{\substack{{\rm card}\{p\}=l+1 \\ {\rm card}\{q\} = l-1}}
\sum_{j=1}^{l+1}
(-)^{j-1}
c_{\{p \backslash p_j \}} 
c_{\{p_j,q\}}
\psi_{\{ p \}} |-l\rangle
\otimes
\psi_{\{ q \}} |-l\rangle
\end{align}

\noindent with the first sum taken over all sets of integers $\{p_1 > \cdots > p_{l+1} \geq -l\}$ and $\{q_1 > \cdots > q_{l-1} \geq -l\}$, and where we have defined

\begin{align}
c_{\{p_j,q\}} 
=
(-)^{i-1} c_{\{q_1,\ldots,q_{i-1},p_j,q_{i},\ldots,q_{l-1}\}}
\end{align}

\noindent if $q_{i-1} > p_j > q_{i}$ for some $1\leq i \leq l$, and $c_{\{p_j,q\}} = 0$ if $p_j \in \{q\}$. The right hand side of (\ref{pluck-4}) vanishes if and only if

\begin{align}
\sum_{i=1}^{l+1}  
(-)^{i-1}
c_{\{m\backslash m_i\}}
c_{\{m_i, n \}}
=
0
\label{KPpluck1}
\end{align}

\noindent for all sets $\{m_1 > \ldots > m_{l+1} \geq -l\}$ and $\{n_1 > \cdots > n_{l-1} \geq -l\}$. Collectively, these conditions are called the {\it KP Pl\"ucker relations}, and we summarize their significance with the following statement (which we have already proved). 

\setcounter{lemma}{10}
\begin{lemma}
{\rm 
$g_{\psi} \in Cl_{\psi}^{(0)}$ satisfies the CFBI (\ref{KPhelp}) if and only if the expansion coefficients (\ref{pluck-exp}) of $g_{\psi} |0\rangle$ obey the KP Pl\"ucker relations (\ref{KPpluck1}).}
\end{lemma} 
 
\subsection{Determinant solution of KP Pl\"ucker relations}
\label{KPdet}

With the following result, we present a general determinant solution of the KP Pl\"ucker relations (\ref{KPpluck1}). 

\begin{lemma}
{\rm To every set of integers $\{m\} = \{m_1, \ldots, m_l\}$ we associate the coefficient 

\begin{align}
c_{\{m\}} 
=
\det \Big(c_{m_i,j}\Big)_{1 \leq i,j \leq l}
=
\left|
\begin{array}{ccc}
c_{m_1,1}     & \cdots & c_{m_1,l}
\\
\vdots        &        & \vdots
\\
c_{m_{l},1} & \cdots & c_{m_{l},l}
\end{array}
\right|
\label{KPdet1}
\end{align}

\noindent where the matrix entries $c_{i,j}$ are arbitrary constants. These coefficients satisfy the Pl\"ucker relations $(\ref{KPpluck1})$.
}
\end{lemma}

\begin{proof}
{\rm The proof is taken from the paper \cite{ostt}. Define two ordered sets of integers $\{m\}=\{m_1 > \cdots > m_{l+1} \geq -l \}$ and $\{n\}=\{n_1 > \cdots > n_{l-1} \geq -l \}$, and from these sets construct a $2l \times 2l$ matrix $M$ given by

\begin{align}
M\Big(\{m\},\{n\}\Big)
=
\left(
\begin{array}{cccccc}
c_{m_1,1}      & \cdots & c_{m_1,l}     & c_{m_1,1}             & \cdots & c_{m_1,l}             
\\
\vdots         &        & \vdots        & \vdots        &        &\vdots
\\
c_{m_{l+1},1}  & \cdots & c_{m_{l+1},l} & c_{m_{l+1},1}             & \cdots & c_{m_{l+1},l} 
\\
c_{n_1,1}      & \cdots & c_{n_1,l}     & 0     & \cdots & 0
\\
\vdots         &        & \vdots        & \vdots        &        &\vdots
\\
c_{n_{l-1},1}  & \cdots & c_{n_{l-1},l} & 0 & \cdots & 0    
\end{array}
\right)
\label{KPdet2}
\end{align}

\noindent Using Laplace's formula for the expansion of $\det M$ yields 

\begin{align}
\det M
=
\sum_{i=1}^{l+1}
(-)^{i-1}
\left|
\begin{array}{ccc}
c_{m_1,1}           & \cdots & c_{m_1,l}
\\
\vdots              &        & \vdots
\\
\widehat{c_{m_i,1}} & \cdots & \widehat{c_{m_i,l}}
\\
\vdots              &        & \vdots
\\
c_{m_{l+1},1}       & \cdots & c_{m_{l+1},l}
\end{array}
\right|
\left|
\begin{array}{ccc}
c_{m_i,1}     & \cdots & c_{m_i,l}
\\
c_{n_1,1}     & \cdots & c_{n_1,l}
\\
\vdots        &        & \vdots
\\
c_{n_{l-1},1}     & \cdots & c_{n_{l-1},l}
\end{array}
\right|
\label{KPdet3}
\end{align}

\noindent where the $i^{\rm th}$ row has been omitted from the first determinant, and inserted as the first row of the second determinant. Writing these determinants in terms of the coefficients (\ref{KPdet1}), we obtain

\begin{align}
\det M
= 
\sum_{i=1}^{l+1}
(-)^{i-1}
c_{\{m\backslash m_i \}} c_{\{m_i,n\}}
\label{KPdet4}
\end{align}

\noindent which is precisely the left hand side of the KP Pl\"ucker relations $(\ref{KPpluck1})$. All that remains in the proof is to demonstrate that $\det M = 0$. This is achieved by subtracting the last $l$ columns from the first $l$ columns in 
$\det M$ to obtain an equivalent determinant $\det M^{(1)}$, where 

\begin{align}
M^{(1)}\Big(\{m\},\{n\}\Big)
=
\left(
\begin{array}{cccccc}
0      & \cdots & 0     & c_{m_1,1}             & \cdots & c_{m_1,l}             
\\
\vdots         &        & \vdots        & \vdots        &        &\vdots
\\
0  & \cdots & 0 & c_{m_{l+1},1}             & \cdots & c_{m_{l+1},l} 
\\
c_{n_1,1}      & \cdots & c_{n_1,l}     & 0     & \cdots & 0
\\
\vdots         &        & \vdots        & \vdots        &        &\vdots
\\
c_{n_{l-1},1}  & \cdots & c_{n_{l-1},l} & 0 & \cdots & 0    
\end{array}
\right)
\label{KPdet5}
\end{align}

\noindent Writing
$
\det M^{(1)} 
= 
\sum_{\sigma \in S_{2l}}
{\rm sgn}(\sigma)
\prod_{i=1}^{2l}
M^{(1)}_{i,\sigma(i)}
$
and using $(\ref{KPdet5})$, we see that every product in this sum has at least one zero term, proving
that $\det M = \det M^{(1)} = 0$. Hence we have shown that the sum $(\ref{KPdet4})$ is equal to zero, proving that the coefficients $(\ref{KPdet1})$ satisfy the KP Pl\"ucker relations $(\ref{KPpluck1})$. 
}
\end{proof}


\section{Neutral fermions and related definitions}
\label{n-fermi}

\subsection{Neutral fermions}
\label{nfermi}

Consider the infinite set $\{\phi_m\}_{m \in \mathbb{Z}}$, where $m$ ranges over all integers. The elements in this set are called {\it neutral fermions}, and they are defined as linear combinations of the charged fermions (\ref{cfermi1}). Explicitly, we set   

\begin{align}
\phi_m 
= 
\psi_m + (-)^m \psis_{-m} 
\label{nfermi1}
\end{align}

\noindent for all $m \in \mathbb{Z}$.\footnote{In \cite{jm1}, the neutral fermions were defined as $\phi_m = \frac{1}{\sqrt{2}}(\psi_m+(-)^m\psis_{-m})$. We have deliberately omitted the factor $\frac{1}{\sqrt{2}}$ from our definition, in order to obtain the correct normalization for polynomials which we later discuss.} From (\ref{nfermi1}) and the charged fermion anticommutation relations (\ref{cfermi1}), we see that the neutral fermions satisfy

\begin{align}
\left[ \phi_m, \phi_n \right]_{+} = 2(-)^m \delta_{m + n, 0}
\label{nfermi2} 
\end{align}

\noindent for all $m,n \in \mathbb{Z}$. As a special case of the equations (\ref{nfermi2}), we obtain 

\begin{align}
\phi_m^2= \delta_{m,0}
\end{align}

\noindent for all $m \in \mathbb{Z}$. For our later convenience, we introduce a second set of fermions $\{\phis_m\}_{m\in\mathbb{Z}}$, defined as

\begin{align}
\phis_m 
= 
(-)^m \phi_{-m} 
=
\psis_{m} + (-)^m \psi_{-m}
\end{align}

\noindent for all $m \in \mathbb{Z}$. This second set of fermions is purely a relabelling of the first set.\footnote{The neutral fermions $\phis_m$ are genuinely different from the second species defined in \cite{jm1}, which are given by $\hat{\phi}_m = \frac{i}{\sqrt{2}}(\psi_m-(-)^m \psis_{-m})$.}

\subsection{Clifford algebra $Cl_{\phi}$}
\label{clB}

Let $Cl_{\phi}$ be the associative subalgebra of $Cl_{\psi}$ generated by $1$ and the neutral fermions $\{\phi_m\}_{m \in \mathbb{Z}}$, modulo the anticommutation relations (\ref{nfermi1}). Considered as a vector space, $Cl_{\phi}$ has the basis

\begin{align}
{\rm Basis}\left(Cl_{\phi}\right)
=
\Big\{
1,\phi_{m_1}\ldots\phi_{m_r}
\Big\}
\end{align}

\noindent where $\{m_1 > \cdots > m_r\}$ ranges over all integers, and the cardinality of this set takes all values $r \geq 1$. We decompose $Cl_{\phi}$ into the direct sum of subalgebras

\begin{align}
Cl_{\phi} 
=
\bigoplus_{i \in \{0,1\}} 
Cl_{\phi}^{(i)}
\label{clB1}
\end{align}

\noindent where $Cl_{\phi}^{(i)}$ is the linear span of all neutral fermion monomials of length $i\ {\rm mod}\ 2$. In this chapter we are mainly interested in the subalgebra $Cl_{\phi}^{(0)}$, which has the basis

\begin{align}
{\rm Basis}\left( Cl_{\phi}^{(0)} \right)
=
\Big\{
1, \phi_{m_1}\ldots \phi_{m_{2r}}
\Big\}
\end{align}

\noindent where $\{m_1 > \cdots > m_{2r} \}$ ranges over all integers, and the cardinality of this set takes all positive even values.


\subsection{Fock representation of $Cl_{\phi}$}
\label{repclB}

Using the annihilation properties (\ref{repclA1}) of the charged fermions, the action of $Cl_{\phi}$ on the vacuum $|0\rangle$ and dual vacuum $\langle 0|$ is given by  

\begin{align}
\phi_m |0\rangle
=
0
,
\quad
\langle 0| \phi_n
=
0
\label{repclB1}
\end{align}

\noindent for all integers $m < 0, n > 0$. The Fock space $\mathcal{F}_{\phi}$ and dual Fock space $\mathcal{F}_{\phi}^{*}$ are the vector spaces generated linearly by the action of $Cl_{\phi}$ on $|0\rangle$ and $\langle 0|$, respectively. Due to the annihilation relations (\ref{repclB1}), they have the bases

\begin{align}
{\rm Basis}\left(\mathcal{F}_{\phi}\right)
&=
\Big\{
|0\rangle,
\phi_{m_1}\ldots\phi_{m_r}|0\rangle
\Big\},
\label{repclB2}
\quad
{\rm Basis}\left(\mathcal{F}_{\phi}^{*}\right)
=
\Big\{
\langle 0|,
\langle 0|\phis_{m_r}\ldots\phis_{m_1}
\Big\}
\end{align}

\noindent where in both cases $\{m_1>\cdots>m_r\geq 0\}$ ranges over all non-negative integers, and the cardinality of this set takes all values $r \geq 1$.  

Using the definition of the Clifford subalgebras (\ref{clB1}), we decompose $\mathcal{F}_{\phi}$ and $\mathcal{F}_{\phi}^{*}$ into the following direct sums of subspaces

\begin{align}
\mathcal{F}_{\phi}
=
\bigoplus_{i \in \{0,1\}}
\mathcal{F}_{\phi}^{(i)}
,
\quad
\mathcal{F}_{\phi}^{*}
=
\bigoplus_{i \in \{0,1\}}
\mathcal{F}_{\phi}^{*(i)}
\label{repclB4}
\end{align}

\noindent where $\mathcal{F}_{\phi}^{(i)}$ and $\mathcal{F}_{\phi}^{*(i)}$ are the subspaces generated linearly by the action of $Cl_{\phi}^{(i)}$ on $|0\rangle$ and $\langle 0|$, respectively. The bases of $\mathcal{F}_{\phi}^{(0)}$ and $\mathcal{F}_{\phi}^{*(0)}$ are given by

\begin{align}
{\rm Basis}\left(\mathcal{F}_{\phi}^{(0)}\right)
&=
\Big\{
|0\rangle,
\phi_{m_1}\ldots\phi_{m_{2r}} |0\rangle
\Big\},
\label{repclB5}
\quad
{\rm Basis}\left(\mathcal{F}_{\phi}^{*(0)}\right)
=
\Big\{
\langle 0|,
\langle 0| \phis_{m_{2r}}\ldots\phis_{m_1}
\Big\}
\end{align}

\noindent where in both cases $\{m_1>\cdots>m_{2r} \geq 0\}$ ranges over all non-negative integers, and the cardinality of this set takes all positive even values.

\begin{remark}
{\rm 
In future sections we will sometimes adopt the notation

\begin{align}
\phi_{\{m\}} = \phi_{m_1}\ldots \phi_{m_r},
\quad
\phis_{\{m\}} = \phis_{m_r} \ldots \phis_{m_1}
\end{align}

\noindent for all ordered sets $\{m\} = \{m_1 > \cdots > m_r\}$ with cardinality $r \geq 1$. We also define $\phi_{\{m\}} = \phis_{\{m\}} = 1$ when $\{m\}$ is empty. For example, using this notation the bases (\ref{repclB5}) can be written as

\begin{align}
{\rm Basis}\left(\mathcal{F}_{\phi}^{(0)}\right)
&=
\Big\{ \phi_{\{m\}} |0\rangle\ \Big|\ {\rm card}\{m\}\ {\rm even} \Big\}
\\
{\rm Basis}\left(\mathcal{F}_{\phi}^{*(0)}\right)
&=
\Big\{ \langle 0| \phis_{\{m\}}\ \Big|\ {\rm card}\{m\}\ {\rm even} \Big\}
\end{align}

\noindent where $\{m\}$ ranges over all ordered, even-cardinality sets of non-negative integers.
}
\end{remark}
  
%
%
%

\subsection{Strict partitions}
\label{spartitions}

A {\it strict partition} $\tilde{\mu} = \{\mu_1 > \cdots > \mu_{l} > \mu_{l+1} = \cdots = 0 \}$ is a set of non-negative integers whose non-zero elements are strictly decreasing and finite in number.\footnote{Throughout the rest of this thesis we shall use a tilde to indicate that a partition is strict.} When represented as a Young diagram they consist of $l$ rows of boxes, such that no two rows have the same length.

\begin{figure}[H]

\begin{center}
\begin{minipage}{4.3in}

\setlength{\unitlength}{0.0003cm}
\begin{picture}(20000,12000)(-10000,10000)

\path(0,22000)(14000,22000)
\path(0,20000)(14000,20000)
\path(0,18000)(12000,18000)
\path(0,16000)(10000,16000)
\path(0,14000)(6000,14000)
\path(0,12000)(4000,12000)
\path(0,10000)(2000,10000)

\path(0,10000)(0,22000)
\path(2000,10000)(2000,22000)
\path(4000,12000)(4000,22000)
\path(6000,14000)(6000,22000)
\path(8000,16000)(8000,22000)
\path(10000,16000)(10000,22000)
\path(12000,18000)(12000,22000)
\path(14000,20000)(14000,22000)

\end{picture}

\end{minipage}
\end{center}

\caption[Young diagram of the strict partition $\tilde{\mu} = \{7,6,5,3,2,1\}$]{Young diagram of the strict partition $\tilde{\mu} = \{7,6,5,3,2,1\}$. Every row of boxes has a different length.}
\end{figure}
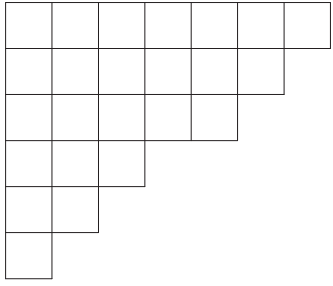

\noindent The elements of the bases (\ref{repclB5}) can be matched with strict partitions. We define $|0\rangle = |\emptyset)$ and $\langle 0| = (\emptyset|$, and for all sets of integers $\{\mu_1 > \cdots > \mu_{2r} \geq 0\}$ we write

\begin{align}
\phi_{\mu_1} \ldots \phi_{\mu_{2r}} |0\rangle
&=
|\mu_1,\ldots,\mu_{2r})
=
|\tilde{\mu})
\\ 
\langle 0| \phis_{\mu_{2r}} \ldots \phis_{\mu_1}
&=
(\mu_1,\ldots,\mu_{2r}|
=
(\tilde{\mu}|
\nonumber
\end{align}

\noindent We let $|\emptyset)$ and $(\emptyset|$ be copies of the empty partition $\emptyset$ as before, and $|\tilde{\mu})$ and $(\tilde{\mu}|$ be copies of the strict partition $\tilde{\mu} = \{\mu_1>\cdots>\mu_{2r} \geq 0\}$. This gives a one-to-one correspondence between the elements of the bases (\ref{repclB5}) and the elements of the set of all strict partitions. Strict partitions of even length are paired to basis vectors with $\mu_{2r} > 0$, while strict partitions of odd length are paired to basis vectors with $\mu_{2r} =0$. This correspondence is useful in later sections, when we encounter functions which are indexed by strict partitions.

%
%

\subsection{Neutral fermion expectation values}
\label{I'}

Since $Cl_{\phi} \subset Cl_{\psi}$, the vacuum expectation value of $g_{\phi} \in Cl_{\phi}$ is inherited from the definition (\ref{I1}). The following result is the neutral fermion analogue of lemma 2. 
%
%

\begin{lemma}
{\rm 
Let $\{m_1,\ldots,m_{2r}\}$ be an arbitrary set of integers, with even cardinality. We claim that

\begin{align}
\langle 
\phi_{m_1} \ldots \phi_{m_{2r}}
\rangle
=
{\rm Pf}
\Big( 
\langle \phi_{m_i} \phi_{m_j} \rangle 
\Big)_{1 \leq i < j \leq 2r} 
\label{BKPpoly1}
\end{align}

\noindent where ${\rm Pf}()$ denotes a Pfaffian.
}
\end{lemma}

\begin{proof} 
{\rm
The case $r=1$ is trivial, since

\begin{align}
\langle \phi_{m_1} \phi_{m_2} \rangle
=
{\rm Pf}
\Big(
\langle \phi_{m_i} \phi_{m_j} \rangle 
\Big)_{1\leq i < j \leq 2}
\end{align}

\noindent Using this case as the basis for induction, we assume there exists $r\geq 2$ such that

\begin{align}
\langle \phi_{m_1} \ldots \phi_{m_{2r-2}} \rangle
=
{\rm Pf}
\Big(
\langle \phi_{m_i} \phi_{m_j} \rangle
\Big)_{1\leq i < j \leq 2r-2}
\label{ass}
\end{align}

\noindent for all sets of integers $\{m_1,\ldots,m_{2r-2}\}$. We define 

\begin{align}
I_1
&=
\langle \phi_{m_1} \ldots \phi_{m_{2r}}\rangle
\\
\nonumber
I_2
&= 
\sum_{k=2}^{2r}
(-)^{k}
\langle \phi_{m_1}\phi_{m_k} \rangle
\langle \phi_{m_2} \ldots 
\widehat{\phi_{m_k}} \ldots 
\phi_{m_{2r}} \rangle
\end{align}

\noindent where $\{m_1,\ldots,m_{2r}\}$ is an arbitrary set of integers, and $\widehat{\phi_{m_k}}$ means the omission of the indicated fermion. By the annihilation properties (\ref{repclB1}) of the fermions, $I_1 = I_2=0$ unless both $m_1 \leq 0$ and $m_1 = \{-m_{k_1},\ldots,-m_{k_l}\}$ for some integers $2 \leq k_1 < \cdots < k_l \leq 2r$. When both these conditions are satisfied it is readily verified that

\begin{align}
I_1
=
I_2
=
\sum_{k \in \{k_1,\ldots,k_l\}}
(-)^{k}
\langle \phi_{m_1} \phi_{m_k} \rangle
\langle 
\phi_{m_2} 
\ldots 
\widehat{\phi_{m_k}} 
\ldots
\phi_{m_{2r}}
\rangle 
\end{align}

\noindent which proves that $I_1=I_2$ for all $m_1$. Therefore we obtain

\begin{align}
I_1
&=
\sum_{k=2}^{2r}
(-)^k
\langle \phi_{m_1} \phi_{m_k} \rangle
\langle \phi_{m_2} \ldots \widehat{\phi_{m_k}}
\ldots \phi_{m_{2r}} \rangle
\label{lasteqn}
\\
&=
\sum_{k=2}^{2r}
(-)^k
\langle \phi_{m_1} \phi_{m_k} \rangle
{\rm Pf}\Big(
\langle \phi_{m_i} \phi_{m_j} \rangle
\Big)_{\substack{2 \leq i < j \leq 2r \\ i,j\not= k}}
=
{\rm Pf}\Big(
\langle \phi_{m_i} \phi_{m_j} \rangle
\Big)_{1 \leq i<j \leq 2r}
\nonumber
\end{align}

\noindent where we have used the assumption (\ref{ass}) to produce the $(2r-2) \times (2r-2)$ Pfaffian in the second line of (\ref{lasteqn}). This completes the proof by induction.
}
\end{proof}

Since $\mathcal{F}_{\phi} \subset \mathcal{F}_{\psi}$ and $\mathcal{F}_{\phi}^{*} \subset \mathcal{F}_{\psi}^{*}$, we inherit a mapping $\mathcal{F}_{\phi}^{*} \times \mathcal{F}_{\phi} \rightarrow \mathbb{C}$ from the bilinear form (\ref{form}). Calculating its action on the strict partition vectors $(\tilde{\mu}| = \langle 0| \phis_{\mu_{2r}}\ldots \phis_{\mu_1}$ and $|\tilde{\nu}) = \phi_{\nu_1}\ldots\phi_{\nu_{2s}}|0\rangle$, we obtain

\begin{align}
\Big\langle
(\tilde{\mu}|,|\tilde{\nu})
\Big\rangle
=
\langle 0| \phis_{\mu_{2r}}\ldots\phis_{\mu_1}
\phi_{\nu_1}\ldots\phi_{\nu_{2s}}|0\rangle
&=
2^{\ell(\tilde{\mu})}
\delta_{r,s}
\prod_{i=1}^{2r}
\delta_{\mu_i,\nu_i}
\label{spart-orth}
=
2^{\ell(\tilde{\mu})}
\delta_{\tilde{\mu},\tilde{\nu}} 
\end{align}

\noindent where $\ell(\tilde{\mu})$ denotes the length of $\tilde{\mu}$. In particular, $\ell(\tilde{\mu})= 2r$ if $\mu_{2r} > 0$, and $\ell(\tilde{\mu}) = 2r-1$ if $\mu_{2r} =0$. The factors of 2 which appear in (\ref{spart-orth}) are due to the anticommutation relations (\ref{nfermi2}), and are characteristic of many calculations in the context of neutral fermions. In analogy with the earlier result (\ref{part-orth}), the bilinear form $\langle,\rangle$ induces orthogonality between the strict partition elements of $\mathcal{F}_{\phi}^{(0)}$ and $\mathcal{F}_{\phi}^{*(0)}$.  

\subsection{Lie algebra $B_{\infty}$}
\label{binfty}

Let $B_{\infty} \subset Cl_{\phi}^{(0)}$ be the vector space whose elements $Y \in B_{\infty}$ are of the form

\begin{align}
Y
=
\sum_{i,j \in \mathbb{Z}}
b_{i,j} :\phi_i \phi_j:
+
\kappa
\label{Binfty1}
\end{align}


\noindent where we have defined the normal ordering $:\phi_i \phi_j: = \phi_i \phi_j - \langle 0| \phi_i \phi_j |0\rangle$, and where the coefficients satisfy $b_{i,j}=0$ for $|i+j|$ sufficiently large, with $\kappa \in \mathbb{C}$.\footnote{In \cite{jm1}, $B_{\infty}$ was defined as the subset of $A_{\infty}$ which is invariant under a certain automorphism $\sigma_0$. The vector space (\ref{Binfty1}) was denoted $B'_{\infty}$ and shown to be isomorphic to $B_{\infty} \subset A_{\infty}$. Two different classes of $\tau$-function were obtained for the BKP hierarchy, the first corresponding with $B_{\infty}$, and the second with $B'_{\infty}$. In this thesis we only consider the second class of $\tau$-function, and have abbreviated $B'_{\infty} = B_{\infty}$ since there is no potential for confusion.} Since $:\phi_i \phi_j: = -:\phi_j \phi_i:$ and $:\phi_i \phi_i: = 0$ for all $i,j \in \mathbb{Z}$, we can equivalently write the elements of this vector space as

\begin{align}
Y
=
\sum_{i<j}
\Big(b_{i,j} - b_{j,i}\Big) :\phi_i \phi_j:
+
\kappa
\end{align}

\noindent where the sum is now over all integers $i,j$ such that $i < j$.

\begin{lemma}
{\rm 
The vector space $B_{\infty}$ becomes a Lie algebra when it is equipped with the commutator as Lie bracket.
}
\end{lemma}

\begin{proof}
{\rm
We prove the closure of $B_{\infty}$ under commutation. Let $Y$ be given by (\ref{Binfty1}) and define

\begin{align}
Y_1
=
\sum_{i,j \in \mathbb{Z}}
b^{(1)}_{i,j} :\phi_i \phi_j:
+
\kappa_1
\end{align}


\noindent where $b^{(1)}_{i,j} = 0$ for $|i+j|$ sufficiently large. By direct calculation, we obtain

\begin{align}
[Y,Y_1]
=
\sum_{i,j \in \mathbb{Z}}
b^{(2)}_{i,j} :\phi_i \phi_j:
+
\kappa_2
\label{Binfty2}
\end{align}


\noindent where for all $i,j \in \mathbb{Z}$ we have defined

\begin{align}
b^{(2)}_{i,j}
=
2
\sum_{k\in\mathbb{Z}}
(-)^k
\Big(b_{i,k}-b_{k,i}\Big)
\Big(b^{(1)}_{-k,j}-b^{(1)}_{j,-k}\Big)
\label{Binfty3}
\end{align}

\noindent and where

\begin{align}
\kappa_2
&=
2
\left(
\sum_{k<0}
-
\sum_{k>0}
\right)
(-)^k
\Big(b_{0,k}-b_{k,0}\Big)
\Big(b^{(1)}_{-k,0}-b^{(1)}_{0,-k}\Big)
\label{Binfty4}
\\
&+
4
\left(
\sum_{i>0,j<0}
-
\sum_{i<0,j>0}
\right)
(-)^{i+j}
\Big(
b_{i,-j}b^{(1)}_{-i,j}
+
b^{(1)}_{i,-j} b_{j,-i}
\Big)
\nonumber
%
\end{align}

\noindent Clearly the right hand side of equation (\ref{Binfty2}) is also an element of $B_{\infty}$, proving the closure of the set under commutation.
}
\end{proof}

%
%
%
%
%
%
%
%

\subsection{$B_{\infty}$ Heisenberg subalgebra}
\label{bheisen}

Adopting the notation of exercise 14.15 in chapter 14 of \cite{kac}, we define operators $\lambda_{m} \in B_{\infty}$ which are given by

\begin{align}
\lambda_m
=
\frac{1}{4} \sum_{j \in \mathbb{Z}}
(-)^j \phi_{(-j-m)} \phi_j
\label{bheisen1}
\end{align}


\noindent for all $m \in \tilde{\mathbb{Z}}$, where $\tilde{\mathbb{Z}}=\{\ldots,-3,-1,1,3,\ldots\}$ is the set of odd integers.\footnote{Throughout the remainder of this chapter, a tilde is also used to denote a set containing odd elements.} Together with the central element 1, the operators $\{\lambda_{m}\}_{m \in \tilde{\mathbb{Z}}}$ generate a Heisenberg subalgebra. The closure of this set follows from the commutation relation


\begin{align}
\left[ \lambda_{m}, \lambda_{n} \right] 
= 
\frac{m}{2}\delta_{m+n, 0} 
\label{bheisen2}
\end{align}

\noindent for all $m,n \in \tilde{\mathbb{Z}}$.\footnote{The relation (\ref{bheisen2}) can be proved by noticing that $\lambda_m,\lambda_n$ are obtained from $Y,Y_1$ by setting $b_{i,j}=\frac{1}{4}(-)^j\delta_{i+j+m,0},\ b^{(1)}_{i,j}=\frac{1}{4}(-)^j\delta_{i+j+n,0}$ and $\kappa = \kappa_1 = 0$. Substituting these values into (\ref{Binfty3}) and (\ref{Binfty4}), we obtain the desired commutator (\ref{bheisen2}) by virtue of (\ref{Binfty2}).} The Heisenberg generators (\ref{bheisen1}) also have a simple commutation relation with the neutral fermions (\ref{nfermi1}), given by

\begin{align}
[\lambda_{m},\phi_n]
=
\phi_{n-m}
\label{bheisen3}
\end{align}

\noindent for all $m \in \tilde{\mathbb{Z}}$ and $n \in \mathbb{Z}$.

\subsection{BKP evolution operators}
\label{BKPev}

We introduce the Hamiltonian

\begin{align}
\lambda\{\tilde{t}\}
=
\sum_{\nodd}
t_n \lambda_n
\label{BKPev1}
\end{align}

\noindent where $\{\tilde{t}\} = \{t_1,t_3,t_5,\ldots\}$ is an infinite set of free variables with odd indices, and define the generating function

\begin{align}
\Phi(k)
=
\sum_{i\in \mathbb{Z}}
\phi_i k^i
\label{BKPev2}
\end{align}

\noindent where $k$ is an indeterminate. Using these definitions and the equation (\ref{bheisen3}), we obtain the commutation relation

\begin{align}
[\lambda\{\tilde{t}\}, \Phi(k)]
=
\sum_{\nodd}
\sum_{i \in \mathbb{Z}}
[\lambda_n,\phi_i]
t_n k^i
=
\left(
\sum_{\nodd}
t_n k^n
\right)
\Phi(k)
\label{BKPev3}
\end{align}

\noindent which, in turn, implies that

\begin{align}
e^{\lambda\{\tilde{t}\}} \Phi(k)
=
\exp\left(\sum_{\nodd} t_n k^n \right)
\Phi(k) e^{\lambda\{\tilde{t}\}}
\label{BKPev4}
\end{align}

\noindent In analogy with the first part of the chapter, the operator $e^{\lambda\{\tilde{t}\}}$ is called a {\it BKP evolution operator.} It plays an essential role in constructing solutions of the BKP hierarchy of partial differential equations.

\subsection{Schur $Q$-polynomials}
\label{scharacter}

For all $m \geq 0$, the {\it one-row Schur $Q$-polynomial}
$\mathcal{Q}_m\{\tilde{t}\}$ in the infinite set of odd variables $\{\tilde{t}\}$ is defined by

\begin{align}
\mathcal{Q}_m\{\tilde{t}\}
=
{\rm Coeff}_{k^m}
\left[
\exp\left( \sum_{\nodd} t_n k^n \right)
\right]
=
\chi_{m}\{t_1,0,t_3,0,t_5,\ldots\}
\label{scharac-1}
\end{align}

\noindent where $\chi_m\{t_1,0,t_3,0,t_5,\ldots\}$ is the one-row Schur polynomial (\ref{KPbil-1}) with all even variables set to zero. We also define $\mathcal{Q}_m\{\tilde{t}\} = 0$ for all $m < 0$. From this, the {\it Schur $Q$-polynomial} \footnote{The Schur $Q$-polynomials were introduced in \cite{sch}, which studied their connection with the projective representations of the symmetric and alternating groups.} $\mathcal{Q}_{\tilde{\mu}}\{\tilde{t}\}$ associated to the strict partition $\tilde{\mu} = \{\mu_1>\cdots>\mu_{2r} \geq 0\}$ is given by

\begin{align}
\mathcal{Q}_{\tilde{\mu}}\{\tilde{t}\}
=
{\rm Pf}\left( 
\mathcal{Q}_{\mu_i}\{\tilde{t}\} 
\mathcal{Q}_{\mu_j}\{\tilde{t}\}
+
2
\sum_{k = 1}^{\mu_j}
(-)^{k}
\mathcal{Q}_{(\mu_i+k)}\{\tilde{t}\}
\mathcal{Q}_{(\mu_j-k)}\{\tilde{t}\}
\right)_{1 \leq i < j \leq 2r}
\label{BKPpoly9}
\end{align}

\noindent where each $\mathcal{Q}_m\{\tilde{t}\}$ represents a one-row Schur $Q$-polynomial. The following result maps the strict partition elements of $\mathcal{F}_{\phi}^{(0)}$ to their corresponding Schur $Q$-polynomials.

\begin{lemma}
{\rm Let $|\tilde{\mu}) = |\mu_1,\ldots,\mu_{2r}) = \phi_{\mu_1}\ldots \phi_{\mu_{2r}} |0\rangle$ be an element of the strict partition basis of $\mathcal{F}_{\phi}^{(0)}$. We claim that 

\begin{align}
\mathcal{Q}_{\tilde{\mu}}\{\tilde{t}\}
=
(\emptyset| e^{\lambda\{\tilde{t}\}} |\tilde{\mu})
\label{nfermi-qschur}
\end{align}

\noindent This result may also be found in, for example, \cite{orl}.

}
\end{lemma}

\begin{proof}
{\rm
Using the definition of the one-row Schur $Q$-polynomials (\ref{scharac-1}) in the commutation relation (\ref{BKPev4}) and then extracting the coefficient of $k^m$ from the resulting equation, we obtain

\begin{align}
e^{\lambda\{\tilde{t}\}}
\phi_m
=
\left(
\sum_{i=0}^{\infty}
\phi_{(m-i)} \mathcal{Q}_i\{\tilde{t}\}
\right)
e^{\lambda\{\tilde{t}\}}
\label{BKPpoly5}
\end{align}

\noindent By definition, we have

\begin{align}
(\emptyset| e^{\lambda\{\tilde{t}\}} |\tilde{\mu})
=
\langle 0| e^{\lambda\{\tilde{t}\}} \phi_{\mu_1} \ldots \phi_{\mu_{2r}} |0\rangle
\label{BKPpoly4}
\end{align}

\noindent and using the commutation relation (\ref{BKPpoly5}) we move the evolution operator $e^{\lambda\{\tilde{t}\}}$ in (\ref{BKPpoly4}) towards the right, obtaining

\begin{align}
(\emptyset| e^{\lambda\{\tilde{t}\}} |\tilde{\mu})
=
\sum_{i_1,\ldots,i_{2r} = 0}^{\infty}
\Big\langle 
\phi_{(\mu_1-i_1)}  
\ldots 
\phi_{(\mu_{2r}-i_{2r})}  
\Big\rangle
\mathcal{Q}_{i_1}\{\tilde{t}\}
\ldots
\mathcal{Q}_{i_{2r}}\{\tilde{t}\}
\label{BKPpoly6}
\end{align}

\noindent where we have used the fact that
$e^{\lambda\{\tilde{t}\}} |0\rangle = |0\rangle$. Applying lemma 13 to the previous vacuum expectation value, we find 

\begin{align}
(\emptyset| e^{\lambda\{\tilde{t}\}} |\tilde{\mu})
=
\sum_{i_1,\ldots,i_{2r}=0}^{\infty}
{\rm Pf} \Big( \langle 
\phi_{(\mu_p-i_p)} \phi_{(\mu_q-i_q)}
\rangle \Big)_{1 \leq p < q \leq 2r}
\mathcal{Q}_{i_1}\{\tilde{t}\} 
\ldots 
\mathcal{Q}_{i_{2r}}\{\tilde{t}\}
\label{BKPpoly7}
\end{align}

\noindent Collecting the one-row Schur $Q$-polynomials inside the Pfaffian, we obtain

\begin{align}
(\emptyset| e^{\lambda\{\tilde{t}\}} |\tilde{\mu})
=
{\rm Pf} \left(
\sum_{i_p=0}^{\infty}
\sum_{i_q=0}^{\infty}
\langle
\phi_{(\mu_p-i_p)} \phi_{(\mu_q-i_q)}
\rangle
\mathcal{Q}_{i_p}\{\tilde{t}\}
\mathcal{Q}_{i_q}\{\tilde{t}\}
\right)_{1 \leq p < q \leq 2r}
\label{BKPpoly8}
\end{align}

\noindent Using the annihilation properties (\ref{repclB1}) and defining $\mathcal{Q}_m\{\tilde{t}\} = 0$ for all $m < 0$, equation (\ref{BKPpoly8}) becomes 

\begin{align}
(\emptyset| e^{\lambda\{\tilde{t}\}} |\tilde{\mu})
&=
{\rm Pf} \left(
\sum_{i_p\in \mathbb{Z}}
\sum_{i_q=0}^{\mu_q}
\langle
\phi_{(\mu_p-i_p)} \phi_{(\mu_q-i_q)}
\rangle
\mathcal{Q}_{i_p}\{\tilde{t}\}
\mathcal{Q}_{i_q}\{\tilde{t}\}
\right)_{1 \leq p < q \leq 2r}
\label{scharacter2}
\\
&
=
{\rm Pf} \left( 
\mathcal{Q}_{\mu_p}\{\tilde{t}\} 
\mathcal{Q}_{\mu_q}\{\tilde{t}\}
+
\sum_{\substack{i_p \in \mathbb{Z} \\ 0 \leq i_q < \mu_q }}
\langle
\phi_{(\mu_p-i_p)} \phi_{(\mu_q-i_q)}
\rangle
\mathcal{Q}_{i_p}\{\tilde{t}\}
\mathcal{Q}_{i_q}\{\tilde{t}\}
\right)_{1 \leq p < q \leq 2r}
\nonumber
\end{align}

\noindent We evaluate the vacuum expectation value inside the previous Pfaffian, using the anticommutation relations (\ref{nfermi2}) and annihilation properties (\ref{repclB1}) to obtain

\begin{align}
\langle
\phi_{(\mu_p-i_p)}
\phi_{(\mu_q-i_q)}
\rangle
=
2
(-)^{\mu_p-i_p}
\delta_{\mu_p-i_p,i_q-\mu_q}
\end{align}

\noindent for all $i_q < \mu_q$. Substituting this result into (\ref{scharacter2}) and using the Kronecker delta to truncate the sum over $i_p$, we find

\begin{align}
&
(\emptyset| e^{\lambda\{\tilde{t}\}} |\tilde{\mu})
\label{BKPpoly10}
=
{\rm Pf}\left( 
\mathcal{Q}_{\mu_p}\{\tilde{t}\} 
\mathcal{Q}_{\mu_q}\{\tilde{t}\}
+
2
\sum_{i_q=0}^{\mu_q-1}
(-)^{\mu_q-i_q}
\mathcal{Q}_{(\mu_p+\mu_q-i_q)}\{\tilde{t}\}
\mathcal{Q}_{i_q}\{\tilde{t}\}
\right)_{1 \leq p < q \leq 2r}
\end{align}

\noindent The Schur $Q$-polynomial (\ref{BKPpoly9}) is recovered by making the change of indices $k=\mu_q-i_q$ in each entry of the Pfaffian (\ref{BKPpoly10}).
}
\end{proof}



\medskip
\noindent
{\bf Corollary.} Let $\{m\} =\{m_1 > \cdots > m_{2r+1} \geq 0\}$ be an ordered set of integers with odd cardinality. Given that $\langle \phi_0 \phi_{n_1}\ldots\phi_{n_{2r+1}} \rangle=\langle \phi_{n_1}\ldots\phi_{n_{2r+1}} \phi_0 \rangle$ for all sets of integers $\{n_1,\ldots,n_{2r+1}\}$, we find

\begin{align}
\langle 1| e^{\lambda\{\tilde{t}\}}
\phi_{m_1}\ldots \phi_{m_{2r+1}} |0\rangle
=
\langle 0|e^{\lambda\{\tilde{t}\}}
\phi_{m_1}\ldots \phi_{m_{2r+1}} \phi_0|0\rangle
\label{cor1}
\end{align}

\noindent where we have made use of the fact that $\langle 1| = \langle 0|\phi_0$. Using the result (\ref{nfermi-qschur}) from the previous lemma in the right hand side of (\ref{cor1}), we obtain

\begin{align}
\langle 1| e^{\lambda\{\tilde{t}\}}
\phi_{m_1}\ldots \phi_{m_{2r+1}} |0\rangle
=
\mathcal{Q}_{\tilde{\mu}}\{\tilde{t}\}
\label{cor2}
\end{align}

\noindent where we have defined the strict partition 

\begin{align}
\tilde{\mu}
=
\left\{
\begin{array}{ll}
\{m_1,\ldots,m_{2r+1},0\},\quad 
& 
m_{2r+1} > 0
\\
\{m_1,\ldots,m_{2r}\},\quad 
& 
m_{2r+1} = 0
\end{array}
\right.
\label{corcor}
\end{align}

\noindent Equation (\ref{cor2}) will prove useful in later calculations.
%
%










\subsection{Schur $Q$-functions}
\label{qschur}

Following section 2 in chapter III of \cite{mac}, the function $q_m\{x\}$ in the infinite set of variables $\{x\}$ is defined as

\begin{align}
q_m\{x\}
=
{\rm Coeff}_{k^m}\left[
\prod_{i=1}^{\infty}
\frac{1+x_i k}{1-x_i k}
\right]
\label{q-com-sym}
\end{align}

\noindent From this definition, the {\it Schur $Q$-function} $Q_{\tilde{\mu}}\{x\}$ associated to the strict partition  $\tilde{\mu} = \{\mu_1> \cdots > \mu_{2r} \geq 0\}$ is given by

\begin{align}
Q_{\tilde{\mu}}\{x\}
=
{\rm Pf}\left(
q_{\mu_i}\{x\} q_{\mu_j}\{x\}
+
2
\sum_{k=1}^{\mu_j}
(-)^{k}
q_{(\mu_i+k)}\{x\} q_{(\mu_j-k)}\{x\}
\right)_{1 \leq i < j \leq 2r}
\label{schurq}
\end{align}

\begin{lemma}
{\rm 
The Schur $Q$-polynomial $\mathcal{Q}_{\tilde{\mu}}\{\tilde{t}\}$ and the Schur $Q$-function $Q_{\tilde{\mu}}\{x\}$ are equal under the change of variables $t_n = \frac{2}{n} \sum_{i=1}^{\infty} x_i^n$ for all $n \in \tilde{\mathbb{N}}$.
}
\end{lemma} 

\begin{proof}
{\rm 
Fixing $t_n = \frac{2}{n} \sum_{i=1}^{\infty} x_i^n$ for all $n \in \tilde{\mathbb{N}}$, we obtain

{\footnotesize
\begin{align}
{\rm Coeff}_{k^m}
\left[
\exp\left(
\sum_{n \in \tilde{\mathbb{N}}}
t_n k^n
\right)
\right]
&=
{\rm Coeff}_{k^m}
\left[
\exp\left(
\sum_{i=1}^{\infty}
\sum_{n \in \tilde{\mathbb{N}}}
\frac{2}{n} (x_i k)^n
\right)
\right]
=
{\rm Coeff}_{k^m}
\left[
\prod_{i=1}^{\infty}
\frac{1+x_i k}{1-x_i k}
\right]
\end{align}
}

\noindent implying that $\mathcal{Q}_m\{\tilde{t}\} = q_m \{x\}$ for all $m \geq 0$. Comparing the definition (\ref{BKPpoly9}) with (\ref{schurq}), we immediately see that $\mathcal{Q}_{\tilde{\mu}}\{\tilde{t}\} = Q_{\tilde{\mu}}\{x\}$ under the prescribed change of variables.
}
\end{proof}

From section 8 in chapter III of \cite{mac}, the Schur $Q$-functions $Q_{\tilde{\mu}}\{x\}$ comprise a basis for symmetric functions in $\{x\}$ which are independent of the even power sums $\frac{1}{2n}\sum_{i=1}^{\infty}x_i^{2n}$. This fact, together with the equality of $\mathcal{Q}_{\tilde{\mu}}\{\tilde{t}\}$ and $Q_{\tilde{\mu}}\{x\}$ under the previous change of variables, proves that the Schur $Q$-polynomials $\mathcal{Q}_{\tilde{\mu}}\{\tilde{t}\}$ are a basis for the set of all polynomials in the odd variables $\{\tilde{t}\}$.

In chapter 3 we will consider Schur $Q$-functions $Q_{\tilde{\mu}}\{x\}$ in finitely many variables $\{x\}=\{x_1,\ldots,x_N\}$. These are obtained from the formulae (\ref{q-com-sym}), (\ref{schurq}) by setting $x_n = 0$ for all $n >N$. In these cases, $Q_{\tilde{\mu}}\{x\} = 0$ if $\ell(\tilde{\mu}) > N$.

\section{BKP hierarchy}
\label{n-BKP}

\subsection{BKP hierarchy in bilinear form}
\label{BKPbil}

The {\it BKP hierarchy} is an infinite set of partial differential equations in the independent variables $\{\tilde{t}\}=\{t_1,t_3,t_5,\ldots\}$. As in the case of the KP hierarchy, the BKP hierarchy derives from a single integral equation, called the {\it BKP bilinear identity}. A function $\tau\{\tilde{t}\}$ which satisfies every differential equation in the hierarchy, or equivalently, satisfies the BKP bilinear identity, is called a {\it BKP $\tau$-function}.

Define the shifted sets of variables

\begin{align}
\{\tilde{t} \pm 2\tilde{\epsilon}_k \}
=
\Big\{
t_1 \pm 2k^{-1},
t_3 \pm \frac{2}{3} k^{-3},
t_5 \pm \frac{2}{5} k^{-5},
\ldots
\Big\}
\end{align}

\noindent where $k$ is a free parameter. 
The BKP bilinear identity is the equation

\begin{align}
\oint
\exp\left(\sum_{n \in \tilde{\mathbb{N}} }(t_n-s_n)k^n\right)
\tau\{\tilde{t} - 2\tilde{\epsilon}_k\}
\tau\{\tilde{s} + 2\tilde{\epsilon}_k\}
\frac{dk}{2\pi i k}
=
\tau\{\tilde{t}\}
\tau\{\tilde{s}\}
\label{BKPbil1}
\end{align}

\noindent where $\{\tilde{t}\}=\{t_1,t_3,t_5,\ldots\}$ and $\{\tilde{s}\} = \{s_1,s_3,s_5,\ldots\}$ are two infinite sets of variables with odd subscripts, and the integration in the $k$-plane is taken around a small contour at $k= \infty$.

As in the case of the KP hierarchy, we will assume $\tau\{\tilde{t}\}$ is a polynomial in its variables. In this special case, $\tau\{\tilde{t}-2\tilde{\epsilon}_k\}$ and $\tau\{\tilde{s}+2\tilde{\epsilon}_k\}$ have singularities in $k$ only at $k=0$. Therefore, for polynomial $\tau$-functions, the BKP bilinear identity becomes

\begin{align}
{\rm Coeff}_{1}
\left[
\exp\left(\sum_{n \in \tilde{\mathbb{N}} } (t_n-s_n)k^n \right)
\tau\{\tilde{t}-2\tilde{\epsilon}_k\}
\tau\{\tilde{s}+2\tilde{\epsilon}_k\}
\right]
=
\tau\{\tilde{t}\}
\tau\{\tilde{s}\}
\label{BKPbil-1}
\end{align}

\noindent where ${\rm Coeff}_1[f(k)]$ denotes the coefficient of $k^0$ in the Laurent series of $f(k)$. We expose the infinitely many differential equations which underly the equation (\ref{BKPbil-1}) by making the substitutions $\{\tilde{t}\} \rightarrow \{\tilde{t}-\tilde{s}\}$ and $\{\tilde{s}\} \rightarrow \{\tilde{t}+\tilde{s}\}$, giving 

\begin{align}
&
{\rm Coeff}_{1}
\left[
\exp{\left( -2 \sum_{n \in \tilde{\mathbb{N}} } s_n k^n \right) }
\tau\{\tilde{t}-\tilde{s}-2\tilde{\epsilon}_k\}
\tau\{\tilde{t}+\tilde{s}+2\tilde{\epsilon}_k\}
\right]
\label{BKPbil2}
=
\tau\{\tilde{t} - \tilde{s}\}
\tau\{\tilde{t} + \tilde{s}\}
\end{align}

\noindent Using the definition (\ref{scharac-1}), the exponential term in (\ref{BKPbil2}) can be replaced with a sum over one-row Schur $Q$-polynomials, producing the equation

\begin{align}
{\rm Coeff}_{1}
\left[
\sum_{m=0}^{\infty}
\mathcal{Q}_m\{-2\tilde{s}\}
k^m
\tau\{\tilde{t}-\tilde{s}-2\tilde{\epsilon}_k\}
\tau\{\tilde{t}+\tilde{s}+2\tilde{\epsilon}_k\}
\right]
=
\tau\{\tilde{t} - \tilde{s}\}
\tau\{\tilde{t} + \tilde{s}\}
\end{align}

\noindent or equivalently,

\begin{align}
\sum_{m=0}^{\infty}
\mathcal{Q}_m \{-2\tilde{s}\}
{\rm Coeff}_{k^{-m}}
\Big[
\tau\{\tilde{t}-\tilde{s}-2\tilde{\epsilon}_k\}
\tau\{\tilde{t}+\tilde{s}+2\tilde{\epsilon}_k\}
\Big]
=
\tau\{\tilde{t} - \tilde{s}\}
\tau\{\tilde{t} + \tilde{s}\}
\label{BKPbil-2}
\end{align}

\noindent In order to progress further we need the following result, which is a simple modification of lemma 7.

\begin{lemma}
{\rm
\begin{align}
\tau\{\tilde{t}-\tilde{s}-2\tilde{\epsilon}_k\}
\tau\{\tilde{t}+\tilde{s}+2\tilde{\epsilon}_k\}
=
\exp\left(
\sum_{n \in \tilde{\mathbb{N}} }
(s_n+\frac{2}{n}k^{-n})
D_n
\right)
\tau\{\tilde{t}\} \cdot \tau\{\tilde{t}\}
\label{BKPbil4}
\end{align}
}
\end{lemma}

\begin{proof}
{\rm
We refer the reader to the proof of lemma 7.
}
\end{proof}

Returning to equation (\ref{BKPbil-2}), we employ the result (\ref{BKPbil4}) to obtain  

{\small
\begin{align}
&
\sum_{m=0}^{\infty}
\mathcal{Q}_m \{-2\tilde{s}\}
{\rm Coeff}_{k^{-m}}
\left[
\exp\left(
\sum_{n \in \tilde{\mathbb{N}} }
(s_n+\frac{2}{n}k^{-n})
D_n
\right)
\right]
\tau\{\tilde{t}\} \cdot \tau\{\tilde{t}\}
=
\tau\{\tilde{t} - \tilde{s}\}
\tau\{\tilde{t} + \tilde{s}\}
\label{BKPbil-7}
\end{align}
}

\noindent Recalling the definition (\ref{scharac-1}) of the one-row Schur $Q$-polynomials once again, equation (\ref{BKPbil-7}) becomes


\begin{align}
\sum_{m=0}^{\infty}
\mathcal{Q}_m\{-2\tilde{s}\}
\mathcal{Q}_m\{2\tilde{\mathbb{D}}\}
\exp\left(
\sum_{n \in \tilde{\mathbb{N}} }
s_n D_n
\right)
\tau\{\tilde{t}\} \cdot \tau\{\tilde{t}\}
=
\tau\{\tilde{t} - \tilde{s}\}
\tau\{\tilde{t} + \tilde{s}\}
\label{BKPbil-6}
\end{align}

\noindent where we have defined the set of operators $\{\tilde{\mathbb{D}}\}=\{D_1,\frac{1}{3}D_3,\frac{1}{5}D_5,\ldots\}$. We notice that the right hand side of (\ref{BKPbil-6}) can be expressed as

\begin{align}
\tau\{\tilde{t}-\tilde{s}\} \tau\{\tilde{t}+\tilde{s}\}
=
\exp\left(
\sum_{n \in \tilde{\mathbb{N}} }
s_n D_n
\right)
\tau\{\tilde{t}\} \cdot \tau\{\tilde{t}\}
\end{align}

\noindent which cancels with the $m=0$ term of the left hand side summation, yielding

\begin{align}
\sum_{m=1}^{\infty}
\mathcal{Q}_m\{-2\tilde{s}\}
\mathcal{Q}_m\{2\tilde{\mathbb{D}}\}
\exp\left(
\sum_{n \in \tilde{\mathbb{N}} }
s_n D_n
\right)
\tau\{\tilde{t}\} \cdot \tau\{\tilde{t}\}
=
0
\label{BKPbil-4}
\end{align}

\noindent For all odd $\{1\leq m_1 < \cdots < m_l\}$ and $\{n_1,\ldots,n_l \geq 1\}$ the coefficient of the monomial $s_{m_1}^{n_1}\ldots s_{m_l}^{n_l}$ on the left hand side of (\ref{BKPbil-4}) must vanish, giving rise to infinitely many consistency equations, which are the differential equations of the BKP hierarchy. 

\begin{example}
{\rm Up to an irrelevant factor the coefficient of $s_3^2$ on the left hand side of (\ref{BKPbil-4}) is equal to 
$(D_1^6-5D_1^3 D_3 -5D_3^2+9D_1 D_5)\tau\{\tilde{t}\}\cdot\tau\{\tilde{t}\}$, implying that

\begin{align}
(
D_1^6-5D_1^3 D_3-5D_3^2+9D_1 D_5)
\tau\{\tilde{t}\}\cdot\tau\{\tilde{t}\}
=
0
\label{BKPbil7}
\end{align}

\noindent which is the BKP equation in bilinear form. Higher equations in the hierarchy are obtained from the coefficients of different monomials.
}
\end{example}

\subsection{Neutral fermion bilinear identity}

We now focus on constructing solutions of the BKP bilinear identity (\ref{BKPbil1}) using the calculus of the neutral fermions $\{\phi_m\}_{m\in \mathbb{Z}}$. The following result is the neutral fermion analogue of theorem 1.

\begin{theorem}
{\rm 
Let $g_{\phi}$ be a finite element of $Cl_{\phi}^{(0)}$ and define

\begin{align}
\tau\{\tilde{t}\}
&=
\langle 0|e^{\lambda\{\tilde{t}\}} g_{\phi} |0\rangle
=
\Big\langle e^{\lambda\{\tilde{t}\}} g_{\phi} \Big\rangle
\label{BKPexpI-1}
=
\Big\langle \phi_0 e^{\lambda\{\tilde{t}\}} g_{\phi} \phi_0 \Big\rangle
\end{align}

\noindent where the equality between these two expectation values follows from the fact that $\langle \phi_0 \phi_{m_1} \ldots \phi_{m_{2r}} \phi_0 \rangle = \langle \phi_{m_1} \ldots \phi_{m_{2r}} \rangle$ for all sets of integers $\{m_1,\ldots,m_{2r}\}$. The polynomial $\tau\{\tilde{t}\}$ satisfies the BKP bilinear identity (\ref{BKPbil1}) if and only if $g_{\phi}$ satisfies the {\it neutral fermion bilinear identity} (NFBI)

\begin{align}
\sum_{i\in \mathbb{Z}}
\phi_i g_{\phi} |0\rangle 
\otimes
\phis_{i} g_{\phi} |0\rangle
=
g_{\phi}  |1\rangle
\otimes
g_{\phi}  |1\rangle
\label{BKPexpI7}
\end{align}
}
\end{theorem}

\begin{proof}
{\rm 
As we did in the proof of theorem 1, we split this proof into three steps. In the first two steps we prove that if (\ref{BKPexpI7}) holds, then $\tau\{\t{t}\}$ as given by (\ref{BKPexpI-1}) satisfies the BKP bilinear identity (\ref{BKPbil1}). In the third step we prove the converse statement.

\medskip
\noindent
{\bf Step 1.} Acting upon the left hand side of $(\ref{BKPexpI7})$ with the tensored dual states  
$\langle 1|e^{\lambda\{\tilde{t}\}} 
\otimes
\langle 1|e^{\lambda\{\tilde{s}\}}
$
we have the result

\begin{align}
\sum_{i\in \mathbb{Z}}
\langle 1|
e^{\lambda\{\tilde{t}\}}
\phi_i g_{\phi}
|0\rangle
\langle 1|
e^{\lambda\{\tilde{s}\}}
\phis_i g_{\phi}
|0\rangle
=
\tau\{\tilde{t}\}
\tau\{\tilde{s}\}
\label{BKPexpI8}
\end{align}

\noindent We convert the sum on the left hand side of (\ref{BKPexpI8}) into a contour integral, using the generating function (\ref{BKPev2}) to write

\begin{align}
\oint
\langle 1|
e^{\lambda\{\tilde{t}\}}
\Phi(k) g_{\phi}
|0\rangle
\langle 1|
e^{\lambda\{\tilde{s}\}}
\Phi(-k) g_{\phi}
|0\rangle
\frac{dk}{2\pi i k}
=
\tau\{\tilde{t}\}
\tau\{\tilde{s}\}
\label{BKPwhat}
\end{align}

\noindent where the contour of integration surrounds the pole at $k=0$. By virtue of the commutation relation (\ref{BKPev4}), it is possible to switch the order of $e^{\lambda\{\tilde{t}\}}$ and $\Phi(k)$, and likewise $e^{\lambda\{\tilde{s}\}}$ and $\Phi(-k)$ in (\ref{BKPwhat}), giving 

\begin{align}
&
\oint
e^{
\sum_{\nodd} (t_n-s_n)k^n
}
\langle 1|
\Phi(k) 
e^{\lambda\{\tilde{t}\}} g_{\phi}
|0\rangle
\langle 1|
\Phi(-k) 
e^{\lambda\{\tilde{s}\}} g_{\phi}
|0\rangle
\frac{dk}{2\pi i k}
=
\tau\{\tilde{t}\}
\tau\{\tilde{s}\}
\label{BKPexpI13}
\end{align}

\medskip
\noindent
{\bf Step 2. (Lemma 18.)} We propose the identity

\begin{align}
\langle 1|\Phi(k)
=
\langle 0|
\exp\Big(-\sum_{\nodd}\frac{2}{n}k^{-n} \lambda_n\Big)
\label{BKPexpI14}
\end{align}

\begin{proof}
{\rm Using the fact that $\langle 1|= \langle0| \phi_0$ and the definition of the generating function (\ref{BKPev2}), we obtain

\begin{align}
\langle 1|\Phi(k) 
= 
\langle 0|\phi_0 \sum_{i \in \mathbb{Z}} \phi_i k^i 
= 
\langle 0|
+
\sum_{i=1}^{\infty} k^{-i} 
\langle 0| \phi_0 \phi_{-i}  
\label{BKPexpI15}
\end{align}

\noindent where we have used the identity $\phi_0^2=1$ and the annihilation properties (\ref{repclB1}) to truncate the sum. Now consider the expression (\ref{scharac-1}) for the one-row Schur $Q$-polynomial. When the variables $\{\t{t}\}$ are set to $t_n =-2k^{-n}/n$ for all $n \in \t{\mathbb{N}}$, this expression simplifies greatly.  We obtain

\begin{align}
\mathcal{Q}_{m}\{\t{t}\}
\Big|_{t_n = -2k^{-n}/n}
=
\left\{
\begin{array}{ll}
1, & m=0
\\
2(-k)^{-m}, & m \geq 1
\end{array}
\right.
\end{align}

\noindent and substitute this formula into the expression (\ref{BKPpoly9}) for the Schur $Q$-polynomial associated to $\tilde{\mu}=\{\mu_1>\cdots>\mu_{2r}\geq 0\}$, giving

\begin{align}
\mathcal{Q}_{\tilde{\mu}}\{\t{t}\}
\Big|_{t_n=-2k^{-n}/n}
=
2(-k)^{-\mu_1}\delta_{r,1}\delta_{\mu_2,0}
\end{align}

\noindent Using the result of lemma 15, this equation becomes

\begin{align}
\langle 0|
\exp\Big(
-\sum_{\nodd} \frac{2}{n} k^{-n} \lambda_n
\Big)
\phi_{\mu_1}\ldots\phi_{\mu_{2r}}
|0\rangle
=
2
(-k)^{-\mu_1} 
\delta_{r,1}
\delta_{\mu_2,0}
\end{align}

\noindent Finally, due to the orthogonality of strict partition vectors (\ref{spart-orth}), we obtain 

\begin{align}
\langle 0|
\exp\Big(-\sum_{\nodd}\frac{2}{n}k^{-n} \lambda_n\Big)
=
\langle 0|
+
\sum_{i=1}^{\infty} (-k)^{-i} 
\langle 0| \phis_0 \phis_{i}
=
\langle 0|
+
\sum_{i=1}^{\infty} k^{-i}
\langle 0|\phi_0 \phi_{-i}
\end{align}

\noindent Comparing this equation with (\ref{BKPexpI15}), we complete the proof of (\ref{BKPexpI14}).
}
\end{proof} 

Applying (\ref{BKPexpI14}) to equation (\ref{BKPexpI13}), we obtain

\begin{align}
\oint
e^{\sum_{\nodd} (t_n-s_n) k^n}
\langle 0|
e^{\lambda\{\tilde{t}-2\tilde{\epsilon}_k\}}
g_{\phi}
|0\rangle
\langle 0|
e^{\lambda\{\tilde{s}+2\tilde{\epsilon}_k\}}
g_{\phi}
|0\rangle
\frac{dk}{2\pi i k}
=
\tau\{\tilde{t}\}
\tau\{\tilde{s}\}
\label{BKPexpI12}
\end{align}

\noindent Equation (\ref{BKPexpI12}) proves that if (\ref{BKPexpI7}) holds, functions given by (\ref{BKPexpI-1}) satisfy the BKP bilinear identity (\ref{BKPbil1}).

\medskip
\noindent
{\bf Step 3.} For any finite $g_{\phi} \in Cl_{\phi}^{(0)}$, there exist coefficients $\kappa_{\{m\},\{n\}}$ such that

\begin{align}
&
\sum_{i\in\mathbb{Z}}
\phi_i g_{\phi} |0\rangle
\otimes
\phis_i g_{\phi} |0\rangle
-
g_{\phi} |1\rangle
\otimes
g_{\phi} |1\rangle
\label{godknows}
=
\sum_{\{m\},\{n\}}
\kappa_{\{m\},\{n\}}
\phi_{\{m\}} |0\rangle
\otimes
\phi_{\{n\}} |0\rangle
\end{align}

\noindent where the sum is over all sets of integers $\{m\} = \{m_1 > \cdots > m_{2r+1} \geq 0\}$ and $\{n\} = \{n_1 > \cdots > n_{2s+1} \geq 0\}$, whose cardinalities can assume all odd values $(2r+1),(2s+1) \geq 1$. Acting upon both sides of this equation with the tensored dual states $\langle 1|e^{\lambda\{\t{t}\}} \otimes \langle 1|e^{\lambda\{\t{s}\}}$ we find

{\small
\begin{align}
\oint
e^{\sum_{\nodd} (t_n-s_n) k^n}
\langle 0|
e^{\lambda\{\tilde{t}-2\tilde{\epsilon}_k\}}
g_{\phi}
|0\rangle
\langle 0|
e^{\lambda\{\tilde{s}+2\tilde{\epsilon}_k\}}
g_{\phi}
&
|0\rangle
\frac{dk}{2\pi i k}
-
\langle 1|e^{\lambda\{\tilde{t}\}} g_{\phi} |1\rangle
\langle 1|e^{\lambda\{\tilde{s}\}} g_{\phi} |1\rangle
\nonumber
\\
&
=
\sum_{\{m\},\{n\}}
\kappa_{\{m\},\{n\}}
\mathcal{Q}_{\tilde{\mu}}\{\t{t}\}
\mathcal{Q}_{\tilde{\nu}}\{\t{s}\}
\label{linind}
\end{align}
}

\noindent where the left hand side of (\ref{linind}) has already been derived in steps 1 and 2, and the right hand side follows from (\ref{cor2}) with $\tilde{\mu},\tilde{\nu}$ defined by (\ref{corcor}). Assuming that $\tau\{\tilde{t}\}$ as given by (\ref{BKPexpI-1}) satisfies the BKP bilinear identity, we thus obtain

\begin{align}
\sum_{\{m\},\{n\}}
\kappa_{\{m\},\{n\}}
\mathcal{Q}_{\tilde{\mu}}\{\t{t}\}
\mathcal{Q}_{\tilde{\nu}}\{\t{s}\}
=
0
\end{align}

\noindent which can only be true if all of the coefficients $\kappa_{\{m\},\{n\}} = 0$, since the Schur $Q$-polynomials are linearly independent. Substituting this trivial value for the coefficients into (\ref{godknows}), we recover the NFBI (\ref{BKPexpI7}). This completes the proof of the converse statement.
}
\end{proof}

\section{Solutions of the NFBI}
\label{n-sol}

\subsection{Orbit of $O_{\infty}$}
\label{BKPexpI}

\begin{theorem}
{\rm Suppose $g_{\phi}$ is a finite element of $Cl_{\phi}^{(0)}$. Then $g_{\phi}$ solves the neutral fermionic bilinear identity (\ref{BKPexpI7}) if and only if

\begin{align}
g_{\phi}|0\rangle
=
e^{Y_1}\ldots e^{Y_l}|0\rangle 
\label{BKPnfermI-1}
\end{align}

\noindent for some $\{Y_1,\ldots,Y_l\} \in B_{\infty}$. In other words, the solution space of (\ref{BKPexpI7}) is generated by the orbit of the Lie group

\begin{align}
O_{\infty}
=
\Big\{
e^{Y_1}\ldots e^{Y_l}\ 
\Big|\ 
Y_i \in B_{\infty}\ {\rm for\ all}\ 1\leq i \leq l
\Big\}
\end{align}

}
\end{theorem}

\medskip
\noindent
{\it Proof.} As we did in the proof of theorem 2, we split this proof into two steps. In the first step we prove the forward statement, in the second step we prove its converse.

\medskip
\noindent
{\bf Step 1. (Lemma 19.)} Let $|u\rangle$ and $|v\rangle$ be arbitrary state vectors in $\mathcal{F}_{\phi}$, and let $g=e^{Y_1} \ldots e^{Y_l}$ with each $Y_i \in B_{\infty}$. We have

\begin{align}
\sum_{i\in \mathbb{Z}}
\phi_i g_{\phi} |u\rangle 
\otimes
\phis_i g_{\phi} |v\rangle
=
\sum_{i \in \mathbb{Z}}
g_{\phi}\phi_i |u\rangle
\otimes
g_{\phi} \phis_i |v\rangle
\label{BKPexpI2}
\end{align}

\begin{proof}
{\it (Lemma 19.)}
{\rm For $m \geq 0$ and arbitrary $Y \in B_{\infty}$, let $\mathcal{P}_m$ denote the proposition

\begin{align}
(\phi_i \otimes \phis_{i})
\sum_{n=0}^{m}
\binom{m}{n}
Y^n \otimes  Y^{m-n}
=
\sum_{n=0}^{m}
\binom{m}{n}
Y^n  \otimes Y^{m-n} 
(\phi_i \otimes \phis_{i})
\label{BKPexpI3}
\end{align}

\noindent where summation over all integers $i$ is implied. The proposition $\mathcal{P}_0$ is trivial. Furthermore, letting $Y \in B_{\infty}$ be given by (\ref{Binfty1}), by direct calculation we obtain the commutation relations  

\begin{align}
[\phi_i,Y] 
= 
2(-)^i \sum_{j \in \mathbb{Z}} 
(b_{-i,j}-b_{j,-i}) \phi_j,
\quad
[\phis_i,Y] 
=
2 
\sum_{j \in \mathbb{Z}}
(-)^j 
(b_{i,-j}-b_{-j,i}) \phis_j
\end{align}

\noindent Using these commutators in the left hand side of $\mathcal{P}_1$, we obtain

\begin{align} 
(\phi_i \otimes \phis_{i})
(1\otimes Y + Y\otimes 1)
&=
(1 \otimes Y + Y \otimes 1)
(\phi_i \otimes \phis_{i})
\\
&+
2(-)^j(b_{i,-j}-b_{-j,i})
\phi_i \otimes \phis_j 
+
2(-)^i(b_{-i,j}-b_{j,-i})
\phi_j \otimes \phis_{i}
\nonumber
\\
&
=
(1 \otimes Y + Y \otimes 1)
(\phi_i \otimes \phis_{i})
\nonumber
\end{align}

\noindent where summation over all integers $i,j$ is implied. This proves $\mathcal{P}_1$ is true. Using the inductive procedure from the proof of lemma 9, we find that $\mathcal{P}_m$ is true for all $m \geq 0$.\footnote{Since the inductive basis $\mathcal{P}_1$ holds, the proof of (\ref{BKPexpI3}) becomes immediate from the proof of (\ref{KPexpI3}) by substituting $\psi_i \rightarrow \phi_i$, $\psis_i \rightarrow \phis_i$, $X\rightarrow Y$.} By virtue of the proposition (\ref{BKPexpI3}), for any $Y \in B_{\infty}$ we have

\begin{align}
(\phi_i \otimes \phis_{i})
(e^Y \otimes e^Y)
|u\rangle \otimes |v\rangle
=
&
(\phi_i \otimes \phis_{i})
\sum_{m=0}^{\infty}
\frac{1}{m!}
\sum_{n=0}^{m}
\binom{m}{n}
Y^{n}
\otimes
Y^{m-n}
|u\rangle \otimes |v\rangle
\nonumber
\\
=
&
\sum_{m=0}^{\infty}
\frac{1}{m!}
\sum_{n=0}^{m}
\binom{m}{n}
Y^{n}
\otimes
Y^{m-n}
(\phi_i \otimes \phis_{i})
|u\rangle \otimes |v\rangle
\nonumber
\\
=
&
(e^Y \otimes e^Y)(\phi_i \otimes \phis_{i})
|u\rangle \otimes |v\rangle
\end{align}

\noindent Therefore we have proved that

\begin{align}
\sum_{i \in \mathbb{Z}}
\phi_i e^{Y} |u\rangle \otimes \phis_{i} e^{Y} |v\rangle
=
\sum_{i \in \mathbb{Z}}
e^{Y} \phi_i |u\rangle \otimes e^{Y} \phis_{i} |v\rangle
\label{BKPnfermI-2}
\end{align}

\noindent for arbitrary $Y \in B_{\infty}$. Using (\ref{BKPnfermI-2}) $l$ times successively, once for each $e^{Y_i}$ in $g_{\phi}$, we prove $(\ref{BKPexpI2})$.
}
\end{proof}

\medskip
\noindent
{\bf Corollary.} Having established the validity of equation (\ref{BKPexpI2}) we employ a particular case of it, namely when $|u\rangle = |v\rangle = |0\rangle$, which gives

\begin{align}
\sum_{i \in \mathbb{Z}}
\phi_i g_{\phi} |0\rangle
\otimes
\phis_i g_{\phi} |0\rangle
=
\sum_{i \in \mathbb{Z}}
g_{\phi} \phi_i |0\rangle
\otimes
g_{\phi} \phis_i |0\rangle
=
g_{\phi} |1\rangle
\otimes
g_{\phi} |1\rangle
\label{cor}
\end{align}

\noindent where the final equality is due to the fact that for all $i\in \mathbb{Z}^{\times}$ either $\phi_i|0\rangle = 0$ or $\phis_i|0\rangle = 0$. Equation (\ref{cor}) completes the proof that when $g_{\phi}|0\rangle$ is of the form (\ref{BKPnfermI-1}), $g_{\phi}$ satisfies the NFBI (\ref{BKPexpI7}).

\medskip
\noindent
{\bf Step 2. (Lemma 20.)} Let $g_{\phi} \in Cl_{\phi}^{(0)}$ satisfy the NFBI (\ref{BKPexpI7}). Then for suitable $\{Y_1,\ldots,Y_l\} \in B_{\infty}$ we can write $g_{\phi}|0\rangle= e^{Y_1}\ldots e^{Y_l}|0\rangle$.

\begin{proof}
{\it (Lemma 20.)}
{\rm The proof is analogous to the proof of lemma 10. Since $g_{\phi}|0\rangle \in \mathcal{F}_{\phi}^{(0)}$ we can expand it in terms of the basis (\ref{repclB5}), by writing

\begin{align}
g_{\phi}
|0\rangle
&=
c_{\emptyset}|0\rangle 
+
\sum_{m > n \geq 0}
c_{m,n}
\phi_m \phi_n
|0\rangle 
+
g^{(1)}_{\phi}
|0\rangle 
\end{align}

\noindent for some suitable coefficients $c_{\emptyset}$ and $c_{m,n}$, and where all monomials within $g^{(1)}_{\phi} \in Cl_{\phi}^{(0)}$ consist of at least four neutral fermions.\footnote{Throughout the rest of the proof, we will always use $g^{(i)}_{\phi}$ to denote an element of $Cl_{\phi}^{(0)}$ with precisely this property.} From here, we need to consider the cases $c_{\emptyset} \not= 0$ and $c_{\emptyset} = 0$ separately.   

\medskip
\noindent
{\bf Case 1. ($c_{\emptyset} \not= 0$)} We define the elements $Y_1,Y_2$ of $B_{\infty}$ as follows

\begin{align}
Y_1
=
\log c_{\emptyset},
\quad
Y_2
=
\sum_{m>n\geq 0}
c^{(1)}_{m,n} \phi_m \phi_n
\end{align}

\noindent where $c^{(1)}_{m,n} = c_{m,n} / c_{\emptyset}$ for all $m > n \geq 0$. We trivially obtain

\begin{align}
e^{-Y_1} g_{\phi}
|0\rangle
=
|0\rangle
+
\sum_{m > n \geq 0} c^{(1)}_{m,n} \phi_{m} \phi_{n} 
|0\rangle
+
g^{(2)}_{\phi}
|0\rangle  
\label{BKPexpII2}
\end{align}

\noindent where we have defined $g^{(2)}_{\phi} = g^{(1)}_{\phi} / c_{\emptyset}$. Next, we act on equation (\ref{BKPexpII2}) with the operator $e^{-Y_2}$. Since 
$
\Big(
\sum_{m > 0}
c^{(1)}_{m,0} \phi_m \phi_0
\Big)^2
=
0
$, term by term we have

\begin{align}
&
e^{-Y_2}
|0\rangle
=
|0\rangle
-
\sum_{m > n \geq 0} c^{(1)}_{m,n} \phi_m \phi_n
|0\rangle
+
g^{(3)}_{\phi}
|0\rangle
\\
&
e^{-Y_2}
\sum_{m >n \geq 0}
c^{(1)}_{m,n} \phi_m \phi_n
|0\rangle
=
\sum_{m > n \geq 0}
c^{(1)}_{m,n} \phi_m \phi_n
|0\rangle
+
g^{(4)}_{\phi}
|0\rangle
\\
&
e^{-Y_2} g^{(2)}_{\phi}
|0\rangle 
=
g^{(5)}_{\phi}
|0\rangle 
\end{align}

\noindent for some suitable $g^{(3)}_{\phi},g^{(4)}_{\phi}, g^{(5)}_{\phi} \in Cl_{\phi}^{(0)}$. Combining these three results, we obtain

\begin{align}
e^{-Y_2}
e^{-Y_1}
g_{\phi}
|0\rangle 
=
|0\rangle
+
g^{(6)}_{\phi}
|0\rangle
\label{BKPnfermI-4}
\end{align}

\noindent where we have defined $g^{(6)}_{\phi} =g^{(3)}_{\phi}+g^{(4)}_{\phi}+g^{(5)}_{\phi}$. By virtue of equation (\ref{BKPexpI2}) and the fact that $g_{\phi}$ obeys the NFBI (\ref{BKPexpI7}), we have

\begin{align}
e^{-Y_2} e^{-Y_1} g_{\phi}|1\rangle
\otimes
e^{-Y_2} e^{-Y_1} g_{\phi}|1\rangle
&
=
\sum_{i \in \mathbb{Z}}
e^{-Y_2} e^{-Y_1} \phi_i  g_{\phi} |0\rangle
\otimes
e^{-Y_2} e^{-Y_1} \phis_{i} g_{\phi} |0\rangle
\nonumber
\\
&
=
\sum_{i \in \mathbb{Z}}
\phi_i e^{-Y_2} e^{-Y_1} g_{\phi} |0\rangle
\otimes
\phis_{i} e^{-Y_2} e^{-Y_1} g_{\phi} |0\rangle
\label{BKPnfermI-3}
\end{align}

\noindent Substituting the expression (\ref{BKPnfermI-4}) for $e^{-Y_2} e^{-Y_1} g_{\phi}|0\rangle$ into (\ref{BKPnfermI-3}) and using the annihilation properties (\ref{repclB1}), we find 

\begin{align}
&
\sum_{i \geq 0}
\phi_i |0\rangle
\otimes
\phis_{i} g^{(6)}_{\phi} |0\rangle
+
\sum_{i \leq 0}
\phi_i g^{(6)}_{\phi} |0\rangle
\otimes
\phis_{i} |0\rangle
+
\sum_{i \in \mathbb{Z}}
\phi_i g^{(6)}_{\phi} |0\rangle
\otimes
\phis_{i} g^{(6)}_{\phi} |0\rangle
\label{BKPblah}
\\
&
=
|1\rangle
\otimes
g^{(6)}_{\phi} |1\rangle
+
g^{(6)}_{\phi} |1\rangle
\otimes
|1\rangle
+
g^{(6)}_{\phi} |1\rangle
\otimes
g^{(6)}_{\phi} |1\rangle
\nonumber
\end{align}

\noindent We recall that all monomials within $g^{(6)}_{\phi} \in Cl_{\phi}^{(0)}$ consist of at least four neutral fermions. Therefore the first two sums on the left hand side of (\ref{BKPblah}) contain terms which do not appear in the rest of the equation. These terms vanish if and only if
$
\phi_i g^{(6)}_{\phi} |0\rangle
=
0
$ for all $i < 0$. The only possible resolution is that $g^{(6)}_{\phi}|0\rangle = 0$. Substituting this value for $g^{(6)}_{\phi}|0\rangle$ into (\ref{BKPnfermI-4}) we see that $e^{-Y_2} e^{-Y_1} g_{\phi}|0\rangle = |0\rangle$, or equivalently, 
$g_{\phi}|0\rangle = e^{Y_1} e^{Y_2}|0\rangle$. This completes the proof in the case $c_{\emptyset} \not= 0$.

\medskip
\noindent 
{\bf Case 2. ($c_{\emptyset} = 0$)} We begin by stating two identities which we use in the proof. Fix two integers $p > q > 0$ and a set $\{m\} = \{m_1 > \cdots > m_{2r} \geq 0\}$. The first identity reads 

\begin{align}
e^{-\frac{1}{2}\phi_p \phi_q}
e^{-\frac{1}{2} \phis_{p} \phis_{q}}
\phi_{\{m\}}|0\rangle
=
\left\{
\begin{array}{ll}
2(-)^{i+j+1}
\phi_{\{m \backslash m_i,m_j\}}|0\rangle,
&\quad
p = m_i
\\
&\quad
q = m_j
\\
\\
\phi_{\{m\}}|0\rangle
+
\frac{1}{2}
\phi_p \phi_q \phi_{\{m\}}|0\rangle,
&\quad
p \not\in \{m\} 
\\
&\quad
q \not\in \{m\}
\\
\\
\phi_{\{m\}}|0\rangle,
&\quad
{\rm otherwise}
\end{array}
\right.
\label{BKPblah2}
\end{align}

\noindent where we have used the notation $\{m\backslash m_i,m_j\}$ to denote the omission of the $i^{\rm th}$ and $j^{\rm th}$ elements from the set $\{m\}$. The second identity reads 

\begin{align}
e^{-\frac{1}{2}\phi_p \phi_0}
e^{-\phis_{p} \phis_{0}}
\phi_{\{m\}} |0\rangle
=
\left\{
\begin{array}{ll}
2(-)^{i+1}
\phi_{\{m \backslash m_i,m_{2r}\}} |0\rangle,
&\quad
p = m_i
\\
&\quad
m_{2r} = 0
\\
\\
2(-)^i  
\phi_{\{m \backslash m_i\}}
\phi_0
|0\rangle,
&\quad
p = m_i
\\
&\quad
m_{2r} > 0 
\\
\\
\phi_{\{m\}}
|0\rangle
+
\frac{1}{2}
\phi_p
\phi_{\{m \backslash m_{2r} \}}
|0\rangle,
&\quad
p \not\in \{m\}
\\
&\quad
m_{2r} = 0
\\
\\
\phi_{\{m\}}
|0\rangle
-
\frac{1}{2}
\phi_p 
\phi_{\{m\}}
\phi_0
|0\rangle,
&\quad
p \not\in \{m\} 
\\
&\quad
m_{2r} > 0 
\end{array}
\right.
\label{BKPblah3}
\end{align}

\noindent where the meaning of the notations $\{m\backslash m_i, m_{2r}\}$, $\{m\backslash m_i\}$ and $\{m \backslash m_{2r}\}$ should be clear from previous explanations. Returning to the proof, we observe that since $c_{\emptyset} = 0$ we can write

\begin{align}
g_{\phi}|0\rangle
=
\sum_{{\rm card}\{m\} = 2r}
c_{\{m\}} \phi_{\{m\}} |0\rangle
+
g^{(7)}_{\phi} |0\rangle
\label{c=0}
\end{align}

\noindent where the sum is taken over all sets of integers $\{m_1>\cdots>m_{2r} \geq 0\}$ of some fixed cardinality $2r \geq 2$, and all monomials within $g^{(7)}_{\phi} \in Cl_{\phi}^{(0)}$ consist of at least $2r+2$ neutral fermions. Let $c_{\{p\}}$ be a particular non-zero coefficient in the sum (\ref{c=0}), corresponding to the set $\{p\} = \{p_1 > \cdots > p_{2r} \geq 0\}$, and define

\begin{align}
Y_{2i-1}
=
\frac{1}{2}
\left(1+\delta_{p_{2i},0}\right)
\phis_{p_{2i-1}} \phis_{p_{2i}},
\quad
Y_{2i}
=
\frac{1}{2}
\phi_{p_{2i-1}} \phi_{p_{2i}}
\end{align}

\noindent for all $1\leq i \leq r$. Successively applying the identities (\ref{BKPblah2}) and (\ref{BKPblah3}) to $g_{\phi}|0\rangle$, we obtain

\begin{align}
e^{-Y_{2r}} 
e^{-Y_{2r-1}}
\ldots
e^{-Y_2}
e^{-Y_1}
g_{\phi}|0\rangle
=
c^{(2)}_{\emptyset}|0\rangle
+
\sum_{m > n \geq 0}
c^{(2)}_{m,n}
\phi_m \phi_n |0\rangle
+
g^{(8)}_{\phi}|0\rangle
\label{BKPblah4}
\end{align}

\noindent with $c^{(2)}_{\emptyset} = 2^r c_{\{p\}}$ and the remaining coefficients $c^{(2)}_{m,n}$ suitably chosen, and where all monomials within $g^{(8)}_{\phi} \in Cl_{\phi}^{(0)}$ consist of at least four neutral fermions. Since $c^{(2)}_{\emptyset} \not= 0$, we can apply the procedure of case 1 to (\ref{BKPblah4}), ultimately obtaining

\begin{align}
g_{\phi}|0\rangle
=
e^{Y_1}
e^{Y_2}
\ldots
e^{Y_{2r+1}}
e^{Y_{2r+2}}
|0\rangle
\label{BKPblah5}
\end{align}

\noindent where we have defined

\begin{align}
Y_{2r+1}
=
\log c^{(2)}_{\emptyset},
\quad
Y_{2r+2}
=
\sum_{m>n\geq 0}
c^{(2)}_{m,n} / c^{(2)}_{\emptyset}
\phi_m \phi_n |0\rangle
\end{align}

\noindent Since all $\{Y_1,\ldots,Y_{2r+2}\} \in B_{\infty}$, equation (\ref{BKPblah5}) completes the proof in the $c_{\emptyset}=0$ case. We have therefore proved lemma 20 which, in turn, finishes the proof of theorem 4.
}
\end{proof}

\subsection{Schur $Q$-polynomials and the orbit of $O_{\infty}$}

\begin{example}
{\rm As a particular case of theorem 4, we show that every Schur $Q$-polynomial (\ref{BKPpoly9}) is a BKP $\tau$-function. Let $|\tilde{\mu})=|\mu_1,\ldots,\mu_{2r})$ be an arbitrary strict partition equal to the Fock space vector $\phi_{\mu_1}\ldots\phi_{\mu_{2r}}|0\rangle$. Recalling equation (\ref{BKPpoly4}) from the proof of lemma 15, we have

\begin{align}
\mathcal{Q}_{\tilde{\mu}}\{\tilde{t}\}
=
\Big\langle 
e^{\lambda\{\tilde{t}\}}
\phi_{\mu_1}\ldots\phi_{\mu_{2r}}
\Big\rangle
\label{BKPpoly42}
\end{align}

\noindent Defining

\begin{align}
Y_{2i-1}
=
\frac{1}{2}
(1+\delta_{\mu_{2i},0})
\phis_{\mu_{2i-1}} \phis_{\mu_{2i}},
\quad
Y_{2i}
=
\frac{1}{2}\phi_{\mu_{2i-1}} \phi_{\mu_{2i}}
\end{align}

\noindent for all $1\leq i \leq r$ and using the identities (\ref{BKPblah2}) and (\ref{BKPblah3}) from the last subsection, we obtain

\begin{align}
e^{-Y_{2r}} e^{-Y_{2r-1}}
\ldots
e^{-Y_2} e^{-Y_1}
\phi_{\mu_1} \ldots \phi_{\mu_{2r}}|0\rangle
=
2^r|0\rangle
\end{align}

\noindent or equivalently,

\begin{align}
\phi_{\mu_1}\ldots \phi_{\mu_{2r}} |0\rangle
=
2^r e^{Y_1}e^{Y_2}\ldots e^{Y_{2r-1}} e^{Y_{2r}} |0\rangle
\label{orbit}
\end{align}

\noindent Substituting (\ref{orbit}) into (\ref{BKPpoly42}), we find

\begin{align}
\mathcal{Q}_{\tilde{\mu}}\{\tilde{t}\}
=
2^r
\Big\langle
e^{\lambda\{\tilde{t}\}}
e^{Y_1}
\ldots
e^{Y_{2r}}
\Big\rangle
\end{align}

\noindent Hence any Schur $Q$-polynomial can be written as an expectation value of the form (\ref{BKPexpI-1}), with $g_{\phi} \in O_{\infty}$. By theorem 4, the Schur $Q$-polynomials are therefore $\tau$-functions of the BKP hierarchy \cite{nim}, \cite{you}.
}
\end{example}

\subsection{BKP Pl\"ucker relations}
\label{BKPpluck}

In this subsection we solve the NFBI (\ref{BKPexpI7}) from another, more direct perspective. Let $g_{\phi}|0\rangle$ and $g_{\phi}|1\rangle$ be corresponding finite elements of $\mathcal{F}_{\phi}^{(0)}$ and $\mathcal{F}_{\phi}^{(1)}$. Expanding $g_{\phi}|0\rangle$ in terms of the basis (\ref{repclB5}), there exist coefficients $c_{\{m\}}$ such that

\begin{align}
g_{\phi}|0\rangle
=
\sum_{\{m\}} c_{\{m\}} \phi_{\{m\}} |0\rangle
\label{BKPpluck2}
\end{align}

\noindent where the sum is over all sets of integers $\{m\} = \{m_1>\cdots>m_{2r} \geq 0\}$ whose cardinalities take all values $2r \geq 0$. Because $g_{\phi}|0\rangle$ is finite, $c_{\{m\}} = 0$ if ${\rm card}\{m\}$ is sufficiently large. Similarly, we have

\begin{align}
g_{\phi}|1\rangle
=
\sum_{\{m\}} c_{\{m\}} \phi_{\{m\}} |1\rangle
\end{align}

\noindent Using these expansions of $g_{\phi}|0\rangle$ and $g_{\phi}|1\rangle$, we obtain

\begin{align}
&
\sum_{i\in\mathbb{Z}}
\phi_i g_{\phi}|0\rangle \otimes \phis_i g_{\phi}|0\rangle
-
g_{\phi} |1\rangle \otimes g_{\phi} |1\rangle
=
\\
&
\sum_{\{m\},\{n\}}
c_{\{m\}} c_{\{n\}}
\left(
\sum_{i \in \mathbb{Z}}
\phi_i \phi_{\{m\}} |0\rangle
\otimes
\phis_{i} \phi_{\{n\}} |0\rangle
-
\phi_{\{m\}} \phi_0|0\rangle
\otimes
\phi_{\{n\}} \phi_0 |0\rangle
\right)
\nonumber
\end{align}

\noindent where the first sum is over all sets of integers $\{m\} = \{m_1 > \cdots > m_{2r} \geq 0\}$ and $\{n\} = \{n_1 > \cdots > n_{2s} \geq 0\}$, whose cardinalities take all even values $2r,2s \geq 0$. Using the annihilation properties (\ref{repclB1}) of the fermions, we find

\begin{align}
&
\sum_{i\in\mathbb{Z}}
\phi_i g_{\phi}|0\rangle \otimes \phis_i g_{\phi}|0\rangle
-
g_{\phi} |1\rangle \otimes g_{\phi} |1\rangle
=
2
\sum_{\{m\},\{n\}}
c_{\{m\}} c_{\{n\}}
\times
\label{BKPpluck-1}
\\
&
\left(
\sum_{i=1}^{2r}
(-)^{i-1}
\phi_{\{m \backslash m_i\}} |0\rangle
\otimes
\phi_{\{m_i,n\}} |0\rangle
+
\sum_{j=1}^{2s}
(-)^{j-1}
\phi_{\{n_j,m\}} |0\rangle
\otimes
\phi_{\{n\backslash n_j\}} |0\rangle
\right)
\nonumber
\end{align}

\noindent where we have defined $\phi_{\{m_i,n\}} = 0$ if $m_i \in \{n\}$ and $\phi_{\{n_j,m\}} = 0$ if $n_j \in \{m\}$. Changing the indexing sets of the first sum in (\ref{BKPpluck-1}), we obtain the equivalent expression

\begin{align}
&
\sum_{i\in\mathbb{Z}}
\phi_i g_{\phi}|0\rangle \otimes \phis_i g_{\phi}|0\rangle
-
g_{\phi} |1\rangle \otimes g_{\phi} |1\rangle
=
\label{BKPpluck-2}
\\
&
2
\sum_{\{p\},\{q\}}
\phi_{\{p\}} |0\rangle
\otimes
\phi_{\{q\}} |0\rangle
\left(
\sum_{i=1}^{2r-1}
(-)^{i-1}
c_{\{p\backslash p_i\}}
c_{\{p_i,q\}}
+
\sum_{j=1}^{2s-1}
(-)^{j-1}
c_{\{q_j, p\}}
c_{\{q\backslash q_j\}}
\right)
\nonumber
\end{align}

\noindent where the sum is over all sets of integers $\{p\} = \{p_1 > \cdots > p_{2r-1} \geq 0\}$ and $\{q\} = \{q_1 > \cdots > q_{2s-1} \geq 0 \}$ whose cardinalities take all odd values, and where we have defined

\begin{align}
c_{\{p_i,q\}} &= (-)^{j-1} c_{\{q_1,\ldots,q_{j-1},p_i,q_{j},\ldots,q_{2s-1}\}}
\end{align}

\noindent if $q_{j-1} > p_i > q_j$ for some $1\leq j \leq 2s$, and $c_{\{p_i,q\}}=0$ if $p_i \in \{q\}$. A similar definition applies to $c_{\{q_j,p\}}$. The right hand side of (\ref{BKPpluck-2}) vanishes if and only if 

\begin{align}
\sum_{i=1}^{2r-1}
(-)^{i}
c_{\{m \backslash m_i\}}
c_{\{m_i,n\}}
+
\sum_{j=1}^{2s-1}
(-)^{j}
c_{\{n_j, m\}}
c_{\{n\backslash n_j\}}
=
0
\label{BKPpluck3}
\end{align}

\noindent for all sets $\{m_1>\cdots>m_{2r-1}\geq 0\}$ and $\{n_1>\cdots>n_{2s-1}\geq 0\}$. Collectively, these conditions are called the {\it BKP Pl\"ucker relations}, and we summarize their significance with the following statement (which we have already proved).

\setcounter{lemma}{20}
\begin{lemma}
{\rm
$g_{\phi} \in Cl_{\phi}^{(0)}$ satisfies the NFBI (\ref{BKPexpI7}) if and only if the expansion coefficients (\ref{BKPpluck2}) of $g_{\phi} |0\rangle$ obey the BKP Pl\"ucker relations (\ref{BKPpluck3}).
}
\end{lemma}

\subsection{Pfaffian solution of BKP Pl\"ucker relations}
\label{BKPpfaff}

With the following result, we present a general Pfaffian solution of the BKP Pl\"ucker relations (\ref{BKPpluck3}).

\begin{lemma}
{\rm To every even-cardinality set of integers $\{m\} = \{m_1,\ldots,m_{2r}\}$ we associate the coefficient

\begin{align}
c_{\{m\}}
=
{\rm Pf}
\Big( c_{m_i,m_j} \Big)_{1\leq i<j\leq 2r}
=
|m_1,\ldots,m_{2r}|
\label{BKPpfaff1}
\end{align}

\noindent where the matrix entries $c_{i,j}$ are arbitrary constants that satisfy the antisymmetry condition $c_{i,j}=-c_{j,i}$. These coefficients satisfy the BKP Pl\"ucker relations (\ref{BKPpluck3}).
}
\end{lemma}

\begin{proof}
{\rm 
The proof is based on identity $(2.97)$ in section 2.8 of \cite{hir}. Define two ordered sets of integers $\{m\} = \{m_1>\cdots>m_{2r-1}\geq 0\}$ and $\{n\} = \{n_1>\cdots>n_{2s-1}\geq 0\}$, with fixed odd cardinalities. By the definition of the coefficients (\ref{BKPpfaff1}), we obtain

\begin{align}
&
\sum_{i=1}^{2r-1}
(-)^{i}
c_{\{m \backslash m_i\}}
c_{\{m_i,n\}}
\label{BKPpfaff2}
=
\sum_{i=1}^{2r-1}
(-)^i
|m_1,\ldots,\widehat{m_i},\ldots,m_{2r-1}|
|m_i,n_1,\ldots,n_{2s-1}|
\end{align}

\noindent Expanding the second Pfaffian in (\ref{BKPpfaff2}), we have

{\footnotesize
\begin{align}
&
\sum_{i=1}^{2r-1}
(-)^{i}
c_{\{m \backslash m_i\}}
c_{\{m_i,n\}}
\label{BKPpfaff3}
=
\sum_{i=1}^{2r-1}
\sum_{j=1}^{2s-1}
(-)^{i+j}
|m_1,\ldots,\widehat{m_i},\ldots,m_{2r-1}|
|m_i,n_j|
|n_1,\ldots,\widehat{n_j},\ldots,n_{2s-1}|
\end{align}
}

\noindent Similarly, we find that

\begin{align}
& 
\sum_{j=1}^{2s-1}
(-)^{j}
c_{\{n_j, m\}}
c_{\{n\backslash n_j\}}
\label{BKPpfaff4}
=
\sum_{j=1}^{2s-1}
(-)^j
|n_j,m_1,\ldots,m_{2r-1}|
|n_1,\ldots,\widehat{n_j},\ldots,n_{2s-1}|
\end{align}

\noindent Expanding the first Pfaffian in (\ref{BKPpfaff4}), we obtain

{\footnotesize
\begin{align}
& 
\sum_{j=1}^{2s-1}
(-)^{j}
c_{\{n_j, m\}}
c_{\{n\backslash n_j\}}
\label{BKPpfaff5}
=
\sum_{j=1}^{2s-1}
\sum_{i=1}^{2r-1}
(-)^{j+i}
|n_j,m_i|
|m_1,\ldots,\widehat{m_i},\ldots,m_{2r-1}|
|n_1,\ldots,\widehat{n_j},\ldots,n_{2s-1}|
\end{align}
}

\noindent Summing the equations (\ref{BKPpfaff3}) and (\ref{BKPpfaff5}) we observe that the right hand side of the resultant equation vanishes, where we have used the fact that $|m_i,n_j|=-|n_j,m_i|$ for all $1\leq i \leq 2r-1$ and $1\leq j \leq 2s-1$. This shows that the BKP Pl\"ucker relations (\ref{BKPpluck3}) are satisfied.
}
\end{proof}


\section{Conclusion}

Before ending this chapter, we present a brief summary of the material that has been discussed. We especially wish to emphasize those results which find application in the remainder of the thesis.

The Clifford algebras $Cl_{\psi}$ and $Cl_{\phi}$ provide a basic framework for the study of the KP and BKP hierarchies, respectively. We defined Fock representations of these algebras, and constructed partition (\ref{partitions3}) and strict partition (\ref{repclB5})  bases for the respective Fock subspaces $\mathcal{F}_{\psi}^{(0)}$ and $\mathcal{F}_{\phi}^{(0)}$. We demonstrated that the elements of these bases may be identified with Schur (\ref{surprising}) and Schur $Q$-polynomials (\ref{BKPpoly9}), via the respective equations (\ref{fermi-char}) and (\ref{nfermi-qschur}). These polynomials, in turn, form a basis for the solution space of the KP and BKP hierarchies. We will continue to refer to the bases (\ref{partitions3}) and (\ref{repclB5}) and the polynomials (\ref{surprising}) and (\ref{BKPpoly9}) throughout the rest of the thesis, particularly in chapter 3, where they play a prominent role.

The KP and BKP bilinear identities, (\ref{KPbil1}) and (\ref{BKPbil1}) respectively, contain all of the differential equations of their corresponding hierarchies. With theorems 1 and 3 we showed that the solutions of these bilinear identities, the $\tau$-functions, are expressible as fermionic expectation values. The task of solving (\ref{KPbil1}) and (\ref{BKPbil1}) was shown to be equivalent to solving the charged and neutral fermionic bilinear identities, (\ref{KPhelp}) and (\ref{BKPexpI7}), respectively.

Solutions of the CFBI and NFBI may be obtained from two different perspectives. The first perspective was demonstrated with theorems 2 and 4, where we showed that solutions of (\ref{KPhelp}) and (\ref{BKPexpI7}) are given by the orbit of the vacuum under $GL_{\infty}$ and $O_{\infty}$, respectively. The second perspective depends on finding solutions to the Pl\"ucker relations, (\ref{KPpluck1}) and (\ref{BKPpluck3}). Explicit solutions of the Pl\"ucker relations were presented by the formulae (\ref{KPdet1}) and (\ref{BKPpfaff1}), respectively. In the coming chapters we will encounter objects whose expansion coefficients are determinants or Pfaffians. By virtue of lemmas 12 and 22, we will thus be able to connect these objects with solutions of the KP and BKP hierarchies.     


\newpage

\thispagestyle{empty}

\phantom{nothing}


\setcounter{secnumdepth}{1}
\chapter{Overview of quantum inverse scattering method}
\setcounter{example}{0}
\setcounter{remark}{0}
\setcounter{lemma}{0}
\setcounter{theorem}{0}

\setcounter{secnumdepth}{2}

\setcounter{section}{-1}

\section{Introduction}

In the 1970s the Leningrad school developed a quantum version of the technique which had been discovered in \cite{ggkm}. One of the first steps towards this quantization can be found in \cite{zf}. The technique itself became known as the quantum inverse scattering method and it was introduced in \cite{fst}, where it was used in the context of integrable field-theoretical models. During the 1980s the method was extended to the discrete, lattice versions of these models. In this thesis we shall apply the quantum inverse scattering method to the descendents and relatives of another type of discrete model, the XYZ spin-$\frac{1}{2}$ chain, which is closely connected with the eight-vertex model of statistical mechanics \cite{bax3}, \cite{bax2}, \cite{bax4}, \cite{bax1}.

The purpose of this chapter is to provide a brief introduction to the quantum inverse scattering method, in a sufficiently general setting. All of the models that we study later are specializations of the generic model discussed here, and this chapter enables us to unify much of the notation and conventions used throughout the thesis. In section \ref{aba-intro} we discuss the basic aspects common to all quantum integrable models which we study. These include the Hamiltonian $\mathcal{H}$ of the model, and its representation on a lattice of finitely many sites. The complete space of lattice states is denoted by $\mathcal{V}$, and the goal of the quantum inverse scattering method is to construct states within $\mathcal{V}$ which are eigenvectors of $\mathcal{H}$.

In the context of a quantum mechanical model, integrability means that $\mathcal{H}$ belongs to a family of commuting operators. The generating function of these operators is called the transfer matrix $t(u)$, and the quantum inverse scattering method is the technique through which $t(u)$ is constructed. In section \ref{aba-tra}, we review the quantum inverse scattering method for a generalized model. We introduce the $R$-matrix, which is a solution of the Yang-Baxter equation, as well as the $L$-matrix and monodromy matrix, whose entries are operators acting in $\mathcal{V}$. The commutation relations between these entries are given by the intertwining equations. We conclude by expressing the transfer matrix as the trace of the monodromy matrix. All of the objects defined in this section have graphical representations which we also include, since they provide a correspondence with the lattice models of statistical physics.

In section \ref{aba-aba} we describe the algebraic Bethe Ansatz for calculating eigenvectors of the transfer matrix, and derive the system of equations necessary for the success of the Ansatz, known as the Bethe equations. We define the scalar product between two different Bethe eigenvectors, and discuss its graphical representation as a two-dimensional lattice. The definitions appearing in this section are essential to the remainder of the thesis, throughout which we focus on Bethe eigenvectors and scalar products across several different models.

The material presented in this chapter is fairly ubiquitous throughout the literature, and as such we do not adhere to any particular reference. Essentially we will provide gleanings from \cite{fad}, \cite{kbi}, \cite{takh}, which are three standard introductory works on the subject. For more information, the reader is referred to these sources. 

\section{Quantum integrable models}
\label{aba-intro}

\subsection{Discrete one-dimensional models and their space of states $\mathcal{V}$}  

In this thesis we will study several discrete one-dimensional quantum mechanical systems. The playing field for these systems is a one-dimensional integral lattice with finitely many sites. Particles are placed at each lattice site, and every unique way of assigning particles to the lattice is called a {\it configuration}. The system is allowed to adopt any state which is a linear combination of individual lattice configurations.

To place these concepts on a more mathematical foundation, it is convenient to use the language of state vectors. Let $M \geq 1$ denote the number of lattice sites, and associate a {\it local quantum space} $\mathcal{V}_i$ to the $i^{\rm th}$ site for all $1\leq i \leq M$. These vector spaces have the basis

\begin{align}
{\rm Basis}(\mathcal{V}_i)
=
\Big\{|n\rangle_i\ \Big|\ n \in \mathfrak{N} \Big\}
\end{align}

\noindent where $n$ is a number which can take any value in the set $\mathfrak{N}$. The interpretation of $\mathfrak{N}$ and the state vectors $|n\rangle_i$ depends on the system under consideration. 

In chapters 3 and 4 we study systems which obey {\it Bose-Einstein} statistics, and have no limit on the number of particles per site. For these systems, $\mathfrak{N} = \mathbb{N} \cup \{0\}$ and  $|n \rangle_i$ represents the number of particles at the $i^{\rm th}$ lattice site. In this case the integers $n$ are called {\it occupation numbers}. On the other hand, in chapters 5 and 6 we study systems which obey {\it Fermi-Dirac} statistics, and have one spin-$\frac{1}{2}$ particle per site. For these systems, $\mathfrak{N} = \{+\frac{1}{2},-\frac{1}{2}\}=\{\uparrow,\downarrow\}$ and $|n\rangle_i$ represents the spin of a single particle at the $i^{\rm th}$ lattice site.

With the definition of each local space $\mathcal{V}_i$ fixed, we introduce the {\it global quantum space} $\mathcal{V}= \mathcal{V}_1 \otimes \cdots \otimes \mathcal{V}_M$. This vector space has the basis 

\begin{align}
{\rm Basis}(\mathcal{V}) 
=
\Big\{ 
|n_1\rangle_1 \otimes \cdots \otimes |n_M\rangle_M\ 
\Big|\ 
n_1,\ldots, n_M \in \mathfrak{N}
\Big\}
\label{general-V}
\end{align}

\noindent where the numbers $n_1,\ldots,n_M$ can take any value in the set $\mathfrak{N}$. Every basis vector $|n_1\rangle_1 \otimes \cdots \otimes |n_M\rangle_M$ describes an individual lattice configuration. In the Bose-Einstein case, $|n_1\rangle_1 \otimes \cdots \otimes |n_M\rangle_M$ represents a lattice configuration with $n_i$ bosons at the $i^{\rm th}$ site for all $1\leq i \leq M$. In the Fermi-Dirac case, $|n_1\rangle_1 \otimes \cdots \otimes |n_M\rangle_M$ represents a lattice configuration of $M$ fermions, such that the $i^{\rm th}$ fermion has intrinsic spin $n_i$ for all $1\leq i \leq M$. As mentioned previously, the system is allowed to adopt any state in $\mathcal{V}$. 


\subsection{Quantum algebras $\mathcal{A}_i$ and their representation on $\mathcal{V}_i$}

In addition to the vector space (\ref{general-V}), a quantum model is described by a set of commuting algebras $\mathcal{A}_1,\ldots,\mathcal{A}_M$. These algebras are in fact copies of a single algebra, with one copy assigned to each lattice site. For all $1\leq i \leq M$, the algebra $\mathcal{A}_i$ has a representation on the local space $\mathcal{V}_i$, and from this we deduce the action of $\mathcal{A} = \mathcal{A}_1\otimes \cdots \otimes \mathcal{A}_M$ on $\mathcal{V}$.

Let us make some general remarks which categorize the algebras $\mathcal{A}_1,\ldots,\mathcal{A}_M$ for the models under our consideration. In all cases, $\mathcal{A}_i$ is generated by three elements $\{\mathfrak{a}_i^{+},\mathfrak{a}_i^{-},\mathfrak{a}_i^{0}\}$. When these generators act on basis vectors of $\mathcal{V}_i$, both $\mathfrak{a}_i^{\pm}$ produce a new state, while $\mathfrak{a}_i^{0}$ returns the original state. More specifically, we can say that


\begin{align}
\mathfrak{a}^{+}_i |n\rangle_i 
=
a^{+}_i(n) |n+1\rangle_i, 
\quad
\mathfrak{a}^{-}_i |n\rangle_i
=
a^{-}_i(n) |n-1\rangle_i,
\quad
\mathfrak{a}^{0}_i |n\rangle_i
=
a^{0}_i(n)|n\rangle_i
\label{general-action}
\end{align}

\noindent for all $n \in \mathfrak{N}$ and some suitable constants $a^{\pm}_i(n),a^{0}_i(n)$. The interpretation of (\ref{general-action}) is specific to the model at hand. 

When the quantum system obeys Bose-Einstein statistics, $\mathfrak{a}_i^{+}$/$\mathfrak{a}_i^{-}$ play the role of creation/annihilation operators, adding/deleting particles from the $i^{\rm th}$ lattice site. In this case, $a_i^{-}(0) = 0$, and all states in $\mathcal{V}_i$ can be constructed from the action of $\mathfrak{a}_i^{+}$ on the vacuum state $|0\rangle_i$. When the system obeys Fermi-Dirac statistics, $\mathfrak{a}_i^{+}$/$\mathfrak{a}_i^{-}$ play the role of raising/lowering operators, raising/lowering the spin of the $i^{\rm th}$ particle. In this case, $a_i^{+}(\uparrow) = a_i^{-}(\downarrow)= 0$ and all states in $\mathcal{V}_i$ can be constructed from the action of $\mathfrak{a}_i^{-}$ on $|\uparrow\rangle_i$, or the action of $\mathfrak{a}_i^{+}$ on $|\downarrow\rangle_i$. For all physical systems, the state vector $|n\rangle_i$ is an eigenstate of the operator $\mathfrak{a}_i^{0}$.

\subsection{Inner products}

Let us define an inner product $\mathcal{I}_i$ on the local space $\mathcal{V}_i$. Suppose that $|m\rangle_i$ and $|n\rangle_i$ are two basis vectors of $\mathcal{V}_i$. The inner product $\mathcal{I}_i$ between these vectors is defined as

\begin{align}
\mathcal{I}_i \Big( |m\rangle_i, |n\rangle_i \Big)
=
c_i(m) \delta_{m,n}
\label{local-I}
\end{align}

\noindent where $c_i(m)$ denotes a function of the discrete variable $m \in \mathfrak{N}$, which is specific to the model under consideration. Typically, this function is chosen such that the operators $\mathfrak{a}_i^{+}$ and $\mathfrak{a}_i^{-}$ are adjoint. Imposing this condition, we find that

\begin{align}
\mathcal{I}_i \Big(\mathfrak{a}_i^{+} |m\rangle_i, |n\rangle_i \Big)
=
\mathcal{I}_i \Big(|m\rangle_i, \mathfrak{a}_i^{-} |n\rangle_i \Big)
\implies 
a_i^{+}(m) c_i(n) \delta_{m+1,n} = a_i^{-}(n) c_i(m) \delta_{m,n-1}
\label{const-rec}
\end{align}

\noindent which follows from the actions (\ref{general-action}) of $\mathfrak{a}_i^{\pm}$ and the definition (\ref{local-I}) of $\mathcal{I}_i$. Equation (\ref{const-rec}) is trivially satisfied if $m+1 \not= n$, while it leads to the constraint 

\begin{align}
a_i^{+}(m) c_i(m+1) = a_i^{-}(m+1) c_i(m)
\end{align}

\noindent on the function $c_i(m)$ in the case $m+1=n$. The operator $\mathfrak{a}_i^{0}$ is self-adjoint without any further conditions imposed on the function $c_i(m)$, since

\begin{align}
\mathcal{I}_i \Big(
\mathfrak{a}_i^{0} |m\rangle_i,
|n\rangle_i \Big)
=
a_i^{0}(m) c_i(m) \delta_{m,n}
=
a_i^{0}(n) c_i(m) \delta_{m,n}
=
\mathcal{I}_i \Big(
|m\rangle_i,
\mathfrak{a}_i^{0} |n\rangle_i
\Big)
\end{align}

\noindent Now we construct an inner product $\mathcal{I}$ on the global space $\mathcal{V}$, using the definition (\ref{local-I}) of the local inner products $\mathcal{I}_i$. Let $|m\rangle = |m_1\rangle_1\otimes\cdots\otimes |m_M\rangle_M$ and $|n\rangle = |n_1\rangle_1\otimes\cdots\otimes |n_M\rangle_M$ be basis vectors of $\mathcal{V}$. The inner product $\mathcal{I}$ between these vectors is defined as 

\begin{align}
\mathcal{I} \Big( |m\rangle, | n\rangle \Big)
=
\prod_{i=1}^{M} 
\mathcal{I}_i \Big( |m_i\rangle_i,|n_i\rangle_i \Big)
=
\prod_{i=1}^{M}
c_i(m_i)
\delta_{m_i,n_i}
\label{general-inn-prod}
\end{align}

\noindent That is, $\mathcal{I}$ induces orthogonality between the basis vectors of $\mathcal{V}$. The inner product between more general elements of $\mathcal{V}$ can be calculated from the assumption that $\mathcal{I}$ is bilinear. 

\subsection{Dual space of states $\mathcal{V}^{*}$}
\label{2-dual}

Rather than using the inner product notation adopted in the last subsection, a standard procedure is to introduce vector spaces which are dual to those already considered. To this end, let $\mathcal{V}_i^{*}$ denote the dual of $\mathcal{V}_i$. It has the basis 

\begin{align}
{\rm Basis}(\mathcal{V}_i^{*})
=
\Big\{\langle m|_i\ \Big|\ m \in \mathfrak{N} \Big\}
\end{align}

\noindent where the action of each dual state vector $\langle m|_i$ is given by

\begin{align}
\langle m|_i ()
=
\mathcal{I}_i\Big(
|m\rangle_i, \Big)
\label{dual-action}
\end{align}

\noindent The generators of $\mathcal{A}_i$ act on basis elements of $\mathcal{V}_i^{*}$ as follows 

\begin{align}
\langle m|_i \mathfrak{a}^{+}_i 
=
a^{-}_i(m) \langle m-1|_i, 
\quad
\langle m|_i\mathfrak{a}^{-}_i
=
a^{+}_i(m) \langle m+1|_i,
\quad
\langle m|_i \mathfrak{a}^{0}_i
=
a^{0}_i(m)\langle m|_i
\label{general-action*}
\end{align}

\noindent By virtue of (\ref{general-action}) and (\ref{general-action*}), the action (\ref{dual-action}) of the dual space $\mathcal{V}_i^{*}$, and the fact that $\mathfrak{a}_i^{\pm}$ are adjoint (whilst $\mathfrak{a}_i^{0}$ is self-adjoint), we recover the equations

\begin{align}
&
\langle m|_i \mathfrak{a}^{+}_i \Big( |n\rangle_i \Big)
=
\mathcal{I}_i\Big( \mathfrak{a}^{-}_i |m\rangle_i, |n\rangle_i \Big)
=
\mathcal{I}_i\Big( |m\rangle_i, \mathfrak{a}^{+}_i |n\rangle_i \Big)
=
\langle m|_i \Big( \mathfrak{a}^{+}_i |n\rangle_i \Big)
\\
&
\langle m|_i \mathfrak{a}^{-}_i \Big( |n\rangle_i \Big)
=
\mathcal{I}_i\Big( \mathfrak{a}^{+}_i |m\rangle_i, |n\rangle_i \Big)
=
\mathcal{I}_i\Big( |m\rangle_i, \mathfrak{a}^{-}_i |n\rangle_i \Big)
=
\langle m|_i \Big( \mathfrak{a}^{-}_i |n\rangle_i \Big)
\\
&
\langle m|_i \mathfrak{a}^{0}_i \Big( |n\rangle_i \Big)
=
\mathcal{I}_i\Big( \mathfrak{a}^{0}_i |m\rangle_i, |n\rangle_i \Big)
=
\mathcal{I}_i\Big( |m\rangle_i, \mathfrak{a}^{0}_i |n\rangle_i \Big)
=
\langle m|_i \Big( \mathfrak{a}^{0}_i |n\rangle_i \Big)
\end{align}

\noindent These equations ensure that the quantities $\langle m|_i \mathfrak{a}_i^{\pm} |n\rangle_i, \langle m|_i \mathfrak{a}_i^{0} |n\rangle_i$ are well defined without specifying the direction in which the operators $\mathfrak{a}_i^{\pm}, \mathfrak{a}_i^{0}$ act. We refer to these quantities as {\it expectation values} of the operators $\mathfrak{a}_i^{\pm}, \mathfrak{a}_i^{0}$. More generally, expectation values of arbitrary elements of $\mathcal{A}_i$ are unambiguously defined. 

These ideas can be extended to the dual $\mathcal{V}^{*}$ of the global vector space $\mathcal{V}$. Its basis is given by

\begin{align}
{\rm Basis} (\mathcal{V}^{*})
=
\Big\{
\langle m_1|_1 \otimes \cdots \otimes \langle m_M|_M\
\Big|\
m_1,\ldots,m_M \in \mathfrak{N} 
\Big\}
\label{general-vec*}
\end{align}

\noindent where the action of each dual state vector $\langle m| = \langle m_1|_1\otimes\cdots\otimes \langle m_M|_M$ is given by

\begin{align}
\langle m| ( ) 
=
\mathcal{I} \Big( |m\rangle, \Big)
\label{general-duality}
\end{align}

\noindent By this definition, the inner product $\mathcal{I}( |m\rangle, |n\rangle)$ can be written as $\langle m|( |n\rangle )$, or more simply $\langle m|n\rangle$. Expectation values of arbitrary elements of $\mathcal{A} = \mathcal{A}_1 \otimes \cdots \otimes \mathcal{A}_M$ remain well defined, regardless of the direction in which operators act.

\subsection{Hamiltonian $\mathcal{H}$}

The physical interactions of a quantum mechanical system are described by its {\it Hamiltonian} $\mathcal{H} \in {\rm End}(\mathcal{V})$. In the models which we study, $\mathcal{H}$ is an algebraic combination of the operators $\mathfrak{a}_i^{\pm},\mathfrak{a}_i^{0}$. It incorporates the interaction of the $i^{\rm th}$ lattice site with its nearest neighbours, the $(i-1)^{\rm th}$ and $(i+1)^{\rm th}$ lattice sites, for all $1\leq i \leq M$. Periodicity is imposed, meaning that the $1^{\rm st}$ and $M^{\rm th}$ sites are considered nearest neighbours. 

To be more explicit, we give some examples of the types of Hamiltonians which we will encounter. In chapters 3 and 4 we will study models with Hamiltonians of the form

\begin{align}
\mathcal{H}
=
\sum_{i=1}^{M}
\Big( 
\mathfrak{a}_i^{+} \mathfrak{a}_{i+1}^{-} 
+
\mathfrak{a}_i^{-} \mathfrak{a}_{i+1}^{+}
+
\Delta
\mathfrak{a}_i^{0}
\Big)
\label{ham-34}
\end{align}

\noindent while in chapters 5 and 6 we will study models with Hamiltonians of the form

\begin{align}
\mathcal{H}
=
\sum_{i=1}^{M}
\Big( 
\mathfrak{a}_i^{+} \mathfrak{a}_{i+1}^{-} 
+
\mathfrak{a}_i^{-} \mathfrak{a}_{i+1}^{+}
+
\Delta
\mathfrak{a}_i^{0} \mathfrak{a}_{i+1}^{0}
\Big)
\label{ham-56}
\end{align}

\noindent with $\Delta$ a constant, and where we assume the periodicity $\mathfrak{a}_{M+1}^{\pm} = \mathfrak{a}_1^{\pm}, \mathfrak{a}_{M+1}^{0} = \mathfrak{a}_1^{0}$ in both (\ref{ham-34}) and (\ref{ham-56}).  

An important goal in the study of a particular quantum mechanical model is to calculate the spectrum of its Hamiltonian $\mathcal{H}$. That is, one wishes to find state vectors $|\Psi\rangle \in \mathcal{V}$ which are eigenvectors of $\mathcal{H}$, satisfying      

\begin{align}
\mathcal{H}|\Psi\rangle
=
\mathcal{E}_{\Psi}|\Psi\rangle
\label{eigen-eqn1}
\end{align}

\noindent and to compute the corresponding eigenvalues $\mathcal{E}_{\Psi}$. In accomplishing such a task, one is commonly said to have solved the model. All the models studied in this thesis are {\it exactly solvable}, meaning that their Hamiltonian $\mathcal{H}$ belongs to a family of commuting operators. The quantum inverse scattering method/algebraic Bethe Ansatz are techniques which diagonalize the entire family of commuting operators simultaneously. We shall devote the remainder of this chapter to describing these techniques.

\section{Quantum inverse scattering method}
\label{aba-tra}

\subsection{$R$-matrix and Yang-Baxter equation}

The quantum inverse scattering approach to solving a given quantum integrable model relies on an $n^2 \times n^2$ matrix called an {\it $R$-matrix}, where $n \geq 2$. The value of $n$ and the entries of the $R$-matrix are specific to the model under consideration, however we can make three remarks which apply universally. {\bf 1.} The $R$-matrix depends on two parameters called {\it rapidities}, which we typically write as $u$ and $v$,\footnote{In later chapters we will also use $x$ and $y$ for the rapidities. In those cases all the theory established here still applies, if one simply replaces $u$ and $v$ with $x$ and $y$, respectively.} {\bf 2.} The $R$-matrix is an element of ${\rm End}(\mathcal{V}_a \otimes \mathcal{V}_b)$, where $\mathcal{V}_a$ and $\mathcal{V}_b$ are copies of $\mathbb{C}^n$, and are called {\it auxiliary vector spaces,} {\bf 3.} The $R$-matrix is a solution of the {\it Yang-Baxter equation,} which will be described in detail below. 

In this thesis we will focus on the case $n=2$, and models which have $4\times 4$ $R$-matrices of the form

\begin{align}
R_{ab}(u,v)
&=
\label{general-Rmat}
\left(
\begin{array}{cccc}
R^{++}_{++}(u,v) & R^{++}_{+-}(u,v) & R^{+-}_{++}(u,v) & R^{+-}_{+-}(u,v)
\\
R^{++}_{-+}(u,v) & R^{++}_{--}(u,v) & R^{+-}_{-+}(u,v) & R^{+-}_{--}(u,v)
\\
R^{-+}_{++}(u,v) & R^{-+}_{+-}(u,v) & R^{--}_{++}(u,v) & R^{--}_{+-}(u,v)
\\
R^{-+}_{-+}(u,v) & R^{-+}_{--}(u,v) & R^{--}_{-+}(u,v) & R^{--}_{--}(u,v)
\end{array}
\right)_{ab}
\\
\nonumber
\\
&=
\left(
\begin{array}{cccc}
a_{+}(u,v) & 0 & 0 & 0
\\
0 & b_{+}(u,v) & c_{+}(u,v) & 0
\\
0 & c_{-}(u,v) & b_{-}(u,v) & 0
\\
0 & 0 & 0 & a_{-}(u,v)
\end{array}
\right)_{ab}
\nonumber
\end{align}

\noindent where the entries are functions of the rapidities $u,v$ which are specific to the model under consideration. We have placed the subscript $ab$ on the $R$-matrix to denote the fact that it is an element of ${\rm End}(\mathcal{V}_a\otimes \mathcal{V}_b)$, where $\mathcal{V}_a,\mathcal{V}_b$ are copies of $\mathbb{C}^2$. The $R$-matrices (\ref{general-Rmat}) are solutions of the {\it Yang-Baxter equation}

\begin{align}
R_{ab}(u,v)R_{ac}(u,w)R_{bc}(v,w)
=
R_{bc}(v,w)R_{ac}(u,w)R_{ab}(u,v)
\label{general-YB}
\end{align}

\noindent which is an identity acting in the tensor product $\mathcal{V}_a \otimes \mathcal{V}_b \otimes \mathcal{V}_c$ of three auxiliary spaces, for general values of the rapidities $u,v,w$. The Yang-Baxter equation is a strong restriction on the functions $a_{\pm},b_{\pm},c_{\pm}$ which are entries of the $R$-matrix (\ref{general-Rmat}). The full strength is revealed when (\ref{general-YB}) is written in component notation, giving

\begin{align}
R^{i_1 k_1}_{i_2 k_2}(u,v) R^{k_1 j_1}_{i_3 k_3}(u,w) R^{k_2 j_2}_{k_3 j_3}(v,w)
=
R^{i_2 k_2}_{i_3 k_3}(v,w) R^{i_1 k_1}_{k_3 j_3}(u,w) R^{k_1 j_1}_{k_2 j_2}(u,v)
\label{general-YB-comp}
\end{align}

\noindent where all indices take values in $\{+1,-1\}$, with each of $\{i_1,i_2,i_3,j_1,j_2,j_3\}$ held fixed, while $\{k_1,k_2,k_3\}$ are summed. We see that (\ref{general-YB-comp}) gives rise to $2^6$ scalar equations involving the functions $a_{\pm},b_{\pm},c_{\pm}$, one corresponding to each configuration of the indices $\{i_1,i_2,i_3,j_1,j_2,j_3\}$. 

\begin{example}
{\rm 
Setting $\{i_1,i_2,i_3,j_1,j_2,j_3\} = \{-,+,-,-,-,+\}$ in (\ref{general-YB-comp}), we recover the equation

\begin{align}
R^{-+}_{+-}(u,v) R^{+-}_{-+}(u,w) R^{--}_{++}(v,w)
&+
R^{--}_{++}(u,v) R^{--}_{--}(u,w) R^{+-}_{-+}(v,w)
\label{general-YB-example-pt1}
\\
&=
R^{+-}_{-+}(v,w) R^{--}_{++}(u,w) R^{--}_{--}(u,v)
\nonumber
\end{align}

\noindent where some terms within the summation have vanished due to their corresponding $R$-matrix entries being zero. Substituting the functions which comprise the $R$-matrix entries into (\ref{general-YB-example-pt1}), we obtain

\begin{align}
c_{-}(u,v) c_{+}(u,w) b_{-}(v,w)
+
b_{-}(u,v) a_{-}(u,w) c_{+}(v,w)
=
c_{+}(v,w) b_{-}(u,w) a_{-}(u,v)
\label{general-YB-example-pt2}
\end{align}
}
\end{example}

\subsection{Graphical representation of $R$-matrix}

It is possible to represent the elements of the $R$-matrix (\ref{general-Rmat}) graphically, a procedure which leads to an elegant diagrammatic interpretation of the Yang-Baxter equation (\ref{general-YB-comp}). This graphical correspondence is realized by matching each non-zero element of (\ref{general-Rmat}) with a {\it vertex}, as shown in figure 2.1.

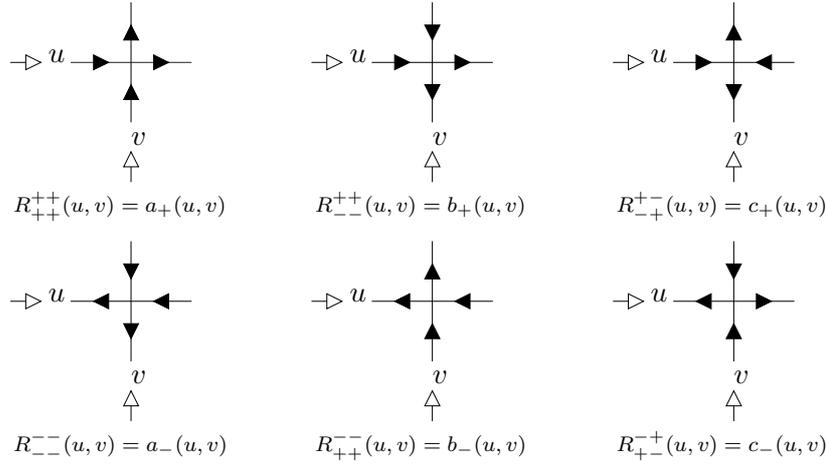
\begin{figure}[H]
\begin{center}
\begin{minipage}{4.3in}

\setlength{\unitlength}{0.0004cm}
\begin{picture}(20000,15000)(-4500,-13000)

\path(-2000,0000)(2000,0000)
\blacken\path(-1250,250)(-1250,-250)(-750,0)(-1250,250)
\blacken\path(750,250)(750,-250)(1250,0)(750,250)
\put(-2750,0){$u$}
\path(-4000,0)(-3000,0)
\whiten\path(-3500,250)(-3500,-250)(-3000,0)(-3500,250)
\path(0000,-2000)(0000,2000)
\blacken\path(-250,-1250)(250,-1250)(0,-750)(-250,-1250)
\blacken\path(-250,750)(250,750)(0,1250)(-250,750)
\put(-3900,-5000){\scriptsize$R^{++}_{++}(u,v) = a_{+}(u,v)$}
\put(0,-2750){$v$}
\path(0,-4000)(0,-3000)
\whiten\path(-250,-3500)(250,-3500)(0,-3000)(-250,-3500)

\path(8000,0000)(12000,0000)
\blacken\path(8750,250)(8750,-250)(9250,0)(8750,250)
\blacken\path(10750,250)(10750,-250)(11250,0)(10750,250)
\put(7250,0){$u$}
\path(6000,0)(7000,0)
\whiten\path(6500,250)(6500,-250)(7000,0)(6500,250)
\path(10000,-2000)(10000,2000)
\blacken\path(9750,-750)(10250,-750)(10000,-1250)(9750,-750)
\blacken\path(9750,1250)(10250,1250)(10000,750)(9750,1250)
\put(6100,-5000){\scriptsize$R^{++}_{--}(u,v)= b_{+}(u,v)$}
\put(10000,-2750){$v$}
\path(10000,-4000)(10000,-3000)
\whiten\path(9750,-3500)(10250,-3500)(10000,-3000)(9750,-3500)

\path(18000,0000)(22000,0000)
\blacken\path(18750,250)(18750,-250)(19250,0)(18750,250)
\blacken\path(21250,250)(21250,-250)(20750,0)(21250,250)
\put(17250,0){$u$}
\path(16000,0)(17000,0)
\whiten\path(16500,250)(16500,-250)(17000,0)(16500,250)
\path(20000,-2000)(20000,2000)
\blacken\path(19750,-750)(20250,-750)(20000,-1250)(19750,-750)
\blacken\path(19750,750)(20250,750)(20000,1250)(19750,750)
\put(16100,-5000){\scriptsize$R^{+-}_{-+}(u,v)= c_{+}(u,v)$}
\put(20000,-2750){$v$}
\path(20000,-4000)(20000,-3000)
\whiten\path(19750,-3500)(20250,-3500)(20000,-3000)(19750,-3500)

\path(-2000,-8000)(2000,-8000)
\blacken\path(-750,-7750)(-750,-8250)(-1250,-8000)(-750,-7750)
\blacken\path(1250,-7750)(1250,-8250)(750,-8000)(1250,-7750)
\put(-2750,-8000){$u$}
\path(-4000,-8000)(-3000,-8000)
\whiten\path(-3500,-7750)(-3500,-8250)(-3000,-8000)(-3500,-7750)
\path(0000,-10000)(0000,-6000)
\blacken\path(-250,-8750)(250,-8750)(0,-9250)(-250,-8750)
\blacken\path(-250,-6750)(250,-6750)(0,-7250)(-250,-6750)
\put(-3900,-13000){\scriptsize$R^{--}_{--}(u,v)= a_{-}(u,v)$}
\put(0,-10750){$v$}
\path(0,-12000)(0,-11000)
\whiten\path(-250,-11500)(250,-11500)(0,-11000)(-250,-11500)

\path(8000,-8000)(12000,-8000)
\blacken\path(9250,-7750)(9250,-8250)(8750,-8000)(9250,-7750)
\blacken\path(11250,-7750)(11250,-8250)(10750,-8000)(11250,-7750)
\put(7250,-8000){$u$}
\path(6000,-8000)(7000,-8000)
\whiten\path(6500,-7750)(6500,-8250)(7000,-8000)(6500,-7750)
\path(10000,-10000)(10000,-6000)
\blacken\path(9750,-9250)(10250,-9250)(10000,-8750)(9750,-9250)
\blacken\path(9750,-7250)(10250,-7250)(10000,-6750)(9750,-7250)
\put(6100,-13000){\scriptsize$R^{--}_{++}(u,v)= b_{-}(u,v)$}
\put(10000,-10750){$v$}
\path(10000,-12000)(10000,-11000)
\whiten\path(9750,-11500)(10250,-11500)(10000,-11000)(9750,-11500)

\path(18000,-8000)(22000,-8000)
\blacken\path(19250,-7750)(19250,-8250)(18750,-8000)(19250,-7750)
\blacken\path(20750,-7750)(20750,-8250)(21250,-8000)(20750,-7750)
\put(17250,-8000){$u$}
\path(16000,-8000)(17000,-8000)
\whiten\path(16500,-7750)(16500,-8250)(17000,-8000)(16500,-7750)
\path(20000,-10000)(20000,-6000)
\blacken\path(19750,-9250)(20250,-9250)(20000,-8750)(19750,-9250)
\blacken\path(19750,-6750)(20250,-6750)(20000,-7250)(19750,-6750)
\put(16100,-13000){\scriptsize$R^{-+}_{+-}(u,v)= c_{-}(u,v)$}
\put(20000,-10750){$v$}
\path(20000,-12000)(20000,-11000)
\whiten\path(19750,-11500)(20250,-11500)(20000,-11000)(19750,-11500)

\end{picture}

\end{minipage}
\end{center}

\caption[Six vertices of the generalized $R$-matrix]{Six vertices of the generalized $R$-matrix. Each non-zero entry of (\ref{general-Rmat}) is matched with a vertex.} 
\end{figure}

Each vertex in figure 2.1 is the intersection of a horizontal line with two black arrows attached, and a vertical line with two black arrows attached. The horizontal line is considered to have the rapidity $u$ {\it flowing through it,} in the direction indicated by the horizontal white arrow. Similarly, the vertical line is considered to have the rapidity $v$ flowing through it, in the direction indicated by the vertical white arrow. When a black arrow points in the direction of variable flow it is assigned the value $+1$, and conversely when a black arrow points opposite the direction of variable flow it is assigned the value $-1$.

In any given line, the black arrow nearest to the external white arrow is called {\it incoming}, since it precedes the intersection point of the vertex. The black arrow farthest from the external white arrow is called {\it outgoing}, since it succeeds the intersection point of the vertex. A line thus gives rise to an ordered pair of values $(i,j) \in \{+1,-1\}$, where $i$ is the value assigned to the incoming arrow, and $j$ is the value assigned to the outgoing arrow. The $R$-matrix element $R^{i_1 j_1}_{i_2 j_2}(u,v)$ is matched with the vertex having horizontal line values $(i_1,j_1)$ and vertical line values $(i_2,j_2)$. 

Using these graphical definitions, the Yang-Baxter equation (\ref{general-YB-comp}) may be written in the form shown in figure 2.2.

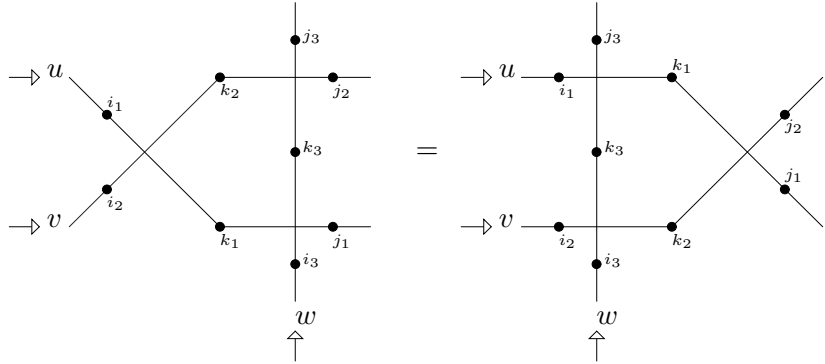
\begin{figure}[H]
\begin{center}
\begin{minipage}{4.3in}

\setlength{\unitlength}{0.0004cm}
\begin{picture}(20000,12500)(-2000,-4000)

\path(0,0)(5000,5000)
\put(1250,1250){\circle*{300}}
\put(1250,650){\tiny{$i_2$}}
\put(5000,5000){\circle*{300}}
\put(5000,4400){\tiny{$k_2$}}
\put(8750,5000){\circle*{300}}
\put(8750,4400){\tiny{$j_2$}}
\path(-2000,0)(-1000,0)
\whiten\path(-1250,250)(-1250,-250)(-1000,0)(-1250,250)
\put(-750,0){$v$}
\path(5000,0)(0,5000)
\put(1250,3750){\circle*{300}}
\put(1250,4000){\tiny{$i_1$}}
\put(5000,0){\circle*{300}}
\put(5000,-600){\tiny{$k_1$}}
\put(8750,0){\circle*{300}}
\put(8750,-600){\tiny{$j_1$}}
\path(-2000,5000)(-1000,5000)
\whiten\path(-1250,5250)(-1250,4750)(-1000,5000)(-1250,5250)
\put(-750,5000){$u$}
\path(5000,5000)(10000,5000)
\path(5000,0)(10000,0)
\path(7500,-2500)(7500,7500)
\put(7500,-1250){\circle*{300}}
\put(7750,-1250){\tiny{$i_3$}}
\put(7500,2500){\circle*{300}}
\put(7750,2500){\tiny{$k_3$}}
\put(7500,6250){\circle*{300}}
\put(7750,6250){\tiny{$j_3$}}
\put(7500,-3250){$w$}
\path(7500,-4500)(7500,-3500)
\whiten\path(7250,-3750)(7750,-3750)(7500,-3500)(7250,-3750)


\put(11500,2300){$=$}

\path(13000,5000)(14000,5000)
\whiten\path(13750,5250)(13750,4750)(14000,5000)(13750,5250)
\put(14250,5000){$u$}
\path(15000,5000)(20000,5000)
\put(16250,5000){\circle*{300}}
\put(16250,4400){\tiny{$i_1$}}
\path(20000,5000)(25000,0)
\put(20000,5000){\circle*{300}}
\put(20000,5250){\tiny{$k_1$}}
\put(23750,1250){\circle*{300}}
\put(23750,1600){\tiny{$j_1$}}
\path(13000,0)(14000,0)
\whiten\path(13750,250)(13750,-250)(14000,0)(13750,250)
\put(14250,0){$v$}
\path(15000,0)(20000,0)
\put(16250,0){\circle*{300}}
\put(16250,-600){\tiny{$i_2$}}
\path(20000,0)(25000,5000)
\put(20000,0){\circle*{300}}
\put(20000,-600){\tiny{$k_2$}}
\put(23750,3750){\circle*{300}}
\put(23750,3250){\tiny{$j_2$}}
\path(17500,-4500)(17500,-3500)
\whiten\path(17250,-3750)(17750,-3750)(17500,-3500)(17250,-3750)
\put(17500,-3250){$w$}
\path(17500,-2500)(17500,7500)
\put(17500,-1250){\circle*{300}}
\put(17750,-1250){\tiny{$i_3$}}
\put(17500,2500){\circle*{300}}
\put(17750,2500){\tiny{$k_3$}}
\put(17500,6250){\circle*{300}}
\put(17750,6250){\tiny{$j_3$}}

\end{picture}

\end{minipage}
\end{center}

\caption[Graphical depiction of the Yang-Baxter equation]{Graphical depiction of the Yang-Baxter equation.}
\end{figure} 

\noindent Both sides of this equation should be interpreted as three conjoined vertices. There is a vertex at the intersection of the $(u,v)$ lines, another at the intersection of the $(u,w)$ lines, and yet another at the intersection of the $(v,w)$ lines. The joining of these vertices is shorthand for multiplication of the three corresponding $R$-matrix elements. The absence of black arrows in this picture is a notational convenience, which requires some explanation.

Examining figure 2.2, we see that a conjoined trio of vertices possesses 6 {\it external} line segments, which have been labelled $\{i_1,i_2,i_3,j_1,j_2,j_3\}$. Each $i$ or $j$ represents an undisclosed black arrow, which is held fixed on both sides of the equation. By leaving these black arrows unspecified we can write (\ref{general-YB-comp}) as a single equation, when in fact it implies $2^6$ equations, one corresponding to each of the $2^6$ external configurations. 

The 3 {\it internal} line segments, which have been labelled $\{k_1,k_2,k_3\}$, have a different meaning. Each $k$ is summed over a black arrow that points with the variable flow, and a black arrow that points opposite the variable flow. By omitting black arrows from these points we imply summation over $2^3$ terms on each side of the equation. It should be noted that many terms in this summation vanish, since any vertex not shown in figure 2.1 is by definition equal to zero.

\subsection{$L$-matrix and local intertwining equation}

Another fundamental object in the quantum inverse scattering method is the $n\times n$ {\it $L$-matrix}. As before, the value of $n$ and the entries of the $L$-matrix are specific to the model under consideration, but we can make three universal remarks. {\bf 1.} The $L$-matrix depends on a single rapidity $u$, {\bf 2.} The $L$-matrix is an element of ${\rm End}(\mathcal{V}_a)$, where $\mathcal{V}_a$ is a copy of $\mathbb{C}^n$, and its entries are elements of the $m^{\rm th}$ quantum algebra $\mathcal{A}_m$, {\bf 3.} The $L$-matrix satisfies the {\it local intertwining equation}, which will be described in detail below.

In this thesis we restrict our attention to models with $2\times 2$ $L$-matrices of the form

\begin{align} 
L_{am}(u)
=
\left(
\begin{array}{cc}
L^{++}_m(u) & L^{+-}_m(u) \\
L^{-+}_m(u) & L^{--}_m(u)
\end{array}
\right)_a 
\label{general-Lmat}
\end{align}

\noindent where the entries depend on $u$ and are elements of $\mathcal{A}_m$, specific to the model under consideration. We have placed the subscript $a$ on the $L$-matrix to denote the fact that it is an element of ${\rm End}(\mathcal{V}_a)$, where $\mathcal{V}_a$ is a copy of $\mathbb{C}^2$. The $L$-matrix (\ref{general-Lmat}) satisfies the relation

\begin{align}
R_{ab}(u,v) L_{am}(u) L_{bm}(v)
=
L_{bm}(v) L_{am}(u) R_{ab}(u,v)
\label{general-intertwining-L}
\end{align}

\noindent which is an identity acting in the tensor product $\mathcal{V}_a \otimes \mathcal{V}_b$ of two auxiliary spaces, for general values of the rapidities $u,v$. The $R$-matrix $R_{ab}(u,v)$, given by (\ref{general-Rmat}), is said to {\it intertwine} the $L$-matrices $L_{am}(u)$ and $L_{bm}(v)$. For this reason we refer to (\ref{general-intertwining-L}) as the {\it local intertwining equation.} It is a local equation insofar as the entries of the $L$-matrices act only at the $m^{\rm th}$ site in the model. Writing (\ref{general-intertwining-L}) in component notation we obtain

\begin{align}
R^{i_1 k_1}_{i_2 k_2}(u,v) L^{k_1 j_1}_m(u) L^{k_2 j_2}_m(v)
=
L^{i_2 k_2}_m(v) L^{i_1 k_1}_m(u) R^{k_1 j_1}_{k_2 j_2}(u,v)
\label{general-intertwining-L-comp}
\end{align}

\noindent where all indices take values in $\{+1,-1\}$, with each of $\{i_1,i_2,j_1,j_2\}$ held fixed, while $\{k_1,k_2\}$ are summed. Hence (\ref{general-intertwining-L-comp}) gives rise to $2^4$ commutation relations involving the operators $L^{++}_m(u),L^{+-}_m(u),L^{-+}_m(u),L^{--}_m(u)$, one corresponding to each configuration of the indices $\{i_1,i_2,j_1,j_2\}$.

\begin{example}
{\rm 
Setting $\{i_1,i_2,j_1,j_2\}=\{+,+,+,-\}$ in (\ref{general-intertwining-L}), we recover the commutation relation

\begin{align}
R^{++}_{++}(u,v) L^{++}_m(u) L^{+-}_m(v)
=
L^{+-}_m(v) L^{++}_m(u) R^{++}_{--}(u,v)
+
L^{++}_m(v) L^{+-}_m(u) R^{-+}_{+-}(u,v)
\label{general-intertwining-L-example}
\end{align}

\noindent where some terms within the summation have vanished due to their corresponding $R$-matrix entries being zero. Substituting the functions which comprise the $R$-matrix entries into (\ref{general-intertwining-L-example}), we obtain

\begin{align}
a_{+}(u,v) L^{++}_m(u) L^{+-}_m(v)
=
b_{+}(u,v) L^{+-}_m(v) L^{++}_m(u)
+
c_{-}(u,v) L^{++}_m(v) L^{+-}_m(u)
\end{align}
}
\end{example}

\begin{remark}
{\rm 
Let us specialize, for the moment, to models obeying Bose-Einstein statistics. An important property of the $L$-matrix (\ref{general-Lmat}) is the action of its entries on the local vacuum states $|0\rangle_m$ and $\langle 0|_m$. In all of the bosonic models which we study, these vacuum states will be eigenvectors of $L^{++}_m(u)$ and $L^{--}_m(u)$, giving rise to the equations

\begin{align}
L^{++}_m(u)|0\rangle_m &= \alpha_m(u) |0\rangle_m, 
\quad 
\langle 0|_m L^{++}_m(u) = \alpha_m(u) \langle 0|_m
\label{eigL1}
\\
L^{--}_m(u)|0\rangle_m &= \delta_m(u) |0\rangle_m,
\quad
\langle 0|_m L^{--}_m(u) = \delta_m(u) \langle 0|_m
\label{eigL2}
\end{align}

\noindent where $\alpha_m(u)$ and $\delta_m(u)$ are functions of $u$ which are specific to the model under consideration. Furthermore, $L^{+-}_m(u)$ and $L^{-+}_m(u)$ will play the role of creation and annihilation operators, giving rise to the equations

\begin{align}
L^{+-}_m(u)|0\rangle_m & \not= 0,
\quad
\langle 0|_m L^{+-}_m(u) = 0
\label{creatL}
\\
L^{-+}_m(u) |0\rangle_m & = 0,
\quad
\langle 0|_m L^{-+}_m(u) \not= 0 
\label{anniL}
\end{align}

\noindent Analogous statements also apply to the models that we study which obey Fermi-Dirac statistics. In those cases, the equations (\ref{eigL1})--(\ref{anniL}) still hold, but with $|0\rangle_m,\langle 0|_m$ replaced by $|\uparrow\rangle_m,\langle \uparrow|_m$. 
}
\end{remark}

\subsection{Graphical representation of $L$-matrix}

The objects introduced in the previous subsection admit the following graphical description. 
We identify each element of the matrix 
$L_{am}(u)$ with a vertex, as shown in figure 2.3.

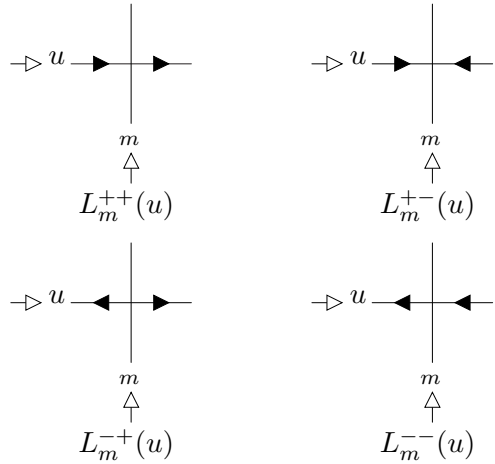
\begin{figure}[H]
\begin{center}
\begin{minipage}{4.3in}

\setlength{\unitlength}{0.0004cm}
\begin{picture}(20000,15000)(-9000,-13000)

\path(-2000,0000)(2000,0000)
\blacken\path(-1250,250)(-1250,-250)(-750,0)(-1250,250)
\blacken\path(750,250)(750,-250)(1250,0)(750,250)
\put(-2750,0){$u$}
\path(-4000,0)(-3000,0)
\whiten\path(-3500,250)(-3500,-250)(-3000,0)(-3500,250)
\path(0000,-2000)(0000,2000)
\put(-1750,-5000)
{$L^{++}_m(u)$}
\put(-400,-2700){\scriptsize$m$}
\path(0,-4000)(0,-3000)
\whiten\path(-250,-3500)(250,-3500)(0,-3000)(-250,-3500)

\path(8000,0000)(12000,0000)
\blacken\path(8750,250)(8750,-250)(9250,0)(8750,250)
\blacken\path(11250,250)(11250,-250)(10750,0)(11250,250)
\put(7250,0){$u$}
\path(6000,0)(7000,0)
\whiten\path(6500,250)(6500,-250)(7000,0)(6500,250)
\path(10000,-2000)(10000,2000)
\put(8250,-5000)
{$L^{+-}_m(u)$}
\put(9600,-2700){\scriptsize$m$}
\path(10000,-4000)(10000,-3000)
\whiten\path(9750,-3500)(10250,-3500)(10000,-3000)(9750,-3500)

\path(-2000,-8000)(2000,-8000)
\blacken\path(-750,-7750)(-750,-8250)(-1250,-8000)(-750,-7750)
\blacken\path(750,-7750)(750,-8250)(1250,-8000)(750,-7750)
\put(-2750,-8000){$u$}
\path(-4000,-8000)(-3000,-8000)
\whiten\path(-3500,-7750)(-3500,-8250)(-3000,-8000)(-3500,-7750)
\path(0000,-10000)(0000,-6000)
\put(-1750,-13000)
{$L^{-+}_m(u)$}
\put(-400,-10700){\scriptsize$m$}
\path(0,-12000)(0,-11000)
\whiten\path(-250,-11500)(250,-11500)(0,-11000)(-250,-11500)

\path(8000,-8000)(12000,-8000)
\blacken\path(9250,-7750)(9250,-8250)(8750,-8000)(9250,-7750)
\blacken\path(11250,-7750)(11250,-8250)(10750,-8000)(11250,-7750)
\put(7250,-8000){$u$}
\path(6000,-8000)(7000,-8000)
\whiten\path(6500,-7750)(6500,-8250)(7000,-8000)(6500,-7750)
\path(10000,-10000)(10000,-6000)
\put(8250,-13000)
{$L^{--}_m(u)$}
\put(9600,-10700){\scriptsize$m$}
\path(10000,-12000)(10000,-11000)
\whiten\path(9750,-11500)(10250,-11500)(10000,-11000)(9750,-11500)

\end{picture}

\end{minipage}
\end{center}

\caption[Four vertices of the generalized $L$-matrix]{Four vertices of the generalized $L$-matrix. Each entry of (\ref{general-Lmat}) is matched with a vertex.}
\end{figure}

Each vertex in figure 2.3 is the intersection of a horizontal line with two black arrows attached, and a blank vertical line. 
The variable $u$ flows through the horizontal line in the direction indicated, and the vertical line is marked with $m$, to indicate the $m^{\rm th}$ quantum space. On the horizontal line, when a black arrow points with the orientation it is assigned the value $+1$, and when a black arrow points against the orientation it is assigned the value $-1$. The vertical line has no values associated to it, and for the moment, it serves only to partition the horizontal line. 
The $L$-matrix element 
$L^{ij}_m(u)$ is matched with the vertex having horizontal line values $(i,j)$.

Using these graphical definitions, the local intertwining equation (\ref{general-intertwining-L-comp}) may be written in the form shown in figure 2.4. 

\begin{figure}[H]

\begin{center}
\begin{minipage}{4.3in}

\setlength{\unitlength}{0.0004cm}
\begin{picture}(20000,12500)(-2000,-4000)

\path(0,0)(5000,5000)
\put(1250,1250){\circle*{300}}
\put(1250,650){\tiny{$i_2$}}
\put(5000,5000){\circle*{300}}
\put(5000,4400){\tiny{$k_2$}}
\put(8750,5000){\circle*{300}}
\put(8750,4400){\tiny{$j_2$}}
\path(-2000,0)(-1000,0)
\whiten\path(-1250,250)(-1250,-250)(-1000,0)(-1250,250)
\put(-750,0){$v$}
\path(5000,0)(0,5000)
\put(1250,3750){\circle*{300}}
\put(1250,4000){\tiny{$i_1$}}
\put(5000,0){\circle*{300}}
\put(5000,-600){\tiny{$k_1$}}
\put(8750,0){\circle*{300}}
\put(8750,-600){\tiny{$j_1$}}
\path(-2000,5000)(-1000,5000)
\whiten\path(-1250,5250)(-1250,4750)(-1000,5000)(-1250,5250)
\put(-750,5000){$u$}
\path(5000,5000)(10000,5000)
\path(5000,0)(10000,0)
\path(7500,-2500)(7500,7500)

\put(7200,-3200){\scriptsize$m$}
\path(7500,-4500)(7500,-3500)
\whiten\path(7250,-3750)(7750,-3750)(7500,-3500)(7250,-3750)


\put(11500,2300){$=$}

\path(13000,5000)(14000,5000)
\whiten\path(13750,5250)(13750,4750)(14000,5000)(13750,5250)
\put(14250,5000){$u$}
\path(15000,5000)(20000,5000)
\put(16250,5000){\circle*{300}}
\put(16250,4400){\tiny{$i_1$}}
\path(20000,5000)(25000,0)
\put(20000,5000){\circle*{300}}
\put(20000,5250){\tiny{$k_1$}}
\put(23750,1250){\circle*{300}}
\put(23750,1600){\tiny{$j_1$}}
\path(13000,0)(14000,0)
\whiten\path(13750,250)(13750,-250)(14000,0)(13750,250)
\put(14250,0){$v$}
\path(15000,0)(20000,0)
\put(16250,0){\circle*{300}}
\put(16250,-600){\tiny{$i_2$}}
\path(20000,0)(25000,5000)
\put(20000,0){\circle*{300}}
\put(20000,-600){\tiny{$k_2$}}
\put(23750,3750){\circle*{300}}
\put(23750,3250){\tiny{$j_2$}}
\put(17200,-3200){\scriptsize$m$}
\path(17500,-4500)(17500,-3500)
\whiten\path(17250,-3750)(17750,-3750)(17500,-3500)(17250,-3750)

\path(17500,-2500)(17500,7500)

\end{picture}

\end{minipage}
\end{center}

\caption[Graphical depiction of the local intertwining equation]{Graphical depiction of the local intertwining equation.}

\end{figure}
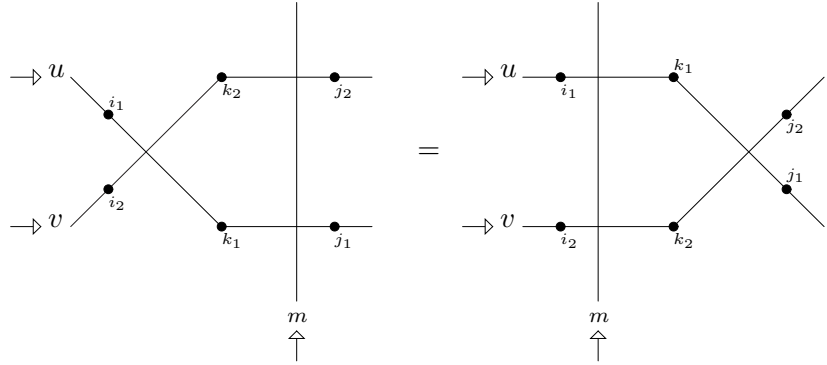

\noindent Both sides of this equation should be interpreted as three conjoined vertices. There is an $R$-matrix vertex at the intersection of the $(u,v)$ lines, and $L$-matrix vertices at the remaining two intersection points. The joining of these vertices is shorthand for multiplication of the three corresponding matrix elements. In this multiplication, the $L$-matrix elements are ordered from the one closest the vertical white arrow to the one farthest.

We label the horizontal external line segments by $\{i_1,i_2,j_1,j_2\}$. Each $i$ or $j$ represents an undisclosed black arrow, which is held fixed on both sides of the equation. The internal horizontal line segments have been labelled $\{k_1,k_2\}$. Each $k$ is summed over a black arrow that points with the orientation, and a black arrow that points against the orientation. 


\subsection{Monodromy matrix and global intertwining equation}

The {\it monodromy matrix} is defined as an $M$-fold product of $L$-matrices, where the product is taken over each site of the model. It is given explicitly by\footnote{In Chapter 5 we will define the monodromy matrix as $T_a(u) = L_{a1}(u)\ldots L_{aM}(u)$. This reversal of the quantum space ordering is merely a convenience and the results of this chapter hold for either definition.}

\begin{align}
T_a(u)
=
L_{aM}(u)\ldots L_{a1}(u)
=
\lprod_{m=1}^{M} L_{am}(u)
\label{general-monodromy1}
\end{align}

\noindent where $L_{am}(u)$ is the $L$-matrix (\ref{general-Lmat}) of the model, acting at the $m^{\rm th}$ site. For the models under our consideration, which have $2\times 2$ $L$-matrices, the monodromy matrix has the form

\begin{align}
T_a(u)
=
\left(
\begin{array}{cc}
A(u) & B(u)
\\
C(u) & D(u)
\end{array}
\right)_a
\label{general-monodromy2}
\end{align}

\noindent where we have placed the subscript $a$ on the monodromy matrix to denote the fact that it is an element of ${\rm End}(\mathcal{V}_a)$, with $\mathcal{V}_a$ a copy of $\mathbb{C}^2$. 

The entries $A(u),B(u),C(u),D(u)$ in (\ref{general-monodromy2}) are the operators which result from performing the multiplication (\ref{general-monodromy1}) of $L$-matrices. They are dependent on $u$, since each $L$-matrix in the product depends on $u$, and they are sums of $2^{M-1}$ monomials. These monomials, in turn, are products of $M$ local operators, one acting at each site of the model. Therefore, in general, the entries of the monodromy matrix are complicated elements of $\mathcal{A}_1\otimes\cdots\otimes\mathcal{A}_M$. For notational convenience, it is conventional to suppress the ${\rm End}(\mathcal{V}_1\otimes\cdots\otimes\mathcal{V}_M)$ dependence of these entries.

\begin{lemma}
\label{2-lem1}
{\rm 
The monodromy matrix satisfies the equation 

\begin{align}
R_{ab}(u,v) T_{a}(u) T_{b}(v)
=
T_{b}(v) T_{a}(u) R_{ab}(u,v)
\label{general-intertwining-T}
\end{align}

\noindent which is an identity acting in the tensor product $\mathcal{V}_a\otimes \mathcal{V}_b$ of two auxiliary spaces, for general values of the rapidities $u,v$. We refer to this as the {\it global intertwining equation}, since the $R$-matrix $R_{ab}(u,v)$ now intertwines the monodromy matrices $T_a(u), T_b(v)$ which act over all sites in the model.
}
\end{lemma}

\begin{proof} 
The result (\ref{general-intertwining-T}) is a corollary of the local intertwining relation 
(\ref{general-intertwining-L}). Using the definition (\ref{general-monodromy1}) of the monodromy matrix, we write

\begin{align}
R_{ab}(u,v) T_a(u) T_b(v)
&=
R_{ab}(u,v) 
\left(\lprod_{m=1}^{M} L_{am}(u)\right) 
\left(\lprod_{m=1}^{M} L_{bm}(v)\right)
\label{global-proof1}
\\
&=
R_{ab}(u,v)
\lprod_{m=1}^{M}
\Big(L_{am}(u) L_{bm}(v)\Big)
\nonumber
\end{align}

\noindent where we have changed the ordering of the $L$-matrices by commuting those which act in different spaces. Applying (\ref{general-intertwining-L}) $M$ times successively to the right hand side of (\ref{global-proof1}), we obtain 

\begin{align}
R_{ab}(u,v) T_a(u) T_b(v)
&=
\lprod_{m=1}^{M}
\Big(L_{bm}(v) L_{am}(u)\Big)
R_{ab}(u,v)
\\
&=
\left(\lprod_{m=1}^{M} L_{bm}(v) \right)
\left(\lprod_{m=1}^{M} L_{am}(u) \right)
R_{ab}(u,v)
\nonumber
\\
&=
T_b(v) T_a(u) R_{ab}(u,v)
\nonumber
\end{align}

\noindent where we have restored the original ordering of the $L$-matrices to complete the proof.
\end{proof}

As we have done with earlier equations, we can write (\ref{general-intertwining-T}) in component notation, obtaining

\begin{align}
R^{i_1 k_1}_{i_2 k_2}(u,v) T^{k_1 j_1}(u) T^{k_2 j_2}(v)
=
T^{i_2 k_2}(v) T^{i_1 k_1}(u) R^{k_1 j_1}_{k_2 j_2}(u,v)
\label{general-intertwining-T-comp}
\end{align}

\noindent where all indices take values in $\{+1,-1\}$, with each of $\{i_1,i_2,j_1,j_2\}$ held fixed, while $\{k_1,k_2\}$ are summed. Hence (\ref{general-intertwining-T-comp}) gives rise to $2^4$ commutation relations involving the operators $A(u),B(u),C(u),D(u)$, one corresponding to each configuration of the indices $\{i_1,i_2,j_1,j_2\}$.

\begin{example}
{\rm
We will list four of the commutation relations contained in (\ref{general-intertwining-T-comp}), since they are used later in this chapter. Setting $\{i_1,i_2,j_1,j_2\} = \{+,+,-,-\}$ in (\ref{general-intertwining-T-comp}), we obtain
 
\begin{align}
R^{++}_{++}(u,v) T^{+-}(u) T^{+-}(v)
&=
T^{+-}(v) T^{+-}(u) R^{--}_{--}(u,v)
\nonumber
\\
\implies
a_{+}(u,v)
B(u)B(v)
&=
a_{-}(u,v)
B(v)B(u)
\label{general-BB}
\end{align}


\noindent Setting $\{i_1,i_2,j_1,j_2\} = \{-,-,+,+\}$ in (\ref{general-intertwining-T-comp}), we obtain

\begin{align}
R^{--}_{--}(u,v) T^{-+}(u) T^{-+}(v)
&=
T^{-+}(v) T^{-+}(u) R^{++}_{++}(u,v)
\nonumber
\\
\implies
a_{-}(u,v)
C(u)C(v)
&=
a_{+}(u,v)
C(v)C(u)
\label{general-CC}
\end{align}  

\noindent Setting $\{i_1,i_2,j_1,j_2\} = \{+,+,-,+\}$ in (\ref{general-intertwining-T-comp}), we obtain

\begin{align}
R^{++}_{++}(u,v) T^{+-}(u) T^{++}(v)
&=
T^{+-}(v) T^{++}(u) R^{+-}_{-+}(u,v)
+
T^{++}(v) T^{+-}(u) R^{--}_{++}(u,v)
\nonumber
\\
\implies
a_{+}(u,v) B(u) A(v)
&=
c_{+}(u,v) B(v) A(u)
+
b_{-}(u,v) A(v) B(u)
\label{general-AB}
\end{align}

\noindent Finally, setting $\{i_1,i_2,j_1,j_2\} = \{-,+,-,-\}$ in (\ref{general-intertwining-T-comp}), we obtain

\begin{align}
R^{-+}_{+-}(u,v) T^{+-}(u) T^{--}(v)
+
R^{--}_{++}(u,v) T^{--}(u) T^{+-}(v)
&=
T^{+-}(v) T^{--}(u) R^{--}_{--}(u,v)
\nonumber
\\
\implies
c_{-}(u,v) B(u) D(v)
+
b_{-}(u,v) D(u) B(v)
&=
a_{-}(u,v) B(v) D(u)
\label{general-DB}
\end{align}
}
\end{example}

\begin{lemma}
\label{2-lem2}
{\rm 
Let us consider, once again, models which obey Bose-Einstein statistics. Defining the global vacua

\begin{align}
|0\rangle = |0\rangle_1 \otimes \cdots \otimes |0\rangle_M,
\quad
\langle 0| = \langle 0|_1 \otimes \cdots \otimes \langle 0|_M
\end{align}

\noindent we find that these states are eigenvectors of the operators $A(u)$ and $D(u)$, giving rise to the equations

\begin{align}
A(u)|0\rangle &= \alpha(u) |0\rangle = \prod_{m=1}^{M} \alpha_m(u) |0\rangle, 
\quad
\langle 0| A(u) = \alpha(u) \langle 0| = \prod_{m=1}^{M} \alpha_m(u) \langle 0|  
\label{eigT1}
\\
D(u)|0\rangle &= \delta(u) |0\rangle = \prod_{m=1}^{M} \delta_m(u) |0\rangle, 
\quad
\langle 0| D(u) = \delta(u) \langle 0| = \prod_{m=1}^{M} \delta_m(u) \langle 0|  
\label{eigT2}
\end{align}

\noindent where $\alpha_m(u)$ and $\delta_m(u)$ are the eigenvalues as defined in (\ref{eigL1}) and (\ref{eigL2}), respectively. In addition, $B(u)$ and $C(u)$ play the role of global creation and annihilation operators, giving rise to the equations

\begin{align}
B(u) |0\rangle &\not= 0,
\quad
\langle 0| B(u) = 0
\label{annirule1}
\\
C(u) |0\rangle & = 0,
\quad
\langle 0|C(u) \not= 0
\label{annirule2}
\end{align}

\noindent Similar equations apply to models with Fermi-Dirac statistics, by replacing the global vacua $|0\rangle,\langle 0|$ in (\ref{eigT1})--(\ref{annirule2}) with the spin-up states

\begin{align} 
|\Uparrow\rangle 
= 
|\uparrow\rangle_1 \otimes \cdots \otimes |\uparrow\rangle_M,
\quad
\langle \Uparrow|
=
\langle \uparrow|_1 \otimes \cdots \otimes \langle \uparrow|_M
\end{align}

}
\end{lemma}

\begin{proof} 
These equations follow immediately from the definition of the monodromy matrix (\ref{general-monodromy1}), and from the actions (\ref{eigL1})--(\ref{anniL}) of the $L$-matrix entries on the local vacuum states. For example,

\begin{align}
A(u) |0\rangle 
&=
\left( 
\begin{array}{cc}
1 & 0
\end{array}
\right)
\left(
\begin{array}{cc}
L^{++}_M |0\rangle_M & L^{+-}_M |0\rangle_M
\\
L^{-+}_M |0\rangle_M & L^{--}_M |0\rangle_M
\end{array}
\right)
\ldots
\left(
\begin{array}{cc}
L^{++}_1 |0\rangle_1 & L^{+-}_1 |0\rangle_1
\\
L^{-+}_1 |0\rangle_1 & L^{--}_1 |0\rangle_1
\end{array}
\right)
\left(
\begin{array}{c}
1 \\ 0
\end{array}
\right)
\nonumber
\\
&=
\left( 
\begin{array}{cc}
1 & 0
\end{array}
\right)
\left(
\begin{array}{cr}
\alpha_M |0\rangle_M & L^{+-}_M |0\rangle_M
\\
0 & \delta_M |0\rangle_M
\end{array}
\right)
\ldots
\left(
\begin{array}{cr}
\alpha_1 |0\rangle_1 & L^{+-}_1 |0\rangle_1
\\
0 & \delta_1 |0\rangle_1
\end{array}
\right)
\left(
\begin{array}{c}
1 \\ 0
\end{array}
\right)\nonumber
\\
&=
\prod_{m=1}^{M} \alpha_m(u)
|0\rangle_1 \otimes \cdots \otimes |0\rangle_M
\end{align}

\noindent The remaining equations are proved analogously.  
\end{proof}

\subsection{Graphical representation of monodromy matrix}

Using the previous diagrammatic conventions for $L$-matrix elements, we identify each element of the matrix 
$T_a(u)$ with a {\it string of vertices}, as shown in figure 2.5.

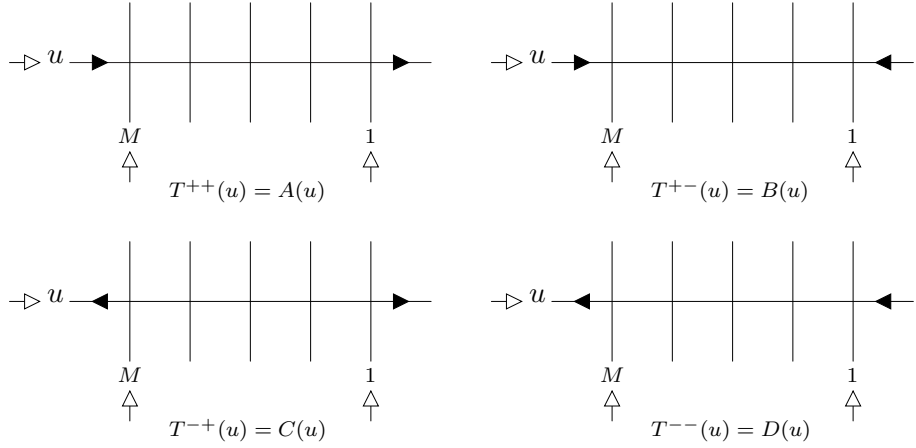
\begin{figure}[H]

\begin{center}
\begin{minipage}{4.3in}

\setlength{\unitlength}{0.0004cm}
\begin{picture}(20000,15000)(-2500,-13000)

\path(-2000,0000)(10000,0000)
\blacken\path(-1250,250)(-1250,-250)(-750,0)(-1250,250)
\blacken\path(8750,250)(8750,-250)(9250,0)(8750,250)
\put(-2750,0){$u$}
\path(-4000,0)(-3000,0)
\whiten\path(-3500,250)(-3500,-250)(-3000,0)(-3500,250)
\path(0000,-2000)(0000,2000)
\put(-400,-2700){\scriptsize$M$}
\path(0,-4000)(0,-3000)
\whiten\path(-250,-3500)(250,-3500)(0,-3000)(-250,-3500)
\path(2000,-2000)(2000,2000)
\path(4000,-2000)(4000,2000)
\path(6000,-2000)(6000,2000)
\path(8000,-2000)(8000,2000)
\put(7800,-2700){\scriptsize$1$}
\path(8000,-4000)(8000,-3000)
\whiten\path(7750,-3500)(8250,-3500)(8000,-3000)(7750,-3500)
\put(1300,-4500)
{\scriptsize$T^{++}(u) = A(u)$}

\path(14000,0000)(26000,0000)
\blacken\path(14750,250)(14750,-250)(15250,0)(14750,250)
\blacken\path(25250,250)(25250,-250)(24750,0)(25250,250)
\put(13250,0){$u$}
\path(12000,0)(13000,0)
\whiten\path(12500,250)(12500,-250)(13000,0)(12500,250)
\path(16000,-2000)(16000,2000)
\put(15600,-2700){\scriptsize$M$}
\path(16000,-4000)(16000,-3000)
\whiten\path(15750,-3500)(16250,-3500)(16000,-3000)(15750,-3500)
\path(18000,-2000)(18000,2000)
\path(20000,-2000)(20000,2000)
\path(22000,-2000)(22000,2000)
\path(24000,-2000)(24000,2000)
\put(23800,-2700){\scriptsize$1$}
\path(24000,-4000)(24000,-3000)
\whiten\path(23750,-3500)(24250,-3500)(24000,-3000)(23750,-3500)
\put(17300,-4500)
{\scriptsize$T^{+-}(u)=B(u)$}

\path(-2000,-8000)(10000,-8000)
\blacken\path(-750,-7750)(-750,-8250)(-1250,-8000)(-750,-7750)
\blacken\path(8750,-7750)(8750,-8250)(9250,-8000)(8750,-7750)
\put(-2750,-8000){$u$}
\path(-4000,-8000)(-3000,-8000)
\whiten\path(-3500,-7750)(-3500,-8250)(-3000,-8000)(-3500,-7750)
\path(0000,-10000)(0000,-6000)
\put(-400,-10700){\scriptsize$M$}
\path(0,-12000)(0,-11000)
\whiten\path(-250,-11500)(250,-11500)(0,-11000)(-250,-11500)
\path(2000,-10000)(2000,-6000)
\path(4000,-10000)(4000,-6000)
\path(6000,-10000)(6000,-6000)
\path(8000,-10000)(8000,-6000)
\put(7800,-10700){\scriptsize$1$}
\path(8000,-12000)(8000,-11000)
\whiten\path(7750,-11500)(8250,-11500)(8000,-11000)(7750,-11500)
\put(1300,-12500)
{\scriptsize$T^{-+}(u)=C(u)$}

\path(14000,-8000)(26000,-8000)
\blacken\path(15250,-7750)(15250,-8250)(14750,-8000)(15250,-7750)
\blacken\path(25250,-7750)(25250,-8250)(24750,-8000)(25250,-7750)
\put(13250,-8000){$u$}
\path(12000,-8000)(13000,-8000)
\whiten\path(12500,-7750)(12500,-8250)(13000,-8000)(12500,-7750)
\path(16000,-10000)(16000,-6000)
\put(15600,-10700){\scriptsize$M$}
\path(16000,-12000)(16000,-11000)
\whiten\path(15750,-11500)(16250,-11500)(16000,-11000)(15750,-11500)
\path(18000,-10000)(18000,-6000)
\path(20000,-10000)(20000,-6000)
\path(22000,-10000)(22000,-6000)
\path(24000,-10000)(24000,-6000)
\put(23800,-10700){\scriptsize$1$}
\path(24000,-12000)(24000,-11000)
\whiten\path(23750,-11500)(24250,-11500)(24000,-11000)(23750,-11500)
\put(17300,-12500)
{\scriptsize$T^{--}(u)=D(u)$}
\end{picture}

\end{minipage}
\end{center}

\caption[Four vertex-strings of the monodromy matrix]{Four vertex-strings of the monodromy matrix. Each entry of (\ref{general-monodromy2}) is matched with a string of $L$-matrix vertices.}
\end{figure}


\noindent These diagrams are interpreted as $M$ conjoined $L$-matrix vertices which represent, from left to right, the multiplication of the corresponding $L$-matrix elements. The horizontal internal line segments are summed over black arrows that point with the orientation, and black arrows that point against the orientation, so each string of vertices implies a sum over $2^{M-1}$ terms. The monodromy matrix element 
$T^{ij}(u)$ is matched with the string of vertices having external horizontal line values $(i,j)$. 

The global intertwining equation (\ref{general-intertwining-T-comp}) has the diagrammatic form

\begin{figure}[H]

\begin{center}
\begin{minipage}{4.3in}

\setlength{\unitlength}{0.00028cm}
\begin{picture}(20000,12500)(-500,-4000)

\path(0,0)(5000,5000)
\put(1250,1250){\circle*{300}}
\put(1250,550){\tiny{$i_2$}}
\put(5000,5000){\circle*{300}}
\put(5000,4200){\tiny$k_2$}
\path(-2000,0)(-1000,0)
\whiten\path(-1250,250)(-1250,-250)(-1000,0)(-1250,250)
\put(-850,0){$v$}
\path(5000,0)(0,5000)
\put(1250,3750){\circle*{300}}
\put(1250,4000){\tiny{$i_1$}}
\put(5000,0){\circle*{300}}
\put(5000,-800){\tiny$k_1$}
\path(-2000,5000)(-1000,5000)
\whiten\path(-1250,5250)(-1250,4750)(-1000,5000)(-1250,5250)
\put(-850,5000){$u$}
\path(5000,5000)(17500,5000)
\put(16250,5000){\circle*{300}}
\put(16250,4300){\tiny{$j_2$}}
\path(5000,0)(17500,0)
\put(16250,0){\circle*{300}}
\put(16250,-700){\tiny{$j_1$}}
\path(7500,-2500)(7500,7500)
\put(7100,-3400){\scriptsize$M$}
\path(7500,-4750)(7500,-3750)
\whiten\path(7250,-4000)(7750,-4000)(7500,-3750)(7250,-4000)
\path(10000,-2500)(10000,7500)
\path(12500,-2500)(12500,7500)
\path(15000,-2500)(15000,7500)
\put(14800,-3400){\scriptsize$1$}
\path(15000,-4750)(15000,-3750)
\whiten\path(14750,-4000)(15250,-4000)(15000,-3750)(14750,-4000)


\put(18000,2300){$=$}

\path(19500,5000)(20500,5000)
\whiten\path(20250,5250)(20250,4750)(20500,5000)(20250,5250)
\put(20650,5000){$u$}
\path(21500,5000)(34000,5000)
\put(22750,5000){\circle*{300}}
\put(22750,4300){\tiny{$i_1$}}
\path(34000,5000)(39000,0)
\put(34000,5000){\circle*{300}}
\put(34000,5300){\tiny$k_1$}
\put(37750,1250){\circle*{300}}
\put(37750,200){\tiny{$j_1$}}
\path(19500,0)(20500,0)
\whiten\path(20250,250)(20250,-250)(20500,0)(20250,250)
\put(20650,0){$v$}
\path(21500,0)(34000,0)
\put(22750,0){\circle*{300}}
\put(22750,-800){\tiny{$i_2$}}
\path(34000,0)(39000,5000)
\put(34000,0){\circle*{300}}
\put(34000,-800){\tiny$k_2$}
\put(37750,3750){\circle*{300}}
\put(37750,3150){\tiny{$j_2$}}
\put(23600,-3400){\scriptsize$M$}
\path(24000,-4750)(24000,-3750)
\whiten\path(23750,-4000)(24250,-4000)(24000,-3750)(23750,-4000)
\path(24000,-2500)(24000,7500)
\path(26500,-2500)(26500,7500)
\path(29000,-2500)(29000,7500)
\put(31300,-3400){\scriptsize$1$}
\path(31500,-4750)(31500,-3750)
\whiten\path(31250,-4000)(31750,-4000)(31500,-3750)(31250,-4000)
\path(31500,-2500)(31500,7500)

\end{picture}

\end{minipage}
\end{center}

\caption[Graphical depiction of the global intertwining equation]{Graphical depiction of the global intertwining equation. This follows by repeated application of the local intertwining equation, which moves the $R$-matrix vertex attached on the left until it emerges from the right.}
\end{figure}
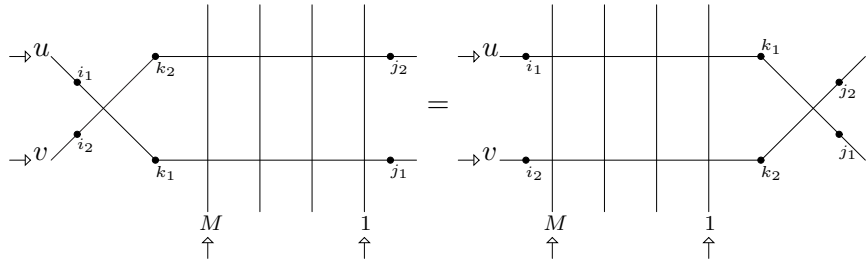

\noindent Both sides of this equation contain an $R$-matrix vertex at the intersection of the $(u,v)$ lines, and $L$-matrix vertices at the remaining $2M$ intersection points. The joining of these vertices is shorthand for multiplication of the corresponding matrix elements. In this multiplication, the $L$-matrix vertices are ordered from those closest the vertical white arrows to those farthest.

We write the external line segments as $\{i_1,i_2,j_1,j_2\}$. Each $i$ or $j$ represents an undisclosed black arrow, which is held fixed on both sides of the equation. 
For simplicity we have not labelled the horizontal internal line segments, which are summed over black arrows that point with the orientation, and black arrows that point against the orientation. 

\subsection{Transfer matrix and quantum trace identities}
 
In this subsection we describe the reconstruction of the Hamiltonian $\mathcal{H}$ in terms of a set of commuting operators. This reconstruction allows us to study the spectrum of $\mathcal{H}$ jointly with another operator, the {\it transfer matrix.}   

\begin{lemma}
{\rm 
Define the transfer matrix $t(u)$ as the trace of the monodromy matrix, taken in the auxiliary space $\mathcal{V}_a$. In other words, let

\begin{align}
t(u) = {\rm tr}_a T_a(u) = A(u)+D(u)
\label{general-transfer}
\end{align}

\noindent Then for all $u,v \in \mathbb{C}$ the transfer matrices $t(u),t(v)$ satisfy

\begin{align}
[t(u),t(v)] = 0
\label{tran-com}
\end{align}
}
\end{lemma}

\begin{proof}

Consider the global intertwining equation (\ref{general-intertwining-T}). Multiplying this equation from the left by the inverse of the $R$-matrix (\ref{general-Rmat}) and taking the trace over $\mathcal{V}_a$ and $\mathcal{V}_b$, we obtain

\begin{align}
{\rm tr}_a T_a(u) 
{\rm tr}_b T_b(v)
&=
{\rm tr}_a
{\rm tr}_b
\Big(
R_{ab}^{-1}(u,v)
T_b(v) T_a(u)
R_{ab}(u,v)
\Big)
\label{tran-proof}
\\
&=
{\rm tr}_a
{\rm tr}_b
\Big(
T_b(v) T_a(u)
R_{ab}(u,v)
R_{ab}^{-1}(u,v)
\Big)
\nonumber
\\
&=
{\rm tr}_b T_b(v)
{\rm tr}_a T_a(u)
\nonumber
\end{align}

\noindent where the second line of (\ref{tran-proof}) follows from the cyclicity of the trace. Recalling the definition (\ref{general-transfer}) of the transfer matrix, the final line of (\ref{tran-proof}) completes the proof.

\end{proof}

The transfer matrix can be viewed as a generating function of the conserved quantities of a given quantum integrable model. To see this, one expands $t(u)$ in powers of $u$ or $e^u$ (depending on the particular model) to obtain

\begin{align}
t(u) = \sum_{n} \mathfrak{t}_n u^n,\quad
{\rm or} \quad
t(u) = \sum_{n} \mathfrak{t}_n e^{nu}
\label{tran-gf}
\end{align}

\noindent where each $\mathfrak{t}_n$ is an element of $\mathcal{A}_1 \otimes \cdots \otimes \mathcal{A}_M$. Substituting the generating function (\ref{tran-gf}) into the commutation relation (\ref{tran-com}), it follows that the quantities $\mathfrak{t}_m,\mathfrak{t}_n$ satisfy

\begin{align}
[\mathfrak{t}_m,\mathfrak{t}_n] = 0, \quad
{\rm for\ all}\ m,n
\end{align}

\noindent meaning that they are in involution. An {\it integrable} quantum model is one whose Hamiltonian $\mathcal{H}$ may be expressed as an algebraic combination of the operators $\mathfrak{t}_n$, via a {\it quantum trace identity.} Whilst the specifics of this reconstruction depend on the model being studied, we always recover the equation

\begin{align}
[\mathcal{H},t(u)] = 0, \quad {\rm for\ all}\ u \in \mathbb{C}
\label{tran-ham}
\end{align}

\noindent as a consequence. This leads us to the following important result.

\begin{lemma}
{\rm
Let $|\Psi\rangle \in \mathcal{V}$ be an eigenvector of $\mathcal{H}$ with non-degenerate eigenvalue $\mathcal{E}_{\Psi}$. Then $|\Psi\rangle$ is also an eigenvector of $t(u)$.
}
\end{lemma}

\begin{proof}
Using the commutation relation (\ref{tran-ham}), we have

\begin{align}
\mathcal{H} t(u) |\Psi\rangle
=
t(u) \mathcal{H} |\Psi\rangle
=
\mathcal{E}_{\Psi} t(u) |\Psi\rangle
\end{align}

\noindent Hence $t(u) |\Psi\rangle$ is an eigenstate of $\mathcal{H}$ with eigenvalue $\mathcal{E}_{\Psi}$. Since this eigenvalue is non-degenerate, it follows that $t(u) |\Psi\rangle = \tau_{\Psi}(u) |\Psi\rangle$ for some scalar function $\tau_{\Psi}(u)$. This proves that $|\Psi\rangle$ is also an eigenvector of $t(u)$. 

\end{proof}

By virtue of lemma 4, we see that all non-degenerate eigenvectors of $\mathcal{H}$ are also eigenvectors of $t(u)$. Hence the non-degenerate spectrum of $\mathcal{H}$ can be recovered by studying the spectrum of $t(u)$. This will be the focus of the next section.

\section{Algebraic Bethe Ansatz}
\label{aba-aba}
  
\subsection{Construction of the Bethe eigenvectors}

Our goal in this subsection is to find vectors $|\Psi\rangle \in \mathcal{V}$ and $\langle \Psi| \in \mathcal{V}^{*}$ such that

\begin{align}
t(u)
|\Psi\rangle =
\tau_{\Psi}(u) |\Psi\rangle,
\quad
\langle\Psi| 
t(u)
=
\tau_{\Psi}(u)
\langle\Psi|
\label{eigeneqn}
\end{align}

\noindent where $t(u)$ is the transfer matrix (\ref{general-transfer}) and $\tau_{\Psi}(u)$ are suitable scalar functions. This goal is achieved by making an educated guess at the form of the eigenvectors, known as the {\it algebraic Bethe Ansatz.} In the following theorem, the eigenvectors $|\Psi\rangle$ and $\langle \Psi |$ are constructed using the off-diagonal entries of the monodromy matrix (\ref{general-monodromy2}). The action of $t(u) = A(u) + D(u)$ on the proposed eigenvectors can then be calculated using the commutation relations (\ref{general-intertwining-T-comp}).
 
\begin{theorem}
{\rm 
Suppose that the entries of the $R$-matrix (\ref{general-Rmat}) satisfy\footnote{The models studied in chapters 3,4,5 have $R$-matrices which obey (\ref{Rmat-cond}). In chapter 6, we study models which require a slightly separate treatment.} 

\begin{align}
a_{\pm}(u,v) = a(u,v),\ \   
b_{\pm}(u,v) = -b_{\pm}(v,u),\ \   
c_{\pm}(u,v) = c(u,v) = c(v,u)
\label{Rmat-cond}
\end{align}
 
\noindent Then the state vectors\footnote{Throughout the theorem, we specialize to models with Bose-Einstein statistics. The results obtained apply equally to fermionic models, by replacing all instances of $|0\rangle, \langle 0|$ with $|\Uparrow\rangle, \langle \Uparrow|$.}

\begin{align}
|\Psi\rangle 
&= 
B(v_1) \ldots B(v_N) |0\rangle 
= 
\rprod_{n=1}^{N} B(v_n) |0\rangle
\label{Bet1}
\\
\langle\Psi| 
&= 
\langle 0|C(v_N)\ldots C(v_1)
=
\langle 0| \lprod_{n=1}^{N} C(v_n)
\label{Bet2}
\end{align}

\noindent are solutions of the eigenvector equations (\ref{eigeneqn}), with eigenvalues $\tau_{\Psi}(u)$ given by

\begin{align}
\tau_{\Psi}(u)
=
\alpha(u)
\prod_{n=1}^{N}
\frac{a(v_n,u)}{b_{-}(v_n,u)}
+
\delta(u)
\prod_{n=1}^{N}
\frac{a(u,v_n)}{b_{-}(u,v_n)}
\label{eigenpsi}
\end{align} 

\noindent provided that the variables $\{v_1,\ldots,v_N\}$ satisfy the system of coupled equations

\begin{align}
\alpha(v_i)
\displaystyle{\prod_{\substack{n\not= i \\ n=1}}^{N}
a(v_n,v_i)
+
(-)^N
\delta(v_i)
\prod_{\substack{n \not= i \\ n=1}}^{N}
a(v_i,v_n)
=
0
}
\label{Beteq}
\end{align}

\noindent for all $1 \leq i \leq N$. We refer to (\ref{Bet1}) and (\ref{Bet2}) as the {\it Bethe eigenvectors} of the model, and to the constraints (\ref{Beteq}) as the {\it Bethe equations.}
}
\end{theorem}

\begin{proof}
Following the procedure given in chapter VII of \cite{kbi}, we shall prove the theorem for (\ref{Bet1}), but omit the proof for (\ref{Bet2}) as it is very similar. Acting on the vector (\ref{Bet1}) with the transfer matrix $t(u)$, we obtain

\begin{align}
t(u)|\Psi\rangle
=
\Big(A(u)+D(u)\Big)|\Psi\rangle
=
A(u) \prod_{n=1}^{N} B(v_n) |0\rangle
+
D(u) \prod_{n=1}^{N} B(v_n) |0\rangle
\label{proof-step1}
\end{align}

\noindent where the products in (\ref{proof-step1}) are left unordered, because when $a_{+}(v_i,v_j) = a_{-}(v_i,v_j)$ the commutation relation (\ref{general-BB}) yields

\begin{align}
[B(v_i),B(v_j)] = 0,
\quad {\rm for\ all}\ 1 \leq i,j \leq N
\end{align} 

\noindent In order to show that $|\Psi\rangle$ is an eigenvector of $t(u)$, we must calculate the two terms on the right hand side of (\ref{proof-step1}). To this end, let $\mathcal{P}_N$ denote the proposition  

\begin{align}
A(u) \prod_{n=1}^{N} B(v_n) |0\rangle
&=
\alpha(u)
\prod_{n=1}^{N}
\Big[
\frac{a(v_n,u)}{b_{-}(v_n,u)}
B(v_n)
\Big]
|0\rangle
\label{proof-step2}
\\
&-
\sum_{i=1}^{N}
\alpha(v_i)
\frac{c(v_i,u)}{b_{-}(v_i,u)}
B(u)
\prod_{\substack{n\not=i \\ n=1}}^{N}
\Big[
\frac{a(v_n,v_i)}{b_{-}(v_n,v_i)}
B(v_n)
\Big]
|0\rangle
\nonumber
\end{align}

\noindent which we will prove for general $N \geq 1$. To begin, we use the commutation relation (\ref{general-AB}) with $u \rightarrow v_1, v \rightarrow u$ and equation (\ref{eigT1}) to calculate

\begin{align}
A(u)B(v_1) |0\rangle
=
\alpha(u)
\frac{a(v_1,u)}{b_{-}(v_1,u)}
B(v_1)|0\rangle
-
\alpha(v_1)
\frac{c(v_1,u)}{b_{-}(v_1,u)}
B(u)|0\rangle
\end{align}


\noindent where we have used the fact that $a_{+}(v_1,u) = a(v_1,u),  c_{+}(v_1,u) = c(v_1,u)$. This establishes that $\mathcal{P}_1$ is true. Now suppose $\mathcal{P}_{m-1}$ is true for some integer $m \geq 2$. Once again, using the commutation relation (\ref{general-AB}) we find that

\begin{align}
A(u) \prod_{n=1}^{m} B(v_n) |0\rangle
&=
\frac{a(v_1,u)}{b_{-}(v_1,u)}
B(v_1) A(u) \prod_{n=2}^{m} B(v_n) |0\rangle
\label{m-1}
\\
&-
\frac{c(v_1,u)}{b_{-}(v_1,u)}
B(u) A(v_1) \prod_{n=2}^{m} B(v_n) |0\rangle
\nonumber
\end{align}

\noindent Since $\mathcal{P}_{m-1}$ holds, we are able to explicitly calculate the terms on the right hand side of (\ref{m-1}). Substituting $\mathcal{P}_{m-1}$ into (\ref{m-1}) we obtain

{\small
\begin{align}
A(u) \prod_{n=1}^{m} B(v_n) |0\rangle
&=
\frac{a(v_1,u)}{b_{-}(v_1,u)}B(v_1)
\alpha(u) 
\prod_{n=2}^{m}
\Big[
\frac{a(v_n,u)}{b_{-}(v_n,u)}
B(v_n)
\Big] 
|0\rangle
\label{2and4}
\\
&
-
\frac{a(v_1,u)}{b_{-}(v_1,u)}B(v_1)
\sum_{i=2}^{m}
\alpha(v_i)
\frac{c(v_i,u)}{b_{-}(v_i,u)}
B(u)
\prod_{\substack{n\not=i \\ n=2}}^{m}
\Big[
\frac{a(v_n,v_i)}{b_{-}(v_n,v_i)}
B(v_n)
\Big]
|0\rangle
\nonumber
\\
&
-
\frac{c(v_1,u)}{b_{-}(v_1,u)} B(u)
\alpha(v_1) 
\prod_{n=2}^{m}
\Big[
\frac{a(v_n,v_1)}{b_{-}(v_n,v_1)}
B(v_n)
\Big] 
|0\rangle
\nonumber
\\
&
+
\frac{c(v_1,u)}{b_{-}(v_1,u)} B(u)
\sum_{i=2}^{m}
\alpha(v_i)
\frac{c(v_i,v_1)}{b_{-}(v_i,v_1)}
B(v_1)
\prod_{\substack{n\not=i \\ n=2}}^{m}
\Big[
\frac{a(v_n,v_i)}{b_{-}(v_n,v_i)}
B(v_n)
\Big] 
|0\rangle
\nonumber 
\end{align}
}

\noindent Now consider the single Yang-Baxter equation (\ref{general-YB-example-pt2}), as given in example 1. Recalling the assumptions (\ref{Rmat-cond}) and setting $u \rightarrow v_1, v \rightarrow v_i, w \rightarrow u$ in this equation, it becomes

\begin{align}
-
b_{-}(v_i,v_1) a(v_1,u) c(v_i,u) 
+
c(v_i,v_1) c(v_1,u) b_{-}(v_i,u) 
=
c(v_i,u) b_{-}(v_1,u) a(v_1,v_i)
\label{littleYB}
\end{align}

\noindent where we have used $b_{-}(v_1,v_i) = -b_{-}(v_i,v_1),\ c(v_1,v_i) = c(v_i,v_1)$ to change the order of $v_1,v_i$ on the left hand side. By virtue of (\ref{littleYB}) we are able to combine the second and fourth term on the right hand side of (\ref{2and4}), which yields

\begin{align}
A(u)\prod_{n=1}^{m} B(v_n) |0\rangle
&=
\alpha(u) 
\prod_{n=1}^{m}
\Big[
\frac{a(v_n,u)}{b_{-}(v_n,u)}
B(v_n)
\Big] 
|0\rangle
\\
&
-
\alpha(v_1)
\frac{c(v_1,u)}{b_{-}(v_1,u)} 
B(u) 
\prod_{n=2}^{m}
\Big[
\frac{a(v_n,v_1)}{b_{-}(v_n,v_1)}
B(v_n) 
\Big]
|0\rangle
\nonumber
\\
&
-
\sum_{i=2}^{m}
\alpha(v_i)
\frac{c(v_i,u)}{b_{-}(v_i,u)}B(u)
\prod_{\substack{n\not=i \\ n=1}}^{m}
\Big[
\frac{a(v_n,v_i)}{b_{-}(v_n,v_i)}
B(v_n)
\Big] |0\rangle
\nonumber
\end{align}

\noindent proving that $\mathcal{P}_m$ is true. Therefore by induction $\mathcal{P}_N$ is true for arbitrary $N \geq 1$. By analogous arguments, which use the commutation relation (\ref{general-DB}) and the equation (\ref{eigT2}), we are also able to show that 

\begin{align}
D(u) \prod_{n=1}^{N} B(v_n) |0\rangle
&=
\delta(u)
\prod_{n=1}^{N}
\Big[
\frac{a(u,v_n)}{b_{-}(u,v_n)}
B(v_n)
\Big]
|0\rangle
\label{proof-step3}
\\
&-
\sum_{i=1}^{N}
\delta(v_i)
\frac{c(u,v_i)}{b_{-}(u,v_i)}
B(u)
\prod_{\substack{n\not=i \\ n=1}}^{N}
\Big[
\frac{a(v_i,v_n)}{b_{-}(v_i,v_n)}
B(v_n)
\Big]
|0\rangle
\nonumber
\end{align}

\noindent for arbitrary $N \geq 1$. Summing the equations (\ref{proof-step2}) and (\ref{proof-step3}), we find that $|\Psi\rangle$ is an eigenvector of $t(u)$ if and only if their sub-leading terms cancel via the equations

\begin{align}
\alpha(v_i)
\frac{c(v_i,u)}{b_{-}(v_i,u)}
\prod_{\substack{n\not=i \\ n=1}}^{N}
\frac{a(v_n,v_i)}{b_{-}(v_n,v_i)}
+
\delta(v_i)
\frac{c(u,v_i)}{b_{-}(u,v_i)}
\prod_{\substack{n\not=i \\ n=1}}^{N}
\frac{a(v_i,v_n)}{b_{-}(v_i,v_n)}
=
0
\label{pre-Beteq}
\end{align}

\noindent for all $1 \leq i \leq N$. Cancelling factors which are common to both terms in (\ref{pre-Beteq}), we obtain the Bethe equations (\ref{Beteq}). Furthermore, summing the leading terms in (\ref{proof-step2}) and (\ref{proof-step3}), we recover the eigenvalue (\ref{eigenpsi}). 

\end{proof} 

\subsection{Scalar product}

When studying a quantum integrable model, aside from calculating the spectrum of its Hamiltonian $\mathcal{H}$, another important problem is the calculation of its {\it scalar product} $S_N(\{u\}_N,\{v\}_N)$. The scalar product is a function of the $2N$ variables $\{u\}_N = \{u_1,\ldots,u_N\}$, $\{v\}_N = \{v_1,\ldots,v_N\}$ given by  

\begin{align}
S_N\Big(\{u\}_N,\{v\}_N\Big)
=
\langle 0|
\lprod_{m=1}^{N} C(u_m)
\rprod_{n=1}^{N} B(v_n)
|0\rangle
\label{scalar}
\end{align}

\noindent for Bose-Einstein models, and by

\begin{align}
S_N\Big(\{u\}_N,\{v\}_N\Big)
=
\langle \Uparrow|
\lprod_{m=1}^{N} C(u_m)
\rprod_{n=1}^{N} B(v_n)
|\Uparrow\rangle
\label{scalar2}
\end{align}

\noindent for Fermi-Dirac models. For simplicity, in the remainder of this section we specialize to the former case, though our treatment may be equally applied to the latter.
 
In general the variables $\{u\}_N$ and $\{v\}_N$ are kept free. Specializing to the case where $\{u\}_N$ and $\{v\}_N$ are solutions of the Bethe equations, the scalar product expresses the action of a dual eigenstate $\langle \Psi_u|$ on another eigenstate $|\Psi_v\rangle$. If $\{u\}_N$ and $\{v\}_N$ are different solutions of the Bethe equations, this action is trivially zero. This follows from the fact that 

{\small
\begin{align} 
\tau_{\Psi_u}(z)
S_N\Big(\{u\}_N,\{v\}_N\Big)
&=
\langle 0|
\lprod_{m=1}^{N} C(u_m)
t(z)
\rprod_{n=1}^{N} B(v_n)
|0\rangle
\label{beth-orth}
=
\tau_{\Psi_v}(z)
S_N\Big(\{u\}_N,\{v\}_N\Big)
\end{align}
}

\noindent and since the eigenvalues $\tau_{\Psi_{u}}(z),\tau_{\Psi_{v}}(z)$ in (\ref{beth-orth}) are assumed to be different, the only possible resolution is that $S_N(\{u\}_N,\{v\}_N) = 0$. On the other hand, when $\{u\}_N$ and $\{v\}_N$ are equal to the {\it same} solution of the Bethe equations, $S_N(\{v\}_N,\{v\}_N)$ is non-zero and used in the normalization of other physical entities, such as correlation functions.

In this thesis we will calculate the scalar product in a variety of different models.\footnote{In chapters 3 and 4 we will calculate $S_N(\{u\}_N,\{v\}_N)$ when the variables $\{u\}_N, \{v\}_N$ are free. In chapters 5 and 6 we will consider an intermediate case, when one set of variables $\{u\}_N$ is free whilst the other, $\{v\}_N$, satisfies the Bethe equations.} A universal technique for achieving this is to use the commutation relations (\ref{general-intertwining-T-comp}) and the annihilation rules (\ref{annirule1}), (\ref{annirule2}) to manipulate the operators appearing in (\ref{scalar}). Unfortunately, this is a complicated approach which generally does not lead to a compact expression. Our approach will be to refrain from the commutation relations (\ref{general-intertwining-T-comp}) as much as possible, preferring simpler techniques which pertain to each individual model.    

\subsection{Graphical representation of scalar product}

In this subsection we specialize to models whose basis vectors are orthonormal. That is, we consider models which satisfy

\begin{align}
\mathcal{I}\Big(
|m\rangle, |n\rangle
\Big)
=
\prod_{i=1}^{M}
\delta_{m_i,n_i}
\end{align}

\noindent for all basis vectors $|m\rangle = |m_1\rangle_1 \otimes \cdots \otimes |m_M\rangle_M$ and $|n\rangle = |n_1\rangle_1 \otimes \cdots \otimes |n_M\rangle_M$ in $\mathcal{V}$. Comparing with equation (\ref{general-inn-prod}), this corresponds to the case $c_i(m_i) = 1$ for all $1\leq i \leq M$ and $m_i \in \mathfrak{N}$. For such models, the scalar product $S_N(\{u\}_N,\{v\}_N)$ can be graphically represented as a lattice. In order to demonstrate this, we prepare some notations.

For all $0\leq i \leq N$, let $\beta_i,\gamma_i$ denote vectors $(\beta_{i,1},\ldots,\beta_{i,M}), (\gamma_{i,1},\ldots,\gamma_{i,M}) \in \mathfrak{N}^M$ and define

\begin{align}
&|\beta_i\rangle=
|\beta_{i,1}\rangle_1 \otimes \cdots \otimes |\beta_{i,M}\rangle_M,
\quad
\langle \beta_i| =
\langle \beta_{i,1} |_1 \otimes \cdots \otimes \langle \beta_{i,M}|_M
\\
&|\gamma_i\rangle = 
|\gamma_{i,1}\rangle_1 \otimes \cdots \otimes |\gamma_{i,M}\rangle_M,
\quad
\langle \gamma_i| =
\langle \gamma_{i,1} |_1 \otimes \cdots \otimes \langle \gamma_{i,M}|_M
\nonumber
\end{align}

\noindent to be their corresponding states in $\mathcal{V}$ and $\mathcal{V}^{*}$. For all $0\leq i,j \leq N$, we insert the complete sets of states $\sum_{\gamma_i} |\gamma_i \rangle \langle \gamma_i |$ and $\sum_{\beta_j} |\beta_j \rangle \langle \beta_j |$ into (\ref{scalar}), yielding

\begin{align}
S_N\Big(
\{u\}_N,\{v\}_N
\Big)
=
\sum_{\substack{
\gamma_0,\ldots,\gamma_N \\ \beta_0,\ldots, \beta_N }}
\delta_{\gamma_N,\vec{0}}\ 
\lprod_{i=1}^{N}
\langle \gamma_{i}|  C(u_i) |\gamma_{i-1}\rangle
\delta_{\gamma_0,\beta_0}
\rprod_{j=1}^{N}
\langle \beta_{j-1}| B(v_j) |\beta_j \rangle
\delta_{\beta_N,\vec{0}}
\label{scalar-explode}
\end{align}

\noindent where a term in the sum is equal to zero unless $\beta_N = \gamma_N = \vec{0}$ and $\beta_0 = \gamma_0$. Using the diagrammatic conventions discussed earlier in the chapter we identify each term in (\ref{scalar-explode}) with a string of vertices, as shown in the following figures.

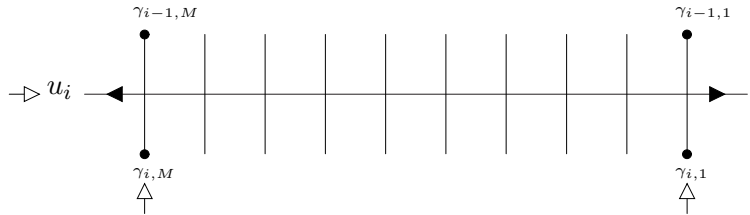
\begin{figure}[H]

\begin{center}
\begin{minipage}{4.3in}

\setlength{\unitlength}{0.0004cm}
\begin{picture}(20000,7000)(-4500,-3000)

\path(-2000,0000)(20000,0000)
\blacken\path(-750,250)(-750,-250)(-1250,0)(-750,250)
\blacken\path(18750,250)(18750,-250)(19250,0)(18750,250)
\put(-3250,0){$u_i$}
\path(-4500,0)(-3500,0)
\whiten\path(-4000,250)(-4000,-250)(-3500,0)(-4000,250)
\path(0000,-2000)(0000,2000)
\put(0,-2000){\circle*{300}}
\put(-400,-2700){\tiny{$\gamma_{i,M}$}}
\put(0,2000){\circle*{300}}
\put(-400,2700){\tiny{$\gamma_{i-1,M}$}}
\path(0,-4000)(0,-3000)
\whiten\path(-250,-3500)(250,-3500)(0,-3000)(-250,-3500)
\path(2000,-2000)(2000,2000)
\path(4000,-2000)(4000,2000)
\path(6000,-2000)(6000,2000)
\path(8000,-2000)(8000,2000)
\path(10000,-2000)(10000,2000)
\path(12000,-2000)(12000,2000)
\path(14000,-2000)(14000,2000)
\path(16000,-2000)(16000,2000)
\path(18000,-2000)(18000,2000)
\put(18000,-2000){\circle*{300}}
\put(17600,-2700){\tiny{$\gamma_{i,1}$}}
\put(18000,2000){\circle*{300}}
\put(17600,2700){\tiny{$\gamma_{i-1,1}$}}
\path(18000,-4000)(18000,-3000)
\whiten\path(17750,-3500)(18250,-3500)(18000,-3000)(17750,-3500)

\end{picture}

\end{minipage}
\end{center}

\caption[Vertex string for $\langle \gamma_i | C(u_i) |\gamma_{i-1} \rangle$]{Vertex string for $\langle \gamma_i | C(u_i) |\gamma_{i-1} \rangle$.}
\end{figure}

\begin{figure}[H]

\begin{center}
\begin{minipage}{4.3in}

\setlength{\unitlength}{0.0004cm}
\begin{picture}(20000,7000)(-4500,-3000)

\path(-2000,0000)(20000,0000)
\blacken\path(-1250,250)(-1250,-250)(-750,0)(-1250,250)
\blacken\path(19250,250)(19250,-250)(18750,0)(19250,250)
\put(-3250,0){$v_j$}
\path(-4500,0)(-3500,0)
\whiten\path(-4000,250)(-4000,-250)(-3500,0)(-4000,250)
\path(0000,-2000)(0000,2000)
\put(0,-2000){\circle*{300}}
\put(-400,-2700){\tiny{$\beta_{j-1,M}$}}
\put(0,2000){\circle*{300}}
\put(-400,2700){\tiny{$\beta_{j,M}$}}
\path(0,-4200)(0,-3200)
\whiten\path(-250,-3700)(250,-3700)(0,-3200)(-250,-3700)
\path(2000,-2000)(2000,2000)
\path(4000,-2000)(4000,2000)
\path(6000,-2000)(6000,2000)
\path(8000,-2000)(8000,2000)
\path(10000,-2000)(10000,2000)
\path(12000,-2000)(12000,2000)
\path(14000,-2000)(14000,2000)
\path(16000,-2000)(16000,2000)
\path(18000,-2000)(18000,2000)
\put(18000,-2000){\circle*{300}}
\put(17600,-2700){\tiny{$\beta_{j-1,1}$}}
\put(18000,2000){\circle*{300}}
\put(17600,2700){\tiny{$\beta_{j,1}$}}
\path(18000,-4200)(18000,-3200)
\whiten\path(17750,-3700)(18250,-3700)(18000,-3200)(17750,-3700)

\end{picture}

\end{minipage}
\end{center}

\caption[Vertex string for $\langle \beta_{j-1} | B(v_j) |\beta_j \rangle$]{Vertex string for $\langle \beta_{j-1} | B(v_j) |\beta_j \rangle$.}
\end{figure}
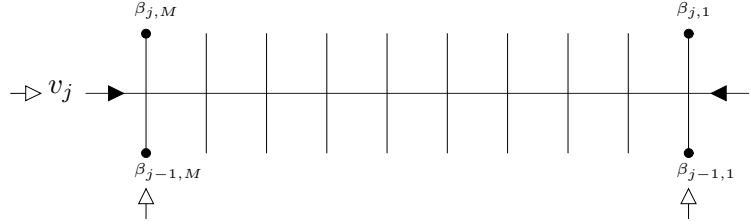

\noindent Attaching these strings of vertices along identified indices, we arrive at a lattice representation of $S_N(\{u\}_N,\{v\}_N)$ as shown in figure 2.9.

\begin{figure}[H]

\begin{center}
\begin{minipage}{4.3in}

\setlength{\unitlength}{0.0004cm}
\begin{picture}(20000,20000)(-4500,-3000)

\path(-2000,0)(20000,0)
\put(-3250,0){$u_N$}
\path(-4500,0)(-3500,0)
\whiten\path(-3750,250)(-3750,-250)(-3500,0)(-3750,250)
\blacken\path(-750,250)(-750,-250)(-1250,0)(-750,250)
\blacken\path(18750,250)(18750,-250)(19250,0)(18750,250)

\path(-2000,2000)(20000,2000)
\path(-4500,2000)(-3500,2000)
\whiten\path(-3750,2250)(-3750,1750)(-3500,2000)(-3750,2250)
\blacken\path(-750,2250)(-750,1750)(-1250,2000)(-750,2250)
\blacken\path(18750,2250)(18750,1750)(19250,2000)(18750,2250)

\path(-2000,4000)(20000,4000)
\path(-4500,4000)(-3500,4000)
\whiten\path(-3750,4250)(-3750,3750)(-3500,4000)(-3750,4250)
\blacken\path(-750,4250)(-750,3750)(-1250,4000)(-750,4250)
\blacken\path(18750,4250)(18750,3750)(19250,4000)(18750,4250)

\path(-2000,6000)(20000,6000)
\put(-3250,6000){$u_1$}
\path(-4500,6000)(-3500,6000)
\whiten\path(-3750,6250)(-3750,5750)(-3500,6000)(-3750,6250)
\blacken\path(-750,6250)(-750,5750)(-1250,6000)(-750,6250)
\blacken\path(18750,6250)(18750,5750)(19250,6000)(18750,6250)

\path(-2000,8000)(20000,8000)
\put(-3250,8000){$v_1$}
\path(-4500,8000)(-3500,8000)
\whiten\path(-3750,8250)(-3750,7750)(-3500,8000)(-3750,8250)
\blacken\path(-1250,8250)(-1250,7750)(-750,8000)(-1250,8250)
\blacken\path(19250,8250)(19250,7750)(18750,8000)(19250,8250)

\path(-2000,10000)(20000,10000)
\path(-4500,10000)(-3500,10000)
\whiten\path(-3750,10250)(-3750,9750)(-3500,10000)(-3750,10250)
\blacken\path(-1250,10250)(-1250,9750)(-750,10000)(-1250,10250)
\blacken\path(19250,10250)(19250,9750)(18750,10000)(19250,10250)

\path(-2000,12000)(20000,12000)
\path(-4500,12000)(-3500,12000)
\whiten\path(-3750,12250)(-3750,11750)(-3500,12000)(-3750,12250)
\blacken\path(-1250,12250)(-1250,11750)(-750,12000)(-1250,12250)
\blacken\path(19250,12250)(19250,11750)(18750,12000)(19250,12250)

\path(-2000,14000)(20000,14000)
\put(-3250,14000){$v_N$}
\path(-4500,14000)(-3500,14000)
\whiten\path(-3750,14250)(-3750,13750)(-3500,14000)(-3750,14250)
\blacken\path(-1250,14250)(-1250,13750)(-750,14000)(-1250,14250)
\blacken\path(19250,14250)(19250,13750)(18750,14000)(19250,14250)


\path(0,-2000)(0,16000)
\path(0,-3500)(0,-2500)
\whiten\path(-250,-2750)(250,-2750)(0,-2500)(-250,-2750)
\put(0,-1000){\circle*{300}}
\put(250,-1000){\tiny0}
\put(0,7000){\circle*{300}}
\put(0,15000){\circle*{300}}
\put(250,15000){\tiny0}

\path(2000,-2000)(2000,16000)
\path(2000,-3500)(2000,-2500)
\whiten\path(1750,-2750)(2250,-2750)(2000,-2500)(1750,-2750)
\put(2000,-1000){\circle*{300}}
\put(2250,-1000){\tiny0}
\put(2000,7000){\circle*{300}}
\put(2000,15000){\circle*{300}}
\put(2250,15000){\tiny0}

\path(4000,-2000)(4000,16000)
\path(4000,-3500)(4000,-2500)
\whiten\path(3750,-2750)(4250,-2750)(4000,-2500)(3750,-2750)
\put(4000,-1000){\circle*{300}}
\put(4250,-1000){\tiny0}
\put(4000,7000){\circle*{300}}
\put(4000,15000){\circle*{300}}
\put(4250,15000){\tiny0}

\path(6000,-2000)(6000,16000)
\path(6000,-3500)(6000,-2500)
\whiten\path(5750,-2750)(6250,-2750)(6000,-2500)(5750,-2750)
\put(6000,-1000){\circle*{300}}
\put(6250,-1000){\tiny0}
\put(6000,7000){\circle*{300}}
\put(6000,15000){\circle*{300}}
\put(6250,15000){\tiny0}

\path(8000,-2000)(8000,16000)
\path(8000,-3500)(8000,-2500)
\whiten\path(7750,-2750)(8250,-2750)(8000,-2500)(7750,-2750)
\put(8000,-1000){\circle*{300}}
\put(8250,-1000){\tiny0}
\put(8000,7000){\circle*{300}}
\put(8000,15000){\circle*{300}}
\put(8250,15000){\tiny0}

\path(10000,-2000)(10000,16000)
\path(10000,-3500)(10000,-2500)
\whiten\path(9750,-2750)(10250,-2750)(10000,-2500)(9750,-2750)
\put(10000,-1000){\circle*{300}}
\put(10250,-1000){\tiny0}
\put(10000,7000){\circle*{300}}
\put(10000,15000){\circle*{300}}
\put(10250,15000){\tiny0}

\path(12000,-2000)(12000,16000)
\path(12000,-3500)(12000,-2500)
\whiten\path(11750,-2750)(12250,-2750)(12000,-2500)(11750,-2750)
\put(12000,-1000){\circle*{300}}
\put(12250,-1000){\tiny0}
\put(12000,7000){\circle*{300}}
\put(12000,15000){\circle*{300}}
\put(12250,15000){\tiny0}

\path(14000,-2000)(14000,16000)
\path(14000,-3500)(14000,-2500)
\whiten\path(13750,-2750)(14250,-2750)(14000,-2500)(13750,-2750)
\put(14000,-1000){\circle*{300}}
\put(14250,-1000){\tiny0}
\put(14000,7000){\circle*{300}}
\put(14000,15000){\circle*{300}}
\put(14250,15000){\tiny0}

\path(16000,-2000)(16000,16000)
\path(16000,-3500)(16000,-2500)
\whiten\path(15750,-2750)(16250,-2750)(16000,-2500)(15750,-2750)
\put(16000,-1000){\circle*{300}}
\put(16250,-1000){\tiny0}
\put(16000,7000){\circle*{300}}
\put(16000,15000){\circle*{300}}
\put(16250,15000){\tiny0}

\path(18000,-2000)(18000,16000)
\path(18000,-3500)(18000,-2500)
\whiten\path(17750,-2750)(18250,-2750)(18000,-2500)(17750,-2750)
\put(18000,-1000){\circle*{300}}
\put(18250,-1000){\tiny0}
\put(18000,7000){\circle*{300}}
\put(18000,15000){\circle*{300}}
\put(18250,15000){\tiny0}

\end{picture}

\end{minipage}
\end{center}

\caption[Graphical depiction of the scalar product]{Graphical depiction of the scalar product.}
\end{figure}
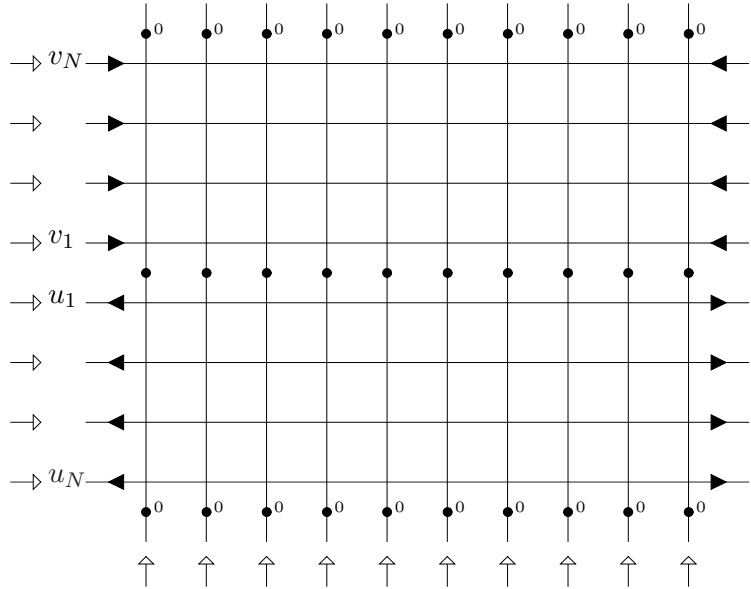

\noindent The horizontal lines in this lattice should be interpreted as strings of $L$-matrix vertices, corresponding to monodromy matrix elements. The lowest $N$ horizontal lines, through which the variables $u_i$ are flowing, have external line values $(-1,+1)$. Therefore, they represent the monodromy matrix elements $\langle \gamma_i| T^{-+}(u_i) |\gamma_{i-1}\rangle = \langle \gamma_i| C(u_i) |\gamma_{i-1}\rangle$. The highest $N$ horizontal lines, through which the variables $v_j$ are flowing, have external line values $(+1,-1)$. Therefore, they represent the monodromy matrix elements $\langle \beta_{j-1}| T^{+-}(v_j) |\beta_j\rangle = \langle \beta_{j-1}| B(v_j) |\beta_j\rangle$. All horizontal internal line segments are summed over black arrows that point with the orientation, and black arrows that point against the orientation. 

The vertical external line segments have been frozen to zero, to represent the fact that the vectors $\beta_N = \gamma_N = \vec{0}$. Conversely, the vertical internal line segments are summed over all elements of the set $\mathfrak{N}$. For simplicity, we have omitted all internal labels. 

\section{Conclusion}

The definitions presented in this chapter will be used ubiquitously throughout the rest of the thesis. We shall consider a number of quantum integrable models, and in each case we list their quantum algebras $\mathcal{A}_1,\ldots,\mathcal{A}_M$ and define representations of these algebras on the vector space $\mathcal{V}$. We also state the Hamiltonian $\mathcal{H}$ of each model under our consideration, but this is mainly for completeness and it is {\it not} our aim to study their spectra in any detail.

The most essential material in this chapter is the description of the quantum inverse scattering method/algebraic Bethe Ansatz. Indeed, we shall apply these techniques to every model that we encounter, listing its $R$-matrix and $L$-matrix, and constructing its Bethe eigenvectors. The graphical conventions of this chapter are also prevalent throughout the remainder of our work, particularly in chapter 5 and 6. 

Finally, let us remark that scalar products are of key interest in our later studies. We will calculate the scalar product of almost every model under our consideration, studying their role as $\tau$-functions in chapters 3,5 and as generating functions of plane partitions in chapters 3,4.


\newpage

\thispagestyle{empty}

\phantom{nothing}


\chapter{Bosonic models and plane partitions}
\setcounter{theorem}{0}
\setcounter{lemma}{0}
\setcounter{example}{0}
\setcounter{definition}{0}
\setcounter{section}{-1}

\section{Introduction}

In this chapter we study two closely related quantum integrable models which are solvable by the algebraic Bethe Ansatz. We will discuss the relationship of these models with the hierarchies discussed in chapter 1, and with plane partitions, which are classical combinatorial objects. Similarly to chapter 1, every result in the context of the first model is mirrored by an analogous result in the context of the second, and accordingly we have split this chapter into two parallel parts.

The first part of the chapter considers the phase model, which was introduced in \cite{bik1} and subsequently studied in \cite{bik2} as the limiting case $q\rightarrow \infty$ of the more general $q$-boson model. In section \ref{phase-intro} we give the space of states $\mathcal{V}$ of the model, the Hamiltonian $\mathcal{H}$, and review the construction of its eigenvectors using the algebraic Bethe Ansatz. 

An explicit expression for the phase model Bethe eigenvectors was found by N~M~Bogoliubov in \cite{bog1}. To perform this calculation, Bogoliubov defined a simple correspondence between the basis vectors of $\mathcal{V}$ and partitions. Under this correspondence, the Bethe eigenvectors can be written as sums over partitions which are weighted by Schur functions. In section \ref{phase-BE} we reproduce these results while appealing to the charged fermion calculus discussed in chapter 1. We map the basis vectors of $\mathcal{V}$ to partitions in the Fock space $\mathcal{F}_{\psi}$ and calculate the image of the Bethe eigenvectors under this map. At the level of the vector space $\mathcal{V}$, a Bethe eigenvector is constructed by the action of $B$-operators acting on the vacuum state. We find that at the level of the Fock space $\mathcal{F}_{\psi}$, a Bethe eigenvector lies in the orbit of the Fock vacuum under $GL_{\infty}$.

The connection of the phase model with plane partitions was also discovered in \cite{bog1}. Using the diagrammatic interpretation of the monodromy matrix operators, as we described in chapter 2, Bogoliubov was able to show that the phase model scalar product is a generating function for plane partitions within a box of size $N\times N\times M$. In section \ref{phase-bpp} we give another proof of this result, taking the perspective of A~Okounkov and N~Reshetikhin in \cite{or}, where it was observed that the diagonal slices of an arbitrary plane partition form a sequence of interlacing partitions. The proof involves showing that monodromy matrix operators act on a partition state to generate a sum of partitions which interlace with the original. We end the section by writing the phase model scalar product in the form of a KP $\tau$-function, that is, as an expectation value of charged fermionic operators.       

In section \ref{phase-pp} we study the phase model as $M \rightarrow \infty$, which is the infinite lattice limit. We prove that in this limit the action of a $B$-operator on an arbitrary vector in $\mathcal{V}$ is equivalent to the action of a KP half-vertex operator on the image state in $\mathcal{F}_{\psi}$. This is achieved by showing that KP half-vertex operators act on a partition state in $\mathcal{F}_{\psi}$ to generate a sum of interlacing partitions. Working at the level of fermionic operators, the infinite lattice scalar product is readily evaluated. In fact, because KP half-vertex operators have simple commutation relations, we find that the scalar product factorizes into product form. We thus obtain a fermionic construction of MacMahon's generating function for plane partitions, first proposed in \cite{orv} and explained in detail in \cite{fwz3}. 

In the second part of the chapter we repeat the calculations of the first part, but in the context of the $q\rightarrow i$ limit of the $q$-boson model, where $i = \sqrt{-1}$. For brevity we call this the $i$-boson model, and to the best of our knowledge it has not been studied in the literature, in its own right. In section \ref{ib-intro} we give the space of states $\tilde{\mathcal{V}}$ of the model, the Hamiltonian $\tilde{\mathcal{H}}$, and describe the construction of its eigenvectors using the algebraic Bethe Ansatz.\footnote{In the second part of the chapter, we distinguish all operators and spaces from their direct counterparts in the first part by use of a tilde.}

After this, our attention turns to evaluating the Bethe eigenvectors of the $i$-boson model. There exists a correspondence between the basis vectors of $\tilde{\mathcal{V}}$ and strict partitions. Under this correspondence, a Bethe eigenvector can be written as a sum over strict partitions which are weighted by Schur $Q$-functions. In section \ref{ib-BE} we make these notions precise by using the neutral fermion calculus discussed in chapter 1. We map basis vectors of $\tilde{\mathcal{V}}$ to strict partitions in the Fock space $\mathcal{F}_{\phi}$, and calculate the image of the Bethe eigenvectors under this map. At the level of the vector space $\tilde{\mathcal{V}}$, a Bethe eigenvector is constructed by the action of $B$-operators acting on the vacuum state. We find that at the level of the Fock space $\mathcal{F}_{\phi}$, a Bethe eigenvector lies in the orbit of the Fock vacuum under $O_{\infty}$.

In section \ref{ib-bpp} we establish a connection between the $i$-boson model and strict plane partitions. Specifically, we prove that the scalar product of the $i$-boson model is a generating function for strict plane partitions within a box of size $N\times N\times M$. This time around, the diagonal slices of an arbitrary strict plane partition form a sequence of interlacing strict partitions. Hence the proof involves showing that monodromy matrix operators act on a strict partition state to generate a sum of strict partitions which interlace with the original. We end the section by writing the $i$-boson model scalar product in the form of a BKP $\tau$-function, that is, as an expectation value of neutral fermionic operators.

Finally, in section \ref{ib-pp} we study the $i$-boson model as $M\rightarrow\infty$. We prove that in this limit the action of a $B$-operator on an arbitrary vector in $\tilde{\mathcal{V}}$ is equivalent to the action of a BKP half-vertex operator on the image state in $\mathcal{F}_{\phi}$. This is achieved by showing that BKP half-vertex operators act on a strict partition state in $\mathcal{F}_{\phi}$ to generate a sum of interlacing strict partitions. As in the case of section \ref{phase-pp}, the BKP half-vertex operators have simple commutation relations, meaning that the scalar product factorizes into product form. We thus obtain a fermionic construction of the generating function for strict plane partitions, which first appeared in \cite{fw1} and later in \cite{fwz3}.

The essence of this chapter is the observation that free fermions appear in the Bethe eigenvectors of two bosonic integrable models. We hope that this result is indicative of a deeper correspondence between the concerned classical and quantum models.

\section{Phase model}
\label{phase-intro}

\subsection{Space of states $\mathcal{V}$ and inner product $\mathcal{I}$}

Following the procedure outlined in the previous chapter, we construct a vector space $\mathcal{V}$ which provides the framework for study of the phase model. Consider a one-dimensional integral lattice, consisting of $M+1$ sites. A configuration of this lattice corresponds to placing $n_i \geq 0$ bosons at the $i^{\rm th}$ site, for all $0 \leq i \leq M$. The vector space $\mathcal{V}$ is defined as the linear span of all lattice configurations. 

Mathematically, we represent a configuration by a tensor product of state vectors  $|n_i\rangle_i$, where $n_i$ is the $i^{\rm th}$ occupation number in that particular configuration. It follows that $\mathcal{V}$ has the basis

\begin{align}
{\rm Basis}\left(\mathcal{V}\right)
=
\Big\{
|n\rangle
=
|n_0\rangle_0 \otimes |n_1\rangle_1 \otimes \cdots \otimes |n_M\rangle_M
\Big\}
\label{vec1}
\end{align}

\noindent where $\{n_0,n_1,\ldots,n_M \}$ range over all non-negative integers. The inner product $\mathcal{I}$ between two basis vectors $|m\rangle = |m_0\rangle_0\otimes \cdots \otimes |m_M\rangle_M$ and $|n\rangle = |n_0\rangle_0 \otimes \cdots \otimes |n_M\rangle_M$ is defined as 

\begin{align}
\mathcal{I}\Big(|m\rangle,|n\rangle\Big)
=
\prod_{i=0}^{M} \delta_{m_i,n_i}
\label{innp}
\end{align}

\noindent which corresponds to setting $c_i(m_i) = 1$ for all $0 \leq i \leq M$, $m_i \geq 0$ in equation (\ref{general-inn-prod}). We also define a space of states $\mathcal{V}^{*}$ that is dual to $\mathcal{V}$. We write

\begin{align}
{\rm Basis}\left(\mathcal{V}^{*}\right)
=
\Big\{
\langle m|
=
\langle m_0|_0 \otimes \langle m_1|_1 \otimes \cdots \otimes \langle m_M|_M
\Big\}
\label{vec2}
\end{align}

\noindent where once again $\{m_0,m_1,\ldots,m_M \}$ range over all non-negative integers. The action of a basis vector $\langle m| = \langle m_0|_0 \otimes \cdots \otimes \langle m_M|_M \in \mathcal{V}^{*}$ on $|n\rangle = |n_0\rangle_0 \otimes \cdots \otimes |n_M\rangle_M \in \mathcal{V}$ is given by

\begin{align}
\langle m|n\rangle 
= 
\mathcal{I}\Big(|m\rangle, |n\rangle \Big)
\label{dualact}
\end{align}

\noindent Both the vector spaces (\ref{vec1}) and (\ref{vec2}) are infinite-dimensional, since there is no upper bound enforced on the occupation numbers $m_i$ and $n_i$.

\subsection{Phase algebra}

The phase algebra is defined in \cite{bog1}, \cite{bik2}. It is generated by $\{\phi,\phid,\mathcal{N},\pi\}$ which satisfy the relations

\begin{align}
[\phi,\phid]=\pi,
\quad
[\mathcal{N},\phi]=-\phi,
\quad
[\mathcal{N},\phid]=\phid,
\quad
\phi\pi
=
\pi\phid
=
0
\label{phase-alg}
\end{align}

\noindent This algebra is the $q\rightarrow\infty$ case of the $q$-boson algebra (\ref{realqb}), discussed in the next chapter. We will consider $M+1$ copies of the phase algebra, generated by $\{\phi_0,\phid_0,\mathcal{N}_0,\pi_0\}$ through to $\{\phi_M,\phid_M,\mathcal{N}_M,\pi_M\}$.\footnote{The generators $\phi_i$ should not be confused with the neutral fermions of chapter 1.} Employing the language of chapter 2, we denote these algebras by $\mathcal{A}_0,\ldots,\mathcal{A}_M$ with $\mathfrak{a}_i^{+} = \phid_i,\mathfrak{a}_i^{-} = \phi_i$ and where $\mathcal{N}_i,\pi_i$ are both of type $\mathfrak{a}_i^{0}$. Different copies of the phase algebra are assumed to commute, giving rise to the equations

\begin{align}
[\phi_i,\phid_j]=\delta_{i,j}\pi_i,
\quad
[\mathcal{N}_i,\phi_j]=-\delta_{i,j}\phi_i,
\quad
[\mathcal{N}_i,\phid_j]=\delta_{i,j}\phid_i,
\quad
\phi_i\pi_i
=
\pi_i\phid_i
=
0
\label{phase-alg2}
\end{align}

\noindent for all $0\leq i,j \leq M$. 

%
%

\subsection{Representations of phase algebras}
\label{phaserep}

Following \cite{bog1}, \cite{bik2}, we fix representations of the phase algebras on the vector space $\mathcal{V}$. We begin with a synopsis of the role played by each operator. Firstly $\phi_i$ is an annihilation operator, removing particles from the $i^{\rm th}$ lattice site. Conversely $\phid_i$ is a creation operator, adding particles to the $i^{\rm th}$ lattice site. The operator $\mathcal{N}_i$ counts, but does not change, the number of particles at the $i^{\rm th}$ lattice site. Finally $\pi_i$ is a {\it vacuum projector}, leaving the $i^{\rm th}$ site unchanged if it is unoccupied by particles, but annihilating any non-empty $i^{\rm th}$ state vector.  

Let us more precisely define these representations of $\mathcal{A}_0,\ldots,\mathcal{A}_M$ on $\mathcal{V}$. The operator $\phi_i$ acts on the $i^{\rm th}$ state vector in a basis element. If the $i^{\rm th}$ occupation number is zero $\phi_i$ annihilates the basis element, otherwise it lowers the $i^{\rm th}$ occupation number by one. This is described by the equation

\begin{align}
\phi_i |n_0\rangle_0 \otimes \cdots \otimes |n_M\rangle_M
=
\left\{
\begin{array}{ll}
0, & n_i = 0
\\
\\
|n_0\rangle_0 
\otimes \cdots \otimes 
|n_i-1\rangle_i 
\otimes \cdots \otimes 
|n_M\rangle_M,
& n_i \geq 1
\end{array}
\right.
\label{p1}
\end{align}

\noindent The operator $\phid_i$ acts on the $i^{\rm th}$ state vector in a basis element and raises the $i^{\rm th}$ occupation number by one. This is described by the equation

\begin{align}
\phid_i |n_0\rangle_0 \otimes \cdots \otimes |n_M\rangle_M
=
|n_0\rangle_0 
\otimes \cdots \otimes 
|n_i+1\rangle_i 
\otimes \cdots \otimes 
|n_M\rangle_M
\label{pd1}
\end{align}

%

\noindent Every basis element of $\mathcal{V}$ is an eigenvector of the operator $\mathcal{N}_i$ with eigenvalue equal to the $i^{\rm th}$ occupation number. This is described by the equation 

\begin{align}
\mathcal{N}_i |n_0\rangle_0 \otimes \cdots \otimes |n_M\rangle_M
=
n_i |n_0\rangle_0 \otimes \cdots \otimes |n_M\rangle_M
\label{n1}
\end{align}

\noindent Lastly, the operator $\pi_i$ acts on the $i^{\rm th}$ state vector in a basis element. If the $i^{\rm th}$ occupation number is zero $\pi_i$ acts identically, otherwise it annihilates the basis element. This is described by the equation

\begin{align}
\pi_i |n_0\rangle_0 \otimes \cdots \otimes |n_M\rangle_M
=
\left\{
\begin{array}{ll}
|n_0\rangle_0 \otimes \cdots \otimes |n_M\rangle_M,
&
n_i = 0
\\
\\
0,
&
n_i \geq 1
\end{array}
\right.
\label{proj1}
\end{align}

\noindent The set of definitions (\ref{p1})--(\ref{proj1}) provide a faithful representation of the phase algebras $\mathcal{A}_0,\ldots,\mathcal{A}_M$. Assuming that all operators act linearly, equations (\ref{p1})--(\ref{proj1}) completely determine the action of $\{\phi_i,\phid_i,\mathcal{N}_i,\pi_i\}$ on the vector space $\mathcal{V}$. Also, from the definition of the inner product (\ref{innp}) we find that

\begin{align}
\mathcal{I}\Big(\phi_i|m\rangle,|n\rangle\Big)
=
\mathcal{I}\Big(|m\rangle,\phid_i|n\rangle\Big)
\end{align}

\noindent which shows that $\phi_i,\phid_i$ are adjoint operators, while $\mathcal{N}_i,\pi_i$ are clearly self-adjoint.

Following subsection \ref{2-dual} in the previous chapter, we fix appropriate actions for $\mathcal{A}_0,\ldots,\mathcal{A}_M$ on the basis elements of the dual space $\mathcal{V}^{*}$. In short, the roles of $\phi_i,\phid_i$ get interchanged, while $\mathcal{N}_i,\pi_i$ behave as before. More explicitly, the operator $\phid_i$ acts on the $i^{\rm th}$ state vector in a dual basis element. If the $i^{\rm th}$ occupation number is zero $\phid_i$ annihilates the dual basis element, otherwise it lowers the $i^{\rm th}$ occupation number by one. This is described by the equation

\begin{align}
\langle n_0|_0 \otimes \cdots \otimes \langle n_M|_M \phid_i
=
\left\{
\begin{array}{ll}
0, & n_i = 0
\\
\\
\langle n_0|_0 
\otimes \cdots \otimes 
\langle n_i-1|_i 
\otimes \cdots \otimes 
\langle n_M|_M,
& n_i \geq 1
\end{array}
\right.
\label{pd2}
\end{align}

\noindent The operator $\phi_i$ acts on the $i^{\rm th}$ state vector in a dual basis element and raises the $i^{\rm th}$ occupation number by one. This is described by the equation
 
\begin{align}
\langle n_0|_0 \otimes \cdots \otimes \langle n_M|_M \phi_i
=
\langle n_0|_0 
\otimes \cdots \otimes 
\langle n_i+1|_i 
\otimes \cdots \otimes 
\langle n_M|_M
\label{p2}
\end{align}



\noindent Every basis element of $\mathcal{V}^{*}$ is an eigenvector of the operator $\mathcal{N}_i$ with eigenvalue equal to the $i^{\rm th}$ occupation number. This is described by the equation

\begin{align}
\langle n_0|_0 \otimes \cdots \otimes \langle n_M|_M \mathcal{N}_i
=
n_i \langle n_0|_0 \otimes \cdots \otimes \langle n_M|_M
\label{n2}
\end{align}

\noindent Finally, the operator $\pi_i$ acts on the $i^{\rm th}$ state vector in a dual basis element. If the $i^{\rm th}$ occupation number is zero $\pi_i$ acts identically, otherwise it annihilates the dual basis element. This is described by the equation

\begin{align}
\langle n_0|_0 \otimes \cdots \otimes \langle n_M|_M \pi_i
=
\left\{
\begin{array}{ll}
\langle n_0|_0 
\otimes \cdots \otimes 
\langle n_M|_M,
& n_i = 0
\\
\\
0,
& n_i \geq 1
\end{array}
\right.
\label{proj2}
\end{align}

\noindent The set of definitions (\ref{pd2})--(\ref{proj2}) provide the dual representation of the algebras $\mathcal{A}_0,\ldots,\mathcal{A}_M$. Assuming that all operators act linearly, the equations (\ref{pd2})--(\ref{proj2}) completely determine the action of $\{\phi_i,\phid_i,\mathcal{N}_i,\pi_i\}$ on the dual vector space $\mathcal{V}^{*}$.

\subsection{Calculation of $\langle m|n\rangle$}

In the interest of self-consistency, in this subsection we check that (\ref{dualact}) is actually obeyed. By virtue of the equation (\ref{pd1}), it is possible to write any basis element of $\mathcal{V}$ in terms of the operators $\phid_i$ acting on the vacuum state. Explicitly speaking, we have

\begin{align}
|n\rangle
=
|n_0\rangle_0 \otimes \cdots \otimes |n_M\rangle_M
&=
(\phid_0)^{n_0}\ldots (\phid_M)^{n_M}
|0\rangle_0 \otimes \cdots \otimes |0\rangle_M
\label{phase-genstat}
\\
&=
(\phid_0)^{n_0}\ldots (\phid_M)^{n_M}
|0\rangle
\nonumber
\end{align}

\noindent Similarly, the equation (\ref{p2}) makes it possible to write any basis element of $\mathcal{V}^{*}$ in terms of the operators $\phi_i$ acting on the dual vacuum state. We find that

\begin{align}
\langle m|
=
\langle m_0|_0 \otimes \cdots \otimes \langle m_M|_M
&=
\langle 0|_0 \otimes \cdots \otimes \langle 0|_M
(\phi_0)^{m_0}\ldots (\phi_M)^{m_M}
\label{phase-dgenstat}
\\
&=
\langle 0|
(\phi_0)^{m_0}\ldots (\phi_M)^{m_M}
\nonumber
\end{align}

\noindent Using equations (\ref{phase-genstat}) and (\ref{phase-dgenstat}) and the fact that the vacuum expectation value of any element in $\mathcal{A} = \mathcal{A}_0 \otimes \cdots \otimes \mathcal{A}_M$ is unambiguously defined, we have

\begin{align}
\langle m|n\rangle
=
\langle 0|(\phi_0)^{m_0}\ldots (\phi_M)^{m_M}
(\phid_0)^{n_0}\ldots (\phid_M)^{n_M} |0\rangle
\label{phase-ip1}
\end{align}

\noindent From the commutation relations (\ref{phase-alg2}) and the fact that $\phi_i|0\rangle = \langle 0|\phid_i = 0$ for all $0\leq i \leq M$, we calculate (\ref{phase-ip1}) explicitly to obtain

\begin{align}
\langle m|n\rangle
=
\prod_{i=0}^{M}
\delta_{m_i,n_i}
\label{phase-isom}
\end{align}

\noindent in agreement with equation (\ref{dualact}).

\subsection{Hamiltonian $\mathcal{H}$}

The Hamiltonian of the phase model is given by

\begin{align}
\mathcal{H}
=
-\frac{1}{2}
\sum_{i=0}^{M}
\left(
\phid_i \phi_{i+1} + \phi_i \phid_{i+1}
\right)
+
\bar{\mathcal{N}} 
\label{phase-ham}
\end{align}

\noindent with $\bar{\mathcal{N}} = \sum_{i=0}^{M} \mathcal{N}_i$ and where the periodicity $\phi_{M+1} = \phi_0$ and $\phid_{M+1} = \phid_0$ is imposed. The problem of finding eigenvectors $|\Psi\rangle$ of this Hamiltonian can be solved using the quantum inverse scattering method/algebraic Bethe Ansatz. In the forthcoming subsections we recover $\mathcal{H}$ from the transfer matrix of the model and construct the Bethe eigenvectors, using the terminologies which were described in chapter 2.  

\subsection{$L$-matrix and local intertwining equation}


The $R$-matrix for the phase model depends on two indeterminates $x,y$ and acts in the tensor product $\mathcal{V}_a \otimes \mathcal{V}_b$, where $\mathcal{V}_a,\mathcal{V}_b$ are copies of $\mathbb{C}^2$. It is given by

\begin{align}
R_{ab}(x,y)
=
\left(
\begin{array}{cccc}
x &               0                 &              0                   & 0
\\
0 &              0              &                x^{\frac{1}{2}} y^{\frac{1}{2}}             & 0
\\
0 &               x^{\frac{1}{2}} y^{\frac{1}{2}}                 & x-y  & 0
\\
0 &               0                 &              0                   & x
\end{array}
\right)_{ab}
\label{phase-R}
\end{align}

\noindent and corresponds to the $a_{\pm}(x,y) = x,\ b_{+}(x,y) = 0,\ b_{-}(x,y) = x-y,\ c_{\pm}(x,y) = x^{\frac{1}{2}} y^{\frac{1}{2}}$ case of (\ref{general-Rmat}). The $L$-matrix for the phase model depends on a single indeterminate $x$, and acts in the space $\mathcal{V}_a$. Its entries are operators acting at the $m^{\rm th}$ lattice site, and identically everywhere else. It has the form

\begin{align}
L_{am}(x)
=
\left(
\begin{array}{cc}
x^{-\frac{1}{2}} & \phid_m
\\
\phi_m              & x^{\frac{1}{2}}
\end{array}
\right)_{a}
\label{phase-L}
\end{align}

\noindent Using these definitions, the local intertwining equation is given by

\begin{align}
R_{ab}(x,y) L_{am}(x) L_{bm}(y)
=
L_{bm}(y) L_{am}(x) R_{ab}(x,y)
\label{phase-intL}
\end{align}

\noindent This is a $4\times 4$ matrix equation, which gives rise to sixteen scalar identities. Each of these identities may be verified by direct calculation. The $L$-matrix and $R$-matrix of the phase model may be found in \cite{bog1},\cite{bik2}, which use slightly different parametrizations from those that we have adopted. 

\subsection{Monodromy matrix and global intertwining equation}

The monodromy matrix is an $(M+1)$-fold product of the $L$-matrices (\ref{phase-L}), taken in the auxiliary space ${\rm End}(\mathcal{V}_a)$. It has the form 

\begin{equation}
T_{a}(x) = L_{aM}(x)\ldots L_{a0}(x)
=
\left(
\begin{array}{cc}
A(x) & B(x)
\\
C(x) & D(x)
\end{array}
\right)_a
\label{phase-T}
\end{equation}

\noindent where $A(x), B(x), C(x), D(x)$ are elements of $\mathcal{A}_0\otimes \cdots \otimes \mathcal{A}_M$. The monodromy matrix satisfies the global intertwining equation

\begin{equation}
R_{ab}(x,y) T_a(x) T_b(y)
=
T_b(y) T_a(x) R_{ab}(x,y)
\label{phase-intT}
\end{equation}

\noindent which follows immediately from the local intertwining equation (\ref{phase-intL}).\footnote{See lemma \ref{2-lem1} in chapter 2.} The identity (\ref{phase-intT}) gives sixteen commutation relations between the $A(x),B(x),C(x),D(x)$ operators, but for our purposes we will only require two. These are the equations

\begin{align}
[B(x),B(y)] = [C(x),C(y)] = 0
\label{phasebc}
\end{align}

\noindent and they are necessary to show that the Bethe eigenvectors are symmetric in their rapidity variables. 

\subsection{Recovering $\mathcal{H}$ from the transfer matrix}

Let $t(x) = {\rm tr}_a T_a(x) = A(x)+D(x)$ be the transfer matrix of the phase model. The Hamiltonian (\ref{phase-ham}) may be recovered via the equation

\begin{align}
\mathcal{H}
=
\frac{1}{2}
\left[
x^2
\frac{d}{dx} \Big(x^{-(M+1)/2} t(x) \Big)
\right]_{x\rightarrow \infty}
-
\frac{1}{2}
\left[
\frac{d}{dx} \Big(x^{(M+1)/2} t(x) \Big)\right]_{x \rightarrow 0}
+
\bar{\mathcal{N}}
\end{align}

\noindent and using the fact that $[\bar{\mathcal{N}},t(x)] = 0$, it follows that $[\mathcal{H},t(x)]=0$. Hence the eigenvectors of $\mathcal{H}$ may be found by studying the eigenvectors of $t(x)$.

\subsection{Bethe Ansatz for the eigenvectors}

As was explained in theorem 1 of the previous chapter, the eigenvectors of the transfer matrix $t(x)$ are given by

\begin{align}
|\Psi\rangle = B(y_1)\ldots B(y_N) |0\rangle,
\quad
\langle \Psi| = \langle 0| C(y_N)\ldots C(y_1)
\label{3-1-bethe}
\end{align}

\noindent where the variables $\{y_1,\ldots, y_N\}$ are assumed to obey the Bethe equations (\ref{Beteq}). For the present model, we have $a(y_i,y_j)=y_i,\ \alpha(y_i) = y_i^{-(M+1)/2},\ \delta(y_i) = y_i^{(M+1)/2}$. Substituting these expressions into (\ref{Beteq}), the Bethe equations for the phase model read

\begin{align}
y_i^{M+N}
=
(-)^{N-1}
\prod_{\substack{ j\not=i \\ j=1}}^{N}
y_j
\label{phase-eig}
\end{align}

\noindent for all $1\leq i \leq N$. Although these equations are necessary for the success of the Bethe Ansatz, in our subsequent analysis we will study the vectors (\ref{3-1-bethe}) {\it without} imposing the restrictions (\ref{phase-eig}). We shall continue to call the objects (\ref{3-1-bethe}) Bethe eigenvectors, despite the fact that the Bethe equations are superfluous in our calculations.

\section{Calculation of phase model Bethe eigenvectors}
\label{phase-BE}

In this section we derive an explicit expression for the Bethe eigenvectors (\ref{3-1-bethe}). Although the result that we obtain first appeared in \cite{bog1}, our derivation has some new features. In particular, we map the Bethe eigenvectors to the charged fermionic Fock space of chapter 1, and show that they lie in the $GL_{\infty}$ orbit of the Fock vacuum.

\subsection{The maps $\mathcal{M}_{\psi}$ and $\mathcal{M}_{\psi}^{*}$}

\begin{definition}
{\rm 
Let $|n\rangle = |n_0\rangle_0 \otimes \cdots \otimes |n_M\rangle_M$ and $\langle n| = \langle n_0|_0 \otimes \cdots \otimes \langle n_M|_M$ be basis elements of $\mathcal{V}$ and $\mathcal{V}^{*}$, respectively, and define

\begin{align}
\Sigma_0 = \sum_{j=0}^{M} n_j
\end{align}

\noindent From this, let $|\nu) = |\nu_1,\ldots,\nu_{\Sigma_0})$ and $( \nu|=( \nu_1,\ldots,\nu_{\Sigma_0}|$ be partitions in the Fock spaces $\mathcal{F}_{\psi}^{(0)}$ and $\mathcal{F}_{\psi}^{*(0)}$ with $n_i$ parts equal to $i$ for all $0 \leq i \leq M$. That is, we let

\begin{align}
|\nu) &= |M^{n_M},\ldots,1^{n_1},0^{n_0}) = |M^{n_M},\ldots,1^{n_1})
\label{part1} 
\\
(\nu| &= (M^{n_M},\ldots,1^{n_1},0^{n_0}| =
(M^{n_M},\ldots,1^{n_1}|
\label{part2}
\end{align} 

\noindent We define linear maps $\mathcal{M}_{\psi}:\mathcal{V} \rightarrow \mathcal{F}_{\psi}^{(0)}$ and $\mathcal{M}_{\psi}^{*} :\mathcal{V}^{*} \rightarrow \mathcal{F}_{\psi}^{*(0)}$ whose actions are given by

\begin{align}
\mathcal{M}_{\psi} |n\rangle 
=
|\nu),
\quad
\langle n| \mathcal{M}_{\psi}^{*}  
=
( \nu|
\label{3-2-map}
\end{align}

\noindent Notice that these mappings are {\it not} one-to-one since they are insensitive to the value of $n_0$, which only appears as trivial information in the corresponding partition $\nu$. Furthermore, they are isometric in the sense that $\langle m|n\rangle =\langle \langle m|\mathcal{M}_{\psi}^{*},\mathcal{M}_{\psi} |n\rangle \rangle$ for all $\langle m| \in \mathcal{V}^{*}, |n\rangle \in \mathcal{V}$ which satisfy $m_0 = n_0$. From equation (\ref{3-2-map}), the action of $\mathcal{M}_{\psi},\mathcal{M}_{\psi}^{*}$ on any element of $\mathcal{V},\mathcal{V}^{*}$ can be calculated using linearity. 

The maps $\mathcal{M}_{\psi}$ and $\mathcal{M}_{\psi}^{*}$ are motivated by section V of \cite{bog1}, which discusses the same correspondence between basis elements of $\mathcal{V}$ and partitions. 
}
\end{definition} 

\begin{example} 
{\rm
Fix $M=4$ and $|n\rangle = |2\rangle_0 \otimes |3\rangle_1 \otimes |0\rangle_2 \otimes |2\rangle_3 \otimes |1\rangle_4$. We have $\Sigma_0 = 2+3+0+2+1=8$, and we let 

\begin{align}
|\nu) = |\nu_1,\ldots,\nu_8) = |4^1,3^2,2^0,1^3,0^2) = |4,3,3,1,1,1)
\end{align}

\noindent Then $\mathcal{M}_{\psi}|n\rangle  = |\nu)$. This correspondence is also shown in the figure below.
}
\end{example}

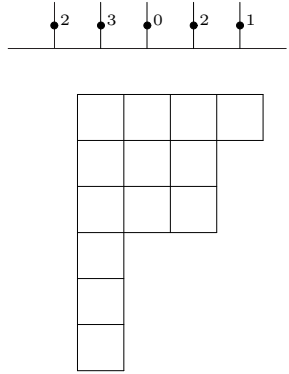
\begin{figure}[H]

\begin{center}

\begin{minipage}{4.3in}

\setlength{\unitlength}{0.0003cm}
\begin{picture}(20000,14000)(-15000,11000)

\path(-3000,24000)(9000,24000)

\path(-1000,24000)(-1000,26000)
\put(-1000,25000){\circle*{300}}
\put(-750,25000){\tiny2}

\path(1000,24000)(1000,26000)
\put(1000,25000){\circle*{300}}
\put(1250,25000){\tiny3}

\path(3000,24000)(3000,26000)
\put(3000,25000){\circle*{300}}
\put(3250,25000){\tiny0}

\path(5000,24000)(5000,26000)
\put(5000,25000){\circle*{300}}
\put(5250,25000){\tiny2}

\path(7000,24000)(7000,26000)
\put(7000,25000){\circle*{300}}
\put(7250,25000){\tiny1}

\path(0,22000)(8000,22000)
\path(0,20000)(8000,20000)
\path(0,18000)(6000,18000)
\path(0,16000)(6000,16000)
\path(0,14000)(2000,14000)
\path(0,12000)(2000,12000)
\path(0,10000)(2000,10000)

\path(0,10000)(0,22000)
\path(2000,10000)(2000,22000)
\path(4000,16000)(4000,22000)
\path(6000,16000)(6000,22000)
\path(8000,20000)(8000,22000)

\end{picture}

\end{minipage}
\end{center}

\caption[Mapping of 
$|n\rangle =|2\rangle_0\otimes|3\rangle_1\otimes|0\rangle_2\otimes|2\rangle_3\otimes|1\rangle_4$ to $|\nu) =|4,3,3,1,1,1)$]{Mapping of 
$|n\rangle =|2\rangle_0\otimes|3\rangle_1\otimes|0\rangle_2\otimes|2\rangle_3\otimes|1\rangle_4$ to $|\nu)=|4,3,3,1,1,1)$. The top part of the figure represents the state vector $|n\rangle$, while the lower part represents the Young diagram of $\nu$. Each occupation number $n_i$ manifests itself as $n_i$ rows of boxes of length $i$ in the Young diagram.}

\end{figure}

\subsection{Admissible basis elements}

\begin{definition}
{\rm
Let $|m\rangle = |m_0\rangle_0 \otimes \cdots \otimes |m_M\rangle_M$ and $|n\rangle = |n_0\rangle_0 \otimes \cdots \otimes |n_M\rangle_M$ be basis elements of $\mathcal{V}$. Define the partial sums of occupation numbers

\begin{align}
\Sigma^{m}_i = \sum_{j=i}^{M} m_j,
\quad
\Sigma^{n}_i = \sum_{j=i}^{M} n_j
\end{align}

\noindent for all $0 \leq i \leq M$. We say that $|m\rangle$ is {\it admissible} to $|n\rangle$, and write $|m\rangle \triangleright |n\rangle$, if and only if

\begin{align}
0 \leq & \left(\Sigma^{m}_{i} - \Sigma^{n}_{i}\right) \leq 1
\quad {\rm for\ all}\quad 1 \leq i \leq M
\\
{\rm and}
\quad & \left(\Sigma^{m}_0 - \Sigma^{n}_0 \right) = 1 
\nonumber
\end{align}

\noindent This definition extends in an obvious way to the basis elements of $\mathcal{V}^{*}$. For $\langle m| = \langle m_0|_0\otimes \cdots \otimes \langle m_M|_M$ and $\langle n| = \langle n_0|_0 \otimes \cdots \otimes \langle n_M|_M$, we say that $\langle m|$ is admissible to $\langle n|$, and write $\langle n| \triangleleft \langle m|$, if and only if the above condition on the occupation numbers is satisfied.
}
\end{definition}

\begin{example}
{\rm
Fix $M=4$, and let 

\begin{align}
|m\rangle 
&= 
|2\rangle_0 \otimes |4\rangle_1 \otimes |0\rangle_2 \otimes |1\rangle_3 \otimes |2\rangle_4
\\
|n\rangle 
&= 
|2\rangle_0 \otimes |3\rangle_1 \otimes |0\rangle_2 \otimes |2\rangle_3 \otimes |1\rangle_4
\end{align}

\noindent The partial sums are given by

\begin{align}
\Big\{\Sigma^{m}_0,\Sigma^{m}_1,\Sigma^{m}_2,\Sigma^{m}_3,\Sigma^{m}_4 \Big\}
&
=
\{9,7,3,3,2\}
\\
\Big\{\Sigma^{n}_0,\Sigma^{n}_1,\Sigma^{n}_2,\Sigma^{n}_3,\Sigma^{n}_4\Big\}
&
=
\{8,6,3,3,1\}
\end{align}

\noindent and we find that

\begin{align}
\left(\Sigma^{m}_i - \Sigma^{n}_i\right)
=
\left\{
\begin{array}{ll}
1,\quad & i=0
\\
1,\quad & i=1
\\
0,\quad & i=2
\\
0,\quad & i=3
\\
1,\quad & i=4
\end{array}
\right.
\end{align}

\noindent Therefore, these two basis vectors satisfy the condition $|m\rangle \triangleright |n\rangle$.
}
\end{example}

\subsection{Interlacing partitions}

\begin{definition}
{\rm
Let $|\mu) = |\mu_1,\ldots,\mu_{l+1})$ and $|\nu) = |\nu_1,\ldots,\nu_l)$ be two partitions in $\mathcal{F}_{\psi}^{(0)}$. We say that $|\mu)$ {\it interlaces} $|\nu)$, and write $|\mu) \succ |\nu)$, if and only if

\begin{align}
\mu_i \geq \nu_i \geq \mu_{i+1}
\label{3-2-interlace}
\end{align}

\noindent for all $1 \leq i \leq l$. Similarly, for the two partitions $(\mu| = ( \mu_1,\ldots,\mu_{l+1}|$ and $( \nu| = ( \nu_1,\ldots,\nu_l|$ in $\mathcal{F}_{\psi}^{*(0)}$ we say that $(\mu|$ interlaces $(\nu|$, and write $( \nu| \prec ( \mu|$, if and only if the above condition on the partition parts is satisfied.\footnote{The relationship (\ref{3-2-interlace}) between two partitions is ubiquitous in the literature, albeit under different nomenclature. For example, the condition $\mu \succ \nu$ is equivalent to saying that the skew diagram $\mu/\nu$ forms a horizontal strip, which is the terminology preferred in \cite{mac} and most other references.}
}
\end{definition} 

\begin{example} 
{\rm
The partitions 

\begin{align}
|\mu)
=
|\mu_1,\ldots,\mu_7)
&=
|4,4,3,1,1,1,1)
\\
|\nu)
=
|\nu_1,\ldots,\nu_6)
&=
|4,3,3,1,1,1)
\end{align}

\noindent obey $\mu_i \geq \nu_i \geq \mu_{i+1}$ for all $1 \leq i \leq 6$. Therefore, we have $|\mu) \succ |\nu)$. This example is further illustrated by the figure below.
}
\end{example}

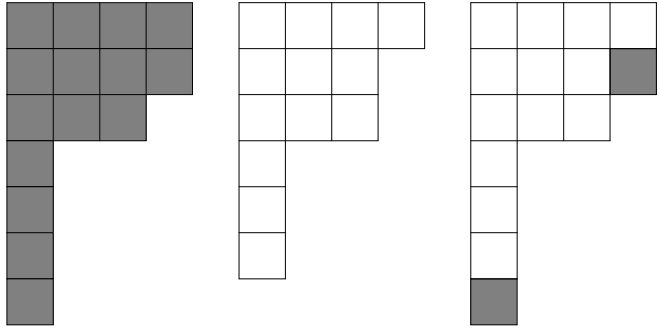
\begin{figure}[H]

\centering

\begin{minipage}{9.4cm}

\setlength{\unitlength}{0.0003cm}
\begin{picture}(28000,16000)(-21000,8000)

\shade\path(-20000,22000)(-12000,22000)(-12000,20000)(-20000,20000)(-20000,22000)
\shade\path(-20000,20000)(-12000,20000)(-12000,18000)(-20000,18000)(-20000,20000)
\shade\path(-20000,18000)(-14000,18000)(-14000,16000)(-20000,16000)(-20000,18000)
\shade\path(-20000,16000)(-18000,16000)(-18000,14000)(-20000,14000)(-20000,16000)
\shade\path(-20000,14000)(-18000,14000)(-18000,12000)(-20000,12000)(-20000,14000)
\shade\path(-20000,12000)(-18000,12000)(-18000,10000)(-20000,10000)(-20000,12000)
\shade\path(-20000,10000)(-18000,10000)(-18000,8000)(-20000,8000)(-20000,10000)

\path(-18000,16000)(-18000,22000)
\path(-16000,16000)(-16000,22000)
\path(-14000,18000)(-14000,22000)

\path(-10000,22000)(-2000,22000)
\path(-10000,20000)(-2000,20000)
\path(-10000,18000)(-4000,18000)
\path(-10000,16000)(-4000,16000)
\path(-10000,14000)(-8000,14000)
\path(-10000,12000)(-8000,12000)
\path(-10000,10000)(-8000,10000)

\path(-10000,10000)(-10000,22000)
\path(-8000,10000)(-8000,22000)
\path(-6000,16000)(-6000,22000)
\path(-4000,16000)(-4000,22000)
\path(-2000,20000)(-2000,22000)

\shade\path(6000,20000)(8000,20000)(8000,18000)(6000,18000)(6000,20000)
\shade\path(0,10000)(2000,10000)(2000,8000)(0,8000)(0,10000)

\path(0,22000)(8000,22000)
\path(0,20000)(8000,20000)
\path(0,18000)(6000,18000)
\path(0,16000)(6000,16000)
\path(0,14000)(2000,14000)
\path(0,12000)(2000,12000)
\path(0,10000)(2000,10000)

\path(0,10000)(0,22000)
\path(2000,10000)(2000,22000)
\path(4000,16000)(4000,22000)
\path(6000,16000)(6000,22000)
\path(8000,20000)(8000,22000)


\end{picture}

\end{minipage}

\caption[Interlacing partitions $|\mu) = |4,4,3,1,1,1,1)$ and $|\nu) = |4,3,3,1,1,1)$]{Interlacing partitions $|\mu) = |4,4,3,1,1,1,1)$ and $|\nu) = |4,3,3,1,1,1)$. The grey Young diagram represents $\mu$, while the white Young diagram represents $\nu$. When the smaller Young diagram is stacked on the larger one, the visible grey boxes comprise the skew diagram $\mu/\nu$. Because the partitions interlace, $\mu/\nu$ is a horizontal strip, \cite{mac}.}

\end{figure}

\subsection{Admissible basis vectors map to interlacing partitions}

\begin{lemma}
{\rm
Let $|m\rangle = |m_0\rangle_0 \otimes \cdots \otimes |m_M\rangle_M$ and $|n\rangle = |n_0\rangle_0 \otimes \cdots \otimes |n_M\rangle_M$ be basis elements of $\mathcal{V}$, and let

\begin{align}
|\mu) = \mathcal{M}_{\psi} |m\rangle ,\quad
|\nu) = \mathcal{M}_{\psi} |n\rangle 
\end{align}

\noindent be their corresponding partitions in $\mathcal{F}_{\psi}^{(0)}$. Then if $|m\rangle$ is admissible to $|n\rangle$, the partition $|\mu)$ interlaces $|\nu)$. That is,

\begin{align}
|m\rangle \triangleright |n\rangle 
\implies
|\mu) \succ |\nu)
\label{3-2-lem1}
\end{align}

\noindent Similarly, at the level of the dual spaces $\mathcal{V}^{*}$ and $\mathcal{F}_{\psi}^{*(0)}$ we have

\begin{align}
\langle n| \triangleleft \langle m|
\implies 
(\nu| \prec (\mu|
\label{3-2-lem1.5}
\end{align}
}
\end{lemma}

\begin{proof}
Due to the assumption $|m\rangle \triangleright |n\rangle$ we know that $\Sigma^m_0-\Sigma^n_0 =1$, which implies that the corresponding partitions $|\mu)$ and $|\nu)$ have $\Sigma^n_0+1$ and $\Sigma^n_0$ parts, respectively. The proof is achieved by showing that

\begin{align}
\mu_i \geq \nu_i \geq \mu_{i+1},
\quad {\rm for\ all}\ 
1\leq i \leq \Sigma^n_0
\label{intcond}
\end{align}

\noindent We will demonstrate this fact by contradiction. Using the definition (\ref{3-2-map}) of the map $\mathcal{M}_{\psi}$, we write

\begin{align}
|\mu) = |M^{m_M},\ldots,1^{m_1},0^{m_0}),
\quad
|\nu) = |M^{n_M},\ldots,1^{n_1},0^{n_0})
\label{3-2-parts}
\end{align}

\noindent from which we recover the inequalities

\begin{align}
&
\Sigma_{\mu_j+1}^m < j \leq \Sigma_{\mu_j}^m
\quad {\rm for\ all}\ 1 \leq j \leq \Sigma_0^m
\label{ineq1}
\\
&
\Sigma_{\nu_j+1}^n < j \leq \Sigma_{\nu_j}^n
\quad {\rm for\ all}\ 1 \leq j \leq \Sigma_0^n
\label{ineq2}
\end{align}

\noindent Now suppose that $\nu_i > \mu_i$ for some $1 \leq i \leq \Sigma_0^n$. Then from (\ref{ineq1}) we have $\Sigma_{\nu_i}^m \leq \Sigma_{\mu_i+1}^m < i$, while from (\ref{ineq2}) we have $i \leq \Sigma_{\nu_i}^n$. Together these imply that $\Sigma^{m}_{\nu_i} -\Sigma^{n}_{\nu_i} < 0$, which contradicts the assumption $|m\rangle \triangleright |n\rangle$.

Alternatively, suppose that $\mu_{i+1} > \nu_i$ for some $1\leq i \leq \Sigma_0^n$. Then from (\ref{ineq1}) we have $i+1 \leq \Sigma_{\mu_{(i+1)}}^m$, while from (\ref{ineq2}) we have $\Sigma_{\mu_{(i+1)}}^n \leq \Sigma_{\nu_i+1}^n < i$. Together these imply that $1<\Sigma_{\mu_{(i+1)}}^m-\Sigma_{\mu_{(i+1)}}^n$, which once again contradicts the assumption $|m\rangle \triangleright |n\rangle$. 

Hence we see that to avoid any contradiction the sequence of inequalities (\ref{intcond}) must hold. The following figure provides an example of admissible basis vectors and interlacing partitions.






\end{proof}

\begin{figure}[H]

\centering
\begin{minipage}{4.3in}

\setlength{\unitlength}{0.0003cm}
\begin{picture}(20000,19000)(-13000,6000)

\put(-6000,24500){\scriptsize$|n\rangle$}
\put(-6000,22500){\scriptsize$|m\rangle$}
\path(-3000,24000)(15000,24000)

\path(-1000,22000)(-1000,26000)
\put(-1000,25000){\circle*{300}}
\put(-750,25000){\tiny0}
\put(-1000,23000){\circle*{300}}
\put(-750,23000){\tiny0}

\path(1000,22000)(1000,26000)
\put(1000,25000){\circle*{300}}
\put(1250,25000){\tiny2}
\put(1000,23000){\circle*{300}}
\put(1250,23000){\tiny3}

\path(3000,22000)(3000,26000)
\put(3000,25000){\circle*{300}}
\put(3250,25000){\tiny1}
\put(3000,23000){\circle*{300}}
\put(3250,23000){\tiny1}

\path(5000,22000)(5000,26000)
\put(5000,25000){\circle*{300}}
\put(5250,25000){\tiny1}
\put(5000,23000){\circle*{300}}
\put(5250,23000){\tiny0}

\path(7000,22000)(7000,26000)
\put(7000,25000){\circle*{300}}
\put(7250,25000){\tiny1}
\put(7000,23000){\circle*{300}}
\put(7250,23000){\tiny1}

\path(9000,22000)(9000,26000)
\put(9000,25000){\circle*{300}}
\put(9250,25000){\tiny0}
\put(9000,23000){\circle*{300}}
\put(9250,23000){\tiny0}

\path(11000,22000)(11000,26000)
\put(11000,25000){\circle*{300}}
\put(11250,25000){\tiny1}
\put(11000,23000){\circle*{300}}
\put(11250,23000){\tiny1}

\path(13000,22000)(13000,26000)
\put(13000,25000){\circle*{300}}
\put(13250,25000){\tiny1}
\put(13000,23000){\circle*{300}}
\put(13250,23000){\tiny2}

\shade\path(0,8000)(2000,8000)(2000,6000)(0,6000)(0,8000)
\shade\path(6000,16000)(8000,16000)(8000,14000)(6000,14000)(6000,16000)
\shade\path(8000,18000)(10000,18000)(10000,16000)(8000,16000)(8000,18000)
\shade\path(10000,18000)(12000,18000)(12000,16000)(10000,16000)(10000,18000)
\shade\path(12000,20000)(14000,20000)(14000,18000)(12000,18000)(12000,20000)

\path(0,22000)(14000,22000)
\path(0,20000)(14000,20000)
\path(0,18000)(12000,18000)
\path(0,16000)(8000,16000)
\path(0,14000)(6000,14000)
\path(0,12000)(4000,12000)
\path(0,10000)(2000,10000)
\path(0,8000)(2000,8000)

\path(0,8000)(0,22000)
\path(2000,8000)(2000,22000)
\path(4000,12000)(4000,22000)
\path(6000,14000)(6000,22000)
\path(8000,16000)(8000,22000)
\path(10000,18000)(10000,22000)
\path(12000,18000)(12000,22000)
\path(14000,20000)(14000,22000)


\shade\path(14000,8000)(16000,8000)(16000,6000)(14000,6000)(14000,8000)
\put(17000,7000){\scriptsize$\mu$}

\path(14000,10000)(16000,10000)(16000,8000)(14000,8000)(14000,10000)
\put(17000,9000){\scriptsize$\nu$}

\end{picture}

\end{minipage}

\caption[Admissible basis vectors and corresponding interlacing partitions]{Admissible basis vectors and corresponding interlacing partitions. The two rows of numbers represent admissible basis vectors. The top row of occupation numbers gives rise to the white Young diagram, while the bottom row gives rise to the grey Young diagram. When these Young diagrams are stacked, the skew diagram is a horizontal strip.}

\end{figure}
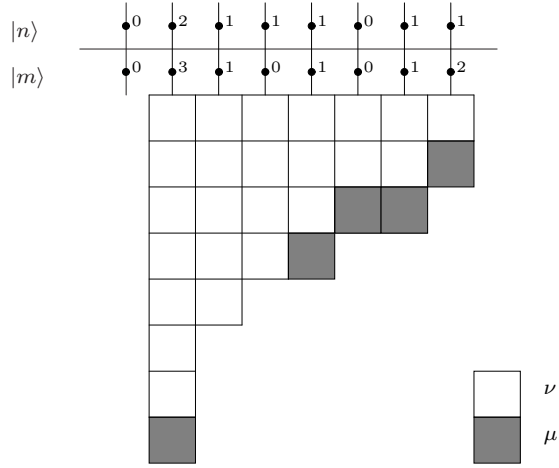

\subsection{Calculation of $\mathbb{B}(x) | n \rangle$}

\begin{lemma}
\label{3-lem2}
{\rm
Define $\mathbb{B}(x) = x^{\frac{M}{2}} B(x)$ and let $|n\rangle = |n_0\rangle_0 \otimes \cdots \otimes |n_M\rangle_M$ be an arbitrary basis vector of $\mathcal{V}$. The action of $\mathbb{B}(x)$ on $|n\rangle$ is given by

\begin{align}
\mathbb{B}(x) |n\rangle
=
\sum_{|m\rangle \triangleright |n\rangle}
\prod_{i=1}^{M}
x^{i(m_i-n_i)} 
|m\rangle
\label{3-2-lem2}
\end{align}

\noindent where the sum is over all basis vectors $|m\rangle = |m_0\rangle_0 \otimes \cdots \otimes |m_M\rangle_M$ which are admissible to $|n\rangle$.
}
\end{lemma}

\begin{proof}
Let $\langle m| = \langle m_0|_0 \otimes \cdots \otimes \langle m_M|_M$ be an arbitrary basis vector of $\mathcal{V}^{*}$. We begin by writing the $B$-operator as a contraction on the auxiliary space $\mathcal{V}_a$, as follows 

\begin{align}
B(x) 
=
\left(
\begin{array}{cc}
1 & 0 
\end{array}
\right)_a
\left(
\begin{array}{cc}
A(x) & B(x)
\\
C(x) & D(x)
\end{array}
\right)_a
\left(
\begin{array}{c}
0 
\\
1
\end{array}
\right)_a
=  
\uparrow^{*}_a
T_a(x)
\downarrow_a
\end{align}

\noindent which leads to the equation

\begin{align}
\langle m| B(x) |n\rangle
=
\uparrow^{*}_a
\otimes
\langle m|
T_a(x)
|n\rangle
\otimes
\downarrow_a
=
\uparrow^{*}_a
\otimes
\langle m|
L_{aM}(x) \ldots L_{a0}(x)
|n\rangle
\otimes
\downarrow_a
\end{align}

%

\noindent By commuting operators and vectors which reside in different spaces we find that

\begin{align}
\langle m|
B(x)
|n\rangle
=
\uparrow^{*}
L^{(M)}(x)
\ldots
L^{(0)}(x)
\downarrow
\end{align}

\noindent where we have dropped the redundant subscripts $a$, and have defined the modified $L$-matrices

\begin{align}
L^{(i)}(x)
=
\left(
\begin{array}{cc}
 \langle m_i|_i x^{-\frac{1}{2}} |n_i\rangle_i
&
\langle m_i|_i\phid_i |n_i\rangle_i
\\
\langle m_i|_i \phi_i |n_i\rangle_i
&
 \langle m_i|_i x^{\frac{1}{2}} |n_i\rangle_i
\end{array}
\right)
\end{align}

\noindent for all $0 \leq i \leq M$. Calculating the entries within these matrices explicitly, we obtain

\begin{align}
L^{(i)}(x)
=
\left\{
\begin{array}{cl}
\left(
\begin{array}{cc}
x^{-\frac{1}{2}} & 0
\\
0 & x^{\frac{1}{2}}
\end{array}
\right)
&
\quad\quad
m_i = n_i
\\
\\
\left(
\begin{array}{cc}
0 & 1
\\
0 & 0
\end{array}
\right)
&
\quad\quad
m_i = n_i+1
\\
\\
\left(
\begin{array}{cc}
0 & 0
\\
1 & 0
\end{array}
\right)
&
\quad\quad
m_i+1 = n_i
\\
\\
\left(
\begin{array}{cc}
0 & 0 
\\
0 & 0
\end{array}
\right)
&
\quad\quad
{\rm otherwise}
\end{array}
\right.
\label{3-2-Lmat}
\end{align}

\noindent Using the expression (\ref{3-2-Lmat}) for $L^{(i)}(x)$, we find that

\begin{align}
\langle m|B(x)|n\rangle
=
\uparrow^{*}
L^{(M)}(x)
\ldots
L^{(0)}(x)
\downarrow
=
0
\label{3-2-case1}
\end{align}

\noindent when $|m\rangle \notad |n\rangle$. In the case when $|m\rangle \triangleright |n\rangle$, let $\{p_1 < \cdots < p_r\}$ be the set of all integers $p$ such that $m_p = n_p+1$. Similarly, let $\{q_1 < \cdots < q_s\}$ be the set of all integers $q$ such that $m_q+1=n_q$. The admissibility relation means that necessarily $s=r-1$ and 

\begin{align}
p_i < q_i < p_{i+1},
\quad
{\rm for\ all}\ 1\leq i \leq r-1
\label{3-2-order}
\end{align} 

\noindent By virtue of the ordering (\ref{3-2-order}) and the expression (\ref{3-2-Lmat}) for $L^{(i)}(x)$, we obtain  

\begin{align}
&
\langle m| B(x) |n\rangle
=
\uparrow^{*}
L^{(M)}(x)
\ldots
L^{(0)}(x)
\left(
\begin{array}{cc}
0 & 0
\\
1 & 0
\end{array}
\right)
\uparrow
=
\\
&
\uparrow^{*}
\lprod_{i=1}^{r}
\left[
\left(
\begin{array}{cc}
x^{-\frac{1}{2}} & 0
\\
0 & x^{\frac{1}{2}}
\end{array}
\right)^{q_i-p_i-1}
\left(
\begin{array}{cc}
0 & 1
\\
0 & 0 
\end{array}
\right)
\left(
\begin{array}{cc}
x^{-\frac{1}{2}} & 0
\\
0 & x^{\frac{1}{2}}
\end{array}
\right)^{p_i-q_{(i-1)}-1}
\left(
\begin{array}{cc}
0 & 0
\\
1 & 0
\end{array}
\right)
\right]
\uparrow
\nonumber
\end{align}

\noindent where we have defined $q_0=-1$ and $q_r=M+1$. Calculating this matrix product explicitly, we find

\begin{align}
\langle m| B(x) |n\rangle
=
x^{-\frac{M}{2}}
\prod_{i=1}^{r-1}
x^{p_i-q_i}
x^{p_r}
=
x^{-\frac{M}{2}}
\prod_{i=1}^{M}
x^{i(m_i-n_i)}
\label{3-2-case2}
\end{align}

\noindent Combining the equations (\ref{3-2-case1}) and (\ref{3-2-case2}) into a single case, we have

\begin{align}
x^{\frac{M}{2}}
\langle m| B(x) |n\rangle
=
\left\{
\begin{array}{ll}
\displaystyle{\prod_{i=1}^{M}}
x^{i(m_i-n_i)},
&
\quad |m\rangle \triangleright |n\rangle
\\
\\
0, 
&
\quad
{\rm otherwise}
\end{array}
\right.
\end{align}

\noindent The result (\ref{3-2-lem2}) follows from the orthonormality (\ref{phase-isom}) of the basis vectors of $\mathcal{V}$, and from the definition of $\mathbb{B}(x)$.
\end{proof}

\subsection{Calculation of $\langle n| \mathbb{C}(x)$}

\begin{lemma}
\label{3-lem3}
{\rm
Define $\mathbb{C}(x) = x^{\frac{M}{2}} C(1/x)$ and let $\langle n| = \langle n_0|_0 \otimes \cdots \otimes \langle n_M|_M$ be an arbitrary basis vector of $\mathcal{V}^{*}$. The action of $\mathbb{C}(x)$ on $\langle n|$ is given by

\begin{align}
\langle n| \mathbb{C}(x)
=
\sum_{\langle n| \triangleleft \langle m|}
\prod_{i=1}^{M}
x^{i(m_i-n_i)}
\langle m|
\label{3-2-lem3}
\end{align}

\noindent where the sum is over all basis vectors $\langle m| = \langle m_0|_0 \otimes \cdots \otimes \langle m_M|_M$ which are admissible to $\langle n|$.
}
\end{lemma}

\begin{proof}
A simple modification of the proof of lemma 2.
\end{proof}

%
%

\subsection{Calculation of $\mathcal{M}_{\psi} \mathbb{B}(x) |n\rangle$ and $\langle n| \mathbb{C}(x) \mathcal{M}_{\psi}^{*}$}

%

Let $|n\rangle$ and $\langle n|$ be arbitrary basis vectors of $\mathcal{V}$ and $\mathcal{V}^{*}$, respectively, and let $|\nu)$ and $(\nu|$ be their corresponding partitions, given by equations (\ref{part1}) and (\ref{part2}). Furthermore, define $l = \ell(\nu)$ to be the number of non-zero parts in the partition $\nu$. Using the definition (\ref{3-2-map}) of the maps $\mathcal{M}_{\psi}$ and $\mathcal{M}^{*}_{\psi}$, the expressions (\ref{3-2-lem2}) and (\ref{3-2-lem3}) and the result of lemma 1, we obtain

\begin{align}
\mathcal{M}_{\psi} \mathbb{B}(x) |n\rangle 
&=
\sum_{\nu \prec \mu \subseteq [l+1,M]}
x^{|\mu|-|\nu|}
|\mu)
\label{acespa}
\\
\langle n| \mathbb{C}(x) \mathcal{M}_{\psi}^{*}
&=
\sum_{ \nu \prec \mu \subseteq [l+1,M]}
x^{|\mu|-|\nu|}
(\mu|
\label{kinspa}
\end{align}

\noindent Both sums are over all partitions $\mu$ which interlace with $\nu$, and whose Young diagrams are contained in the rectangle $[l+1,M]$.

\subsection{Skew Schur functions}

For an arbitrary pair of partitions $\mu,\nu$ and an indeterminate $x$, the single variable {\it skew Schur function} $s_{\mu/\nu}(x)$ is given by

\begin{align}
s_{\mu / \nu} (x) 
=
\left\{
\begin{array}{ll}
x^{|\mu|-|\nu|}, & \mu \succ \nu
\\
\\
0,                   & {\rm otherwise}
\end{array}
\right.
\label{schuridentity2}
\end{align}

\noindent In the case $\nu = \emptyset$ we have $s_{\mu/\nu}(x) = s_{\mu}(x)$, where $s_{\mu}(x)$ is the ordinary Schur function in a single variable $x$. The skew Schur function satisfies the identity

\begin{align}
s_{\mu}\{x_1,\ldots,x_n\}
=
\sum_{\nu \subseteq [n-1,\infty]}
s_{\mu / \nu}(x_n)
s_{\nu}\{x_1,\ldots,x_{n-1}\}
\label{skew-schur-id}
\end{align}

\noindent where the sum is taken over all partitions $\nu$ whose lengths satisfy $\ell(\nu) \leq n-1$, and $s_{\mu}\{x_1,\ldots,x_n\}$ and $s_{\nu}\{x_1,\ldots,x_{n-1}\}$ are Schur functions in $n$ and $n-1$ variables, respectively.\footnote{For more information on skew Schur functions, the reader is referred to section 5 of chapter I in \cite{mac}.}

\subsection{Calculation of $\mathcal{M}_{\psi} \mathbb{B}(x_1)\ldots \mathbb{B}(x_N) |0\rangle$}

The purpose of the previous subsection was to provide the equation (\ref{skew-schur-id}), which we now use to calculate the phase model Bethe eigenvectors explicitly.

\begin{lemma}
{\rm Let $\{x_1,\ldots,x_N\}$ be a finite set of variables. We claim that

\begin{align}
\mathcal{M}_{\psi}
\mathbb{B}(x_1)
\ldots
\mathbb{B}(x_N)
|0\rangle
=
\sum_{\mu \subseteq [N,M]}
s_{\mu}\{x_1,\ldots,x_N\}
|\mu)
\label{p-lem4-0}
\end{align}

\noindent where $s_{\mu}\{x_1,\ldots,x_N\}$ is the Schur function in $N$ variables (\ref{schurfunct}), and the sum is over all partitions $\mu$ whose Young diagrams are contained in the rectangle $[N,M]$. This result was originally obtained in \cite{bog1}.
}
\end{lemma}

\begin{proof}
{\rm We begin by specializing equation (\ref{acespa}) to the case $|n\rangle = |0\rangle$, to obtain

\begin{align}
\mathcal{M}_{\psi}  \mathbb{B}(x) |0\rangle 
=
\sum_{\emptyset \prec \mu \subseteq [1,M] }
s_{\mu/\emptyset}(x)
|\mu)
=
\sum_{\mu \subseteq [1,M] }
s_{\mu}(x)
|\mu)
\label{p-lem4-1}
\end{align}

\noindent where we have used the equation (\ref{schuridentity2}) for the skew Schur function, and the definition $\ell(\emptyset) = 0$. We use equation (\ref{p-lem4-1}) as the basis for induction, and assume that 

\begin{align}
\mathcal{M}_{\psi} \mathbb{B}(x_1) \ldots \mathbb{B}(x_{N-1}) |0\rangle 
=
\sum_{\nu \subseteq [N-1,M] }
s_{\nu}\{x_1,\ldots,x_{N-1}\}
|\nu)
\end{align}

\noindent for some $N \geq 2$. In terms of the basis vectors of $\mathcal{V}$, this assumption is written as

\begin{align}
\mathbb{B}(x_1)\ldots \mathbb{B}(x_{N-1}) |0\rangle
=
\sum_{|n\rangle | \Sigma_0 = N-1}
s_{\nu}\{x_1,\ldots,x_{N-1}\}
|n\rangle
\label{p-lem4-2}
\end{align}

\noindent where the sum is over all basis vectors $|n\rangle$ whose occupation numbers satisfy the condition $\sum_{i=0}^{M} n_i = N-1$, and $\nu$ is the partition corresponding to each $|n\rangle$. Acting on (\ref{p-lem4-2}) with the composition of operators $\mathcal{M}_{\psi} \circ \mathbb{B}(x_N)$ and using the fact that the $B$-operators commute (\ref{phasebc}), we obtain

\begin{align}
\mathcal{M}_{\psi}
\mathbb{B}(x_1) \ldots \mathbb{B}(x_N) |0\rangle
\label{p-lem4-3}
=
\sum_{\nu \subseteq [N-1,M] }
s_{\nu}\{x_1,\ldots,x_{N-1}\}
\sum_{\nu \prec \mu \subseteq [N,M] }
s_{\mu/\nu}(x_N)
|\mu)
\end{align}

\noindent Since $s_{\mu/\nu}(x_N) = 0$ if $\mu \not\succ \nu$, we may alter the sums appearing in (\ref{p-lem4-3}), yielding

\begin{align}
\mathcal{M}_{\psi}
\mathbb{B}(x_1) \ldots \mathbb{B}(x_N) |0\rangle
&
=
\sum_{\mu \subseteq [N,M] }
\sum_{\nu \subseteq [N-1,M] }
s_{\mu / \nu}(x_N)
s_{\nu}\{x_1,\ldots,x_{N-1}\}
|\mu)
\\
&
=
\sum_{\mu \subseteq [N,M]}
\sum_{\nu \subseteq [N-1,\infty] }
s_{\mu / \nu}(x_N)
s_{\nu}\{x_1,\ldots,x_{N-1}\}
|\mu)
\nonumber
\end{align}

\noindent where the final equality holds since every part of $\mu$ is less than or equal to $M$, and therefore $s_{\mu/\nu}(x_N) = 0$ if any part of $\nu$ is greater than $M$. Using the identity (\ref{skew-schur-id}) we evaluate the sum over $\nu$ explicitly, producing the equation (\ref{p-lem4-0}). Therefore by induction the result (\ref{p-lem4-0}) must hold for arbitrary $N \geq 1$.
}
\end{proof}

%
%
%

\subsection{Calculation of $\langle 0| \mathbb{C}(x_N)\ldots \mathbb{C}(x_1) \mathcal{M}_{\psi}^{*} $}

By following essentially the same steps that were used in the previous subsection, we can also derive the expression

\begin{align}
\langle 0| \mathbb{C}(x_N) \ldots \mathbb{C}(x_1)
\mathcal{M}_{\psi}^{*}
=
\sum_{\mu \subseteq [N,M]}
s_{\mu}\{x_1,\ldots,x_N\}
(\mu|
\label{eigenvec-phase*}
\end{align}

\noindent for the dual Bethe eigenvectors. As before, this sum is taken over all partitions $\mu$ whose Young diagrams are contained in the rectangle $[N,M]$.

\subsection{Charged fermionic expression for Bethe eigenvectors}

The first goal of this subsection is to show that $\mathcal{M}_{\psi}$ maps the phase model Bethe eigenvectors $\mathbb{B}(x_1)\ldots \mathbb{B}(x_N) |0\rangle$ to vectors $g_{\psi}|0\rangle \in \mathcal{F}_{\psi}^{(0)}$ which satisfy the charged fermion bilinear identity (\ref{KPhelp}). In order to do this, we make some definitions. For all integers $i \geq -N$ and $ 1\leq j \leq N$ we define 

\begin{align}
c_{i,j}\{x\}
=
\left\{
\begin{array}{ll}
h_{i+j}\{x\},\quad & -N \leq i < M
\\
\\
0,\quad & {i \geq M}
\end{array}
\right.
\end{align}

\noindent where $h_{i+j}\{x\}$ is a complete symmetric function (\ref{complete}) in the finite set of variables $\{x\} = \{x_1,\ldots,x_N\}$. Using this expression for the functions $c_{i,j}\{x\}$, for all ordered sets of integers $\{m\} = \{m_1 > \cdots > m_N \geq -N\}$ let us also define the coefficients  

\begin{align}
c_{\{m\}}\{x\}
=
\det \Big( c_{m_i,j}\{x\} \Big)_{1 \leq i, j \leq N}
\label{phase-coeff}
\end{align}

\noindent Since $c_{i,j}\{x\} = 0$ if $i \geq M$, the coefficient $c_{\{m\}}\{x\}$ vanishes if $m_1 \geq M$. Now suppose that $\mu = \{\mu_1,\ldots,\mu_N\}$ is the partition formed by setting $\mu_i = m_i+i$ for all $1 \leq i \leq N$. By the definition of the Schur functions (\ref{schurfunct}), it follows that

\begin{align}
c_{\{m\}}\{x\} 
=
\left\{
\begin{array}{ll}
s_{\mu}\{x\}, & \mu \subseteq [N,M]
\\
\\
0, & \mu \not\subseteq [N,M]
\end{array}
\right.
\label{phase-coeff3}
\end{align}

\noindent Returning to the expression (\ref{p-lem4-0}) for the Bethe eigenvectors, we can use the coefficients (\ref{phase-coeff3}) to write

\begin{align}
\mathcal{M}_{\psi} 
\mathbb{B}(x_1)\ldots \mathbb{B}(x_N) |0\rangle
=
\sum_{{\rm card}\{m\} = N} c_{\{m\}}\{x\} \psi_{m_1}\ldots \psi_{m_N} |-N\rangle
\label{phase-coeff2}
\end{align}

\noindent where the sum is over all sets of integers $\{m\} = \{m_1 > \cdots > m_N \geq -N\}$, and we have made the identification $|\mu)=\psi_{m_1}\ldots\psi_{m_N}|-N\rangle$. Thanks to their determinant form (\ref{phase-coeff}) and lemma 12 of chapter 1, we see that the coefficients $c_{\{m\}}\{x\}$ satisfy the KP Pl\"ucker relations (\ref{KPpluck1}). This implies that the right hand side of (\ref{phase-coeff2}) satisfies the charged fermion bilinear identity (\ref{KPhelp}), as we intended to show.

The second goal of this subsection is to express $\mathcal{M}_{\psi} \mathbb{B}(x_1)\ldots \mathbb{B}(x_N) |0\rangle$ in the orbit of the Fock vacuum  under $GL_{\infty}$. We know that this is possible using theorem 2 of chapter 1, and the fact that the right hand side of (\ref{phase-coeff2}) satisfies the CFBI. We begin by writing the formula (\ref{p-lem4-0}) for $\mathcal{M}_{\psi} \mathbb{B}(x_1)\ldots \mathbb{B}(x_N) |0\rangle$ in terms of the canonical $\mathcal{F}_{\psi}^{(0)}$ basis (\ref{repclA5}), yielding

\begin{align}
\mathcal{M}_{\psi}
\mathbb{B}(x_1) \ldots \mathbb{B}(x_N) 
|0\rangle
=
|0\rangle 
+ 
\sum_{\substack{1 \leq m \leq M \\ 1 \leq n \leq N}}
(-)^{n-1}
s_{\{m,1^{n-1}\}}\{x\}
\psi_{m-1} \psis_{-n} |0\rangle
+
g_{\psi}^{(1)}|0\rangle
\label{phase-coeff4}
\end{align}

\noindent where $s_{\{m,1^{n-1}\}}\{x\}$ is the Schur function associated to the partition with one part of size $m$ and $n-1$ parts of size 1, and we assume that all monomials within $g_{\psi}^{(1)} \in Cl_{\psi}^{(0)}$ consist of at least two $(+1)$ and two $(-1)$ fermions. Because the right hand side of (\ref{phase-coeff4}) obeys the CFBI, we can use the method adopted in the proof of lemma 10 in chapter 1 to obtain

\begin{align}
\mathcal{M}_{\psi}
\mathbb{B}(x_1) \ldots \mathbb{B}(x_N) |0\rangle
=
\exp\left( 
\sum_{\substack{1 \leq m \leq M \\ 1 \leq n \leq N}}
(-)^{n-1}
s_{\{m,1^{n-1}\}}\{x\}
\psi_{m-1} \psis_{-n}
\right)
|0\rangle
\label{phase-fermi}
\end{align}

\noindent This result explicitly places $\mathcal{M}_{\psi} \mathbb{B}(x_1)\ldots \mathbb{B}(x_N) |0\rangle$ in the vacuum orbit of $GL_{\infty}$. 

Finally, let us remark that all of these results can be extended to the dual Bethe eigenvectors $\langle 0|\mathbb{C}(x_N)\ldots \mathbb{C}(x_1)\mathcal{M}_{\psi}^{*}$. For example, by completely analogous reasoning it is possible to show that

\begin{align}
\langle 0|
\mathbb{C}(x_N) \ldots \mathbb{C}(x_1)
\mathcal{M}_{\psi}^{*}
=
\langle 0|
\exp\left(
\sum_{\substack{1 \leq m \leq M \\ 1 \leq n \leq N}}
(-)^{n-1}
s_{\{m,1^{n-1}\}}\{x\}
\psi_{-n} \psis_{m-1}
\right)
\end{align}

\section{Scalar product, boxed plane partitions}
\label{phase-bpp}

\subsection{Plane partitions}

\begin{definition}
{\rm 
A {\it plane partition} $\pi$ is a set of non-negative integers $\pi(i,j)$ which satisfy

\begin{align}
\pi(i,j) \geq \pi(i+1,j),
\quad 
\pi(i,j) \geq \pi(i,j+1)
\label{pp-def}
\end{align}

\noindent for all integers $i,j \geq 1$, as well as the finiteness condition

\begin{align}
\lim_{i \rightarrow \infty} \pi(i,j)
= 
\lim_{j \rightarrow \infty} \pi(i,j) 
= 
0
\end{align}

\noindent An {\it $M$-boxed plane partition} is a set of non-negative integers $\pi(i,j)$ satisfying the above properties, as well as the supplementary condition

\begin{align}
0 \leq \pi(i,j) \leq M
\end{align}

\noindent for all integers $i,j \geq 1$, where $M \geq 1$ is some fixed positive integer.

Plane partitions are two-dimensional analogues of ordinary partitions. They are pictorially represented in one of two ways. The first way uses the notion of a {\it tableau}, whereby the non-negative integer $\pi(i,j)$ is placed in the coordinate-labelled box $(i,j)$ for all $i,j \geq 1$.

\begin{figure}[H]
\begin{center}
\begin{minipage}{3in}

\setlength{\unitlength}{0.0012cm}
\renewcommand{\dashlinestretch}{30}
\begin{picture}(4800, 2000)(-2000, 1000)

\path(0000,0600)(0600,0600)

\path(0000,1200)(1800,1200)
\path(0000,1800)(2400,1800)
\path(0000,2400)(3000,2400)
\path(0000,3000)(0000,0600)
\path(0000,3000)(3000,3000)
\path(0600,3000)(0600,0600)
\path(1200,3000)(1200,1200)
\path(1800,3000)(1800,1200)

\path(2400,3000)(2400,1800)
\path(3000,3000)(3000,2400)

%

\put(0300,0900){1}
\put(0300,1500){2}
\put(0300,2100){3}
\put(0300,2700){4}
\put(0900,1500){1}
\put(0900,2100){2}
\put(0900,2700){2}
\put(1500,1500){1}
\put(1500,2100){1}
\put(1500,2700){1}
\put(2100,2100){1}
\put(2100,2700){1}
\put(2700,2700){1}


\end{picture}
\end{minipage}
\end{center}
\caption[Tableau representation of a plane partition]{Tableau representation of a plane partition. An integer $\pi(i,j)$ is assigned to each box $(i,j)$. The top row of the tableau corresponds to the boxes $(1,n)$, where $1 \leq n \leq 5$, while the left-most column corresponds to the boxes $(m,1)$, where $1 \leq m \leq 4$.}
 
\end{figure}

The second way is by stacking a column of cubes of height $\pi(i,j)$ over the coordinate-labelled square $(i,j)$ for all $i,j \geq 1$. When viewed in its three-dimensional representation, an $M$-boxed plane partition has columns of cubes which are maximally of height $M$.

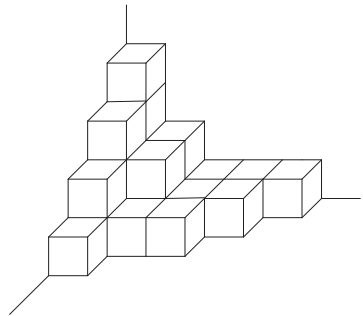
\begin{figure}[H]
\begin{center}
\begin{minipage}{2.5in}
\setlength{\unitlength}{0.00085cm}
\renewcommand{\dashlinestretch}{30}
\begin{picture}(4800, 5000)(-1000, 0)

\path(3900,1500)(4500,1500)
\path(1800,4200)(2400,4200)
\path(1800,4200)(1500,3900)
\path(2400,4200)(2100,3900)
\path(1500,3900)(2100,3900)
\path(1500,3900)(1500,3300)
\path(2100,3900)(2100,3300)
\path(1500,3300)(2100,3312)
\path(2400,4200)(2400,3600)
\path(2400,3600)(2100,3300)
\path(1500,3300)(1200,3000)
\path(2100,3300)(1800,3000)
\path(1200,3000)(1800,3000)
\path(1200,3000)(1200,2400)
\path(1800,3000)(1800,2400)
\path(1200,2400)(1800,2400)
\path(1200,2400)(0900,2100)
\path(1800,2400)(1500,2100)
\path(0900,2100)(1500,2100)
\path(0900,2100)(0900,1500)
\path(1500,2100)(1500,1500)
\path(0900,1500)(1500,1500)
\path(0900,1500)(0600,1200)
\path(1500,1500)(1200,1200)
\path(0600,1200)(1200,1200)
\path(0600,1200)(0600,0600)
\path(1200,1200)(1200,0600)
\path(0600,0600)(1200,0600)
\path(0600,0600)(0000,0000)
\path(2100,3300)(2100,2700)
\path(2100,2700)(1800,2400)
\path(1800,2400)(1800,1800)
\path(1800,1800)(1500,1500)
\path(1500,1500)(1500,0900)
\path(1500,0900)(1200,0600)
\path(2400,3600)(2400,3000)
\path(2400,3000)(2100,2700)
\path(2400,3000)(3000,3000)
\path(2100,2700)(2700,2700)
\path(3000,3000)(2700,2700)
\path(1800,2400)(2400,2400)
\path(2700,2700)(2400,2400)
\path(2400,2400)(2400,1800)
\path(1800,1800)(2400,1800)
\path(2400,1800)(2100,1500)
\path(1500,1500)(2100,1500)
\path(2100,1500)(2100,0900)
\path(1500,0900)(2100,0900)
\path(3000,3000)(3000,2400)
\path(2700,2700)(2700,2100)
\path(3000,2400)(2400,1800)
\path(3000,2400)(4500,2400)
\path(3600,2400)(2700,1500)
\path(2700,2100)(3300,2100)
\path(2400,1800)(3000,1812)
\path(2100,1500)(2700,1500)
\path(2700,1500)(2700,0900)
\path(2100,0900)(2700,0900)
\path(4200,2400)(3900,2100)
\path(4500,2400)(4800,2400)
\path(4800,2400)(4500,2100)
\path(3300,2100)(4500,2100)
\path(3000,1800)(3000,1200)
\path(3900,2100)(3900,1500)
\path(4500,2100)(4500,1500)
\path(4800,2400)(4800,1800)
\path(4800,1800)(4500,1500)
\path(4800,1800)(5400,1800)
\path(3900,2100)(3600,1800)
\path(3000,1800)(3600,1800)
\path(3600,1800)(3600,1200)
\path(3000,1200)(3600,1200)
\path(3900,1500)(3600,1200)
\path(3000,1200)(2700,0900)
\path(1800,4800)(1800,4200)
\end{picture}
\end{minipage}
\end{center}
\caption[Three-dimensional representation of a plane partition]{Three-dimensional representation of a plane partition. Columns of cubes of height $\pi(i,j)$ are stacked above each square $(i,j)$. This plane partition is 4-boxed.}
\end{figure}
 
}
\end{definition}

\subsection{Diagonal slices of plane partitions}

\begin{definition}
{\rm 
Let $\pi$ be an arbitrary plane partition. For $i \geq 0$ define the partitions $|\pi_i) \in \mathcal{F}_{\psi}^{(0)}$ whose parts are given by

\begin{align}
(\pi_{i})_j = \pi(j,i+j)
\end{align}

\noindent for all $j \geq 1$. Similarly for $i \leq 0$ define the partitions $(\pi_i| \in \mathcal{F}_{\psi}^{*(0)}$ whose parts are given by

\begin{align}
(\pi_i)_j = \pi(-i+j,j)
\end{align}

\noindent for all $j \geq 1$. The partitions $|\pi_i)$ and $( \pi_i|$ are called the {\it diagonal slices} of the plane partition $\pi$.
}
\end{definition}

%
%

\begin{figure}[H]

\centering
\begin{minipage}{4.2in}

\setlength{\unitlength}{0.00024cm}
\begin{picture}(30000,28000)(-13000,-20000)

\path(0,0)(2000,-2000)
\path(2000,2000)(8000,-4000)
\path(4000,4000)(12000,-4000)
\path(6000,6000)(16000,-4000)
\path(8000,8000)(18000,-2000)

\path(0,0)(8000,8000)
\path(2000,-2000)(10000,6000)
\path(6000,-2000)(12000,4000)
\path(8000,-4000)(14000,2000)
\path(12000,-4000)(16000,0)
\path(16000,-4000)(18000,-2000)

\put(8000,6000){4}
\put(6000,4000){3}
\put(4000,2000){2}
\put(2000,0){1}

\put(6000,0){1}
\put(8000,2000){2}
\put(10000,4000){2}

\put(8000,-2000){1}
\put(10000,0){1}
\put(12000,2000){1}

\put(12000,-2000){1}
\put(14000,0){1}

\put(16000,-2000){1}


\path(-4000,-12000)(-6000,-14000)(-4000,-16000)(-2000,-14000)(-4000,-12000)
\put(-4000,-14000){1}
\put(-4500,-17000){\scriptsize$\pi_{-3}$}

\path(0,-10000)(-2000,-12000)(0,-14000)(2000,-12000)(0,-10000)
\put(0,-12000){2}
\put(-500,-15000){\scriptsize$\pi_{-2}$}

\path(4000,-8000)(2000,-10000)(6000,-14000)(4000,-16000)(2000,-14000)(6000,-10000)(4000,-8000)
\put(4000,-10000){3}
\put(4000,-14000){1}
\put(3500,-17000){\scriptsize$\pi_{-1}$}

\path(8000,-6000)(6000,-8000)(10000,-12000)(6000,-16000)(8000,-18000)(10000,-16000)(6000,-12000)(10000,-8000)(8000,-6000)
\put(8000,-8000){4}
\put(8000,-12000){2}
\put(8000,-16000){1}
\put(7500,-19000){\scriptsize$\pi_{0}$}

\path(12000,-8000)(10000,-10000)(14000,-14000)(12000,-16000)(10000,-14000)(14000,-10000)(12000,-8000)
\put(12000,-10000){2}
\put(12000,-14000){1}
\put(11500,-17000){\scriptsize$\pi_{1}$}

\path(16000,-10000)(14000,-12000)(18000,-16000)(16000,-18000)(14000,-16000)(18000,-12000)(16000,-10000)
\put(16000,-12000){1}
\put(16000,-16000){1}
\put(15500,-19000){\scriptsize$\pi_{2}$}

\path(20000,-12000)(18000,-14000)(20000,-16000)(22000,-14000)(20000,-12000)
\put(20000,-14000){1}
\put(19500,-17000){\scriptsize$\pi_{3}$}

\path(24000,-14000)(22000,-16000)(24000,-18000)(26000,-16000)(24000,-14000)
\put(24000,-16000){1}
\put(23500,-19000){\scriptsize$\pi_{4}$}

\end{picture}

\end{minipage}

\caption[Diagonal slices of a plane partition]{Diagonal slices of a plane partition. Each column of boxes represents a regular partition. For example, the column labelled $\pi_0$ represents the partition $\{4,2,1\}$.}

\end{figure}
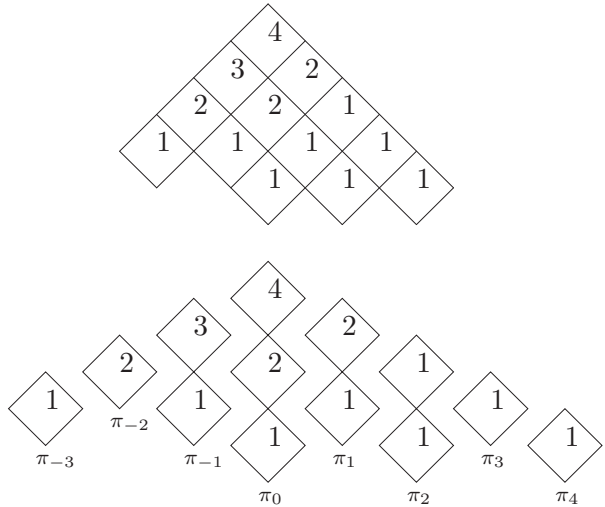

\begin{lemma}
{\rm 
Let $|\pi_{i})$ and $( \pi_{i}|$ be the diagonal slices of the arbitrary plane partition $\pi$. Then we have 

\begin{align}
( \pi_{i-1}| \prec ( \pi_{i}| \ \ {\rm for\ all} \ i \leq 0,
\quad\quad
|\pi_{i}) \succ |\pi_{i+1}) \ \ {\rm for\ all} \ i \geq 0
\end{align}

}
\end{lemma}

\begin{proof} 
This observation is due to Okounkov and Reshetikhin in \cite{or}, where it was used to define and study the Schur stochastic process. The proof is immediate from the definition (\ref{3-2-interlace}) of interlacing partitions and the defining property (\ref{pp-def}) of plane partitions. Since $\pi$ is a plane partition we have 

\begin{align}
\pi(-i+j,j) &\geq \pi(-i+1+j,j) \geq \pi(-i+j+1,j+1)
\ \ {\rm for\ all}\ i \leq 0,\ j\geq 1 
\nonumber
\\
&\implies
(\pi_{i})_j \geq (\pi_{i-1})_j \geq (\pi_{i})_{j+1} 
\ \ {\rm for\ all}\ i \leq 0,\ j\geq 1
\end{align}

\noindent proving that $(\pi_{i-1}| \prec (\pi_{i}|$ for all $i \leq 0$. Similarly we find 

\begin{align}
\pi(j,i+j) &\geq \pi(j,i+1+j) \geq \pi(j+1,i+j+1)
\ \ {\rm for\ all}\ i \geq 0,\ j \geq 1
\nonumber
\\
&\implies 
(\pi_{i})_j \geq (\pi_{i+1})_j \geq (\pi_{i})_{j+1}
\ \ {\rm for\ all}\ i \geq 0,\ j \geq 1
\end{align}

\noindent proving that $|\pi_{i}) \succ |\pi_{i+1})$ for all $i \geq 0$.

\end{proof}

\subsection{Generating $M$-boxed plane partitions}

In this subsection we reproduce the result of \cite{bog1}, where it was shown that the scalar product of the phase model on $M+1$ sites is a generating function for $M$-boxed plane partitions. This correspondence may be realized by iterating the $|n\rangle = |0\rangle$ case of equation (\ref{acespa}) $N$ times, giving

\begin{align}
\mathcal{M}_{\psi}
\mathbb{B}(x_1)
\ldots
\mathbb{B}(x_N)
|0\rangle
=
\sum_{[N,M] \supseteq \pi_{0} \succ \cdots \succ \pi_{N}= \emptyset}\ 
\prod_{i=1}^{N}
x_i^{|\pi_{i-1}|-|\pi_{i}|}
|\pi_{0})
\end{align}

\noindent where the sum is over all interlacing partitions $\{\pi_{0} \succ \cdots \succ \pi_{N}\}$ which are subject to $\pi_{0} \subseteq [N,M]$ and $\pi_{N}=\emptyset$. Similarly, one can iterate the $\langle n| = \langle 0|$ case of (\ref{kinspa}) $N$ times, giving

\begin{align}
\langle 0|
\mathbb{C}(x_N)
\ldots
\mathbb{C}(x_1)
\mathcal{M}_{\psi}^{*}
=
\sum_{\emptyset = \pi_{-N} \prec \cdots \prec \pi_{0} \subseteq [N,M]}\ 
\prod_{i=1}^{N}
x_i^{|\pi_{-i+1}|-|\pi_{-i}|}
(\pi_{0}|
\end{align}

\noindent where the sum is over all interlacing partitions $\{\pi_{-N} \prec \cdots \prec \pi_{0}\}$ which are subject to $\pi_{0} \subseteq [N,M]$ and $\pi_{-N} = \emptyset$. Due to the isometry of the maps (\ref{3-2-map}) and the orthonormality (\ref{part-orth}) of partition states, we thus obtain

\begin{align}
\langle 0|
\mathbb{C}(x_N)
\ldots
\mathbb{C}(x_1)
\mathbb{B}(y_1)
\ldots
\mathbb{B}(y_N)
|0\rangle
=
\sum_{\pi \subseteq [N,N,M]}
A_{\pi}\Big(\{x\},\{y\}\Big)
\label{phase-gf}
\end{align}


\noindent where the sum is over all plane partitions $\pi$ which fit inside the box of dimension $N \times N \times M$, and where we have defined the weighting factor

\begin{align}
A_{\pi}\Big(\{x\},\{y\}\Big)
=
\prod_{i=1}^{N}
x_i^{|\pi_{-i+1}|-|\pi_{-i}|}
y_i^{|\pi_{i-1}|-|\pi_{i}|}
\label{a-fact}
\end{align}

\noindent which depends on the diagonal slices of $\pi$. From equation (\ref{phase-gf}), we see that the scalar product is a generating function of $M$-boxed plane partitions. A closed form expression for this generating function can be obtained by using the formulae (\ref{p-lem4-0}) and (\ref{eigenvec-phase*}) for the Bethe eigenvectors to show that

\begin{align}
&
\langle 0| \mathbb{C}(x_N)\ldots \mathbb{C}(x_1)
\mathbb{B}(y_1)\ldots \mathbb{B}(y_N) |0\rangle
=
\sum_{\mu \subseteq [N,M]}
s_{\mu}\{x\}
s_{\mu}\{y\}
\label{phase-JT}
\\
&
=
\frac{
\displaystyle{
\sum_{M-1 \geq m_1 > \cdots > m_N \geq -N}
}
\det
\left(
x_i^{m_j+N}
\right)_{1 \leq i,j \leq N}
\det
\left(
y_j^{m_i+N}
\right)_{1 \leq i,j \leq N}
}{
\displaystyle{
\prod_{1\leq i<j \leq N}
}
(x_i-x_j)(y_i-y_j)
}
\nonumber
\end{align}

\noindent where we have used the Jacobi-Trudi identity for Schur functions\footnote{See section 3 of chapter I in \cite{mac}.}

\begin{align}
s_{\mu}\{x\} 
=
\det\Big( 
h_{\mu_i-i+j}\{x\}
\Big)_{1\leq i, j \leq N}
=
\frac{
\det\Big(
x_i^{\mu_j-j+N}
\Big)_{1\leq i,j \leq N}
}
{\displaystyle{\prod_{1 \leq i< j \leq N}}(x_i-x_j)}
\label{jac-trud}
\end{align}  

\noindent in conjunction with the definition $m_i=\mu_i-i$ for all $1 \leq i \leq N$. Using the Cauchy-Binet identity\footnote{See subsection 5.2.5 of the thesis.} to convert the sum in the numerator of (\ref{phase-JT}) into a single determinant, we obtain

\begin{align}
&
\langle 0| \mathbb{C}(x_N)\ldots \mathbb{C}(x_1)
\mathbb{B}(y_1)\ldots \mathbb{B}(y_N)|0\rangle
\label{phase-JT2}
=
\frac{\displaystyle{
\det\left(
\sum_{m=0}^{M+N-1}
(x_i y_j)^m
\right)_{1\leq i,j \leq N}
}}{\displaystyle{
\prod_{1\leq i<j \leq N}
(x_i-x_j)(y_i-y_j)
}}
\\
&
=
\frac{\displaystyle{
\det\Big(
\left(1-(x_i y_j)^{M+N}\right) \Big/ \left(1-x_i y_j\right)
\Big)_{1\leq i,j \leq N}
}}
{\displaystyle{
\prod_{1\leq i<j \leq N}
(x_i-x_j)(y_i-y_j)
}}
\nonumber
\end{align}

\noindent Equating the right hand sides of (\ref{phase-gf}) and (\ref{phase-JT2}), we have proved that

\begin{align}
\sum_{\pi \subseteq [N,N,M]}
A_{\pi}\Big(\{x\},\{y\}\Big)
=
\frac{\displaystyle{
\det\Big(
\left(1-(x_i y_j)^{M+N}\right) \Big/ \left(1-x_i y_j\right)
\Big)_{1\leq i,j \leq N}
}}
{\displaystyle{
\prod_{1\leq i<j \leq N}
(x_i-x_j)(y_i-y_j)
}}
\label{bogs}
\end{align}

\noindent which matches the evaluation of this generating function in \cite{bog1}.

\subsection{Scalar product as a power-sum specialized KP $\tau$-function}

We now demonstrate that the phase model scalar product is a specialization of a KP $\tau$-function. The specialization is achieved by setting the KP time variables to power sums in the phase model rapidities. Our starting point is the equation

\begin{align}
\langle 0|
\exp\left(\sum_{m=1}^{\infty} t_m H_m\right)
=
\sum_{\mu} \chi_{\mu}\{t\} (\mu|
\label{allstates}
\end{align}

\noindent whose sum is over all partitions $\mu$, which follows from lemma 4 in chapter 1 and the orthonormality (\ref{part-orth}) of partitions. Defining $t_m = \frac{1}{m} \sum_{n=1}^{N} x_n^m$ for all $m \geq 1$, equation (\ref{allstates}) becomes

\begin{align}
\langle 0| 
\exp\left(
\sum_{m=1}^{\infty}
\sum_{n=1}^{N} \frac{1}{m} x_n^m H_m
\right)
=
\sum_{\mu \subseteq [N,\infty]}
s_{\mu}\{x\}
(\mu|
\label{allstates2} 
\end{align}

\noindent where the sum is over all partitions $\mu$ with maximal length $N$. Equating the right hand sides of (\ref{p-lem4-0}) and (\ref{phase-fermi}) and using the identity (\ref{allstates2}), we find

\begin{align}
\langle 0| 
\exp\left(
\sum_{m=1}^{\infty} \sum_{n=1}^{N}\frac{1}{m} x_n^m H_m
\right)
\exp X\{y\} 
|0\rangle
=
\sum_{\mu \subseteq [N,M]}
s_{\mu}\{x\} s_{\mu}\{y\}
\label{yay1}
\end{align}

\noindent where $X\{y\} \in A_{\infty}$ is defined as

\begin{align}
X\{y\}
=
\sum_{\substack{1 \leq m \leq M \\ 1 \leq n \leq N}}
(-)^{n-1}
s_{\{m,1^{n-1}\}}\{y\}
\psi_{m-1} \psis_{-n}
\end{align}

%

\noindent Now consider the polynomial KP $\tau$-function $\tau\{t\} = \langle e^{H\{t\}} e^{X\{y\}} \rangle$. Comparing the first line of (\ref{phase-JT}) with equation (\ref{yay1}), we conclude that 

\begin{align}
\tau\{t\}
&=
\langle 0| 
\exp\left(
\sum_{m=1}^{\infty} t_m H_m
\right)
\exp X\{y\} 
|0\rangle
=
\langle 0|
\mathbb{C}(x_N)
\ldots
\mathbb{C}(x_1)
\mathbb{B}(y_1)
\ldots
\mathbb{B}(y_N)
|0\rangle
\end{align}

\noindent under the power-sum specialization $t_m = \frac{1}{m}\sum_{n=1}^{N} x_n^m$ for all $m \geq 1$. This connection between plane partition generating functions and KP $\tau$-functions was suggested in \cite{fwz3}, albeit in the context of plane partitions whose column heights are unrestricted. The result of this subsection is at the level of $M$-boxed plane partitions, and it specializes to the result of \cite{fwz3} in the limit $M\rightarrow\infty$. 

\section{Phase model on an infinite lattice}
\label{phase-pp}

In this section we study the action of the monodromy matrix operators $\mathbb{B}(x)$ and $\mathbb{C}(x)$ when the number of lattice sites becomes infinite. Our main result is lemma 6, showing that in the limit $M\rightarrow\infty$ the operators $\mathbb{B}(x)$ and $\mathbb{C}(x)$ acquire equivalent actions to the half-vertex operators $\Gamma_{-}(x)$ and $\Gamma_{+}(x)$ from KP theory. This result rests basically on the works \cite{fwz3},\cite{mntt}, \cite{or},\cite{orv} which studied the actions of these half-vertex operators. 

\subsection{Calculation of 
$\mathcal{M}_{\psi}\mathbb{B}(x)|n\rangle$ and 
$\langle n|\mathbb{C}(x)\mathcal{M}_{\psi}^{*}$ as $M \rightarrow \infty$}


\begin{lemma}
\label{3-lem6}
{\rm 
Consider the infinite lattice limit of the phase model, which is obtained by taking $M \rightarrow \infty$. Let $|n\rangle = \otimes_{j=0}^{\infty} |n_j\rangle_j$ and $\langle n| = \otimes_{j=0}^{\infty} \langle n_j|_j$ be basis vectors of $\mathcal{V}$ and $\mathcal{V}^{*}$, respectively, in this limit.\footnote{When considering such basis vectors, we always assume there exists some integer $I$ such that $n_i=0$ for all $i>I$. This is necessary to ensure that the vectors contain a finite amount of non-trivial information.} In addition, let $|\nu)$ and $(\nu|$ be the image states of these basis vectors under the mappings (\ref{3-2-map}). We claim that   

\begin{align}
\mathcal{M}_{\psi}
\Big[
\lim_{M\rightarrow\infty}
\mathbb{B}(x)
|n\rangle
\Big]
=
\Gamma_{-}(x)
|\nu),
\quad
\Big[
\lim_{M\rightarrow\infty}
\langle n|
\mathbb{C}(x)
\Big]
\mathcal{M}_{\psi}^{*}
=
(\nu|
\Gamma_{+}(x)
\label{p-lem6-1}
\end{align}

\noindent where we have defined the KP half-vertex operators\footnote{We use the terminology {\it half-vertex operator} in reference to the fact that $\Gamma_{-}(x),\Gamma_{+}(x)$ each constitute one half of a charged fermion vertex operator, \cite{jm1}.}

\begin{align}
\Gamma_{-}(x)
=
\exp\left(\sum_{n=1}^{\infty} \frac{x^n}{n} H_{-n} \right),
\quad
\Gamma_{+}(x)
=
\exp\left(\sum_{n=1}^{\infty} \frac{x^{n}}{n} H_n \right)
\label{hvert-def}
\end{align}

\noindent and $H_{-n},H_{n}$ denote the Heisenberg generators (\ref{aheisen1}).
}
\end{lemma} 

\begin{proof}

We split the proof into two steps. In the first step, we show that (\ref{p-lem6-1}) is equivalent to the statement (\ref{p-lem6-3}). In the second step we prove (\ref{p-lem6-3}) using the calculus of charged free fermions.

\medskip
\noindent
{\bf Step 1.}\ Taking the $M \rightarrow \infty$ limit of equations (\ref{acespa}) and (\ref{kinspa}), we obtain

\begin{align}
\mathcal{M}_{\psi}
\Big[
\lim_{M\rightarrow\infty}
\mathbb{B}(x)
|n\rangle
\Big]
&=
\sum_{\mu \succ \nu}
x^{|\mu|-|\nu|}
|\mu)
\\
\Big[
\lim_{M\rightarrow\infty}
\langle n|
\mathbb{C}(x)
\Big]
\mathcal{M}_{\psi}^{*}
&=
\sum_{\mu \succ \nu}
x^{|\mu|-|\nu|}
(\mu|
\end{align}

\noindent where the sums are over all partitions $\mu$ which interlace with $\nu$, whose parts now have no size restriction. The equations (\ref{p-lem6-1}) are therefore equivalent to the statements

\begin{align}
\Gamma_{-}(x)|\nu)
=
\sum_{\mu \succ \nu}
x^{|\mu|-|\nu|}
|\mu),
\quad
(\nu| \Gamma_{+}(x)
=
\sum_{\mu \succ \nu}
x^{|\mu|-|\nu|}
(\mu|
\label{p-lem6-2}
\end{align}

\noindent which are entirely at the level of charged free fermions. Due to the orthonormality (\ref{part-orth}) of partition states, equations (\ref{p-lem6-2}) may be presented in the alternative form

\begin{align}
(\mu| \Gamma_{-}(x)
=
\sum_{\nu \prec \mu}
x^{|\mu|-|\nu|}
(\nu|,
\quad
\Gamma_{+}(x) |\mu)
=
\sum_{\nu \prec \mu}
x^{|\mu|-|\nu|}
|\nu)
\label{p-lem6-3}
\end{align}

\noindent where the sums are now over partitions $\nu$ such that $\nu \prec \mu$. We will find it convenient to prove (\ref{p-lem6-3}), as opposed to (\ref{p-lem6-2}). The point is that the sums in the former are finite, whereas the sums in the latter are infinite and inherently more difficult to handle. When we succeed in showing (\ref{p-lem6-3}), we will have achieved the proof of (\ref{p-lem6-1}).

\medskip
\noindent
{\bf Step 2.}\ Consider the length $l$ partitions

\begin{align}
( \mu| = 
\langle -l|\psis_{m_l} \ldots \psis_{m_1},
\quad
|\mu) = 
\psi_{m_1} \ldots \psi_{m_l} |-l\rangle
\label{p-lem6-5}
\end{align}

\noindent where $\{m_1 > \cdots > m_l > -l\}$, and the elements of the partitions are given by $\mu_i = m_i +i$ for all $1\leq i \leq l$. In order to prove (\ref{p-lem6-3}), we must calculate $(\mu|\Gamma_{-}(x)$ and $\Gamma_{+}(x)|\mu)$. To progress in this direction, we require the commutation relations

\begin{align}
\Gamma_{-}(x)
\left(\sum_{n=0}^{\infty} \psis_{(i-n)} x^n\right)
=
\psis_i \Gamma_{-}(x),
\quad
\Gamma_{+}(x) \psi_i
=
\left(\sum_{n=0}^{\infty} \psi_{(i-n)} x^{n}\right)
\Gamma_{+}(x)
\label{p-lem6-4}
\end{align}

\noindent which are derived following the arguments presented in subsection 1.1.8.\footnote{Setting $t_n = x^{n}/n$ for all $n \geq 1$ in (\ref{KPexpI12}), we obtain
$
\Gamma_{+}(x) \Psi(k)
=
\frac{1}{1-xk}
\Psi(k)
\Gamma_{+}(x)
$. Extracting the coefficients of $k^i$ from this equation, we prove the second commutation relation in (\ref{p-lem6-4}). The first commutation relation may be proved similarly.} Applying the relations (\ref{p-lem6-4}) repeatedly to the partitions (\ref{p-lem6-5}), we find 

\begin{align}
( \mu| \Gamma_{-}(x)
&=
\langle -l|
\left(
\sum_{i_{l} = 0}^{\infty} \psis_{(m_{l}-i_{l})} x^{i_{l}}
\right)
\ldots
\left(
\sum_{i_1 = 0}^{\infty} \psis_{(m_1-i_1)} x^{i_1}
\right)
\label{p-lem6-6}
\\
\Gamma_{+}(x) |\mu)
&=
\left(
\sum_{i_1 = 0}^{\infty} \psi_{(m_1-i_1)} x^{i_1}
\right)
\ldots
\left(
\sum_{i_{l} = 0}^{\infty} \psi_{(m_l-i_l)} x^{i_l}
\right)
|-l\rangle
\label{p-lem6-7}
\end{align}

\noindent where we have used the fact that $\langle -l|\Gamma_{-}(x) = \langle -l|$ and $\Gamma_{+}(x) |-l\rangle = |-l\rangle$. Now for arbitrary integers $m > n$ we have the identities

\begin{align}
\left(
\sum_{i=0}^{\infty}
\psis_{(n-i)} x^{i}
\right)
\left(
\sum_{j=0}^{\infty}
\psis_{(m-j)} x^{j}
\right)
&=
\left(
\sum_{i=0}^{\infty}
\psis_{(n-i)} x^{i}
\right)
\left(
\sum_{j=0}^{m-n-1}
\psis_{(m-j)} x^{j}
\right)
\\
\left(
\sum_{i=0}^{\infty}
\psi_{(m-i)} x^{i}
\right)
\left(
\sum_{j=0}^{\infty}
\psi_{(n-j)} x^{j}
\right)
&=
\left(
\sum_{i=0}^{m-n-1}
\psi_{(m-i)} x^{i}
\right)
\left(
\sum_{j=0}^{\infty}
\psi_{(n-j)} x^{j}
\right)
\end{align}

\noindent which, when substituted into (\ref{p-lem6-6}) and (\ref{p-lem6-7}), lead to the truncated equations

\begin{align}
( \mu| \Gamma_{-}(x)
&=
\langle -l|
\left(
\sum_{i_l=0}^{m_l-m_{\bar{l}}-1}
\psis_{(m_l-i_l)} x^{i_l}
\right)
\ldots
\left(
\sum_{i_1=0}^{m_1-m_2-1}
\psis_{(m_1-i_1)} x^{i_1}
\right)
\label{p-lem6-8}
\\
\Gamma_{+}(x) |\mu)
&=
\left(
\sum_{i_1=0}^{m_1-m_2-1}
\psi_{(m_1-i_1)} x^{i_1}
\right)
\ldots
\left(
\sum_{i_l=0}^{m_l-m_{\bar{l}}-1}
\psi_{(m_l-i_l)} x^{i_l}
\right)
|-l\rangle
\label{p-lem6-9}
\end{align}

\noindent where we have defined $\bar{l} = l+1$ and $m_{\bar{l}} = -\bar{l}$. The indices in the sums (\ref{p-lem6-8}) and (\ref{p-lem6-9}) can then be modified to produce the equations

{\small
\begin{align}
( \mu |\Gamma_{-}(x)
&=
\langle -l|
\left(
\sum_{m_l \geq n_l > m_{\bar{l}}}
\psis_{n_l} x^{m_l-n_l}
\right)
\ldots
\left(
\sum_{m_1 \geq n_1 > m_2}
\psis_{n_1} x^{m_1-n_1}
\right)
\label{p-lem6-10}
=
\sum_{\nu \prec \mu}
x^{|\mu|-|\nu|}
( \nu|
\\
\Gamma_{+}(x) |\mu)
&=
\left(
\sum_{m_1 \geq n_1 > m_2}
\psi_{n_1} x^{m_1-n_1}
\right)
\ldots
\left(
\sum_{m_l \geq n_l > m_{\bar{l}}}
\psi_{n_l} x^{m_l-n_l}
\right)
|-l\rangle
\label{p-lem6-11}
= 
\sum_{\nu \prec \mu}
x^{|\mu|-|\nu|}
|\nu)
\end{align}
}

\noindent where we have defined partitions $(\nu| = \langle -l|\psis_{n_l}\ldots \psis_{n_1}$ and $|\nu) = \psi_{n_1}\ldots \psi_{n_l}|-l\rangle$, which correspond to the ordered set $\{n_1 > \cdots > n_l \geq -l\}$ via $\nu_i = n_i+i$ for all $1\leq i \leq l$. Notice that the final equalities of (\ref{p-lem6-10}) and (\ref{p-lem6-11}) follow from the relationship

\begin{align}
m_i \geq n_i > m_{i+1} 
\ \ {\rm for\ all}\ 1 \leq i \leq l
\implies
\mu_i \geq \nu_i \geq \mu_{i+1} 
\ \ {\rm for\ all}\ 1 \leq i \leq l
\end{align}

\noindent as well as the fact $|\mu|-|\nu| = \sum_{i=1}^{l}(m_i-n_i)$. The equations (\ref{p-lem6-10}) and (\ref{p-lem6-11}) complete the proof of (\ref{p-lem6-3}).

\end{proof}

\subsection{Generating plane partitions of arbitrary size}

In the previous section we demonstrated that the phase model scalar product on a lattice of size $M+1$ generates $M$-boxed plane partitions. Accordingly, we expect that in the limit $M \rightarrow \infty$ the scalar product will generate plane partitions whose column heights are arbitrarily large, giving rise to the equation

\begin{align}
\lim_{M \rightarrow \infty}
\langle 0|
\mathbb{C}(x_N)
\ldots
\mathbb{C}(x_1)
\mathbb{B}(y_1)
\ldots
\mathbb{B}(y_N)
|0\rangle
=
\sum_{\pi \subseteq [N,N,\infty]}
A_{\pi}\Big( \{x\},\{y\} \Big)
\label{before0}
\end{align}

\noindent where the sum is over all plane partitions $\pi$ which fit inside the box of dimension $N\times N \times \infty$, with $A_{\pi}(\{x\},\{y\})$ given by (\ref{a-fact}). On the other hand, using the result of lemma 6 we are able to write

{\small
\begin{align}
\lim_{M\rightarrow\infty}
\langle 0|
\mathbb{C}(x_N)
\ldots
\mathbb{C}(x_1)
\mathbb{B}(y_1)
\ldots
\mathbb{B}(y_N)
|0\rangle
=
\label{before}
(\emptyset|
\Gamma_{+}(x_N)
\ldots
\Gamma_{+}(x_1)
\Gamma_{-}(y_1)
\ldots
\Gamma_{-}(y_N)
|\emptyset)
\end{align}
}

\noindent which lends itself to immediate evaluation, owing to simple commutation relations between the KP half-vertex operators. Explicitly speaking, using the definition (\ref{hvert-def}) of $\Gamma_{-}(y)$ and $\Gamma_{+}(x)$ we find that

\begin{align}
&
\sum_{m=1}^{\infty} \sum_{n=1}^{\infty}
\frac{x^m y^n}{mn}
[H_m,H_{-n}]
=
\sum_{m=1}^{\infty} \sum_{n=1}^{\infty}
\frac{x^m y^n}{mn}
m\delta_{m,n}
=
\sum_{m=1}^{\infty}
\frac{(xy)^m}{m}
\label{hvert-com}
\\
\implies
&
\Gamma_{+}(x) \Gamma_{-}(y)
=
\exp\left(
\sum_{m=1}^{\infty}\frac{(xy)^m}{m}
\right)
\Gamma_{-}(y) \Gamma_{+}(x)
=
\frac{1}{1-xy}
\Gamma_{-}(y) \Gamma_{+}(x)
\nonumber
\end{align}  

\noindent Employing the commutation relation (\ref{hvert-com}) repeatedly in (\ref{before}) and using the fact that $\Gamma_{+}(x)|\emptyset) = |\emptyset)$, $(\emptyset| \Gamma_{-}(y) = (\emptyset|$, we obtain

\begin{align}
\lim_{M\rightarrow\infty}
\langle 0|
\mathbb{C}(x_N)
\ldots
\mathbb{C}(x_1)
\mathbb{B}(y_1)
\ldots
\mathbb{B}(y_N)
|0\rangle
=
\prod_{i,j=1}^{N}
\frac{1}{1-x_i y_j}
\label{after}
\end{align}

\noindent Comparing the equations (\ref{before0}) and (\ref{after}), we have proved that 

\begin{align}
\sum_{\pi \subseteq [N,N,\infty]}
A_{\pi}\Big( \{x\},\{y\} \Big)
=
\prod_{i,j=1}^{N}
\frac{1}{1-x_i y_j}
\end{align}

\noindent which is a much simpler evaluation of this generating function than in the finite case (\ref{bogs}). Let us remark that this calculation could also have been performed using the first line of (\ref{phase-JT}) and the identity

\begin{align}
\sum_{\mu \subseteq [N,\infty]}
s_{\mu}\{x\} s_{\mu}\{y\}
=
\prod_{i,j =1}^{N} \frac{1}{1-x_i y_j}
\end{align}

\noindent from section 4, chapter I of \cite{mac}. The proof which we have given is independent of the properties of symmetric functions. As a final observation, let us specialize the variables $\{x_1,\ldots,x_N\}$ and $\{y_1,\ldots,y_N\}$ to 

\begin{align}
x_i = y_i = z^{i-\frac{1}{2}} 
\ \ {\rm for\ all}\ 1\leq i \leq N
\end{align}

\noindent giving rise to the equation

\begin{align}
\sum_{\pi \subseteq [N,N,\infty]}
z^{|\pi|}
=
\prod_{i,j=1}^{N}
\frac{1}{1-z^{i+j-1}}
\end{align}

\noindent where $|\pi|$ is the {\it weight} of the plane partition $\pi$, defined equal to the sum of all its entries $\pi(i,j)$. Taking the limit $N\rightarrow\infty$ we obtain

\begin{align}
\sum_{\pi}
z^{|\pi|}
=
\prod_{i=1}^{\infty}
\frac{1}{(1-z^i)^i}
\label{macmahon}
\end{align}

\noindent where the sum is now over plane partitions $\pi$ of completely arbitrary dimension. Equation (\ref{macmahon}) is a famous generating function for plane partitions, originally found by MacMahon \cite{macm}. The idea of deriving this generating function using charged fermions is due to \cite{or},\cite{orv} and was explained in detail in \cite{fwz3}.

\section{$i$-boson model}
\label{ib-intro}

\subsection{Space of states $\tilde{\mathcal{V}}$ and inner product $\tilde{\mathcal{I}}$}

It is necessary to introduce a space of states $\tilde{\mathcal{V}}$ which is a subspace of $\mathcal{V}$, defined in section 3.1. In brief, a lattice configuration in $\tilde{\mathcal{V}}$ can have an unlimited number of particles occupying the $0^{\rm th}$ site, but the remaining $M$ sites are occupied by at most one particle. We represent this mathematically by writing

\begin{align} 
{\rm Basis}(\tilde{\mathcal{V}})
=
\Big\{ |\tilde{n}\rangle =
|n_0\rangle_0 \otimes |n_1\rangle_1 \otimes \cdots \otimes |n_M\rangle_M
\ 
\Big|
\ 
0 \leq n_1,\ldots,n_M \leq 1
\Big\}
\end{align}

\noindent where $n_0$ ranges over all non-negative integers, while the remaining occupation numbers are constrained by $0 \leq n_1,\ldots, n_M \leq 1$. The inner product between two basis vectors $|\tilde{m}\rangle = |m_0\rangle_0 \otimes \cdots \otimes |m_M\rangle_M$ and $|\tilde{n}\rangle = |n_0\rangle_0 \otimes \cdots \otimes |n_M\rangle_M$ is defined to be

\begin{align}
\tilde{\mathcal{I}}\Big(
|\tilde{m}\rangle, |\tilde{n}\rangle
\Big)
=
2^{m_0}
\theta\Big(0 \leq m_0 \leq 1\Big)
\delta_{m_0,n_0}
\prod_{j=1}^{M} 
2^{-m_j}
\delta_{m_j,n_j}
\label{Itilde}
\end{align}

\noindent where $\theta(x)$ is the Boolean function, with $\theta(x) =1$ if $x$ is true and $\theta(x) = 0$ if $x$ is false. This corresponds to setting

\begin{align}
c_0(m_0) 
=
2^{m_0}
\theta\Big(0 \leq m_0 \leq 1\Big),
\quad
c_j(m_j)
=
2^{-m_j}
\ \ {\rm for\ all}\ 1\leq j \leq M
\end{align}

\noindent in equation (\ref{general-inn-prod}). The dual space of states $\tilde{\mathcal{V}}^{*}$ has the basis

\begin{align}
{\rm Basis}(\tilde{\mathcal{V}}^{*})
=
\Big\{ \langle \tilde{m}| =
\langle m_0|_0 \otimes \langle m_1|_1 \otimes \cdots \otimes \langle m_M|_M
\ 
\Big|
\ 
0 \leq m_1,\ldots,m_M \leq 1
\Big\}
\end{align}

\noindent where $m_0$ again ranges over all non-negative integers, while the remaining occupation numbers are constrained by $0 \leq m_1,\ldots, m_M \leq 1$. The dual space of states $\tilde{\mathcal{V}}^{*}$ is prescribed the action 

\begin{align}
\langle \tilde{m}| \tilde{n}\rangle
=
\tilde{\mathcal{I}}\Big(
|\tilde{m}\rangle, |\tilde{n}\rangle
\Big)
\label{ib-isom0}
\end{align}

\noindent for all $\langle \tilde{m}| = \langle m_0|_0 \otimes \cdots \otimes \langle m_M|_M \in \tilde{\mathcal{V}}^{*}$ and $|\tilde{n}\rangle = |n_0\rangle_0 \otimes \cdots \otimes |n_M\rangle_M \in \tilde{\mathcal{V}}$.

\subsection{$i$-boson algebra}

Let us define the {\it $i$-boson algebra.} It is generated by $\{\varphi,\varphid,\mathcal{N}\}$ which satisfy the commutation relations

\begin{align}
[\varphi,\varphid]= (-)^{\mathcal{N}},
\quad
[\mathcal{N},\varphi]=-\varphi,
\quad
[\mathcal{N},\varphid]=\varphid
\label{ib-alg}
\end{align}

\noindent This algebra is the $q=i=\sqrt{-1}$ case of the $q$-boson algebra (\ref{realqb}), discussed in the next chapter. As we did for the phase model, we consider $M+1$ copies of the $i$-boson algebra, generated by $\{\varphi_0,\varphid_0,\mathcal{N}_0\}$ through to $\{\varphi_M,\varphid_M,\mathcal{N}_M\}$. Recalling the conventions of chapter 2, we denote these algebras by $\tilde{\mathcal{A}}_0,\ldots,\tilde{\mathcal{A}}_M$ with generators $\mathfrak{a}_i^{+} = \varphid_i,\mathfrak{a}_i^{-} = \varphi_i,\mathfrak{a}_i^{0} = \mathcal{N}_i$. Different copies of the $i$-boson algebra are assumed to commute, giving rise to the equations

\begin{align}
[\varphi_i,\varphid_j]= \delta_{i,j}(-)^{\mathcal{N}_i},
\quad
[\mathcal{N}_i,\varphi_j]=-\delta_{i,j} \varphi_i,
\quad
[\mathcal{N}_i,\varphid_j]= \delta_{i,j} \varphid_i
\label{ib-alg2}
\end{align}

\noindent for all $0 \leq i,j \leq M$.

\subsection{Representations of $i$-boson algebras}
\label{ibrep}

In direct analogy with the first part of the chapter, we fix representations of the $i$-boson algebras (\ref{ib-alg2}) on the vector space $\mathcal{V}$ (\ref{vec1}). These representations have much in common with those of the phase algebras (\ref{phase-alg2}). The operator $\varphi_j$ acts as an annihilator, removing particles from the $j^{\rm th}$ lattice site, and $\varphid_j$ acts as a creation operator, adding particles to the $j^{\rm th}$ lattice site. What is different is that these operators also produce certain factors, which are necessary to correctly represent the $i$-boson algebras. These factors also conspire to ensure that $\tilde{\mathcal{V}} \subset \mathcal{V}$ is closed under the action of $\tilde{\mathcal{A}}_0,\ldots,\tilde{\mathcal{A}}_M$, as we demonstrate below.

We begin by constructing representations for the algebras $\tilde{\mathcal{A}}_1,\ldots,\tilde{\mathcal{A}}_M$. For all $1\leq j \leq M$, the operator $\varphi_j$ has the action  

\begin{align}
\varphi_j |n_0\rangle_0 \otimes \cdots \otimes |n_M\rangle_M
=
\left\{
\begin{array}{ll}
0, & n_j = 0
\\
\\
\frac{1}{\sqrt{2}}
|n_0\rangle_0 
\otimes \cdots \otimes 
|n_j-1\rangle_j 
\otimes \cdots \otimes 
|n_M\rangle_M,
& n_j \geq 1
\end{array}
\right.
\label{vp1}
\end{align}

\noindent while for all $1 \leq j \leq M$ the operator $\varphid_j$ has the action

\begin{align}
\varphid_j |n_0\rangle_0 \otimes \cdots \otimes |n_M\rangle_M
=
\frac{1+(-)^{n_j}}{\sqrt{2}}
|n_0\rangle_0 
\otimes \cdots \otimes 
|n_j+1\rangle_j 
\otimes \ldots \otimes 
|n_M\rangle_M
\label{vpd1}
\end{align}

\noindent We construct a slightly different representation for the algebra $\tilde{\mathcal{A}}_0$. The operator $\varphi_0$ has the action

\begin{align}
\varphi_0 
|n_0\rangle_0 \otimes \cdots \otimes |n_M\rangle_M
=
\frac{1-(-)^{n_0}}{\sqrt{2}}
|n_0-1\rangle_0 
\otimes 
|n_1\rangle_1 
\otimes \cdots \otimes
|n_M\rangle_M
\label{vp0}
\end{align}

\noindent while the operator $\varphid_0$ has the action

\begin{align}
\varphid_0 
|n_0\rangle_0 \otimes \cdots \otimes |n_M\rangle_M
=
\frac{1}{\sqrt{2}}
|n_0+1\rangle_0 
\otimes 
|n_1\rangle_1 
\otimes \cdots \otimes
|n_M\rangle_M
\label{vpd0}
\end{align} 

%

\noindent As before, for all $0\leq j \leq M$ the action of $\mathcal{N}_j$ is given by 

\begin{align}
\mathcal{N}_j
|n_0\rangle_0 \otimes \cdots \otimes |n_M\rangle_M
=
n_j
|n_0\rangle_0 \otimes \cdots \otimes |n_M\rangle_M
\label{ib-N1}
\end{align}

\noindent It is straightforward to check that (\ref{vp1})--(\ref{ib-N1}) faithfully represent the $i$-boson algebras $\tilde{\mathcal{A}}_0,\ldots,\tilde{\mathcal{A}}_M$. Assuming all operators act linearly, the equations (\ref{vp1})--(\ref{ib-N1}) completely determine the action of $\{\varphi_j,\varphid_j,\mathcal{N}_j\}$ on $\mathcal{V}$. Studying (\ref{vpd1}), we see that for all $1\leq j \leq M$ the operator $\varphid_j$ annihilates any state with $n_j=1$. It follows that $\tilde{\mathcal{V}}$ is closed under the action of the $i$-boson algebras. In addition, from the definition of the inner product (\ref{Itilde}) we see that

\begin{align}
\tilde{\mathcal{I}}\Big(
\varphi_j |\tilde{m}\rangle,
|\tilde{n}\rangle
\Big)
=
\tilde{\mathcal{I}}\Big(
|\tilde{m}\rangle,
\varphid_j |\tilde{n}\rangle
\Big)
\end{align}

\noindent for all $|\tilde{m}\rangle,|\tilde{n}\rangle \in \tilde{\mathcal{V}}$. Hence $\varphi_j,\varphid_j$ are adjoint operators on $\tilde{\mathcal{V}}$ for all $0 \leq j \leq M$, while $\mathcal{N}_j$ continues to be self-adjoint.

We again follow subsection \ref{2-dual} to deduce appropriate actions for $\tilde{\mathcal{A}}_0,\ldots,\tilde{\mathcal{A}}_M$ on the dual space $\mathcal{V}^{*}$. We firstly consider the algebras $\tilde{\mathcal{A}}_1,\ldots,\tilde{\mathcal{A}}_M$. For all $1\leq j \leq M$, the operator $\varphid_j$ has the action

\begin{align}
\langle n_0|_0 \otimes \cdots \otimes \langle n_M|_M \varphid_j
=
\left\{
\begin{array}{ll}
0, & n_j = 0
\\
\\
\frac{1}{\sqrt{2}}
\langle n_0|_0 
\otimes \cdots \otimes 
\langle n_j-1|_j 
\otimes \cdots \otimes 
\langle n_M|_M,
& n_j \geq 1
\end{array}
\right.
\label{vpd2}
\end{align}

\noindent while for all $1 \leq j \leq M$ the operator $\varphi_j$ has the action

\begin{align}
\langle n_0|_0 \otimes \cdots \otimes \langle n_M|_M \varphi_j
=
\frac{1+(-)^{n_j}}{\sqrt{2}}
\langle n_0|_0 
\otimes \cdots \otimes 
\langle n_j+1|_j 
\otimes \cdots \otimes 
\langle n_M|_M
\label{vp2}
\end{align}

\noindent As before, $\tilde{\mathcal{A}}_0$ is assigned its own representation. The operator $\varphid_0$ has the action

\begin{align}
\langle n_0|_0 \otimes \cdots \otimes \langle n_M|_M
\varphid_0
=
\frac{1-(-)^{n_0}}{\sqrt{2}}
\langle n_0-1|_0 \otimes \langle n_1|_1 \otimes \cdots \otimes \langle n_M|_M
\label{vpd00}
\end{align}

\noindent while the operator $\varphi_0$ has the action

\begin{align}
\langle n_0|_0 \otimes \cdots \otimes \langle n_M|_M
\varphi_0
=
\frac{1}{\sqrt{2}}
\langle n_0+1|_0 \otimes \langle n_1|_1 \otimes \cdots \otimes \langle n_M|_M
\label{vp00}
\end{align}

\noindent Finally, for all $0 \leq j \leq M$ the action of $\mathcal{N}_j$ is given by

\begin{align}
\langle n_0|_0 \otimes \cdots \otimes \langle n_M|_M
\mathcal{N}_j
=
n_j
\langle n_0|_0 \otimes \cdots \otimes \langle n_M|_M
\label{ib-N2}
\end{align}

\noindent The set of definitions (\ref{vpd2})--(\ref{ib-N2}) provide the dual representation of the algebras $\tilde{\mathcal{A}}_0,\ldots,\tilde{\mathcal{A}}_M$. Assuming that all operators act linearly, the equations (\ref{vpd2})--(\ref{ib-N2}) completely determine the action of $\{\varphi_j,\varphid_j,\mathcal{N}_j\}$ on the dual vector space $\mathcal{V}^{*}$.

\subsection{Hamiltonian $\tilde{\mathcal{H}}$}

The Hamiltonian of the $i$-boson model is given by

\begin{align}
\tilde{\mathcal{H}}
=
-\frac{1}{2}
\sum_{j=0}^{M}
\left(
\varphid_j \varphi_{j+1} + \varphi_j \varphid_{j+1}
\right) 
+
\bar{\mathcal{N}}
\label{ib-ham}
\end{align}

\noindent where the periodicity $\varphi_{M+1} = \varphi_0$ and $\varphid_{M+1} = \varphid_0$ is imposed. This Hamiltonian is deduced simply by taking the $q\rightarrow i$ limit of the $q$-boson Hamiltonian (\ref{qb-ham}). We now describe the quantum inverse scattering/algebraic Bethe Ansatz scheme for finding eigenvectors $|\Psi\rangle \in \tilde{\mathcal{V}}$ of (\ref{ib-ham}).
 

\subsection{$L$-matrix and local intertwining equation}

The $R$-matrix for the $i$-boson model depends on two indeterminates $x,y$ and acts in the tensor product $\mathcal{V}_a \otimes \mathcal{V}_b$, where $\mathcal{V}_a,\mathcal{V}_b$ are copies of $\mathbb{C}^{2}$. It is given by

\begin{align}
\tilde{R}_{ab}(x,y)
=
\left(
\begin{array}{cccc}
x+y &               0                 &              0                   & 0
\\
0 &              y-x              &                2x^{\frac{1}{2}} y^{\frac{1}{2}}             & 0
\\
0 &               2x^{\frac{1}{2}} y^{\frac{1}{2}}                 &  x-y  & 0
\\
0 &               0                 &              0                   & x+y
\end{array}
\right)_{ab}
\label{ib-R}
\end{align}

\noindent and corresponds to the $a_{\pm}(x,y) = x+y,\ b_{\pm}(x,y) = \pm(y-x),\ c_{\pm}(x,y) = 2x^{\frac{1}{2}} y^{\frac{1}{2}}$ case of (\ref{general-Rmat}). The $L$-matrix for the $i$-boson model depends on a single indeterminate $x$, and acts in the space $\mathcal{V}_a$. Its entries are operators acting at the $m^{\rm th}$ lattice site, and identically everywhere else. It has the form

\begin{align}
\tilde{L}_{am}(x)
=
\left(
\begin{array}{cc}
x^{-\frac{1}{2}} & 2^{\frac{1}{2}} \varphid_m
\\
2^{\frac{1}{2}} \varphi_m              & x^{\frac{1}{2}}
\end{array}
\right)_{a}
\label{ib-L}
\end{align}

\noindent Using these definitions, the local intertwining equation is given by 

\begin{align}
\tilde{R}_{ab}(x,y) \tilde{L}_{am}(x) \tilde{L}_{bm}(y)
=
\tilde{L}_{bm}(y) \tilde{L}_{am}(x) \tilde{R}_{ab}(x,y)
\label{ib-intL}
\end{align}

\noindent This is a $4\times 4$ matrix equation, which gives rise to sixteen scalar identities. Each of these identities may be verified by direct calculation, and by using the commutation relations (\ref{ib-alg2}) where appropriate.

\subsection{Monodromy matrix and global intertwining equation}

The monodromy matrix is an $(M+1)$-fold product of the $L$-matrices (\ref{ib-L}), taken in the auxiliary space ${\rm End}(\mathcal{V}_a)$. It has the form

\begin{align}
\tilde{T}_a(x)
=
\tilde{L}_{aM}(x)
\ldots
\tilde{L}_{a0}(x)
=
\left(
\begin{array}{cc}
\tilde{A}(x) & \tilde{B}(x)
\\
\tilde{C}(x) & \tilde{D}(x)
\end{array}
\right)_a
\end{align}

\noindent where $\tilde{A}(x),\tilde{B}(x),\tilde{C}(x),\tilde{D}(x)$ are elements of $\tilde{\mathcal{A}}_0 \otimes \cdots \otimes \tilde{\mathcal{A}}_M$. The monodromy matrix satisfies the global intertwining equation

\begin{align}
\tilde{R}_{ab}(x,y)
\tilde{T}_a(x)
\tilde{T}_b(y)
=
\tilde{T}_b(y)
\tilde{T}_a(x)
\tilde{R}_{ab}(x,y)
\label{ib-glob}
\end{align}

\noindent the proof of which is immediate from the local intertwining equation (\ref{ib-intL}). The identity (\ref{ib-glob}) gives sixteen commutation relations between the monodromy matrix operators $\tilde{A}(x),\tilde{B}(x),\tilde{C}(x),\tilde{D}(x)$, but for our purposes we will only require two. These are the equations

\begin{align}
[\tilde{B}(x), \tilde{B}(y)]
=
[\tilde{C}(x), \tilde{C}(y)]
=
0
\label{bbcc}
\end{align}

\noindent and they are necessary to show that the Bethe eigenvectors are symmetric in their rapidity variables.

\subsection{Recovering $\tilde{\mathcal{H}}$ from the transfer matrix}

Let $\tilde{t}(x) = {\rm tr}_a \tilde{T}_a(x) = \tilde{A}(x)+\tilde{D}(x)$ be the transfer matrix of the $i$-boson model. The Hamiltonian (\ref{ib-ham}) may be recovered via the equation

\begin{align}
\tilde{\mathcal{H}}
=
\frac{1}{4}\left[
x^2 \frac{d}{dx}\Big(
x^{-(M+1)/2}\tilde{t}(x)
\Big)
\right]_{x\rightarrow \infty}
-
\frac{1}{4}\left[
\frac{d}{dx}\Big(
x^{(M+1)/2}\tilde{t}(x)
\Big)
\right]_{x\rightarrow 0}
+
\bar{\mathcal{N}}
\label{tibham}
\end{align}

\noindent from which it follows that $[\tilde{\mathcal{H}},\tilde{t}(x)] = 0$. Hence the eigenvectors of $\tilde{\mathcal{H}}$ may be found by studying the eigenvectors of $\tilde{t}(x)$.

\subsection{Bethe Ansatz for the eigenvectors}

As was explained in theorem 1 of the previous chapter, the eigenvectors of the transfer matrix $\tilde{t}(x)$ are given by

\begin{align}
|\Psi\rangle
=
\tilde{B}(y_1)
\ldots
\tilde{B}(y_N)
|0\rangle,
\quad
\langle \Psi|
=
\langle 0|
\tilde{C}(y_N)
\ldots
\tilde{C}(y_1)
\label{ibbethe}
\end{align}

\noindent where the variables $\{y_1,\ldots,y_N\}$ are assumed to obey the Bethe equations (\ref{Beteq}). For the present model, $a(y_i,y_j) = y_i+y_j,\ \alpha(y_i) = y_i^{-(M+1)/2},\ \delta(y_i) = y_i^{(M+1)/2}$. Substituting these expressions into (\ref{Beteq}), the Bethe equations for the $i$-boson model have the decoupled form

\begin{align}
y_i^{M+1} = (-)^{N+1}
\label{ib-beteq}
\end{align}

\noindent for all $1\leq i \leq N$. The equations (\ref{ib-beteq}) may be trivially solved, and reflect the inherent simplicity of the model under consideration. In the next section we turn to a more rigorous examination of the eigenvectors (\ref{ibbethe}), in which the trivial Bethe equations (\ref{ib-beteq}) are not required. We remark that the vectors (\ref{ibbethe}) are genuinely elements of $\tilde{\mathcal{V}}$ and $\tilde{\mathcal{V}}^{*}$, owing to the closure of these spaces under the action of the algebras (\ref{ib-alg2}) and the fact that $|0\rangle \in \tilde{\mathcal{V}}, \langle 0| \in \tilde{\mathcal{V}}^{*}$.  

\section{Calculation of $i$-boson model Bethe eigenvectors}
\label{ib-BE}

In this section we essentially repeat the calculations of section 3.2, but now in the context of the $i$-boson model. Our main result is that the $i$-boson model Bethe eigenvectors can be mapped to the neutral fermionic Fock space of chapter 1, and under this mapping they lie in the $O_{\infty}$ orbit of the Fock vacuum.

\subsection{The maps $\mathcal{M}_{\phi}$ and $\mathcal{M}_{\phi}^{*}$}

Let us begin by introducing analogues of the maps presented in subsection 3.2.1. Observing that the basis elements of $\tilde{\mathcal{V}} \subset \mathcal{V}$ and $\tilde{\mathcal{V}}^{*} \subset \mathcal{V}^{*}$ correspond with {\it strict} partitions under the maps $\mathcal{M}_{\psi}$ and $\mathcal{M}_{\psi}^{*}$, we are motivated to make the following definition.

\begin{definition}
{\rm
Let $|\tilde{n}\rangle = |n_0\rangle_0 \otimes \cdots \otimes |n_M\rangle_M$ and $\langle \tilde{n}| = \langle n_0|_0 \otimes \cdots \otimes \langle n_M|_M$ be basis elements of $\tilde{\mathcal{V}}$ and $\tilde{\mathcal{V}}^{*}$, respectively, and define

\begin{align}
\Sigma_1 
= 
\sum_{j=1}^{M} n_j
\end{align}

\noindent From this, let $|\tilde{\nu})$ and $(\tilde{\nu}|$ be the strict partitions in $\mathcal{F}_{\phi}^{(0)}$ and $\mathcal{F}_{\phi}^{*(0)}$ with one part equal to $j$ if $n_j = 1$, for all $1\leq j \leq M$. That is, we let 

\begin{align}
|\tilde{\nu})
&=
|M^{n_M},\ldots,1^{n_1}),
\quad
\Sigma_1\ {\rm even}
\quad
\quad
|\tilde{\nu})
=
|M^{n_M},\ldots,1^{n_1},0),
\quad
\Sigma_1\ {\rm odd}
\label{spart1}
\\
(\tilde{\nu}|
&=
(M^{n_M},\ldots,1^{n_1}|,
\quad
\Sigma_1\ {\rm even}
\quad
\quad
(\tilde{\nu}|
=
(M^{n_M},\ldots,1^{n_1},0|,
\quad
\Sigma_1\ {\rm odd}
\label{spart2}
\end{align}

\noindent We define linear maps $\mathcal{M}_{\phi}:\tilde{\mathcal{V}} \rightarrow \mathcal{F}_{\phi}^{(0)}$ and $\mathcal{M}_{\phi}^{*}:\tilde{\mathcal{V}}^{*} \rightarrow \mathcal{F}_{\phi}^{*(0)}$ whose actions are given by

\begin{align}
\mathcal{M}_{\phi} |\tilde{n}\rangle 
= 
2^{-\ell(\tilde{\nu})} |\tilde{\nu}),
\quad
\langle \tilde{n} | \mathcal{M}_{\phi}^{*} 
= 
2^{-\ell(\tilde{\nu})}(\tilde{\nu}|
\label{ib-maps}
\end{align} 

\noindent Since these mappings do not depend on the value of $n_0$, they are not one-to-one. We also remark that the maps (\ref{ib-maps}) are non-isometric, in the sense that 

\begin{align}
\langle \tilde{m} | \tilde{n} \rangle
\not=
\Big\langle
\langle \tilde{m}|
\mathcal{M}_{\phi}^{*},
\mathcal{M}_{\phi} 
|\tilde{n} \rangle 
\Big\rangle
\end{align}

\noindent To show this, let $(\tilde{\mu}|$ and $|\tilde{\nu})$ be the strict partitions corresponding with the respective basis vectors $\langle \tilde{m}|$ and $|\tilde{n}\rangle$ under the maps (\ref{ib-maps}). Using the orthogonality relation (\ref{spart-orth}) we find that

\begin{align}
\Big\langle
\langle \tilde{m} | \mathcal{M}_{\phi}^{*},
\mathcal{M}_{\phi} |\tilde{n} \rangle 
\Big\rangle
=
2^{-\ell(\tilde{\mu})} 2^{-\ell(\tilde{\nu})} 
\Big\langle (\tilde{\mu}|, |\tilde{\nu}) \Big\rangle
=
2^{-\ell(\tilde{\mu})} \delta_{\tilde{\mu},\tilde{\nu}}
=
\prod_{j=1}^{M} 2^{-m_j} \delta_{m_j,n_j}
\label{ib-isom}
\end{align}

\noindent Comparing  (\ref{ib-isom0}) with (\ref{ib-isom}), we see that in general $\langle \tilde{m} |\tilde{n} \rangle \not= \langle \langle \tilde{m}| \mathcal{M}_{\phi}^{*},\mathcal{M}_{\phi}|\tilde{n}\rangle \rangle$. The only difference between these two quantities is that (\ref{ib-isom0}) is sensitive to the value of $m_0,n_0$, whereas (\ref{ib-isom}) is not. Hence the mappings (\ref{ib-maps}) essentially filter out all information from the $0^{\rm th}$ site of the lattice. 


}
\end{definition}
   
\subsection{Calculation of $ \tilde{\mathbb{B}}(x) |\tilde{n} \rangle$}

\begin{lemma}
\label{3-lem7}
{\rm Define $\tilde{\mathbb{B}}(x) = x^{\frac{M}{2}} \tilde{B}(x)$ and let $|\tilde{n}\rangle = |n_0\rangle_0 \otimes \cdots \otimes |n_M\rangle_M$ be an arbitrary basis vector of $\tilde{\mathcal{V}}$. The action of $\tilde{\mathbb{B}}(x)$ on $|\tilde{n}\rangle$ is given by

\begin{align}
\tilde{\mathbb{B}}(x)
|\tilde{n}\rangle
=
\sum_{|\tilde{m}\rangle \triangleright |\tilde{n}\rangle}
\prod_{i=1}^{M}
2^{\delta_{(m_i-n_i),1}}
x^{i(m_i-n_i)}
|\tilde{m}\rangle
\label{ib-lem7-0}
\end{align}

\noindent where the sum is over all basis vectors $|\tilde{m}\rangle = |m_0\rangle_0 \otimes \cdots \otimes |m_M\rangle_M \in \tilde{\mathcal{V}}$ which are admissible to $|\tilde{n}\rangle$.
}
\end{lemma}

\begin{proof}
We defer the proof of this equation to the next chapter, where we obtain a similar result (\ref{qb-lem6-0}) in the context of the $q$-boson model. Specializing (\ref{qb-lem6-0}) to the value $q=i$, we obtain (\ref{ib-lem7-0}) as a corollary.
\end{proof}

\subsection{Calculation of $\langle \tilde{n}| \tilde{\mathbb{C}}(x)$} 

\begin{lemma}
\label{3-lem8}
{\rm
Define $\tilde{\mathbb{C}}(x) = x^{\frac{M}{2}} \tilde{C}(1/x)$ and let $\langle \tilde{n}| = \langle n_0|_0 \otimes \cdots \otimes \langle n_M|_M$ be an arbitrary basis vector of $\tilde{\mathcal{V}}^{*}$. The action of $\tilde{\mathbb{C}}(x)$ on $\langle \tilde{n}|$ is given by

\begin{align}
\langle \tilde{n}|
\tilde{\mathbb{C}}(x)
=
\sum_{\langle \tilde{n}| \triangleleft \langle \tilde{m}|}
\prod_{i=1}^{M}
2^{\delta_{(m_i-n_i),1}}
x^{i(m_i-n_i)}
\langle \tilde{m}|
\label{ib-lem8-0}
\end{align}

\noindent where the sum is over all basis vectors $\langle \tilde{m}| = \langle m_0|_0 \otimes \cdots \otimes \langle m_M|_M \in \tilde{\mathcal{V}}^{*}$ which are admissible to $\langle \tilde{n}|$.
}
\end{lemma}

\begin{proof}
Again, we defer the proof to the next chapter. Specializing the later result (\ref{qb-lem7-0}) to the value $q=i$, we obtain (\ref{ib-lem8-0}) as a corollary.
\end{proof}

\subsection{Calculation of $\mathcal{M}_{\phi} \tilde{\mathbb{B}}(x)|\tilde{n}\rangle $ and $\langle \tilde{n}| \tilde{\mathbb{C}}(x) \mathcal{M}_{\phi}^{*}$}

Let $|\tilde{n}\rangle$ and $\langle \tilde{n}|$ be arbitrary basis vectors of $\tilde{\mathcal{V}}$ and $\tilde{\mathcal{V}}^{*}$, respectively, and let $|\tilde{\nu})$ and $(\tilde{\nu}|$ be their corresponding strict partitions, given by equations (\ref{spart1}) and (\ref{spart2}). We fix $l = \ell(\tilde{\nu})$ to be the number of non-zero parts of the strict partition $\tilde{\nu}$. For any two strict partitions $\tilde{\mu},\tilde{\nu}$ let us also define $\#(\tilde{\mu}|\tilde{\nu})$ to be the number of parts in $\tilde{\mu}$ which are not in $\tilde{\nu}$. Using the definition (\ref{ib-maps}) of the maps $\mathcal{M}_{\phi}$ and $\mathcal{M}_{\phi}^{*}$, the expressions (\ref{ib-lem7-0}) and (\ref{ib-lem8-0}) and the result of lemma 1, we obtain

\begin{align}
\mathcal{M}_{\phi}
\tilde{\mathbb{B}}(x)
|\tilde{n}\rangle
&=
\sum_{\tilde{\nu} \prec \tilde{\mu} \subseteq [l+1,M]}
2^{\#(\tilde{\mu}|\tilde{\nu})-\ell(\tilde{\mu})}
x^{|\tilde{\mu}|-|\tilde{\nu}|}
|\tilde{\mu})
\label{quespa}
\\
\langle \tilde{n}|
\tilde{\mathbb{C}}(x)
\mathcal{M}_{\phi}^{*}
&=
\sum_{ \tilde{\nu} \prec \tilde{\mu} \subseteq [l+1,M]}
2^{\#(\tilde{\mu}|\tilde{\nu})-\ell(\tilde{\mu})}
x^{|\tilde{\mu}|-|\tilde{\nu}|}
(\tilde{\mu}|
\label{jacspa}
\end{align}

\noindent Both sums are over all strict partitions $\tilde{\mu}$ which interlace with $\tilde{\nu}$, and whose Young diagrams are contained within the rectangle $[l+1,M]$. 





\subsection{Skew Schur $Q$-functions}

Before we progress to the calculation of the $i$-boson model Bethe eigenvectors, we present some formulae from the theory of symmetric functions.\footnote{For more information on this material, see section 8 in chapter III of \cite{mac}.} For an arbitrary pair of strict partitions $\tilde{\mu},\tilde{\nu}$ and an indeterminate $x$, the single variable {\it skew Schur $Q$-function} $Q_{\tilde{\mu}/\tilde{\nu}}(x)$ is given by

\begin{align}
Q_{\tilde{\mu} / \tilde{\nu}} (x)
=
\left\{
\begin{array}{ll}
2^{\#(\tilde{\mu}|\tilde{\nu})} x^{|\tilde{\mu}|-|\tilde{\nu}|}, 
& 
\tilde{\mu} \succ \tilde{\nu}
\\
\\
0, & {\rm otherwise}
\end{array}
\right.
\label{ib-skew}
\end{align}

\noindent In the case $\tilde{\nu} = \emptyset$ we have $Q_{\tilde{\mu}/\tilde{\nu}}(x) = Q_{\tilde{\mu}}(x)$, where $Q_{\tilde{\mu}}(x)$ is the ordinary Schur $Q$-function in a single variable $x$. The skew Schur $Q$-function satisfies the identity

\begin{align}
Q_{\tilde{\mu}} \{x_1,\ldots,x_n\}
=
\sum_{\tilde{\nu} \subseteq [n-1,\infty]} 
Q_{\tilde{\mu}/\tilde{\nu}} (x_n)
Q_{\tilde{\nu}}\{x_1,\ldots,x_{n-1}\}
\label{ib-skewid}
\end{align}

\noindent where the sum is taken over all strict partitions $\tilde{\nu}$ whose lengths satisfy $\ell(\tilde{\nu}) \leq n-1$, and $Q_{\tilde{\mu}}\{x_1,\ldots,x_n\}$ and $Q_{\tilde{\nu}}\{x_1,\ldots,x_{n-1}\}$ are Schur $Q$-functions in $n$ and $n-1$ variables, respectively.

\subsection{Calculation of $\mathcal{M}_{\phi} \tilde{\mathbb{B}}(x_1) \ldots \tilde{\mathbb{B}}(x_N) |0\rangle $}

\begin{lemma}
{\rm
Let $\{x_1,\ldots,x_N\}$ be a finite set of variables. We claim that

\begin{align}
\mathcal{M}_{\phi}
\tilde{\mathbb{B}}(x_1)
\ldots
\tilde{\mathbb{B}}(x_N) 
|0\rangle
=
\sum_{\tilde{\mu} \subseteq [N,M]}
2^{-\ell(\tilde{\mu})}
Q_{\tilde{\mu}}\{x_1,\ldots,x_N\}
|\tilde{\mu})
\label{ib-ind5}
\end{align}

\noindent where $Q_{\tilde{\mu}}\{x_1,\ldots,x_N\}$ is the Schur $Q$-function in $N$ variables (\ref{schurq}), and the sum is over all strict partitions $\tilde{\mu}$ whose Young diagrams are contained in the rectangle $[N,M]$.
}
\end{lemma}

\begin{proof}

Taking the special case $|\tilde{n}\rangle = |0\rangle$ of equation (\ref{quespa}) we obtain

\begin{align}
\mathcal{M}_{\phi}
\tilde{\mathbb{B}}(x) |0\rangle
=
\sum_{\emptyset \prec \tilde{\mu} \subseteq [1,M]}
2^{-\ell(\tilde{\mu})}
Q_{\tilde{\mu}/\emptyset}(x)
|\tilde{\mu})
=
\sum_{\tilde{\mu} \subseteq [1,M]}
2^{-\ell(\tilde{\mu})}
Q_{\tilde{\mu}}(x)
|\tilde{\mu})
\label{ib-ind1}
\end{align}

\noindent where we have used the equation (\ref{ib-skew}) for the skew Schur $Q$-function, and the definition $\ell(\emptyset) = 0$. We use equation (\ref{ib-ind1}) as the basis for induction, and assume that

\begin{align}
\mathcal{M}_{\phi}
\tilde{\mathbb{B}}(x_1) 
\ldots 
\tilde{\mathbb{B}}(x_{N-1}) 
|0\rangle
=
\sum_{\tilde{\nu} \subseteq [N-1,M]}
2^{-\ell(\tilde{\nu})}
Q_{\tilde{\nu}}\{x_1,\ldots,x_{N-1}\}
|\tilde{\nu})
\end{align}

\noindent for some $N \geq 2$. In terms of the basis vectors of $\tilde{\mathcal{V}}$, this assumption is written as

\begin{align}
\tilde{\mathbb{B}}(x_1) 
\ldots 
\tilde{\mathbb{B}}(x_{N-1}) 
|0\rangle
=
\sum_{|\tilde{n}\rangle | \Sigma_0 = N-1}
Q_{\tilde{\nu}}\{x_1,\ldots,x_{N-1}\}
|\tilde{n}\rangle
\label{ib-ind2}
\end{align}

\noindent where the sum is over all basis vectors $|\tilde{n}\rangle$ such that $\sum_{j=0}^{M} n_j$ $= N-1$, and $\tilde{\nu}$ is the strict partition corresponding to each $|\tilde{n}\rangle$. Acting on (\ref{ib-ind2}) with the composition of operators $\mathcal{M}_{\phi} \circ \tilde{\mathbb{B}}(x_N)$ and using the fact that the $B$-operators commute (\ref{bbcc}), we obtain

\begin{align}
&\mathcal{M}_{\phi} 
\tilde{\mathbb{B}}(x_1) \ldots \tilde{\mathbb{B}}(x_N)
|0\rangle
\label{ib-ind3}
=
\sum_{\tilde{\nu} \subseteq [N-1,M]}
Q_{\tilde{\nu}}\{x_1,\ldots,x_{N-1}\}
\sum_{\tilde{\nu} \prec \tilde{\mu} \subseteq [N,M]}
2^{-\ell(\tilde{\mu})}
Q_{\tilde{\mu}/\tilde{\nu}}(x_N)
|\tilde{\mu})
\end{align}

\noindent Since $Q_{\tilde{\mu}/\tilde{\nu}}(x_N) = 0$ if $\tilde{\mu} \not\succ \tilde{\nu}$, we may alter the sums appearing in (\ref{ib-ind3}), yielding

\begin{align}
\mathcal{M}_{\phi} 
\tilde{\mathbb{B}}(x_1) \ldots \tilde{\mathbb{B}}(x_N)
|0\rangle
\nonumber
&=
\sum_{\tilde{\mu} \subseteq [N,M]}
2^{-\ell(\tilde{\mu})}
\sum_{\tilde{\nu} \subseteq [N-1,M]}
Q_{\tilde{\mu}/\tilde{\nu}}(x_N)
Q_{\tilde{\nu}}\{x_1,\ldots,x_{N-1}\}
|\tilde{\mu})
\\
&=
\sum_{\tilde{\mu} \subseteq [N,M]}
2^{-\ell(\tilde{\mu})}
\sum_{\tilde{\nu} \subseteq [N-1,\infty]}
Q_{\tilde{\mu}/\tilde{\nu}}(x_N)
Q_{\tilde{\nu}}\{x_1,\ldots,x_{N-1}\}
|\tilde{\mu})
\end{align}

\noindent where the final equality holds since the leading part of $\tilde{\mu}$ is less than or equal to $M$, and therefore $Q_{\tilde{\mu}/\tilde{\nu}}(x_N) =0$ if the leading part of $\tilde{\nu}$ exceeds $M$. Using the identity (\ref{ib-skewid}) we evaluate the sum over $\tilde{\nu}$ explicitly, producing the equation (\ref{ib-ind5}). Therefore by induction the result (\ref{ib-ind5}) must hold for arbitrary $N \geq 1$.

\end{proof} 

\subsection{Calculation of $ \langle 0| \tilde{\mathbb{C}}(x_N) \ldots \tilde{\mathbb{C}}(x_1) \mathcal{M}_{\phi}^{*}$}

By following essentially the same steps that were used in the previous subsection, we can also derive the expression

\begin{align}
\langle 0| 
\tilde{\mathbb{C}}(x_N) \ldots \tilde{\mathbb{C}}(x_1) 
\mathcal{M}_{\phi}^{*}
=
\sum_{\tilde{\mu} \subseteq [N,M] }
2^{-\ell(\tilde{\mu})}
Q_{\tilde{\mu}}\{x_1,\ldots,x_N\}
(\tilde{\mu}|
\label{ib-ind6}
\end{align}

\noindent for the dual Bethe eigenvectors. As before, this sum is taken over all strict partitions $\tilde{\mu}$ whose Young diagrams are contained in the rectangle $[N,M]$.

\subsection{Neutral fermionic expression for Bethe eigenvectors}

In analogy with subsection 3.2.11, we proceed to show that $\mathcal{M}_{\phi}$ maps the $i$-boson model Bethe eigenvectors $\tilde{\mathbb{B}}(x_1)\ldots \tilde{\mathbb{B}}(x_N)|0\rangle$ to vectors $g_{\phi}|0\rangle \in \mathcal{F}_{\phi}^{(0)}$ which satisfy the neutral fermion bilinear identity (\ref{BKPexpI7}). In order to do this, we prepare some definitions. For all integers $m>n\geq 0$ we define

\begin{align}
c_{m,n}\{x\}
=
\left\{
\begin{array}{ll}
2^{\delta_{n,0}}
\Big(
\frac{1}{4}
q_m\{x\} q_n\{x\}
+
\frac{1}{2}
\displaystyle{
\sum_{k=1}^{n}
}
(-)^{k}
q_{m+k}\{x\} q_{n-k}\{x\}
\Big),
& 
m \leq M
\\
\\
0,
& 
m > M
\end{array}
\right.
\end{align}

\noindent where $q_m\{x\}$ is the function (\ref{q-com-sym}) in the variables $\{x\} = \{x_1,\ldots,x_N\}$, as defined in chapter 1. Using this expression for the functions $c_{m,n}\{x\}$, for all strict partitions $\tilde{\mu} = \{\mu_1> \cdots > \mu_{2r} \geq 0\}$ let us also define the coefficients

\begin{align}
c_{\tilde{\mu}}\{x\}
=
{\rm Pf}\Big(
c_{\mu_i,\mu_j}\{x\}
\Big)_{1\leq i<j \leq 2r}
\label{ib-coeff2}
\end{align}

\noindent Since $c_{m,n}\{x\} = 0$ if $m > M$, the coefficient $c_{\tilde{\mu}}\{x\}$ vanishes when $\mu_1 > M$. By the definition of the Schur $Q$-functions (\ref{schurq}), it follows that

\begin{align}
c_{\tilde{\mu}}\{x\}
=
\left\{
\begin{array}{ll}
2^{-\ell(\tilde{\mu})}
Q_{\tilde{\mu}}\{x\},
&
\tilde{\mu} \subseteq [N,M]
\\
\\
0, 
& 
\tilde{\mu} \not\subseteq [N,M]
\end{array}
\right.
\label{ib-coeff1}
\end{align}

\noindent Returning to the expression (\ref{ib-ind5}) for the Bethe eigenvectors, we can use the coefficients (\ref{ib-coeff1}) to write

\begin{align}
\mathcal{M}_{\phi}
\tilde{\mathbb{B}}(x_1)\ldots \tilde{\mathbb{B}}(x_N) |0\rangle
=
\sum_{\tilde{\mu}}
c_{\tilde{\mu}}\{x\}
\phi_{\mu_1}\ldots \phi_{\mu_{2r}} |0\rangle
\label{ib-coeff3}
\end{align}

\noindent where the sum is over strict partitions $\tilde{\mu} = \{\mu_1 >\cdots > \mu_{2r} \geq 0\}$ with $r$ taking all non-negative values, and where we have made the identification $|\tilde{\mu}) = \phi_{\mu_1}\ldots \phi_{\mu_{2r}}|0\rangle$.  Owing to their Pfaffian form (\ref{ib-coeff2}) and lemma 22 of chapter 1, we see that the coefficients $c_{\tilde{\mu}}\{x\}$ satisfy the BKP Pl\"ucker relations (\ref{BKPpluck3}). This implies that the right hand side of (\ref{ib-coeff3}) satisfies the neutral fermion bilinear identity (\ref{BKPexpI7}), as we intended to show.

Having established the preceding result, we now express $\mathcal{M}_{\phi} \tilde{\mathbb{B}}(x_1)\ldots \tilde{\mathbb{B}}(x_N) |0\rangle$ in the orbit of the Fock vacuum under $O_{\infty}$. We know that this is possible using theorem 4 of chapter 1, and the fact that the right hand side of (\ref{ib-coeff3}) satisfies the NFBI. We begin by expanding $\mathcal{M}_{\phi} \tilde{\mathbb{B}}(x_1)\ldots \tilde{\mathbb{B}}(x_N) |0\rangle$ in the $\mathcal{F}_{\phi}^{(0)}$ basis (\ref{repclB5}), yielding

\begin{align}
\mathcal{M}_{\phi} 
\tilde{\mathbb{B}}(x_1)\ldots \tilde{\mathbb{B}}(x_N) |0\rangle
=
|0\rangle
+
\sum_{0 \leq n < m \leq M}
2^{\delta_{n,0}-2}
Q_{\{m,n\}}\{x\}
\phi_m \phi_n
|0\rangle
+
g^{(1)}_{\phi}
|0\rangle
\label{ib-coeff4}
\end{align}

\noindent where $Q_{\{m,n\}}\{x\}$ is the Schur $Q$-function associated to the strict partition with one part of size $m$, and another part of size $n$. As usual we assume that all monomials within $g_{\phi}^{(1)} \in Cl_{\phi}^{(0)}$ consist of at least four neutral fermions. Because the right hand side of (\ref{ib-coeff4}) obeys the NFBI, we can use the method adopted in the proof of lemma 20 in chapter 1 to obtain

\begin{align}
\mathcal{M}_{\phi} 
\tilde{\mathbb{B}}(x_1)\ldots \tilde{\mathbb{B}}(x_N) |0\rangle
=
\exp\left(
\sum_{0 \leq n < m \leq M}
2^{\delta_{n,0}-2}
Q_{\{m,n\}}\{x\}
\phi_m \phi_n
\right)
|0\rangle
\label{ib-neut}
\end{align}

\noindent This result explicitly places $\mathcal{M}_{\phi} \tilde{\mathbb{B}}(x_1)\ldots \tilde{\mathbb{B}}(x_N) |0\rangle$ in the vacuum orbit of $O_{\infty}$.

Finally, let us remark that all of these results can be extended to the dual Bethe eigenvectors $\langle 0|\tilde{\mathbb{C}}(x_N)\ldots \tilde{\mathbb{C}}(x_1) \mathcal{M}_{\phi}^{*}$. For example, by completely analogous reasoning it is possible to show that

\begin{align}
\langle 0|\tilde{\mathbb{C}}(x_N)\ldots \tilde{\mathbb{C}}(x_1) \mathcal{M}_{\phi}^{*}
=
\langle 0|
\exp\left(
\sum_{0 \leq n < m \leq M}
2^{\delta_{n,0}-2}
Q_{\{m,n\}}\{x\}
\phis_{n} \phis_{m}
\right)
\end{align}

\section{Scalar product, boxed strict plane partitions}
\label{ib-bpp}

\subsection{Strict plane partitions}

\begin{definition}
{\rm
A {\it strict plane partition} $\tilde{\pi}$ is a set of non-negative integers $\pi(i,j)$ which satisfy

\begin{align}
&
{\pi}(i,j) \geq {\pi}(i+1,j),
\quad
{\pi}(i,j) \geq {\pi}(i,j+1)
\\
&
{\pi}(i,j) > {\pi}(i+1,j+1)
\label{ppsupple}
\end{align}

\noindent for all integers $i,j \geq 1$, as well as the finiteness condition

\begin{align}
\lim_{i \rightarrow \infty}
{\pi}(i,j)
=
\lim_{j \rightarrow \infty}
{\pi}(i,j)
=
0
\end{align}

\noindent That is, strict plane partitions obey all of the axioms of ordinary plane partitions, plus the additional constraint (\ref{ppsupple}) which imposes strictness on every diagonal. Similarly, an {\it $M$-boxed strict plane partition} is a set of non-negative integers ${\pi}(i,j)$ satisfying the above properties, as well as the supplementary condition

\begin{align}
0 \leq {\pi}(i,j) \leq M
\end{align}

\noindent for all integers $i,j \geq 1$, and where $M \geq 1$ is some fixed positive integer. 

To illustrate this definition, we now give the two and three-dimensional diagrams of an exemplary strict plane partition. 

\begin{figure}[H]
\begin{center}
\begin{minipage}{3in}

\setlength{\unitlength}{0.0012cm}
\renewcommand{\dashlinestretch}{30}
\begin{picture}(4800, 2000)(-2000, 1000)

\path(0000,0600)(0600,0600)

\path(0000,1200)(1800,1200)
\path(0000,1800)(1800,1800)
\path(0000,2400)(3000,2400)
\path(0000,3000)(3000,3000)
\path(0000,3000)(0000,0600)
\path(0600,3000)(0600,0600)
\path(1200,3000)(1200,1200)
\path(1800,3000)(1800,1200)
\path(2400,3000)(2400,2400)
\path(3000,3000)(3000,2400)

%

\put(0300,0900){1}
\put(0300,1500){2}
\put(0300,2100){3}
\put(0300,2700){4}
\put(0900,1500){1}
\put(0900,2100){2}
\put(0900,2700){2}
\put(1500,1500){1}
\put(1500,2100){1}
\put(1500,2700){1}
\put(2100,2700){1}
\put(2700,2700){1}


\end{picture}
\end{minipage}
\end{center}
\caption[Tableau representation of a strict plane partition]{Tableau representation of a strict plane partition. The integers on the diagonals of this tableau are strictly decreasing. For example, the central diagonal $\{\pi(1,1)=4, \pi(2,2) =2, \pi(3,3) = 1\}$ is a strictly decreasing sequence.}
 
\end{figure}

\begin{figure}[H]
\begin{center}
\begin{minipage}{2.5in}
\setlength{\unitlength}{0.00085cm}\renewcommand{\dashlinestretch}{30}
\begin{picture}(4800, 4000)(-1500, 500)

%
\path(3912,1512)(3312,1512)
\path(1812,4212)(2412,4212)
\path(1812,4212)(1512,3912)
\path(2412,4212)(2112,3912)
\path(1512,3912)(2112,3912)
\path(1512,3912)(1512,3312)
\path(2112,3912)(2112,3312)
\path(1512,3312)(2112,3312)
\path(2412,4212)(2412,3612)
\path(2412,3612)(2112,3312)
\path(1512,3312)(1212,3012)
\path(2112,3312)(1812,3012)
\path(1212,3012)(1812,3012)
\path(1212,3012)(1212,2412)
\path(1812,3012)(1812,2412)
\path(1212,2412)(1812,2412)
\path(1212,2412)(0912,2112)
\path(1812,2412)(1512,2112)
\path(0912,2112)(1512,2112)
\path(0912,2112)(0912,1512)
\path(1512,2112)(1512,1512)
\path(0912,1512)(1512,1512)
\path(0912,1512)(0612,1212)
\path(1512,1512)(1212,1212)
\path(0612,1212)(1212,1212)
\path(0612,1212)(0612,0612)
\path(1212,1212)(1212,0612)
\path(0612,0612)(1212,0612)
\path(0612,0612)(0012,0012)
\path(2112,3312)(2112,2712)
\path(2112,2712)(1812,2412)
\path(1812,2412)(1812,1812)
\path(1812,1812)(1512,1512)
\path(1512,1512)(1512,0912)
\path(1512,0912)(1212,0612)
\path(2412,3612)(2412,3012)
\path(2412,3012)(2112,2712)
\path(2412,3012)(3012,3012)
\path(2112,2712)(2712,2712)
\path(3012,3012)(2712,2712)
\path(1812,2412)(2412,2412)
\path(2712,2712)(2412,2412)
\path(2412,2412)(2412,1812)
\path(1812,1812)(2412,1812)
\path(2412,1812)(2112,1512)
\path(1512,1512)(2112,1512)
\path(2112,1512)(2112,0912)
\path(1512,0912)(2112,0912)
\path(3012,3012)(3012,2412)
\path(2712,2712)(2712,2112)
\path(3012,2412)(2412,1812)
\path(3012,2412)(4512,2412)
\path(3612,2412)(2712,1512)
\path(2712,2112)(3312,2112)
\path(2412,1812)(3012,1812)
\path(2112,1512)(2712,1512)
\path(2712,1512)(2712,0912)
\path(2112,0912)(2712,0912)
\path(4212,2412)(3912,2112)
\path(4512,2412)(4812,2412)
\path(4812,2412)(4512,2112)
\path(3312,2112)(4512,2112)
\path(3012,1812)(3012,1212)
\path(3912,2112)(3912,1512)
\path(4512,2112)(4512,1512)
\path(4812,2412)(4812,1812)
\path(4812,1812)(4512,1512)
\path(4812,1812)(5412,1812)
\path(3012,1212)(2712,0912)
\path(3912,1512)(4512,1512)
\path(3312,2112)(3312,1512)
\path(3312,1512)(3012,1212)
\path(1812,4812)(1812,4212)
\end{picture}

\end{minipage}
\end{center}

\caption[Three-dimensional representation of a strict plane partition]{Three-dimensional representation of a strict plane partition. This plane partition has six connected horizontal plateaux. The strictness imposed on the diagonals means that all connected horizontal plateaux are one square wide.}

\end{figure}
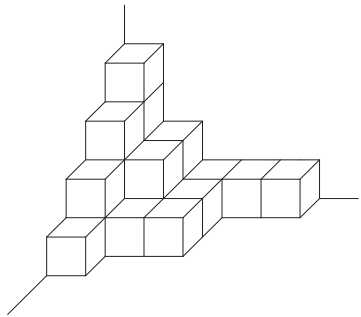
}
\end{definition}

\subsection{Diagonal slices of strict plane partitions}

\begin{definition}
{\rm
Let $\tilde{\pi}$ be an arbitrary strict plane partition. For $i \geq 0$ define the strict partitions $|\tilde{\pi}_{i}) \in \mathcal{F}_{\phi}^{(0)}$ whose elements are given by

\begin{align}
(\tilde{\pi}_{i})_j = {\pi}(j,i+j)
\end{align}

\noindent for all $j \geq 1$. Similarly for $i \leq 0$ define the strict partitions $(\tilde{\pi}_{i}| \in \mathcal{F}_{\phi}^{*(0)}$ whose elements are given by

\begin{align}
(\tilde{\pi}_{i})_j = {\pi}(-i+j,j)
\end{align}

\noindent for all $j \geq 1$. The strict partitions $|\tilde{\pi}_{i})$ and $( \tilde{\pi}_{i}|$ comprise the diagonal slices of the strict plane partition $\tilde{\pi}$.
}
\end{definition}

\begin{lemma}
{\rm Let $|\tilde{\pi}_{i})$ and $( \tilde{\pi}_{i}|$ be the diagonal slices of the strict plane partition $\tilde{\pi}$. Then we have 

\begin{align}
( \tilde{\pi}_{i-1}| \prec ( \tilde{\pi}_{i}|\ \ {\rm for\ all}\ i\leq 0,
\quad\quad
|\tilde{\pi}_{i}) \succ |\tilde{\pi}_{i+1}) \ \ {\rm for\ all}\ i \geq 0
\end{align}
}
\end{lemma}

\begin{proof}
{\rm This is a trivial corollary of the result in subsection 3.3.2. The diagonal slices of $\tilde{\pi}$ are (strict) partitions, and must therefore interlace by lemma 5.
}
\end{proof}

\subsection{Connected elements, paths in strict plane partitions} 

\begin{definition}
{\rm
The element ${\pi}(i,j)$ of $\pi$ is considered to be {\it connected} with both the elements ${\pi}(i+1,j),{\pi}(i,j+1)$ and, assuming they exist, with ${\pi}(i-1,j),{\pi}(i,j-1)$.\footnote{If $\pi(i,j)$ is on the edge of the plane partition, one or both of $\pi(i-1,j),\pi(i,j-1)$ may not exist.} We indicate that two elements are connected by writing, for example, $\pi(i,j) \sim \pi(i,j+1)$. A set $\mathcal{S}$ of more than two elements in ${\pi}$ is connected if, for any two $\pi(i_0,j_0),\pi(i_n,j_n) \in \mathcal{S}$, there exists a subset $\{\pi(i_1,j_1),\ldots,\pi(i_{n-1},j_{n-1})\} \subset \mathcal{S}$ such that $\pi(i_k,j_k) \sim \pi(i_{k+1},j_{k+1})$ for all $0 \leq k \leq n-1$.}
\end{definition}

\begin{definition}
{\rm 
Let $\tilde{\pi}$ be an arbitrary strict plane partition. A {\it path} in $\tilde{\pi}$ is a set of connected elements of $\tilde{\pi}$ which all have the same numerical value. When viewed in the standard three-dimensional representation, paths in a strict plane partition are connected horizontal plateaux which are maximally one square wide. We let $p(\tilde{\pi})$ denote the number of paths possessed by $\tilde{\pi}$. This definition of paths within a strict plane partition was originally given in \cite{fw1}.
}
\end{definition}

\begin{definition}
{\rm
Let $\tilde{\pi}$ be a strict plane partition living inside the box $[N,N,M]$, with diagonal slices $\{\emptyset = \tilde{\pi}_{-N} \prec \cdots \prec \tilde{\pi}_{-1} \prec \tilde{\pi}_{0} \succ \tilde{\pi}_{1} \succ \cdots \succ \tilde{\pi}_{N} = \emptyset \}$. We associate to this strict plane partition the weighting $B_{\tilde{\pi}}(\{x\},\{y\})$ given by

\begin{align}
B_{\tilde{\pi}}\Big(\{x\},\{y\}\Big)
=
2^{p(\tilde{\pi})}
\prod_{i=1}^{N}
x_i^{|\tilde{\pi}_{-i+1}|-|\tilde{\pi}_{-i}|}
y_i^{|\tilde{\pi}_{i-1}|-|\tilde{\pi}_{i}|}
\label{B-def}
\end{align}

\noindent where $p(\tilde{\pi})$ is the number of paths in $\tilde{\pi}$.

\begin{figure}[H]

\begin{center}
\begin{minipage}{4.3in}

\setlength{\unitlength}{0.00022cm}
\begin{picture}(30000,28000)(-15000,-20000)

\path(0,0)(2000,-2000)
\path(2000,2000)(8000,-4000)
\path(4000,4000)(12000,-4000)
\path(6000,6000)(16000,-4000)
\path(8000,8000)(18000,-2000)

\path(0,0)(8000,8000)
\path(2000,-2000)(10000,6000)
\path(6000,-2000)(12000,4000)
\path(8000,-4000)(14000,2000)
\path(12000,-4000)(16000,0)
\path(16000,-4000)(18000,-2000)

\put(7700,6000){4}
\put(5700,4000){3}
\put(3700,2000){2}
\put(1700,0){1}

\put(5700,0){1}
\put(7700,2000){2}
\put(9700,4000){2}

\put(7700,-2000){1}
\put(9700,0){1}
\put(11700,2000){2}

\put(11700,-2000){1}
\put(13700,0){1}

\put(15700,-2000){1}


\path(8000,-6000)(6000,-8000)(8000,-10000)(10000,-8000)(8000,-6000)
\put(7700,-8000){4}

\path(7000,-13000)(9000,-15000)
\path(9000,-11000)(13000,-15000)
\path(11000,-9000)(15000,-13000)
\path(11000,-9000)(7000,-13000)
\path(13000,-11000)(9000,-15000)
\path(15000,-13000)(13000,-15000)
\put(8700,-13000){2}
\put(10700,-11000){2}
\put(12700,-13000){2}

\path(5000,-17000)(9000,-21000)
\path(7000,-15000)(13000,-21000)
\path(11000,-15000)(17000,-21000)
\path(15000,-15000)(19000,-19000)
\path(5000,-17000)(7000,-15000)
\path(7000,-19000)(11000,-15000)
\path(9000,-21000)(15000,-15000)
\path(13000,-21000)(17000,-17000)
\path(17000,-21000)(19000,-19000)
\put(6700,-17000){1}
\put(8700,-19000){1}
\put(10700,-17000){1}
\put(12700,-19000){1}
\put(14700,-17000){1}
\put(16700,-19000){1}

\path(5000,-9000)(3000,-11000)(5000,-13000)(7000,-11000)(5000,-9000)
\put(4700,-11000){3}

\path(3000,-13000)(1000,-15000)(3000,-17000)(5000,-15000)(3000,-13000)
\put(2700,-15000){2}

\path(0000,-16000)(-2000,-18000)(0,-20000)(2000,-18000)(0,-16000)
\put(-300,-18000){1}

\end{picture}

\end{minipage}
\end{center}

\caption{Splitting a strict plane partition into its constituent paths.}

\end{figure}
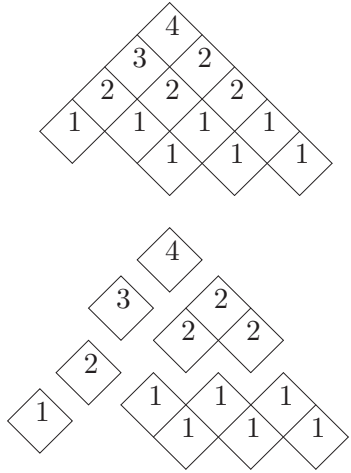

}
\end{definition}

\begin{lemma}
{\rm 
Let $\tilde{\pi}$ be a strict plane partition as described in definition 11. Then

\begin{align}
2^{-\ell(\tilde{\pi}_0)}
\prod_{i=1}^{N}
2^{\#(\tilde{\pi}_{-i+1}|\tilde{\pi}_{-i})}
\prod_{j=1}^{N}
2^{\#(\tilde{\pi}_{j-1}|\tilde{\pi}_j)}
=
2^{p(\tilde{\pi})}
\label{ib-lem10-1}
\end{align}
}
\end{lemma}

\begin{proof}
The proof is well illustrated by an example, so we consider the strict plane partition $\tilde{\pi}$ drawn in figure 3.9. We see that $\tilde{\pi}$ lives inside the box $[5,5,4]$, and its diagonal slices are given by

\begin{align}
&\tilde{\pi}_{-5} = \emptyset 
&& \phantom{\tilde{\pi}_{0} = \{3,3,3\}} 
\quad\quad\quad\quad\quad \tilde{\pi}_{1} = \{2,1\} \\ \nonumber
&\tilde{\pi}_{-4} = \emptyset 
&& \phantom{\tilde{\pi}_{0} = \{3,3,3\}}
\quad\quad\quad\quad\quad \tilde{\pi}_{2} = \{2,1\} \\ \nonumber
&\tilde{\pi}_{-3} = \{1\} 
&& \tilde{\pi}_{0} = \{4,2,1\} 
\quad\quad\quad\quad\quad \tilde{\pi}_{3} = \{1\} \\ \nonumber
&\tilde{\pi}_{-2} = \{2\} 
&& \phantom{\tilde{\pi}_{0} = \{3,3,3\}} 
\quad\quad\quad\quad\quad \tilde{\pi}_{4} = \{1\} \\ \nonumber
&\tilde{\pi}_{-1} = \{3,1\} 
&& \phantom{\tilde{\pi}_{0} = \{3,3,3\}} 
\quad\quad\quad\quad\quad \tilde{\pi}_{5} = \emptyset
\end{align}

\noindent Using these strict partitions, we evaluate

\begin{align}
&2^{\#(\tilde{\pi}_{-4}|\tilde{\pi}_{-5})} = 2^0 
&& 2^{\#(\tilde{\pi}_{0}|\tilde{\pi}_{1})} = 2^1
\\
& 2^{\#(\tilde{\pi}_{-3}|\tilde{\pi}_{-4})} = 2^1
&& 2^{\#(\tilde{\pi}_{1}|\tilde{\pi}_{2})} = 2^0
\nonumber
\\
& 2^{\#(\tilde{\pi}_{-2}|\tilde{\pi}_{-3})} = 2^1
&& 2^{\#(\tilde{\pi}_{2}|\tilde{\pi}_{3})} = 2^1
\nonumber
\\
& 2^{\#(\tilde{\pi}_{-1}|\tilde{\pi}_{-2})} = 2^2
&& 2^{\#(\tilde{\pi}_{3}|\tilde{\pi}_{4})} = 2^0
\nonumber
\\
& 2^{\#(\tilde{\pi}_{0}|\tilde{\pi}_{-1})} = 2^2
&& 2^{\#(\tilde{\pi}_{4}|\tilde{\pi}_{5})} = 2^1
\nonumber
\end{align}

\noindent Let us consider these factors, progressing from the extremal diagonal slices of $\tilde{\pi}$ towards its central slice $\tilde{\pi}_0$. We obtain a factor of 2 for every path which begins in $\tilde{\pi}$ and does {\it not} intersect the central diagonal. Paths that {\it do} intersect the central diagonal are assigned a factor of $2^2$, which is a double counting. We cure this double counting by dividing by 2 for every element in $\tilde{\pi}_0$, that is, by dividing by $2^{\ell(\tilde{\pi}_0)} = 2^3$. The result is

\begin{align}
2^{-\ell(\tilde{\pi}_0)}
\prod_{i=1}^{5}
2^{\#(\tilde{\pi}_{-i+1}|\tilde{\pi}_{-i})}
\prod_{j=1}^{5}
2^{\#(\tilde{\pi}_{j-1}|\tilde{\pi}_j)}
=
2^{6}
=
2^{p(\tilde{\pi})}
\end{align}

\noindent as required. It is clear that this method extends to arbitrary strict plane partitions.

\end{proof}

\subsection{Generating $M$-boxed strict plane partitions}

We now derive an analogue of the result given in subsection 3.3.3, this time relating to the scalar product of the $i$-boson model on $M+1$ sites. The result is that the scalar product between the image Bethe eigenstates (\ref{ib-ind5}) and (\ref{ib-ind6}) is a generating function for $M$-boxed strict plane partitions. This correspondence may be realized by iterating the $|\tilde{n}\rangle = |0\rangle$ case of equation (\ref{quespa}) $N$ times, giving

\begin{align}
& \mathcal{M}_{\phi}
\tilde{\mathbb{B}}(x_1)
\ldots
\tilde{\mathbb{B}}(x_N)
|0\rangle
=
\sum_{
[N,M]
\supseteq
\tilde{\pi}_{0}
\succ
\cdots
\succ
\tilde{\pi}_{N}
=
\emptyset
}
\
2^{-\ell(\tilde{\pi}_0)}
\prod_{i=1}^{N}
2^{\#(\tilde{\pi}_{i-1}|\tilde{\pi}_{i})}
x_i^{|\tilde{\pi}_{i-1}|-|\tilde{\pi}_{i}|}
|\tilde{\pi}_0)
\end{align}

\noindent where the sum is over all interlacing strict partitions $\{\tilde{\pi}_{0} \succ \cdots \succ \tilde{\pi}_{N}\}$ subject to $\tilde{\pi}_{0} \subseteq [N,M]$ and $\tilde{\pi}_{N} = \emptyset$. Similarly, one can iterate the $\langle \tilde{n}| = \langle 0|$ case of (\ref{jacspa}) $N$ times, giving

{\small
\begin{align}
& \langle 0|
\tilde{\mathbb{C}}(x_N)
\ldots
\tilde{\mathbb{C}}(x_1)
\mathcal{M}_{\phi}^{*}
=
\sum_{
\emptyset
=
\tilde{\pi}_{-N}
\prec
\cdots
\prec
\tilde{\pi}_{0}
\subseteq
[N,M]
}
\
2^{-\ell(\tilde{\pi}_0)} 
\prod_{i=1}^{N}
2^{\#(\tilde{\pi}_{-i+1}|\tilde{\pi}_{-i})}
x_i^{|\tilde{\pi}_{-i+1}|-|\tilde{\pi}_{-i}|}
(\tilde{\pi}_0|
\end{align}
}

\noindent where the sum is over all interlacing strict partitions $\{\tilde{\pi}_{-N} \prec \cdots \prec \tilde{\pi}_{0}\}$ subject to $\tilde{\pi}_0 \subseteq [N,M]$ and $\tilde{\pi}_{-N} = \emptyset$. By the definition (\ref{B-def}) of $B_{\tilde{\pi}}(\{x\},\{y\})$, the result (\ref{ib-lem10-1}) of lemma 11 and the orthogonality (\ref{spart-orth}) of strict partition states, we thus obtain

\begin{align}
\Big\langle
\langle 0|
\tilde{\mathbb{C}}(x_N)
\ldots
\tilde{\mathbb{C}}(x_1)
\mathcal{M}_{\phi}^{*},
\mathcal{M}_{\phi}
\tilde{\mathbb{B}}(y_1)
\ldots
\tilde{\mathbb{B}}(y_N)
|0\rangle
\Big\rangle
=
\sum_{\tilde{\pi} \subseteq [N,N,M]}
B_{\tilde{\pi}}\Big(\{x\},\{y\}\Big)
\label{thebeg5}
\end{align}

\noindent where the sum is taken over all strict plane partitions $\tilde{\pi}$ which fit inside the box of dimension $N \times N \times M$. Hence the scalar product between the image Bethe eigenstates (\ref{ib-ind5}) and (\ref{ib-ind6}) is a generating function for $M$-boxed strict plane partitions. This generating function is evaluated explicitly by using the equations (\ref{ib-ind5}), (\ref{ib-ind6}) and the orthogonality relation (\ref{spart-orth}) to give

\begin{align}
\Big\langle
\langle 0|
\tilde{\mathbb{C}}(x_N)
\ldots
\tilde{\mathbb{C}}(x_1)
\mathcal{M}_{\phi}^{*},
\mathcal{M}_{\phi}
\tilde{\mathbb{B}}(y_1)
\ldots
\tilde{\mathbb{B}}(y_N)
|0\rangle
\Big\rangle
=
\sum_{\tilde{\mu} \subseteq [N,M]}
2^{-\ell(\tilde{\mu})}
Q_{\tilde{\mu}}\{x\}
Q_{\tilde{\mu}}\{y\}
\label{thebeg4}
\end{align}

\noindent where $Q_{\tilde{\mu}}\{x\}, Q_{\tilde{\mu}}\{y\}$ denote Schur $Q$-functions. Comparing equations (\ref{thebeg5}) and (\ref{thebeg4}), we have proved the result

\begin{align}
\sum_{\tilde{\pi} \subseteq [N,N,M]}
B_{\tilde{\pi}}\Big(\{x\},\{y\}\Big)
=
\sum_{\tilde{\mu} \subseteq [N,M]}
2^{-\ell(\tilde{\mu})}
Q_{\tilde{\mu}}\{x\}
Q_{\tilde{\mu}}\{y\}
\label{spp-exp3}
\end{align}

\subsection{Scalar product as a power-sum specialized BKP $\tau$-function}

In this subsection we demonstrate that the scalar product (\ref{thebeg4}) is a specialization of a BKP $\tau$-function. The specialization is achieved by setting the BKP time variables to power sums in the $i$-boson model rapidities. This result parallels the one obtained in subsection 3.3.4, in the context of KP $\tau$-functions. We begin with the equation

\begin{align}
\langle 0|
\exp\left( \sum_{m \in \tilde{\mathbb{N}}} t_m \lambda_m \right)
=
\sum_{\tilde{\mu}}
2^{-\ell(\tilde{\mu})} 
\mathcal{Q}_{\tilde{\mu}} \{\tilde{t}\}
(\tilde{\mu}|
\label{thebeg}
\end{align}

\noindent whose sum is over all strict partitions $\tilde{\mu}$, which follows from lemma 15 in chapter 1 and the orthogonality (\ref{spart-orth}) of strict partitions. Defining $t_m = \frac{2}{m} \sum_{n=1}^{N} x_n^m$ for all $m \in \tilde{\mathbb{N}}$, equation (\ref{thebeg}) becomes

\begin{align}
\langle 0|
\exp\left( \sum_{m \in \tilde{\mathbb{N}}} \sum_{n=1}^{N} 
\frac{2}{m} x_n^m \lambda_m \right)
=
\sum_{\tilde{\mu} \subseteq [N, \infty]}
2^{-\ell(\tilde{\mu})}  
Q_{\tilde{\mu}} \{x\}
(\tilde{\mu}|
\label{thebeg2}
\end{align}

\noindent where the sum is over all strict partitions $\tilde{\mu}$ with maximal length $N$. Equating the right hand sides of (\ref{ib-ind5}) and (\ref{ib-neut}) and using the identity (\ref{thebeg2}), we find

\begin{align}
\langle 0|
\exp\left(
\sum_{m\in\tilde{\mathbb{N}}}
\sum_{n=1}^{N}
\frac{2}{m}
x_n^m
\lambda_m
\right)
\exp Y\{y\}
|0\rangle
=
\sum_{\tilde{\mu} \subseteq [N,M]}
2^{-\ell(\tilde{\mu})}
Q_{\tilde{\mu}}\{x\}
Q_{\tilde{\mu}}\{y\}
\label{thebeg3} 
\end{align}

\noindent where $Y\{y\} \in B_{\infty}$ is defined as

\begin{align}
Y\{y\}
=
\sum_{0 \leq n < m \leq M}
2^{\delta_{n,0}-2}
Q_{\{m,n\}}\{y\}
\phi_m \phi_n
\end{align}

\noindent Now consider the polynomial BKP $\tau$-function $\tau\{\tilde{t}\} = \langle e^{\lambda\{\tilde{t}\}} e^{Y\{y\}} \rangle$. Comparing the equations (\ref{thebeg4}) and (\ref{thebeg3}) we find that

\begin{align}
\tau\{\tilde{t}\}
&=
\langle 0| 
\exp\left(\sum_{m\in\tilde{\mathbb{N}}} t_m \lambda_m \right)
\exp Y\{y\} |0\rangle
\\
&
=
\Big\langle
\langle 0|
\tilde{\mathbb{C}}(x_N)
\ldots
\tilde{\mathbb{C}}(x_1)
\mathcal{M}_{\phi}^{*},
\mathcal{M}_{\phi}
\tilde{\mathbb{B}}(y_1)
\ldots
\tilde{\mathbb{B}}(y_N)
|0\rangle
\Big\rangle
\nonumber
\end{align}

\noindent under the power-sum specialization $t_m = \frac{2}{m} \sum_{n=1}^{N} x_n^m$ for all $m \in \tilde{\mathbb{N}}$. This connection between the generating function of strict plane partitions and BKP $\tau$-functions first appeared in \cite{fwz3}, but in the context of strict plane partitions whose column heights are unrestricted. The result of this subsection is at the level of $M$-boxed strict plane partitions, and it specializes to the result of \cite{fwz3} in the limit $M\rightarrow\infty$.

\section{$i$-boson model on an infinite lattice}
\label{ib-pp}

This section is the $i$-boson model analogue of the earlier section 3.4 on the phase model. We study the action of the monodromy matrix operators $\tilde{\mathbb{B}}(x)$ and $\tilde{\mathbb{C}}(x)$ when the number of lattice sites becomes infinite. Our main result is lemma 12, showing that in the limit $M \rightarrow \infty$ the operators $\tilde{\mathbb{B}}(x)$ and $\tilde{\mathbb{C}}(x)$ acquire equivalent actions to the half-vertex operators $\tilde{\Gamma}_{-}(x)$ and $\tilde{\Gamma}_{+}(x)$ from BKP theory. The actions of these half-vertex operators were studied briefly in \cite{fw1}, and in more detail in \cite{fwz3}.  

\subsection{Calculation of $\mathcal{M}_{\phi} \tilde{\mathbb{B}}(x)|\tilde{n}\rangle$ and $\langle \tilde{n}| \tilde{\mathbb{C}}(x) \mathcal{M}_{\phi}^{*}$ as $M \rightarrow \infty$}

\begin{lemma}
\label{3-lem11}
{\rm 
Consider the infinite lattice limit of the $i$-boson model, obtained by taking $M \rightarrow \infty$. Let $|\tilde{n}\rangle = \otimes_{j=0}^{\infty} |n_j\rangle_j$ and $\langle \tilde{n}| = \otimes_{j=0}^{\infty} \langle n_j|_j$ be basis vectors of $\tilde{\mathcal{V}}$ and $\tilde{\mathcal{V}}^{*}$, respectively, in this limit. In addition, let $2^{-\ell(\tilde{\nu})}|\tilde{\nu})$ and $2^{-\ell(\tilde{\nu})}(\tilde{\nu}|$ be the image states of these basis vectors under the mappings (\ref{ib-maps}). We claim that

\begin{align}
\mathcal{M}_{\phi}
\Big[
\lim_{M \rightarrow \infty}
\tilde{\mathbb{B}}(x)
|\tilde{n}\rangle
\Big]
=
2^{-\ell(\tilde{\nu})}
\tilde{\Gamma}_{-}(x)
|\tilde{\nu}),
\quad
\Big[
\lim_{M \rightarrow \infty}
\langle \tilde{n}|
\tilde{\mathbb{C}}(x)
\Big]
\mathcal{M}_{\phi}^{*}
=
2^{-\ell(\tilde{\nu})}
(\tilde{\nu}|
\tilde{\Gamma}_{+}(x)
\label{ib-lem11-1}
\end{align}

\noindent where we have defined the BKP half-vertex operators\footnote{Once again we use the half-vertex nomenclature, since $\tilde{\Gamma}_{-}(x),\tilde{\Gamma}_{+}(x)$ each comprise one half of a neutral fermion vertex operator, \cite{jm1}.}

\begin{align}
\tilde{\Gamma}_{-}(x)
=
\exp\left(
\sum_{n \in \tilde{\mathbb{N}}}
\frac{2}{n} x^{n} \lambda_{-n}
\right),
\quad
\tilde{\Gamma}_{+}(x)
=
\exp\left( 
\sum_{n \in \tilde{\mathbb{N}}}
\frac{2}{n} x^{n} \lambda_n
\right)
\label{bkp-half}
\end{align}

\noindent and $\lambda_{-n},\lambda_{n}$ denote the Heisenberg generators (\ref{bheisen1}).
}
\end{lemma}

\begin{proof}

We split the proof into two steps. In the first step, we show that (\ref{ib-lem11-1}) is equivalent to the statement (\ref{ib-lem11-2}). In the second step we prove (\ref{ib-lem11-2}) using the calculus of neutral free fermions.

\medskip
\noindent
{\bf Step 1.}\ Taking the $M \rightarrow \infty$ limit of equations (\ref{quespa}) and (\ref{jacspa}), we obtain

\begin{align}
\mathcal{M}_{\phi}
\Big[\lim_{M \rightarrow \infty}
\tilde{\mathbb{B}}(x) |\tilde{n}\rangle 
\Big]
&=
\sum_{\tilde{\mu} \succ \tilde{\nu}}
2^{\#(\tilde{\mu}|\tilde{\nu}) - \ell(\tilde{\mu})}
x^{|\tilde{\mu}|-|\tilde{\nu}|}
|\tilde{\mu})
\\
\Big[\lim_{M \rightarrow \infty}
\langle \tilde{n}| \tilde{\mathbb{C}}(x)
\Big]
\mathcal{M}_{\phi}^{*}
&=
\sum_{\tilde{\mu} \succ \tilde{\nu}}
2^{\#(\tilde{\mu}|\tilde{\nu}) - \ell(\tilde{\mu})}
x^{|\tilde{\mu}|-|\tilde{\nu}|}
(\tilde{\mu}|
\end{align}

\noindent where the sums are over all strict partitions $\tilde{\mu}$ which interlace with $\tilde{\nu}$, whose parts now have no size restriction. The equations (\ref{ib-lem11-1}) are therefore equivalent to the statements

\begin{align}
\tilde{\Gamma}_{-}(x) |\tilde{\nu})
&=
\sum_{\tilde{\mu} \succ \tilde{\nu}}
2^{\#(\tilde{\mu}|\tilde{\nu}) 
- 
\ell(\tilde{\mu})
+
\ell(\tilde{\nu})}
x^{|\tilde{\mu}|-|\tilde{\nu}|}
|\tilde{\mu})
\label{ib-lem11-3}
\\
(\tilde{\nu}| \tilde{\Gamma}_{+}(x) 
&=
\sum_{\tilde{\mu} \succ \tilde{\nu}}
2^{\#(\tilde{\mu}|\tilde{\nu}) 
- 
\ell(\tilde{\mu})
+
\ell(\tilde{\nu})}
x^{|\tilde{\mu}|-|\tilde{\nu}|}
(\tilde{\mu}|
\label{ib-lem11-4}
\end{align}

\noindent which are entirely at the level of neutral free fermions. Due to the orthogonality (\ref{spart-orth}) of strict partition states, equations (\ref{ib-lem11-3}) and (\ref{ib-lem11-4}) may be presented in the alternative form

\begin{align}
(\tilde{\mu}|
\tilde{\Gamma}_{-}(x)
=
\sum_{\tilde{\nu} \prec \tilde{\mu}}
2^{\#(\tilde{\mu}|\tilde{\nu})}
x^{|\tilde{\mu}|-|\tilde{\nu}|}
(\tilde{\nu}|, 
\quad
\tilde{\Gamma}_{+}(x)
|\tilde{\mu})
=
\sum_{\tilde{\nu} \prec \tilde{\mu}}
2^{\#(\tilde{\mu}|\tilde{\nu})}
x^{|\tilde{\mu}|-|\tilde{\nu}|}
|\tilde{\nu})
\label{ib-lem11-2}
\end{align}

\noindent where the sums are now over strict partitions $\tilde{\nu}$ such that $\tilde{\nu} \prec \tilde{\mu}$. The sums in (\ref{ib-lem11-2}) are finite, whereas those in (\ref{ib-lem11-3}) and (\ref{ib-lem11-4}) are infinite. As we mentioned earlier, it is thus easiest to prove (\ref{ib-lem11-2}), and this will in turn establish the equations (\ref{ib-lem11-1}).

\medskip
\noindent
{\bf Step 2.}\ Consider the even-length strict partitions

\begin{align}
(\tilde{\mu}|
=
\langle 0| \phis_{\mu_{2r}} \ldots \phis_{\mu_1},
\quad
|\tilde{\mu})
=
\phi_{\mu_1} \ldots \phi_{\mu_{2r}} |0\rangle
\label{ib-lem11-6}
\end{align}

\noindent where we assume that $\{\mu_1 > \cdots > \mu_{2r} > 0\}$. In order to prove (\ref{ib-lem11-2}), we must calculate $(\tilde{\mu}| \tilde{\Gamma}_{-}(x)$ and $\tilde{\Gamma}_{+}(x)|\tilde{\mu})$. To achieve this we require the commutation relations

\begin{align}
\tilde{\Gamma}_{-}(x)
\left(
\phis_i
+
2 \sum_{n=1}^{\infty}
\phis_{(i-n)} x^n
\right)
=
\phis_i
\tilde{\Gamma}_{-}(x),
\quad
\tilde{\Gamma}_{+}(x)
\phi_i
=
\left( 
\phi_i
+
2 \sum_{n=1}^{\infty}
\phi_{(i-n)} x^{n}
\right)
\tilde{\Gamma}_{+}(x)
\label{ib-lem11-5}
\end{align}

\noindent which are derived following the arguments presented in subsection 1.4.8.\footnote{Setting $t_n=\frac{2}{n}x^{n}$ for all $n \in \tilde{\mathbb{N}}$ in (\ref{BKPev4}), we obtain $\tilde{\Gamma}_{+}(x) \Phi(k) = \frac{1+xk}{1-xk} \Phi(k) \tilde{\Gamma}_{+}(x)$. Extracting the coefficients of $k^i$ from this equation, we prove the second commutation relation in (\ref{ib-lem11-5}). The first commutation relation may be proved similarly.} Applying the relations (\ref{ib-lem11-5}) repeatedly to the strict partitions (\ref{ib-lem11-6}), we obtain
 
\begin{align}
( \tilde{\mu} | \tilde{\Gamma}_{-}(x)
&=
\langle 0|
\left(
\phis_{\mu_{2r}}
+
2 \sum_{i_{2r}=1}^{\infty}
\phis_{(\mu_{2r}-i_{2r})} x^{i_{2r}}
\right)
\ldots
\left(
\phis_{\mu_1}
+
2 \sum_{i_1 =1}^{\infty}
\phis_{(\mu_1-i_1)} x^{i_1}
\right)
\nonumber
\\
&=
\langle 0|
\left(
\sum_{i_{2r}=0}^{\infty}
(2-\delta_{i_{2r},0})
\phis_{(\mu_{2r}-i_{2r})}
x^{i_{2r}}
\right)
\ldots
\left(
\sum_{i_1=0}^{\infty}
(2-\delta_{i_1,0})
\phis_{(\mu_1-i_1)}
x^{i_1}
\right)
\label{ib-trunc1}
\end{align}

\noindent where we have used the fact that $\langle 0| \tilde{\Gamma}_{-}(x) = \langle 0|$, and

\begin{align}
\tilde{\Gamma}_{+}(x)|\tilde{\mu})
&=
\left(
\phi_{\mu_1}
+
2 \sum_{i_1=1}^{\infty}
\phi_{(\mu_1-i_1)} x^{i_1}
\right)
\ldots
\left(
\phi_{\mu_{2r}}
+
2 \sum_{i_{2r}=1}^{\infty}
\phi_{(\mu_{2r}-i_{2r})} x^{i_{2r}}
\right)
|0\rangle
\nonumber
\\
&=
\left(
\sum_{i_1=0}^{\infty}
(2-\delta_{i_1,0})
\phi_{(\mu_1-i_1)}
x^{i_1}
\right)
\ldots
\left(
\sum_{i_{2r}=0}^{\infty}
(2-\delta_{i_{2r},0})
\phi_{(\mu_{2r}-i_{2r})}
x^{i_{2r}}
\right)
|0\rangle
\label{ib-trunc2}
\end{align}

\noindent where we have used the fact that $\tilde{\Gamma}_{+}(x) |0\rangle = |0\rangle$. The equation (\ref{ib-trunc1}) contains infinite sums which can be truncated by means of the identity

\begin{align}
&\left(
\sum_{i=0}^{\infty}
(2-\delta_{i,0}) \phis_{(n-i)} x^i
\right)
\left(
\sum_{j=0}^{\infty}
(2-\delta_{j,0}) \phis_{(m-j)} x^j
\right)
\label{ib-trunc3}
\\
=&
\left(
\sum_{i=0}^{\infty}
(2-\delta_{i,0}) \phis_{(n-i)} x^i
\right)
\left(
\sum_{j=0}^{m-n}
(2-\delta_{j,0}-\delta_{j,m-n}) \phis_{(m-j)} x^j
\right)
\nonumber
\end{align}


\noindent which holds for all integers $m > n > 0$, while the sums in (\ref{ib-trunc2}) may be truncated by means of the identity

\begin{align}
&\left(
\sum_{i=0}^{\infty}
(2-\delta_{i,0})
\phi_{(m-i)}
x^{i}
\right)
\left(
\sum_{j=0}^{\infty}
(2-\delta_{j,0})
\phi_{(n-j)}
x^{j}
\right)
\label{ib-trunc4}
\\
=&
\left(
\sum_{i=0}^{m-n}
(2-\delta_{i,0}-\delta_{i,m-n})
\phi_{(m-i)}
x^{i}
\right)
\left(
\sum_{j=0}^{\infty}
(2-\delta_{j,0})
\phi_{(n-j)}
x^{j}
\right)
\nonumber
\end{align}


\noindent which also holds for all integers $m > n > 0$. 
%
Substituting the identity (\ref{ib-trunc3}) into (\ref{ib-trunc1}) we obtain

\begin{align}
(\tilde{\mu}|
\tilde{\Gamma}_{-}(x)
=
\langle 0|
&
\left(
\sum_{i_{2r}=0}^{\mu_{2r}}
(2-\delta_{i_{2r},0})
\phis_{(\mu_{2r}-i_{2r})}
x^{i_{2r}}
\right)
\label{ib-trunc5}
\\
\times
&
\left(
\sum_{i_{\widetilde{2r}}=0}^{\mu_{\widetilde{2r}}-\mu_{2r}}
(2-\delta_{i_{\widetilde{2r}},0}-\delta_{i_{\widetilde{2r}},\mu_{\widetilde{2r}}-\mu_{2r}}) 
\phis_{(\mu_{\widetilde{2r}}-i_{\widetilde{2r}})}
x^{i_{\widetilde{2r}}}
\right)
\ldots
\nonumber
\\
\times
&
\left(
\sum_{i_1=0}^{\mu_1-\mu_2}
(2-\delta_{i_1,0}-\delta_{i_1,\mu_1-\mu_2}) 
\phis_{(\mu_1-i_1)}
x^{i_1}
\right)
\nonumber
\end{align}

\noindent where we have defined $\widetilde{2r} = 2r-1$, and used the annihilation properties (\ref{repclB1}) to truncate the left-most sum. Furthermore, substituting (\ref{ib-trunc4}) into (\ref{ib-trunc2}) gives

\begin{align}
\tilde{\Gamma}_{+}(x)
|\tilde{\mu})
=
&
\left(
\sum_{i_1=0}^{\mu_1-\mu_2}
(2-\delta_{i_1,0}-\delta_{i_1,\mu_1-\mu_2}) 
\phi_{(\mu_1-i_1)}
x^{i_1}
\right)
\ldots
\label{ib-trunc6}
\\
\times
&
\left(
\sum_{i_{\widetilde{2r}}=0}^{\mu_{\widetilde{2r}}-\mu_{2r}}
(2-\delta_{i_{\widetilde{2r}},0}-\delta_{i_{\widetilde{2r}},\mu_{\widetilde{2r}}-\mu_{2r}}) 
\phi_{(\mu_{\widetilde{2r}}-i_{\widetilde{2r}})}
x^{i_{\widetilde{2r}}}
\right)
\nonumber
\\
\times
&
\left(
\sum_{i_{2r}=0}^{\mu_{2r}}
(2-\delta_{i_{2r},0})
\phi_{(\mu_{2r}-i_{2r})}
x^{i_{2r}}
\right)
|0\rangle
\nonumber
\end{align}

\noindent where we have again set $\widetilde{2r} = 2r-1$, and used the annihilation properties (\ref{repclB1}) to truncate the right-most sum. The indices in (\ref{ib-trunc5}) can then be modified to produce the equation

\begin{align}
(\tilde{\mu}| \tilde{\Gamma}_{-}(x)
=
\langle 0|
& 
\left(
\sum_{\mu_{2r} \geq \nu_{2r} \geq 0}
(2-\delta_{\mu_{2r},\nu_{2r}})
\phis_{\nu_{2r}}
x^{\mu_{2r}-\nu_{2r}}
\right)
\label{ib-trunc7}
\\
\times
&
\left(
\sum_{\mu_{\widetilde{2r}} \geq \nu_{\widetilde{2r}} \geq \mu_{2r}}
(2-\delta_{\mu_{\widetilde{2r}},\nu_{\widetilde{2r}}} - \delta_{\mu_{2r},\nu_{\widetilde{2r}}})
\phis_{\nu_{\widetilde{2r}}}
x^{\mu_{\widetilde{2r}}-\nu_{\widetilde{2r}}}
\right)
\ldots
\nonumber
\\
\times
&
\left(
\sum_{\mu_1 \geq \nu_1 \geq \mu_2}
(2-\delta_{\mu_1,\nu_1}-\delta_{\mu_2,\nu_1})
\phis_{\nu_1}
x^{\mu_1-\nu_1}
\right)
=
\sum_{\tilde{\nu} \prec \tilde{\mu}}
2^{\#(\tilde{\mu}|\tilde{\nu})}
x^{|\tilde{\mu}|-|\tilde{\nu}|}
(\tilde{\nu}|
\nonumber
\end{align}

\noindent where we have defined the strict partitions $(\tilde{\nu}| = \langle 0| \phis_{\nu_{2r}} \ldots \phis_{\nu_1}$, while the indices in (\ref{ib-trunc6}) can be modified to produce the equation 

\begin{align}
\tilde{\Gamma}_{+}(x)
|\tilde{\mu})
=
&
\left(
\sum_{\mu_1 \geq \nu_1 \geq \mu_2}
(2-\delta_{\mu_1,\nu_1}-\delta_{\mu_2,\nu_1}) 
\phi_{\nu_1}
x^{\mu_1-\nu_1}
\right)
\ldots
\label{ib-trunc8}
\\
\times
&
\left(
\sum_{\mu_{\widetilde{2r}} \geq \nu_{\widetilde{2r}} \geq \mu_{2r}}
(2-\delta_{\mu_{\widetilde{2r}},\nu_{\widetilde{2r}}} - \delta_{\mu_{2r},\nu_{\widetilde{2r}}})
\phi_{\nu_{\widetilde{2r}}}
x^{\mu_{\widetilde{2r}}-\nu_{\widetilde{2r}}}\right)
\nonumber
\\
\times
&
\left(
\sum_{\mu_{2r} \geq \nu_{2r} \geq 0}
(2-\delta_{\mu_{2r},\nu_{2r}})
\phi_{\nu_{2r}}
x^{\mu_{2r}-\nu_{2r}}
\right)
|0\rangle
=
\sum_{\tilde{\nu} \prec \tilde{\mu}}
2^{\#(\tilde{\mu}|\tilde{\nu})}
x^{|\tilde{\mu}|-|\tilde{\nu}|}
|\tilde{\nu})
\nonumber
\end{align}

\noindent where we have defined the strict partitions $|\tilde{\nu}) = \phi_{\nu_1}\ldots \phi_{\nu_{2r}} |0\rangle$. Notice that the final equality in (\ref{ib-trunc7}) and (\ref{ib-trunc8}) follows from the fact that the summation variables satisfy

\begin{align}
\mu_i \geq \nu_i \geq \mu_{i+1}\ \ 
{\rm for\ all}\ 1\leq i \leq \widetilde{2r},
\quad
\mu_{2r} \geq \nu_{2r} \geq 0
\end{align}

\noindent as well as the fact that $|\tilde{\mu}|-|\tilde{\nu}| = \sum_{i=1}^{2r} (\mu_i-\nu_i)$. It is also straightforward to check that the correct factors $2^{\#(\tilde{\mu}|\tilde{\nu})}$ are recovered from the sums in (\ref{ib-trunc7}) and (\ref{ib-trunc8}). This completes the proof of (\ref{ib-lem11-2}) for even-length strict partitions $\tilde{\mu}$. The proof for odd-length strict partitions starts by acting on the states

\begin{align}
(\tilde{\mu}| = \langle 1|\phis_{\mu_{2r-1}}\ldots \phis_{\mu_1},
\quad
|\tilde{\mu}) = \phi_{\mu_1}\ldots \phi_{\mu_{2r-1}} |1\rangle
\end{align}

\noindent with $\{\mu_1 > \cdots > \mu_{2r-1} > 0\}$, and is achieved following precisely the same procedure as above.   

\end{proof}

\subsection{Generating strict plane partitions of arbitrary size}

In the previous section we demonstrated that the scalar product (\ref{thebeg5}) on a lattice of size $M+1$ generates $M$-boxed strict plane partitions. Accordingly, we expect that in the limit $M\rightarrow\infty$ it will generate strict plane partitions whose column heights are arbitrarily large, giving rise to the equation

\begin{align}
\lim_{M \rightarrow\infty}
\Big\langle
\langle 0|
\tilde{\mathbb{C}}(x_N)
\ldots
\tilde{\mathbb{C}}(x_1)
\mathcal{M}_{\phi}^{*},
\mathcal{M}_{\phi}
\tilde{\mathbb{B}}(y_1)
\ldots
\tilde{\mathbb{B}}(y_N)
|0\rangle
\Big\rangle
=
\sum_{\tilde{\pi} \subseteq [N,N,\infty]}
B_{\tilde{\pi}}\Big(\{x\},\{y\}\Big)
\label{spp-exp1}
\end{align}

\noindent where the sum is over all strict plane partitions $\tilde{\pi}$ which fit inside the box of dimension $N \times N \times \infty$, with $B_{\tilde{\pi}}(\{x\},\{y\})$ given by (\ref{B-def}). On the other hand, using the result of lemma 12 we are able to write

{\footnotesize
\begin{align}
\lim_{M\rightarrow\infty}
\Big\langle
\langle 0|
\tilde{\mathbb{C}}(x_N)
\ldots
\tilde{\mathbb{C}}(x_1)
\mathcal{M}_{\phi}^{*},
\mathcal{M}_{\phi}
\tilde{\mathbb{B}}(y_1)
&\ldots
\tilde{\mathbb{B}}(y_N)
|0\rangle
\Big\rangle
=
\label{bhv-com2}
(\emptyset| 
\tilde{\Gamma}_{+}(x_N)
\ldots
\tilde{\Gamma}_{+}(x_1)
\tilde{\Gamma}_{-}(y_1)
\ldots
\tilde{\Gamma}_{-}(y_N)
|\emptyset)
\end{align}
}

\noindent which lends itself to immediate evaluation, owing to simple commutation relations between the BKP half-vertex operators. Explicitly speaking, using the definition (\ref{bkp-half}) of $\tilde{\Gamma}_{-}(y)$ and $\tilde{\Gamma}_{+}(x)$ we find that

\begin{align}
& 
4
\sum_{m \in \tilde{\mathbb{N}}}
\sum_{n \in \tilde{\mathbb{N}}}
\frac{x^m y^n}{mn} [\lambda_m,\lambda_{-n}]
=
2
\sum_{m \in \tilde{\mathbb{N}}}
\sum_{n \in \tilde{\mathbb{N}}}
\frac{x^m y^n}{mn} m \delta_{m,n}
=
2
\sum_{m \in \tilde{\mathbb{N}}}
\frac{(xy)^m}{m}
\label{bhv-com}
\\
\implies
&
\tilde{\Gamma}_{+}(x)
\tilde{\Gamma}_{-}(y)
=
\exp\left(
2
\sum_{m \in \tilde{\mathbb{N}}}
\frac{(xy)^m}{m}
\right)
\tilde{\Gamma}_{-}(y)
\tilde{\Gamma}_{+}(x)
=
\frac{1+xy}{1-xy}
\tilde{\Gamma}_{-}(y)
\tilde{\Gamma}_{+}(x)
\nonumber
\end{align}

\noindent Employing the commutation relation (\ref{bhv-com}) repeatedly in (\ref{bhv-com2}), we obtain

\begin{align}
\lim_{M\rightarrow\infty}
\Big\langle
\langle 0|
\tilde{\mathbb{C}}(x_N)
\ldots
\tilde{\mathbb{C}}(x_1)
\mathcal{M}_{\phi}^{*},
\mathcal{M}_{\phi}
\tilde{\mathbb{B}}(y_1)
\ldots
\tilde{\mathbb{B}}(y_N)
|0\rangle
\Big\rangle
=
\prod_{i,j=1}^{N}
\frac{1+x_i y_j}{1-x_i y_j}
\label{spp-exp2}
\end{align}

\noindent Comparing equations (\ref{spp-exp1}) and (\ref{spp-exp2}), we have proved that

\begin{align}
\sum_{\tilde{\pi} \subseteq [N,N,\infty]}
B_{\tilde{\pi}}\Big(\{x\},\{y\}\Big)
=
\prod_{i,j=1}^{N} 
\frac{
1+x_i y_j
}{
1-x_i y_j
}
\end{align}

\noindent which is a simpler evaluation of this generating function than in the finite case (\ref{spp-exp3}). This calculation could also have been performed using (\ref{thebeg4}) and the identity

\begin{align}
\sum_{\tilde{\mu} \subseteq [N,\infty]}
2^{-\ell(\tilde{\mu})}
Q_{\tilde{\mu}}\{x\}
Q_{\tilde{\mu}}\{y\}
=
\prod_{i,j=1}^{N}
\frac{1+x_i y_j}{1-x_i y_j}
\end{align}

\noindent from section 8 in chapter III of \cite{mac}. We believe our proof is more fundamental, since it is independent of the properties of symmetric functions. As a final observation, let us specialize the variables $\{x_1,\ldots,x_N\}$ and $\{y_1,\ldots,y_N\}$ to 

\begin{align}
x_i = 
y_i = z^{i-\frac{1}{2}}\ \
{\rm for\ all}\ 1\leq i \leq N
\end{align}

\noindent giving rise to the equation

\begin{align}
\sum_{\tilde{\pi} \subseteq [N,N,\infty]}
2^{p(\tilde{\pi})}
z^{|\tilde{\pi}|}
=
\prod_{i,j=1}^{N}
\frac{
1+z^{i+j-1}
}{
1-z^{i+j-1}
}
\end{align}

\noindent where $p(\tilde{\pi})$ and $|\tilde{\pi}|$ are the number of paths and weight of the strict plane partition $\tilde{\pi}$, respectively. Taking the limit $N \rightarrow \infty$ we obtain

\begin{align}
\sum_{\tilde{\pi}}
2^{p(\tilde{\pi})}
z^{|\tilde{\pi}|}
=
\prod_{i=1}^{\infty}
\frac{
(1+z^i)^i
}{
(1-z^i)^i
}
\label{spp-exp4}
\end{align}

\noindent where the sum is over strict plane partitions $\tilde{\pi}$ of completely arbitrary dimension. The generating function (\ref{spp-exp4}) first appeared in \cite{fw1}. It was used independently in \cite{vul1} to study the shifted Schur process, extending the earlier work of \cite{or}.

\section{Conclusion}

In this chapter we studied the Bethe eigenvectors of the phase and $i$-boson models. Both models admit a representation on the vector space $\mathcal{V}$, and we showed that the basis elements of $\mathcal{V}$ can be mapped quite naturally to partitions in $\mathcal{F}_{\psi}^{(0)},\mathcal{F}_{\phi}^{(0)}$. Our main observation was that under these maps, the Bethe eigenvectors of these models lie in the orbit of the vacuum under $GL_{\infty}$ and $O_{\infty}$, respectively. This proved that the corresponding scalar products are power-sum specializations of KP and BKP $\tau$-functions, respectively.

The other key results in this chapter are the lemmas \ref{3-lem6} and \ref{3-lem11}, which pertain to the infinite lattice limit of the phase and $i$-boson models. We found that when $M\rightarrow\infty$, the action of a $B$-operator on a general state maps to the action of a charged/neutral half-vertex operator on the image state. Owing to the elementary commutations between the half-vertex operators, we could easily evaluate the scalar products of these models in the $M\rightarrow \infty$ limit. Hence we obtained new proofs of the generating functions for ordinary and strict plane partitions.

This chapter raises several questions which might lead to interesting research in the future. We list two such questions below.

{\bf 1.} {\it Why do the Bethe eigenvectors of the phase and $i$-boson models lead to solutions of the KP and BKP hierarchies?} At a superficial level we would expect these two areas of integrability to be unrelated, and yet we have considerable evidence to indicate that this is not the case. Ultimately, we desire a more fundamental explanation for this link between classical and quantum integrable models. Let us also remark that our work complements that of \cite{zup}, where it was shown that one-point boundary correlation functions of the phase model are $\tau$-functions of the 2-Toda hierarchy \cite{tak1}. It would be worthwhile to repeat these calculations in the context of the $i$-boson model.

{\bf 2.} {\it Is it possible to obtain fermionic proofs of the generating functions for various symmetry classes of plane partitions?} There are many different symmetry classes of plane partitions, whose enumerations are in factorized form.\footnote{See, for example, chapter 6 of \cite{bre}.} It is possible that the techniques of this chapter could be extended to constructing these restricted plane partitions, and to calculating their generating functions.


\newpage

\thispagestyle{empty}

\phantom{nothing}


\chapter{$q$-boson model and Hall-Littlewood plane partitions}
\setcounter{lemma}{0}
\setcounter{theorem}{0}
\setcounter{remark}{0}
\setcounter{example}{0}
\setcounter{definition}{0}

\setcounter{section}{-1}

\section{Introduction}

In chapter 3 we studied the $q\rightarrow\infty$ and $q\rightarrow i$ limits of the $q$-boson model. In the $q\rightarrow\infty$ case we reviewed the work of \cite{bog1}, before showing that the Bethe eigenvectors map to elements of $\mathcal{F}_{\psi}$ which satisfy the charged fermionic bilinear identity. We obtained an analogous result in the context of the $q\rightarrow i$ limit, where the Bethe eigenvectors map to elements of $\mathcal{F}_{\phi}$ which satisfy the neutral fermionic bilinear identity. The aim of this chapter is to provide a fermionic description of the $q$-boson model itself.

The $q$-boson model was introduced in \cite{bb} by applying the Primakov-Holstein transformation to the $L$-matrix of the spin-$\frac{1}{2}$ XXZ model. The bosons which appear in this model generate a $q$-deformed Heisenberg algebra, which has several different representations on the vector space discussed in section \ref{phase-intro}. In section \ref{qb-intro} we discuss one such representation on $\mathcal{V}$ \cite{tsi}, give the $q$-boson model Hamiltonian $\mathcal{H}$, and construct its eigenvectors using the algebraic Bethe Ansatz.

After this we will find it necessary to introduce a set of $t$-fermions, where $t$ is a deformation parameter,\footnote{Throughout this chapter the parameters $q$ and $t$ play the same role, and they are related via the equation $t = q^{-2}$. Generally we will work in terms of $t$, which is the parameter used in the study of Hall-Littlewood functions.} which generalize the charged fermions of chapter 1. These fermions originally appeared in the papers \cite{jin1} and \cite{jin2} by N~Jing, where they were used in the context of Hall-Littlewood functions, and their connection with the $q$-boson model was proposed by P~Sulkowski in \cite{sul}. In section \ref{qb-fermi} we study the algebra generated by the $t$-fermions, deriving several useful identities. We define a representation of this algebra on the $t$-deformed Fock space $\mathcal{F}_{\psi}(t)$, and calculate inner products between the partition elements of $\mathcal{F}_{\psi}(t)$. We also state a $t$-deformed version of the KP half-vertex operator, which is used later in the chapter.  

An explicit expression for the $q$-boson model Bethe eigenvectors was found by N~Tsilevich in \cite{tsi}. Tsilevich extended the earlier work of Bogoliubov, writing the Bethe eigenvectors as sums over partitions which are weighted by Hall-Littlewood functions. In section \ref{qb-BE} we reproduce this result and transfer it to the language of the $t$-deformed fermions, as was suggested in \cite{sul}. We map the basis vectors of $\mathcal{V}$ to partitions in the Fock space $\mathcal{F}_{\psi}(t)$ and calculate the image of the Bethe eigenvectors under this map. 

The remainder of the chapter consists of original work. In section \ref{qb-bpp} we extend the results of chapter 3, by showing that the $q$-boson model scalar product is a generating function for plane partitions inside a box of size $N\times N\times M$. Within this generating function, each plane partition is assigned a weight which depends on the deformation parameter $t$. All the weights collapse to 1 in the $t\rightarrow 0\ (q\rightarrow \infty)$ limit, giving rise to Bogoliubov's generating function as discussed in section \ref{phase-bpp}. In the $t\rightarrow -1\ (q\rightarrow i)$ limit, all weights assigned to non-strict plane partitions collapse to 0, giving rise to the generating function discussed in section \ref{ib-bpp}. The $t$-weighted generating function for plane partitions was originally found by M~Vuleti\'{c} in \cite{vul2}, using purely combinatorial arguments.

The most interesting result of the chapter is given in section \ref{qb-pp}, which considers the $M \rightarrow \infty$ limit of the $q$-boson model. In this limit, we prove that the action of a $B$-operator on an arbitrary vector in $\mathcal{V}$ is equivalent to the action of a $t$-deformed half-vertex operator on the image state in $\mathcal{F}_{\psi}(t)$.\footnote{This connection was also noticed in \cite{sul}, by identifying the basis vectors of $\mathcal{V}$ and partitions in $\mathcal{F}_{\psi}(t)$ with Hall-Littlewood functions. Our derivation is distinct from that of \cite{sul}, in that it relies solely on the calculus of the $t$-deformed fermions.} This is an extension of the results obtained in chapter 3, and allows the infinite lattice scalar product to be easily evaluated. The $t$-deformed half-vertex operators have simple commutation relations, meaning that the scalar product once again factorizes into product form. We thus obtain a fermionic proof of Vuleti\'{c}'s generating function, first proposed in \cite{fw2}.

\section{$q$-boson model}
\label{qb-intro}

In this section we gather together a number of preliminary results pertaining to the $q$-boson model. We mainly follow \cite{bik2} and \cite{tsi}. The sections 3.1 and 3.5 from the previous chapter can be viewed as specializations of the material presented here.

\subsection{Space of states $\mathcal{V}$ and inner product $\mathcal{I}_t$}

Like the phase model in the previous chapter, the $q$-boson model consists of a lattice of $M+1$ sites which can each be occupied by an unlimited number of particles. For this reason, its space of states $\mathcal{V}$ is the same as that defined in subsection 3.1.1. For any two basis vectors $|m\rangle = |m_0\rangle_0 \otimes \cdots \otimes |m_M\rangle_M, |n\rangle = |n_0\rangle_0 \otimes \cdots \otimes |n_M\rangle_M$ we define a bilinear inner product $\mathcal{I}_t$ given by

\begin{align}
\mathcal{I}_t
\Big(
|m\rangle,|n\rangle
\Big)
=
\frac{
[m_0]_t!
}
{\prod_{i=1}^{M} 
[m_i]_t!
}\ 
\displaystyle{\prod_{i=0}^{M}} \delta_{m_i,n_i}
\label{qb-ip1}
\end{align}

\noindent where we have adopted the notations

\begin{align}
[n]_t! = \prod_{i=1}^{n} (1-t^i)\ \ 
{\rm for\ all}\ n \geq 1,
\quad
{\rm and}
\quad
[0]_t! = 1
\end{align}

\noindent The inner product between more general states of $\mathcal{V}$ is deduced using bilinearity. Here $t = q^{-2}$ is the deformation parameter characteristic of the model. By setting $t=0\ (q \rightarrow \infty)$ we recover the inner product (\ref{innp}) used for the phase model. Also, letting the basis vectors in (\ref{qb-ip1}) be elements of $\tilde{\mathcal{V}}$ and setting $t=-1\ (q=i)$, we recover the inner product (\ref{Itilde}) used for the $i$-boson model. 

The dual space of states $\mathcal{V}^{*}$ acts according to

\begin{align}
\langle m|n\rangle
=
\mathcal{I}_t\Big(
|m\rangle, |n\rangle
\Big)
\label{qb-ip2}
\end{align}

\noindent for all $\langle m| = \langle m_0|_0 \otimes \cdots \otimes \langle m_M|_M \in \mathcal{V}^{*}$ and $|n\rangle = |n_0\rangle_0 \otimes \cdots \otimes |n_M\rangle_M \in \mathcal{V}$.    

\subsection{$q$-boson algebra}

The $q$-boson algebra is generated by $\{b,\bd,\mathcal{N}\}$ which satisfy the commutation relations

\begin{align}
[b,\bd] = q^{-2\mathcal{N}}=t^{\mathcal{N}},
\quad
[\mathcal{N},b] = -b,
\quad
[\mathcal{N},\bd] = \bd
\label{realqb}
\end{align}

\noindent where we have again identified $q^{-2} = t$, and now retain this definition throughout the entire chapter without further comment. This algebra collapses to the phase algebra (\ref{phase-alg}) in the limit $q\rightarrow\infty$, and to the $i$-boson algebra (\ref{ib-alg}) in the limit $q\rightarrow i$. As in the previous chapter, we consider $M+1$ copies of the $q$-boson algebra, generated by $\{b_0,\bd_0,\mathcal{N}_0\}$ through to $\{b_M,\bd_M,\mathcal{N}_M\}$. Adopting the labelling system of chapter 2, we denote these algebras by $\mathcal{A}_0,\ldots,\mathcal{A}_M$ with $\mathfrak{a}_i^{+} = \bd_i,\mathfrak{a}_i^{-} = b_i,\mathfrak{a}_i^{0} = \mathcal{N}_i$. As usual different copies are commuting, giving rise to the equations

\begin{align}
[b_i,\bd_j] = \delta_{i,j}t^{\mathcal{N}_i},
\quad
[\mathcal{N}_i,b_j] = -\delta_{i,j}b_i,
\quad
[\mathcal{N}_i,\bd_j] = \delta_{i,j} \bd_i
\label{qb-alg}
\end{align}

\noindent for all $0 \leq i,j \leq M$.

\subsection{Representations of $q$-boson algebras}

In this subsection, following \cite{tsi}, we fix representations of the $q$-boson algebras on the vector space $\mathcal{V}$. Quite generally speaking, $b_i$ plays the role of an annihilation operator, removing particles from the $i^{\rm th}$ lattice site. Conversely, $\bd_i$ is a creation operator, adding particles to the $i^{\rm th}$ lattice site. In order to represent the $q$-boson algebras correctly, these operators must also produce accompanying factors that depend on $t$. It transpires that there is a certain amount of freedom in choosing these factors. We will make one choice for the representation of $\mathcal{A}_1,\ldots,\mathcal{A}_M$ and a different choice for the representation of $\mathcal{A}_0$, as we describe below.  


Firstly we consider the algebras $\mathcal{A}_1,\ldots,\mathcal{A}_M$. For all $1\leq i \leq M$, the operator $b_i$ has the action

{\small
\begin{align}
&
b_i |n_0\rangle_0 \otimes \cdots \otimes |n_M\rangle_M
\label{b1}
=
\left\{
\begin{array}{ll}
0, 
& 
n_i = 0
\\ \\
(1-t)^{-\frac{1}{2}}
|n_0\rangle_0 
\otimes \cdots \otimes 
|n_i-1\rangle_i 
\otimes \cdots \otimes 
|n_M\rangle_M,
&
n_i \geq 1
\end{array}
\right.
\end{align}
}

\noindent while for all $1\leq i \leq M$ the operator $\bd_i$ has the action

\begin{align}
\bd_i |n_0\rangle_0 \otimes \cdots \otimes |n_M\rangle_M
=
\frac{
1-t^{n_i+1}
}{(1-t)^{\frac{1}{2}}}
|n_0\rangle_0 
\otimes \cdots \otimes 
|n_i+1\rangle_i 
\otimes \cdots \otimes 
|n_M\rangle_M
\label{bd1}
\end{align}

\noindent Secondly we consider the algebra $\mathcal{A}_0$, for which the representation is slightly different. The operator $b_0$ has the action 

\begin{align}
b_0 |n_0\rangle_0 \otimes \cdots \otimes |n_M\rangle_M
=
\frac{
1-t^{n_0}
}{(1-t)^{\frac{1}{2}}}
|n_0-1\rangle_0 \otimes |n_1\rangle_1  \otimes \cdots \otimes |n_M\rangle_M
\end{align}

\noindent while the operator $\bd_0$ has the action

\begin{align}
\bd_0 |n_0\rangle_0 \otimes \cdots \otimes |n_M\rangle_M
=
(1-t)^{-\frac{1}{2}}
|n_0+1\rangle_0 \otimes |n_1\rangle_1  \otimes \cdots \otimes |n_M\rangle_M
\label{bd0}
\end{align}

\noindent For all $0 \leq i \leq M$ the action of $\mathcal{N}_i$ is given by 

\begin{align}
\mathcal{N}_i 
|n_0\rangle_0 \otimes \cdots \otimes |n_M\rangle_M
=
n_i |n_0\rangle_0 \otimes \cdots \otimes |n_M\rangle_M
\label{numb1}
\end{align}

\noindent The set of definitions (\ref{b1})--(\ref{numb1}) faithfully represent the algebras $\mathcal{A}_0,\ldots,\mathcal{A}_M$. The representations of the phase algebras in subsection \ref{phaserep} are obtained by setting $t=0$. Similarly, the representations of the $i$-boson algebras in subsection \ref{ibrep} are obtained by setting $t=-1$.

By virtue of the definitions (\ref{qb-ip1}) and (\ref{b1})--(\ref{bd0}), we notice that for all $|m\rangle,|n\rangle \in \mathcal{V}$ and $0\leq i \leq M$ the operators $b_i,\bd_i$ satisfy the equation

\begin{align}
\mathcal{I}_t\Big(
b_i |m\rangle, |n\rangle
\Big)
=
\mathcal{I}_t\Big(
|m\rangle, \bd_i |n\rangle
\Big)
\end{align}

\noindent meaning that they are adjoint. Once again, the operator $\mathcal{N}_i$ is self-adjoint. Hence we can immediately deduce actions for $\{b_i,\bd_i,\mathcal{N}_i\}$ on the dual space $\mathcal{V}^{*}$, as follows. For all $1\leq i \leq M$, the operator $\bd_i$ has the action

{\small
\begin{align}
&
\langle n_0|_0 \otimes \cdots \otimes \langle n_M|_M \bd_i
\label{bd2}
=
\left\{
\begin{array}{ll}
0, & n_i = 0
\\
\\
(1-t)^{-\frac{1}{2}}
\langle n_0|_0 
\otimes \cdots \otimes 
\langle n_i-1|_i 
\otimes \cdots \otimes 
\langle n_M|_M,
& 
n_i \geq 1
\end{array}
\right.
\end{align}
}

\noindent while for all $1\leq i \leq M$ the operator $b_i$ has the action

\begin{align}
\langle n_0|_0 \otimes \cdots \otimes \langle n_M|_M b_i
=
\frac{
1-t^{n_i+1}
}{(1-t)^{\frac{1}{2}}}
\langle n_0|_0 
\otimes \cdots \otimes 
\langle n_i+1|_i 
\otimes \cdots \otimes 
\langle n_M|_M
\label{b2}
\end{align}

\noindent As before, the algebra $\mathcal{A}_0$ is prescribed its own representation. The operator $\bd_0$ has the action

\begin{align}
\langle n_0|_0 \otimes \cdots \otimes \langle n_M|_M 
\bd_0
=
\frac{
1-t^{n_0}
}{(1-t)^{\frac{1}{2}}}
\langle n_0-1|_0 \otimes \langle n_1|_1 \otimes \cdots \otimes \langle n_M|_M
\end{align}

\noindent while the operator $b_0$ has the action

\begin{align}
\langle n_0|_0 \otimes \cdots \otimes \langle n_M|_M 
b_0
=
(1-t)^{-\frac{1}{2}}
\langle n_0+1|_0 \otimes \langle n_1|_1 \otimes \cdots \otimes \langle n_M|_M
\label{b0}
\end{align}

\noindent For all $0 \leq i \leq M$ the action of $\mathcal{N}_i$ is given by

\begin{align}
\langle n_0|_0 \otimes \cdots \otimes \langle n_M|_M 
\mathcal{N}_i
=
n_i
\langle n_0|_0 \otimes \cdots \otimes \langle n_M|_M
\label{numb2}
\end{align}

\noindent The set of definitions (\ref{bd2})--(\ref{numb2}) provide the dual representation of the algebras $\mathcal{A}_0,\ldots,\mathcal{A}_M$.

\subsection{Hamiltonian $\mathcal{H}$}

The Hamiltonian of the $q$-boson model is given by

\begin{align}
\mathcal{H}
=
-\frac{1}{2}
\sum_{i=0}^{M}
\left(
\bd_i b_{i+1} + b_i \bd_{i+1}
\right)
+
\bar{\mathcal{N}}
\label{qb-ham}
\end{align}

\noindent where the periodicity $b_{M+1} = b_0$ and $\bd_{M+1} = \bd_0$ is imposed. The Hamiltonians (\ref{phase-ham}) and (\ref{ib-ham}) are trivially recovered by taking the limits $t\rightarrow 0$ and $t\rightarrow -1$. It follows that the eigenvectors of these earlier Hamiltonians can be recovered by constructing the eigenvectors of (\ref{qb-ham}) directly, as we do in the remainder of this section.    

\subsection{$L$-matrix and local intertwining equation}

The $R$-matrix for the $q$-boson model depends on two indeterminates $x,y$ and acts in the tensor product $\mathcal{V}_a \otimes \mathcal{V}_b$, where $\mathcal{V}_a, \mathcal{V}_b$ are copies of $\mathbb{C}^2$. It is given by

\begin{align}
R_{ab}(x,y,t)
=
\left(
\begin{array}{cccc}
x-ty &               0                 &              0                   & 0
\\
0 &              t(x-y)              &                (1-t) x^{\frac{1}{2}} y^{\frac{1}{2}}             & 0
\\
0 &              (1-t) x^{\frac{1}{2}} y^{\frac{1}{2}}                 &  x-y  & 0
\\
0 &               0                 &              0                   & x-ty
\end{array}
\right)_{ab}
\label{qb-R}
\end{align}

\noindent and corresponds to choosing $a_{\pm}(x,y) = x-ty,\ b_{+}(x,y)=t(x-y),\ b_{-}(x,y) = x-y,\ c_{\pm}(x,y) = (1-t)x^{\frac{1}{2}} y^{\frac{1}{2}}$ in (\ref{general-Rmat}). The $L$-matrix for the $q$-boson model depends on a single indeterminate $x$, and acts in the space $\mathcal{V}_a$. Its entries are operators acting at the $m^{\rm th}$ lattice site, and identically everywhere else. It has the form

\begin{align}
L_{am}(x,t)
=
\left(
\begin{array}{cc}
x^{-\frac{1}{2}} & (1-t)^{\frac{1}{2}} \bd_m
\\
(1-t)^{\frac{1}{2}} b_m  & x^{\frac{1}{2}}
\end{array}
\right)_{a}
\label{qb-L}
\end{align}

\noindent Notice that both the $R$-matrix (\ref{qb-R}) and $L$-matrix (\ref{qb-L}) collapse to their phase model counterpart in the limit $t\rightarrow 0$, and to their $i$-boson model counterpart in the limit $t \rightarrow -1$. Returning to the parent model, the local intertwining equation has the usual form

\begin{align}
R_{ab}(x,y,t) L_{am}(x,t) L_{bm}(y,t)
=
L_{bm}(y,t) L_{am}(x,t) R_{ab}(x,y,t)
\label{qb-intL}
\end{align}

\noindent This is a $4\times 4$ matrix equation, which gives rise to sixteen scalar identities. Each of these identities may be verified by direct calculation, and by using the commutation relations (\ref{qb-alg}) where appropriate.

\subsection{Monodromy matrix and global intertwining equation}

The monodromy matrix is an $(M+1)$-fold product of the $L$-matrices (\ref{qb-L}), taken in the auxiliary space ${\rm End}(\mathcal{V}_a)$. It has the form

\begin{align}
T_a(x,t)
=
L_{aM}(x,t)
\ldots
L_{a0}(x,t)
=
\left(
\begin{array}{cc}
A(x,t) & B(x,t)
\\
C(x,t) & D(x,t)
\end{array}
\right)_a
\end{align}

\noindent where $A(x,t), B(x,t), C(x,t), D(x,t)$ are elements of $\mathcal{A}_0\otimes\cdots\otimes\mathcal{A}_M$. The monodromy matrix satisfies the global intertwining equation

\begin{align}
R_{ab}(x,y,t)
T_a(x,t)
T_b(y,t)
=
T_b(y,t)
T_a(x,t)
R_{ab}(x,y,t)
\label{qb-intT}
\end{align}

\noindent the proof of which is immediate from the local intertwining relation (\ref{qb-intL}). The equation (\ref{qb-intT}) contains sixteen commutation relations between the monodromy matrix operators $A(x,t),B(x,t)$,$C(x,t),D(x,t)$, but for our purposes we will only require two. These are the equations

\begin{align}
[B(x,t),B(y,t)] = [C(x,t),C(y,t)] = 0
\label{qb-bbcc}
\end{align} 

\noindent and they are necessary to show that the Bethe eigenvectors are symmetric in their rapidity variables.

\subsection{Recovering $\mathcal{H}$ from the transfer matrix}

Let ${\rm tr}_a T_a = A(x,t)+D(x,t)$ be the transfer matrix of the $q$-boson model. The Hamiltonian (\ref{qb-ham}) may be recovered via the equation

{\small
\begin{align}
\mathcal{H}
=
\frac{1}{2(1-t)}\left[
x^2 \frac{d}{dx}\Big(
x^{-(M+1)/2}{\rm tr}_a T_a
\Big)
\right]_{x\rightarrow \infty}
-
\frac{1}{2(1-t)}\left[
\frac{d}{dx}\Big(
x^{(M+1)/2}{\rm tr}_a T_a
\Big)
\right]_{x\rightarrow 0}
+
\bar{\mathcal{N}}
\end{align}
}

\noindent from which it follows that $[\mathcal{H},{\rm tr}_a T_a]=0$. Hence the eigenvectors of $\mathcal{H}$ may be found by studying the eigenvectors of the transfer matrix.

\subsection{Bethe Ansatz for the eigenvectors}

The eigenvectors of the transfer matrix ${\rm tr}_a T_a$ are obtained via the Ansatz

\begin{align}
|\Psi\rangle = B(y_1,t)\ldots B(y_N,t) |0\rangle,
\quad
\langle \Psi| = \langle 0| C(y_N,t)\ldots C(y_1,t)
\label{4-1-bethe}
\end{align}

\noindent where the variables $\{y_1,\ldots, y_N\}$ are assumed to obey the Bethe equations (\ref{Beteq}). For the present model, $a(y_i,y_j)=y_i-ty_j,\ \alpha(y_i) = y_i^{-(M+1)/2},\ \delta(y_i) = y_i^{(M+1)/2}$. Substituting these expressions into (\ref{Beteq}), the Bethe equations for the $q$-boson model read

\begin{align}
y_i^{M+1}
=
\prod_{\substack{j\not= i \\ j=1}}^{N}
\frac{ty_i-y_j}{y_i-ty_j}
\label{qb-eig}
\end{align}

\noindent for all $1\leq i \leq N$. As was the case in the previous chapter, we now proceed towards an explicit evaluation of the eigenvectors (\ref{4-1-bethe}) {\it without} assuming the Bethe equations (\ref{qb-eig}). 

\section{Charged $t$-fermions and related definitions}
\label{qb-fermi}


Before we begin our calculation of the Bethe eigenvectors, we digress briefly to discuss a $t$-deformed species of fermions and its corresponding Fock space $\mathcal{F}_{\psi}(t)$. These fermions were introduced in \cite{jin1}, and appeared in a slightly modified form in \cite{jin2}. The material of this section is mostly taken from \cite{fw2}, and will be necessary when we map the Bethe eigenvectors to elements of $\mathcal{F}_{\psi}(t)$.

\subsection{Charged $t$-fermions}

Following \cite{jin2}, consider two infinite sets $\{\psi_m(t)\}_{m \in \mathbb{Z}}$ and $\{\psis_m(t)\}_{m \in \mathbb{Z}}$, where $m$ runs over all integers. The elements in these sets are called {\it charged $t$-fermions} and they satisfy the anticommutation relations

\begin{align}
\psi_m \psi_n + \psi_n \psi_m &= 
t \psi_{(m+1)} \psi_{(n-1)} + t \psi_{(n+1)} \psi_{(m-1)}
\label{a-c1}
\\
\psis_m \psis_n + \psis_n \psis_m &= 
t \psis_{(m-1)} \psis_{(n+1)} + t \psis_{(n-1)} \psis_{(m+1)}
\label{a-c2}
\\
\psi_m \psis_n + \psis_n \psi_m &= 
t \psi_{(m-1)} \psis_{(n-1)} + t \psis_{(n+1)} \psi_{(m+1)} 
+ 
(1-t)^2 \delta_{m,n}
\label{a-c3}
\end{align}

\noindent for all $m,n \in \mathbb{Z}$.\footnote{To reduce notational complexity, we will often abbreviate $\psi_m(t) = \psi_m,\psis_m(t) = \psis_m$ throughout this chapter. This notation is not to be confused with the charged fermions of chapter 1, and the reader should assume that all fermions appearing in this chapter obey the deformed anticommutation relations (\ref{a-c1})--(\ref{a-c3}).} In these equations $t \in \complex$ plays the role of a deformation parameter, and for simplicity we will always assume that $|t|<1$. The charged free fermions of section 1.1 are recovered as the $t=0$ specialization of equations (\ref{a-c1})--(\ref{a-c3}).

\subsection{Clifford algebra $Cl_{\psi}(t)$ and identities}

The Clifford algebra $Cl_{\psi}(t)$ is the associative algebra generated by 1 and the charged $t$-fermions $\{\psi_m(t)\}_{m\in\mathbb{Z}}$ and $\{\psis_m(t)\}_{m\in\mathbb{Z}}$, modulo the equations (\ref{a-c1})--(\ref{a-c3}). For the purposes of calculation, the $t$-deformed anticommutation relations can be rather cumbersome. For this reason, we will prove several identities which make the algebra $Cl_{\psi}(t)$ easier to handle.\footnote{These identities were originally proved in \cite{fw2}.}

\begin{lemma}
{\rm 
For all $m,n \in \mathbb{Z}$, we have

\begin{align}
\psis_m \psi_n
=
t \psi_{(n-1)} \psis_{(m-1)}
+
(t^2-1)
\sum_{i=0}^{\infty}
\psi_{(n+i)} \psis_{(m+i)} t^i
+
(1-t) \delta_{m,n}
\label{qb-lem1-0}
\end{align}
}
\end{lemma}

\begin{proof} 
Rearranging the anticommutation relation (\ref{a-c3}), we have

\begin{align}
\psis_m \psi_n
=
t \psi_{(n-1)} \psis_{(m-1)}
-
\psi_n \psis_m
+
t \psis_{(m+1)} \psi_{(n+1)}
+
(1-t)^2 \delta_{m,n}
\label{qb-lem1-1}
\end{align}

\noindent and repeating this rearrangement to replace the term $t\psis_{(m+1)} \psi_{(n+1)}$ in (\ref{qb-lem1-1}), we recover

\begin{align}
\psis_m \psi_n
&=
t \psi_{(n-1)} \psis_{(m-1)}
+
(t^2-1) \psi_n \psis_m
-
t \psi_{(n+1)} \psis_{(m+1)}
+
t^2 \psis_{(m+2)} \psi_{(n+2)}
\nonumber
\\
&
+
(1-t)^2(1+t) \delta_{m,n}
\end{align}

\noindent Iterating this substitution procedure infinitely, we arrive at the equation

\begin{align}
\psis_m \psi_n
&=
t \psi_{(n-1)} \psis_{(m-1)}
+
(t^2-1)
\sum_{i=0}^{\infty}
\psi_{(n+i)} \psis_{(m+i)} t^i
+
(1-t)^2
\sum_{i=0}^{\infty} t^i \delta_{m,n}
\end{align}

\noindent and the proof is achieved by the geometric series identity $\sum_{i=0}^{\infty} t^i = \frac{1}{1-t}$.
\end{proof}

\begin{lemma} 
{\rm 
For arbitrary $m,n \in \mathbb{Z}$, we have

\begin{align}
\psi_m \psis_n
=
t \psis_{(n+1)} \psi_{(m+1)}
+
(t^2-1)
\sum_{i=0}^{\infty}
\psis_{(n-i)} \psi_{(m-i)} t^i
+
(1-t) \delta_{m,n}
\label{qb-lem2-1}
\end{align}
}
\end{lemma}

\begin{proof} 
Analogous to the proof of lemma 1.
\end{proof} 

%
%
%
%
%
 
\begin{lemma}
{\rm 
For arbitrary $m \in \mathbb{Z}$ and $n \geq 0$, we propose the identity

\begin{align}
\psi_{(m-n)} \psi_{m}
+
(1-t)
\sum_{i=1}^{n}
\psi_{(m-n+i)} \psi_{(m-i)}
=
t
\psi_{(m+1)} \psi_{(m-n-1)}
\label{qb-lem3-1}
\end{align}
}
\end{lemma}

\begin{proof} 
Let $\mathcal{P}_n$ denote the proposition (\ref{qb-lem3-1}). Using the anticommutation relation (\ref{a-c1}) we obtain the equations

\begin{align}
&
\psi_m \psi_m
=
t \psi_{(m+1)} \psi_{(m-1)}
\\
&
\psi_{(m-1)} \psi_{m}
+
(1-t)
\psi_{m} \psi_{(m-1)}
=
t \psi_{(m+1)} \psi_{(m-2)}
\end{align}

\noindent which prove that $\mathcal{P}_0$ and $\mathcal{P}_1$ are true. For $n \geq 2$, we rearrange the left hand side of the proposition $\mathcal{P}_{n}$ to give 

\begin{align}
&
\psi_{(m-n)} \psi_{m}
+
(1-t)
\sum_{i=1}^{n}
\psi_{(m-n+i)} \psi_{(m-i)}
\label{qb-lem3-2}
\\
=
&
\psi_{(m-n)} \psi_{m}
-
t \psi_{(m-n+1)} \psi_{(m-1)}
+
(1-t) \psi_{m} \psi_{(m-n)}
\nonumber
\\
+
&
\left(
\psi_{(m-n+1)} \psi_{(m-1)}
+
(1-t)
\sum_{i=1}^{n-2}
\psi_{(m-n+1+i)} \psi_{(m-1-i)}
\right)
\nonumber
\end{align}

\noindent Assuming that the proposition $\mathcal{P}_{n-2}$ is true, the parenthesized term in (\ref{qb-lem3-2}) is equal to $t \psi_{m} \psi_{(m-n)}$, and we recover

\begin{align}
&
\psi_{(m-n)} \psi_{m}
+
(1-t)
\sum_{i=1}^{n}
\psi_{(m-n+i)} \psi_{(m-i)}
\label{qb-lem3-3}
\\
=
&
\psi_{(m-n)} \psi_{m}
-
t \psi_{(m-n+1)} \psi_{(m-1)}
+
\psi_{m} \psi_{(m-n)}
\nonumber
\end{align}

\noindent Finally, applying the anticommutation relation (\ref{a-c1}) to the right hand side of (\ref{qb-lem3-3}), we prove that $\mathcal{P}_n$ is true. Therefore $\mathcal{P}_{n-2}$ true $\implies \mathcal{P}_n$ true, and by induction the proposition (\ref{qb-lem3-1}) holds for all $n \geq 0$.
\end{proof}

\subsection{Fock representations of $Cl_{\psi}(t)$}

As we did in section 1.1, we introduce a vacuum vector $|0\rangle$ and dual vacuum vector $\langle 0|$. We define actions of $Cl_{\psi}(t)$ on these vacuum states by setting

\begin{align}
\psi_m(t) |0\rangle = \psis_n(t) |0\rangle = 0,
\quad
\langle 0|\psis_m(t) = \langle 0|\psi_n(t) = 0
\label{qb-anni}
\end{align}

\noindent for all integers $m < 0, n \geq 0$. The $t$-deformed Fock space $\mathcal{F}_{\psi}(t)$ and its dual $\mathcal{F}^{*}_{\psi}(t)$ are the vector spaces generated linearly by the action of $Cl_{\psi}(t)$ on $|0\rangle$ and $\langle 0|$, respectively.

\begin{lemma}
{\rm 
For all $l \geq 1$ we define the charged vacuum states

\begin{align}
|-l\rangle
=
\psis_{-l}(t)\ldots\psis_{-1}(t)|0\rangle,
\quad
\langle -l|
=
\langle 0|\psi_{-1}(t)\ldots\psi_{-l}(t)
\end{align}

\noindent and propose the identities

\begin{align}
\psi_{m}(t) |-l\rangle
=
\left\{
\begin{array}{ll}
0, & {m < -l}
\\
|-l+1\rangle, & m =-l
\end{array}
\right.
\quad\ {\rm and}\ \quad
\langle -l| \psis_m(t)
=
\left\{
\begin{array}{ll}
0, & {m < -l}
\\
\langle -l+1|, & m=-l
\end{array}
\right.
\label{qb-lem4-1}
\end{align}
}
\end{lemma}

\begin{proof} 
We prove only the first of the propositions in (\ref{qb-lem4-1}), as the proof of the second is completely analogous. Let us denote this first proposition by $\mathcal{P}_l$. In the case $m < -1$ we can use the identity (\ref{qb-lem2-1}) and the annihilation properties (\ref{qb-anni}) to show that $\psi_m |-1\rangle = \psi_m \psis_{-1}|0\rangle = 0$. Furthermore, when $m=-1$ we use the anticommutation relation (\ref{a-c3}) to obtain

\begin{align}
\psi_{-1} \psis_{-1}|0\rangle
&=
\Big\{ 
(1-t)^2 + t\psi_{-2} \psis_{-2} + t \psis_0 \psi_0
\Big\}
|0\rangle
\nonumber
\\
&
=
\Big\{
(1+t)(1-t)^2 +t\psi_{-2} \psis_{-2} +t^2\psi_{-1} \psis_{-1} + t^2 \psis_1 \psi_1
\Big\}
|0\rangle
\label{curious}
\end{align}

\noindent and by application of (\ref{qb-lem2-1}) we have $\psi_{-2} \psis_{-2} |0\rangle = (1-t) |0\rangle$, while by (\ref{qb-lem1-0}) we see that $\psis_1 \psi_1 |0\rangle =(1-t)|0\rangle$. Substituting these results into (\ref{curious}), we obtain 

\begin{align}
\psi_{-1} \psis_{-1}|0\rangle
=
t^2\psi_{-1} \psis_{-1} |0\rangle
+
(1-t^2)|0\rangle
\end{align}

\noindent and therefore $\psi_{-1}|-1\rangle = |0\rangle$. Hence we have shown that $\mathcal{P}_1$ is true. Now assume that $\mathcal{P}_l$ is true for some $l\geq 1$. In the case $m < -(l+1)$ we use the identity (\ref{qb-lem2-1}) to write

\begin{align}
\psi_m |-l-1\rangle
&=
\psi_m \psis_{-(l+1)} |-l\rangle
\\
&
=
\Big\{
t\psis_{-l} \psi_{(m+1)}
+
(t^2-1)
\sum_{i=0}^{\infty}
\psis_{-(i+l+1)} \psi_{(m-i)}
t^i
\Big\}
|-l\rangle
=
0
\nonumber
\end{align}

\noindent where every term on the right hand side vanishes, because $\mathcal{P}_l$ holds. In the case $m=-(l+1)$ we again use identity (\ref{qb-lem2-1}) to write

\begin{align}
\psi_{-(l+1)} \psis_{-(l+1)} |-l\rangle
&
=
\Big\{
t \psis_{-l} \psi_{-l}
+
(t^2-1)
\sum_{i=0}^{\infty}
\psis_{-(i+l+1)} \psi_{-(i+l+1)}
t^i
+ 
(1-t)
\Big\}
|-l\rangle
\nonumber
\\
&=
|-l\rangle
\end{align}

\noindent where, again, the final equality holds because $\mathcal{P}_l$ is true. This establishes the identity $\psi_{-(l+1)} |-l-1\rangle = |-l\rangle$. Hence $\mathcal{P}_l$ true $\implies \mathcal{P}_{l+1}$ true, and the proof of (\ref{qb-lem4-1}) is complete by induction.
\end{proof}

\subsection{Partitions}

In direct analogy with section 1.1, we now identify elements of the deformed Fock spaces $\mathcal{F}_{\psi}(t)$ and $\mathcal{F}^{*}_{\psi}(t)$ with partitions. The correspondence is essentially the same as equation (\ref{cfermi-part}), except that all charged fermions are replaced with their $t$-deformed counterparts. Explicitly, we write 

\begin{align}  
\psi_{m_1}(t)\ldots\psi_{m_l}(t)|-l\rangle
=
|\mu_1,\ldots,\mu_l),
\quad
\langle -l| \psis_{m_l}(t)\ldots\psis_{m_1}(t)
=
(\mu_1,\ldots,\mu_l|
\end{align}

\noindent where $\mu_i = m_i+i$ for all $1\leq i \leq l$. The following result is an extension of (\ref{part-orth}), and evaluates the form $\langle,\rangle$ between $t$-deformed partitions.

\begin{lemma}
{\rm
Let $\mu$ be an arbitrary partition having $p_i(\mu) \geq 0$ parts of size $i$, for all $i \geq 1$. We associate to this partition a function $b_{\mu}(t)$, defined as

\begin{align}
b_{\mu}(t)
=
\prod_{i=1}^{\infty}
\left(
\prod_{j=1}^{p_i(\mu)}
(1-t^j)
\right)
=
\prod_{i=1}^{\infty}
[p_i(\mu)]_t!
\label{bmu}
\end{align} 

\noindent This product is actually finite, since there must exist an $I \geq 1$ such that $p_i(\mu) = 0$ for all $i>I$. Let $(\mu| \in \mathcal{F}_{\psi}^{*}(t)$ and $|\nu) \in \mathcal{F}_{\psi}(t)$ be two arbitrary partitions, given by

\begin{align}
(\mu| = \langle -l |\psis_{m_l}(t) \ldots \psis_{m_1}(t),
\quad
|\nu) = \psi_{n_1}(t)\ldots\psi_{n_k}(t)|-k\rangle
\end{align}

\noindent We claim that

\begin{align}
\Big\langle (\mu|, |\nu) \Big\rangle
=
\langle -l |\psis_{m_l}(t)\ldots \psis_{m_1}(t)
\psi_{n_1}(t)\ldots \psi_{n_k}(t) |-k\rangle
=
b_{\mu}(t) \delta_{\mu,\nu}
\label{innerproduct}
\end{align}

\noindent where $\delta_{\mu,\nu} =1$ if $\mu = \nu$, and $\delta_{\mu,\nu}=0$ if $\mu$ and $\nu$ are different.
}
\end{lemma}

\begin{proof}

We take the proof from \cite{fw2}. Using the identity (\ref{qb-lem1-0}) and the annihilation properties (\ref{qb-anni}) of the charged $t$-fermions, it follows that $\langle(\mu|,|\nu)\rangle=0$ if $m_1\not=n_1$. In the case when $m_1=n_1$, we assume that $\{m_1,\ldots,m_s\}$ are nearest neighbours for some $1 \leq s \leq l$.\footnote{This is equivalent to assuming that the first $s$ parts of $\mu$ have the same size.} That is, we fix

\begin{align} 
m_i = m_{(i+1)}+1, \ {\rm for\ all}\ 1 \leq i \leq s-1
\end{align}

\noindent but take $m_{s} > m_{(s+1)}+1$. Commuting the central pair of $t$-fermions $\psis_{m_1} \psi_{m_1}$ using (\ref{qb-lem1-0}) and annihilating terms with (\ref{qb-anni}), we obtain 

\begin{align}
\Big\langle (\mu|, |\nu) \Big\rangle
&=
(1-t) 
\langle -l| \psis_{m_l} \ldots \psis_{m_2}
\psi_{n_2} \ldots \psi_{n_k} |-k\rangle
\label{qb-lem5-1}
\\ 
&+
t 
\langle -l| \psis_{m_l} \ldots \psis_{m_2} \psi_{m_2}
\psis_{m_2} \psi_{n_2} \ldots \psi_{n_k} |-k\rangle
\nonumber
\end{align}

\noindent where we have recalled that $m_2 = m_1 -1$. Iterating this calculation on the second term on the right hand side of (\ref{qb-lem5-1}), we ultimately find

\begin{align}
\Big\langle (\mu|, |\nu)\Big\rangle
=
(1-t^s)
\langle -l| \psis_{m_l} \ldots \psis_{m_2}
\psi_{n_2} \ldots \psi_{n_k} |-k\rangle
\end{align}

\noindent and we have reduced, by two, the number of fermions appearing in the expectation value. Repeating this overall procedure, we find that $\langle(\mu|,|\nu)\rangle=0$ if $m_i \not= n_i$ for any $1\leq i \leq s$. In the case when $m_i = n_i$ for all $1 \leq i \leq s$, we obtain 

\begin{align}
\Big\langle ( \mu|,|\nu )\Big\rangle
=
\prod_{i=1}^{s} 
(1-t^i)
\langle -l| \psis_{m_l}\ldots\psis_{m_{(s+1)}}
\psi_{n_{(s+1)}}\ldots \psi_{n_k} |-k\rangle
\end{align}

\noindent and we have acquired a factor of $\prod_{i=1}^{s} (1-t^i)$ when the first $s$ parts of $\mu$ have the same size, as desired. Finally we can see that $\langle (\mu|,|\nu)\rangle = 0$ unless $l=k$ and $m_i = n_i$ for all $1\leq i \leq l$, or equivalently, $\mu = \nu$. If this condition holds, we clearly have $\langle (\mu|,|\nu)\rangle = b_{\mu}(t)$.

\end{proof}

\subsection{$t$-deformed Heisenberg algebra}

In this subsection, once again following \cite{jin2}, we give $t$-deformed analogues of the operators (\ref{aheisen1}). We define 

\begin{align}
H_m(t) 
=
\frac{1}{1-t}
\sum_{i\in \mathbb{Z}}
\psi_i(t) \psis_{(i+m)}(t)
\label{t-heis-def1}
\end{align}

\noindent for all integers $m > 0$, and

\begin{align}
H_m(t)
=
\frac{1}{(1-t)(1-t^{|m|})}
\sum_{i\in \mathbb{Z}}
\psi_i(t) \psis_{(i+m)}(t)
\label{t-heis-def2} 
\end{align}

\noindent for all integers $m<0$. Using the $t$-anticommutation relations (\ref{a-c1})--(\ref{a-c3}), it is possible to show that these operators obey the commutation relation\footnote{See section III of \cite{jin2} for a detailed proof of (\ref{t-heis}).} 

\begin{align}
[H_m(t),H_n(t)]
=
\frac{m}{1-t^{|m|}}
\delta_{m+n,0}
\label{t-heis}
\end{align}

\noindent for all $m,n \in \mathbb{Z}^{\times}$. Extending the identities (\ref{aheisen3}), we also have the commutation relations

\begin{align}
[H_m(t),\psi_n(t)]
=
\psi_{n-m}(t),
\quad
[H_m(t),\psis_n(t)]
=
-\psis_{m+n}(t)
\label{t-heis-2}
\end{align}

\noindent Lastly, we state the annihilation identities

\begin{align}
H_m(t) |-l\rangle =0,
\quad
\langle -l | H_{-m}(t) = 0
\label{t-heis-anni}
\end{align}

\noindent which are true for all $l \geq 0$ and $m \geq 1$. The relations (\ref{t-heis-anni}) are proved by using the definitions (\ref{t-heis-def1}) and (\ref{t-heis-def2}), the annihilation properties (\ref{qb-anni}), as well as the algebraic relations (\ref{a-c1})--(\ref{a-c3}). For brevity, we omit this proof.

\subsection{$t$-deformed half-vertex operators}

We introduce the Hamiltonians

\begin{align}
H_{\pm}(x,t)
=
\sum_{n=1}^{\infty}
\frac{1-t^n}{n}
x^{n}
H_{\pm n}(t)
\end{align}

\noindent where $x$ is an indeterminate. We also define $t$-analogues of the generating functions (\ref{KPexpI9}), given by

\begin{align}
\Psi(k) = \sum_{i\in\mathbb{Z}} \psi_i(t)k^i,
\quad
\Psis(k) = \sum_{i\in\mathbb{Z}} \psis_i(t) k^{-i}
\end{align}

\noindent Using these definitions and the commutation relations (\ref{t-heis-2}), we obtain

{\small
\begin{align}
[H_{+}(x,t),\Psi(k)]
&=
\sum_{n=1}^{\infty}
\sum_{i \in \mathbb{Z}}
[H_n(t),\psi_i(t)]
\frac{1-t^n}{n} x^{n}
k^i
=
\Psi(k)
\sum_{n=1}^{\infty}
\frac{1-t^n}{n} (xk)^n
\\
[\Psis(k),H_{-}(x,t)]
&=
\sum_{n=1}^{\infty}
\sum_{i \in \mathbb{Z}}
[\psis_i(t),H_{-n}(t)]
\frac{1-t^n}{n} x^n
k^{-i}
=
\Psis(k)
\sum_{n=1}^{\infty}
\frac{1-t^n}{n} \left(\frac{x}{k}\right)^n
\end{align} 
}

\noindent which, in turn, imply that

\begin{align}
e^{H_{+}(x,t)} \Psi(k)
&=
\frac{1-txk}{1-xk}
\Psi(k) e^{H_{+}(x,t)}
\label{t-ev1}
\\
\Psis(k) e^{H_{-}(x,t)}
&=
\frac{k-tx}{k-x}
e^{H_{-}(x,t)} \Psis(k)
\label{t-ev2}
\end{align}

\noindent where we have used the formal power series identity 

\begin{align}
\sum_{n=1}^{\infty}\frac{1-t^n}{n}z^n = \log\left(\frac{1-tz}{1-z}\right)
\end{align}

\noindent Defining $\Gamma_{\pm}(x,t) = e^{H_{\pm}(x,t)}$ and eliminating the generating parameter $k$ from the equations (\ref{t-ev1}) and (\ref{t-ev2}), we obtain 


\begin{align}
\Gamma_{+}(x,t) \psi_i
&=
\left( \psi_i +(1-t) \sum_{n=1}^{\infty} \psi_{(i-n)} x^{n} \right) 
\Gamma_{+}(x,t)
\label{t-ev3}
\\
\psis_i \Gamma_{-}(x,t)
&=
\Gamma_{-}(x,t)
\left(\psis_i + (1-t) \sum_{n=1}^{\infty} \psis_{(i-n)} x^{n} \right)
\label{t-ev4}
\end{align}

\noindent for all $i \in \mathbb{Z}$. The operators $\Gamma_{\pm}(x,t)$ are $t$-generalizations of $\Gamma_{\pm}(x)$ as defined in subsection 3.4.1. They each constitute one half of the $t$-fermion vertex operators in \cite{jin2}, and for this reason we call them {\it $t$-deformed half-vertex operators.} 

Equations (\ref{t-ev3}) and (\ref{t-ev4}) play an important role when we consider the $q$-boson model on an infinite lattice, in section 4.5. Essentially they extend the equations (\ref{p-lem6-4}) from the last chapter to arbitrary $t$ values. 

\section{Calculation of Bethe eigenvectors}
\label{qb-BE}

\subsection{The maps $\mathcal{M}_{\psi}(t)$ and $\mathcal{M}_{\psi}^{*}(t)$}

We begin by defining the maps $\mathcal{M}_{\psi}(t)$ and $\mathcal{M}_{\psi}^{*}(t)$, which take the basis vectors of $\mathcal{V}$ and $\mathcal{V}^{*}$ to partitions in the deformed Fock spaces $\mathcal{F}_{\psi}(t)$ and $\mathcal{F}^{*}_{\psi}(t)$, respectively. These maps are the natural $t$-extension of those defined in section 3.2.
 
\begin{definition}
{\rm 
Let $|n\rangle = |n_0\rangle_0 \otimes \cdots \otimes |n_M\rangle_M$ and $\langle n| = \langle n_0|_0 \otimes \cdots \otimes \langle n_M|_M$ be basis elements of $\mathcal{V}$ and $\mathcal{V}^{*}$, respectively, and define 

\begin{align}
\Sigma_0 = \sum_{j=0}^{M} n_j
\end{align} 

\noindent From this, let $|\nu) = |\nu_1,\ldots,\nu_{\Sigma_0})$ and $( \nu|=( \nu_1,\ldots,\nu_{\Sigma_0}|$ be the partitions in $\mathcal{F}_{\psi}(t)$ and $\mathcal{F}^{*}_{\psi}(t)$ with $n_i$ parts equal to $i$ for all $0 \leq i \leq M$. That is, we let

\begin{align}
|\nu) &= |M^{n_M},\ldots,1^{n_1},0^{n_0}) = |M^{n_M},\ldots,1^{n_1}) 
\\
(\nu| &= (M^{n_M},\ldots,1^{n_1},0^{n_0}| =
(M^{n_M},\ldots,1^{n_1}|
\end{align} 

\noindent We define linear maps $\mathcal{M}_{\psi}(t):\mathcal{V} \rightarrow \mathcal{F}_{\psi}(t)$ and $\mathcal{M}_{\psi}^{*}(t):\mathcal{V}^{*} \rightarrow \mathcal{F}^{*}_{\psi}(t)$ whose actions are given by

\begin{align}
\mathcal{M}_{\psi}(t) |n\rangle
=
\frac{1}{b_{\nu}(t)}
|\nu),
\quad
\langle n|\mathcal{M}_{\psi}^{*}(t) 
=
\frac{1}{b_{\nu}(t)}
( \nu|
\label{qb-def1-1}
\end{align}

\noindent where $b_{\nu}(t)$ denotes the factor (\ref{bmu}) assigned to the partition $\nu$. These maps are motivated by the orthogonality relation (\ref{innerproduct}), from which we see that

\begin{align}
\Big\langle
\langle m| \mathcal{M}_{\psi}^{*}(t),
\mathcal{M}_{\psi}(t) |n\rangle
\Big\rangle
=
\frac{\langle (\mu|,|\nu) \rangle}{b_{\mu}(t) b_{\nu}(t)}
=
\frac{\delta_{\mu,\nu}}{b_{\mu}(t)}
=
\prod_{i=1}^{M}\frac{\delta_{m_i,n_i}}{[m_i]_t!}
\end{align}

\noindent In other words, the maps (\ref{qb-def1-1}) preserve the inner product (\ref{qb-ip1}) on all sites $\{1,\ldots, M\}$, but project out all information from the $0^{\rm th}$ site.


}
\end{definition}

\subsection{Calculation of $\mathbb{B}(x,t) | n \rangle$}

\begin{lemma}
{\rm Define $\mathbb{B}(x,t)=x^{\frac{M}{2}}B(x,t)$ and let $|n\rangle = |n_0\rangle_0 \otimes \cdots \otimes |n_M\rangle_M$ be an arbitrary basis vector of $\mathcal{V}$. The action of $\mathbb{B}(x,t)$ on $|n\rangle$ is given by

\begin{align}
\mathbb{B}(x,t) |n\rangle
=
\sum_{|m\rangle \triangleright |n\rangle}
\prod_{i=1}^{M}
x^{i(m_i-n_i)}
\Big(
1
-
\delta_{m_i,n_i+1}
t^{m_i}
\Big)
|m\rangle
\label{qb-lem6-0}
\end{align}

\noindent where the sum is over all basis vectors $|m\rangle = |m_0\rangle_0 \otimes \cdots \otimes |m_M\rangle_M$ which are admissible to $|n\rangle$. In the case $t=0\ (q\rightarrow \infty)$, this equation clearly reduces to lemma \ref{3-lem2} from chapter 3. Furthermore, when acting on basis vectors $|\tilde{n}\rangle \in \tilde{\mathcal{V}}$ and with $t=-1\ (q = i)$, the weighting factor in (\ref{qb-lem6-0}) vanishes if $m_i = 2,n_i=1$ for any $1\leq i \leq M$. In this case, the right hand side of (\ref{qb-lem6-0}) becomes a sum over admissible basis vectors $|\tilde{m}\rangle \in \tilde{\mathcal{V}}$, and we recover lemma \ref{3-lem7} from chapter 3.

}
\end{lemma}

\begin{proof}
We proceed along similar lines to the proof of lemma 2 in chapter 3. Let $\langle m| = \langle m_0|_0 \otimes \cdots \otimes \langle m_M|_M$ be an arbitrary basis vector of $\mathcal{V}^{*}$. We write the $B$-operator as a contraction on the auxiliary space $\mathcal{V}_a$, as follows

\begin{align}
B(x,t)
=
\left(
\begin{array}{cc}
1 & 0
\end{array}
\right)_a
\left(
\begin{array}{cc}
A(x,t) & B(x,t)
\\
C(x,t) & D(x,t)
\end{array}
\right)_a
\left(
\begin{array}{c}
0
\\
1
\end{array}
\right)_a
=
\uparrow_a^{*}
T_a(x,t)
\downarrow_a
\end{align}

\noindent which leads to the equation

\begin{align}
\langle m| B(x,t) |n\rangle
&=
\uparrow_a^{*}
\otimes
\langle m|
T_a(x,t)
|n\rangle
\otimes
\downarrow_a
=
\uparrow_a^{*}
\otimes
\langle m|
L_{aM}(x,t)
\ldots
L_{a0}(x,t)
|n\rangle
\otimes
\downarrow_a
\end{align}

\noindent By commuting operators and vectors which reside in different spaces we obtain

\begin{align}
\langle m | B(x,t) | n \rangle
=
\uparrow^{*}
L^{(M)}(x,t)
\ldots
L^{(0)}(x,t)
\downarrow
\end{align}

\noindent where we have dropped the redundant subscripts $a$, and have defined the modified $L$-matrices

\begin{align}
L^{(i)}(x,t)
=
\left(
\begin{array}{cc}
\langle m_i|_i x^{-\frac{1}{2}} |n_i\rangle_i
& 
\langle m_i|_i (1-t)^{\frac{1}{2}} \bd_i |n_i\rangle_i
\\
\langle m_i|_i (1-t)^{\frac{1}{2}} b_i |n_i\rangle_i
&
\langle m_i|_i x^{\frac{1}{2}} |n_i\rangle_i
\end{array}
\right)
\end{align}

\noindent for all $0 \leq i \leq M$. Calculating the entries within these matrices explicitly yields

\begin{align}
\label{qb-lem6-1}
[m_i]_t!
L^{(i)}(x,t)
=
\left\{
\begin{array}{rl}
\left(
\begin{array}{cc}
x^{-\frac{1}{2}} & 0
\\
0 & x^{\frac{1}{2}}
\end{array}
\right)
&
\quad m_i = n_i
\\
\\
(1-t^{m_i})
\left(
\begin{array}{cc}
0 & 1
\\
0 & 0
\end{array}
\right)
&
\quad m_i = n_i+1
\\
\\
\left(
\begin{array}{cc}
0 & 0
\\
1 & 0
\end{array}
\right)
&
\quad m_i+1 = n_i
\\
\\
\left(
\begin{array}{cc}
0 & 0
\\
0 & 0
\end{array}
\right)
& 
\quad {\rm otherwise}
\end{array}
\right.
\end{align}

\noindent for all $1\leq i\leq M$, as well as

\begin{align}
\label{qb-lem6-2}
\frac{L^{(0)}(x,t)}
{
[m_0]_t!
}
=
\left\{
\begin{array}{rl}
\left(
\begin{array}{cc}
x^{-\frac{1}{2}} & 0
\\
0 & x^{\frac{1}{2}}
\end{array}
\right)
&
\quad m_0 = n_0
\\
\\
\left(
\begin{array}{cc}
0 & 1
\\
0 & 0
\end{array}
\right)
&
\quad m_0 = n_0+1
\\
\\
(1-t^{n_0})
\left(
\begin{array}{cc}
0 & 0
\\
1 & 0
\end{array}
\right)
&
\quad m_0+1 = n_0
\\
\\
\left(
\begin{array}{cc}
0 & 0
\\
0 & 0
\end{array}
\right)
& 
\quad {\rm otherwise}
\end{array}
\right.
\end{align}

\noindent Using the expressions (\ref{qb-lem6-1}) and (\ref{qb-lem6-2}) for $L^{(i)}(x,t)$ and $L^{(0)}(x,t)$ respectively, we see that

\begin{align}
\langle m|B(x,t)|n\rangle
=
\uparrow^{*}
L^{(M)}(x,t)
\ldots
L^{(0)}(x,t)
\downarrow
=
0
\label{qb-lem6-3}
\end{align}

\noindent when $|m\rangle \notad |n\rangle$. In the case $|m\rangle \triangleright |n\rangle$, we follow essentially the same argument that was used to calculate $\uparrow^{*} L^{(M)}(x)\ldots L^{(0)}(x) \downarrow$ when proving lemma 2 in chapter 3. The calculation of $\uparrow^{*} L^{(M)}(x,t)\ldots L^{(0)}(x,t)\downarrow$ deviates only up to overall factors depending on $t$, and we easily surmise that 

\begin{align}
\uparrow^{*}
L^{(M)}(x,t)
\ldots
L^{(0)}(x,t)
&
\downarrow
=
\label{qb-lem6-4}
\frac{
x^{-\frac{M}{2}}[m_0]_t!  
\displaystyle{\prod_{i=1}^{M}}
x^{i(m_i-n_i)}
\Big(
1-\delta_{m_i,n_i+1} t^{m_i}
\Big)
}
{\displaystyle{\prod_{i=1}^{M}} [m_i]_t!}
\end{align}

\noindent when $|m\rangle \triangleright |n\rangle$. Combining the equations (\ref{qb-lem6-3}) and (\ref{qb-lem6-4}) into a single case, we have

\begin{align}
x^{\frac{M}{2}}
\langle m| B(x,t) |n\rangle
=
\left\{
\begin{array}{ll}
\frac{
[m_0]_t!
\displaystyle{ \prod_{i=1}^{M}}
x^{i(m_i-n_i)}
\Big(
1-\delta_{m_i,n_i+1} t^{m_i}
\Big)
}
{\displaystyle{\prod_{i=1}^{M}} [m_i]_t!},
&
\quad
|m\rangle \triangleright |n\rangle
\\
\\
0, & \quad {\rm otherwise}
\end{array}
\right.
\end{align}

\noindent The result (\ref{qb-lem6-0}) follows from the orthogonality (\ref{qb-ip1}) of the basis vectors of $\mathcal{V}$, and from the definition of $\mathbb{B}(x,t)$.
\end{proof}

%
%

\subsection{Calculation of $\langle n| \mathbb{C}(x,t)$}

\begin{lemma} 
{\rm Define $\mathbb{C}(x,t) = x^{\frac{M}{2}} C(1/x,t)$ and let $\langle n| = \langle n_0|_0 \otimes \cdots \otimes \langle n_M|_M$ be an arbitrary basis vector of $\mathcal{V}^{*}$. The action of $\mathbb{C}(x,t)$ on $\langle n|$ is given by

\begin{align}
\langle n|\mathbb{C}(x,t)
=
\sum_{\langle n| \triangleleft \langle m|}
\prod_{i=1}^{M}
x^{i(m_i-n_i)}
\Big(
1-\delta_{m_i,n_i+1}t^{m_i}
\Big)
\langle m|
\label{qb-lem7-0}
\end{align}

\noindent where the sum is over all basis vectors $\langle m| = \langle m_0|_0 \otimes \cdots \otimes \langle m_M|_M$ which are admissible to $\langle n|$. Once again, we notice that this result reduces to lemma \ref{3-lem3} of chapter 3 in the case $t=0\ (q\rightarrow \infty)$, and to lemma \ref{3-lem8} of chapter 3 in the case $t=-1\ (q = i)$.
}
\end{lemma}

\begin{proof}
A simple modification of the proof of lemma 6.
\end{proof}

\subsection{Calculation of $\mathcal{M}_{\psi}(t) \mathbb{B}(x,t) |n\rangle$ and $\langle n| \mathbb{C}(x,t) \mathcal{M}_{\psi}^{*}(t)$}

Throughout the rest of the chapter we will require the function $p_{\mu / \nu}(t)$,\footnote{To translate to the notation of chapter III in \cite{mac}, we remark that $p_{\mu/\nu}(t)=\varphi_{\mu / \nu}(t)$ and $p_{\nu/\mu}(t) = \psi_{\mu/\nu}(t)$.} which compares the part multiplicities of the partitions $\mu,\nu$ and returns
   
\begin{align}
p_{\mu / \nu}(t)
=
\prod_{i=1}^{\infty}
\Big(
1
-
\delta_{p_i(\mu),p_i(\nu)+1}
t^{p_i(\mu)}
\Big)
\label{pfunct}
\end{align}

\noindent Let $|n\rangle$ and $\langle n|$ be arbitrary basis vectors of $\mathcal{V}$ and $\mathcal{V}^{*}$ respectively, and let $|\nu)$ and $(\nu|$ be their corresponding partitions, given by equation (\ref{qb-def1-1}). We also fix $l = \ell(\nu)$. Using the definition of the maps (\ref{qb-def1-1}), the expressions (\ref{qb-lem6-0}) and (\ref{qb-lem7-0}) and the relationship between admissible basis vectors and interlacing partitions (lemma 1 of chapter 3), we obtain

\begin{align}
\mathcal{M}_{\psi}(t)
\mathbb{B}(x,t) |n\rangle
&=
\sum_{\nu \prec \mu \subseteq [l+1,M]}
x^{|\mu|-|\nu|}
\frac{p_{\mu/\nu}(t)}{b_{\mu}(t)}
|\mu)
\label{sum1}
\\
\langle n| \mathbb{C}(x,t) \mathcal{M}_{\psi}^{*}(t)
&=
\sum_{\nu \prec \mu \subseteq [l+1,M]}
x^{|\mu|-|\nu|}
\frac{p_{\mu/\nu}(t)}{b_{\mu}(t)}
(\mu|
\label{sum2}
\end{align}

\noindent Both sums are over all partitions $\mu$ which interlace with $\nu$, and whose Young diagrams are contained in the rectangle $[l+1,M]$. These equations may be written in the equivalent form

\begin{align}
\mathcal{M}_{\psi}(t)
\mathbb{B}(x,t) |n\rangle
&=
\sum_{\nu \prec \mu \subseteq [l+1,M]}
x^{|\mu|-|\nu|}
\frac{p_{\nu/\mu}(t)}{b_{\nu}(t)}
|\mu)
\label{sum3}
\\
\langle n| \mathbb{C}(x,t) \mathcal{M}_{\psi}^{*}(t)
&=
\sum_{\nu \prec \mu \subseteq [l+1,M]}
x^{|\mu|-|\nu|}
\frac{p_{\nu/\mu}(t)}{b_{\nu}(t)}
(\mu|
\label{sum4}
\end{align}

\noindent where the weighting factors in the sums (\ref{sum1}) and (\ref{sum2}) have been adjusted using the identity\footnote{See equation (5.12) in chapter III of \cite{mac}.}

\begin{align}
\frac{p_{\mu/\nu}(t)}{p_{\nu/\mu}(t)}
=
\frac{b_{\mu}(t)}{b_{\nu}(t)}
\end{align}

\subsection{Hall-Littlewood functions}

Let $\{x_1,\ldots,x_n\}$ be free variables, and $t$ an additional parameter. Following chapter III of \cite{mac}, the {\it Hall-Littlewood function} $P_{\mu}(\{x_1,\ldots,x_n\},t)$ associated to the partition $\mu$ is defined as

\begin{align}
&
P_{\mu}(\{x_1,\ldots,x_n\}, t)
=
\frac{1}{v_{\mu}(t)}
\sum_{\sigma \in S_n}
\sigma\left(
x_1^{\mu_1}\ldots x_n^{\mu_n} \prod_{1 \leq i < j \leq n} \frac{x_i-tx_j}{x_i-x_j}
\right)
\label{hl-def}
\end{align}

\noindent where the function $v_{\mu}(t)$ is given by

\begin{align}
v_{\mu}(t)
=
\prod_{i=1}^{\infty}
\prod_{j=1}^{p_i(\mu)}
\left(\frac{1-t^j}{1-t}\right)
\label{vmu-def}
\end{align}

\noindent The Hall-Littlewood function $P_{\mu}(\{x\},t)$ specializes to the Schur function $s_{\mu}\{x\}$ in the limit $t\rightarrow 0$. Also, for all strict partitions $\tilde{\mu}$ the function $P_{\tilde{\mu}}(\{x\},t)$ specializes to the Schur $Q$-function $2^{-\ell(\tilde{\mu})}Q_{\tilde{\mu}}\{x\}$ by setting $t=-1$. Hence the material of chapter 3 is recovered by suitable specializations of the results stated below.

For an arbitrary pair of partitions $\mu,\nu$ and indeterminates $x,t$ the single variable {\it skew Hall-Littlewood function} $P_{\mu / \nu}(x,t)$ is given by\footnote{This definition is consistent with equation $(5.14')$ in chapter III of \cite{mac}, if one replaces $p_{\nu/\mu}(t)$ with $\psi_{\mu/\nu}(t)$.}

\begin{align}
P_{\mu / \nu}(x,t)
=
\left\{
\begin{array}{ll}
x^{|\mu|-|\nu|}
p_{\nu / \mu}(t),
&
\mu \succ \nu
\\
\\
0,
&
{\rm otherwise}
\end{array}
\right.
\label{skew-hl}
\end{align}

\noindent In the case $\nu = \emptyset$ we have $P_{\mu/\nu}(x,t) = P_{\mu}(x,t)$, where $P_{\mu}(x,t)$ is the ordinary Hall-Littlewood function in a single variable $x$. The skew Hall-Littlewood function satisfies the identity\footnote{See equation $(5.5')$ in chapter III of \cite{mac}.}

\begin{align}
P_{\mu}(
\{x_1,\ldots,x_n\},t
)
=
\sum_{\nu \subseteq [n-1,\infty]}
P_{\mu / \nu}(x_n,t)
P_{\nu}(
\{x_1,\ldots,x_{n-1}\},t
)
\label{skew-hl-id}
\end{align}

\noindent where the sum is over all partitions $\nu$ with length $\ell(\nu) \leq n-1$, and $P_{\mu}(\{x_1,\ldots,x_n\},t)$ and $P_{\nu}(\{x_1,\ldots,x_{n-1}\},t)$ are Hall-Littlewood functions in $n$ and $n-1$ variables, respectively.

\subsection{Calculation of $\mathcal{M}_{\psi}(t) \mathbb{B}(x_1,t) \ldots \mathbb{B}(x_N,t) |0\rangle$}

Equipped with the necessary symmetric function identities, we are now able to calculate the $q$-boson model Bethe eigenvectors explicitly.

\begin{lemma}
{\rm Let $\{x_1,\ldots,x_N\}$ be a finite set of variables and $t$ an extra parameter. We claim that

\begin{align}
\mathcal{M}_{\psi}(t)
\mathbb{B}(x_1,t)
\ldots
\mathbb{B}(x_N,t)
|0\rangle
=
\sum_{\mu \subseteq [N,M]}
P_{\mu}(
\{x_1,\ldots,x_N\},t
)
|\mu)
\label{hl-result}
\end{align}

\noindent where $P_{\mu}(\{x_1,\ldots,x_N\},t)$ is the Hall-Littlewood function in $N$ variables (\ref{hl-def}), and the sum is over all partitions $\mu$ whose Young diagrams are contained in the rectangle $[N,M]$. This formula was originally proved in \cite{tsi}.
}
\end{lemma}

\begin{proof}

We begin by specializing equation (\ref{sum3}) to the case $|n\rangle = |0\rangle$, to obtain

\begin{align}
\mathcal{M}_{\psi}(t)
\mathbb{B}(x,t) |0\rangle
=
\sum_{\emptyset \prec \mu \subseteq [1,M]}
P_{\mu/\emptyset}(x,t) |\mu)
=
\sum_{\mu \subseteq [1,M]}
P_{\mu}(x,t) |\mu)
\label{hl-ind}
\end{align}

\noindent where we have used the equation (\ref{skew-hl}) for the skew Hall-Littlewood function, and the definitions $b_{\emptyset}(t) =1, \ell(\emptyset)=0$. We use equation (\ref{hl-ind}) as the basis for induction, and assume that

\begin{align}
\mathcal{M}_{\psi}(t)
\mathbb{B}(x_1,t)
\ldots
\mathbb{B}(x_{N-1},t)
|0\rangle
=
\sum_{\nu \subseteq [N-1,M]}
P_{\nu}
(
\{x_1,\ldots,x_{N-1}\},t
)
|\nu)
\end{align}

\noindent for some $N\geq 2$. In terms of the basis vectors of $\mathcal{V}$, this assumption is written as

\begin{align}
\mathbb{B}(x_1,t)
\ldots
\mathbb{B}(x_{N-1},t)
|0\rangle
=
\sum_{|n\rangle | \Sigma_0=N-1}
P_{\nu}(
\{x_1,\ldots,x_{N-1}\},t
)
b_{\nu}(t)
|n\rangle
\label{hl-ind2}
\end{align}

\noindent where the sum is over all basis vectors $|n\rangle$ whose occupation numbers satisfy the condition $\sum_{i=0}^{M}n_i = N-1$, and $\nu$ is the partition corresponding to each $|n\rangle$. Acting on (\ref{hl-ind2}) with the composition of operators $\mathcal{M}_{\psi}(t)\circ \mathbb{B}(x_N,t)$ and using the fact that the $B$-operators commute (\ref{qb-bbcc}), we obtain  

{\small
\begin{align}
\mathcal{M}_{\psi}(t)
\mathbb{B}(x_1,t)
\ldots
\mathbb{B}(x_N,t)
|0\rangle
=
\label{hl-ind3}
\sum_{\nu \subseteq [N-1,M] }
P_{\nu}
(
\{x_1,\ldots,x_{N-1}\},t
)
\sum_{\nu \prec \mu \subseteq [N,M]}
P_{\mu / \nu}(x_N,t)
|\mu)
\end{align}
}

\noindent Since $P_{\mu/\nu}(x_N,t)= 0$ if $\mu \not\succ \nu$, we may alter the sums appearing in (\ref{hl-ind3}), yielding

{\small
\begin{align}
\mathcal{M}_{\psi}(t)
\mathbb{B}(x_1,t)
\ldots
\mathbb{B}(x_N,t)
|0\rangle
&=
\sum_{\mu \subseteq [N,M]}
\sum_{\nu \subseteq [N-1,M]}
P_{\mu / \nu}(x_N,t)
P_{\nu}(
\{x_1,\ldots,x_{N-1}\},t
)
|\mu)
\nonumber
\\
&=
\sum_{\mu \subseteq [N,M]}
\sum_{\nu \subseteq [N-1,\infty]}
P_{\mu / \nu}(x_N,t)
P_{\nu}(
\{x_1,\ldots,x_{N-1}\},t
)
|\mu)
\end{align}
}

\noindent where the final equality holds since every part of $\mu$ is less than or equal to $M$, and therefore $P_{\mu/\nu}(x_N,t) = 0$ if any part of $\nu$ is greater than $M$. Using the identity (\ref{skew-hl-id}) we evaluate the sum over $\nu$ explicitly, producing the equation (\ref{hl-result}). Therefore by induction the result (\ref{hl-result}) must hold for arbitrary $N \geq 1$. 

\end{proof}

\subsection{Calculation of $\langle 0| \mathbb{C}(x_N,t)\ldots\mathbb{C}(x_1,t) \mathcal{M}_{\psi}^{*}(t)$}

By following essentially the same steps that were used in the previous subsection, we can also derive the expression

\begin{align} 
\langle 0| \mathbb{C}(x_N,t)\ldots\mathbb{C}(x_1,t)
\mathcal{M}_{\psi}^{*}(t)
=
\sum_{\mu \subseteq [N,M]}
P_{\mu}
(
\{x_1,\ldots,x_N\},t
)
(\mu|
\label{hl-result2}
\end{align}

\noindent for the dual Bethe eigenvectors. As before, this sum is taken over all partitions $\mu$ whose Young diagrams are contained in the rectangle $[N,M]$.

\section{Scalar product, weighted plane partitions}
\label{qb-bpp}

\subsection{Levels of paths within plane partitions}

\begin{definition}
{\rm 
Let $\pi$ be a plane partition. The element $\pi(i,j)$ is said to be at level $l$ if 

\begin{align}
\pi(i,j)
=
\cdots
=
\pi(i+l-1,j+l-1)
>
\pi(i+l,j+l)
\end{align}

\noindent for some $l \geq 1$. A {\it path at level $l$} is a set of connected elements in $\pi$ which have the same numerical value and the same level $l$. We let $p_l(\pi)$ denote the number of paths in $\pi$ at level $l$. This definition of levels within a plane partition originally appeared in \cite{vul2}.

\begin{figure}[H]

\begin{center}
\begin{minipage}{4.3in}

\setlength{\unitlength}{0.00025cm}
\begin{picture}(30000,33000)(-13000,-26000)

\path(-2000,-2000)(2000,-6000)
\path(0,0)(4000,-4000)
\path(2000,2000)(10000,-6000)
\path(4000,4000)(14000,-6000)
\path(6000,6000)(16000,-4000)
\path(8000,8000)(18000,-2000)

\path(-2000,-2000)(8000,8000)
\path(0000,-4000)(10000,6000)
\path(2000,-6000)(12000,4000)
\path(8000,-4000)(14000,2000)
\path(10000,-6000)(16000,0)
\path(14000,-6000)(18000,-2000)

\put(8000,6000){3}
\put(6000,4000){3}
\put(4000,2000){3}
\put(2000,0){3}
\put(0,-2000){1}

\put(2000,-4000){1}
\put(4000,-2000){1}
\put(6000,0){3}
\put(8000,2000){3}
\put(10000,4000){3}

\put(8000,-2000){3}
\put(10000,0){3}
\put(12000,2000){3}

\put(10000,-4000){2}
\put(12000,-2000){2}
\put(14000,0){2}

\put(14000,-4000){2}
\put(16000,-2000){2}


\blacken\path(8000,-8000)(6000,-10000)(8000,-12000)(10000,-10000)(8000,-8000)
\put(8000,-10000){\color{white}3}

\shade\path(6000,-12000)(8000,-14000)(10000,-12000)(12000,-14000)(8000,-18000)(4000,-14000)(6000,-12000)
\path(6000,-16000)(8000,-14000)(10000,-16000)
\put(6000,-14000){\color{white}3}
\put(8000,-16000){\color{white}3}
\put(10000,-14000){\color{white}3}

\path(0,-20000)(2000,-22000)
\path(2000,-18000)(8000,-24000)
\path(4000,-16000)(10000,-22000)
\path(12000,-20000)(10000,-18000)
\path(14000,-18000)(12000,-16000)
\path(0,-20000)(4000,-16000)
\path(2000,-22000)(6000,-18000)
\path(6000,-22000)(12000,-16000)
\path(8000,-24000)(14000,-18000)
\put(2000,-20000){3}
\put(4000,-18000){3}
\put(6000,-20000){3}
\put(8000,-22000){3}
\put(10000,-20000){3}
\put(12000,-18000){3}

\path(12000,-22000)(16000,-18000)
\path(14000,-24000)(20000,-18000)
\path(18000,-24000)(22000,-20000)
\path(12000,-22000)(14000,-24000)
\path(14000,-20000)(18000,-24000)
\path(16000,-18000)(20000,-22000)
\path(20000,-18000)(22000,-20000)
\put(14000,-22000){2}
\put(16000,-20000){2}
\put(18000,-22000){2}
\put(20000,-20000){2}

\shade\path(20000,-16000)(18000,-14000)(16000,-16000)(18000,-18000)(20000,-16000)
\put(18000,-16000){\color{white}2}

\path(-2000,-24000)(2000,-28000)
\path(0,-22000)(4000,-26000)
\path(4000,-22000)(6000,-24000)
\path(-2000,-24000)(0,-22000)
\path(0,-26000)(4000,-22000)
\path(2000,-28000)(6000,-24000)
\put(0,-24000){1}
\put(2000,-26000){1}
\put(4000,-24000){1}

\blacken\path(-10000,-8000)(-9000,-8000)(-9000,-7000)(-10000,-7000)(-10000,-8000)
\put(-8000,-8000){level 3}

\shade\path(-10000,-10000)(-9000,-10000)(-9000,-9000)(-10000,-9000)(-10000,-10000)
\put(-8000,-10000){level 2}

\path(-10000,-12000)(-9000,-12000)(-9000,-11000)(-10000,-11000)(-10000,-12000)
\put(-8000,-12000){level 1}

\end{picture}

\end{minipage}
\end{center}

\caption{Paths at various levels within a plane partition.}

\end{figure}
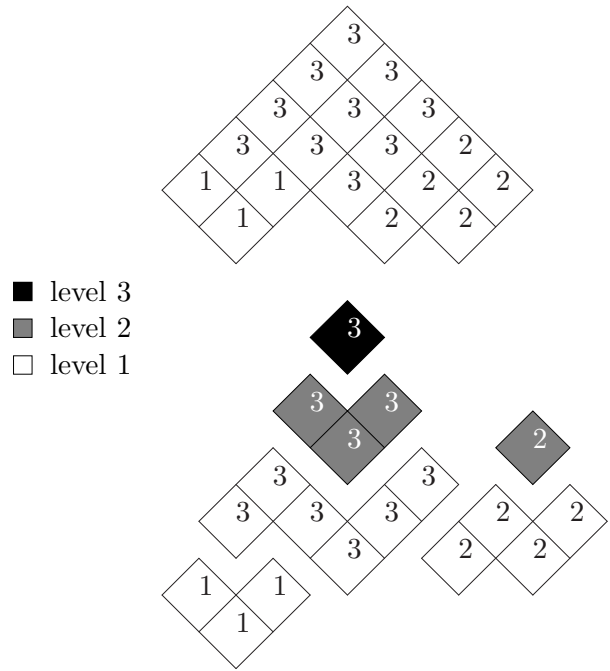
}
\end{definition}

\subsection{Path-weighted plane partitions}

In this subsection we assign a $t$-weighting to plane partitions, which was originally introduced in \cite{vul2}. In the limit $t=0$ the weighting becomes 1 for all plane partitions. In the limit $t=-1$ the weighting collapses to zero for all plane partitions which contain paths at level 2, or equivalently, for all plane partitions which are non-strict.

\begin{definition}
{\rm
Let $\pi$ be a plane partition living inside the box $[N,N,M]$, with diagonal slices $\{\emptyset = \pi_{-N} \prec \cdots \prec \pi_{-1} \prec \pi_{0} \succ \pi_{1} \succ \cdots \succ \pi_{N} = \emptyset\}$. We associate to this plane partition the weighting $A_{\pi}(\{x\},\{y\},t)$, given by

\begin{align}
A_{\pi}\Big(
\{x\},\{y\},t
\Big)
=
\prod_{i=1}^{N}
\Big(
1-t^i
\Big)^{p_i(\pi)}
x_i^{|\pi_{-i+1}|-|\pi_{-i}|}
y_i^{|\pi_{i-1}|-|\pi_{i}|}
\label{apit}
\end{align}

\noindent where $p_i(\pi)$ is the number of paths in $\pi$ at level $i$. Notice that $A_{\pi}(\{x\},\{y\},t)$ only takes contributions from paths whose levels are less than or equal to $N$, but in fact paths at greater levels cannot exist, since $\pi \subseteq [N,N,M]$.

}
\end{definition}

\begin{lemma}
{\rm Let $\pi$ be a plane partition as described in definition 3. Then

\begin{align}
\frac{1}{b_{\pi_{0}}(t)}
\prod_{i=1}^{N} 
p_{\pi_{-i+1}/\pi_{-i}}(t) 
\prod_{j=1}^{N}
p_{\pi_{j-1}/\pi_{j}}(t)
=
\prod_{i=1}^{N} 
\Big( 1-t^i \Big)^{p_i(\pi)}
\label{qb-lem8-1}
\end{align}

}
\end{lemma}

\begin{proof}
The proof is best illustrated by an example, so we consider the plane partition drawn in figure 4.1. This plane partition lives inside the box $[5,5,3]$, and its diagonal slices are given by

\begin{align}
&\pi_{-5} = \emptyset && \phantom{\pi_{0} = \{3,3,3\}} \quad\quad\quad\quad \pi_{1} = \{3,3,2\} \\
\nonumber
&\pi_{-4} = \{1\} && \phantom{\pi_{0} = \{3,3,3\}} \quad\quad\quad\quad \pi_{2} = \{3,2\} \\
\nonumber
&\pi_{-3} = \{3,1\} && \pi_{0} = \{3,3,3\} \quad\quad\quad\quad \pi_{3} = \{2,2\} \\
\nonumber
&\pi_{-2} = \{3,1\} && \phantom{\pi_{0} = \{3,3,3\}} \quad\quad\quad\quad \pi_{4} = \{2\} \\
\nonumber
&\pi_{-1} = \{3,3\} && \phantom{\pi_{0} = \{3,3,3\}} \quad\quad\quad\quad \pi_{5} = \emptyset
\end{align}

\noindent Using these partitions and the definition (\ref{pfunct}), we evaluate

\begin{align}
&p_{\pi_{-4}/\pi_{-5}}(t) = 1-t 
&& p_{\pi_{0}/\pi_{1}}(t) = 1-t^3
\\
\nonumber
&p_{\pi_{-3}/\pi_{-4}}(t) = 1-t
&& p_{\pi_{1}/\pi_{2}}(t) = 1-t^2
\\
\nonumber
&p_{\pi_{-2}/\pi_{-3}}(t) = 1
&& p_{\pi_{2}/\pi_{3}}(t) = 1-t
\\
\nonumber
&p_{\pi_{-1}/\pi_{-2}}(t) = 1-t^2
&& p_{\pi_{3}/\pi_{4}}(t) = 1-t^2
\\
\nonumber
&p_{\pi_{0}/\pi_{-1}}(t) = 1-t^3
&& p_{\pi_{4}/\pi_{5}}(t) = 1-t
\end{align}

\noindent Multiplying all of these terms together, we obtain a factor of $(1-t^i)$ for every level $i$ path which {\it does not} intersect the central diagonal. All level $i$ paths which {\it do} intersect the central diagonal obtain a factor of $(1-t^i)^2$. This double counting is cured by dividing by $b_{\pi_{0}}(t) = (1-t)(1-t^2)(1-t^3)$. The result is

\begin{align}
\frac{1}{b_{\pi_{0}}(t)}
\prod_{i=1}^{5} 
p_{\pi_{-i+1}/\pi_{-i}}(t) 
\prod_{j=1}^{5}
p_{\pi_{j-1}/\pi_{j}}(t)
&=
(1-t)^3(1-t^2)^2(1-t^3)
\nonumber
\\
&=
\prod_{i=1}^{5} \Big(1-t^i\Big)^{p_i(\pi)}
\end{align}

\noindent as required. This method can be easily extended to an arbitrary plane partition.
 
\end{proof}

\subsection{Generating $M$-boxed path-weighted plane partitions}

Aided by the results of the previous subsections, we are now able to calculate the generating function for $M$-boxed plane partitions with the prescribed weighting (\ref{apit}). We will see that this generating function is intrinsically related to the scalar product of the $q$-boson model. We start by iterating the $|n\rangle = |0\rangle$ case of equation (\ref{sum1}) $N$ times successively, which gives

\begin{align}
&
\mathcal{M}_{\psi}(t)
\mathbb{B}(x_1,t)
\ldots
\mathbb{B}(x_N,t)
|0\rangle
=
\sum_{[N,M] \supseteq \pi_{0} \succ \cdots \succ \pi_{N} = \emptyset}
\frac{1}{b_{\pi_{0}}(t)}
\prod_{i=1}^{N}
p_{\pi_{i-1}/\pi_{i}}(t)
x_i^{|\pi_{i-1}|-|\pi_{i}|}
|\pi_{0})
\end{align}

\noindent where the sum is over all interlacing partitions $\{\pi_{0} \succ \cdots \succ \pi_{N}\}$ which are subject to $\pi_0 \subseteq [N,M]$ and $\pi_{N} = \emptyset$. Similarly, one can iterate the $\langle n| = \langle 0|$ case of (\ref{sum2}) $N$ times successively, obtaining

{\small
\begin{align}
\langle 0|
\mathbb{C}(x_N,t)
\ldots
\mathbb{C}(x_1,t)
\mathcal{M}_{\psi}^{*}(t)
=
\sum_{\emptyset = \pi_{-N} \prec \cdots \prec \pi_{0} \subseteq [N,M]}
\frac{1}{b_{\pi_0}(t)}
\prod_{i=1}^{N}
p_{\pi_{-i+1}/\pi_{-i}}(t)
x_i^{|\pi_{-i+1}|-|\pi_{-i}|}
(\pi_{0}|
\end{align}
}

\noindent where the sum is over all interlacing partitions $\{\pi_{-N} \prec \cdots \prec \pi_{0}\}$ which are subject to $\pi_0 \subseteq [N,M]$ and $\pi_{-N} = \emptyset$. By the definition (\ref{apit}) of $A_{\pi}(\{x\},\{y\},t)$, the result (\ref{qb-lem8-1}) of lemma 9 and the orthogonality relation (\ref{innerproduct}), we therefore obtain

{\small
\begin{align}
\Big\langle
\langle 0|
\mathbb{C}(x_N,t)
\ldots
\mathbb{C}(x_1,t)
\mathcal{M}_{\psi}^{*}(t),
\mathcal{M}_{\psi}(t)
\mathbb{B}(y_1,t)
&\ldots
\mathbb{B}(y_N,t)
|0\rangle
\Big\rangle
\label{hl-genf1}
=
\sum_{\pi \subseteq [N,N,M]}
A_{\pi}\Big(
\{x\},\{y\},t
\Big)
\end{align}
}

\noindent where the sum is taken over all plane partitions $\pi$ which fit inside the box $[N,N,M]$. Hence the scalar product between the image Bethe eigenstates (\ref{hl-result}) and (\ref{hl-result2}) is a generating function for $M$-boxed path-weighted plane partitions. This generating function is evaluated explicitly by using the equations (\ref{hl-result}), (\ref{hl-result2}) and the orthogonality relation (\ref{innerproduct}) to give

{\footnotesize
\begin{align}
\Big\langle
\langle 0|
\mathbb{C}(x_N,t)
\ldots
\mathbb{C}(x_1,t)
\mathcal{M}_{\psi}^{*}(t),
\mathcal{M}_{\psi}(t)
\mathbb{B}(y_1,t)
\ldots
\mathbb{B}(y_N,t)
|0\rangle
\Big\rangle
\label{hl-genf2}
=
\sum_{\mu \subseteq [N,M]}
b_{\mu}(t)
P_{\mu}
(
\{x\},t
)
P_{\mu}
(
\{y\},t
)
\end{align}
}

\noindent Comparing equations (\ref{hl-genf1}) and (\ref{hl-genf2}), we have proved the result

\begin{align}
\sum_{\pi \subseteq [N,N,M]}
A_{\pi}\Big(
\{x\},\{y\},t
\Big)
=
\sum_{\mu \subseteq [N,M]}
b_{\mu}(t)
P_{\mu}
(
\{x\},t
)
P_{\mu}
(
\{y\},t
)
\label{tsp-fin}
\end{align}

\noindent We remark that the finite generating functions (\ref{bogs}) and (\ref{spp-exp3}) in the previous chapter are obtained from (\ref{tsp-fin}) by setting $t=0$ and $t=-1$, respectively.

\section{$q$-boson model on an infinite lattice}
\label{qb-pp}

In this section we elaborate upon results which were obtained in \cite{fw2}. Most of our attention centres on proving theorem 1, which is the $t$-deformation of lemma \ref{3-lem6} in the previous chapter. An independent derivation of theorem 1, using the properties of Hall-Littlewood functions, can be found in \cite{sul}.

\subsection{$\mathcal{M}_{\psi}(t) \mathbb{B}(x,t)|n\rangle$ and $\langle n|\mathbb{C}(x,t) \mathcal{M}_{\psi}^{*}(t)$ as $M\rightarrow\infty$}

\begin{theorem}
{\rm 
Consider the infinite lattice limit of the $q$-boson model, obtained by taking $M\rightarrow\infty$. Let $|n\rangle = \otimes_{i=0}^{\infty} |n_i\rangle_i$ and $\langle n|=\otimes_{i=0}^{\infty} \langle n_i|_i$ be basis vectors of $\mathcal{V}$ and $\mathcal{V}^{*}$, respectively, in this limit. Similarly, let $\frac{1}{b_{\nu}(t)}|\nu)$ and $\frac{1}{b_{\nu}(t)}(\nu|$ be the image states of these basis vectors under the mappings (\ref{qb-def1-1}). We claim that

\begin{align}
\mathcal{M}_{\psi}(t)
\left[
\lim_{M\rightarrow \infty}
\mathbb{B}(x,t)
|n\rangle
\right]
&=
\frac{1}{b_{\nu}(t)}
\Gamma_{-}(x,t)
|\nu)
\label{qb-the-1}
\\
\left[
\lim_{M \rightarrow \infty}
\langle n| \mathbb{C}(x,t)
\right]
\mathcal{M}_{\psi}^{*}(t)
&=
\frac{1}{b_{\nu}(t)}
(\nu| 
\Gamma_{+}(x,t)
\label{qb-the-2}
\end{align}

\noindent where $\Gamma_{\pm}(x,t)$ denote the $t$-deformed half-vertex operators

\begin{align}
\Gamma_{\pm}(x,t) 
= 
\exp 
\left( 
\sum_{n=1}^{\infty} \frac{1-t^n}{n} x^{n} H_{\pm n}(t) 
\right)
\end{align}

\noindent which were studied in subsection 4.2.6.
}
\end{theorem}

%

\begin{proof}
We split the proof into four steps. In the first step, we translate the equations (\ref{qb-the-1}), (\ref{qb-the-2}) to the equivalent statements (\ref{nottheone}), (\ref{theone}) at the level of the $t$-deformed Fock spaces. Thereafter we focus on proving (\ref{theone}), since (\ref{nottheone}) follows by direct analogy. In the second step we define the function $f_0$ and use it to express (\ref{theone}) in the alternative form (\ref{qb-the-5}). The third step contains several identities and a useful change of notation. In the fourth step we prove (\ref{qb-the-5}) using induction.

\medskip
\noindent
{\bf Step 1.}\ Taking the $M \rightarrow \infty$ limit of equations (\ref{sum3}) and (\ref{sum4}), we obtain 

\begin{align}
\mathcal{M}_{\psi}(t)
\left[
\lim_{M\rightarrow\infty}
\mathbb{B}(x,t)
|n\rangle
\right]
&=
\sum_{\mu \succ \nu }
x^{|\mu|-|\nu|}
\frac{p_{\nu / \mu}(t)}{b_{\nu}(t)}
|\mu)
\\
\left[
\lim_{M\rightarrow\infty}
\langle n|
\mathbb{C}(x,t)
\right]
\mathcal{M}_{\psi}^{*}(t)
&=
\sum_{\mu \succ \nu}
x^{|\mu|-|\nu|}
\frac{p_{\nu / \mu}(t)}{b_{\nu}(t)}
(\mu|
\end{align}

\noindent where the sums are over all partitions $\mu$ which interlace $\nu$, but whose parts have no size restriction. The equations (\ref{qb-the-1}) and (\ref{qb-the-2}) are therefore equivalent to the statements

\begin{align}
\Gamma_{-}(x,t)
|\nu)
&=
\sum_{\mu \succ \nu}
x^{|\mu|-|\nu|}
p_{\nu / \mu}(t)
|\mu)
\label{qb-the-3}
\\
(\nu|
\Gamma_{+}(x,t)
&=
\sum_{\mu \succ \nu}
x^{|\mu|-|\nu|}
p_{\nu / \mu}(t)
(\mu|
\label{qb-the-4}
\end{align}

\noindent Due to the orthogonality (\ref{innerproduct}) of partition states, equations (\ref{qb-the-3}) and (\ref{qb-the-4}) may be presented in the alternative form

\begin{align}
(\mu| \Gamma_{-}(x,t)
=
\sum_{\nu \prec \mu}
x^{|\mu|-|\nu|}
p_{\nu/\mu}(t)
\frac{ b_{\mu}(t)}{b_{\nu}(t)}
(\nu|
=
\sum_{\nu \prec \mu}
x^{|\mu|-|\nu|}
p_{\mu/\nu}(t)
(\nu|
\label{nottheone}
\\
\Gamma_{+}(x,t) |\mu)
=
\sum_{\nu \prec \mu}
x^{|\mu|-|\nu|}
p_{\nu/\mu}(t)
\frac{ b_{\mu}(t)}{b_{\nu}(t)}
|\nu)
=
\sum_{\nu \prec \mu}
x^{|\mu|-|\nu|}
p_{\mu/\nu}(t)
|\nu)
\label{theone}
\end{align}

\noindent where the sums are now over all partitions $\nu$ such that $\nu \prec \mu$. In contrast with (\ref{qb-the-3}) and (\ref{qb-the-4}), equations (\ref{nottheone}) and (\ref{theone}) contain only finite sums, which simplifies their analysis. We will give an explicit proof of (\ref{theone}). The proof of (\ref{nottheone}) is omitted, but as it is of such a similar nature we will claim it as a corollary of (\ref{theone}).

\medskip
\noindent
{\bf Step 2. (Definition 4.)}\ To every set $\{m\} = \{m_1 > \cdots > m_l > -l\}$ we associate a unique collection of integers $\{s\} = \{0 = s_0 < s_1 < \cdots < s_r < s_{r+1} = l\}$ such that the subsets 

\begin{align}
\mathcal{S}_k
=
\{m_{(s_k+1)} > \cdots > m_{s_{(k+1)}}\}
\end{align} 

\noindent are comprised of nearest neighbours, for all $0 \leq k \leq r$, with $r$ minimized. We call this the {\it nearest neighbour partitioning} of $\{m\}$.

Let $\{m\}=\{m_1 >\cdots > m_l > -l\}$ and $\{n\} = \{n_1 > \cdots > n_l \geq -l\}$ be two arbitrary sets of integers, and fix $n_0 = \infty$. For all $0 \leq j \leq r$, we define the functions $f_j(\{m\},\{n\},t)$ by the equation  

\begin{align}
f_j\Big(\{m\},\{n\},t\Big)
=
\prod_{k=j}^{r}
\Big(
1
-
\theta(n_{s_k} > m_{(s_k+1)}+1)
\theta(m_{s_{(k+1)}} > n_{s_{(k+1)}})
t^{\Delta_{k}}
\Big)
\label{qb-def4-1}
\end{align}

\noindent where $\{s\}$ is the set of nearest neighbour points associated to $\{m\}$, $\Delta_k = s_{k+1}-s_k$, and $\theta(z)$ is the Boolean function

\begin{align}
\theta(z) =
\left\{
\begin{array}{ll}
1, & z\ {\rm true}
\\
0, & z\ {\rm false}
\end{array}
\right.
\end{align}

\medskip
\noindent
{\bf (Lemma 10.)}\ Let $\mu = \{\mu_1 \geq \cdots \geq \mu_l > 0\}$ and $\nu= \{\nu_1 \geq \cdots \geq \nu_l \geq 0\}$ be partitions whose $\mathcal{F}_{\psi}(t)$ equivalents are given by

\begin{align}
|\mu)
=
\psi_{m_1}(t)\ldots\psi_{m_l}(t) |-l\rangle,
\quad
|\nu)
=
\psi_{n_1}(t)\ldots\psi_{n_l}(t)|-l\rangle
\end{align}

\noindent where $m_i=\mu_i-i,n_i = \nu_i-i$ for all $1\leq i \leq l$. Then if $\mu \succ \nu$, we have

\begin{align}
p_{\mu / \nu}(t)
=
f_0\Big(
\{m\},\{n\},t
\Big)
\label{qb-lem9-1}
\end{align}

\begin{proof} {\it (Lemma 10.)}\/\ Since $\mu \succ \nu$, we know that $m_i \geq n_i > m_{i+1}$ for all $1\leq i \leq l$. It follows that the inequality $m_i > n_i$ is only allowed if $i \in \{s\}$. Hence all differences between the sets $\{m\},\{n\}$ occur at the points $\{s\}$. 

Now we notice that each $\mathcal{S}_k \subset \{m\}$ corresponds to $\Delta_k$ parts in $\mu$ of the same size. The necessary and sufficient condition for there to be one less part in $\nu$ is that $m_{s_{(k+1)}} > n_{s_{(k+1)}}$ and $n_{s_k} > m_{(s_k+1)}+1$. In such a case $f_0(\{m\},\{n\},t)$ returns a factor of $(1-t^{\Delta_k})$, which is precisely the same factor returned by $p_{\mu/\nu}(t)$. Iterating this argument across all $0\leq k \leq r$, we obtain the equality (\ref{qb-lem9-1}). 

To further clarify the proof, we present a short example. Let the partitions $\mu$ and $\nu$ be given by 

\begin{align}
\mu 
&= 
\{\mu_1,\mu_2,\mu_3,\mu_4,\mu_5,\mu_6,\mu_7\}
=
\{7,6,6,6,4,4,2\}
\\ 
\nu
&= 
\{\nu_1,\nu_2,\nu_3,\nu_4,\nu_5,\nu_6,\nu_7\}
=
\{7,6,6,4,4,3,1\}
\end{align}

\noindent These partitions satisfy $\mu \succ \nu$. Furthermore, using the definition (\ref{pfunct}) we obtain 

\begin{align}
p_{\mu/\nu}(t)
=
\prod_{i=1}^{7}  
\Big( 1 - \delta_{p_i(\mu),p_i(\nu)+1} t^{p_i(\mu)} \Big)
=
(1-t^3)(1-t)
\label{qb-lem9-2}
\end{align}

\noindent Now let $\{m\}$ and $\{n\}$ be the sets formed by fixing, respectively, $m_i = \mu_i - i$ and $n_i = \nu_i - i$ for all $1\leq i \leq 7$. We find that

\begin{align}
\{m\} &= \{m_1,m_2,m_3,m_4,m_5,m_6,m_7\} = \{6,4,3,2,-1,-2,-5\}
\\
\{n\} &= \{n_1,n_2,n_3,n_4,n_5,n_6,n_7\} = \{6,4,3,0,-1,-3,-6\}
\end{align}

\noindent Each of $\{m_1\},\{m_2>m_3>m_4\},\{m_5>m_6\},\{m_7\}$ are nearest neighbours, and this is the smallest possible decomposition of $\{m\}$ into such subsets. Hence the set $\{s\}$ of nearest neighbour points associated to $\{m\}$ is given by

\begin{align}
\{s\} = \{s_0,s_1,s_2,s_3,s_4\} = \{0,1,4,6,7\}
\end{align}

\noindent Finally, setting $n_0 = \infty$ and using the definition (\ref{qb-def4-1}) we obtain

\begin{align}
&
f_0\Big(\{m\},\{n\},t\Big)
=
\Big(
1-\theta(n_0>m_1+1) \theta(m_1>n_1) t
\Big)
\\
&
\times
\Big(
1-\theta(n_1>m_2+1) \theta(m_4>n_4) t^{3}
\Big)
\Big(
1-\theta(n_4>m_5+1) \theta(m_6>n_6) t^{2}
\Big)
\nonumber
\\
&
\times
\Big(
1-\theta(n_6>m_7+1) \theta(m_7>n_7) t
\Big)
=
(1-t^3)(1-t)
\nonumber
\end{align}

\noindent which is in agreement with (\ref{qb-lem9-2}). 


\end{proof}

\noindent By virtue of lemma 10, we are able to write (\ref{theone}) in the equivalent form

\begin{align}
\Gamma_{+}(x,t) \psi_{m_1}\ldots \psi_{m_l} |-l\rangle
=
\sum_{\nu \prec \mu}
x^{|\mu|-|\nu|}
f_{0}\Big(\{m\},\{n\},t\Big)
\psi_{n_1}\ldots \psi_{n_l} |-l\rangle
\label{qb-the-5}
\end{align}

\noindent where we have assumed that $\ell(\mu) = l$ and defined $m_i = \mu_i-i,n_i = \nu_i-i$ for all $1\leq i \leq l$. We now proceed to calculate the left hand side of (\ref{qb-the-5}), aiming to show that it evaluates to the sum on the right hand side. In order to achieve this we need several identities, which are introduced in the next step.  

%

\medskip
\noindent
{\bf Step 3. (Identity 1.)} Let $x$ be an indeterminate and $m$ an arbitrary integer. We have

\begin{align} 
& \ \ \ 
\left(
\sum_{n=0}^{\infty}
\psi_{(m-n)} x^{n+1}
\right)
\left(
\psi_{m}
+
(1-t)
\sum_{n=1}^{\infty}
\psi_{(m-n)} x^n
\right)
\\
&
=
\sum_{n=0}^{\infty}
\left(
\psi_{(m-n)} \psi_m
+
(1-t)
\sum_{i=1}^{n}
\psi_{(m-n+i)} \psi_{(m-i)}
\right) x^{n+1}
=
t \psi_{(m+1)}
\sum_{n=1}^{\infty}
\psi_{(m-n)} x^n
\nonumber
\end{align}

\noindent where the first equality follows trivially by collecting the coefficient of $x^{n+1}$ for all $n \geq 0$, and the second equality holds due to lemma 3. 

\medskip
\noindent
{\bf (Identity 2.)} Let $x$ be an indeterminate, $m$ an arbitrary integer, and fix another integer $s \geq 1$. We obtain

{\footnotesize
\begin{align}
&\ 
\left(
\psi_{(m+1)}
+
(1-t^s)
\sum_{i=1}^{\infty}
\psi_{(m+1-i)} x^i
\right)
\left(
\psi_m
+
(1-t)
\sum_{i=1}^{\infty}
\psi_{(m-i)} x^i
\right)
\\
&
=
\psi_{(m+1)} 
\left(
\psi_m
+
(1-t)
\sum_{i=1}^{\infty}
\psi_{(m-i)} x^i
\right)
+
(1-t^s)
\left(
\sum_{i=0}^{\infty}
\psi_{(m-i)} x^{i+1}
\right)
\left(
\psi_m
+
(1-t)
\sum_{i=1}^{\infty}
\psi_{(m-i)} x^i
\right)
\nonumber
\\
&
=
\psi_{(m+1)} 
\left(
\psi_m
+
(1-t^{s+1})
\sum_{i=1}^{\infty}
\psi_{(m-i)} x^i
\right)
\nonumber
\end{align}
}

\noindent where the second line follows trivially by expanding the first, and the final line holds by application of identity 1.

\medskip
\noindent
{\bf (Identity 3.)} Let $x$ be an indeterminate, $m$ an arbitrary integer, and fix two more integers $n \geq 2$, $s\geq 1$. We find

\begin{align}
&
\left(
\psi_{(m+n)}
+
(1-t^s)
\sum_{i=1}^{\infty}
\psi_{(m+n-i)} x^i
\right)
\left(
\psi_m
+
(1-t)
\sum_{i=1}^{\infty}
\psi_{(m-i)} x^i
\right)
\\
=
&
\left(
\psi_{(m+n)}
+
(1-t^s)
\sum_{i=1}^{n-1}
\psi_{(m+n-i)} x^i
\right)
\left(
\psi_m
+
(1-t)
\sum_{i=1}^{\infty}
\psi_{(m-i)} x^i
\right)
\nonumber
\\
&
+
(1-t^s) x^{n-1} 
\left(
\sum_{i=0}^{\infty}
\psi_{(m-i)} x^{i+1}
\right)
\left(
\psi_m
+
(1-t)
\sum_{i=1}^{\infty}
\psi_{(m-i)} x^i
\right)
\nonumber
\\
=
&
\left(
\psi_{(m+n)}
+
(1-t^s)
\sum_{i=1}^{n-2}
\psi_{(m+n-i)} x^i
\right)
\left(
\psi_m
+
(1-t)
\sum_{i=1}^{\infty}
\psi_{(m-i)} x^i
\right)
\nonumber
\\
&
+
(1-t^s) x^{n-1} \psi_{(m+1)}
\left(
\psi_m 
+
\sum_{i=1}^{\infty}
\psi_{(m-i)} x^i
\right) 
\nonumber
\end{align}

\noindent where the second line follows trivially by expanding the first, and the final line holds by application of identity 1.

\medskip
\noindent
{\bf (Change of notation.)} We introduce the notation

\begin{align}
\Psi_m(x,t^s)
=
\psi_m
+
(1-t^s)
\sum_{i=1}^{\infty}
\psi_{(m-i)} x^i
\label{change}
\end{align}

\noindent which allows us to abbreviate identity 2 as follows

\begin{align}
\Psi_{(m+1)}(x,t^s)
\Psi_m (x,t)
=
\psi_{(m+1)}
\Psi_m (x,t^{s+1})
\label{t-trunc1}
\end{align}

\noindent and identity 3 as follows

\begin{align}
\Psi_{(m+n)}(x,t^s)
\Psi_m (x,t)
&=
\left(
\psi_{(m+n)}
+
(1-t^s)
\sum_{i=1}^{n-2}
\psi_{(m+n-i)} x^i
\right)
\Psi_m (x,t)
\nonumber
\\
&+
(1-t^s) x^{n-1}
\psi_{(m+1)}
\Psi_m (x,t^{\infty})
\label{t-trunc2}
\end{align} 

\noindent where we have defined $t^{\infty}=0$, which is perfectly sensible given the assumption $|t|<1$. The equations (\ref{t-trunc1}) and (\ref{t-trunc2}) are essential to the calculations of the final step.

\medskip
\noindent
{\bf Step 4.} Let us return to the proof of (\ref{qb-the-5}). Our starting point is the commutation relation (\ref{t-ev3}), which can be expressed in the more succinct form   

\begin{align}
\Gamma_{+}(x,t)
\psi_m
&=
\Psi_m(x,t)
\Gamma_{+}(x,t)
\label{t-ev5}
\end{align}

\noindent using the notation (\ref{change}). Applying the identity (\ref{t-ev5}) repeatedly to the left hand side of (\ref{qb-the-5}), we find that

\begin{align}
\Gamma_{+}(x,t)
\psi_{m_1}\ldots \psi_{m_l} |-l\rangle
&=
\Psi_{m_1}(x,t) \ldots \Psi_{m_l}(x,t) 
\Gamma_{+}(x,t) |-l\rangle
\\
&=
\Psi_{m_1}(x,t)\ldots \Psi_{m_l}(x,t) |-l\rangle
\nonumber
\end{align}

\noindent where the final line follows from the fact that $\Gamma_{+}(x,t) |-l\rangle = e^{H_{+}(x,t)} |-l\rangle = |-l\rangle$, which is a simple consequence of the annihilation relation (\ref{t-heis-anni}). Hence we see that (\ref{qb-the-5}) is equivalent to the proposition

\begin{align}
\Psi_{m_1}(x,t)\ldots \Psi_{m_l}(x,t) |-l\rangle
=
\sum_{\nu \prec \mu}
x^{|\mu|-|\nu|}
f_{0}\Big(\{m\},\{n\},t\Big)
\psi_{n_1}\ldots \psi_{n_l} |-l\rangle
\label{qb-the-6}
\end{align} 

\noindent In order to prove (\ref{qb-the-6}), fix $\{s\}=\{0=s_0<s_1<\cdots<s_r<s_{r+1} = l\}$ as the set arising from the nearest neighbour partitioning of $\{m\}$. We let $\mathcal{P}^{(1)}_j$ denote the proposition  

\begin{align}
\psi_{m_1}\ldots \psi_{m_{s_j}}
\rprod_{k=\bar{s}_j}^{l}
\Psi_{m_k}(x,t)
|-l\rangle
&=
\sum_{ \substack{\nu \prec \mu \\ [\nu]_{s_j} = [\mu]_{s_j}} }
x^{|\mu|-|\nu|}
f_j\Big(
\{m\},\{n\},t
\Big)
\rprod_{k=1}^{l}
\psi_{n_k}
|-l\rangle
\label{doub-prop1}
\end{align}

\noindent and similarly let $\mathcal{P}^{(2)}_j$ denote the proposition 

{\footnotesize
\begin{align}
\psi_{m_1}\ldots \psi_{m_{s_j}}
\Psi_{m_{\bar{s}_j}}(x,0)
\rprod_{k=\bar{s}_j+1}^{l}
\Psi_{m_k}(x,t)
|-l\rangle
\label{doub-prop2}
=
\sum_{ \substack{\nu \prec \mu \\ [\nu]_{s_j} = [\mu]_{s_j}} }
x^{|\mu|-|\nu|}
f_{j+1}\Big(
\{m\},\{n\},t
\Big)
\rprod_{k=1}^{l}
\psi_{n_k}
|-l\rangle
\end{align}
}

\noindent where we have abbreviated $\bar{s}_j = s_j+1$ for convenience, and where in both cases the sum is over all partitions $\nu$ which satisfy 

\begin{align}
\nu_i = \mu_i,\ {\rm for\ all}\ 1 \leq i \leq s_j,
\quad\quad
\mu_i \geq \nu_i \geq \mu_{i+1},\ {\rm for\ all}\ \bar{s}_j \leq i \leq l
\end{align}

\noindent Our aim is to prove that $\mathcal{P}_0^{(1)}$ is true, because equation (\ref{doub-prop1}) specializes to (\ref{qb-the-6}) in the case $j=0$. Firstly we prove the propositions $\mathcal{P}^{(1)}_r$ and $\mathcal{P}^{(2)}_r$. By definition the integers $\{m_{\bar{s}_r} > \cdots > m_l\}$ are nearest neighbours, so we may use the identity (\ref{t-trunc1}) repeatedly to obtain

\begin{align}
\psi_{m_1}\ldots \psi_{m_{s_r}}
\Psi_{m_{\bar{s}_r}}(x,t^{\delta+1})
\rprod_{k=\bar{s}_r+1}^{l}
\Psi_{m_{k}}(x,t)
|-l\rangle
\label{doub-prop3}
=
\rprod_{k=1}^{l-1}
\psi_{m_{k}}
\Psi_{m_l}(x,t^{\delta+\Delta_{r}})
|-l\rangle
\end{align}

\noindent where we have defined $\Delta_r = s_{r+1} - s_r = l - s_r$, and at this stage $\delta$ is unspecified. Now due to the annihilation properties (\ref{qb-lem4-1}), we obtain the truncation

\begin{align}
\Psi_{m_l}(x,t^{\delta+\Delta_r})
|-l\rangle
=
\left(
\psi_{m_l}
+
(1-t^{\delta+\Delta_r})
\sum_{m_l > n_l \geq -l}
\psi_{n_l}
x^{m_l-n_l}
\right)
|-l\rangle
\end{align}

\noindent Substituting this result into (\ref{doub-prop3}), we take the limit $\delta\rightarrow 0$ to recover

\begin{align}
\psi_{m_1}\ldots \psi_{m_{s_r}}
\rprod_{k=\bar{s}_r}^{l}
\Psi_{m_k}(x,t)
|-l\rangle
&=
\sum_{ \substack{\nu \prec \mu \\ [\nu]_{s_r} = [\mu]_{s_r}} }
x^{|\mu|-|\nu|}
f_r\Big(
\{m\},\{n\},t
\Big)
\rprod_{k=1}^{l}
\psi_{n_k}
|-l\rangle
\label{doub-prop4}
\end{align}

\noindent and the limit $\delta \rightarrow \infty$ to recover 

\begin{align}
\psi_{m_1}\ldots \psi_{m_{s_r}}
\Psi_{m_{\bar{s}_r}}(x,0)
\rprod_{k=\bar{s}_r+1}^{l}
\Psi_{m_k}(x,t)
|-l\rangle
=
\label{doub-prop5}
\sum_{ \substack{\nu \prec \mu \\ [\nu]_{s_r} = [\mu]_{s_r}} }
x^{|\mu|-|\nu|}
f_{r+1}
\rprod_{k=1}^{l}
\psi_{n_k}
|-l\rangle
\end{align} 


\noindent where we have used the definition (\ref{qb-def4-1}) of $f_r(\{m\},\{n\},t)$ to produce (\ref{doub-prop4}), and fixed $f_{r+1} = 1$ to produce (\ref{doub-prop5}). Therefore $\mathcal{P}^{(1)}_r$ and $\mathcal{P}^{(2)}_r$ are true. Now assume that $\mathcal{P}^{(1)}_j$ and $\mathcal{P}^{(2)}_j$ are true for some $1 \leq j \leq r$, and adopt the notation $\tilde{\jmath} = j-1$.  Since the integers $\{m_{\bar{s}_{\tilde{\jmath}}} > \cdots > m_{s_j}\}$ are nearest neighbours, we may use the identity (\ref{t-trunc1}) repeatedly to show that 

{\footnotesize
\begin{align}
\psi_{m_1}\ldots \psi_{m_{s_{\tilde{\jmath}}}}
\Psi_{m_{\bar{s}_{\tilde{\jmath}}}}(x,t^{\delta+1})
\rprod_{k=\bar{s}_{\tilde{\jmath}}+1}^{l}
\Psi_{m_k}(x,t)
|-l\rangle
=
\label{doub-prop6}
\rprod_{k=1}^{s_j-1}
\psi_{m_k}
\Psi_{m_{s_j}}(x,t^{\delta+\Delta_{\tilde{\jmath}}})
\rprod_{k=\bar{s}_j}^{l}
\Psi_{m_k}(x,t)
|-l\rangle
\end{align}
}

\noindent By definition the integers $m_{s_j}$ and $m_{\bar{s}_j}$ are {\it not} nearest neighbours, so we may apply the identity (\ref{t-trunc2}) to the right hand side of (\ref{doub-prop6}), giving

{\small
\begin{align}
\quad
&
\psi_{m_1}\ldots \psi_{m_{s_{\tilde{\jmath}}}}
\Psi_{m_{\bar{s}_{\tilde{\jmath}}}}(x,t^{\delta+1})
\rprod_{k=\bar{s}_{\tilde{\jmath}}+1}^{l}
\Psi_{m_k}(x,t)
|-l\rangle
=
\label{doub-prop7}
\\
\quad
&
\rprod_{k=1}^{s_j-1}
\psi_{m_k}
\left\{
\Big(
\psi_{m_{s_j}} 
+ 
(1-t^{\delta+\Delta_{\tilde{\jmath}}}) 
\sum_{ m_{s_j} > n_{s_j} > m_{\bar{s}_j}+1}
\psi_{n_{s_j}}
x^{m_{s_j}-n_{s_j}}
\Big)
\right.
\rprod_{k=\bar{s}_j}^{l}
\Psi_{m_k}(x,t)
|-l\rangle
\nonumber
\\
&
\quad\quad\quad\quad\quad\quad\ \ 
\left.
+
(1-t^{\delta+\Delta_{\tilde{\jmath}}})
x^{m_{s_j}-m_{\bar{s}_j}-1}
\psi_{(m_{\bar{s}_j}+1)}
\Psi_{m_{\bar{s}_j}}(x,0)
\rprod_{k=\bar{s}_j+1}^{l}
\Psi_{m_k}(x,t)
|-l\rangle
\right\}
&
\nonumber
\end{align}
}

\noindent Applying the assumptions $\mathcal{P}^{(1)}_j$ and $\mathcal{P}^{(2)}_j$ to (\ref{doub-prop7}), we take the limit $\delta\rightarrow0$ to recover 

\begin{align}
\psi_{m_1}\ldots \psi_{m_{s_{\tilde{\jmath}}}}
\rprod_{k=\bar{s}_{\tilde{\jmath}}}^{l}
\Psi_{m_k}(x,t)
|-l\rangle
&
=
\sum_{ \substack{\nu \prec \mu \\ [\nu]_{s_{\tilde{\jmath}}} = [\mu]_{s_{\tilde{\jmath}}}} }
x^{|\mu|-|\nu|}
f_{\tilde{\jmath}}\Big(
\{m\},\{n\},t
\Big)
\rprod_{k=1}^{l}
\psi_{n_k}
|-l\rangle
\end{align}

\noindent and the limit $\delta\rightarrow\infty$ to recover

{\footnotesize
\begin{align}
\psi_{m_1}\ldots \psi_{m_{s_{\tilde{\jmath}}}}
\Psi_{m_{\bar{s}_{\tilde{\jmath}}}}(x,0)
\rprod_{k=\bar{s}_{\tilde{\jmath}}+1}^{l}
\Psi_{m_k}(x,t)
|-l\rangle
=
\sum_{ \substack{\nu \prec \mu \\ [\nu]_{s_{\tilde{\jmath}}} = [\mu]_{s_{\tilde{\jmath}}}} }
x^{|\mu|-|\nu|}
f_{j}\Big(
\{m\},\{n\},t
\Big)
\rprod_{k=1}^{l}
\psi_{n_k}
|-l\rangle
\end{align}
}

\noindent Hence $\mathcal{P}^{(1)}_j,\mathcal{P}^{(2)}_j$ true $\implies \mathcal{P}^{(1)}_{\tilde{\jmath}},\mathcal{P}^{(2)}_{\tilde{\jmath}}$ true, proving the propositions (\ref{doub-prop1}) and (\ref{doub-prop2}) in general by induction. Therefore $\mathcal{P}^{(1)}_0$ holds, completing the proof of the theorem.

\end{proof} 

\subsection{Generating path-weighted plane partitions of arbitrary size}

In the previous section we demonstrated that the scalar product (\ref{hl-genf1}) on a lattice of size $M+1$ generates $M$-boxed path-weighted plane partitions. Accordingly, we expect that in the limit $M \rightarrow \infty$ it will generate path-weighted plane partitions whose column heights are arbitrarily large, giving rise to the equation

{\footnotesize
\begin{align}
\lim_{M \rightarrow \infty}
\Big\langle
\langle 0|
\mathbb{C}(x_N,t)
\ldots
\mathbb{C}(x_1,t)
\mathcal{M}_{\psi}^{*}(t),
\mathcal{M}_{\psi}(t)
\mathbb{B}(y_1,t)
\ldots
\mathbb{B}(y_N,t)
|0\rangle
\Big\rangle
\label{tsp-0}
=
\sum_{\pi \subseteq [N,N,\infty]}
A_{\pi}
\Big(
\{x\},\{y\},t
\Big)
\end{align}
}

\noindent where the sum is over all plane partitions $\pi$ which fit inside the box of dimension $N \times N \times \infty$, with $A_{\pi}(\{x\},\{y\},t)$ given by (\ref{apit}). On the other hand, using the result of theorem 1 we are able to write

\begin{align}
\lim_{M \rightarrow \infty}
\Big\langle
\langle 0|
\mathbb{C}(x_N,t)
\ldots
\mathbb{C}(x_1,t)
\mathcal{M}_{\psi}^{*}(t),
&
\mathcal{M}_{\psi}(t)
\mathbb{B}(y_1,t)
\ldots
\mathbb{B}(y_N,t)
|0\rangle
\Big\rangle
\label{tsp-1}
\\
=
(\emptyset|
&
\Gamma_{+}(x_N,t)
\ldots
\Gamma_{+}(x_1,t)
\Gamma_{-}(y_1,t)
\ldots
\Gamma_{-}(y_N,t)
|\emptyset)
\nonumber
\end{align}

\noindent which lends itself to immediate evaluation. This is because the $t$-deformed half-vertex operators, like their counterparts in the previous chapter, obey a simple commutation rule. To derive this rule, we use the commutator (\ref{t-heis}) to obtain the equation

\begin{align}
&
\sum_{m=1}^{\infty}
\sum_{n=1}^{\infty}
\frac{(1-t^m)(1-t^n) x^m y^n}{mn}
[H_m(t),H_{-n}(t)]
\\
=&
\sum_{m=1}^{\infty}
\sum_{n=1}^{\infty}
\frac{(1-t^m)(1-t^n) x^m y^n}{mn}
\frac{m \delta_{m,n}}{1-t^m}
=
\sum_{m=1}^{\infty}
\frac{(1-t^m) x^m y^m}{m}
\nonumber
\end{align}

\noindent which, in turn, implies that

\begin{align}
\Gamma_{+}(x,t) \Gamma_{-}(y,t)
&=
\exp\left(
\sum_{m=1}^{\infty}
\frac{(1-t^m) (xy)^m}{m}
\right)
\Gamma_{-}(y,t) \Gamma_{+}(x,t)
\label{thv-com}
\\
&=
\frac{1-txy}{1-xy}
\Gamma_{-}(y,t) \Gamma_{+}(x,t)
\nonumber
\end{align}

\noindent Employing the commutation relation (\ref{thv-com}) repeatedly in (\ref{tsp-1}) yields

{\small
\begin{align}
\lim_{M \rightarrow \infty}
\Big\langle
\langle 0|
\mathbb{C}(x_N,t)
\ldots
\mathbb{C}(x_1,t)
\mathcal{M}_{\psi}^{*}(t),
\mathcal{M}_{\psi}(t)
\mathbb{B}(y_1,t)
\ldots
\mathbb{B}(y_N,t)
|0\rangle
\Big\rangle
\label{tsp-2}
=
\prod_{i,j=1}^{N}
\frac{1-t x_i y_j}{1- x_i y_j}
\end{align}
}

\noindent Comparing equations (\ref{tsp-0}) and (\ref{tsp-2}), we have proved that

\begin{align}
\sum_{\pi \subseteq [N,N,\infty]}
A_{\pi}\Big(\{x\},\{y\},t\Big)
=
\prod_{i,j=1}^{N}
\frac{1-t x_i y_j}{1- x_i y_j}
\label{whatwhat}
\end{align}

\noindent which is a simpler evaluation of this generating function than in the finite case (\ref{tsp-fin}). As we observed in the previous chapter, this type of calculation could be performed using a symmetric function identity. This continues to be the case here, and we remark that (\ref{whatwhat}) could also be obtained using (\ref{tsp-fin}) and the identity

\begin{align}
\sum_{\mu \subseteq [N,\infty]}
b_{\mu}(t) P_{\mu}(\{x\},t) P_{\mu}(\{y\},t)
=
\prod_{i,j=1}^{N} 
\frac{1-tx_i y_j}{1-x_i y_j}
\end{align}

\noindent from section 4, chapter III of \cite{mac}. As we have done in the previous chapter, let us specialize the variables $\{x_1,\ldots,x_N\}$ and $\{y_1,\ldots,y_N\}$ to 

\begin{align}
x_i= y_i = z^{i-\frac{1}{2}}\ \ {\rm for\ all}\ 1\leq i \leq N
\end{align}

\noindent giving rise to the equation

\begin{align}
\sum_{\pi \subseteq [N,N,\infty]}
\prod_{i=1}^{N}
\Big(1-t^i\Big)^{p_i(\pi)}
z^{|\pi|}
=
\prod_{i,j=1}^{N}
\frac{1-tz^{i+j-1}}{1-z^{i+j-1}}
\end{align}

\noindent where $|\pi|$ is the weight of the plane partition $\pi$, and $p_i(\pi)$ is the number of paths in $\pi$ at level $i$. Taking the limit $N \rightarrow \infty$ we obtain

\begin{align}
\sum_{\pi}
\prod_{i=1}^{\infty}
\Big(1-t^i\Big)^{p_i(\pi)}
z^{|\pi|}
=
\prod_{i=1}^{\infty}
\frac{(1-t z^i)^i}{(1-z^i)^i}
\label{vulgf}
\end{align}

\noindent where the sum is now over plane partitions of completely arbitrary dimension. This result specializes to the generating function (\ref{macmahon}) by setting $t=0$, and to (\ref{spp-exp4}) by setting $t=-1$. The generating function (\ref{vulgf}) first appeared in \cite{vul2}, where it was proved using combinatorial methods. The fermionic proof which we have described was the key result of \cite{fw2}.

\section{Conclusion}

In this chapter we studied the Bethe eigenvectors of the $q$-boson model. We gave a representation of the $q$-boson algebra on the vector space $\mathcal{V}$, which collapses to the previously encountered representations of the phase and $i$-boson algebras in the respective limits $q\rightarrow\infty$ and $q\rightarrow i$. We defined a map taking basis elements of $\mathcal{V}$ to partitions in the deformed Fock space $\mathcal{F}_{\psi}(t)$, and calculated the image of the Bethe eigenvectors under this map. However, since the underlying fermions are more complicated than in the previous chapter, we were unable to connect the finite lattice scalar product with a solution of a classical hierarchy.

On the other hand, in the infinite lattice limit of the $q$-boson model we were able to obtain a result which logically extends the material of chapter 3. We refer to theorem 1, which shows that when $M \rightarrow \infty$ the action of a $B$-operator on a general state maps to the action of a $t$-deformed half-vertex operator on the image state. Since the $t$-deformed half-vertex operators obey simple commutation relations, we could evaluate the scalar product in product form in the $M\rightarrow\infty$ limit. This provided a new proof of Vuleti\'c's path-weighted generating function for plane partitions. 


We now list two questions which arise from this work, which are worthy of further investigation.

{\bf 1.} {\it Does there exist a $t$-deformed hierarchy admitting Hall-Littlewood polynomials as $\tau$-functions?} This question does not seem to have been properly addressed in the literature. The closing remarks of \cite{jin2} discuss this possibility, but in the context of a slightly different $t$-deformed Clifford algebra than the one presented in this chapter. Assuming the existence of such a hierarchy, we would expect that individual Hall-Littlewood polynomials and the finite lattice $q$-boson scalar product should be valid solutions.

{\bf 2.} {\it Can we extend our procedure to calculating the generating functions for other weighted plane partitions?} A more general generating function for plane partitions, related to the Macdonald polynomials, was obtained in \cite{vul2}. Furthermore, a more general species of deformed fermions appeared in \cite{jin3}, where they were used once again in the context of Macdonald polynomials. It was realized in \cite{fw2} that the generating function for Macdonald-type plane partitions could be obtained using the fermions of \cite{jin3}, but an explicit proof is yet to be presented.


\newpage

\thispagestyle{empty}

\phantom{nothing}


\chapter{XXZ model and the KP hierarchy}
\newtheorem{korepin}{ }
\newtheorem{propertylist}{ }
\newtheorem{korepin2}{ }
\newtheorem{propertylist2}{ }
\newtheorem{propertylist3}{ }
\newtheorem{propertylist4}{ }
\newtheorem{propertylist5}{ }

\newcommand{\ttop}[2]{\genfrac{}{}{0pt}{}{#1}{#2}}

\def\brack{[}
\def\no{\nonumber}
\def\ni{\noindent}
\def\proofend{\ensuremath{\square}}
\def\pr{'}
\def\beqa{\begin{eqnarray}}
\def\eeqa{\end{eqnarray}}
\def\ba{\begin{array}}
\def\ea{\end{array}}
\def\gl{\begin{swabfamily}gl\end{swabfamily}}
\def\psis{\psi^{*}}
\def\Psis{\Psi^{*}}
\def\union{\mathop{\bigcup}}
\def\vac{|\mbox{vac}\rangle}
\def\cav{\langle\mbox{vac}|}
\def\dprod{\mathop{\prod{\mkern-29.5mu}{\mathbf\longleftarrow}}}
\def\rprod{\mathop{\prod{\mkern-28.0mu}{\mathbf\longrightarrow}}}
\def\r{\rangle}
\def\l{\langle}
\def\a{\alpha}
\def\b{\beta}
\def\hb{\hat\beta}
\def\d{\delta}
\def\g{\gamma}
\def\e{\epsilon}
\def\tg{\operatorname{tg}}
\def\ctg{\operatorname{ctg}}
 \def\sh{\operatorname{sh}}
 \def\ch{\operatorname{ch}}
\def\cth{\operatorname{cth}}
 \def\th{\operatorname{th}}
\def\eps{\varepsilon}
 \def\la{\lambda}
\def\tla{\tilde{\lambda}}
\def\Gh{\widehat{\Gamma}}
\def\tmu{\tilde{\mu}}
\def\s{\sigma}
\def\sul{\sum\limits}
\def\pl{\prod\limits}
\def\lt({\left(}
\def\rt){\right)}
\def\const{{\rm const}}
\def\argum{\{\mu_j\},\{\la_k\}}
\def\umarg{\{\la_k\},\{\mu_j\}}
\def\prodmu #1{\prod\limits_{j #1 k} \sinh(\mu_k-\mu_j)}
\def\prodla #1{\prod\limits_{j #1 k} \sinh(\lambda_k-\lambda_j)}
\def\tr{\operatorname{tr}}
\def\Res{\operatorname{Res}}
\def\det{\operatorname{det}}
\def\ivac{\langle\Omega|}
\def\fvac{|\Omega\rangle}
\def\fdirac{|0\rangle}
\def\idirac{\langle0|}
\def\psis{\psi^{*}}
\def\Psis{\Psi^{*}}
\def\lprod{\mathop{\prod{\mkern-29.5mu}{\mathbf\longleftarrow}}}
\def\rprod{\mathop{\prod{\mkern-28.0mu}{\mathbf\longrightarrow}}}
\def\complex{\mathbb{C}}
\def\integer{\mathbb{Z}}

\setcounter{section}{-1}
\setcounter{lemma}{0}
\setcounter{theorem}{0}
\setcounter{remark}{0}
\setcounter{definition}{0}

\section{Introduction}

The one-dimensional Heisenberg magnet was introduced in 1928 \cite{hei} and first solved in 1931 \cite{bet}, marking the invention of the Bethe Ansatz. It is a model for a one-dimensional lattice of spin-$\frac{1}{2}$ fermions with nearest neighbour interactions, and its Hamiltonian contains three parameters $J_x,J_y,J_z$ which describe the anisotropy of the system.\footnote{See, for example, chapter II of \cite{kbi}.} When these three parameters are different, it is called the XYZ model. In this chapter we will consider the case of partial isotropy, with $J_x=J_y\not=J_z$, which is known as the XXZ model or XXZ spin-$\frac{1}{2}$ chain. Despite its extensive history, the XXZ model continues to be of great interest to researchers in the field of quantum integrable models. In this chapter we will describe several original contributions which we have made to the study of this model, outlining in particular its apparent connection with the KP hierarchy.

In section \ref{xxz-intro} we introduce the basics of the XXZ model, including its space of states $\mathcal{V}$, Hamiltonian $\mathcal{H}$, and review the construction of its eigenvectors using the algebraic Bethe Ansatz. Following chapter 2, we also provide graphical representations for the entries of the $R$-matrix and the monodromy matrix.

One of the most fundamental quantities pertaining to the XXZ model is the domain wall partition function. This object acquires its name because it is equal to the partition function of the six-vertex model \cite{bax1}, under domain wall boundary conditions. In \cite{kor}, V~E~Korepin found a set of conditions on the partition function which determine it uniquely. These conditions were subsequently solved by A~G~Izergin in \cite{ize}, where a determinant expression for the partition function was obtained. We reproduce these results in section \ref{xxz-pf}, working from both an algebraic and a graphical perspective. We conclude the section with the first of our new results, showing that the partition function is a power-sum specialization of a KP $\tau$-function \cite{fwz4}.


The scalar product is another object of essential interest in the XXZ model. It is a function of two sets of auxiliary rapidities $\{u\}_N$ and $\{v\}_N$, and its evaluation depends on the restrictions imposed on these rapidities. When $\{u\}_N$ and $\{v\}_N$ are unrestricted, the scalar product can be expressed as a complicated sum over a product of two determinants \cite{kbi}. On the other hand, when $\{u\}_N$ and $\{v\}_N$ are equal and satisfy the Bethe equations, the scalar product has a compact determinant expression proposed by M~Gaudin \cite{gau} and proved by Korepin in \cite{kor}. In this chapter we are interested in the intermediate case when $\{v\}_N$ satisfies the Bethe equations, but $\{u\}_N$ is unrestricted. Although these conditions are weaker than in Gaudin's case, the scalar product remains expressible as a determinant, as was discovered by N~A~Slavnov in \cite{sla}. This determinant formula has been crucial to the work of N~Kitanine et al. on correlation functions, see for example \cite{kmst}, \cite{kmst2}, \cite{kmt}. 

In section \ref{xxz-sp} we outline a new proof of the Slavnov scalar product formula, which is inspired by the Izergin-Korepin procedure for the partition function. Our proof is based on a sequence of incremental scalar products, $S_0$ through to $S_N$, where $S_0$ is equal to the partition function up to a multiplicative factor, and $S_N$ is the actual scalar product. These scalar products were defined and calculated in \cite{kmt}, but this earlier proof was less elementary and relied on the Drinfel'd twist technique of \cite{ms}. We end the section with a new result which extends that of \cite{fwz4}, showing that the Slavnov scalar product is a power-sum specialization of a KP $\tau$-function \cite{fwz5}.     

Motivated by chapters 3 and 4, in section \ref{xxz-BE} we derive an explicit expression for the XXZ model Bethe eigenvectors. That is, we write the Bethe eigenvectors as sums over elementary spin states, and calculate the coefficients within these sums. We find that the coefficients can be expressed as objects which generalize determinants. We name these objects weighted determinants, since they are determinant-like sums over permutations, only each term in the sum is weighted with a factor that depends on pairs of elements in the permutation. It transpires that Hall-Littlewood functions are weighted determinants, which allows us to compare our form for the XXZ eigenvectors with Tsilevich's form for the $q$-boson eigenvectors, as was given in section \ref{qb-BE}.

\section{XXZ spin-$\frac{1}{2}$ chain}
\label{xxz-intro}

In this section we introduce the basics of the XXZ spin-$\frac{1}{2}$ chain, describing its solution via the quantum inverse scattering method/algebraic Bethe Ansatz. The notations and conventions that we adopt are basically consistent with those of\cite{kmt}. For a more general introduction, the reader is referred to chapters VI and VII of the book \cite{kbi}.

\subsection{Space of states $\mathcal{V}$ and inner product $\mathcal{I}$}

The finite length XXZ spin-$\frac{1}{2}$ chain consists of a one-dimensional lattice with $M$ sites. As we mentioned in chapter 2, each site is populated by a single spin-$\frac{1}{2}$ fermion. To each site $m$ we therefore associate a two-dimensional vector space $\mathcal{V}_m$ with the basis

\begin{align}
{\rm Basis}(\mathcal{V}_m)
=
\Big\{ \uparrow_m,\downarrow_m \Big\}
\end{align}

\noindent where for convenience we have adopted the notations 

\begin{align}
\uparrow_m\  
=
\Big(
\begin{array}{c}
1 \\ 0
\end{array}
\Big)_m,
\quad 
\downarrow_m\ 
=
\Big(
\begin{array}{c}
0 \\ 1
\end{array}
\Big)_m
\label{V_i}
\end{align}

\noindent Physically speaking, $\uparrow_m$ and $\downarrow_m$ represent the spin eigenstates of a spin-$\frac{1}{2}$ fermion at site $m$. The global vector space $\mathcal{V} = \mathcal{V}_1 \otimes \cdots \otimes \mathcal{V}_M$ has the basis

\begin{align}
{\rm Basis}(\mathcal{V})
=
\left\{
|\lambda\rangle
=
\bigotimes_{m \in \lambda } \uparrow_m
\bigotimes_{m \not\in \lambda } \downarrow_m
\right\}
\label{elem}
\end{align}

\noindent where $\lambda = \{\lambda_1,\ldots,\lambda_l\}$ ranges over all strict partitions $\{M \geq \lambda_1 > \cdots > \lambda_l > 0\}$ and all lengths $ 0 \leq l \leq M$ are allowed. The partition $\lambda = \{M,\ldots,1\}$ corresponds with the state in which all spins are up, while $\lambda = \emptyset$ corresponds with the state in which all spins are down. Since they play an important role in our calculations, we prescribe these states their own special notation, by defining

\begin{align}
|\Uparrow_M\rangle = \bigotimes_{m=1}^{M} \uparrow_m,
\quad
|\Downarrow_M\rangle = \bigotimes_{m=1}^{M} \downarrow_m
\end{align}

\noindent We will also make use of the notation

\begin{align}
|\Uparrow_{N/M}\rangle
=
\bigotimes_{1 \leq m \leq N}
\uparrow_m
\bigotimes_{N < m \leq M}
\downarrow_m,
\quad\quad
|\Downarrow_{N/M}\rangle
=
\bigotimes_{1 \leq m \leq N}
\downarrow_m
\bigotimes_{N < m \leq M}
\uparrow_m
\end{align}

\noindent for states whose first $N$ spins are up (down), with all the remaining spins being down (up), respectively. As always, we fix a bilinear inner product $\mathcal{I}$ acting on $\mathcal{V}$. Its action is given by 

\begin{align}
\mathcal{I}\Big( 
|\lambda\rangle,|\mu\rangle
\Big)
=
\delta_{\lambda,\mu}
\end{align}

\noindent for all basis vectors $|\lambda\rangle, |\mu\rangle$. In other words, $\mathcal{I}$ induces orthonormality between the elements of the basis (\ref{elem}).

As was explained in chapter 2, it is advantageous to define vector spaces dual to those already introduced. To each site $m$ we associate the dual vector space $\mathcal{V}_m^{*}$ with the basis

\begin{align}
{\rm Basis}(\mathcal{V}_m^{*})
=
\Big\{ 
\uparrow_m^{*},\downarrow_m^{*}
\Big\}
\end{align}

\noindent where we have adopted the notations

\begin{align}
\uparrow^{*}_m\ 
=
\left(
\begin{array}{cc}
1 & 0 
\end{array}
\right)_m,
\quad 
\downarrow^{*}_m\ 
=
\left(
\begin{array}{cc}
0 & 1
\end{array} 
\right)_m
\label{V_i*}
\end{align}

\noindent From this, we construct the dual space of states $\mathcal{V}^{*} = \mathcal{V}_1^{*} \otimes \cdots \otimes \mathcal{V}_M^{*}$ whose basis is given by

\begin{align}
{\rm Basis}(\mathcal{V}^{*})
=
\left\{
\langle \lambda| = 
\bigotimes_{m \in \lambda} \uparrow_m^{*}
\bigotimes_{m \not\in \lambda} \downarrow_m^{*}
\right\}
\label{elem*}
\end{align}

\noindent where, as before, $\lambda = \{\lambda_1,\ldots,\lambda_l\}$ ranges over all strict partitions which satisfy $\{M \geq \lambda_1 > \cdots > \lambda_l > 0\}$ and all lengths $0 \leq l \leq M$ are allowed. Similarly to above, we also introduce the notations

\begin{align}
\langle \Uparrow_M| = \bigotimes_{m=1}^{M} \uparrow^*_m,
\quad 
\langle \Downarrow_M| = \bigotimes_{m=1}^{M} \downarrow^*_m
\end{align}

\noindent for the dual total spin up/down states, and

\begin{align}
\langle \Uparrow_{N/M}|
=
\bigotimes_{1 \leq m \leq N}
\uparrow^{*}_m
\bigotimes_{N < m \leq M}
\downarrow^{*}_m,
\quad\quad
\langle \Downarrow_{N/M}|
=
\bigotimes_{1 \leq m \leq N}
\downarrow^{*}_m
\bigotimes_{N < m \leq M}
\uparrow^{*}_m
\end{align}

\noindent for the dual partial spin up/down states. Finally, we fix an action of $\mathcal{V}^{*}$ on $\mathcal{V}$ by defining

\begin{align}
\langle \lambda|\mu \rangle
=
\mathcal{I}\Big(
|\lambda \rangle, |\mu\rangle
\Big)
\end{align}

\noindent for all basis vectors $\langle \lambda| \in \mathcal{V}^{*}$ and $|\mu\rangle \in \mathcal{V}$. Notice that this definition is consistent with treating $\langle \lambda|\mu\rangle$ as a product of the matrices (\ref{V_i}) and (\ref{V_i*}).

%
%
%
%
%
%

%


\subsection{$sl_q(2)$ algebra}

The algebra which underpins the theory of the XXZ model is a $q$-deformation of the Lie algebra $sl(2)$. It is denoted $sl_q(2)$ and generated by $\{\sigma^{+},\sigma^{-},\sigma^{z}\}$ which satisfy the commutation relations

\begin{align}
[\sigma^{+},\sigma^{-}]
=
\frac{
q^{\sigma^{z}/2} - q^{-\sigma^{z}/2}
}
{q^{1/2}-q^{-1/2}},
\quad
[\sigma^{z},\sigma^{+}]
=
2\sigma^{+},
\quad
[\sigma^{z},\sigma^{-}]
=
-2\sigma^{-}
\label{slq2-1}
\end{align}

\noindent By taking the limit $q\rightarrow 1$, one recovers the commutation relations of $sl(2)$. As in previous chapters, we will consider $M$ copies of $sl_q(2)$, generated by $\{\sigma_1^{+},\sigma_1^{-},\sigma_1^{z}\}$ through to $\{\sigma_M^{+},\sigma_M^{-},\sigma_M^{z}\}$. In accordance with our chapter 2 conventions, we denote these algebras by $\mathcal{A}_1,\ldots,\mathcal{A}_M$ with $\mathfrak{a}_m^{+} = \sigma_m^{+}, \mathfrak{a}_m^{-} = \sigma_m^{-},\mathfrak{a}_m^{0} = \sigma_m^{z}$. Different copies of $sl_q(2)$ commute, giving rise to the equations

\begin{align} 
[\sigma_m^{+},\sigma_n^{-}]
=
\delta_{m,n}
\frac{q^{\sigma_m^{z}/2}-q^{-\sigma_m^{z}/2}}
{q^{1/2}-q^{-1/2}},
\quad
[\sigma_m^{z},\sigma_n^{+}]
=
2\delta_{m,n}\sigma_m^{+},
\quad
[\sigma_m^{z},\sigma_n^{-}]
=
-2\delta_{m,n} \sigma_m^{-}
\label{slq2-2}
\end{align} 

\noindent for all $1\leq m,n \leq M$.

\subsection{Representations of $sl_q(2)$ algebras}

It is possible to provide representations of the algebras $\mathcal{A}_1,\ldots,\mathcal{A}_M$ by identifying each of the generators $\{\sigma_m^{+},\sigma_m^{-},\sigma_m^{z}\}$ with a $2\times 2$ matrix. In particular, let us define the Pauli matrices

\begin{align}
\sigma_m^{x}
=
\left(
\begin{array}{rr}
0 & 1 
\\
1 & 0 
\end{array}
\right)_m,
\quad
\sigma_m^{y}
=
\left(
\begin{array}{rr}
0 & -i 
\\
i & 0 
\end{array}
\right)_m,
\quad
\sigma_m^{z}
=
\left(
\begin{array}{rr}
1 & 0 
\\
0 & -1
\end{array}
\right)_m
\label{pauli}
\end{align}

\noindent with $i = \sqrt{-1}$, and the spin-changing matrices

\begin{align}
\sigma_m^{+}
=
\frac{1}{2}
(\sigma_m^{x} + i \sigma_m^{y})
=
\left(
\begin{array}{cc}
0 & 1
\\
0 & 0 
\end{array}
\right)_m,
\quad
\sigma_m^{-}
=
\frac{1}{2}
(\sigma_m^{x}-i\sigma_m^{y})
=
\left(
\begin{array}{cc}
0 & 0
\\
1 & 0
\end{array}
\right)_m
\label{spin-chang} 
\end{align}

\noindent where in all cases the subscript $m$ is used to indicate that the matrices belong to the algebra $\mathcal{A}_m$. It is easy to show that the relations (\ref{slq2-2}) are faithfully represented with $\sigma^{z}_m$ as defined by (\ref{pauli}) and $\sigma^{\pm}_m$ as defined by (\ref{spin-chang}).

By virtue of these identifications and the interpretation (\ref{V_i}) of $\uparrow_m, \downarrow_m$ we automatically obtain actions of $\mathcal{A}_1,\ldots,\mathcal{A}_M$ on $\mathcal{V}$. For all $1\leq m \leq M$ we have

\begin{align}
\sigma_m^{+} \uparrow_m = 0,
\ 
\sigma_m^{+} \downarrow_m = \uparrow_m,
\quad
\sigma_m^{-} \uparrow_m = \downarrow_m,
\ 
\sigma_m^{-} \downarrow_m = 0,
\quad
\sigma_m^{z} \uparrow_m = \uparrow_m,
\ 
\sigma_m^{z} \downarrow_m = -\downarrow_m
\label{action1}
\end{align}

\noindent Similarly, using the interpretation (\ref{V_i*}) of $\uparrow_m^{*}, \downarrow_m^{*}$ we deduce actions of $\mathcal{A}_1,\ldots,\mathcal{A}_M$ on $\mathcal{V}^{*}$. For all $1\leq m \leq M$ we obtain  


\begin{align}
\uparrow_m^{*} \sigma_m^{+} = \downarrow_m^{*},
\ 
\downarrow_m^{*} \sigma_m^{+} = 0,
\quad
\uparrow_m^{*} \sigma_m^{-} = 0,
\ 
\downarrow_m^{*} \sigma_m^{-} = \uparrow_m^{*},
\quad
\uparrow_m^{*} \sigma_m^{z} = \uparrow_m^{*},
\ 
\downarrow_m^{*} \sigma_m^{z} = -\downarrow_m^{*}  
\label{action2}
\end{align}

\noindent Notice that the actions (\ref{action1}) and (\ref{action2}) are consistent with those of the most general context, that is, with equations (\ref{general-action}) and (\ref{general-action*}) from chapter 2. The correspondence is realized by setting $\uparrow_m = |\frac{1}{2} \rangle_m,\ \downarrow_m = |-\frac{1}{2}\rangle_m$, $\uparrow_m^{*} = \langle \frac{1}{2}|_m,\ \downarrow_m^{*} = \langle -\frac{1}{2}|_m$.

\subsection{Hamiltonian $\mathcal{H}$}

The Hamiltonian of the finite length XXZ spin-$\frac{1}{2}$ chain is given by 

\begin{equation}
\mathcal{H}
=
\sum_{m=1}^{M} 
\Big(
\sigma_m^x \sigma_{m+1}^x
+
\sigma_m^y \sigma_{m+1}^y
+
\Delta (\sigma_m^z \sigma_{m+1}^z-1)
\Big)
\label{hamiltonian}
\end{equation}

\noindent where $\Delta = \frac{1}{2}(q^{1/2}+q^{-1/2})$ is the anisotropy parameter of the model, and the periodicity conditions $\sigma_{M+1}^{x}=\sigma_1^{x},\sigma_{M+1}^{y}=\sigma_1^{y},\sigma_{M+1}^{z}=\sigma_1^{z}$ 
are assumed. Alternatively, the Hamiltonian (\ref{hamiltonian}) may be expressed as

\begin{align}
\mathcal{H}
=
\sum_{m=1}^{M} 
\Big(
2
\sigma_m^{+} \sigma_{m+1}^{-}
+
2
\sigma_m^{-} \sigma_{m+1}^{+}
+
\Delta (\sigma_m^z \sigma_{m+1}^z-1)
\Big)
\end{align}

\noindent where we assume $\sigma_{M+1}^{\pm} = \sigma_{1}^{\pm}$. This is the form which corresponds with the general Hamiltonian (\ref{ham-56}) discussed in chapter 2. Once again, we follow the quantum inverse scattering/algebraic Bethe Ansatz procedure for finding the eigenvectors $|\Psi\rangle \in \mathcal{V}$ of $\mathcal{H}$.

\subsection{$R$-matrix, crossing symmetry, Yang-Baxter equation}

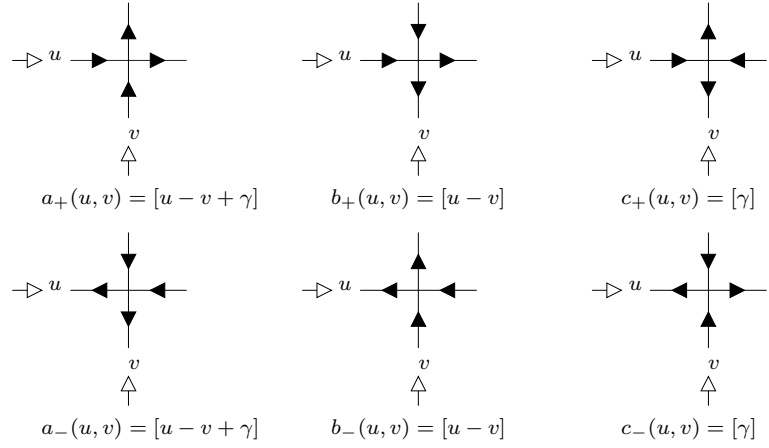
\begin{figure}[H]
\begin{center}
\begin{minipage}{4.3in}

\setlength{\unitlength}{0.00038cm}
\begin{picture}(20000,14000)(-4500,-12000)

\path(-2000,0000)(2000,0000)
\blacken\path(-1250,250)(-1250,-250)(-750,0)(-1250,250)
\blacken\path(750,250)(750,-250)(1250,0)(750,250)
\put(-2750,0){\scriptsize{$u$}}
\path(-4000,0)(-3000,0)
\whiten\path(-3500,250)(-3500,-250)(-3000,0)(-3500,250)
\path(0000,-2000)(0000,2000)
\blacken\path(-250,-1250)(250,-1250)(0,-750)(-250,-1250)
\blacken\path(-250,750)(250,750)(0,1250)(-250,750)
\put(-3000,-5000){\scriptsize$a_{+}(u,v)=[u-v+\gamma]$}
\put(0,-2750){\scriptsize{$v$}}
\path(0,-4000)(0,-3000)
\whiten\path(-250,-3500)(250,-3500)(0,-3000)(-250,-3500)

\path(8000,0000)(12000,0000)
\blacken\path(8750,250)(8750,-250)(9250,0)(8750,250)
\blacken\path(10750,250)(10750,-250)(11250,0)(10750,250)
\put(7250,0){\scriptsize{$u$}}
\path(6000,0)(7000,0)
\whiten\path(6500,250)(6500,-250)(7000,0)(6500,250)
\path(10000,-2000)(10000,2000)
\blacken\path(9750,-750)(10250,-750)(10000,-1250)(9750,-750)
\blacken\path(9750,1250)(10250,1250)(10000,750)(9750,1250)
\put(7000,-5000){\scriptsize$b_{+}(u,v)=[u-v]$}
\put(10000,-2750){\scriptsize{$v$}}
\path(10000,-4000)(10000,-3000)
\whiten\path(9750,-3500)(10250,-3500)(10000,-3000)(9750,-3500)

\path(18000,0000)(22000,0000)
\blacken\path(18750,250)(18750,-250)(19250,0)(18750,250)
\blacken\path(21250,250)(21250,-250)(20750,0)(21250,250)
\put(17250,0){\scriptsize{$u$}}
\path(16000,0)(17000,0)
\whiten\path(16500,250)(16500,-250)(17000,0)(16500,250)
\path(20000,-2000)(20000,2000)
\blacken\path(19750,-750)(20250,-750)(20000,-1250)(19750,-750)
\blacken\path(19750,750)(20250,750)(20000,1250)(19750,750)
\put(17000,-5000){\scriptsize$c_{+}(u,v)=[\gamma]$}
\put(20000,-2750){\scriptsize{$v$}}
\path(20000,-4000)(20000,-3000)
\whiten\path(19750,-3500)(20250,-3500)(20000,-3000)(19750,-3500)

\path(-2000,-8000)(2000,-8000)
\blacken\path(-750,-7750)(-750,-8250)(-1250,-8000)(-750,-7750)
\blacken\path(1250,-7750)(1250,-8250)(750,-8000)(1250,-7750)
\put(-2750,-8000){\scriptsize{$u$}}
\path(-4000,-8000)(-3000,-8000)
\whiten\path(-3500,-7750)(-3500,-8250)(-3000,-8000)(-3500,-7750)
\path(0000,-10000)(0000,-6000)
\blacken\path(-250,-8750)(250,-8750)(0,-9250)(-250,-8750)
\blacken\path(-250,-6750)(250,-6750)(0,-7250)(-250,-6750)
\put(-3000,-13000){\scriptsize$a_{-}(u,v)=[u-v+\gamma]$}
\put(0,-10750){\scriptsize{$v$}}
\path(0,-12000)(0,-11000)
\whiten\path(-250,-11500)(250,-11500)(0,-11000)(-250,-11500)

\path(8000,-8000)(12000,-8000)
\blacken\path(9250,-7750)(9250,-8250)(8750,-8000)(9250,-7750)
\blacken\path(11250,-7750)(11250,-8250)(10750,-8000)(11250,-7750)
\put(7250,-8000){\scriptsize{$u$}}
\path(6000,-8000)(7000,-8000)
\whiten\path(6500,-7750)(6500,-8250)(7000,-8000)(6500,-7750)
\path(10000,-10000)(10000,-6000)
\blacken\path(9750,-9250)(10250,-9250)(10000,-8750)(9750,-9250)
\blacken\path(9750,-7250)(10250,-7250)(10000,-6750)(9750,-7250)
\put(7000,-13000){\scriptsize$b_{-}(u,v)=[u-v]$}
\put(10000,-10750){\scriptsize{$v$}}
\path(10000,-12000)(10000,-11000)
\whiten\path(9750,-11500)(10250,-11500)(10000,-11000)(9750,-11500)

\path(18000,-8000)(22000,-8000)
\blacken\path(19250,-7750)(19250,-8250)(18750,-8000)(19250,-7750)
\blacken\path(20750,-7750)(20750,-8250)(21250,-8000)(20750,-7750)
\put(17250,-8000){\scriptsize{$u$}}
\path(16000,-8000)(17000,-8000)
\whiten\path(16500,-7750)(16500,-8250)(17000,-8000)(16500,-7750)
\path(20000,-10000)(20000,-6000)
\blacken\path(19750,-9250)(20250,-9250)(20000,-8750)(19750,-9250)
\blacken\path(19750,-6750)(20250,-6750)(20000,-7250)(19750,-6750)
\put(17000,-13000){\scriptsize$c_{-}(u,v)=[\gamma]$}
\put(20000,-10750){\scriptsize{$v$}}
\path(20000,-12000)(20000,-11000)
\whiten\path(19750,-11500)(20250,-11500)(20000,-11000)(19750,-11500)

\end{picture}

\end{minipage}
\end{center}

\caption[Six vertices associated to the XXZ $R$-matrix]{Six vertices associated to the XXZ $R$-matrix. Each entry of the $R$-matrix (\ref{Rmat1}) is matched with a vertex.} 

\label{R6v}
\end{figure}

The $R$-matrix corresponding to the XXZ spin-$\frac{1}{2}$ chain is given by

\begin{align}
R_{ab}(u,v)
=
\left(
\begin{array}{cccc}
[u-v+\gamma] & 0                          & 0              & 0    \\
0                                        & [u-v]    & [\gamma]            & 0    \\
0                                        & [\gamma]             & [u-v]  & 0    \\
0                            & 0                          & 0              & [u-v+\gamma] 
\end{array}
\right)_{ab}
\label{Rmat1}
\end{align}

\noindent where we have defined $[u]=2\sinh u$. The $R$-matrix is an element of ${\rm End}(\mathcal{V}_a\otimes \mathcal{V}_b)$. The variables $u,v$ are rapidities associated to the vector spaces $\mathcal{V}_a,\mathcal{V}_b$, and $\gamma$ is the crossing parameter, related to $q$ via the equation $q=e^{2\gamma}$. Recalling the conventions of chapter 2 we identify the entries of (\ref{Rmat1}) with vertices, as shown in figure \ref{R6v}.

The entries in the $R$-matrix (\ref{Rmat1}) are parametrized in terms of trigonometric functions. An alternative parametrization is obtained by defining 

\begin{align}
x=e^{2 u},\quad
y=e^{2 v}, \quad 
q=e^{2 \gamma}
\label{cov}
\end{align}

\noindent Letting $e^{u+v+\gamma} R_{ab}(u,v)=R'_{ab}(x,y)$ under the change of variables (\ref{cov}), we find
  
\begin{align}
R'_{ab}(x,y)
=
\left(
\begin{array}{cccc}
qx-y & 0                          & 0              & 0    \\
0                                        & q^{\frac{1}{2}}(x-y)    & x^{\frac{1}{2}} y^{\frac{1}{2}}(q-1)            & 0    \\
0                                        & x^{\frac{1}{2}} y^{\frac{1}{2}}(q-1)             & q^{\frac{1}{2}}(x-y)  & 0    \\
0                            & 0                          & 0              & qx-y 
\end{array}
\right)_{ab}
\label{Rmat2}
\end{align}

\noindent Either of the parametrizations (\ref{Rmat1}) or (\ref{Rmat2}) may be used in the quantum inverse scattering approach to the XXZ spin-$\frac{1}{2}$ chain. We will generally prefer to use (\ref{Rmat1}), since its matrix entries are neater. However the parametrization (\ref{Rmat2}) is necessary to produce polynomial KP $\tau$-functions. We reconcile this problem by performing all calculations in terms of (\ref{Rmat1}), and then making the simple change of variables (\ref{cov}) in those places where we discuss KP $\tau$-functions. 

\begin{lemma} 
{\rm
Define $\bar{u} = u+\gamma$ for all rapidities $u$. The $R$-matrix has the {\it crossing symmetry} property

\begin{align}
R_{ab}(u,v)
=
-\sigma_b^y R_{ba}(v,\bar{u})^{{\rm t}_b} \sigma_b^y
\label{cross}
\end{align}

\noindent where $\sigma_b^y$ is the second of the Pauli matrices (\ref{pauli}) acting in $\mathcal{V}_b$, and ${\rm t}_b$ denotes transposition in the space ${\rm End}(\mathcal{V}_b)$. 
}
\end{lemma}

\begin{proof} 
We have

\begin{align}
R_{ba}(v,\bar{u})^{{\rm t}_b}
=
\left(
\begin{array}{cccc}
[v-\bar{u}+\gamma] & 0 & 0 & [\gamma]
\\
0 & [v-\bar{u}] & 0 & 0
\\
0 & 0 & [v-\bar{u}] & 0
\\ 
\phantom{.}[\gamma] & 0 & 0 & [v-\bar{u}+\gamma]
\end{array}
\right)_{ba}
\end{align}

\noindent which leads to the equation

\begin{align}
-\sigma_b^{y} R_{ba}(v,\bar{u})^{{\rm t}_b} \sigma_b^{y}
=
\left(
\begin{array}{cccc}
[\bar{u}-v] & 0 & 0 & 0
\\
0 & [\bar{u}-v-\gamma] & [\gamma] & 0
\\
0 & [\gamma] & [\bar{u}-v-\gamma] & 0
\\
0 & 0 & 0 & [\bar{u}-v]
\end{array}
\right)_{ba}
\end{align}

\noindent Finally, using the definition $\bar{u} = u+\gamma$ and the fact that $R_{ba}(u,v) = R_{ab}(u,v)$, we prove (\ref{cross}). The crossing symmetry relation $(\ref{cross})$ will be essential in the calculations of later sections.

\end{proof}

\begin{lemma}
{\rm The $R$-matrix obeys the {\it Yang-Baxter equation} 

\begin{equation}
R_{ab}(u,v) R_{ac}(u,w) R_{bc}(v,w)
=
R_{bc}(v,w) R_{ac}(u,w) R_{ab}(u,v)
\label{YB}
\end{equation}

\noindent which holds in ${\rm End}(\mathcal{V}_a \otimes \mathcal{V}_b \otimes \mathcal{V}_c)$ for all $u,v,w$.
}
\end{lemma}

\begin{proof}
This is one of the most essential identities in quantum integrable models, \cite{jim}. It can be verified by direct calculation, but for simplicity we omit these details.
\end{proof}

\subsection{$L$-matrix and local intertwining equation}

The $L$-matrix for the XXZ model depends on a single indeterminate $u$, and acts in the space $\mathcal{V}_a$. Its entries are operators acting at the $m^{\rm th}$ lattice site, and identically everywhere else. It has the form

\begin{align}
L_{am}(u)
=
\left(
\begin{array}{cc}
[u+\frac{\gamma}{2}\sigma_m^{z}]
&
[\gamma] \sigma_m^{-}
\\
\phantom{.}
[\gamma] \sigma_m^{+}
&
[u-\frac{\gamma}{2} \sigma_m^{z}]
\end{array}
\right)_a
\label{xxz-Lmat}
\end{align}

\noindent where we have defined, as before, $[u] = 2\sinh u$. Using the definition of the $R$-matrix (\ref{Rmat1}) and the $L$-matrix (\ref{xxz-Lmat}), the local intertwining equation is given by

\begin{align}
R_{ab}(u,v) L_{am}(u) L_{bm}(v)
=
L_{bm}(v) L_{am}(u) R_{ab}(u,v)
\label{xxz-Lint}
\end{align}

\noindent This is $4\times 4$ matrix equation, which gives rise to sixteen scalar identities. Each of these identities may be checked by direct calculation, and by using the commutation relations (\ref{slq2-2}).

Specializing to the spin-$\frac{1}{2}$ representation of $\{\sigma_m^{+},\sigma_m^{-},\sigma_m^{z}\}$ given by equations (\ref{pauli}) and (\ref{spin-chang}), we find that the $L$-matrix (\ref{xxz-Lmat}) takes the form

\begin{align}
L_{am}(u)
=
\left(
\begin{array}{cccc}
[u+\frac{\gamma}{2}] & 0 & 0 & 0
\\
0 & [u-\frac{\gamma}{2}] & [\gamma] & 0
\\
0 & [\gamma] & [u-\frac{\gamma}{2}] & 0
\\
0 & 0 & 0 & [u+\frac{\gamma}{2}]
\end{array}
\right)_{am}
=
R_{am}(u,\gamma/2)
\end{align}

\noindent from which we see that it is equal to the $R$-matrix $R_{am}(u,w_m)$ with $w_m = \frac{\gamma}{2}$. In this representation, the local intertwining equation (\ref{xxz-Lint}) becomes

\begin{align}
R_{ab}(u,v) R_{am}(u,\gamma/2) R_{bm}(v,\gamma/2)
=
R_{bm}(v,\gamma/2) R_{am}(u,\gamma/2) R_{ab}(u,v)
\end{align}

\noindent which is simply a corollary of the Yang-Baxter equation (\ref{YB}).

\subsection{Monodromy matrix and global intertwining equation}

In the last subsection we observed that the XXZ $L$-matrix $L_{am}(u)$ is equal to the $R$-matrix $R_{am}(u,w_m)$ under the specialization $w_m =\frac{\gamma}{2}$. Using this observation, it is convenient to construct the monodromy matrix as an ordered product of the $R$-matrices $R_{am}(u,w_m)$, without restricting the variables $w_m$. That is, we define

\begin{equation}
T_{a}(u,\{w\}_M)
=
R_{a1}(u,w_1) \ldots R_{aM}(u,w_M)
\label{Tmat}
\end{equation}

\begin{figure}[H]

\begin{center}
\begin{minipage}{4.3in}

\setlength{\unitlength}{0.0004cm}
\begin{picture}(20000,15000)(-2500,-13000)

\path(-2000,0000)(10000,0000)
\blacken\path(-1250,250)(-1250,-250)(-750,0)(-1250,250)
\blacken\path(8750,250)(8750,-250)(9250,0)(8750,250)
\put(-2750,0){$u$}
\path(-4000,0)(-3000,0)
\whiten\path(-3500,250)(-3500,-250)(-3000,0)(-3500,250)
\path(0000,-2000)(0000,2000)
\put(-400,-2700){\scriptsize$w_1$}
\path(0,-4000)(0,-3000)
\whiten\path(-250,-3500)(250,-3500)(0,-3000)(-250,-3500)
\path(2000,-2000)(2000,2000)
\path(4000,-2000)(4000,2000)
\path(6000,-2000)(6000,2000)
\path(8000,-2000)(8000,2000)
\put(7800,-2700){\scriptsize$w_M$}
\path(8000,-4000)(8000,-3000)
\whiten\path(7750,-3500)(8250,-3500)(8000,-3000)(7750,-3500)
\put(1500,-4500)
{$A(u,\{w\}_M)$}

\path(14000,0000)(26000,0000)
\blacken\path(14750,250)(14750,-250)(15250,0)(14750,250)
\blacken\path(25250,250)(25250,-250)(24750,0)(25250,250)
\put(13250,0){$u$}
\path(12000,0)(13000,0)
\whiten\path(12500,250)(12500,-250)(13000,0)(12500,250)
\path(16000,-2000)(16000,2000)
\put(15600,-2700){\scriptsize$w_1$}
\path(16000,-4000)(16000,-3000)
\whiten\path(15750,-3500)(16250,-3500)(16000,-3000)(15750,-3500)
\path(18000,-2000)(18000,2000)
\path(20000,-2000)(20000,2000)
\path(22000,-2000)(22000,2000)
\path(24000,-2000)(24000,2000)
\put(23800,-2700){\scriptsize$w_M$}
\path(24000,-4000)(24000,-3000)
\whiten\path(23750,-3500)(24250,-3500)(24000,-3000)(23750,-3500)
\put(17500,-4500)
{$B(u,\{w\}_M)$}

\path(-2000,-8000)(10000,-8000)
\blacken\path(-750,-7750)(-750,-8250)(-1250,-8000)(-750,-7750)
\blacken\path(8750,-7750)(8750,-8250)(9250,-8000)(8750,-7750)
\put(-2750,-8000){$u$}
\path(-4000,-8000)(-3000,-8000)
\whiten\path(-3500,-7750)(-3500,-8250)(-3000,-8000)(-3500,-7750)
\path(0000,-10000)(0000,-6000)
\put(-400,-10700){\scriptsize$w_1$}
\path(0,-12000)(0,-11000)
\whiten\path(-250,-11500)(250,-11500)(0,-11000)(-250,-11500)
\path(2000,-10000)(2000,-6000)
\path(4000,-10000)(4000,-6000)
\path(6000,-10000)(6000,-6000)
\path(8000,-10000)(8000,-6000)
\put(7800,-10700){\scriptsize$w_M$}
\path(8000,-12000)(8000,-11000)
\whiten\path(7750,-11500)(8250,-11500)(8000,-11000)(7750,-11500)
\put(1500,-12500)
{$C(u,\{w\}_M)$}

\path(14000,-8000)(26000,-8000)
\blacken\path(15250,-7750)(15250,-8250)(14750,-8000)(15250,-7750)
\blacken\path(25250,-7750)(25250,-8250)(24750,-8000)(25250,-7750)
\put(13250,-8000){$u$}
\path(12000,-8000)(13000,-8000)
\whiten\path(12500,-7750)(12500,-8250)(13000,-8000)(12500,-7750)
\path(16000,-10000)(16000,-6000)
\put(15600,-10700){\scriptsize$w_1$}
\path(16000,-12000)(16000,-11000)
\whiten\path(15750,-11500)(16250,-11500)(16000,-11000)(15750,-11500)
\path(18000,-10000)(18000,-6000)
\path(20000,-10000)(20000,-6000)
\path(22000,-10000)(22000,-6000)
\path(24000,-10000)(24000,-6000)
\put(23800,-10700){\scriptsize$w_M$}
\path(24000,-12000)(24000,-11000)
\whiten\path(23750,-11500)(24250,-11500)(24000,-11000)(23750,-11500)
\put(17500,-12500)
{$D(u,\{w\}_M)$}

\end{picture}

\end{minipage}

\end{center}

\caption[Four vertex-strings of the XXZ monodromy matrix]{Four vertex-strings of the XXZ monodromy matrix. Each entry of (\ref{Tmat2}) is matched with a string of $R$-matrix vertices.}
\label{6vT}

\end{figure}
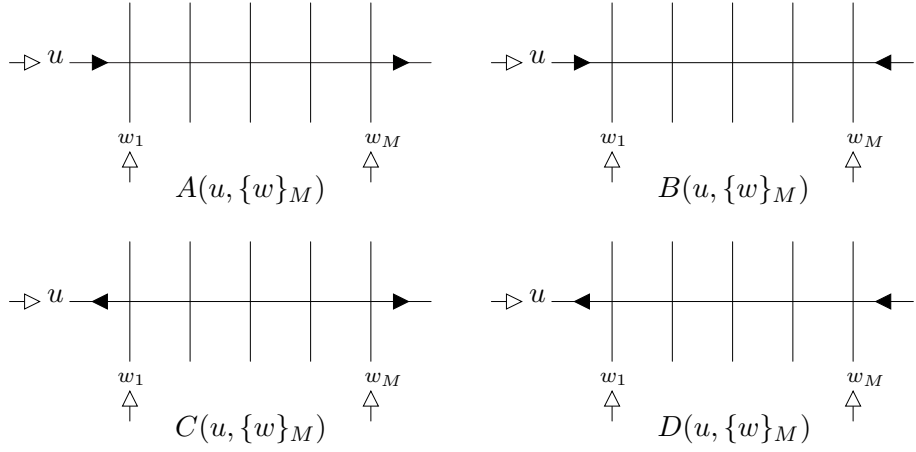

\noindent The variables $\{w_1,\ldots,w_M\}$ are called {\it inhomogeneities} and the usual monodromy matrix is recovered by setting $w_m = \frac{\gamma}{2}$ for all $1\leq m \leq M$. It turns out that the inclusion of the variables $\{w_1,\ldots,w_M\}$ simplifies many later calculations. As usual, the contribution from the space ${\rm End}(\mathcal{V}_a)$ can be exhibited explicitly by defining

\begin{equation}
T_{a}(u,\{w\}_M)
=
\left(
\begin{array}{cc}
A(u,\{w\}_M) & B(u,\{w\}_M) \\
C(u,\{w\}_M) & D(u,\{w\}_M)
\end{array}
\right)_a
\label{Tmat2}
\end{equation}

\noindent where the matrix entries are all operators acting in $\mathcal{V}=\mathcal{V}_1\otimes\cdots\otimes \mathcal{V}_M$. When writing the monodromy matrix and its entries, it is conventional to display dependence only on the variable associated to $\mathcal{V}_a$, letting $T_{a}(u,\{w\}_M) = T_{a}(u)$, $A(u,\{w\}_M) = A(u)$, $B(u,\{w\}_M) = B(u)$, $C(u,\{w\}_M) = C(u)$, $D(u,\{w\}_M) = D(u)$. The diagrammatic version of these operators is given in figure \ref{6vT}.

Owing to the Yang-Baxter equation (\ref{YB}) and the definition of the monodromy matrix (\ref{Tmat}), we obtain the global intertwining equation 

\begin{equation}
R_{ab}(u,v) T_{a}(u) T_{b}(v)
=
T_{b}(v) T_{a}(u) R_{ab}(u,v)
\label{int1}
\end{equation}

\noindent Explicitly writing out the entries of the monodromy matrix, we find that

\begin{eqnarray}
R_{ab}(u,v)
&
\left(
\begin{array}{cccc}
A(u) A(v) & A(u) B(v) & B(u) A(v) & B(u) B(v) \\
A(u) C(v) & A(u) D(v) & B(u) C(v) & B(u) D(v) \\
C(u) A(v) & C(u) B(v) & D(u) A(v) & D(u) B(v) \\
C(u) C(v) & C(u) D(v) & D(u) C(v) & D(u) D(v)
\end{array}
\right)_{ab}
&
\label{int2}
\\
=
&
\left(
\begin{array}{cccc}
A(v) A(u) & B(v) A(u) & A(v) B(u) & B(v) B(u) \\
C(v) A(u) & D(v) A(u) & C(v) B(u) & D(v) B(u) \\
A(v) C(u) & B(v) C(u) & A(v) D(u) & B(v) D(u) \\
C(v) C(u) & D(v) C(u) & C(v) D(u) & D(v) D(u)
\end{array}
\right)_{ab}
&
R_{ab}(u,v)
\nonumber
\end{eqnarray}

\noindent This equation contains sixteen commutation relations between the entries of the monodromy matrix. In the forthcoming calculations we will require

\begin{align}
&B(u) B(v) = B(v) B(u) \label{BB} \\ 
&[\gamma] B(u) D(v) + [u-v] D(u) B(v) = [u-v+\gamma] B(v) D(u) \label{DB} \\ 
&C(u) C(v) = C(v) C(u) \label{CC} \\ 
&[u-v+\gamma] C(u) D(v) = [u-v] D(v) C(u) + [\gamma] C(v) D(u) \label{CD}
\\
& D(u) D(v) = D(v) D(u) \label{DD} 
\end{align}

\noindent which are obtained by multiplying the matrices in (\ref{int2}) and equating components. 

\begin{lemma}
{\rm 
Let $\{u\}_{L} = \{u_1,\ldots,u_L\}$ and $\{w\}_M = \{w_1,\ldots,w_M\}$ be two sets of variables, with cardinalities $L,M \geq 1$. To each variable $u_l$ we associate an auxiliary space $\mathcal{V}_{a_l}$, while to each variable $w_m$ we associate a quantum space $\mathcal{V}_{m}$. Define

\begin{align}
T\Big(\{u\}_{L},\{w\}_{M}\Big)
=
T_{a_L}(u_L,\{w\}_M)
\ldots
T_{a_1}(u_1,\{w\}_M)
\label{doublemon}
\end{align}

\noindent We claim that

\begin{align}
T\Big(\{u\}_{L},\{w\}_{M}\Big)
=
(-)^{LM}
\overline{T}_1(w_1,\{\bar{u}\}_L)
\ldots
\overline{T}_M(w_M,\{\bar{u}\}_L)
\label{doublemon2}
\end{align}

\noindent where for all $1 \leq m \leq M$ we have defined

\begin{align}
\overline{T}_m(w_m,\{\bar{u}\}_L)
=
\left(
\begin{array}{rr}
D(w_m,\{\bar{u}\}_L) & -B(w_m,\{\bar{u}\}_L)
\\
-C(w_m,\{\bar{u}\}_L) & A(w_m,\{\bar{u}\}_L)
\end{array}
\right)_m
\label{doublemon!}
\end{align}

\noindent with $\{\bar{u}\}_L = \{u_1+\gamma,\ldots,u_L+\gamma\}$.
}
\end{lemma}

\begin{proof}
 
By the definition (\ref{doublemon}), we have

\begin{align}
&
T\Big(\{u\}_{L},\{w\}_M\Big)
=
\\
&
\Big(
R_{a_L 1}(u_L,w_1)
\ldots
R_{a_L M}(u_L,w_M)
\Big)
\ldots
\Big(
R_{a_1 1}(u_1,w_1)
\ldots
R_{a_1 M}(u_1,w_M)
\Big)
\nonumber
\end{align}

\noindent Commuting $R$-matrices which act in different spaces leads to the equation

\begin{align}
&
T\Big(\{u\}_{L},\{w\}_M\Big)
=
\label{every}
\\
&
\Big(
R_{a_L 1}(u_L,w_1)
\ldots
R_{a_1 1}(u_1,w_1)
\Big)
\ldots
\Big(
R_{a_L M}(u_L,w_M)
\ldots
R_{a_1 M}(u_1,w_M)
\Big)
\nonumber
\end{align}

\noindent and using the crossing symmetry relation (\ref{cross}) on every $R$-matrix in (\ref{every}) we obtain

\begin{align}
T\Big(\{u\}_{L},\{w\}_M\Big)
=
(-)^{LM}
&
\sigma_1^{y}
\Big(
R_{1 a_L}(w_1,\bar{u}_L)^{{\rm t}_1}
\ldots
R_{1 a_1}(w_1,\bar{u}_1)^{{\rm t}_1}
\Big)
\sigma_1^{y}
\label{every2}
\\
\times
\ldots
\times 
&
\sigma_M^{y}
\Big(
R_{M a_L}(w_M,\bar{u}_L)^{{\rm t}_M}
\ldots
R_{M a_1}(w_M,\bar{u}_1)^{{\rm t}_M}
\Big)
\sigma_M^{y}
\nonumber
\end{align}

\noindent Using a standard identity of matrix transposition we may reverse the order of the $R$-matrices in (\ref{every2}), yielding

\begin{align}
T\Big(\{u\}_{L},\{w\}_M\Big)
=
(-)^{LM}
&
\sigma_1^{y}
\Big(
R_{1 a_1}(w_1,\bar{u}_1)
\ldots
R_{1 a_L}(w_1,\bar{u}_L)
\Big)^{{\rm t}_1}
\sigma_1^{y}
\label{every3}
\\
\times
\ldots
\times
&
\sigma_M^{y}
\Big(
R_{M a_1}(w_M,\bar{u}_1)
\ldots
R_{M a_L}(w_M,\bar{u}_L)
\Big)^{{\rm t}_M}
\sigma_M^{y}
\nonumber
\end{align}

\noindent Finally we replace each parenthesized term in (\ref{every3}) with its corresponding monodromy matrix, which gives 

\begin{align}
T\Big(\{u\}_{L},\{w\}_M\Big)
=
(-)^{LM}
\sigma_1^{y}
T_1(w_1,\{\bar{u}\}_L)^{{\rm t}_1}
\sigma_1^{y}
\ldots
\sigma_M^{y}
T_M(w_M,\{\bar{u}\}_L)^{{\rm t}_M}
\sigma_M^{y}
\label{every4}
\end{align}

\noindent Letting the monodromy matrices in (\ref{every4}) be written in the form

\begin{align}
T_m(w_m,\{\bar{u}\}_L)
=
\left(
\begin{array}{cc}
A(w_m,\{\bar{u}\}_L) & B(w_m,\{\bar{u}\}_L)
\\
C(w_m,\{\bar{u}\}_L) & D(w_m,\{\bar{u}\}_L)
\end{array}
\right)_m
\end{align}

\noindent for all $1 \leq m \leq M$ and contracting on the quantum spaces $\mathcal{V}_1,\ldots,\mathcal{V}_M$, we recover the result (\ref{doublemon2}).

\end{proof}

\begin{lemma} 
{\rm 
The spin-up states $|\Uparrow_M\rangle$ and $\langle \Uparrow_M|$ are eigenvectors of the diagonal elements of the monodromy matrix. Explicitly, we have

\begin{align}
A(u,\{w\}_M)|\Uparrow_M\rangle
&=
\prod_{j=1}^{M} [u-w_j+\gamma] |\Uparrow_M\rangle,
\quad
D(u,\{w\}_M)|\Uparrow_M\rangle
=
\prod_{j=1}^{M} [u-w_j] |\Uparrow_M\rangle
\label{diagonal1}
\\
\langle \Uparrow_M | A(u,\{w\}_M)
&=
\prod_{j=1}^{M} [u-w_j+\gamma] \langle \Uparrow_M|,
\quad
\langle \Uparrow_M | D(u,\{w\}_M)
=
\prod_{j=1}^{M} [u-w_j] \langle \Uparrow_M | 
\label{diagonal1*}
\end{align}

\noindent In addition, the spin-down states $|\Downarrow_M\rangle$ and $\langle \Downarrow_M|$ are eigenvectors of the diagonal elements of the monodromy matrix. Explicitly, we have

\begin{align}
A(u,\{w\}_M)|\Downarrow_M\rangle
&=
\prod_{j=1}^{M} [u-w_j] |\Downarrow_M\rangle,
\quad
D(u,\{w\}_M)|\Downarrow_M\rangle
=
\prod_{j=1}^{M} [u-w_j+\gamma] |\Downarrow_M\rangle
\label{diagonal2}
\\
\langle \Downarrow_M | A(u,\{w\}_M)
&=
\prod_{j=1}^{M} [u-w_j] \langle \Downarrow_M|,
\quad
\langle \Downarrow_M | D(u,\{w\}_M)
=
\prod_{j=1}^{M} [u-w_j+\gamma] \langle \Downarrow_M | 
\label{diagonal2*}
\end{align}

}
\end{lemma}

\begin{proof}
See lemma \ref{2-lem2} in chapter 2.
\end{proof}

\subsection{Recovering $\mathcal{H}$ from the transfer matrix}

Let $t(u,\{w\}_M) = A(u,\{w\}_M)+D(u,\{w\}_M)$ denote the transfer matrix of the XXZ model. The Hamiltonian (\ref{hamiltonian}) is recovered via the formula

\begin{align}
\mathcal{H}
=
[\gamma]
\left.
\frac{d}{du}
\log \breve{t}(u)
\right|_{u = \frac{\gamma}{2}}
\quad
{\rm where}\quad
\breve{t}(u)
=
\left.
\frac{
t(u,\{w\}_M)
}{
\prod_{j=1}^{M}[u-w_j+\gamma]
}
\right|_{w_1=\cdots = w_M =\frac{\gamma}{2}}
\end{align}

\noindent Therefore all eigenvectors of $t(u,\{w\}_M)$ are also eigenvectors of $\mathcal{H}$. Our attention turns, therefore, to finding vectors $|\Psi\rangle \in \mathcal{V}$ satisfying

\begin{align}
\Big(
A(u,\{w\}_M)+D(u,\{w\}_M)
\Big)
|\Psi\rangle = 
\tau_{\Psi}(u,\{w\}_M) |\Psi\rangle
\label{eigenvector}
\end{align}

\noindent for some suitable constants $\tau_{\Psi}(u,\{w\}_M)$.

\subsection{Bethe Ansatz for the eigenvectors}

As we explained in a general setting in chapter 2, the eigenvectors of $t(u,\{w\}_M)$ are given by the Ansatz 

\begin{equation}
|\Psi\rangle = B(v_1,\{w\}_M)\ldots B(v_N,\{w\}_M)|\Uparrow_M\rangle
\label{bethe1}
\end{equation}

\noindent where we assume that $N \leq M$, since we annihilate the state $|\Uparrow_M\rangle$ when acting with more $B$-operators than the number of sites in the spin chain. Similarly, we construct eigenvectors of $t(u,\{w\}_M)$ in the dual space of states via the Ansatz 

\begin{align}
\langle \Psi|
=
\langle \Uparrow_M|
C(v_N,\{w\}_M)
\ldots
C(v_1,\{w\}_M)
\label{bethe3}
\end{align}

\noindent where we again restrict $N \leq M$. To ensure that (\ref{bethe1}) and (\ref{bethe3}) are genuine eigenvectors, the variables $\{v_1,\ldots,v_N\}$ are required to satisfy the Bethe equations. Using the commutation relations (\ref{int2}) and the actions (\ref{diagonal1}), it is possible to show that $|\Psi\rangle,\langle \Psi|$ are eigenvectors of $t(u,\{w\}_M)$ if and only if 

\begin{equation}
\prod_{j=1}^{M}\frac{ [v_i-w_j+\gamma]}{[v_i-w_j]}
=
\prod_{j \not= i}^{N}
\frac{[v_i-v_j + \gamma]}{[v_i - v_j - \gamma]}
\label{bethe2}
\end{equation}

\noindent for all $1\leq i \leq N$.\footnote{See theorem 1 in chapter 2. Notice that for the present model

\begin{align}
a(v_i,v_j) = [v_i-v_j+\gamma],
\quad
\alpha(v_i,\{w\}_M) = \prod_{j=1}^{M}[v_i-w_j+\gamma],
\quad
\delta(v_i,\{w\}_M) = \prod_{j=1}^{M}[v_i-w_j]
\end{align}

\noindent which, when substituted into (\ref{Beteq}), produce the equations (\ref{bethe2}).}
Later in the chapter we will find an explicit expression for the Bethe eigenvectors (\ref{bethe1}) and (\ref{bethe3}) by expanding them in terms of the bases (\ref{elem}) and (\ref{elem*}), whereby we have 

\begin{align}
\prod_{i=1}^{N} B(v_i,\{w\}_M)|\Uparrow_M\rangle
&=
\sum_{\lambda | \ell(\lambda) = M-N} 
b_{\lambda} \Big(\{v\}_N,\{w\}_M\Big) |\lambda\rangle
\label{b}
\\
\langle \Uparrow_M | \prod_{i=1}^{N} C(v_i,\{w\}_M)
&=
\sum_{\lambda | \ell(\lambda) = M-N}
c_{\lambda} \Big(\{v\}_N,\{w\}_M\Big) \langle \lambda|
\label{c*}
\end{align}

\noindent with both sums taken over all strict partitions $\lambda$ of $M-N$ integers which satisfy $\{M \geq \lambda_1 > \cdots > \lambda_{M-N} \geq 1\}$. The focus of section 5.4 will be the calculation of the coefficients $b_{\lambda}(\{v\}_N,\{w\}_M)$ and $c_{\lambda}(\{v\}_N,\{w\}_M)$.

\section{Domain wall partition function}
\label{xxz-pf}

In this section we will study $Z_N$, the domain wall partition function of the XXZ spin-$\frac{1}{2}$ chain. This quantity acquires its name because it is equal to the partition function of the six-vertex model under domain wall boundary conditions. As we will see in sections 5.3 and 5.4, the calculation of $Z_N$ is essential for the explicit evaluation of more complicated objects within the XXZ model, such as its scalar product and Bethe eigenvectors. 

\subsection{Definition of $Z_N(\{v\}_N,\{w\}_N)$}

Let $\{v\}_N = \{v_1,\ldots,v_N\}$ and $\{w\}_N = \{w_1,\ldots,w_N\}$ be two sets of variables. The domain wall partition function has the algebraic definition

\begin{align}
Z_N\Big(\{v\}_N,\{w\}_N\Big)
=
\langle \Downarrow_N | \prod_{i=1}^{N} B(v_i,\{w\}_N) |\Uparrow_N \rangle
\label{pf-xxz}
\end{align}

\noindent where the ordering of the $B$-operators is irrelevant, since from equation (\ref{BB}) we know that they commute. 

The relationship of $Z_N(\{v\}_N,\{w\}_N)$ to the coefficients in equation (\ref{b}) is easily exposed. Consider the Bethe eigenstate (\ref{b}) in the case when $M=N$, which corresponds to acting with the same number of $B$-operators as the number of sites on the spin chain. In this situation we obtain

\begin{align}
\prod_{i=1}^{N}
B(v_i,\{w\}_N)
|\Uparrow_N\rangle
=
b_{\emptyset}\Big(
\{v\}_N,\{w\}_N
\Big)
|\Downarrow_N\rangle
\end{align}

\noindent from which we see that $b_{\emptyset}(\{v\}_N,\{w\}_N) = Z_N(\{v\}_N,\{w\}_N)$.  Hence the domain wall partition function is, in some sense, the simplest non-trivial example amongst the coefficients $b_{\lambda}(\{v\}_N,\{w\}_M)$.  

\subsection{Graphical representation of partition function}

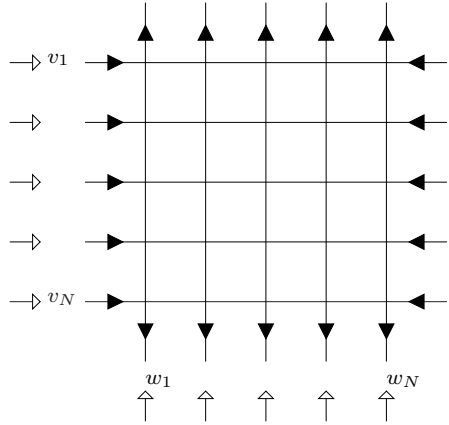
\begin{figure}[H]

\begin{center}
\begin{minipage}{4.3in}

\setlength{\unitlength}{0.0004cm}
\begin{picture}(20000,12500)(-10000,-3000)

\path(-2000,0)(10000,0)
\put(-3250,0){\scriptsize$v_N$}
\path(-4500,0)(-3500,0)
\whiten\path(-3750,250)(-3750,-250)(-3500,0)(-3750,250)
\blacken\path(-1250,250)(-1250,-250)(-750,0)(-1250,250)
\blacken\path(9250,250)(9250,-250)(8750,0)(9250,250)

\path(-2000,2000)(10000,2000)
\path(-4500,2000)(-3500,2000)
\whiten\path(-3750,2250)(-3750,1750)(-3500,2000)(-3750,2250)
\blacken\path(-1250,2250)(-1250,1750)(-750,2000)(-1250,2250)
\blacken\path(9250,2250)(9250,1750)(8750,2000)(9250,2250)

\path(-2000,4000)(10000,4000)
\path(-4500,4000)(-3500,4000)
\whiten\path(-3750,4250)(-3750,3750)(-3500,4000)(-3750,4250)
\blacken\path(-1250,4250)(-1250,3750)(-750,4000)(-1250,4250)
\blacken\path(9250,4250)(9250,3750)(8750,4000)(9250,4250)

\path(-2000,6000)(10000,6000)
\path(-4500,6000)(-3500,6000)
\whiten\path(-3750,6250)(-3750,5750)(-3500,6000)(-3750,6250)
\blacken\path(-1250,6250)(-1250,5750)(-750,6000)(-1250,6250)
\blacken\path(9250,6250)(9250,5750)(8750,6000)(9250,6250)

\path(-2000,8000)(10000,8000)
\put(-3250,8000){\scriptsize$v_1$}
\path(-4500,8000)(-3500,8000)
\whiten\path(-3750,8250)(-3750,7750)(-3500,8000)(-3750,8250)
\blacken\path(-1250,8250)(-1250,7750)(-750,8000)(-1250,8250)
\blacken\path(9250,8250)(9250,7750)(8750,8000)(9250,8250)


\path(0,-2000)(0,10000)
\put(0,-2750){\scriptsize$w_1$}
\path(0,-4000)(0,-3000)
\whiten\path(-250,-3250)(250,-3250)(0,-3000)(-250,-3250)
\blacken\path(-250,-750)(250,-750)(0,-1250)(-250,-750)
\blacken\path(-250,8750)(250,8750)(0,9250)(-250,8750)

\path(2000,-2000)(2000,10000)
\path(2000,-4000)(2000,-3000)
\whiten\path(1750,-3250)(2250,-3250)(2000,-3000)(1750,-3250)
\blacken\path(1750,-750)(2250,-750)(2000,-1250)(1750,-750)
\blacken\path(1750,8750)(2250,8750)(2000,9250)(1750,8750)

\path(4000,-2000)(4000,10000)
\path(4000,-4000)(4000,-3000)
\whiten\path(3750,-3250)(4250,-3250)(4000,-3000)(3750,-3250)
\blacken\path(3750,-750)(4250,-750)(4000,-1250)(3750,-750)
\blacken\path(3750,8750)(4250,8750)(4000,9250)(3750,8750)

\path(6000,-2000)(6000,10000)
\path(6000,-4000)(6000,-3000)
\whiten\path(5750,-3250)(6250,-3250)(6000,-3000)(5750,-3250)
\blacken\path(5750,-750)(6250,-750)(6000,-1250)(5750,-750)
\blacken\path(5750,8750)(6250,8750)(6000,9250)(5750,8750)

\path(8000,-2000)(8000,10000)
\put(8000,-2750){\scriptsize$w_N$}
\path(8000,-4000)(8000,-3000)
\whiten\path(7750,-3250)(8250,-3250)(8000,-3000)(7750,-3250)
\blacken\path(7750,-750)(8250,-750)(8000,-1250)(7750,-750)
\blacken\path(7750,8750)(8250,8750)(8000,9250)(7750,8750)

\end{picture}

\end{minipage}
\end{center}

\caption[Domain wall partition function of the six-vertex model]{Domain wall partition function of the six-vertex model. The top row of upward pointing arrows corresponds with the state vector $|\Uparrow_N\rangle$. The bottom row of downward pointing arrows corresponds with the dual state vector $\langle \Downarrow_N|$. Each horizontal lattice line corresponds to multiplication by a $B$-operator.}

\label{pf6v}
\end{figure}

Using the graphical conventions described in the previous section, the domain wall partition function may be represented as the $N\times N$ lattice shown in figure \ref{pf6v}.

\subsection{Conditions on $Z_N(\{v\}_N,\{w\}_N)$}

In \cite{kor} Korepin showed that the function $Z_N(\{v\}_N,\{w\}_N)$ satisfies a set of four conditions which determine it uniquely. We reproduce these facts below, with the following two lemmas.

\begin{lemma}
{\rm Let us adopt the shorthand $Z_N = Z_N(\{v\}_N,\{w\}_N)$. For all $N \geq 2$ we claim that 

\begin{korepin}
{\rm
$Z_N$ is symmetric in the $\{ w \}_N$ variables.
}
\end{korepin}

\begin{korepin}
{\rm
$Z_N$ is a trigonometric polynomial of degree $N-1$ in the rapidity variable $v_N$.
}
\end{korepin}

\begin{korepin}
{\rm
Setting $v_N = w_N-\gamma$, $Z_{N}$ satisfies the recursion relation 
\begin{align}
Z_N\Big|_{v_N=w_N-\gamma}
&=
[\gamma]
\prod_{i=1}^{N-1}
[v_i-w_N][w_N-w_i-\gamma]
Z_{N-1}
\label{recursion}
\end{align}

\noindent where $Z_{N-1}$ is the domain wall partition function on a square lattice of size $N-1$.
}
\end{korepin}

In addition, we have the supplementary condition 

\begin{korepin}
{\rm The partition function on the $1\times 1$ lattice is given by $Z_{1} = [\gamma]$.
}
\end{korepin}
}
\end{lemma}

\begin{proof}

\begin{propertylist}
{\rm 
We write the domain wall partition function in the form

\begin{align}
Z_N\Big(\{v\}_N,\{w\}_N\Big)
=
\langle \Uparrow_N^a|
\otimes
\langle \Downarrow_N|
T\Big(
\{v\}_N,\{w\}_N
\Big)
|\Uparrow_N\rangle
\otimes
|\Downarrow_N^a\rangle
\end{align}

\noindent with $T(\{v\}_N,\{w\}_N)$ given by (\ref{doublemon}), and where we have defined the auxiliary states

\begin{align}
\langle \Uparrow_N^a|
=
\bigotimes_{i=1}^{N} 
\uparrow_{a_i}^{*},
\quad
|\Downarrow_N^a\rangle
=
\bigotimes_{i=1}^{N}
\downarrow_{a_i}
\end{align}

\noindent Using the expressions (\ref{doublemon2}), (\ref{doublemon!}) for $T(\{v\}_N,\{w\}_N)$ and contracting on the quantum spaces $\mathcal{V}_1,\ldots,\mathcal{V}_N$ gives

\begin{align}
Z_N\Big(
\{v\}_N,\{w\}_N
\Big)
=
\langle \Uparrow_N^a|
C(w_1,\{\bar{v}\}_N)
\ldots
C(w_N,\{\bar{v}\}_N)
|\Downarrow_N^a\rangle
\label{zequiv}
\end{align}

\noindent The diagrammatic interpretation of the equivalence (\ref{zequiv}) is shown in figure \ref{zequiv2}.


\begin{figure}[H]

\begin{center}
\begin{minipage}{4.3in}

\setlength{\unitlength}{0.000325cm}
\begin{picture}(20000,12500)(-4000,-3000)

\path(-2000,0)(10000,0)
\put(-3250,0){\scriptsize$v_N$}
\path(-4500,0)(-3500,0)
\whiten\path(-3750,250)(-3750,-250)(-3500,0)(-3750,250)
\blacken\path(-1250,250)(-1250,-250)(-750,0)(-1250,250)
\blacken\path(9250,250)(9250,-250)(8750,0)(9250,250)

\path(-2000,2000)(10000,2000)
\path(-4500,2000)(-3500,2000)
\whiten\path(-3750,2250)(-3750,1750)(-3500,2000)(-3750,2250)
\blacken\path(-1250,2250)(-1250,1750)(-750,2000)(-1250,2250)
\blacken\path(9250,2250)(9250,1750)(8750,2000)(9250,2250)

\path(-2000,4000)(10000,4000)
\path(-4500,4000)(-3500,4000)
\whiten\path(-3750,4250)(-3750,3750)(-3500,4000)(-3750,4250)
\blacken\path(-1250,4250)(-1250,3750)(-750,4000)(-1250,4250)
\blacken\path(9250,4250)(9250,3750)(8750,4000)(9250,4250)

\path(-2000,6000)(10000,6000)
\path(-4500,6000)(-3500,6000)
\whiten\path(-3750,6250)(-3750,5750)(-3500,6000)(-3750,6250)
\blacken\path(-1250,6250)(-1250,5750)(-750,6000)(-1250,6250)
\blacken\path(9250,6250)(9250,5750)(8750,6000)(9250,6250)

\path(-2000,8000)(10000,8000)
\put(-3250,8000){\scriptsize$v_1$}
\path(-4500,8000)(-3500,8000)
\whiten\path(-3750,8250)(-3750,7750)(-3500,8000)(-3750,8250)
\blacken\path(-1250,8250)(-1250,7750)(-750,8000)(-1250,8250)
\blacken\path(9250,8250)(9250,7750)(8750,8000)(9250,8250)


\path(0,-2000)(0,10000)
\put(0,-2750){\scriptsize$w_1$}
\path(0,-4000)(0,-3000)
\whiten\path(-250,-3250)(250,-3250)(0,-3000)(-250,-3250)
\blacken\path(-250,-750)(250,-750)(0,-1250)(-250,-750)
\blacken\path(-250,8750)(250,8750)(0,9250)(-250,8750)

\path(2000,-2000)(2000,10000)
\path(2000,-4000)(2000,-3000)
\whiten\path(1750,-3250)(2250,-3250)(2000,-3000)(1750,-3250)
\blacken\path(1750,-750)(2250,-750)(2000,-1250)(1750,-750)
\blacken\path(1750,8750)(2250,8750)(2000,9250)(1750,8750)

\path(4000,-2000)(4000,10000)
\path(4000,-4000)(4000,-3000)
\whiten\path(3750,-3250)(4250,-3250)(4000,-3000)(3750,-3250)
\blacken\path(3750,-750)(4250,-750)(4000,-1250)(3750,-750)
\blacken\path(3750,8750)(4250,8750)(4000,9250)(3750,8750)

\path(6000,-2000)(6000,10000)
\path(6000,-4000)(6000,-3000)
\whiten\path(5750,-3250)(6250,-3250)(6000,-3000)(5750,-3250)
\blacken\path(5750,-750)(6250,-750)(6000,-1250)(5750,-750)
\blacken\path(5750,8750)(6250,8750)(6000,9250)(5750,8750)

\path(8000,-2000)(8000,10000)
\put(8000,-2750){\scriptsize$w_N$}
\path(8000,-4000)(8000,-3000)
\whiten\path(7750,-3250)(8250,-3250)(8000,-3000)(7750,-3250)
\blacken\path(7750,-750)(8250,-750)(8000,-1250)(7750,-750)
\blacken\path(7750,8750)(8250,8750)(8000,9250)(7750,8750)


\put(12000,3750){$=$}


\path(18000,0)(30000,0)
\put(16250,0){\scriptsize$w_1$}
\path(15000,0)(16000,0)
\whiten\path(15750,250)(15750,-250)(16000,0)(15750,250)
\blacken\path(19250,250)(19250,-250)(18750,0)(19250,250)
\blacken\path(28750,250)(28750,-250)(29250,0)(28750,250)

\path(18000,2000)(30000,2000)
\path(15000,2000)(16000,2000)
\whiten\path(15750,2250)(15750,1750)(16000,2000)(15750,2250)
\blacken\path(19250,2250)(19250,1750)(18750,2000)(19250,2250)
\blacken\path(28750,2250)(28750,1750)(29250,2000)(28750,2250)

\path(18000,4000)(30000,4000)
\path(15000,4000)(16000,4000)
\whiten\path(15750,4250)(15750,3750)(16000,4000)(15750,4250)
\blacken\path(19250,4250)(19250,3750)(18750,4000)(19250,4250)
\blacken\path(28750,4250)(28750,3750)(29250,4000)(28750,4250)

\path(18000,6000)(30000,6000)
\path(15000,6000)(16000,6000)
\whiten\path(15750,6250)(15750,5750)(16000,6000)(15750,6250)
\blacken\path(19250,6250)(19250,5750)(18750,6000)(19250,6250)
\blacken\path(28750,6250)(28750,5750)(29250,6000)(28750,6250)

\path(18000,8000)(30000,8000)
\put(16250,8000){\scriptsize$w_N$}
\path(15000,8000)(16000,8000)
\whiten\path(15750,8250)(15750,7750)(16000,8000)(15750,8250)
\blacken\path(19250,8250)(19250,7750)(18750,8000)(19250,8250)
\blacken\path(28750,8250)(28750,7750)(29250,8000)(28750,8250)


\path(20000,-2000)(20000,10000)
\put(20000,-2750){\scriptsize$\bar{v}_1$}
\path(20000,-4000)(20000,-3000)
\whiten\path(19750,-3250)(20250,-3250)(20000,-3000)(19750,-3250)
\blacken\path(19750,-1250)(20250,-1250)(20000,-750)(19750,-1250)
\blacken\path(19750,9250)(20250,9250)(20000,8750)(19750,9250)

\path(22000,-2000)(22000,10000)
\path(22000,-4000)(22000,-3000)
\whiten\path(21750,-3250)(22250,-3250)(22000,-3000)(21750,-3250)
\blacken\path(21750,-1250)(22250,-1250)(22000,-750)(21750,-1250)
\blacken\path(21750,9250)(22250,9250)(22000,8750)(21750,9250)

\path(24000,-2000)(24000,10000)
\path(24000,-4000)(24000,-3000)
\whiten\path(23750,-3250)(24250,-3250)(24000,-3000)(23750,-3250)
\blacken\path(23750,-1250)(24250,-1250)(24000,-750)(23750,-1250)
\blacken\path(23750,9250)(24250,9250)(24000,8750)(23750,9250)

\path(26000,-2000)(26000,10000)
\path(26000,-4000)(26000,-3000)
\whiten\path(25750,-3250)(26250,-3250)(26000,-3000)(25750,-3250)
\blacken\path(25750,-1250)(26250,-1250)(26000,-750)(25750,-1250)
\blacken\path(25750,9250)(26250,9250)(26000,8750)(25750,9250)

\path(28000,-2000)(28000,10000)
\put(28000,-2750){\scriptsize$\bar{v}_N$}
\path(28000,-4000)(28000,-3000)
\whiten\path(27750,-3250)(28250,-3250)(28000,-3000)(27750,-3250)
\blacken\path(27750,-1250)(28250,-1250)(28000,-750)(27750,-1250)
\blacken\path(27750,9250)(28250,9250)(28000,8750)(27750,9250)

\end{picture}

\end{minipage}
\end{center}

\caption[Equivalent expressions for the domain wall partition function]{Equivalent expressions for the domain wall partition function. The diagram on the left represents a stacking of the operators $B(v_i,\{w\}_N)$. The diagram on the right represents a stacking of the operators $C(w_i,\{\bar{v}\}_N)$.}

\label{zequiv2}
\end{figure}
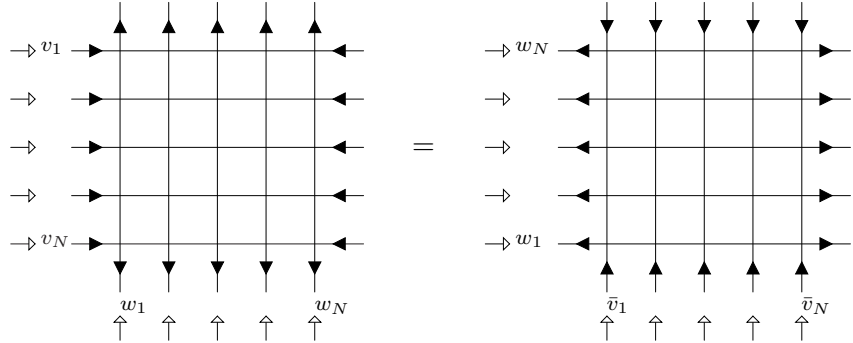

\noindent Thanks to equation (\ref{CC}) the $C(w_i,\{\bar{v}\}_N)$ operators all commute, proving that $Z_N(\{v\}_N,\{w\}_N)$ is symmetric in $\{w\}_N$. 
}
\end{propertylist}

\begin{propertylist}
{\rm 
By inserting the set of states $\sum_{n=1}^{N} \sigma_n^{+} |\Downarrow_N\rangle \langle \Downarrow_N| \sigma_n^{-}$ after the first $B$-operator appearing in (\ref{pf-xxz}), we obtain the expansion

\begin{align}
Z_N\Big(\{v\}_N,\{w\}_N\Big)
&=
\sum_{n=1}^{N}
\langle \Downarrow_N|
B(v_N,\{w\}_N)
\sigma^{+}_n |\Downarrow_N\rangle
\langle \Downarrow_N| \sigma^{-}_n
\prod_{i=1}^{N-1} B(v_i,\{w\}_N)
|\Uparrow_N\rangle
\label{pfexp-xxz}
\end{align}

\noindent in which all dependence on $v_N$ appears in the first factor within the sum. Hence we proceed to calculate $\langle \Downarrow_N| B(v_N,\{w\}_N) \sigma_n^{+} |\Downarrow_N\rangle$ for all $1 \leq n \leq N$, as shown below. 

%


\begin{figure}[H]

\begin{center}
\begin{minipage}{4.3in}

\setlength{\unitlength}{0.00035cm}
\begin{picture}(20000,6000)(-3000,-3000)

\path(-2000,0)(10000,0)
\put(-3250,0){\scriptsize$v_N$}
\path(-4500,0)(-3500,0)
\whiten\path(-3750,250)(-3750,-250)(-3500,0)(-3750,250)
\blacken\path(-1250,250)(-1250,-250)(-750,0)(-1250,250)
\blacken\path(9250,250)(9250,-250)(8750,0)(9250,250)


\path(0,-2000)(0,2000)
\put(0,-2750){\scriptsize$w_1$}
\path(0,-4000)(0,-3000)
\whiten\path(-250,-3250)(250,-3250)(0,-3000)(-250,-3250)
\blacken\path(-250,-750)(250,-750)(0,-1250)(-250,-750)
\blacken\path(-250,1250)(250,1250)(0,750)(-250,1250)

\path(2000,-2000)(2000,2000)
\path(2000,-4000)(2000,-3000)
\whiten\path(1750,-3250)(2250,-3250)(2000,-3000)(1750,-3250)
\blacken\path(1750,-750)(2250,-750)(2000,-1250)(1750,-750)
\blacken\path(1750,1250)(2250,1250)(2000,750)(1750,1250)

\path(4000,-2000)(4000,2000)
\put(4000,-2750){\scriptsize$w_n$}
\path(4000,-4000)(4000,-3000)
\whiten\path(3750,-3250)(4250,-3250)(4000,-3000)(3750,-3250)
\blacken\path(3750,-750)(4250,-750)(4000,-1250)(3750,-750)
\blacken\path(3750,750)(4250,750)(4000,1250)(3750,750)

\path(6000,-2000)(6000,2000)
\path(6000,-4000)(6000,-3000)
\whiten\path(5750,-3250)(6250,-3250)(6000,-3000)(5750,-3250)
\blacken\path(5750,-750)(6250,-750)(6000,-1250)(5750,-750)
\blacken\path(5750,1250)(6250,1250)(6000,750)(5750,1250)

\path(8000,-2000)(8000,2000)
\put(8000,-2750){\scriptsize$w_N$}
\path(8000,-4000)(8000,-3000)
\whiten\path(7750,-3250)(8250,-3250)(8000,-3000)(7750,-3250)
\blacken\path(7750,-750)(8250,-750)(8000,-1250)(7750,-750)
\blacken\path(7750,1250)(8250,1250)(8000,750)(7750,1250)


\put(12000,-250){$=$}


\path(18000,0)(30000,0)
\put(16750,0){\scriptsize$v_N$}
\path(15500,0)(16500,0)
\whiten\path(16250,250)(16250,-250)(16500,0)(16250,250)
\blacken\path(18750,250)(18750,-250)(19250,0)(18750,250)
\blacken\path(20750,250)(20750,-250)(21250,0)(20750,250)
\blacken\path(22750,250)(22750,-250)(23250,0)(22750,250)
\blacken\path(25250,250)(25250,-250)(24750,0)(25250,250)
\blacken\path(27250,250)(27250,-250)(26750,0)(27250,250)
\blacken\path(29250,250)(29250,-250)(28750,0)(29250,250)


\path(20000,-2000)(20000,2000)
\put(20000,-2750){\scriptsize$w_1$}
\path(20000,-4000)(20000,-3000)
\whiten\path(19750,-3250)(20250,-3250)(20000,-3000)(19750,-3250)
\blacken\path(19750,-750)(20250,-750)(20000,-1250)(19750,-750)
\blacken\path(19750,1250)(20250,1250)(20000,750)(19750,1250)

\path(22000,-2000)(22000,2000)
\path(22000,-4000)(22000,-3000)
\whiten\path(21750,-3250)(22250,-3250)(22000,-3000)(21750,-3250)
\blacken\path(21750,-750)(22250,-750)(22000,-1250)(21750,-750)
\blacken\path(21750,1250)(22250,1250)(22000,750)(21750,1250)

\path(24000,-2000)(24000,2000)
\put(24000,-2750){\scriptsize$w_n$}
\path(24000,-4000)(24000,-3000)
\whiten\path(23750,-3250)(24250,-3250)(24000,-3000)(23750,-3250)
\blacken\path(23750,-750)(24250,-750)(24000,-1250)(23750,-750)
\blacken\path(23750,750)(24250,750)(24000,1250)(23750,750)

\path(26000,-2000)(26000,2000)
\path(26000,-4000)(26000,-3000)
\whiten\path(25750,-3250)(26250,-3250)(26000,-3000)(25750,-3250)
\blacken\path(25750,-750)(26250,-750)(26000,-1250)(25750,-750)
\blacken\path(25750,1250)(26250,1250)(26000,750)(25750,1250)

\path(28000,-2000)(28000,2000)
\put(28000,-2750){\scriptsize$w_N$}
\path(28000,-4000)(28000,-3000)
\whiten\path(27750,-3250)(28250,-3250)(28000,-3000)(27750,-3250)
\blacken\path(27750,-750)(28250,-750)(28000,-1250)(27750,-750)
\blacken\path(27750,1250)(28250,1250)(28000,750)(27750,1250)

\end{picture}

\end{minipage}
\end{center}

\caption[Peeling away the bottom row of the partition function]{Peeling away the bottom row of the partition function. The diagram on the left represents $\langle \Downarrow_N |B(v_N,\{w\}_N) \sigma_n^{+} |\Downarrow_N \rangle$, with the internal black arrows being summed over all configurations. The diagram on the right represents the only surviving configuration.}

\label{zequiv3}
\end{figure}

\noindent The right hand side of figure \ref{zequiv3} is simply a product of vertices. Replacing each vertex with its corresponding trigonometric weight, we conclude that

\begin{align}
\langle \Downarrow_N|
B(v_N,\{w\}_N)
\sigma_n^{+} |\Downarrow_N\rangle
=
\prod_{1\leq i<n} [v_N-w_i]
[\gamma]
\prod_{n < i \leq N} [v_N-w_i+\gamma]
\label{zequiv4}
\end{align}

\noindent Substituting (\ref{zequiv4}) into the expansion (\ref{pfexp-xxz}) gives 

\begin{align}
&
Z_N\Big( \{v\}_N,\{w\}_N \Big)
=
\label{PFexp1}
\\
&
[\gamma]
\sum_{n=1}^{N}
\prod_{1\leq i<n}
[v_N-w_i]
\prod_{n < i \leq N}
[v_N-w_i+\gamma]
\langle \Downarrow_N|\sigma_n^{-}
\prod_{i=1}^{N-1} B(v_i,\{w\}_N)
|\Uparrow_N\rangle
\nonumber
\end{align}

\noindent From this equation we see that every term in $Z_N(\{v\}_N,\{w\}_N)$ contains a product of exactly $N-1$ trigonometric functions with argument $v_N$. Therefore $Z_N(\{v\}_N,\{w\}_N)$ is a trigonometric polynomial of degree $N-1$ in the variable $v_N$. 
}
\end{propertylist}

\begin{propertylist}
{\rm
We start from the expansion (\ref{PFexp1}) of the domain wall partition function, and set $v_N=w_N-\gamma$. This causes all terms in the summation over $1\leq n \leq N$ to collapse to zero except the $n=N$ term, giving

\begin{align}
Z_N\Big(\{v\}_N,\{w\}_N\Big) \Big|_{v_N=w_N-\gamma}
&=
[\gamma]
\prod_{i=1}^{N-1}
[w_N-w_i-\gamma]
\langle \Downarrow_N|
\sigma_N^{-}
\prod_{i=1}^{N-1} B(v_i,\{w\}_N)
|\Uparrow_N\rangle
\label{PFexp3}
\end{align}

\noindent We then consider the graphical representation of $\langle \Downarrow_N | \sigma_N^{-} \prod_{i=1}^{N-1} B(v_i,\{w\}_N) |\Uparrow_N\rangle$, as shown below.

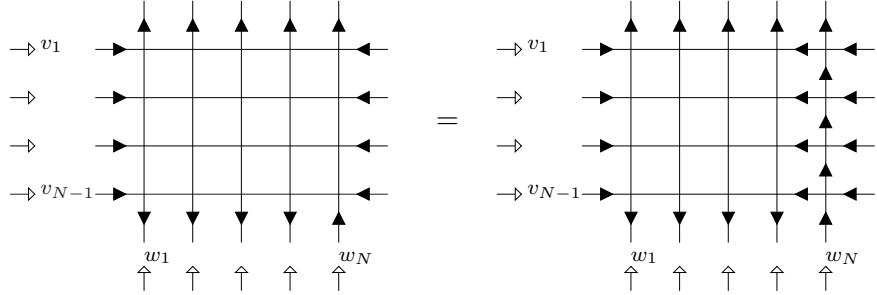
\begin{figure}[H]

\begin{center}
\begin{minipage}{4.3in}

\setlength{\unitlength}{0.000325cm}
\begin{picture}(20000,12500)(-4000,-3000)

\path(-2000,0)(10000,0)
\put(-4250,0){\scriptsize$v_{N-1}$}
\path(-5500,0)(-4500,0)
\whiten\path(-4750,250)(-4750,-250)(-4500,0)(-4750,250)
\blacken\path(-1250,250)(-1250,-250)(-750,0)(-1250,250)
\blacken\path(9250,250)(9250,-250)(8750,0)(9250,250)

\path(-2000,2000)(10000,2000)
\path(-5500,2000)(-4500,2000)
\whiten\path(-4750,2250)(-4750,1750)(-4500,2000)(-4750,2250)
\blacken\path(-1250,2250)(-1250,1750)(-750,2000)(-1250,2250)
\blacken\path(9250,2250)(9250,1750)(8750,2000)(9250,2250)

\path(-2000,4000)(10000,4000)
\path(-5500,4000)(-4500,4000)
\whiten\path(-4750,4250)(-4750,3750)(-4500,4000)(-4750,4250)
\blacken\path(-1250,4250)(-1250,3750)(-750,4000)(-1250,4250)
\blacken\path(9250,4250)(9250,3750)(8750,4000)(9250,4250)

\path(-2000,6000)(10000,6000)
\put(-4250,6000){\scriptsize$v_1$}
\path(-5500,6000)(-4500,6000)
\whiten\path(-4750,6250)(-4750,5750)(-4500,6000)(-4750,6250)
\blacken\path(-1250,6250)(-1250,5750)(-750,6000)(-1250,6250)
\blacken\path(9250,6250)(9250,5750)(8750,6000)(9250,6250)


\path(0,-2000)(0,8000)
\put(0,-2750){\scriptsize$w_1$}
\path(0,-4000)(0,-3000)
\whiten\path(-250,-3250)(250,-3250)(0,-3000)(-250,-3250)
\blacken\path(-250,-750)(250,-750)(0,-1250)(-250,-750)
\blacken\path(-250,6750)(250,6750)(0,7250)(-250,6750)

\path(2000,-2000)(2000,8000)
\path(2000,-4000)(2000,-3000)
\whiten\path(1750,-3250)(2250,-3250)(2000,-3000)(1750,-3250)
\blacken\path(1750,-750)(2250,-750)(2000,-1250)(1750,-750)
\blacken\path(1750,6750)(2250,6750)(2000,7250)(1750,6750)

\path(4000,-2000)(4000,8000)
\path(4000,-4000)(4000,-3000)
\whiten\path(3750,-3250)(4250,-3250)(4000,-3000)(3750,-3250)
\blacken\path(3750,-750)(4250,-750)(4000,-1250)(3750,-750)
\blacken\path(3750,6750)(4250,6750)(4000,7250)(3750,6750)

\path(6000,-2000)(6000,8000)
\path(6000,-4000)(6000,-3000)
\whiten\path(5750,-3250)(6250,-3250)(6000,-3000)(5750,-3250)
\blacken\path(5750,-750)(6250,-750)(6000,-1250)(5750,-750)
\blacken\path(5750,6750)(6250,6750)(6000,7250)(5750,6750)

\path(8000,-2000)(8000,8000)
\put(8000,-2750){\scriptsize$w_N$}
\path(8000,-4000)(8000,-3000)
\whiten\path(7750,-3250)(8250,-3250)(8000,-3000)(7750,-3250)
\blacken\path(7750,-1250)(8250,-1250)(8000,-750)(7750,-1250)
\blacken\path(7750,6750)(8250,6750)(8000,7250)(7750,6750)


\put(12000,2750){$=$}


\path(18000,0)(30000,0)
\put(15750,0){\scriptsize$v_{N-1}$}
\path(14500,0)(15500,0)
\whiten\path(15250,250)(15250,-250)(15500,0)(15250,250)
\blacken\path(18750,250)(18750,-250)(19250,0)(18750,250)
\blacken\path(27250,250)(27250,-250)(26750,0)(27250,250)
\blacken\path(29250,250)(29250,-250)(28750,0)(29250,250)

\path(18000,2000)(30000,2000)
\path(14500,2000)(15500,2000)
\whiten\path(15250,2250)(15250,1750)(15500,2000)(15250,2250)
\blacken\path(18750,2250)(18750,1750)(19250,2000)(18750,2250)
\blacken\path(27250,2250)(27250,1750)(26750,2000)(27250,2250)
\blacken\path(29250,2250)(29250,1750)(28750,2000)(29250,2250)

\path(18000,4000)(30000,4000)
\path(14500,4000)(15500,4000)
\whiten\path(15250,4250)(15250,3750)(15500,4000)(15250,4250)
\blacken\path(18750,4250)(18750,3750)(19250,4000)(18750,4250)
\blacken\path(27250,4250)(27250,3750)(26750,4000)(27250,4250)
\blacken\path(29250,4250)(29250,3750)(28750,4000)(29250,4250)

\path(18000,6000)(30000,6000)
\put(15750,6000){\scriptsize$v_1$}
\path(14500,6000)(15500,6000)
\whiten\path(15250,6250)(15250,5750)(15500,6000)(15250,6250)
\blacken\path(18750,6250)(18750,5750)(19250,6000)(18750,6250)
\blacken\path(27250,6250)(27250,5750)(26750,6000)(27250,6250)
\blacken\path(29250,6250)(29250,5750)(28750,6000)(29250,6250)


\path(20000,-2000)(20000,8000)
\put(20000,-2750){\scriptsize$w_1$}
\path(20000,-4000)(20000,-3000)
\whiten\path(19750,-3250)(20250,-3250)(20000,-3000)(19750,-3250)
\blacken\path(19750,-750)(20250,-750)(20000,-1250)(19750,-750)
\blacken\path(19750,6750)(20250,6750)(20000,7250)(19750,6750)

\path(22000,-2000)(22000,8000)
\path(22000,-4000)(22000,-3000)
\whiten\path(21750,-3250)(22250,-3250)(22000,-3000)(21750,-3250)
\blacken\path(21750,-750)(22250,-750)(22000,-1250)(21750,-750)
\blacken\path(21750,6750)(22250,6750)(22000,7250)(21750,6750)

\path(24000,-2000)(24000,8000)
\path(24000,-4000)(24000,-3000)
\whiten\path(23750,-3250)(24250,-3250)(24000,-3000)(23750,-3250)
\blacken\path(23750,-750)(24250,-750)(24000,-1250)(23750,-750)
\blacken\path(23750,6750)(24250,6750)(24000,7250)(23750,6750)

\path(26000,-2000)(26000,8000)
\path(26000,-4000)(26000,-3000)
\whiten\path(25750,-3250)(26250,-3250)(26000,-3000)(25750,-3250)
\blacken\path(25750,-750)(26250,-750)(26000,-1250)(25750,-750)
\blacken\path(25750,6750)(26250,6750)(26000,7250)(25750,6750)

\path(28000,-2000)(28000,8000)
\put(28000,-2750){\scriptsize$w_N$}
\path(28000,-4000)(28000,-3000)
\whiten\path(27750,-3250)(28250,-3250)(28000,-3000)(27750,-3250)
\blacken\path(27750,-1250)(28250,-1250)(28000,-750)(27750,-1250)
\blacken\path(27750,750)(28250,750)(28000,1250)(27750,750)
\blacken\path(27750,2750)(28250,2750)(28000,3250)(27750,2750)
\blacken\path(27750,4750)(28250,4750)(28000,5250)(27750,4750)
\blacken\path(27750,6750)(28250,6750)(28000,7250)(27750,6750)

\end{picture}

\end{minipage}
\end{center}

\caption[Peeling away the right-most column of the partition function]{Peeling away the right-most column of the partition function. The diagram on the left hand side represents $\langle \Downarrow_N | \sigma_N^{-} \prod_{i=1}^{N-1} B(v_i,\{w\}_N) |\Uparrow_N\rangle$, with the internal black arrows being summed over all configurations. The diagram on the right contains all surviving configurations.}

\label{zequiv5}
\end{figure}

\noindent The right hand side of figure \ref{zequiv5} represents the $(N-1)\times (N-1)$ domain wall partition function, multiplied by a column of vertices. Replacing these vertices with their trigonometric weights, we find that 

\begin{align}
\langle \Downarrow_N|
\sigma_N^{-} 
\prod_{i=1}^{N-1} B(v_i,\{w\}_N) 
|\Uparrow_N\rangle
=
\prod_{i=1}^{N-1}
[v_i-w_N]
Z_{N-1}\Big(\{v\}_{N-1},\{w\}_{N-1}\Big)
\label{zequiv6}
\end{align}

\noindent Substituting (\ref{zequiv6}) into (\ref{PFexp3}) produces the required recursion relation (\ref{recursion}).

}
\end{propertylist}

\begin{propertylist}
{\rm
Specializing the definition (\ref{pf-xxz}) to the case $N=1$ gives

\begin{align} 
Z_1(v_1,w_1) 
=
\langle \Downarrow_1 |
B(v_1,w_1)
| \Uparrow_1 \rangle
=
\uparrow_{a_1}^{*} \otimes \downarrow_1^{*}
R_{a_1 1}(v_1,w_1)
\uparrow_1 \otimes \downarrow_{a_1}
=
[\gamma]
\end{align}

\noindent as required. Alternatively, the lattice representation of $Z_1$ is simply the top right vertex in figure \ref{R6v}, whose weight is equal to $[\gamma]$.  
} 
\end{propertylist}
\end{proof}

\begin{lemma}
\label{uniqueness}
{\rm 
Let $\{\breve{Z}_N\}_{N \in \mathbb{N}}$ denote a set of functions $\breve{Z}_N(\{v\}_N,\{w\}_N)$ which satisfy the four conditions of the previous lemma. Then $\breve{Z}_N = Z_N$ for all $N \geq 1$. In other words, the conditions imposed on the domain wall partition function determine it uniquely. 

%
%
%
%
%
%

}
\end{lemma}

\begin{proof}
From condition {\bf 4} on $\breve{Z}_1,Z_1$ we know that $\breve{Z}_1=Z_1$. Hence we may assume that $\breve{Z}_{N-1} = Z_{N-1}$ for some $N \geq 2$. Using this assumption together with condition {\bf 3} on $\breve{Z}_N,Z_N$ yields 


\begin{align}
\breve{Z}_N
\Big|_{v_N = w_N-\gamma}
&=
[\gamma] \prod_{i=1}^{N-1} [v_i-w_N] [w_N-w_i-\gamma]
\breve{Z}_{N-1}
\\
&=
[\gamma] \prod_{i=1}^{N-1} [v_i-w_N] [w_N-w_i-\gamma]
Z_{N-1}
=
Z_{N} \Big|_{v_N=w_N-\gamma}
\nonumber
\end{align}

\noindent Condition {\bf 1} on $\breve{Z}_N,Z_N$ states that both are symmetric in the variables $\{w\}_N$. Using this fact in the previous equation, we find that 


\begin{align} 
\breve{Z}_N \Big|_{v_N=w_{i}-\gamma}
=
Z_N \Big|_{v_N=w_{i}-\gamma}
\quad
{\rm for\ all}\ 1 \leq i \leq N
\end{align}

\noindent which proves that $\breve{Z}_N$ and $Z_N$ are equal at $N$ distinct values of $v_N$. By condition {\bf 2}, both functions are trigonometric polynomials of degree $N-1$ in $v_N$, so their equality at $N$ points implies $\breve{Z}_N = Z_N$ everywhere. This completes the proof by induction.

\end{proof}

\subsection{Determinant expression for $Z_N$}

In \cite{ize} Izergin found a function which satisfies the four conditions of the previous subsection, and is therefore equal to the domain wall partition function. We present this formula below.

\begin{lemma}
{\rm For all $N \geq 1$ we define  

\begin{align}
\breve{Z}_N\Big(\{v\}_N,\{w\}_N\Big)
&=
\frac{\displaystyle{\prod_{i,j=1}^{N}} [v_i-w_j+\gamma] [v_i-w_j]}
{\displaystyle{\prod_{1 \leq i < j \leq N}} [v_i-v_j] [w_j-w_i]}
\det\left(\frac{[\gamma]}{[v_i-w_j+\gamma][v_i-w_j]}\right)_{1\leq i,j \leq N}
\nonumber
\\
&=
\frac{
\det\left(
[\gamma] 
\displaystyle{\prod_{k \not= i}^{N} [v_k-w_j+\gamma] [v_k-w_j]}
\right)_{1\leq i,j \leq N}
}
{
\displaystyle{
\prod_{1 \leq i < j \leq N} [v_i-v_j] [w_j-w_i]
}
}
\label{dwpf}
\end{align}

\noindent The functions $\{\breve{Z}_N\}_{N\in \mathbb{N}}$ satisfy the four conditions of lemma 5. Equivalently, the domain wall partition function $Z_N$ is equal to the right hand side of (\ref{dwpf}).
}
\end{lemma}

\begin{proof}
 
%
%

\begin{propertylist2}
{\rm
Permuting $w_m\leftrightarrow w_n$ in $\det\left([\gamma] \prod_{k \not= i}^{N} [v_k-w_j+\gamma] [v_k-w_j]\right)_{1\leq i,j \leq N}$ swaps two columns of the determinant, which introduces a minus sign into the numerator of (\ref{dwpf}). Similarly, permuting $w_m \leftrightarrow w_n$ in $\prod_{1 \leq i<j \leq N} [w_j-w_i]$ introduces a minus sign into the denominator of (\ref{dwpf}). These minus signs cancel, leaving $\breve{Z}_N$ invariant under permutations of the $\{w\}_N$ variables. 
}
\end{propertylist2}

\begin{propertylist2}
{\rm
The numerator of (\ref{dwpf}) is a trigonometric polynomial of degree $2N-2$ in $v_N$, with zeros at the points $v_N=v_n$ for all $1 \leq n \leq N-1$, since such a substitution would render two rows of the determinant equal. The denominator of (\ref{dwpf}) is a trigonometric polynomial of degree $N-1$ in $v_N$, with zeros at the same points. Cancelling these common zeros, $\breve{Z}_N$ is a trigonometric polynomial of degree $N-1$ in $v_N$. 
}
\end{propertylist2}

\begin{propertylist2}
{\rm 
Expanding the determinant in $\breve{Z}_N$ along its $N^{\rm th}$ row, we find

\begin{align}
\breve{Z}_N\Big(\{v\}_N,\{w\}_N\Big)
&=
\sum_{j=1}^{N}
\breve{Z}_{N-1}\Big(\{v\}_{N-1},\{w_1,\ldots, \widehat{w_j},\ldots,w_N\}\Big)
\\
&
\times
\frac{
\displaystyle{
[\gamma]
\prod_{k=1}^{N-1}
[v_k-w_j+\gamma] 
[v_k-w_j]
\prod_{k\not=j}^{N}
[v_N-w_k+\gamma]
[v_N-w_k]
}
}{
\displaystyle{
\prod_{k = 1}^{N-1}
[v_k-v_N]
\prod_{k \not= j}^{N}
[w_j-w_k]
}
}
\nonumber
\end{align}

\noindent where $\widehat{w_j}$ denotes the omission of that variable. Setting $v_N=w_N-\gamma$ in the above expression, all terms in the summation over $1 \leq j \leq N$ vanish except the $j=N$ term, and we obtain

\begin{align}
\breve{Z}_N
\Big|_{v_N=w_N-\gamma}
&=
[\gamma]
\prod_{k=1}^{N-1}
[w_N-w_k-\gamma]
[v_k-w_N]
\breve{Z}_{N-1}
\Big(\{v\}_{N-1},\{w\}_{N-1}\Big)
\end{align} 

\noindent which is the desired recursion relation.
}
\end{propertylist2}

\begin{propertylist2}
{\rm 
From the definition (\ref{dwpf}) it is clear that $\breve{Z}_1=[\gamma]$.
} 
\end{propertylist2}

\end{proof}

\subsection{Partition function as a power-sum specialized KP $\tau$-function}

In this subsection we reproduce the main result of \cite{fwz4} where it was shown that, after a normalization and trivial change of variables, the domain wall partition function is a power-sum specialization of a KP $\tau$-function. We start by defining the function $Z_N'$ as follows

\begin{align}
Z_N'\Big(
\{y\}_N,\{z\}_N
\Big)
=
e^{N^2\gamma}
\prod_{i=1}^{N}
e^{(N-1) (v_i-w_i)}
Z_N\Big(
\{v\}_N,\{w\}_N
\Big)
\label{normaliz}
\end{align}

\noindent where we have made the change of variables $e^{2 v_i} = y_i, e^{-2 w_i} = -z_i, e^{2\gamma}=q$. In contrast to $Z_N$, which is a trigonometric function, $Z_N'$ is a genuine polynomial in its variables $\{y\}_N$ and $\{z\}_N$. Applying the normalization (\ref{normaliz}) to (\ref{dwpf}) and performing the prescribed change of variables, we obtain the explicit formula

\begin{align}
Z_N'\Big(
\{y\}_N,\{z\}_N
\Big)
&=
\frac{\displaystyle{
(q-1)^N
\det\left(
\prod_{k\not=j}^{N}
(1+q y_i z_k)(1+ y_i z_k)
\right)_{1 \leq i,j \leq N}
}}
{\displaystyle{
\prod_{1 \leq i < j \leq N}
(y_i-y_j)(z_j-z_i)
}}
\label{z'}
\end{align}

\noindent In order to progress further, we need to define the {\it elementary symmetric functions} $e_m\{x\}$ in the set of variables $\{x\} = \{x_1,\ldots,x_N\}$. Following section 2 of chapter I in \cite{mac}, they are given by

\begin{align}
e_m\{x\}
=
{\rm Coeff}_{z^m}
\left[ \prod_{i=1}^{N} (1+x_i z) \right]
\label{elemen}
\end{align}

\noindent Using the definition (\ref{elemen}) of the elementary symmetric functions, the entries of the determinant (\ref{z'}) can be expanded to produce the equation

\begin{align}
Z_N'\Big(
\{y\}_N,\{z\}_N
\Big)
=
\frac{\displaystyle{
(q-1)^N\det\left(
\sum_{k=1}^{2N-1}
y_i^{k-1}
e_{k-1}\{q\widehat{z}_j\}\cup\{\widehat{z}_j\}
\right)_{1 \leq i, j \leq N}
}}
{\displaystyle{
\prod_{1 \leq i < j \leq N}
(y_i-y_j)(z_j-z_i)
}}
\label{z'2}
\end{align}

\noindent where $\{q\widehat{z}_j\} \cup \{\widehat{z}_j\}$ denotes the set of $2N$ variables $\{qz_1,\ldots, qz_N, z_1,\ldots, z_N\}$ with the omission of the pair $\{qz_j, z_j\}$. The determinant in (\ref{z'2}) can be manipulated further using the Cauchy-Binet identity, which we state below.

\begin{lemma}
{\rm Fix two positive integers $M \geq N \geq 1$. Let $A$ be an arbitrary $N \times M$ matrix with entries $(A)_{i,j} = A_{i,j}$, and $B$ an arbitrary $M \times N$ matrix with entries $(B)_{i,j} =B_{i,j}$. The Cauchy-Binet identity states that 

\begin{align}
\det \left(\sum_{k=1}^{M} A_{i,k} B_{k,j} \right)_{1 \leq i, j \leq N}
\label{cauch-bin}
=
\sum_{M \geq k_1 > \cdots > k_N \geq 1}
\det\Big(A_{i,k_j}\Big)_{1\leq i,j \leq N}
\det\Big(B_{k_i,j}\Big)_{1 \leq i,j \leq N}
\end{align}

\noindent where the sum is taken over all strict partitions satisfying $\{M \geq k_1 > \cdots > k_N \geq 1\}$.

}
\end{lemma}

\begin{proof}
See section 2 of chapter I in \cite{gan}.
\end{proof}

Applying the Cauchy-Binet identity (\ref{cauch-bin}) to the determinant (\ref{z'2}) gives

\begin{align}
&
Z_N'\Big(
\{y\}_N,\{z\}_N
\Big)
=
\frac{\displaystyle{
(q-1)^N
}}
{\displaystyle{
\prod_{1 \leq i < j \leq N}
(y_i-y_j)(z_j-z_i)
}}
\times
\label{z'cb}
\\
&
\sum_{2N-1 \geq k_1 > \cdots > k_N \geq 1}
\det\Big(
y_i^{k_j-1}
\Big)_{1 \leq i,j \leq N}
\det\Big(
e_{k_i-1}
\{q\widehat{z}_j\}\cup\{\widehat{z}_j\}
\Big)_{1 \leq i,j \leq N}
\nonumber
\end{align}

\noindent We perform a change in summation variables, writing $k_i-1= \mu_{i}-i+N$ for all $1\leq i \leq N$. Making this substitution in (\ref{z'cb}) yields 

\begin{align}
&
Z_N'\Big(
\{y\}_N,\{z\}_N
\Big)
=
\frac{\displaystyle{
(q-1)^N
}}
{\displaystyle{
\prod_{1 \leq i < j \leq N}
(y_i-y_j)(z_j-z_i)
}}
\times
\label{z'3}
\\
&
\sum_{\mu \subseteq [N,N-1] }
\det\Big(
y_i^{\mu_j-j+N}
\Big)_{1\leq i,j \leq N}
\det\Big(
e_{\mu_i-i+N}\{q\widehat{z}_j\}\cup\{\widehat{z}_j\}
\Big)_{1\leq i,j \leq N}
\nonumber
\end{align}

\noindent where the sum is over all partitions $\mu$ whose Young diagrams are contained in the rectangle $N\times (N-1)$. By virtue of the Jacobi-Trudi identity (\ref{jac-trud}), the first determinant in (\ref{z'3}) matches the numerator of a Schur function. Hence we are able to write

%
%
%


\begin{align}
&
Z_N'\Big(
\{y\}_N,\{z\}_N
\Big)
=
\sum_{\mu \subseteq [N,N-1]}
s_{\mu}\{y\}
\zeta_{\mu}(\{z\},q)
\label{z'4}
\end{align}

\noindent where we have defined the functions\footnote{The coefficients $\zeta_{\mu}(\{z\},q)$ allow an even simpler expression than (\ref{zeta}) by cancelling the Vandermonde factor in the denominator, as was done in \cite{las}.}

\begin{align}
\zeta_{\mu}(\{z\},q)
=
\frac{\displaystyle{
(q-1)^N
\det\Big(
e_{\mu_i-i+N}\{q\widehat{z}_j\}\cup\{\widehat{z}_j\}
\Big)_{1\leq i,j \leq N}
}}
{\displaystyle{
\prod_{1 \leq i < j \leq N}
(z_j-z_i)
}}
\label{zeta}
\end{align}

\noindent We have thus shown that $Z'_N$ is equal to   

\begin{align}
\tau_{\mbox{\tiny{PF}}}\{t\}
=
\sum_{\mu \subseteq [N,N-1] }
\chi_{\mu}\{t\} \zeta_{\mu}(\{z\},q)
\label{zeta-tau}
\end{align}

\noindent under the power-sum specialization $t_n = \frac{1}{n} \sum_{i=1}^{N} y_i^n$ for all $n \geq 1$. In the next subsection we show that the polynomial $\tau_{\mbox{\tiny{PF}}}\{t\}$ is a KP $\tau$-function, by writing it as an expectation value of charged fermions.

\subsection{$\tau_{\mbox{\tiny{PF}}}\{t\}$ as an expectation value of charged fermions}

Our aim is to transform the right hand side of (\ref{zeta-tau}) to the fermionic form (\ref{KPexpI1}) of a KP $\tau$-function. To achieve this, we follow the procedure of subsection 3.2.11. For all integers $i \geq -N$ and $1\leq j \leq N$ we define

\begin{align}
c_{i,j}
=
\left\{
\begin{array}{ll}
e_{i+N}\{q\widehat{z}_j\}\cup\{\widehat{z}_j\},
&
-N \leq i < N-1
\\
\\
0,
&
i \geq N-1
\end{array}
\right.
\end{align}

\noindent where $e_{i+N}\{q\widehat{z}_j\}\cup\{\widehat{z}_j\}$ is an elementary symmetric function (\ref{elemen}) in the variables $\{qz_1,\ldots,qz_N,z_1,\ldots,z_N\}$ with $\{qz_j,z_j\}$ omitted. Using this expression, for all ordered sets of integers $\{m\} = \{m_1 > \cdots > m_N \geq -N\}$ we define the coefficients  

\begin{align}
c_{\{m\}}\Big(\{z\},q\Big)
=
\frac{(q-1)^N}{\displaystyle{\prod_{1\leq i < j \leq N} (z_j-z_i)}}
\det\Big(
c_{m_i,j}
\Big)_{1\leq i,j \leq N}
\label{coco}
\end{align}

\noindent Since $c_{i,j} = 0$ if $i \geq N-1$, the coefficient $c_{\{m\}}(\{z\},q)$ vanishes if $m_1 \geq N-1$. Letting $\mu$ be the partition formed by setting $\mu_i = m_i+i$ for all $1\leq i\leq N$, we conclude that

\begin{align}
c_{\{m\}}\Big(\{z\},q\Big)
=
\left\{
\begin{array}{ll}
\zeta_{\mu}(\{z\},q),
&
\mu \subseteq [N,N-1]
\\
\\
0, 
& 
\mu \not\subseteq [N,N-1]
\end{array}
\right.
\end{align}

\noindent where $\zeta_{\mu}(\{z\},q)$ denotes the function (\ref{zeta}). This enables us to write (\ref{zeta-tau}) in the form

\begin{align}
&
\tau_{\mbox{\tiny{PF}}}\{t\}
=
\langle 0|
e^{H\{t\}}
\sum_{{\rm card}\{m\}=N}
c_{\{m\}}\Big(\{z\},q\Big)
\psi_{m_1}\ldots\psi_{m_N}|-N\rangle
\label{pfferm}
\end{align}

\noindent where the sum is over all sets of integers $\{m\} = \{m_1 > \cdots > m_N \geq -N\}$, with the identification $|\mu) = \psi_{m_1}\ldots \psi_{m_N} |-N\rangle$. Expanding in the basis (\ref{repclA5}) of $\mathcal{F}_{\psi}^{(0)}$, we have 

{\footnotesize
\begin{align}
\sum_{{\rm card}\{m\}=N}
c_{\{m\}}\Big(\{z\},q\Big)
\psi_{m_1}\ldots & \psi_{m_N}
|-N\rangle
=
\label{pfferm1}
\zeta_{\emptyset}
|0\rangle
+
\sum_{m=1}^{N-1}
\sum_{n=1}^{N}
(-)^{n-1}
\zeta_{\{m,1^{n-1}\}}
\psi_{m-1}
\psis_{-n}
|0\rangle
+
g_{\psi}^{(1)}|0\rangle
\end{align}
}

\noindent where we have abbreviated $\zeta_{\mu}(\{z\},q) = \zeta_{\mu}$ for all partitions $\mu$, and where all monomials within $g^{(1)}_{\psi} \in Cl_{\psi}^{(0)}$ consist of at least two positive and two negative fermions. Up to an overall constant, the coefficients (\ref{coco}) are determinants. Therefore they satisfy the Pl\"ucker relations (\ref{KPpluck1}) automatically, and the right hand side of (\ref{pfferm1}) must obey the charged fermion bilinear identity (\ref{KPhelp}). Hence we may follow the procedure used to prove lemma 10 in chapter 1 to obtain

{\footnotesize
\begin{align}
\sum_{{\rm card}\{m\}=N}
c_{\{m\}}\Big(\{z\},q\Big)
\psi_{m_1}\ldots\psi_{m_N}
|&-N\rangle
=
\label{pfferm3}
\zeta_{\emptyset}
\exp\left(
\sum_{m=1}^{N-1}
\sum_{n=1}^{N}
(-)^{n-1}
\frac{\zeta_{\{m,1^{n-1}\}}}{\zeta_{\emptyset}}
\psi_{m-1}
\psis_{-n}
\right)
|0\rangle
\end{align}
}

\noindent Substituting (\ref{pfferm3}) into (\ref{pfferm}), we arrive at the equation

\begin{align}
&
\tau_{\mbox{\tiny{PF}}}\{t\}
=
\label{pfferm2}
\zeta_{\emptyset}
\langle 0|
e^{H\{t\}}
\exp\left(
\sum_{m=1}^{N-1}
\sum_{n=1}^{N}
(-)^{n-1}
\frac{\zeta_{\{m,1^{n-1}\}}}{\zeta_{\emptyset}}
\psi_{m-1}
\psis_{-n}
\right)
|0\rangle
\end{align}

\noindent which expresses $\tau_{\mbox{\tiny{PF}}}\{t\}$ in the canonical form of a KP $\tau$-function. The fermionic expression (\ref{pfferm2}) was originally derived in \cite{fwz4}, using a relatively complicated inductive argument. The preceding derivation is simpler in that it only requires lemma 10 of chapter 1.

\section{Scalar products}
\label{xxz-sp}

In this section we define and calculate a sequence of {\it intermediate scalar products} $S_n$, which interpolate between the domain wall partition function and the full scalar product. The domain wall partition function corresponds to the case $n=0$, whereas the full scalar product is given by the case $n=N$. These functions were originally studied in \cite{kmt}, using Drinfel'd twists in the algebraic Bethe Ansatz setting \cite{ms}. Our approach is more elementary, and inspired by the Izergin-Korepin procedure for evaluating the domain wall partition function. The results of this section subsequently appeared in \cite{whe1}.

\subsection{Definition of $S_n(\{u\}_n,\{v\}_N,\{w\}_M)$}
\label{xxz-sp-def}

Let $\{u\}_n = \{u_1,\ldots,u_n\}$, $\{v\}_N = \{v_1,\ldots,v_N\}$, $\{w\}_M = \{w_1,\ldots,w_M\}$ be three sets of variables whose cardinalities satisfy $0 \leq n \leq N$ and $1\leq N \leq M$. We proceed to introduce functions $S_n(\{u\}_n,\{v\}_N,\{w\}_M)$ for all $0 \leq n \leq N$. In the case $n=0$ we define

\begin{align}
S_0 \Big( \{v\}_N, \{w\}_M \Big) 
=
\langle \Downarrow_{N/M}|
\prod_{j=1}^{N} 
B(v_j,\{w\}_M)  
|\Uparrow_M\rangle
\label{s0-xxz}
\end{align}


\noindent where, for conciseness, we have suppressed dependence on the set $\{u\}_0 = \emptyset$. As we will soon show, up to an overall normalization the scalar product $S_0$ is equal to the domain wall partition function $Z_N$. Next, for all $1\leq n \leq N-1$ we define

\begin{align}
S_n \Big( \{u\}_n, \{v\}_N, \{w\}_M \Big)
=
\langle \Downarrow_{\widetilde{N}/M}|
\prod_{i=1}^{n}
C(u_i,\{w\}_M)
\prod_{j=1}^{N}  
B(v_j,\{w\}_M) 
|\Uparrow_M\rangle
\label{sn-xxz}
\end{align}


\noindent where we have adopted the notation $\widetilde{N} = N-n$, which is used frequently hereafter. Finally, for $n=N$ we define

\begin{align}
&
S_N \Big( \{u\}_N, \{v\}_N, \{w\}_M \Big)
=
\langle \Uparrow_M|
\prod_{i=1}^{N}
C(u_i,\{w\}_M)
\prod_{j=1}^{N} 
B(v_j,\{w\}_M) 
|\Uparrow_M\rangle
\label{sN-xxz}
\end{align}

\noindent which represents the usual scalar product. In all cases (\ref{s0-xxz})--(\ref{sN-xxz}) we shall assume that the parameters $\{v\}_N$ obey the Bethe equations (\ref{bethe2}), while the remaining variables $\{u\}_n$ are considered free. Accordingly, we name these objects {\it Bethe scalar products}. It turns out that $\{S_n\}_{0 \leq n \leq N}$ are related by a simple recursion. Hence they provide a convenient way of calculating $S_N$, starting from Izergin's determinant formula (\ref{dwpf}) for $Z_N$.  

\subsection{Graphical representation of scalar products}
\label{xxz-sp-graph}

We now provide lattice representations of the Bethe scalar products $\{S_n\}_{0 \leq n \leq N}$. As in the case of the domain wall partition function, these lattices simplify the calculation of the functions themselves.

\begin{figure}[H]

\begin{center}
\begin{minipage}{4.3in}

\setlength{\unitlength}{0.0003cm}
\begin{picture}(20000,15000)(-9000,-3000)

\put(-7000,5000){\fbox{$v$}}


\path(-2000,0)(20000,0)
\put(-3250,0){\tiny{$N$}}
\path(-4500,0)(-3500,0)
\whiten\path(-3750,250)(-3750,-250)(-3500,0)(-3750,250)
\blacken\path(-1250,250)(-1250,-250)(-750,0)(-1250,250)
\blacken\path(19250,250)(19250,-250)(18750,0)(19250,250)

\path(-2000,2000)(20000,2000)
\path(-4500,2000)(-3500,2000)
\whiten\path(-3750,2250)(-3750,1750)(-3500,2000)(-3750,2250)
\blacken\path(-1250,2250)(-1250,1750)(-750,2000)(-1250,2250)
\blacken\path(19250,2250)(19250,1750)(18750,2000)(19250,2250)

\path(-2000,4000)(20000,4000)
\path(-4500,4000)(-3500,4000)
\whiten\path(-3750,4250)(-3750,3750)(-3500,4000)(-3750,4250)
\blacken\path(-1250,4250)(-1250,3750)(-750,4000)(-1250,4250)
\blacken\path(19250,4250)(19250,3750)(18750,4000)(19250,4250)

\path(-2000,6000)(20000,6000)
\path(-4500,6000)(-3500,6000)
\whiten\path(-3750,6250)(-3750,5750)(-3500,6000)(-3750,6250)
\blacken\path(-1250,6250)(-1250,5750)(-750,6000)(-1250,6250)
\blacken\path(19250,6250)(19250,5750)(18750,6000)(19250,6250)

\path(-2000,8000)(20000,8000)
\path(-4500,8000)(-3500,8000)
\whiten\path(-3750,8250)(-3750,7750)(-3500,8000)(-3750,8250)
\blacken\path(-1250,8250)(-1250,7750)(-750,8000)(-1250,8250)
\blacken\path(19250,8250)(19250,7750)(18750,8000)(19250,8250)

\path(-2000,10000)(20000,10000)
\put(-3250,10000){\tiny$1$}
\path(-4500,10000)(-3500,10000)
\whiten\path(-3750,10250)(-3750,9750)(-3500,10000)(-3750,10250)
\blacken\path(-1250,10250)(-1250,9750)(-750,10000)(-1250,10250)
\blacken\path(19250,10250)(19250,9750)(18750,10000)(19250,10250)


\put(20000,-3000){\fbox{$w$}}


\path(0,-2000)(0,12000)
\put(-250,-2750){\tiny$1$}
\path(0,-4250)(0,-3250)
\whiten\path(-250,-3500)(250,-3500)(0,-3250)(-250,-3500)
\blacken\path(-250,-750)(250,-750)(0,-1250)(-250,-750)
\blacken\path(-250,10750)(250,10750)(0,11250)(-250,10750)

\path(2000,-2000)(2000,12000)
\path(2000,-4250)(2000,-3250)
\whiten\path(1750,-3500)(2250,-3500)(2000,-3250)(1750,-3500)
\blacken\path(1750,-750)(2250,-750)(2000,-1250)(1750,-750)
\blacken\path(1750,10750)(2250,10750)(2000,11250)(1750,10750)

\path(4000,-2000)(4000,12000)
\path(4000,-4250)(4000,-3250)
\whiten\path(3750,-3500)(4250,-3500)(4000,-3250)(3750,-3500)
\blacken\path(3750,-750)(4250,-750)(4000,-1250)(3750,-750)
\blacken\path(3750,10750)(4250,10750)(4000,11250)(3750,10750)

\path(6000,-2000)(6000,12000)
\path(6000,-4250)(6000,-3250)
\whiten\path(5750,-3500)(6250,-3500)(6000,-3250)(5750,-3500)
\blacken\path(5750,-750)(6250,-750)(6000,-1250)(5750,-750)
\blacken\path(5750,10750)(6250,10750)(6000,11250)(5750,10750)

\path(8000,-2000)(8000,12000)
\path(8000,-4250)(8000,-3250)
\whiten\path(7750,-3500)(8250,-3500)(8000,-3250)(7750,-3500)
\blacken\path(7750,-750)(8250,-750)(8000,-1250)(7750,-750)
\blacken\path(7750,10750)(8250,10750)(8000,11250)(7750,10750)

\path(10000,-2000)(10000,12000)
\put(9750,-2750){\tiny$N$}
\path(10000,-4250)(10000,-3250)
\whiten\path(9750,-3500)(10250,-3500)(10000,-3250)(9750,-3500)
\blacken\path(9750,-750)(10250,-750)(10000,-1250)(9750,-750)
\blacken\path(9750,10750)(10250,10750)(10000,11250)(9750,10750)

\path(12000,-2000)(12000,12000)
\put(11750,-2750){\tiny$N+1$}
\path(12000,-4250)(12000,-3250)
\whiten\path(11750,-3500)(12250,-3500)(12000,-3250)(11750,-3500)
\blacken\path(11750,-1250)(12250,-1250)(12000,-750)(11750,-1250)
\blacken\path(11750,10750)(12250,10750)(12000,11250)(11750,10750)

\path(14000,-2000)(14000,12000)
\path(14000,-4250)(14000,-3250)
\whiten\path(13750,-3500)(14250,-3500)(14000,-3250)(13750,-3500)
\blacken\path(13750,-1250)(14250,-1250)(14000,-750)(13750,-1250)
\blacken\path(13750,10750)(14250,10750)(14000,11250)(13750,10750)

\path(16000,-2000)(16000,12000)
\path(16000,-4250)(16000,-3250)
\whiten\path(15750,-3500)(16250,-3500)(16000,-3250)(15750,-3500)
\blacken\path(15750,-1250)(16250,-1250)(16000,-750)(15750,-1250)
\blacken\path(15750,10750)(16250,10750)(16000,11250)(15750,10750)

\path(18000,-2000)(18000,12000)
\put(17750,-2750){\tiny$M$}
\path(18000,-4250)(18000,-3250)
\whiten\path(17750,-3500)(18250,-3500)(18000,-3250)(17750,-3500)
\blacken\path(17750,-1250)(18250,-1250)(18000,-750)(17750,-1250)
\blacken\path(17750,10750)(18250,10750)(18000,11250)(17750,10750)

\end{picture}

\end{minipage}
\end{center}

\caption[Lattice representation of $S_0$]{Lattice representation of $S_0$. The top row of arrows corresponds with the state $|\Uparrow_M\rangle$, while the bottom row of arrows corresponds with the dual state $\langle \Downarrow_{N/M}|$. Each horizontal lattice line represents a $B$-operator $B(v_j,\{w\}_M)$.}
\end{figure}
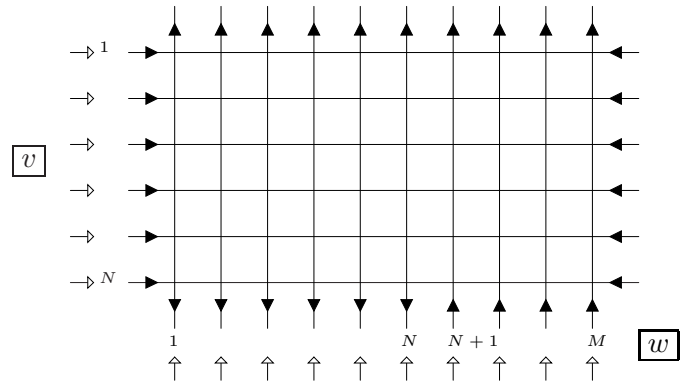

\begin{figure}[H]

\begin{center}
\begin{minipage}{4.3in}

\setlength{\unitlength}{0.0003cm}
\begin{picture}(20000,20000)(-9000,-9000)

\put(-7000,-4000){\fbox{$u$}}


\path(-2000,-6000)(20000,-6000)
\put(-3250,-6000){\tiny$n$}
\path(-4500,-6000)(-3500,-6000)
\whiten\path(-3750,-5750)(-3750,-6250)(-3500,-6000)(-3750,-5750)
\blacken\path(-750,-5750)(-750,-6250)(-1250,-6000)(-750,-5750)
\blacken\path(18750,-5750)(18750,-6250)(19250,-6000)(18750,-5750)

\path(-2000,-4000)(20000,-4000)
\path(-4500,-4000)(-3500,-4000)
\whiten\path(-3750,-3750)(-3750,-4250)(-3500,-4000)(-3750,-3750)
\blacken\path(-750,-3750)(-750,-4250)(-1250,-4000)(-750,-3750)
\blacken\path(18750,-3750)(18750,-4250)(19250,-4000)(18750,-3750)

\path(-2000,-2000)(20000,-2000)
\put(-3250,-2000){\tiny$1$}
\path(-4500,-2000)(-3500,-2000)
\whiten\path(-3750,-1750)(-3750,-2250)(-3500,-2000)(-3750,-1750)
\blacken\path(-750,-1750)(-750,-2250)(-1250,-2000)(-750,-1750)
\blacken\path(18750,-1750)(18750,-2250)(19250,-2000)(18750,-1750)


\put(-7000,5000){\fbox{$v$}}


\path(-2000,0)(20000,0)
\put(-3250,0){\tiny$N$}
\path(-4500,0)(-3500,0)
\whiten\path(-3750,250)(-3750,-250)(-3500,0)(-3750,250)
\blacken\path(-1250,250)(-1250,-250)(-750,0)(-1250,250)
\blacken\path(19250,250)(19250,-250)(18750,0)(19250,250)

\path(-2000,2000)(20000,2000)
\path(-4500,2000)(-3500,2000)
\whiten\path(-3750,2250)(-3750,1750)(-3500,2000)(-3750,2250)
\blacken\path(-1250,2250)(-1250,1750)(-750,2000)(-1250,2250)
\blacken\path(19250,2250)(19250,1750)(18750,2000)(19250,2250)

\path(-2000,4000)(20000,4000)
\path(-4500,4000)(-3500,4000)
\whiten\path(-3750,4250)(-3750,3750)(-3500,4000)(-3750,4250)
\blacken\path(-1250,4250)(-1250,3750)(-750,4000)(-1250,4250)
\blacken\path(19250,4250)(19250,3750)(18750,4000)(19250,4250)

\path(-2000,6000)(20000,6000)
\path(-4500,6000)(-3500,6000)
\whiten\path(-3750,6250)(-3750,5750)(-3500,6000)(-3750,6250)
\blacken\path(-1250,6250)(-1250,5750)(-750,6000)(-1250,6250)
\blacken\path(19250,6250)(19250,5750)(18750,6000)(19250,6250)

\path(-2000,8000)(20000,8000)
\path(-4500,8000)(-3500,8000)
\whiten\path(-3750,8250)(-3750,7750)(-3500,8000)(-3750,8250)
\blacken\path(-1250,8250)(-1250,7750)(-750,8000)(-1250,8250)
\blacken\path(19250,8250)(19250,7750)(18750,8000)(19250,8250)

\path(-2000,10000)(20000,10000)
\put(-3250,10000){\tiny$1$}
\path(-4500,10000)(-3500,10000)
\whiten\path(-3750,10250)(-3750,9750)(-3500,10000)(-3750,10250)
\blacken\path(-1250,10250)(-1250,9750)(-750,10000)(-1250,10250)
\blacken\path(19250,10250)(19250,9750)(18750,10000)(19250,10250)


\put(20000,-9000){\fbox{$w$}}


\path(0,-8000)(0,12000)
\put(-250,-9000){\tiny$1$}
\path(0,-10500)(0,-9500)
\whiten\path(-250,-9750)(250,-9750)(0,-9500)(-250,-9750)
\blacken\path(-250,-6750)(250,-6750)(0,-7250)(-250,-6750)
\blacken\path(-250,10750)(250,10750)(0,11250)(-250,10750)

\path(2000,-8000)(2000,12000)
\path(2000,-10500)(2000,-9500)
\whiten\path(1750,-9750)(2250,-9750)(2000,-9500)(1750,-9750)
\blacken\path(1750,-6750)(2250,-6750)(2000,-7250)(1750,-6750)
\blacken\path(1750,10750)(2250,10750)(2000,11250)(1750,10750)

\path(4000,-8000)(4000,12000)
\put(3750,-9000){\tiny$\widetilde{N}$}
\path(4000,-10500)(4000,-9500)
\whiten\path(3750,-9750)(4250,-9750)(4000,-9500)(3750,-9750)
\blacken\path(3750,-6750)(4250,-6750)(4000,-7250)(3750,-6750)
\blacken\path(3750,10750)(4250,10750)(4000,11250)(3750,10750)

\path(6000,-8000)(6000,12000)
\put(5750,-9000){\tiny$\widetilde{N}+1$}
\path(6000,-10500)(6000,-9500)
\whiten\path(5750,-9750)(6250,-9750)(6000,-9500)(5750,-9750)
\blacken\path(5750,-7250)(6250,-7250)(6000,-6750)(5750,-7250)
\blacken\path(5750,10750)(6250,10750)(6000,11250)(5750,10750)

\path(8000,-8000)(8000,12000)
\path(8000,-10500)(8000,-9500)
\whiten\path(7750,-9750)(8250,-9750)(8000,-9500)(7750,-9750)
\blacken\path(7750,-7250)(8250,-7250)(8000,-6750)(7750,-7250)
\blacken\path(7750,10750)(8250,10750)(8000,11250)(7750,10750)

\path(10000,-8000)(10000,12000)
\path(10000,-10500)(10000,-9500)
\whiten\path(9750,-9750)(10250,-9750)(10000,-9500)(9750,-9750)
\blacken\path(9750,-7250)(10250,-7250)(10000,-6750)(9750,-7250)
\blacken\path(9750,10750)(10250,10750)(10000,11250)(9750,10750)

\path(12000,-8000)(12000,12000)
\path(12000,-10500)(12000,-9500)
\whiten\path(11750,-9750)(12250,-9750)(12000,-9500)(11750,-9750)
\blacken\path(11750,-7250)(12250,-7250)(12000,-6750)(11750,-7250)
\blacken\path(11750,10750)(12250,10750)(12000,11250)(11750,10750)

\path(14000,-8000)(14000,12000)
\path(14000,-10500)(14000,-9500)
\whiten\path(13750,-9750)(14250,-9750)(14000,-9500)(13750,-9750)
\blacken\path(13750,-7250)(14250,-7250)(14000,-6750)(13750,-7250)
\blacken\path(13750,10750)(14250,10750)(14000,11250)(13750,10750)

\path(16000,-8000)(16000,12000)
\path(16000,-10500)(16000,-9500)
\whiten\path(15750,-9750)(16250,-9750)(16000,-9500)(15750,-9750)
\blacken\path(15750,-7250)(16250,-7250)(16000,-6750)(15750,-7250)
\blacken\path(15750,10750)(16250,10750)(16000,11250)(15750,10750)

\path(18000,-8000)(18000,12000)
\put(17750,-9000){\tiny$M$}
\path(18000,-10500)(18000,-9500)
\whiten\path(17750,-9750)(18250,-9750)(18000,-9500)(17750,-9750)
\blacken\path(17750,-7250)(18250,-7250)(18000,-6750)(17750,-7250)
\blacken\path(17750,10750)(18250,10750)(18000,11250)(17750,10750)

\end{picture}

\end{minipage}
\end{center}

\caption[Lattice representation of $S_n$]{Lattice representation of $S_n$. The top row of arrows corresponds with the state $|\Uparrow_M\rangle$, while the bottom row of arrows corresponds with the dual state $\langle \Downarrow_{\widetilde{N}/M}|$. The highest $N$ horizontal lines represent $B$-operators $B(v_j,\{w\}_M)$, while the lowest $n$ horizontal lines represent $C$-operators $C(u_i,\{w\}_M)$.}

\label{lat}

\end{figure}

\begin{figure}[H]

\begin{center}
\begin{minipage}{4.3in}

\setlength{\unitlength}{0.0003cm}
\begin{picture}(20000,18000)(-9000,-9000)

\put(-7000,-3000){\fbox{$u$}}


\path(-2000,-6000)(20000,-6000)
\put(-3250,-6000){\tiny$N$}
\path(-4500,-6000)(-3500,-6000)
\whiten\path(-3750,-5750)(-3750,-6250)(-3500,-6000)(-3750,-5750)
\blacken\path(-750,-5750)(-750,-6250)(-1250,-6000)(-750,-5750)
\blacken\path(18750,-5750)(18750,-6250)(19250,-6000)(18750,-5750)

\path(-2000,-4000)(20000,-4000)
\path(-4500,-4000)(-3500,-4000)
\whiten\path(-3750,-3750)(-3750,-4250)(-3500,-4000)(-3750,-3750)
\blacken\path(-750,-3750)(-750,-4250)(-1250,-4000)(-750,-3750)
\blacken\path(18750,-3750)(18750,-4250)(19250,-4000)(18750,-3750)

\path(-2000,-2000)(20000,-2000)
\path(-4500,-2000)(-3500,-2000)
\whiten\path(-3750,-1750)(-3750,-2250)(-3500,-2000)(-3750,-1750)
\blacken\path(-750,-1750)(-750,-2250)(-1250,-2000)(-750,-1750)
\blacken\path(18750,-1750)(18750,-2250)(19250,-2000)(18750,-1750)

\path(-2000,0)(20000,0)
\put(-3250,0){\tiny$1$}
\path(-4500,0)(-3500,0)
\whiten\path(-3750,250)(-3750,-250)(-3500,0)(-3750,250)
\blacken\path(-750,250)(-750,-250)(-1250,0)(-750,250)
\blacken\path(18750,250)(18750,-250)(19250,0)(18750,250)


\put(-7000,5000){\fbox{$v$}}


\path(-2000,2000)(20000,2000)
\put(-3250,2000){\tiny$N$}
\path(-4500,2000)(-3500,2000)
\whiten\path(-3750,2250)(-3750,1750)(-3500,2000)(-3750,2250)
\blacken\path(-1250,2250)(-1250,1750)(-750,2000)(-1250,2250)
\blacken\path(19250,2250)(19250,1750)(18750,2000)(19250,2250)

\path(-2000,4000)(20000,4000)
\path(-4500,4000)(-3500,4000)
\whiten\path(-3750,4250)(-3750,3750)(-3500,4000)(-3750,4250)
\blacken\path(-1250,4250)(-1250,3750)(-750,4000)(-1250,4250)
\blacken\path(19250,4250)(19250,3750)(18750,4000)(19250,4250)

\path(-2000,6000)(20000,6000)
\path(-4500,6000)(-3500,6000)
\whiten\path(-3750,6250)(-3750,5750)(-3500,6000)(-3750,6250)
\blacken\path(-1250,6250)(-1250,5750)(-750,6000)(-1250,6250)
\blacken\path(19250,6250)(19250,5750)(18750,6000)(19250,6250)

\path(-2000,8000)(20000,8000)
\put(-3250,8000){\tiny$1$}
\path(-4500,8000)(-3500,8000)
\whiten\path(-3750,8250)(-3750,7750)(-3500,8000)(-3750,8250)
\blacken\path(-1250,8250)(-1250,7750)(-750,8000)(-1250,8250)
\blacken\path(19250,8250)(19250,7750)(18750,8000)(19250,8250)


\put(20000,-9000){\fbox{$w$}}


\path(0,-8000)(0,10000)
\put(-250,-8750){\tiny$1$}
\path(0,-10250)(0,-9250)
\whiten\path(-250,-9500)(250,-9500)(0,-9250)(-250,-9500)
\blacken\path(-250,-7250)(250,-7250)(0,-6750)(-250,-7250)
\blacken\path(-250,8750)(250,8750)(0,9250)(-250,8750)

\path(2000,-8000)(2000,10000)
\path(2000,-10250)(2000,-9250)
\whiten\path(1750,-9500)(2250,-9500)(2000,-9250)(1750,-9500)
\blacken\path(1750,-7250)(2250,-7250)(2000,-6750)(1750,-7250)
\blacken\path(1750,8750)(2250,8750)(2000,9250)(1750,8750)

\path(4000,-8000)(4000,10000)
\path(4000,-10250)(4000,-9250)
\whiten\path(3750,-9500)(4250,-9500)(4000,-9250)(3750,-9500)
\blacken\path(3750,-7250)(4250,-7250)(4000,-6750)(3750,-7250)
\blacken\path(3750,8750)(4250,8750)(4000,9250)(3750,8750)

\path(6000,-8000)(6000,10000)
\path(6000,-10250)(6000,-9250)
\whiten\path(5750,-9500)(6250,-9500)(6000,-9250)(5750,-9500)
\blacken\path(5750,-7250)(6250,-7250)(6000,-6750)(5750,-7250)
\blacken\path(5750,8750)(6250,8750)(6000,9250)(5750,8750)

\path(8000,-8000)(8000,10000)
\path(8000,-10250)(8000,-9250)
\whiten\path(7750,-9500)(8250,-9500)(8000,-9250)(7750,-9500)
\blacken\path(7750,-7250)(8250,-7250)(8000,-6750)(7750,-7250)
\blacken\path(7750,8750)(8250,8750)(8000,9250)(7750,8750)

\path(10000,-8000)(10000,10000)
\path(10000,-10250)(10000,-9250)
\whiten\path(9750,-9500)(10250,-9500)(10000,-9250)(9750,-9500)
\blacken\path(9750,-7250)(10250,-7250)(10000,-6750)(9750,-7250)
\blacken\path(9750,8750)(10250,8750)(10000,9250)(9750,8750)

\path(12000,-8000)(12000,10000)
\path(12000,-10250)(12000,-9250)
\whiten\path(11750,-9500)(12250,-9500)(12000,-9250)(11750,-9500)
\blacken\path(11750,-7250)(12250,-7250)(12000,-6750)(11750,-7250)
\blacken\path(11750,8750)(12250,8750)(12000,9250)(11750,8750)

\path(14000,-8000)(14000,10000)
\path(14000,-10250)(14000,-9250)
\whiten\path(13750,-9500)(14250,-9500)(14000,-9250)(13750,-9500)
\blacken\path(13750,-7250)(14250,-7250)(14000,-6750)(13750,-7250)
\blacken\path(13750,8750)(14250,8750)(14000,9250)(13750,8750)

\path(16000,-8000)(16000,10000)
\path(16000,-10250)(16000,-9250)
\whiten\path(15750,-9500)(16250,-9500)(16000,-9250)(15750,-9500)
\blacken\path(15750,-7250)(16250,-7250)(16000,-6750)(15750,-7250)
\blacken\path(15750,8750)(16250,8750)(16000,9250)(15750,8750)

\path(18000,-8000)(18000,10000)
\put(17750,-8750){\tiny$M$}
\path(18000,-10250)(18000,-9250)
\whiten\path(17750,-9500)(18250,-9500)(18000,-9250)(17750,-9500)
\blacken\path(17750,-7250)(18250,-7250)(18000,-6750)(17750,-7250)
\blacken\path(17750,8750)(18250,8750)(18000,9250)(17750,8750)

\end{picture}

\end{minipage}
\end{center}

\caption[Lattice representation of $S_N$]{Lattice representation of $S_N$. The top row of arrows corresponds with the state $|\Uparrow_M\rangle$, while the bottom row of arrows corresponds with the dual state $\langle \Uparrow_M|$. The highest $N$ horizontal lines represent $B$-operators $B(v_j,\{w\}_M)$, while the lowest $N$ horizontal lines represent $C$-operators $C(u_i,\{w\}_M)$.}
\end{figure}
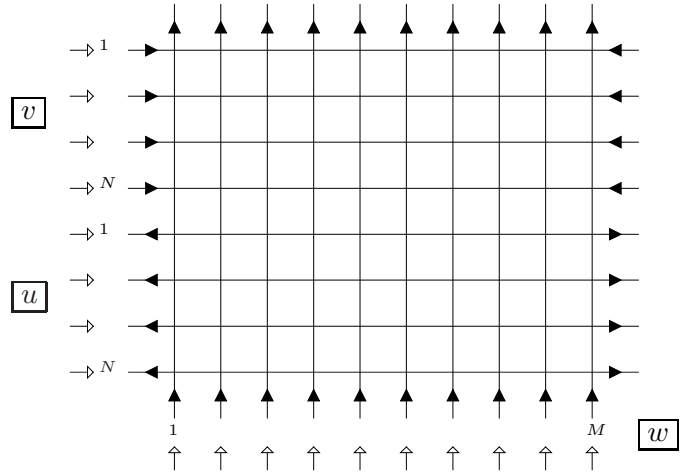

\subsection{Conditions on $S_n(\{u\}_n,\{v\}_N,\{w\}_M)$}

Progressing in the same manner as the previous section, we will show that the Bethe scalar products $S_{n}( \{u\}_n, \{v\}_N, \{w\}_M )$ satisfy a set of properties which determine them uniquely.

\begin{lemma}
\label{sncond}
{\rm 
Let us adopt the usual shorthand $S_n = S_n(\{u\}_n,\{v\}_N,\{w\}_M)$. For all $1 \leq n \leq N$ we claim that 

\begin{propertylist3}
{\rm $S_n$ is symmetric in the variables $\{w_{\widetilde{N}+1},\ldots,w_M\}$.}
\end{propertylist3}

\begin{propertylist3}
{\rm $S_n$ is a trigonometric polynomial of degree $M-1$ in $u_n$, with zeros occurring at the points $u_n = w_i-\gamma$, for all $1 \leq i \leq \widetilde{N}$.}
\end{propertylist3}

\begin{propertylist3}
{\rm Setting $u_n = w_{\widetilde{N}+1}$, $S_n$ satisfies the recursion relation

\begin{align}
&
S_{n}
\Big|_{u_n = w_{\widetilde{N}+1}}
=
\prod_{i=1}^{M}
[w_{\widetilde{N}+1}-w_i+\gamma]
S_{n-1}
\label{sNrec0-xxz}
\end{align}

\noindent where $S_{n-1}$ denotes the Bethe scalar product $S_{n-1}(\{u\}_{n-1},\{v\}_N,\{w\}_M)$.
}
\end{propertylist3}

In addition, we have the supplementary condition 

\begin{propertylist3}
{\rm 
$S_0$ and $Z_N$ are related via the equation

\begin{align}
S_{0} \Big(\{v\}_N,\{w\}_M \Big)
=
\prod_{i=1}^{N} 
\prod_{j=N+1}^{M}
[v_i-w_j]
Z_N \Big( \{v\}_N,\{w\}_{N} \Big)
\label{cond4}
\end{align}

\noindent where we have defined $\{w\}_{N} = \{w_1,\ldots,w_N\}$.}
\end{propertylist3}
}
\end{lemma}

\begin{proof}

\begin{propertylist4}
{\rm We introduce the auxiliary state vectors

\begin{align}
\langle \Uparrow^{a}_N|
=
\bigotimes_{i=1}^{N} \uparrow^{*}_{a_i},
\quad
\langle \Downarrow^{b}_n|
=
\bigotimes_{i=1}^{n} \downarrow^{*}_{b_i},
\quad
|\Downarrow^{a}_N\rangle
=
\bigotimes_{i=1}^{N} \downarrow_{a_i},
\quad
|\Uparrow^{b}_{n}\rangle
=
\bigotimes_{i=1}^{n} \uparrow_{b_i}
\label{auxvec}
\end{align}

\noindent which allow us to write

\begin{align}
&S_n \Big(\{u\}_n,\{v\}_N,\{w\}_M\Big)
=
\\
&
\langle \Downarrow_{\widetilde{N}/M}|
\otimes
\langle\Uparrow_{N}^{a}|
\otimes
\langle\Downarrow_n^{b}| 
T\Big(
\{v\}_N \cup \{u\}_n,\{w\}_M
\Big)
|\Uparrow_n^{b}\rangle
\otimes
|\Downarrow_N^{a}\rangle
\otimes
|\Uparrow_M\rangle
\nonumber
\end{align}

\noindent where we have defined

\begin{align}
&T\Big(\{v\}_N \cup \{u\}_{n},\{w\}_M\Big)
=
\\
&
T_{b_{n}}(u_n,\{w\}_M)\ldots T_{b_{1}}(u_1,\{w\}_M)
T_{a_N}(v_N,\{w\}_M) \ldots T_{a_1}(v_1,\{w\}_M)
\nonumber
\end{align}

\noindent By application of lemma 3 we thus obtain

\begin{align*}
&
T\Big(\{v\}_N \cup \{u\}_{n},\{w\}_M\Big)
=
(-)^{M\widetilde{N}}
\overline{T}_1(w_1,\{\bar{v}\}_N \cup \{\bar{u}\}_n)
\ldots
\overline{T}_M(w_M,\{\bar{v}\}_N \cup \{\bar{u}\}_n)
\end{align*}

\noindent where for all $1 \leq i \leq M$ we have set

\begin{align}
\overline{T}_i(w_i,\{\bar{v}\}_N \cup \{\bar{u}\}_n)
=
\left(
\begin{array}{rr}
D(w_i,\{\bar{v}\}_N \cup \{\bar{u}\}_n)
&
-B(w_i,\{\bar{v}\}_N \cup \{\bar{u}\}_n)
\\
-C(w_i,\{\bar{v}\}_N \cup \{\bar{u}\}_n)
&
A(w_i,\{\bar{v}\}_N \cup \{\bar{u}\}_n)
\end{array}
\right)_i
\end{align}

\noindent with $\{\bar{v}\}_N \cup \{\bar{u}\}_n = \{v_1+\gamma,\ldots,v_N+\gamma,u_1+\gamma,\ldots,u_n+\gamma\}$. Finally, contracting on the quantum spaces $\mathcal{V}_1,\ldots,\mathcal{V}_M$ gives

\begin{align}
&
S_n \Big(\{u\}_n,\{v\}_N,\{w\}_M\Big)
=
(-)^{(M+1)\widetilde{N}}
\times
\label{preceq}
\\
&
\langle \Uparrow_N^a|
\otimes
\langle \Downarrow_n^b|
\prod_{i=1}^{\widetilde{N}}
C(w_i,\{\bar{v}\}_N\cup\{\bar{u}\}_n)
\prod_{j=\widetilde{N}+1}^{M}
D(w_j,\{\bar{v}\}_N\cup\{\bar{u}\}_n)
|\Uparrow_n^b\rangle
\otimes
|\Downarrow_N^a\rangle
\nonumber
\end{align}

\noindent The diagrammatic interpretation of (\ref{preceq}) is shown in figure \ref{altgraph}.

\begin{figure}[H]

\begin{center}
\begin{minipage}{4.3in}

\setlength{\unitlength}{0.0003cm}
\begin{picture}(20000,23000)(-11000,-9000)

\put(-8000,3000){\fbox{$w$}}


\path(-2000,-6000)(18000,-6000)
\put(-4350,-6000){\tiny$1$}
\path(-5500,-6000)(-4500,-6000)
\whiten\path(-4750,-5750)(-4750,-6250)(-4500,-6000)(-4750,-5750)
\blacken\path(-750,-5750)(-750,-6250)(-1250,-6000)(-750,-5750)
\blacken\path(16750,-5750)(16750,-6250)(17250,-6000)(16750,-5750)

\path(-2000,-4000)(18000,-4000)
\path(-5500,-4000)(-4500,-4000)
\whiten\path(-4750,-3750)(-4750,-4250)(-4500,-4000)(-4750,-3750)
\blacken\path(-750,-3750)(-750,-4250)(-1250,-4000)(-750,-3750)
\blacken\path(16750,-3750)(16750,-4250)(17250,-4000)(16750,-3750)

\path(-2000,-2000)(18000,-2000)
\put(-4350,-2000){\tiny$\widetilde{N}$}
\path(-5500,-2000)(-4500,-2000)
\whiten\path(-4750,-1750)(-4750,-2250)(-4500,-2000)(-4750,-1750)
\blacken\path(-750,-1750)(-750,-2250)(-1250,-2000)(-750,-1750)
\blacken\path(16750,-1750)(16750,-2250)(17250,-2000)(16750,-1750)

\path(-2000,0)(18000,0)
\put(-4350,0){\tiny$\widetilde{N}+1$}
\path(-5500,0)(-4500,0)
\whiten\path(-4750,250)(-4750,-250)(-4500,0)(-4750,250)
\blacken\path(-750,250)(-750,-250)(-1250,0)(-750,250)
\blacken\path(17250,250)(17250,-250)(16750,0)(17250,250)

\path(-2000,2000)(18000,2000)
\path(-5500,2000)(-4500,2000)
\whiten\path(-4750,2250)(-4750,1750)(-4500,2000)(-4750,2250)
\blacken\path(-750,2250)(-750,1750)(-1250,2000)(-750,2250)
\blacken\path(17250,2250)(17250,1750)(16750,2000)(17250,2250)

\path(-2000,4000)(18000,4000)
\path(-5500,4000)(-4500,4000)
\whiten\path(-4750,4250)(-4750,3750)(-4500,4000)(-4750,4250)
\blacken\path(-750,4250)(-750,3750)(-1250,4000)(-750,4250)
\blacken\path(17250,4250)(17250,3750)(16750,4000)(17250,4250)

\path(-2000,6000)(18000,6000)
\path(-5500,6000)(-4500,6000)
\whiten\path(-4750,6250)(-4750,5750)(-4500,6000)(-4750,6250)
\blacken\path(-750,6250)(-750,5750)(-1250,6000)(-750,6250)
\blacken\path(17250,6250)(17250,5750)(16750,6000)(17250,6250)

\path(-2000,8000)(18000,8000)
\path(-5500,8000)(-4500,8000)
\whiten\path(-4750,8250)(-4750,7750)(-4500,8000)(-4750,8250)
\blacken\path(-750,8250)(-750,7750)(-1250,8000)(-750,8250)
\blacken\path(17250,8250)(17250,7750)(16750,8000)(17250,8250)

\path(-2000,10000)(18000,10000)
\path(-5500,10000)(-4500,10000)
\whiten\path(-4750,10250)(-4750,9750)(-4500,10000)(-4750,10250)
\blacken\path(-750,10250)(-750,9750)(-1250,10000)(-750,10250)
\blacken\path(17250,10250)(17250,9750)(16750,10000)(17250,10250)

\path(-2000,12000)(18000,12000)
\put(-4350,12000){\tiny$M$}
\path(-5500,12000)(-4500,12000)
\whiten\path(-4750,12250)(-4750,11750)(-4500,12000)(-4750,12250)
\blacken\path(-750,12250)(-750,11750)(-1250,12000)(-750,12250)
\blacken\path(17250,12250)(17250,11750)(16750,12000)(17250,12250)


\path(0,-8000)(0,14000)
\put(-250,-8750){\tiny$\bar{v}_1$}
\path(0,-10500)(0,-9500)
\whiten\path(-250,-9750)(250,-9750)(0,-9500)(-250,-9750)
\blacken\path(-250,-7250)(250,-7250)(0,-6750)(-250,-7250)
\blacken\path(-250,13250)(250,13250)(0,12750)(-250,13250)

\path(2000,-8000)(2000,14000)
\path(2000,-10500)(2000,-9500)
\whiten\path(1750,-9750)(2250,-9750)(2000,-9500)(1750,-9750)
\blacken\path(1750,-7250)(2250,-7250)(2000,-6750)(1750,-7250)
\blacken\path(1750,13250)(2250,13250)(2000,12750)(1750,13250)

\path(4000,-8000)(4000,14000)
\path(4000,-10500)(4000,-9500)
\whiten\path(3750,-9750)(4250,-9750)(4000,-9500)(3750,-9750)
\blacken\path(3750,-7250)(4250,-7250)(4000,-6750)(3750,-7250)
\blacken\path(3750,13250)(4250,13250)(4000,12750)(3750,13250)

\path(6000,-8000)(6000,14000)
\path(6000,-10500)(6000,-9500)
\whiten\path(5750,-9750)(6250,-9750)(6000,-9500)(5750,-9750)
\blacken\path(5750,-7250)(6250,-7250)(6000,-6750)(5750,-7250)
\blacken\path(5750,13250)(6250,13250)(6000,12750)(5750,13250)

\path(8000,-8000)(8000,14000)
\path(8000,-10500)(8000,-9500)
\whiten\path(7750,-9750)(8250,-9750)(8000,-9500)(7750,-9750)
\blacken\path(7750,-7250)(8250,-7250)(8000,-6750)(7750,-7250)
\blacken\path(7750,13250)(8250,13250)(8000,12750)(7750,13250)

\path(10000,-8000)(10000,14000)
\put(9750,-8750){\tiny$\bar{v}_N$}
\path(10000,-10500)(10000,-9500)
\whiten\path(9750,-9750)(10250,-9750)(10000,-9500)(9750,-9750)
\blacken\path(9750,-7250)(10250,-7250)(10000,-6750)(9750,-7250)
\blacken\path(9750,13250)(10250,13250)(10000,12750)(9750,13250)

\path(12000,-8000)(12000,14000)
\put(11750,-8750){\tiny$\bar{u}_1$}
\path(12000,-10500)(12000,-9500)
\whiten\path(11750,-9750)(12250,-9750)(12000,-9500)(11750,-9750)
\blacken\path(11750,-6750)(12250,-6750)(12000,-7250)(11750,-6750)
\blacken\path(11750,12750)(12250,12750)(12000,13250)(11750,12750)

\path(14000,-8000)(14000,14000)
\path(14000,-10500)(14000,-9500)
\whiten\path(13750,-9750)(14250,-9750)(14000,-9500)(13750,-9750)
\blacken\path(13750,-6750)(14250,-6750)(14000,-7250)(13750,-6750)
\blacken\path(13750,12750)(14250,12750)(14000,13250)(13750,12750)

\path(16000,-8000)(16000,14000)
\put(15750,-8750){\tiny$\bar{u}_n$}
\path(16000,-10500)(16000,-9500)
\whiten\path(15750,-9750)(16250,-9750)(16000,-9500)(15750,-9750)
\blacken\path(15750,-6750)(16250,-6750)(16000,-7250)(15750,-6750)
\blacken\path(15750,12750)(16250,12750)(16000,13250)(15750,12750)

\end{picture}

\end{minipage}
\end{center}

\caption[Alternative graphical representation of $S_n$]{Alternative graphical representation of $S_n$. Neglecting an overall minus sign, $S_n$ is equal to this lattice, which is essentially a rotation of figure \ref{lat}. The top row of arrows represents the state $|\Downarrow_N^a\rangle \otimes |\Uparrow_n^b\rangle$, while the bottom row of arrows represents the dual state $\langle \Uparrow_N^a| \otimes \langle \Downarrow_n^b|$. The lowest $\widetilde{N}$ horizontal lines represent $C$-operators, with the remaining lines representing $D$-operators.}

\label{altgraph}

\end{figure}
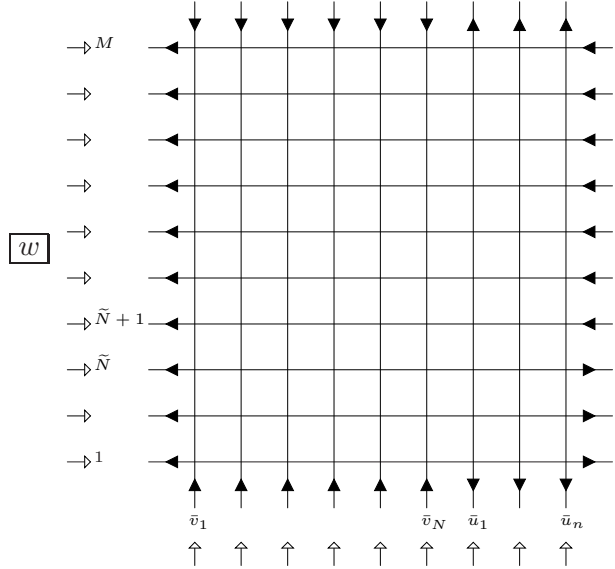

\noindent By virtue of (\ref{DD}) the $D$-operators in (\ref{preceq}) commute, proving that $S_n$ is symmetric in the variables $\{w_{\widetilde{N}+1},\ldots,w_M\}$.
} 
\end{propertylist4}

\begin{propertylist4}
{\rm
Inserting the set of states $\sum_{m > \widetilde{N}}
\sigma_m^{-} |\Downarrow_{\widetilde{N}/M}\rangle \langle \Downarrow_{\widetilde{N}/M}| \sigma_m^{+}$ after the first $C$-operator appearing in (\ref{sn-xxz}), we obtain the expansion

\begin{align}
S_n
\Big(\{u\}_n,\{v\}_N,\{w\}_M\Big) 
=
\sum_{m>\widetilde{N}}
&
\langle \Downarrow_{\widetilde{N}/M}|
C(u_n,\{w\}_M)
\sigma_m^{-} 
|\Downarrow_{\widetilde{N}/M}\rangle
\label{exp100}
\\
\times
&
\langle \Downarrow_{\widetilde{N}/M} |
\sigma_m^{+} 
\prod_{i=1}^{n-1}
C(u_i,\{w\}_M)
\prod_{j=1}^{N}
B(v_j,\{w\}_M)
|\Uparrow_M\rangle
\nonumber
\end{align}

\noindent in which all dependence on $u_n$ appears in the first factor of the sum. We therefore wish to calculate 
$
\langle \Downarrow_{\widetilde{N}/M}|
C(u_n,\{w\}_M)
\sigma_m^{-} 
|\Downarrow_{\widetilde{N}/M}\rangle
$
for all $\widetilde{N} < m \leq M$, and do so by identifying it with the string of vertices shown below.

\begin{figure}[H]

\begin{center}
\begin{minipage}{4.3in}

\setlength{\unitlength}{0.0003cm}
\begin{picture}(20000,13000)(-9000,-9000)

\path(-2000,2000)(20000,2000)
\put(-3250,2000){\tiny$u_n$}
\path(-4500,2000)(-3500,2000)
\whiten\path(-3750,2250)(-3750,1750)(-3500,2000)(-3750,2250)
\blacken\path(-750,2250)(-750,1750)(-1250,2000)(-750,2250)
\blacken\path(18750,2250)(18750,1750)(19250,2000)(18750,2250)


\path(0,000)(0,4000)
\put(-250,-1000){\tiny$1$}
\path(0,-2500)(0,-1500)
\whiten\path(-250,-1750)(250,-1750)(0,-1500)(-250,-1750)
\blacken\path(-250,1250)(250,1250)(0,750)(-250,1250)
\blacken\path(-250,3250)(250,3250)(0,2750)(-250,3250)

\path(2000,000)(2000,4000)
\path(2000,-2500)(2000,-1500)
\whiten\path(1750,-1750)(2250,-1750)(2000,-1500)(1750,-1750)
\blacken\path(1750,1250)(2250,1250)(2000,750)(1750,1250)
\blacken\path(1750,3250)(2250,3250)(2000,2750)(1750,3250)

\path(4000,000)(4000,4000)
\put(3750,-1000){\tiny$\widetilde{N}$}
\path(4000,-2500)(4000,-1500)
\whiten\path(3750,-1750)(4250,-1750)(4000,-1500)(3750,-1750)
\blacken\path(3750,1250)(4250,1250)(4000,750)(3750,1250)
\blacken\path(3750,3250)(4250,3250)(4000,2750)(3750,3250)

\path(6000,000)(6000,4000)
\put(5750,-1000){\tiny$\widetilde{N}+1$}
\path(6000,-2500)(6000,-1500)
\whiten\path(5750,-1750)(6250,-1750)(6000,-1500)(5750,-1750)
\blacken\path(5750,750)(6250,750)(6000,1250)(5750,750)
\blacken\path(5750,2750)(6250,2750)(6000,3250)(5750,2750)

\path(8000,000)(8000,4000)
\path(8000,-2500)(8000,-1500)
\whiten\path(7750,-1750)(8250,-1750)(8000,-1500)(7750,-1750)
\blacken\path(7750,750)(8250,750)(8000,1250)(7750,750)
\blacken\path(7750,2750)(8250,2750)(8000,3250)(7750,2750)

\path(10000,000)(10000,4000)
\path(10000,-2500)(10000,-1500)
\whiten\path(9750,-1750)(10250,-1750)(10000,-1500)(9750,-1750)
\blacken\path(9750,750)(10250,750)(10000,1250)(9750,750)
\blacken\path(9750,2750)(10250,2750)(10000,3250)(9750,2750)

\path(12000,000)(12000,4000)
\put(11750,-1000){\tiny$m$}
\path(12000,-2500)(12000,-1500)
\whiten\path(11750,-1750)(12250,-1750)(12000,-1500)(11750,-1750)
\blacken\path(11750,750)(12250,750)(12000,1250)(11750,750)
\blacken\path(11750,3250)(12250,3250)(12000,2750)(11750,3250)

\path(14000,000)(14000,4000)
\path(14000,-2500)(14000,-1500)
\whiten\path(13750,-1750)(14250,-1750)(14000,-1500)(13750,-1750)
\blacken\path(13750,750)(14250,750)(14000,1250)(13750,750)
\blacken\path(13750,2750)(14250,2750)(14000,3250)(13750,2750)

\path(16000,000)(16000,4000)
\path(16000,-2500)(16000,-1500)
\whiten\path(15750,-1750)(16250,-1750)(16000,-1500)(15750,-1750)
\blacken\path(15750,750)(16250,750)(16000,1250)(15750,750)
\blacken\path(15750,2750)(16250,2750)(16000,3250)(15750,2750)

\path(18000,000)(18000,4000)
\put(17750,-1000){\tiny$M$}
\path(18000,-2500)(18000,-1500)
\whiten\path(17750,-1750)(18250,-1750)(18000,-1500)(17750,-1750)
\blacken\path(17750,750)(18250,750)(18000,1250)(17750,750)
\blacken\path(17750,2750)(18250,2750)(18000,3250)(17750,2750)


\put(20000,-1000){\fbox{$w$}}

\put(-7000,-6250){$=$}

\path(-2000,-6000)(20000,-6000)
\put(-3250,-6000){\tiny$u_n$}
\path(-4500,-6000)(-3500,-6000)
\whiten\path(-3750,-5750)(-3750,-6250)(-3500,-6000)(-3750,-5750)
\blacken\path(-750,-5750)(-750,-6250)(-1250,-6000)(-750,-5750)
\blacken\path(1250,-5750)(1250,-6250)(750,-6000)(1250,-5750)
\blacken\path(3250,-5750)(3250,-6250)(2750,-6000)(3250,-5750)
\blacken\path(5250,-5750)(5250,-6250)(4750,-6000)(5250,-5750)
\blacken\path(7250,-5750)(7250,-6250)(6750,-6000)(7250,-5750)
\blacken\path(9250,-5750)(9250,-6250)(8750,-6000)(9250,-5750)
\blacken\path(11250,-5750)(11250,-6250)(10750,-6000)(11250,-5750)
\blacken\path(12750,-5750)(12750,-6250)(13250,-6000)(12750,-5750)
\blacken\path(14750,-5750)(14750,-6250)(15250,-6000)(14750,-5750)
\blacken\path(16750,-5750)(16750,-6250)(17250,-6000)(16750,-5750)
\blacken\path(18750,-5750)(18750,-6250)(19250,-6000)(18750,-5750)


\path(0,-8000)(0,-4000)
\put(-250,-9000){\tiny$1$}
\path(0,-10500)(0,-9500)
\whiten\path(-250,-9750)(250,-9750)(0,-9500)(-250,-9750)
\blacken\path(-250,-6750)(250,-6750)(0,-7250)(-250,-6750)
\blacken\path(-250,-4750)(250,-4750)(0,-5250)(-250,-4750)

\path(2000,-8000)(2000,-4000)
\path(2000,-10500)(2000,-9500)
\whiten\path(1750,-9750)(2250,-9750)(2000,-9500)(1750,-9750)
\blacken\path(1750,-6750)(2250,-6750)(2000,-7250)(1750,-6750)
\blacken\path(1750,-4750)(2250,-4750)(2000,-5250)(1750,-4750)

\path(4000,-8000)(4000,-4000)
\put(3750,-9000){\tiny$\widetilde{N}$}
\path(4000,-10500)(4000,-9500)
\whiten\path(3750,-9750)(4250,-9750)(4000,-9500)(3750,-9750)
\blacken\path(3750,-6750)(4250,-6750)(4000,-7250)(3750,-6750)
\blacken\path(3750,-4750)(4250,-4750)(4000,-5250)(3750,-4750)

\path(6000,-8000)(6000,-4000)
\put(5750,-9000){\tiny$\widetilde{N}+1$}
\path(6000,-10500)(6000,-9500)
\whiten\path(5750,-9750)(6250,-9750)(6000,-9500)(5750,-9750)
\blacken\path(5750,-7250)(6250,-7250)(6000,-6750)(5750,-7250)
\blacken\path(5750,-5250)(6250,-5250)(6000,-4750)(5750,-5250)

\path(8000,-8000)(8000,-4000)
\path(8000,-10500)(8000,-9500)
\whiten\path(7750,-9750)(8250,-9750)(8000,-9500)(7750,-9750)
\blacken\path(7750,-7250)(8250,-7250)(8000,-6750)(7750,-7250)
\blacken\path(7750,-5250)(8250,-5250)(8000,-4750)(7750,-5250)

\path(10000,-8000)(10000,-4000)
\path(10000,-10500)(10000,-9500)
\whiten\path(9750,-9750)(10250,-9750)(10000,-9500)(9750,-9750)
\blacken\path(9750,-7250)(10250,-7250)(10000,-6750)(9750,-7250)
\blacken\path(9750,-5250)(10250,-5250)(10000,-4750)(9750,-5250)

\path(12000,-8000)(12000,-4000)
\put(11750,-9000){\tiny$m$}
\path(12000,-10500)(12000,-9500)
\whiten\path(11750,-9750)(12250,-9750)(12000,-9500)(11750,-9750)
\blacken\path(11750,-7250)(12250,-7250)(12000,-6750)(11750,-7250)
\blacken\path(11750,-4750)(12250,-4750)(12000,-5250)(11750,-4750)

\path(14000,-8000)(14000,-4000)
\path(14000,-10500)(14000,-9500)
\whiten\path(13750,-9750)(14250,-9750)(14000,-9500)(13750,-9750)
\blacken\path(13750,-7250)(14250,-7250)(14000,-6750)(13750,-7250)
\blacken\path(13750,-5250)(14250,-5250)(14000,-4750)(13750,-5250)

\path(16000,-8000)(16000,-4000)
\path(16000,-10500)(16000,-9500)
\whiten\path(15750,-9750)(16250,-9750)(16000,-9500)(15750,-9750)
\blacken\path(15750,-7250)(16250,-7250)(16000,-6750)(15750,-7250)
\blacken\path(15750,-5250)(16250,-5250)(16000,-4750)(15750,-5250)

\path(18000,-8000)(18000,-4000)
\put(17750,-9000){\tiny$M$}
\path(18000,-10500)(18000,-9500)
\whiten\path(17750,-9750)(18250,-9750)(18000,-9500)(17750,-9750)
\blacken\path(17750,-7250)(18250,-7250)(18000,-6750)(17750,-7250)
\blacken\path(17750,-5250)(18250,-5250)(18000,-4750)(17750,-5250)


\put(20000,-9000){\fbox{$w$}}

\end{picture}

\end{minipage}
\end{center}

\caption[Peeling away the bottom row of $S_n$]{Peeling away the bottom row of $S_n$. The upper diagram represents $\langle \Downarrow_{\widetilde{N}/M}| C(u_n,\{w\}_M) \sigma_m^{-} |\Downarrow_{\widetilde{N}/M}\rangle$, with the internal black arrows being summed over all configurations. The lower diagram represents the only surviving configuration.}

\label{svert100}

\end{figure}
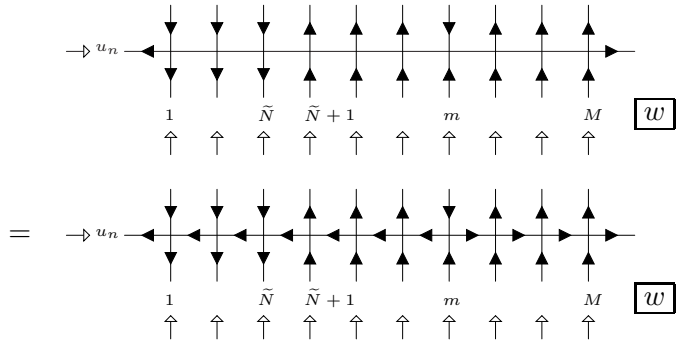

\noindent Replacing each vertex in figure \ref{svert100} with its corresponding trigonometric weight, we find that

{\small
\begin{align}
\langle \Downarrow_{\widetilde{N}/M}|
C(u_n,\{w\}_M)
\sigma_m^{-} 
|\Downarrow_{\widetilde{N}/M}\rangle
=
\label{factor100}
\prod_{i=1}^{\widetilde{N}}
[u_n - w_i +\gamma]
\prod_{\widetilde{N} < i < m}
[u_n -w_i]
[\gamma]
\prod_{m < i \leq M}
[u_n -w_i +\gamma]
\end{align}
}

\noindent Substituting (\ref{factor100}) into the expansion (\ref{exp100}) gives

\begin{align}
S_n
=
[\gamma] \prod_{i=1}^{\widetilde{N}} [u_n-w_i+\gamma]
\sum_{m > \widetilde{N}}
&
\prod_{\widetilde{N} < i < m}
[u_n-w_i]
\prod_{m < i \leq M}
[u_n-w_i+\gamma]
\label{snexp-xxz}
\\
\times
&
\langle \Downarrow_{\widetilde{N}/M}|
\sigma_m^{+}
\prod_{i=1}^{n-1}
C(u_i,\{w\}_M)
\prod_{j=1}^{N}
B(v_j,\{w\}_M)
|\Uparrow_M\rangle
\nonumber
\end{align}

\noindent From the last expression it is apparent that $S_n$ is a trigonometric polynomial of degree $M-1$ in $u_n$. Furthermore, $\widetilde{N}$ of the zeros of this polynomial are contained in the factor $\prod_{i=1}^{\widetilde{N}} [u_n-w_i+\gamma]$. 
}
\end{propertylist4}

\begin{propertylist4}
{\rm
Setting $u_n = w_{\widetilde{N}+1}$ in equation (\ref{snexp-xxz}), all terms in the sum collapse to zero except the term corresponding to $m = \widetilde{N}+1$, and we obtain

\begin{align}
S_n
\Big|_{u_n = w_{\widetilde{N}+1}}
&=
\prod_{i = 1}^{M}
[w_{\widetilde{N}+1}-w_i+\gamma]
\langle \Downarrow_{\widetilde{N}+1/M}|
\prod_{i=1}^{n-1} 
C(u_{i},\{w\}_M)
\prod_{j=1}^{N}
B(v_j,\{w\}_M)
|\Uparrow_M\rangle
\nonumber
\\
&=
\prod_{i=1}^{M}
[w_{\widetilde{N}+1}-w_i+\gamma]
S_{n-1} \Big( \{u\}_{n-1},\{v\}_N,\{w\}_M \Big)
\label{recproof}
\end{align}

\noindent where we have performed the trivial rearrangements $[\gamma] = [w_{\widetilde{N}+1}-w_{\widetilde{N}+1}+\gamma]$ and $\langle \Downarrow_{\widetilde{N}/M}| \sigma_{\widetilde{N}+1}^{+} = \langle \Downarrow_{\widetilde{N}+1/M}|$ to produce the first line of (\ref{recproof}), while the second line follows directly from the definition of $S_{n-1}$. Hence we have proved the recursion relation (\ref{sNrec0-xxz}). The diagrammatic interpretation of this identity is given below.

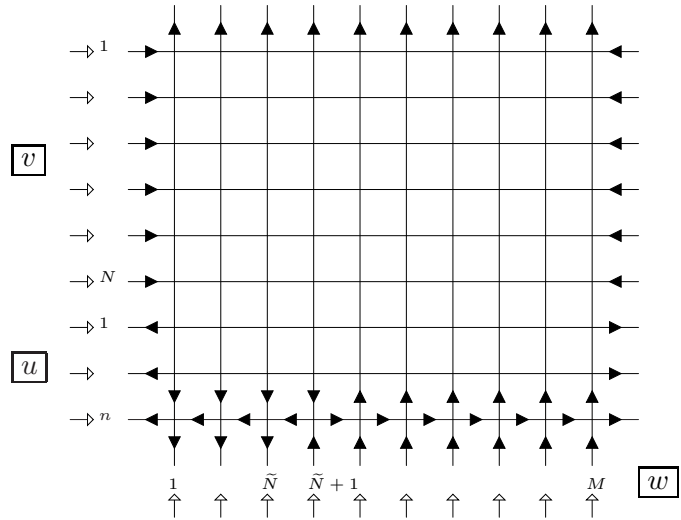
\begin{figure}[H]

\begin{center}
\begin{minipage}{4.3in}

\setlength{\unitlength}{0.0003cm}
\begin{picture}(20000,21000)(-9000,-9000)

\put(-7000,-4000){\fbox{$u$}}


\path(-2000,-6000)(20000,-6000)
\put(-3250,-6000){\tiny$n$}
\path(-4500,-6000)(-3500,-6000)
\whiten\path(-3750,-5750)(-3750,-6250)(-3500,-6000)(-3750,-5750)
\blacken\path(-750,-5750)(-750,-6250)(-1250,-6000)(-750,-5750)
\blacken\path(1250,-5750)(1250,-6250)(750,-6000)(1250,-5750)
\blacken\path(3250,-5750)(3250,-6250)(2750,-6000)(3250,-5750)
\blacken\path(5250,-5750)(5250,-6250)(4750,-6000)(5250,-5750)
\blacken\path(6750,-5750)(6750,-6250)(7250,-6000)(6750,-5750)
\blacken\path(8750,-5750)(8750,-6250)(9250,-6000)(8750,-5750)
\blacken\path(10750,-5750)(10750,-6250)(11250,-6000)(10750,-5750)
\blacken\path(12750,-5750)(12750,-6250)(13250,-6000)(12750,-5750)
\blacken\path(14750,-5750)(14750,-6250)(15250,-6000)(14750,-5750)
\blacken\path(16750,-5750)(16750,-6250)(17250,-6000)(16750,-5750)
\blacken\path(18750,-5750)(18750,-6250)(19250,-6000)(18750,-5750)

\path(-2000,-4000)(20000,-4000)
\path(-4500,-4000)(-3500,-4000)
\whiten\path(-3750,-3750)(-3750,-4250)(-3500,-4000)(-3750,-3750)
\blacken\path(-750,-3750)(-750,-4250)(-1250,-4000)(-750,-3750)
\blacken\path(18750,-3750)(18750,-4250)(19250,-4000)(18750,-3750)

\path(-2000,-2000)(20000,-2000)
\put(-3250,-2000){\tiny$1$}
\path(-4500,-2000)(-3500,-2000)
\whiten\path(-3750,-1750)(-3750,-2250)(-3500,-2000)(-3750,-1750)
\blacken\path(-750,-1750)(-750,-2250)(-1250,-2000)(-750,-1750)
\blacken\path(18750,-1750)(18750,-2250)(19250,-2000)(18750,-1750)


\put(-7000,5000){\fbox{$v$}}


\path(-2000,0)(20000,0)
\put(-3250,0){\tiny$N$}
\path(-4500,0)(-3500,0)
\whiten\path(-3750,250)(-3750,-250)(-3500,0)(-3750,250)
\blacken\path(-1250,250)(-1250,-250)(-750,0)(-1250,250)
\blacken\path(19250,250)(19250,-250)(18750,0)(19250,250)

\path(-2000,2000)(20000,2000)
\path(-4500,2000)(-3500,2000)
\whiten\path(-3750,2250)(-3750,1750)(-3500,2000)(-3750,2250)
\blacken\path(-1250,2250)(-1250,1750)(-750,2000)(-1250,2250)
\blacken\path(19250,2250)(19250,1750)(18750,2000)(19250,2250)

\path(-2000,4000)(20000,4000)
\path(-4500,4000)(-3500,4000)
\whiten\path(-3750,4250)(-3750,3750)(-3500,4000)(-3750,4250)
\blacken\path(-1250,4250)(-1250,3750)(-750,4000)(-1250,4250)
\blacken\path(19250,4250)(19250,3750)(18750,4000)(19250,4250)

\path(-2000,6000)(20000,6000)
\path(-4500,6000)(-3500,6000)
\whiten\path(-3750,6250)(-3750,5750)(-3500,6000)(-3750,6250)
\blacken\path(-1250,6250)(-1250,5750)(-750,6000)(-1250,6250)
\blacken\path(19250,6250)(19250,5750)(18750,6000)(19250,6250)

\path(-2000,8000)(20000,8000)
\path(-4500,8000)(-3500,8000)
\whiten\path(-3750,8250)(-3750,7750)(-3500,8000)(-3750,8250)
\blacken\path(-1250,8250)(-1250,7750)(-750,8000)(-1250,8250)
\blacken\path(19250,8250)(19250,7750)(18750,8000)(19250,8250)

\path(-2000,10000)(20000,10000)
\put(-3250,10000){\tiny$1$}
\path(-4500,10000)(-3500,10000)
\whiten\path(-3750,10250)(-3750,9750)(-3500,10000)(-3750,10250)
\blacken\path(-1250,10250)(-1250,9750)(-750,10000)(-1250,10250)
\blacken\path(19250,10250)(19250,9750)(18750,10000)(19250,10250)


\put(20000,-9000){\fbox{$w$}}


\path(0,-8000)(0,12000)
\put(-250,-9000){\tiny$1$}
\path(0,-10250)(0,-9250)
\whiten\path(-250,-9500)(250,-9500)(0,-9250)(-250,-9500)
\blacken\path(-250,-6750)(250,-6750)(0,-7250)(-250,-6750)
\blacken\path(-250,-4750)(250,-4750)(0,-5250)(-250,-4750)
\blacken\path(-250,10750)(250,10750)(0,11250)(-250,10750)

\path(2000,-8000)(2000,12000)
\path(2000,-10250)(2000,-9250)
\whiten\path(1750,-9500)(2250,-9500)(2000,-9250)(1750,-9500)
\blacken\path(1750,-6750)(2250,-6750)(2000,-7250)(1750,-6750)
\blacken\path(1750,-4750)(2250,-4750)(2000,-5250)(1750,-4750)
\blacken\path(1750,10750)(2250,10750)(2000,11250)(1750,10750)

\path(4000,-8000)(4000,12000)
\put(3750,-9000){\tiny$\widetilde{N}$}
\path(4000,-10250)(4000,-9250)
\whiten\path(3750,-9500)(4250,-9500)(4000,-9250)(3750,-9500)
\blacken\path(3750,-6750)(4250,-6750)(4000,-7250)(3750,-6750)
\blacken\path(3750,-4750)(4250,-4750)(4000,-5250)(3750,-4750)
\blacken\path(3750,10750)(4250,10750)(4000,11250)(3750,10750)

\path(6000,-8000)(6000,12000)
\put(5750,-9000){\tiny$\widetilde{N}+1$}
\path(6000,-10250)(6000,-9250)
\whiten\path(5750,-9500)(6250,-9500)(6000,-9250)(5750,-9500)
\blacken\path(5750,-7250)(6250,-7250)(6000,-6750)(5750,-7250)
\blacken\path(5750,-4750)(6250,-4750)(6000,-5250)(5750,-4750)
\blacken\path(5750,10750)(6250,10750)(6000,11250)(5750,10750)

\path(8000,-8000)(8000,12000)
\path(8000,-10250)(8000,-9250)
\whiten\path(7750,-9500)(8250,-9500)(8000,-9250)(7750,-9500)
\blacken\path(7750,-7250)(8250,-7250)(8000,-6750)(7750,-7250)
\blacken\path(7750,-5250)(8250,-5250)(8000,-4750)(7750,-5250)
\blacken\path(7750,10750)(8250,10750)(8000,11250)(7750,10750)

\path(10000,-8000)(10000,12000)
\path(10000,-10250)(10000,-9250)
\whiten\path(9750,-9500)(10250,-9500)(10000,-9250)(9750,-9500)
\blacken\path(9750,-7250)(10250,-7250)(10000,-6750)(9750,-7250)
\blacken\path(9750,-5250)(10250,-5250)(10000,-4750)(9750,-5250)
\blacken\path(9750,10750)(10250,10750)(10000,11250)(9750,10750)

\path(12000,-8000)(12000,12000)
\path(12000,-10250)(12000,-9250)
\whiten\path(11750,-9500)(12250,-9500)(12000,-9250)(11750,-9500)
\blacken\path(11750,-7250)(12250,-7250)(12000,-6750)(11750,-7250)
\blacken\path(11750,-5250)(12250,-5250)(12000,-4750)(11750,-5250)
\blacken\path(11750,10750)(12250,10750)(12000,11250)(11750,10750)

\path(14000,-8000)(14000,12000)
\path(14000,-10250)(14000,-9250)
\whiten\path(13750,-9500)(14250,-9500)(14000,-9250)(13750,-9500)
\blacken\path(13750,-7250)(14250,-7250)(14000,-6750)(13750,-7250)
\blacken\path(13750,-5250)(14250,-5250)(14000,-4750)(13750,-5250)
\blacken\path(13750,10750)(14250,10750)(14000,11250)(13750,10750)

\path(16000,-8000)(16000,12000)
\path(16000,-10250)(16000,-9250)
\whiten\path(15750,-9500)(16250,-9500)(16000,-9250)(15750,-9500)
\blacken\path(15750,-7250)(16250,-7250)(16000,-6750)(15750,-7250)
\blacken\path(15750,-5250)(16250,-5250)(16000,-4750)(15750,-5250)
\blacken\path(15750,10750)(16250,10750)(16000,11250)(15750,10750)

\path(18000,-8000)(18000,12000)
\put(17750,-9000){\tiny$M$}
\path(18000,-10250)(18000,-9250)
\whiten\path(17750,-9500)(18250,-9500)(18000,-9250)(17750,-9500)
\blacken\path(17750,-7250)(18250,-7250)(18000,-6750)(17750,-7250)
\blacken\path(17750,-5250)(18250,-5250)(18000,-4750)(17750,-5250)
\blacken\path(17750,10750)(18250,10750)(18000,11250)(17750,10750)

\end{picture}

\end{minipage}
\end{center}

\caption[Freezing the last row of the $S_n$ lattice]{Freezing the last row of the $S_n$ lattice. In general, the vertex at the intersection of the $u_n$ and $w_{\widetilde{N}+1}$ lines can be of type $b_{-}(u_n,w_{\widetilde{N}+1})$ or $c_{-}(u_n,w_{\widetilde{N}+1})$. Setting $u_n = w_{\widetilde{N}+1}$ causes all configurations with a $b_{-}(u_n,w_{\widetilde{N}+1})$ vertex to vanish, and we are left with a frozen row of vertices as shown. This row of vertices produces the prefactor in (\ref{sNrec0-xxz}), whilst the remainder of the lattice represents $S_{n-1}$. }
\end{figure}

}
\end{propertylist4}

\begin{propertylist4}
{\rm From the definition of the state vectors (\ref{auxvec}) we obtain 

\begin{align}
&
S_{0} \Big(\{v\}_N,\{w\}_M\Big) 
=
\langle \Downarrow_{N/M}| \otimes \langle \Uparrow_N^{a}|
T\Big( \{v\}_{N},\{w\}_M \Big) 
|\Downarrow_N^{a}\rangle \otimes |\Uparrow_M\rangle  
\end{align}

\noindent with $T(\{v\}_N,\{w\}_M)$ given by (\ref{doublemon}). Using lemma 3 and contracting on the quantum spaces $\mathcal{V}_1,\ldots,\mathcal{V}_M$ gives

\begin{align}
&
S_{0} \Big(\{v\}_N,\{w\}_M\Big) 
=
(-)^{(M+1)N}
\langle \Uparrow_N^{a} |
\prod_{i=1}^{N}
C(w_i,\{\bar{v}\}_N)
\prod_{j=N+1}^{M}
D(w_{j},\{\bar{v}\}_N)
|\Downarrow_N^{a}\rangle
\end{align}

\noindent Now since $|\Downarrow_N^{a}\rangle$ is an eigenvector of the $D$-operators, as can be seen from equation (\ref{diagonal2}), we have

\begin{align}
&
S_{0} \Big(\{v\}_N,\{w\}_M\Big) 
=
(-)^{(M+1)N}
\prod_{i=1}^{N}
\prod_{j=N+1}^{M}
[w_j-\bar{v}_i+\gamma]
\langle \Uparrow_N^{a} |
\prod_{i=1}^{N}
C(w_i,\{\bar{v}\}_N) 
|\Downarrow_N^{a}\rangle
\end{align}

\noindent or equivalently, substituting $\bar{v}_i = v_i+\gamma$ for all $1 \leq i \leq N$ into the previous equation,

\begin{align}
S_{0} \Big(\{v\}_N,\{w\}_M\Big) 
=
\prod_{i=1}^{N}
\prod_{j=N+1}^{M}
[v_i-w_j]
\langle \Uparrow_N^{a} |
\prod_{i=1}^{N}
C(w_i,\{\bar{v}\}_N) 
|\Downarrow_N^{a}\rangle
\label{cond4pf}
\end{align}

\noindent Comparing with the alternative expression (\ref{zequiv}) for the domain wall partition function, equation (\ref{cond4pf}) completes the proof of (\ref{cond4}). A graphical version of this identity is given below.

\begin{figure}[H]

\begin{center}
\begin{minipage}{4.3in}

\setlength{\unitlength}{0.0003cm}
\begin{picture}(20000,15000)(-9000,-3000)

\put(-7000,5000){\fbox{$v$}}


\path(-2000,0)(20000,0)
\put(-3250,0){\tiny$N$}
\path(-4500,0)(-3500,0)
\whiten\path(-3750,250)(-3750,-250)(-3500,0)(-3750,250)
\blacken\path(-1250,250)(-1250,-250)(-750,0)(-1250,250)
\blacken\path(11250,250)(11250,-250)(10750,0)(11250,250)
\blacken\path(13250,250)(13250,-250)(12750,0)(13250,250)
\blacken\path(15250,250)(15250,-250)(14750,0)(15250,250)
\blacken\path(17250,250)(17250,-250)(16750,0)(17250,250)
\blacken\path(19250,250)(19250,-250)(18750,0)(19250,250)

\path(-2000,2000)(20000,2000)
\path(-4500,2000)(-3500,2000)
\whiten\path(-3750,2250)(-3750,1750)(-3500,2000)(-3750,2250)
\blacken\path(-1250,2250)(-1250,1750)(-750,2000)(-1250,2250)
\blacken\path(11250,2250)(11250,1750)(10750,2000)(11250,2250)
\blacken\path(13250,2250)(13250,1750)(12750,2000)(13250,2250)
\blacken\path(15250,2250)(15250,1750)(14750,2000)(15250,2250)
\blacken\path(17250,2250)(17250,1750)(16750,2000)(17250,2250)
\blacken\path(19250,2250)(19250,1750)(18750,2000)(19250,2250)

\path(-2000,4000)(20000,4000)
\path(-4500,4000)(-3500,4000)
\whiten\path(-3750,4250)(-3750,3750)(-3500,4000)(-3750,4250)
\blacken\path(-1250,4250)(-1250,3750)(-750,4000)(-1250,4250)
\blacken\path(11250,4250)(11250,3750)(10750,4000)(11250,4250)
\blacken\path(13250,4250)(13250,3750)(12750,4000)(13250,4250)
\blacken\path(15250,4250)(15250,3750)(14750,4000)(15250,4250)
\blacken\path(17250,4250)(17250,3750)(16750,4000)(17250,4250)
\blacken\path(19250,4250)(19250,3750)(18750,4000)(19250,4250)

\path(-2000,6000)(20000,6000)
\path(-4500,6000)(-3500,6000)
\whiten\path(-3750,6250)(-3750,5750)(-3500,6000)(-3750,6250)
\blacken\path(-1250,6250)(-1250,5750)(-750,6000)(-1250,6250)
\blacken\path(11250,6250)(11250,5750)(10750,6000)(11250,6250)
\blacken\path(13250,6250)(13250,5750)(12750,6000)(13250,6250)
\blacken\path(15250,6250)(15250,5750)(14750,6000)(15250,6250)
\blacken\path(17250,6250)(17250,5750)(16750,6000)(17250,6250)
\blacken\path(19250,6250)(19250,5750)(18750,6000)(19250,6250)

\path(-2000,8000)(20000,8000)
\path(-4500,8000)(-3500,8000)
\whiten\path(-3750,8250)(-3750,7750)(-3500,8000)(-3750,8250)
\blacken\path(-1250,8250)(-1250,7750)(-750,8000)(-1250,8250)
\blacken\path(11250,8250)(11250,7750)(10750,8000)(11250,8250)
\blacken\path(13250,8250)(13250,7750)(12750,8000)(13250,8250)
\blacken\path(15250,8250)(15250,7750)(14750,8000)(15250,8250)
\blacken\path(17250,8250)(17250,7750)(16750,8000)(17250,8250)
\blacken\path(19250,8250)(19250,7750)(18750,8000)(19250,8250)

\path(-2000,10000)(20000,10000)
\put(-3250,10000){\tiny$1$}
\path(-4500,10000)(-3500,10000)
\whiten\path(-3750,10250)(-3750,9750)(-3500,10000)(-3750,10250)
\blacken\path(-1250,10250)(-1250,9750)(-750,10000)(-1250,10250)
\blacken\path(11250,10250)(11250,9750)(10750,10000)(11250,10250)
\blacken\path(13250,10250)(13250,9750)(12750,10000)(13250,10250)
\blacken\path(15250,10250)(15250,9750)(14750,10000)(15250,10250)
\blacken\path(17250,10250)(17250,9750)(16750,10000)(17250,10250)
\blacken\path(19250,10250)(19250,9750)(18750,10000)(19250,10250)


\path(0,-2000)(0,12000)
\put(-250,-2750){\tiny$1$}
\path(0,-4250)(0,-3250)
\whiten\path(-250,-3500)(250,-3500)(0,-3250)(-250,-3500)
\blacken\path(-250,-750)(250,-750)(0,-1250)(-250,-750)
\blacken\path(-250,10750)(250,10750)(0,11250)(-250,10750)

\path(2000,-2000)(2000,12000)
\path(2000,-4250)(2000,-3250)
\whiten\path(1750,-3500)(2250,-3500)(2000,-3250)(1750,-3500)
\blacken\path(1750,-750)(2250,-750)(2000,-1250)(1750,-750)
\blacken\path(1750,10750)(2250,10750)(2000,11250)(1750,10750)

\path(4000,-2000)(4000,12000)
\path(4000,-4250)(4000,-3250)
\whiten\path(3750,-3500)(4250,-3500)(4000,-3250)(3750,-3500)
\blacken\path(3750,-750)(4250,-750)(4000,-1250)(3750,-750)
\blacken\path(3750,10750)(4250,10750)(4000,11250)(3750,10750)

\path(6000,-2000)(6000,12000)
\path(6000,-4250)(6000,-3250)
\whiten\path(5750,-3500)(6250,-3500)(6000,-3250)(5750,-3500)
\blacken\path(5750,-750)(6250,-750)(6000,-1250)(5750,-750)
\blacken\path(5750,10750)(6250,10750)(6000,11250)(5750,10750)

\path(8000,-2000)(8000,12000)
\path(8000,-4250)(8000,-3250)
\whiten\path(7750,-3500)(8250,-3500)(8000,-3250)(7750,-3500)
\blacken\path(7750,-750)(8250,-750)(8000,-1250)(7750,-750)
\blacken\path(7750,10750)(8250,10750)(8000,11250)(7750,10750)

\path(10000,-2000)(10000,12000)
\put(9750,-2750){\tiny$N$}
\path(10000,-4250)(10000,-3250)
\whiten\path(9750,-3500)(10250,-3500)(10000,-3250)(9750,-3500)
\blacken\path(9750,-750)(10250,-750)(10000,-1250)(9750,-750)
\blacken\path(9750,10750)(10250,10750)(10000,11250)(9750,10750)

\path(12000,-2000)(12000,12000)
\put(11750,-2750){\tiny$N+1$}
\path(12000,-4250)(12000,-3250)
\whiten\path(11750,-3500)(12250,-3500)(12000,-3250)(11750,-3500)
\blacken\path(11750,-1250)(12250,-1250)(12000,-750)(11750,-1250)
\blacken\path(11750,750)(12250,750)(12000,1250)(11750,750)
\blacken\path(11750,2750)(12250,2750)(12000,3250)(11750,2750)
\blacken\path(11750,4750)(12250,4750)(12000,5250)(11750,4750)
\blacken\path(11750,6750)(12250,6750)(12000,7250)(11750,6750)
\blacken\path(11750,8750)(12250,8750)(12000,9250)(11750,8750)
\blacken\path(11750,10750)(12250,10750)(12000,11250)(11750,10750)

\path(14000,-2000)(14000,12000)
\path(14000,-4250)(14000,-3250)
\whiten\path(13750,-3500)(14250,-3500)(14000,-3250)(13750,-3500)
\blacken\path(13750,-1250)(14250,-1250)(14000,-750)(13750,-1250)
\blacken\path(13750,750)(14250,750)(14000,1250)(13750,750)
\blacken\path(13750,2750)(14250,2750)(14000,3250)(13750,2750)
\blacken\path(13750,4750)(14250,4750)(14000,5250)(13750,4750)
\blacken\path(13750,6750)(14250,6750)(14000,7250)(13750,6750)
\blacken\path(13750,8750)(14250,8750)(14000,9250)(13750,8750)
\blacken\path(13750,10750)(14250,10750)(14000,11250)(13750,10750)

\path(16000,-2000)(16000,12000)
\path(16000,-4250)(16000,-3250)
\whiten\path(15750,-3500)(16250,-3500)(16000,-3250)(15750,-3500)
\blacken\path(15750,-1250)(16250,-1250)(16000,-750)(15750,-1250)
\blacken\path(15750,750)(16250,750)(16000,1250)(15750,750)
\blacken\path(15750,2750)(16250,2750)(16000,3250)(15750,2750)
\blacken\path(15750,4750)(16250,4750)(16000,5250)(15750,4750)
\blacken\path(15750,6750)(16250,6750)(16000,7250)(15750,6750)
\blacken\path(15750,8750)(16250,8750)(16000,9250)(15750,8750)
\blacken\path(15750,10750)(16250,10750)(16000,11250)(15750,10750)

\path(18000,-2000)(18000,12000)
\put(17750,-2750){\tiny$M$}
\path(18000,-4250)(18000,-3250)
\whiten\path(17750,-3500)(18250,-3500)(18000,-3250)(17750,-3500)
\blacken\path(17750,-1250)(18250,-1250)(18000,-750)(17750,-1250)
\blacken\path(17750,750)(18250,750)(18000,1250)(17750,750)
\blacken\path(17750,2750)(18250,2750)(18000,3250)(17750,2750)
\blacken\path(17750,4750)(18250,4750)(18000,5250)(17750,4750)
\blacken\path(17750,6750)(18250,6750)(18000,7250)(17750,6750)
\blacken\path(17750,8750)(18250,8750)(18000,9250)(17750,8750)
\blacken\path(17750,10750)(18250,10750)(18000,11250)(17750,10750)


\put(20000,-3000){\fbox{$w$}}

\end{picture}

\end{minipage}
\end{center}

\caption[Equivalence between $S_0$ and $Z_N$]{Equivalence between $S_0$ and $Z_N$. The final $M-N$ columns of the $S_0$ lattice must assume the configuration shown. All other configurations vanish. The block of vertices thus obtained corresponds with the prefactor in (\ref{cond4}), whilst the remainder of the lattice represents $Z_N$.}
\end{figure}
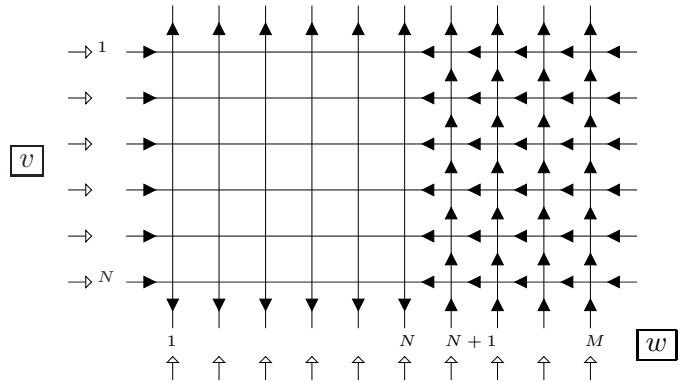

}
\end{propertylist4}

\end{proof}

\subsection{Determinant expression for $S_n(\{u\}_n,\{v\}_N,\{w\}_M)$}

\begin{lemma}
{\rm 
Let us define the functions 

\begin{align}
&
f_i(w)
=
\frac{[\gamma]}{[v_i-w]}
\prod_{k\not=i}^{N} [v_k-w+\gamma]
\label{ffunct}
\\
&
g_i(u)
=
\label{gfunct}
\frac{[\gamma]}{[v_i-u]}
\left(
\prod_{k\not=i}^{N} [v_k-u+\gamma] \prod_{k=1}^{M} [u-w_k+\gamma]
-
\prod_{k\not=i}^{N} [v_k-u-\gamma] \prod_{k=1}^{M}[u-w_k]
\right)
\end{align}

\noindent Using these definitions, we construct the $N \times N$ matrix

\begin{align}
&
\mathcal{M}_n \Big(\{u\}_n,\{v\}_N,\{w\}_M\Big)
=
\label{M}
\left(
\begin{array}{cccccc}
f_1(w_1) & \cdots & f_1(w_{\widetilde{N}}) & 
g_1(u_n) & \cdots & g_1(u_1)
\\
\vdots & & \vdots & \vdots & & \vdots
\\
f_N(w_1) & \cdots & f_N(w_{\widetilde{N}}) &
g_N(u_n) & \cdots & g_N(u_1)
\end{array}
\right)
\end{align}

\noindent Assuming that the parameters $\{v\}_N$ satisfy the Bethe equations (\ref{bethe2}), we have

\begin{align}
\label{sNcalc-xxz}
S_n
=
\frac{
\displaystyle{
\prod_{i=1}^{N} \prod_{j=1}^{M} [v_i-w_j]
\det\mathcal{M}_n \Big(\{u\}_n,\{v\}_N,\{w\}_M\Big)
}
}
{
\displaystyle{
\prod_{i=1}^{n} \prod_{j=1}^{\widetilde{N}}[u_i-w_j]
\prod_{1 \leq i < j \leq n} [u_i-u_j]
\prod_{1 \leq i < j \leq N} [v_i-v_j]
\prod_{1 \leq i < j \leq \widetilde{N}} [w_j-w_i]
}
}
\end{align}

\noindent The expression (\ref{sNcalc-xxz}) for the intermediate Bethe scalar product $S_n$ originally appeared in appendix C of \cite{kmt}.

}
\end{lemma}

\begin{proof}
Firstly, one must show that $S_n$ is uniquely determined by the set of conditions in lemma \ref{sncond}. This is accomplished using similar arguments to those presented in lemma \ref{uniqueness}, and we shall assume this fact {\it a priori.} Hence it will be sufficient to show that the expression (\ref{sNcalc-xxz}) satisfies the list of properties given in lemma \ref{sncond}.

\begin{propertylist5}
{\rm
All dependence of (\ref{sNcalc-xxz}) on the variables $\{w_{\widetilde{N}+1},\ldots,w_M\}$ occurs in the factor $\prod_{i=1}^{N} \prod_{j=1}^{M} [v_i-w_j]$ and in the functions $g_i(u_j)$ in the determinant. Clearly, these terms are invariant under the permutation $w_i \leftrightarrow w_j$ for all $i \not= j$. Hence the expression (\ref{sNcalc-xxz}) is symmetric in $\{w_{\widetilde{N}+1},\ldots,w_M\}$.
}
\end{propertylist5}

\begin{propertylist5} 


{\rm Consider the expression (\ref{gfunct}) for $g_i(u_n)$. Since the variables $\{v\}_N$ satisfy the Bethe equations (\ref{bethe2}), the numerator of $g_i(u_n)$ vanishes in the limit $u_n \rightarrow v_i$. It follows that the pole in (\ref{gfunct}) is removable, and therefore $g_i(u_n)$ is a trigonometric polynomial of degree $M+N-2$ in $u_n$. Using this fact, we see that (\ref{sNcalc-xxz}) is a quotient of trigonometric polynomials in $u_n$. The polynomial in the numerator has degree $M+N-2$, while the polynomial in the denominator has degree $N-1$. We must show that every zero in the denominator is cancelled by a zero in the numerator.

Setting $u_n = u_j$ for $1 \leq j \leq n-1$ causes two columns of the determinant to become equal, producing $n-1$ zeros in the numerator which cancel $n-1$ of the zeros in the denominator. Furthermore since 

\begin{align}
g_i(w_j)
&=
\frac{[\gamma]}{[v_i-w_j]}
\prod_{k \not= i}^{N} [v_k-w_j+\gamma]
\prod_{k = 1}^{M} [w_j-w_k+\gamma]
=
\prod_{k=1}^{M}[w_j-w_k+\gamma]
f_i(w_j)
\label{stuff}
\end{align}

\noindent it follows that by setting $u_n = w_j$ for $1 \leq j \leq \widetilde{N}$, two columns of the determinant are equal up to a multiplicative factor, producing $\widetilde{N}$ zeros in the numerator which cancel $\widetilde{N}$ of the zeros in the denominator. This proves that the expression (\ref{sNcalc-xxz}) is a trigonometric polynomial of degree $M-1$ in $u_n$. 

Finally, since

\begin{align}
g_i(w_j-\gamma)
&=
\frac{-[\gamma]}{[v_i-w_j+\gamma]}
\prod_{k \not= i}^{N} [v_k-w_j]
\prod_{k=1}^{M} [w_j-w_k-\gamma]
\nonumber
\\
&=
-
\prod_{k=1}^{M} [w_j-w_k-\gamma]
\prod_{k=1}^{N} \frac{[v_k-w_j]}{[v_k-w_j+\gamma]}
f_i(w_j)
\end{align}

\noindent we see that by setting $u_n = w_j -\gamma$ for all $1 \leq j \leq \widetilde{N}$, two columns of the determinant are equal up to a multiplicative factor, producing the $\widetilde{N}$ zeros which (\ref{sNcalc-xxz}) requires in order to satisfy property {\bf 2}. 
}
\end{propertylist5}

\begin{propertylist5}
{\rm Using equation (\ref{stuff}) and the definition of the matrix (\ref{M}), it is clear that 

\begin{align}
\det \mathcal{M}_n \Big( \{u\}_n,\{v\}_N,\{w\}_M  \Big)
\Big|_{u_n = w_{\widetilde{N}+1}}
&=
\prod_{k=1}^{M}
[w_{\widetilde{N}+1}-w_k+\gamma]
\label{sNrec-xxz}
\\
&
\times 
\det \mathcal{M}_{n-1} \Big( \{u\}_{n-1}, \{v\}_N, \{w\}_M \Big)
\nonumber
\end{align}

\noindent Furthermore, we notice the trivial product identity

\begin{align}
&
\left.
\left(
\prod_{i=1}^{n} \prod_{j=1}^{\widetilde{N}} [u_i-w_j]
\prod_{1 \leq i<j \leq n} [u_i-u_j]
\prod_{1 \leq i< j \leq \widetilde{N}} [w_j -w_i]
\right)
\right|_{u_n = w_{\widetilde{N}+1}}
=
\label{sNrec2-xxz}
\\
&
\phantom{\Big((}
\prod_{i=1}^{n-1} \prod_{j=1}^{\widetilde{N}+1} [u_i-w_j]
\prod_{1 \leq i<j \leq n-1} [u_i-u_j]
\prod_{1 \leq i < j \leq \widetilde{N}+1}[w_j - w_i]
\nonumber
\end{align}

\noindent Combining the results (\ref{sNrec-xxz}) and (\ref{sNrec2-xxz}), we find that the expression (\ref{sNcalc-xxz}) satisfies the recursion relation (\ref{sNrec0-xxz}).
}
\end{propertylist5}

\begin{propertylist5}
{\rm
Taking the $n=0$ case of (\ref{sNcalc-xxz}) yields

\begin{align}
S_0\Big(
\{v\}_N,\{w\}_M
\Big)
=
\frac{\displaystyle{
\prod_{i=1}^{N}
\prod_{j=1}^{M}
[v_i-w_j]
\det\mathcal{M}_0
}}
{\displaystyle{
\prod_{1\leq i<j \leq N}
[v_i-v_j][w_j-w_i]
}}
\label{s0last}
\end{align}

\noindent with the matrix $\mathcal{M}_0$ given by

\begin{align}
\mathcal{M}_0
&=
\left(
\begin{array}{ccc}
\frac{[\gamma]}{[v_1-w_1]} \prod_{k\not=1} [v_k-w_1+\gamma]
&
\cdots
&
\frac{[\gamma]}{[v_1-w_N]} \prod_{k\not=1} [v_k-w_N+\gamma]
\\
\vdots
&
&
\vdots
\\
\frac{[\gamma]}{[v_N-w_1]} \prod_{k\not=N} [v_k-w_1+\gamma]
&
\cdots
&
\frac{[\gamma]}{[v_N-w_N]} \prod_{k\not=N} [v_k-w_N+\gamma]
\end{array}
\right)
\end{align}

\noindent Comparing (\ref{s0last}) with the determinant expression (\ref{dwpf}) for the domain wall partition function, we see that $S_0 = \prod_{i=1}^{N} \prod_{j=N+1}^{M} [v_i-w_j] Z_N$, as required.
}
\end{propertylist5}

\end{proof}

\subsection{Evaluation of $S_N(\{u\}_N,\{v\}_N,\{w\}_M)$}

Let us now consider the $n=N$ case of equation (\ref{sNcalc-xxz}) in more detail. For purely aesthetic purposes, we simultaneously reverse the order of the columns in the matrix $\mathcal{M}_N$ and the order of the variables in the Vandermonde $\prod_{1\leq i<j \leq N} [u_i-u_j]$. We also take the transpose of the matrix $\mathcal{M}_N$. The formula (\ref{sNcalc-xxz}) is invariant under these transformations, and we obtain 

\begin{align}
&
S_N\Big(\{u\}_N,\{v\}_N,\{w\}_M\Big)
=
\frac{\displaystyle{
[\gamma]^N
\prod_{i=1}^{N} \prod_{j=1}^{M}
[v_i-w_j]
}}
{\displaystyle{
\prod_{1\leq i < j \leq N}
[u_j-u_i][v_i-v_j]
}}
\times
\label{betscal}
\\
&
\det \left(
\frac{\displaystyle{
\prod_{k\not= j}^{N}[v_k-u_i-\gamma]
\prod_{k=1}^{M} [u_i-w_k]
-
\prod_{k\not=j}^{N} [v_k-u_i+\gamma]
\prod_{k=1}^{M} [u_i-w_k+\gamma]
}}
{[u_i-v_j]}
\right)_{1\leq i,j \leq N}
\nonumber
\end{align}

The expression (\ref{betscal}) for the Bethe scalar product was first proved by Slavnov in \cite{sla}. The original proof required a recursion relation between scalar products of dimension $N$ and $N-1$, which can be found in section 3, chapter IX of \cite{kbi}. Although this earlier proof applies universally to quantum models with the $R$-matrix (\ref{Rmat1}), it seems less transparent than the simple Izergin-Korepin proof which we have proposed in the context of the XXZ spin-$\frac{1}{2}$ chain.

\subsection{Scalar product as a power-sum specialized KP $\tau$-function} 

In this subsection we consider the Bethe scalar product $S'_N$, which arises from a normalization and change of variables applied to $S_N$. We will show that, by virtue of its determinant form, $S'_N$ is a specialization of a polynomial KP $\tau$-function \cite{fwz5}. As in the case of the domain wall partition function $Z_N'$, this specialization is achieved by setting the $\tau$-function time variables to power sums of rapidities. The procedure we follow is fairly unremarkable, and requires only elementary determinant operations. However the result itself is certainly interesting, since it provides a link between solutions of a quantum model (Bethe eigenvectors of the XXZ spin-$\frac{1}{2}$ chain) and solutions of a classical hierarchy (KP $\tau$-functions).


Letting $S_N(\{u\}_N,\{v\}_N,\{w\}_M)$ denote the full Bethe scalar product, we define 
{
\small
\begin{align}
S_N'\Big(\{x\}_N,\{y\}_N,\{z\}_M\Big)
&=
e^{N^2\gamma}
\prod_{i=1}^{N}
e^{(M-1)(v_i-u_i)}
\prod_{j=1}^{M}
e^{2N w_j}
S_N\Big( \{u\}_N,\{v\}_N,\{w\}_M \Big)
\label{renormscal}
\end{align}
}

\noindent where we have set $e^{-2 u_i} = x_i, e^{2 v_i} = y_i, e^{2 w_i} = z_i, e^{2\gamma} = q$. Applying this change of variables to (\ref{betscal}) and extracting a factor of $(-)^{N-1}$ from each row of the determinant leads to the explicit formula

{
\small
\begin{align}
&
S_N'\Big(
\{x\}_N,\{y\}_N,\{z\}_M
\Big)
=
\frac{\displaystyle{
(q-1)^N
\prod_{i=1}^{N}
\prod_{j=1}^{M}
(y_i-z_j)
}}
{\displaystyle{
\prod_{1\leq i<j \leq N}
(x_i-x_j)(y_i-y_j)
}}
\times
\label{sNcalc5-xxz}
\\
&
\det\left(
\frac{\displaystyle{
q^{N-1}
\prod_{k\not= j}^{N}
\Big(1-x_i \frac{y_k}{q}\Big)
\prod_{k=1}^{M}
(1-x_i z_k)
-
q^{\frac{M}{2}}
\prod_{k\not=j}^{N}
(1-q x_i y_k)
\prod_{k=1}^{M}
\Big(1 - x_i \frac{z_k}{q}\Big)
}}
{1-x_i y_j}
\right)_{1\leq i,j \leq N}
\nonumber
\end{align}
}

\noindent Expanding the numerator of the determinant in (\ref{sNcalc5-xxz}) in terms of elementary symmetric functions (\ref{elemen}), we obtain the equivalent expression

\begin{align}
&
S_N'\Big(
\{x\}_N,\{y\}_N,\{z\}_M
\Big)
=
\frac{\displaystyle{
(q-1)^N
\prod_{i=1}^{N}
\prod_{j=1}^{M}
(y_i-z_j)
}}
{\displaystyle{
\prod_{1\leq i<j \leq N}
(x_i-x_j)(y_i-y_j)
}}
\times
\label{sNcalc6-xxz}
\\
&
\det\left(
\frac{\displaystyle{
\sum_{k=0}^{M+N-1}
(-x_i)^k
\left[
q^{N-1}
e_k\Big\{\frac{\widehat{y}_j}{q}\Big\}\cup \{z\}
-
q^{\frac{M}{2}}
e_k\{q \widehat{y}_j\}\cup\Big\{\frac{z}{q}\Big\}
\right]
}}
{1-x_i y_j}
\right)_{1\leq i,j \leq N}
\nonumber
\end{align}

\noindent From here, we wish to remove the pole within the determinant in (\ref{sNcalc6-xxz}). This is done using the fact that the rapidities $\{v\}_N$ obey the Bethe equations (\ref{bethe2}). Employing the change of variables $e^{2 v_i} = y_i, e^{2 w_i} = z_i, e^{2\gamma} =q$ in (\ref{bethe2}), the Bethe equations take the form

\begin{align}
\beta_i
=
q^{N-1}
\prod_{j\not=i}^{N}
\Big(1 - \frac{y_j}{q y_i}\Big)
\prod_{j=1}^{M}
\Big(1-\frac{z_j}{y_i}\Big)
-
q^{\frac{M}{2}}
\prod_{j\not=i}^{N}
\Big(1 - \frac{q y_j}{y_i}\Big)
\prod_{j=1}^{M}
\Big(1-\frac{z_j}{q y_i}\Big)
=
0
\label{xxz-bethe1}
\end{align}

\noindent or in terms of elementary symmetric functions

\begin{align}
\beta_i
=
\sum_{k=0}^{M+N-1}
(-y_i)^{-k}
\left[
q^{N-1}
e_k\Big\{\frac{\widehat{y}_i}{q}\Big\}\cup\{z\}
-
q^{\frac{M}{2}}
e_k\{q\widehat{y}_i\}\cup\Big\{\frac{z}{q}\Big\}
\right]
=
0
\label{xxz-bethe2}
\end{align}

\noindent where we have denoted the left hand side of (\ref{xxz-bethe1}) and (\ref{xxz-bethe2}) by $\beta_i$, since these equations correspond with the $i^{\rm th}$ Bethe equation. Using (\ref{xxz-bethe2}), we see that the numerator of the determinant (\ref{sNcalc6-xxz}) collapses to zero when $x_i =1/y_j$. Hence we may remove the pole in this determinant, yielding\footnote{If the arbitrary degree-$m$ polynomial $P(x) = \sum_{k=0}^{m} x^k p_k$ has a zero at the point $x=1/y$, then
$
\frac{P(x)}{1-xy}
=
\sum_{k=0}^{m-1}
x^k
\sum_{l=0}^{k}
y^{k-l} p_l  
$. We use this fact to establish equation (\ref{sNcalc3-xxz}).
}   

\begin{align}
&
S_N'\Big(
\{x\}_N,\{y\}_N,\{z\}_M
\Big)
=
\frac{\displaystyle{
(q-1)^N
\prod_{i=1}^{N}
\prod_{j=1}^{M}
(y_i-z_j)
}}
{\displaystyle{
\prod_{1\leq i<j \leq N}
(x_i-x_j)(y_i-y_j)
}}
\times
\label{sNcalc3-xxz}
\\
&
\det\left(
\sum_{k=0}^{M+N-2}
x_i^k y_j^k
\sum_{l=0}^{k}
(-y_j)^{-l}
\left[
q^{N-1}
e_l\Big\{\frac{\widehat{y}_j}{q}\Big\}\cup \{z\}
-
q^{\frac{M}{2}}
e_l\{q \widehat{y}_j\}\cup\Big\{\frac{z}{q}\Big\}
\right]
\right)_{1 \leq i, j \leq N}
\nonumber
\end{align}

\noindent or more succinctly

{
\small
\begin{align}
&
S_N'\Big(
\{x\}_N,\{y\}_N,\{z\}_M
\Big)
=
\label{sNcalc4-xxz}
\frac{\displaystyle{
(q-1)^N
\prod_{i=1}^{N}
\prod_{j=1}^{M}
(y_i-z_j)
}}
{\displaystyle{
\prod_{1\leq i<j \leq N}
(x_i-x_j)(y_i-y_j)
}}
\det\left(
\sum_{k=0}^{M+N-2}
x_i^k
\Big[y_j^k \beta_j\Big]_+
\right)_{1\leq i,j \leq N}
\end{align}
}

\noindent where $[y_j^k \beta_j]_{+}$ denotes all terms in $y_j^k \beta_j$ which have non-negative degree in $y_j$. Now we may follow the procedure of subsection 5.2.5 to bring (\ref{sNcalc4-xxz}) towards the form of a KP $\tau$-function.  Applying the Cauchy-Binet identity (\ref{cauch-bin}) to the determinant in (\ref{sNcalc4-xxz}), we obtain

\begin{align}
&
S_N'\Big(\{x\}_N,\{y\}_N,\{z\}_M\Big)
=
\frac{\displaystyle{
(q-1)^N
\prod_{i=1}^{N} \prod_{j=1}^{M} (y_i-z_j)
}}
{\displaystyle{
\prod_{1 \leq i < j \leq N}
(x_i-x_j)(y_i-y_j)
}}
\times
\\
&
\sum_{M+N-1 \geq k_1 > \cdots > k_N \geq 1}
\det\Big(
x_i^{k_j-1}
\Big)_{1\leq i,j \leq N}
\det\Big(
\Big[y_j^{k_i-1}\beta_j\Big]_{+}
\Big)_{1\leq i,j \leq N}
\nonumber
\end{align}

\noindent We perform a change in summation variables, writing $k_i-1 = \mu_i-i+N$ for all $1\leq i \leq N$, which gives

\begin{align}
&
S_N'\Big(\{x\}_N,\{y\}_N,\{z\}_M\Big)
=
\frac{\displaystyle{
(q-1)^N
\prod_{i=1}^{N} \prod_{j=1}^{M} (y_i-z_j)
}}
{\displaystyle{
\prod_{1 \leq i < j \leq N}
(x_i-x_j)(y_i-y_j)
}}
\times
\\
&
\sum_{\mu \subseteq [N,M-1]}
\det\Big(
x_i^{\mu_j-j+N}
\Big)_{1\leq i,j \leq N}
\det\Big(
\Big[y_j^{\mu_i-i+N}\beta_j\Big]_{+}
\Big)_{1\leq i,j \leq N}
\nonumber
\end{align}

\noindent where the sum is taken over all partitions $\mu = \{\mu_1 \geq \cdots \geq \mu_N \geq 0 \}$ whose Young diagrams fit inside the rectangle $[N,M-1]$. Finally, by virtue of the Jacobi-Trudi identity (\ref{jac-trud}) we have

\begin{align}
S_N'\Big(
\{x\}_N,\{y\}_N,\{z\}_M
\Big)
=
\sum_{\mu \subseteq [N,M-1]}
s_{\mu}\{x\}
\varsigma_{\mu}(\{y\},\{z\},q)
\end{align}

\noindent with $s_{\mu}\{x\}$ a Schur function in the variables $\{x\} = \{x_1,\ldots,x_N\}$ and where we have defined the function  

\begin{align}
\varsigma_{\mu}(\{y\},\{z\},q)
=
\frac{\displaystyle{
(q-1)^N
\prod_{i=1}^{N} \prod_{j=1}^{M} (y_i-z_j)
\det\Big(\Big[
y_j^{\mu_i-i+N}\beta_j
\Big]_{+}\Big)_{1\leq i,j \leq N}
}}
{\displaystyle{
\prod_{1 \leq i < j \leq N}
(y_i-y_j)
}}
\label{compactcoeff}
\end{align}

\noindent This expression for the coefficients $\varsigma_{\mu}(\{y\},\{z\},q)$ is significantly more compact than the one which appeared in \cite{fwz5}. Written in the form (\ref{compactcoeff}), $\varsigma_{\mu}(\{y\},\{z\},q)$ is not far removed from a Schur function in the variables $\{y\}$. The only difference is the presence of the Bethe variables $\beta_j$ in each entry of the determinant.

We have thus shown that $S'_N(\{x\}_N,\{y\}_N,\{z\}_M)$ is equal to

\begin{align}
\tau_{\mbox{\tiny SP}}\{t\}
\sum_{\mu \subseteq [N,M-1]}
\chi_{\mu}\{t\} \varsigma_{\mu}(\{y\},\{z\},q)
\label{scaltau}
\end{align}

\noindent under the power-sum specialization $t_n = \frac{1}{n} \sum_{i=1}^{N} x_i^n$ for all $n \geq 1$. In the next subsection we show that the polynomial $\tau_{\mbox{\tiny SP}}\{t\}$ is a KP $\tau$-function, by writing it as an expectation value of charged fermions. 

\subsection{$\tau_{\mbox{\tiny SP}}\{t\}$ as an expectation value of charged fermions}

Proceeding in direct analogy with subsection 5.2.6, it is possible to express the right hand side of (\ref{scaltau}) in the canonical form (\ref{KPexpI1}) of a KP $\tau$-function. As usual, this is made possible by the fact that the coefficients $\varsigma_{\mu}(\{y\},\{z\},q)$ are determinants. To avoid needless repetition we omit the details of this calculation and quote only the result, which reads 

%
%
%
%
%
%

\begin{align}
\tau_{\mbox{\tiny SP}}\{t\}
=
\varsigma_{\emptyset}
\langle 0|
e^{H\{t\}}
\exp\left(
\sum_{m=1}^{M-1}
\sum_{n=1}^{N}
(-)^{n-1}
\frac{\varsigma_{\{m,1^{n-1}\}}}{\varsigma_{\emptyset}}
\psi_{m-1}
\psis_{-n}
\right)
|0\rangle
\label{tausp}
\end{align}

\noindent where we have abbreviated $\varsigma_{\mu}(\{y\},\{z\},q) = \varsigma_{\mu}$ for all partitions $\mu$. As was noticed in \cite{fwz5}, equation (\ref{tausp}) allows us to identify the XXZ Bethe eigenvectors with elements of $\mathcal{F}_{\psi}^{(0)}$ which lie in the $GL_{\infty}$ orbit of the Fock vacuum. By writing the scalar product (\ref{renormscal}) in the form

\begin{align}
S_N'\Big(\{x\}_N,\{y\}_N,\{z\}_M\Big)
=
\langle 0|
\prod_{i=1}^{N} \mathbb{C}(x_i,\{z\}_M)
\prod_{j=1}^{N} \mathbb{B}(y_j,\{z\}_M)
|0\rangle
\end{align}

\noindent for some suitably renormalized monodromy matrix operators, we have

\begin{align}
\langle 0|
\prod_{i=1}^{N} \mathbb{C}(x_i,\{z\}_M)
\prod_{j=1}^{N} \mathbb{B}(y_j,\{z\}_M)
|0\rangle
=
\varsigma_{\emptyset}
\left\langle
\prod_{i=1}^{N}
\Big(
e^{\sum_{n=1}^{\infty} \frac{1}{n} x_i^n H_n}
\Big)
e^{X(\{y\},\{z\},q)}
\right\rangle 
\label{suggestive}
\end{align}

\noindent where $X(\{y\},\{z\},q) \in A_{\infty}$ is given by

\begin{align}
X(\{y\},\{z\},q)
=
\sum_{m=1}^{M-1}
\sum_{n=1}^{N}
(-)^{n-1}
\frac{\varsigma_{\{m,1^{n-1}\}}}{\varsigma_{\emptyset}}
\psi_{m-1}
\psis_{-n}
\end{align}

\noindent The identity (\ref{suggestive}) is suggestive of a map from $\mathcal{V}$ to $\mathcal{F}_{\psi}^{(0)}$ under which

\begin{align}
\prod_{j=1}^{N} \mathbb{B}(y_j,\{z\}_M)
|0\rangle
\longmapsto
\varsigma_{\emptyset}
e^{X(\{y\},\{z\},q)} 
|0\rangle
\end{align}

\noindent Such a mapping is reminiscent of those which were discussed in chapter 3, albeit in the context of simpler models. 


\section{Calculation of Bethe eigenvectors}  
\label{xxz-BE}

In this section we calculate the Bethe eigenvector coefficients $b_{\lambda}(\{v\}_N,\{w\}_M)$, as given in equation (\ref{b}). We claim an analogous result for the coefficients $c_{\lambda}(\{v\}_N,\{w\}_M)$ in equation (\ref{c*}), but for the sake of brevity we do not present an explicit proof. 

\subsection{Weighted determinants}

\begin{definition}
{\rm 

Let $\mathcal{W}$ and $\mathcal{M}$ be $M\times M$ matrices and fix an integer $l \leq M$. The {\it weighted determinant} $\overline{{\rm det}}_l(\mathcal{W},\mathcal{M})$ is defined as

\begin{align}
\overline{{\rm det}}_l(\mathcal{W},\mathcal{M})
=
\sum_{\sigma \in S_M}
\left(
\prod_{1\leq i < j \leq l}\mathcal{W}_{\sigma_i,\sigma_j}
\right)
{\rm sgn}(\sigma)
\prod_{i=1}^{M}\mathcal{M}_{i,\sigma_i}
\label{det}
\end{align}

\noindent where the sum is over all permutations $\sigma$ of the integers $\{1,\ldots,M\}$, and the sign of the permutation ${\rm sgn}(\sigma)$ is given by 

\begin{align}
{\rm sgn}(\sigma)
=
\prod_{1\leq i < j \leq M}
{\rm sgn}(\sigma_j-\sigma_i)
\label{sgn}
\end{align}

\noindent We will call $\mathcal{W}$ a {\it weighting matrix,} and the ordinary determinant is recovered by setting its entries to $\mathcal{W}_{i,j}=1$ for all $1\leq i,j \leq M$. Hence equation (\ref{det}) may be viewed as a generalization of the determinant. In a regular determinant, each term in the sum is weighted with the sign (\ref{sgn}) that depends on pairs of elements $\{\sigma_i,\sigma_j\}$ in the permutation. The weighted determinant is different in that it allows the weighting to be a more general function of pairs of elements in the permutation.

}
\end{definition}

\begin{example}
{\rm Let us consider the Hall-Littlewood function $P_{\mu}(\{x\},t)$, as given by the sum (\ref{hl-def}) in chapter 4. By explicitly performing the permutation $\sigma$ of the variables $\{x_1,\ldots,x_N\}$ within this sum, we obtain

\begin{align}
\label{hl}
P_{\mu}(\{x\},t)
=
\frac{\displaystyle{
\sum_{\sigma \in S_N}
\left(\prod_{1\leq i<j \leq N} (x_{\sigma_i}-tx_{\sigma_j})\right)
{\rm sgn}(\sigma)
\prod_{i=1}^{N} x_{\sigma_i}^{\mu_i}}
}{
\displaystyle{v_{\mu}(t)\prod_{1 \leq i<j \leq N}(x_i-x_j) }
}
\end{align}

\noindent where the constant term $v_{\mu}(t)$ is given by (\ref{vmu-def}). We define $N\times N$ matrices $\mathcal{M}_{\mu}\{x\}$ and $\mathcal{W}(\{x\},t)$ whose entries are given by 

\begin{align}
\Big(\mathcal{M}_{\mu}\{x\}\Big)_{i,j} = x_j^{\mu_i},
\quad 
\Big(\mathcal{W}(\{x\},t)\Big)_{i,j}=x_i-tx_j
\end{align}

\noindent for all $1 \leq i,j \leq N$. Then the Hall-Littlewood function $P_{\mu}(\{x\},t)$ can be expressed in the form 
 
\begin{equation}
P_{\mu}(\{x\},t)
=
\frac{
\overline{\det}_N\Big(\mathcal{W}(\{x\},t),\mathcal{M}_{\mu}\{x\}\Big)
}
{\displaystyle{ v_{\mu}(t) \prod_{1\leq i< j \leq N} (x_i-x_j) }}
\label{hl2}
\end{equation}

\noindent which is a ratio of a weighted determinant and the Vandermonde in $\{x_1,\ldots,x_N\}$. One can view (\ref{hl2}) as the Hall-Littlewood analogue of equation (\ref{jac-trud}) for Schur functions. We remark that (\ref{hl2}) in conjunction with (\ref{hl-result}), (\ref{hl-result2}) allows us to write the $q$-boson model Bethe eigenvectors in the form

\begin{align}
&\mathcal{M}_{\psi}(t)
\mathbb{B}(x_1,t)\ldots\mathbb{B}(x_N,t)|0\rangle
=
\sum_{\mu \subseteq [N,M]}
\frac{
\overline{\det}_N\Big(\mathcal{W}(\{x\},t),\mathcal{M}_{\mu}\{x\}\Big)
}
{\displaystyle{ v_{\mu}(t) \prod_{1\leq i< j \leq N} (x_i-x_j) }}
|\mu)
\\
&
\langle 0|
\mathbb{C}(x_N,t)\ldots\mathbb{C}(x_1,t)
\mathcal{M}_{\psi}^{*}(t)
=
\sum_{\mu \subseteq [N,M]}
\frac{
\overline{\det}_N\Big(\mathcal{W}(\{x\},t),\mathcal{M}_{\mu}\{x\}\Big)
}
{\displaystyle{ v_{\mu}(t) \prod_{1\leq i< j \leq N} (x_i-x_j) }}
(\mu|
\end{align}

\noindent The aim of this section is to obtain an analogous result in the context of the XXZ model Bethe eigenvectors. That is, we wish to evaluate the expansion coefficients in (\ref{b}), (\ref{c*}) as weighted determinants divided by a Vandermonde. This proposal is motivated by the preceding equations and the close relationship of the $q$-boson and XXZ models.   


}
\end{example}

\subsection{Isolating coefficients}

Starting from (\ref{b}) and using the orthonormality of the basis vectors (\ref{elem}), we isolate coefficients as follows 

\begin{align}
b_{\lambda}\Big(\{v\}_N,\{w\}_M\Big) = 
\langle \lambda| 
\prod_{i=1}^{N} B(v_i,\{w\}_M) 
|\Uparrow_M\rangle 
\label{step1}
\end{align}

\noindent For the purpose of computation, the formula (\ref{step1}) is not immediately helpful. In order to bring it into a more convenient form, we begin by writing 

\begin{align}
b_{\lambda}\Big( \{v\}_N,\{w\}_M \Big)
=
\langle \lambda| \otimes \langle \Uparrow_N^a|
T\Big(\{v\}_N,\{w\}_M\Big)
|\Downarrow_N^a\rangle \otimes |\Uparrow_M\rangle
\label{step1.5}
\end{align}

\noindent where we have defined 

\begin{align}
T\Big(\{v\}_N,\{w\}_M\Big)
=
T_{a_N}(v_N,\{w\}_M) \ldots T_{a_1}(v_1,\{w\}_M)
\label{double}
\end{align}

\noindent Here the spaces $\mathcal{V}_{a_1},\ldots,\mathcal{V}_{a_N}$ assigned to the monodromy matrices are auxiliary, and we have defined the auxiliary state vectors

\begin{align}
\langle \Uparrow_N^a|
=
\bigotimes_{i=1}^{N} \uparrow_{a_i}^{*},
\quad
|\Downarrow_N^a\rangle
=
\bigotimes_{i=1}^{N}\downarrow_{a_i} 
\end{align}

\noindent Using the result of lemma 3 we may then write

\begin{align}
T\Big(\{v\}_N,\{w\}_M\Big)
=
(-)^{MN}
\overline{T}_1(w_1,\{\bar{v}\}_N)
\ldots
\overline{T}_M(w_M,\{\bar{v}\}_N)
\label{double2}
\end{align}

\noindent where for all $1\leq i\leq M$ we have defined

\begin{align}
\overline{T}_i(w_i,\{\bar{v}\}_N)
=
\left(
\begin{array}{rr}
D(w_i,\{\bar{v}\}_N)
&
-B(w_i,\{\bar{v}\}_N)
\\ 
-C(w_i,\{\bar{v}\}_N)
&
A(w_i,\{\bar{v}\}_N)
\end{array}
\right)_i
\end{align}

\noindent with $\{\bar{v}\}_N = \{v_1+\gamma,\ldots,v_N+\gamma\}$. Substituting the expression (\ref{double2}) into (\ref{step1.5}) and contracting on the quantum spaces $\mathcal{V}_1,\ldots,\mathcal{V}_M$ gives

\begin{align}
b_{\lambda}\Big(\{v\}_N,\{w\}_M\Big)
=
(-)^{(M+1)N}
\langle \Uparrow_N^a|\rprod_{i=1}^{M}O_i(w_i,\{\bar{v}\}_N)|\Downarrow_N^a\rangle
\label{step2}
\end{align}

\noindent where we have recalled the fact $\ell(\lambda) = M-N$ and defined the operators

\begin{align}
O_i(w_i,\{\bar{v}\}_N)
=
\left\{
\begin{array}{ll}
D(w_i,\{\bar{v}\}_N), & i \in \lambda \\
C(w_i,\{\bar{v}\}_N), & i \not\in \lambda
\end{array}
\right.
\end{align}

\noindent Throughout the rest of this section we will set $l=M-N$ for convenience. In terms of this notation, the expression (\ref{step2}) contains $l$ $D$-operators and $N$ $C$-operators. 

\subsection{Eliminating operators} 

The expression (\ref{step2}) is more amenable to calculation than (\ref{step1}). It allows us to employ the strategy that was used in \cite{bpz}, \cite{fp} and \cite{cp} to calculate the one and two-point boundary correlators of the six-vertex model, respectively. Our method is purely algebraic and therefore closest to that of \cite{bpz}, while \cite{fp} took a more graphical perspective.  

The idea is to eliminate the $l$ operators $D(w_{\lambda_i},\{\bar{v}\}_N)$ from (\ref{step2}) by employing the commutation relation (\ref{CD}) repeatedly. This will have the effect of shuffling these $l$ operators to the right until they act on the eigenvector $|\Downarrow_N^a\rangle$. This ultimately leaves us with a domain wall partition function (\ref{zequiv}), whose explicit form is given by lemma 7. To begin in this direction, we use (\ref{CD}) and (\ref{diagonal2}) repeatedly to establish that

{
\footnotesize
\begin{align}
&
D(w_{\lambda},\{\bar{v}\}_N)
\prod_{j \in J}
C(w_j,\{\bar{v}\}_N) |\Downarrow_N^a\rangle
=
\sum_{i \in J\cup\lambda}
\frac{\displaystyle{\prod_{j=1}^{N}[w_i-v_j] \prod_{j \in J} [w_i-w_j-\gamma]}}
{\displaystyle{\prod_{\substack{j \in J\cup\lambda \\ j \not= i}} [w_i-w_j]}}
\prod_{\substack{j \in J\cup\lambda \\ j \not= i}} C(w_j,\{\bar{v}\}_N) |\Downarrow_N^a\rangle
\end{align}
}

\noindent where $J$ is an arbitrary indexing set, and $\lambda$ an arbitrary extra label. By employing this identity $l$ times, once for each operator $D(w_{\lambda_i},\{\bar{v}\}_N)$ in (\ref{step2}), we obtain 

\begin{align}
&
b_{\lambda}\Big(\{v\}_N,\{w\}_M\Big)
=
\sum_{\sigma_1=\lambda_1}^{M}
\cdots
\sum_{\substack{\sigma_{l} = \lambda_{l} \\ \sigma_l \not= \sigma_1,\ldots,\sigma_{l-1} }}^{M} 
\frac{\displaystyle{
\prod_{i=1}^{N} [w_{\sigma_1}-v_i]
\prod_{i= \lambda_1 + 1}^{M} [w_{\sigma_1}-w_i-\gamma]}}
{\displaystyle{ \prod_{\substack{i = \lambda_1 \\ i \not= \sigma_1}}^{M} [w_{\sigma_1}-w_i]}}
\cdots
\nonumber
\\
&
\times
\frac{\displaystyle{
\prod_{i=1}^{N} [w_{\sigma_l}-v_i]
\prod_{\substack{i=\lambda_l+1 \\ i \not= \sigma_1,\ldots,\sigma_{l-1}}}^{M} [w_{\sigma_l}-w_i-\gamma]}}
{\displaystyle{ \prod_{\substack{i=\lambda_l \\ i \not= \sigma_1,\ldots, \sigma_l }}^{M}} [w_{\sigma_l}-w_i]}
\langle \Uparrow_N^a|\prod_{j \not= \sigma_1,\ldots,\sigma_l}^{M} C(w_j,\{\bar{v}\}_N) |\Downarrow_N^a\rangle
\label{wrath}
\end{align}

\noindent The final term in the sum (\ref{wrath}) is the domain wall partition function, given explicitly by

\begin{align}
\label{pf}
\langle \Uparrow_N^a|\prod_{j \not= \sigma_1,\ldots,\sigma_l}^{M} C(w_j,\{\bar{v}\}_N) |\Downarrow_N^a\rangle
=
\frac{\displaystyle{
\det\Big(
[\gamma]\prod_{k\not=i}^{N} [v_k-w_j+\gamma] [v_k-w_j]
\Big)_{\substack{1\leq i\leq N \\ j\not=\sigma_1,\ldots,\sigma_l}}
}}{\displaystyle{
\prod_{1 \leq i<j \leq N} [v_i-v_j]
\prod_{\substack{1 \leq i < j \leq M  \\ i,j \not= \sigma_1,\ldots,\sigma_l}}[w_j-w_i]
}}
\end{align}

\noindent We manipulate the factors which occur in the denominator of the summand of (\ref{wrath}), and in the denominator of the partition function (\ref{pf}). We find that

\begin{align}
&
\phantom{w.}
\displaystyle{
\prod_{\substack{i = \lambda_1 \\ i \not= \sigma_1}}^{M} \frac{1}{[w_{\sigma_1}-w_i]}
\cdots
\prod_{\substack{i = \lambda_l \\ i \not= \sigma_1,\ldots,\sigma_l}}^{M} \frac{1}{[w_{\sigma_l}-w_i]}
\prod_{ \substack{1\leq i<j \leq M \\ i,j \not= \sigma_1,\ldots,\sigma_l}} \frac{1}{[w_j-w_i]}
}
=
\label{denom}
\\
&
\prod_{1\leq i<j \leq l}{\rm sgn}(\sigma_j -\sigma_i)
\frac{\displaystyle{
(-)^{M+\sigma_1}
\prod_{i=1}^{\lambda_1-1} [w_{\sigma_1}-w_{i}]
\ldots
(-)^{M+\sigma_l}
\prod_{i=1}^{\lambda_l-1}[w_{\sigma_l}-w_{i}]
}}
{\displaystyle{
\prod_{1 \leq i < j \leq M}[w_j-w_i]
}}
\nonumber
\end{align}

\noindent Substituting (\ref{denom}) into (\ref{wrath}) gives

\begin{align}
b_{\lambda}\Big(\{v\}_N,\{w\}_M\Big)
&=
\frac{\displaystyle{\sum_{\sigma_1,\ldots,\sigma_l=1}^{M}\ 
\prod_{1\leq i<j \leq l}
{\rm sgn}(\sigma_j-\sigma_i)}}
{\displaystyle{
\prod_{1 \leq i < j \leq N} [v_i-v_j]
\prod_{1 \leq i < j \leq M} [w_j-w_i]}
}
\label{wrath2}
\\
&
\times 
(-)^{M+\sigma_1}
\prod_{i=1}^{N}[w_{\sigma_1}-v_i]
\prod_{i=1}^{\lambda_1-1}[w_{\sigma_1}-w_{i}] 
\prod_{i = \lambda_1+1}^{M} [w_{\sigma_1}-w_i-\gamma]
\ldots
\nonumber
\\
&
\times
(-)^{M+\sigma_l}
\prod_{i=1}^{N}[w_{\sigma_l}-v_i]
\prod_{i=1}^{\lambda_l-1} [w_{\sigma_l}-w_{i}]
\prod_{\substack{i = \lambda_l+1 \\ i \not= \sigma_1,\ldots, \sigma_{l-1} }}^{M} [w_{\sigma_l}-w_i-\gamma]
\nonumber
\\
&
\times
\det
\Big( [\gamma] \prod_{k\not=i}^{N} [v_k-w_j+\gamma][v_k-w_j] \Big)_{\substack{ 1\leq i \leq N \\ j \not= \sigma_1,\ldots,\sigma_l }}
\nonumber
\end{align}

\noindent where the summation indices $\sigma_1,\ldots,\sigma_l$ are assumed to be distinct but are now allowed to range from $1$ to $M$, since all terms corresponding to $\sigma_i < \lambda_i$ vanish trivially on account of the products $\prod_{j=1}^{\lambda_i-1} [w_{\sigma_i}-w_j]$. Manipulating the summand of (\ref{wrath2}) slightly, we obtain 

\begin{align}
b_{\lambda}\Big(\{v\}_N,\{w\}_M\Big)
&=
\frac{\displaystyle{\sum_{\sigma_1,\ldots,\sigma_l=1}^{M}\ 
\prod_{1\leq i<j \leq l}
\Big(
{\rm sgn}(\sigma_j-\sigma_i)
\Big{/}
[w_{\sigma_i}-w_{\sigma_j}+\gamma]
\Big)
}}
{\displaystyle{
\prod_{1 \leq i < j \leq N} [v_i-v_j]
\prod_{1 \leq i < j \leq M} [w_j-w_i]}
}
\label{wrath3}
\\
&
\times 
(-)^{1+\sigma_1}
\prod_{i=1}^{N}[v_i-w_{\sigma_1}]
\prod_{i=1}^{\lambda_1-1}[w_{i}-w_{\sigma_1}] 
\prod_{i = \lambda_1+1}^{M} [w_i-w_{\sigma_1}+\gamma]
\ldots
\nonumber
\\
&
\times
(-)^{l+\sigma_l}
\prod_{i=1}^{N}[v_i-w_{\sigma_l}]
\prod_{i=1}^{\lambda_l-1} [w_{i}-w_{\sigma_l}]
\prod_{i = \lambda_l+1}^{M} [w_i-w_{\sigma_l}+\gamma]
\nonumber
\\
&
\times
(-)^{lN}
\det
\Big( [\gamma] \prod_{k\not=i}^{N} [v_k-w_j+\gamma][v_k-w_j] \Big)_{\substack{ 1\leq i \leq N \\ j \not= \sigma_1,\ldots,\sigma_l }}
\nonumber
\end{align}

\noindent The expression (\ref{wrath3}) allows us to put $b_{\lambda}(\{v\}_N,\{w\}_M)$ in the form of a weighted determinant, as we will show in the next subsection.

\subsection{Weighted determinant expression for $b_{\lambda}(\{v\}_N,\{w\}_M)$}

By expanding the determinant in the summand of (\ref{wrath3}) we arrive at the expression

\begin{align}
&
b_{\lambda}\Big( \{v\}_N,\{w\}_M \Big)
=
\frac{\displaystyle{
\prod_{i=1}^{N} \prod_{j=1}^{M}
[v_i-w_j]
}}
{\displaystyle{
\prod_{1 \leq i < j \leq N}[v_i-v_j]
\prod_{1 \leq i < j \leq M}[w_j-w_i]
}}
\times
\\
&
\sum_{\sigma \in S_M}
\frac{\displaystyle{ 
\prod_{1\leq i < j \leq M}
{\rm sgn}(\sigma_j-\sigma_i) 
}}
{\displaystyle{
\prod_{1 \leq i < j \leq l}[w_{\sigma_i}-w_{\sigma_j}+\gamma]
}}
\prod_{i=1}^{l}
b_{\lambda_i}\left(w_{\sigma_i},\{w\}_M\right)
\prod_{i=1}^{N}
f_i\left(w_{\sigma_{l+i}},\{v\}_N\right)
\nonumber
\end{align}

\noindent where the sum is now taken over all permutations $\sigma \in S_M$, and where we have defined the functions

\begin{align}
f_i\left(w,\{v\}_N\right)
&=
\frac{[\gamma]}{[v_i-w]}
\prod_{k\not=i}^{N}
[v_k-w+\gamma]
\\
b_{\lambda_i}\left(w,\{w\}_M\right)
&=
\prod_{j=1}^{\lambda_i-1} [w_j-w]
\prod_{j=\lambda_i+1}^{M} [w_j-w+\gamma]
\end{align}

\noindent Recalling the definition of the weighted determinant (\ref{det}), we are able to write

\begin{align}
b_{\lambda}\Big(\{v\}_N,\{w\}_M\Big)
=
\frac{\displaystyle{
\prod_{i=1}^{N} \prod_{j=1}^{M}
[v_i-w_j]
\overline{{\rm det}}_l\Big(\mathcal{W}\{w\},\mathcal{B}_{\lambda}\{v\}_N\Big)
}}
{\displaystyle{
\prod_{1\leq i<j \leq N} [v_i-v_j]
\prod_{1 \leq i<j \leq M} [w_j-w_i]
}}
\label{finalc}
\end{align}

\noindent where the components of the $M \times M$ weighting matrix $\mathcal{W}\{w\}$ are given by

\begin{align}
\Big(\mathcal{W}\{w\}\Big)_{i,j} = \frac{1}{[w_i-w_j+\gamma]}
\label{weight}
\end{align}

\noindent and the $M\times M$ matrix $\mathcal{B}_{\lambda}\{v\}_N$, whose form depends only on the strict partition $\lambda$, is defined as

\begin{align}
\mathcal{B}_{\lambda}\{v\}_N
=
\left(
\begin{array}{ccc}
b_{\lambda_1}\left(w_1,\{w\}_M\right) & \cdots & b_{\lambda_1}\left(w_M,\{w\}_M\right) \\
\vdots            &        & \vdots  \\
b_{\lambda_l}\left(w_1,\{w\}_M\right) & \cdots & b_{\lambda_l}\left(w_M,\{w\}_M\right) \\
 f_1\left(w_1,\{v\}_N\right)  & \cdots &  f_1\left(w_M,\{v\}_N\right) \\
\vdots            &        & \vdots \\
 f_N\left(w_1,\{v\}_N\right)  & \cdots &  f_N\left(w_M,\{v\}_N\right)       
\end{array}
\right)
\end{align}

\noindent As a consistency check, we evaluate some special cases of (\ref{finalc}). When $\lambda$ is equal to the staircase partition $\{M,\ldots,1\}$ we have $N=0$, and accordingly we require that $b_{\{M,\ldots,1\}}=1$. We observe that the matrix $\mathcal{B}_{\{M,\ldots,1\}}$ is given by

\begin{align}
\mathcal{B}_{\{M,\ldots,1\}}
=
\left(
\begin{array}{ccc}
0 & 0 & b_M(w_M,\{w\}_M) 
\\
0 & \iddots & 0
\\
b_1(w_1,\{w\}_M) & 0 & 0
\end{array}
\right)
\end{align}

\noindent where all entries which are not on the indicated diagonal are zero. The weighted determinant in (\ref{finalc}) is trivially evaluated in this case, and we obtain $b_{\{M,\ldots,1\}} = 1$, as expected. Similarly, when $\lambda$ is equal to the empty partition $\emptyset$ we have $M=N$, and accordingly we require $b_{\emptyset} = Z_N$. In this case the weighted determinant (\ref{finalc}) collapses to the Izergin-Korepin expression (\ref{dwpf}), as required. In general, the formula (\ref{finalc}) serves to interpolate between these extremal cases.
    
Substituting the coefficients (\ref{finalc}) into (\ref{b}), we can write the Bethe eigenvector $\prod_{i=1}^{N} B(v_i,\{w\}_M) |\Uparrow_M\rangle$ in the form 

\begin{align}
\prod_{i=1}^{N} B(v_i,\{w\}_M)|\Uparrow_M\rangle
&=
\sum_{\lambda|\ell(\lambda) = l}
\frac{\displaystyle{
\prod_{i=1}^{N} \prod_{j=1}^{M}
[v_i-w_j]
\overline{{\rm det}}_l\Big(\mathcal{W}\{w\},\mathcal{B}_{\lambda}\{v\}_N\Big)
}}
{\displaystyle{
\prod_{1\leq i<j \leq N} [v_i-v_j]
\prod_{1 \leq i<j \leq M} [w_j-w_i]}} 
|\lambda\rangle
\label{finalexp1}
\end{align}

\noindent where the sum is taken over all strict partitions $\lambda$ of $l$ integers which satisfy the inequalities $\{M \geq \lambda_1 > \cdots > \lambda_l \geq 1\}$. 

\subsection{Weighted determinant expression for $c_{\lambda}(\{v\}_N,\{w\}_M)$}

By proceeding in an analogous fashion, it is possible to show that

\begin{align}
c_{\lambda} \Big( \{v\}_N,\{w\}_M \Big)
=
\frac{\displaystyle{
\prod_{i=1}^{N} \prod_{j=1}^{M}
[v_i-w_j+\gamma]
\overline{{\rm det}}_l\Big(\mathcal{W}'\{w\},\mathcal{C}_{\lambda}\{v\}_N\Big)
}}
{\displaystyle{
\prod_{1\leq i<j \leq N} [v_i-v_j]
\prod_{1 \leq i<j \leq M} [w_j-w_i]
}}
\label{finalb}
\end{align}

\noindent where the components of the $M \times M$ weighting matrix $\mathcal{W}'\{w\}$ are given by

\begin{align}
\Big(\mathcal{W}'\{w\}\Big)_{i,j}
=
\frac{1}{[w_i-w_j-\gamma]}
\end{align}

\noindent and the $M \times M$ matrix $\mathcal{C}_{\lambda}\{v\}_N$, whose form depends only on the strict partition $\lambda$, is defined as

\begin{align}
\mathcal{C}_{\lambda}\{v\}_N
=
\left(
\begin{array}{ccc}
c_{\lambda_1}\left(w_1,\{w\}_M\right) & \cdots & c_{\lambda_1}\left(w_M,\{w\}_M\right) \\
\vdots            &        & \vdots  \\
c_{\lambda_l}\left(w_1,\{w\}_M\right) & \cdots & c_{\lambda_l}\left(w_M,\{w\}_M\right) \\
f'_1\left(w_1,\{v\}_N\right)  & \cdots & f'_1\left(w_M,\{v\}_N\right) \\
\vdots            &        & \vdots \\
f'_N\left(w_1,\{v\}_N\right)  & \cdots & f'_N\left(w_M,\{v\}_N\right)       
\end{array}
\right)
\end{align}

\noindent with the functions

\begin{align}
f'(w,\{v\}_N)
&=
\frac{[\gamma]}{[v_i-w+\gamma]}
\prod_{k\not= i}^{N}
[v_k-w]
\\
c_{\lambda_i}\left(w,\{w\}_M\right)
&=
\prod_{j=1}^{\lambda_i-1}
[w_j-w]
\prod_{j=\lambda_i+1}^{M}
[w_j-w-\gamma]
\end{align}

\noindent Substituting the coefficients (\ref{finalb}) into (\ref{c*}), we can write the dual Bethe eigenvector $\langle \Uparrow_M|\prod_{i=1}^{N} C(v_i,\{w\}_M)$ in the form 

\begin{align}
\langle \Uparrow_M|
\prod_{i=1}^{N} C(v_i,\{w\}_M)
&=
\sum_{\lambda|\ell(\lambda) = l}
\frac{\displaystyle{
\prod_{i=1}^{N} \prod_{j=1}^{M}
[v_i-w_j+\gamma]
\overline{{\rm det}}_l\Big(\mathcal{W}'\{w\},\mathcal{C}_{\lambda}\{v\}_N\Big)
}}
{\displaystyle{
\prod_{1\leq i<j \leq N} [v_i-v_j]
\prod_{1 \leq i<j \leq M} [w_j-w_i]}} 
\langle\lambda|
\label{finalexp2}
\end{align}

\noindent where the sum is taken over all strict partitions $\lambda$ of $l$ integers which satisfy the inequalities $\{M \geq \lambda_1 > \cdots > \lambda_l \geq 1\}$.

\section{Conclusion}

In this chapter we presented several new results in the context of the XXZ spin-$\frac{1}{2}$ model. Admittedly these results are rather distinct in nature, but we believe that they are each interesting in their own right. We summarize our findings in the following paragraphs.

{\bf 1.} In section \ref{xxz-sp} we gave a new proof of the Slavnov determinant formula for the Bethe scalar product. Our proof relied on evaluating the intermediate Bethe scalar products $S_0$ through to $S_N$, which obey a set of Izergin/Korepin type conditions. We believe this method refines Slavnov's original proof insofar as our recursion relations have a simple graphical representation, whereas the recursion relation used in \cite{sla} can only be derived by complicated arguments, and we feel it presents a more elementary alternative to the method of \cite{kmt}. We hope there is potential to extend our graphical proof to the calculation of scalar products in a variety of lattice models. For example, starting from the domain wall partition functions obtained in \cite{cfk}, it should be possible to apply our method to the calculation of scalar products of the higher spin XXZ models \cite{deg}.

{\bf 2.} By virtue of their determinant form, we were able to show that the partition function and Bethe scalar product are power-sum specializations of KP $\tau$-functions. These provide further examples of quantum mechanical quantities which unexpectedly solve a classical hierarchy. At present we lack a deeper understanding of this classical/quantum relationship, but there is sufficient evidence to suggest that one exists. To achieve this aim, it may be necessary to study the link between these XXZ quantities and the discrete version of the KP hierarchy, as was done in \cite{fs}. In particular, it would be worthwhile to relate the results of \cite{fs} to those of \cite{kri}, in which a family of {\it transfer matrices} were shown to obey a discrete KP equation. The review article \cite{tak2} also provides some insights on a connection with the 2-component KP hierarchy. 

{\bf 3.} In section \ref{xxz-BE} we evaluated the XXZ Bethe eigenvectors by writing them in terms of the elementary spin basis, and evaluating the expansion coefficients. We found that these coefficients can be expressed as a generalization of the determinant, which we called a weighted determinant. Importantly, the Bethe equations were not used in these calculations. If it were possible to incorporate the Bethe equations in some way, it is conceivable that even simpler expressions could be obtained. 



\newpage

\thispagestyle{empty}

\phantom{nothing}


\chapter{Free fermion condition in lattice models}
\newtheorem{property}{ }
\newtheorem{property2}{ }
\newtheorem{property3}{ }
\newtheorem{property4}{ }
\newtheorem{property5}{ }
\newtheorem{property6}{ }
\newtheorem{property7}{ }
\newtheorem{property8}{ }

\setcounter{section}{-1}
\setcounter{lemma}{0}
\setcounter{theorem}{0}
\setcounter{remark}{0}

\section{Introduction}

In the preceding chapters we have noticed the appearance of fermions in the scalar product of the $q$-boson model and its limiting cases, and the XXZ spin-$\frac{1}{2}$ model. The fermionization in sections \ref{phase-pp}, \ref{ib-pp} and \ref{qb-pp} was particularly special, in that it caused the scalar product to factorize into product form. In this chapter we calculate the partition function and scalar product of several different lattice models. We find that these quantities also admit a factorization into product form, a fact which reflects the free fermionic nature of the corresponding models.

The starting point of our studies is the six-vertex model with crossing parameter set to $\gamma = \pi i / 2$. Following the literature, we refer to this as the free fermion point of the six-vertex model. At the free fermion point the Izergin-Korepin formula for the partition function becomes a Cauchy determinant, and therefore factorizes into product form. We review this fact in section \ref{ff-6v}, writing the partition function as an expectation value of KP vertex operators. We will also show that the Slavnov formula for the scalar product admits a similar factorization. This leads to a product expression for the Bethe scalar product at the free fermion point.

In \cite{fel1}, \cite{fel2}, \cite{fel3}, B~U~Felderhof introduced an elliptic eight-vertex model whose weights depend not only on rapidity variables, but on two additional parameters called external fields. In contrast to the eight-vertex model of R~J~Baxter \cite{bax1}, where vertices are invariant under conjugation of state variables, in the Felderhof model no such symmetry exists between vertices, creating an extra degree of anisotropy in the corresponding lattice model. We consider the trigonometric limit of Felderhof's model in section \ref{ff-fel}. In this limit two of the vertices collapse to zero, leading to the model studied by T~Deguchi and A~Akutsu in \cite{da1} which generalizes the free fermion point of the six-vertex model. We are able to calculate the partition function \cite{cfwz} and scalar product of this model, aided by the asymmetry of its vertex weights. Both the partition function and scalar product factorize into product form, indicating that the trigonometric Felderhof model remains free fermionic despite the presence of external fields.

In the remainder of the chapter we turn our attention to a more general class of lattice models. The most fundamental of these was derived from the eight-vertex model by Baxter \cite{bax4}, and is called the eight-vertex solid-on-solid (SOS) model. The weights of the SOS model are parametrized by elliptic functions, and they are graphically represented by squares with a dynamical or height variable attached at their corners. The six-vertex model is recovered as a special case by taking, simultaneously, the trigonometric and heightless limits of the SOS model. In section \ref{ff-sos} we review these facts and define the domain wall partition function of the SOS model. We do this algebraically, by following the quantum inverse scattering method for models with a height parameter, see for example \cite{kit}. As we did for the six-vertex model in section \ref{ff-6v}, we study the $\gamma = \pi i/2$ specialization of the crossing parameter. In this limit the partition function factorizes into product form. This suggests the existence of a free fermion point even in models which possess a height parameter.

Motivated by the results of section \ref{ff-fel}, in section \ref{ff-elda} we introduce external fields into the weights of the SOS model at its free fermion point. This was originally achieved by Deguchi and Akutsu \cite{da2} at the level of trigonometric functions, and extended to an elliptic parametrization in \cite{fwz2}, so we refer to this as the elliptic Deguchi-Akutsu height model. Due to the inherent asymmetry between the weights of this model, we are able to calculate its domain wall partition function using essentially the same techniques that were employed in section \ref{ff-fel}. The partition function is obtained in factorized product form \cite{fwz2}, showing that the elliptic Deguchi-Akutsu height model is free fermionic despite the presence of height and external field parameters.

\section{Free fermion point of six-vertex model}
\label{ff-6v}

\subsection{The limit $\gamma = \pi i/2 $}

The free fermion point of the XXZ/six-vertex model is obtained by setting the crossing parameter to $\gamma = \pi i/2$.\footnote{Throughout the entire chapter, we will reserve the letter $i$ to represent $\sqrt{-1}$.} In this limit the anisotropy parameter $\Delta = \frac{1}{2}(e^{\gamma}+e^{-\gamma})$ vanishes, and the Hamiltonian (\ref{hamiltonian}) for the model becomes

\begin{align}
\mathcal{H}
=
\sum_{j=1}^{M}
\left(
\sigma_j^x \sigma_{j+1}^x
+
\sigma_j^y \sigma_{j+1}^y
\right)
=
2
\sum_{j=1}^{M}
\left(
\sigma_j^{+} \sigma_{j+1}^{-}
+
\sigma_j^{-} \sigma_{j+1}^{+}
\right)
\label{hamiltonian-ff}
\end{align}

\noindent This Hamiltonian incorporates only interactions of the $x$ and $y$ components of the spin, and so it is said to correspond to the XX0 model. In the same limit, the $R$-matrix (\ref{Rmat1}) reads

\begin{align}
R_{ab}(u,v)
=
\left(
\begin{array}{cccc}
\langle u - v\rangle & 0 & 0 & 0 \\
0 & [u - v] & \langle 0 \rangle   & 0 \\
0 & \langle 0 \rangle & [u - v]   & 0 \\
0 & 0 & 0 & \langle u - v \rangle
\end{array}
\right)_{ab}
\label{Rmat-ff}
\end{align}

\noindent where we have employed the notations $[u] =2 \sinh u$, $\langle u \rangle = 2i \cosh u$ and made use of the fact that $[u+\pi i/2] = \langle u\rangle$. We will carry these conventions throughout the next few subsections.

\subsection{Domain wall partition function in the limit $\gamma=\pi i/2$}

Recalling Izergin's determinant expression (\ref{dwpf}) for the domain wall partition function, we set $\gamma = \pi i/2$ to obtain

\begin{align}
\label{IK-ff1}
&
Z_N \Big|_{\gamma=\pi i/2}
=
\frac{\displaystyle{
\langle 0 \rangle^N
\prod_{j,k=1}^{N} \langle v_j-w_k \rangle [v_j-w_k]
}}
{\displaystyle{
\prod_{1 \leq j<k \leq N} 
[v_j-v_k] [w_k-w_j]
}}
\det\left(
\frac{1}{\langle v_j-w_k \rangle [v_j-w_k]}
\right)_{1 \leq j,k \leq N}
\end{align}

\noindent The determinant in (\ref{IK-ff1}) is in Cauchy form, and obeys the factorization\footnote{By using the identity $\langle v \rangle [v] = i [2v]$ we can express (\ref{Cauchy}) in the more standard form

\begin{align}
\det\left(
\frac{1}{[2v_j-2w_k]}
\right)_{1 \leq j, k \leq N}
=
\frac{
\displaystyle{
\prod_{1\leq j<k \leq N} 
[2v_j-2v_k]
[2w_k-2w_j]
}
}
{
\displaystyle{
\prod_{j,k=1}^{N} [2v_j-2w_k]
}
}
\end{align}
}  

\begin{align}
\det\left(
\frac{1}{\langle v_j-w_k \rangle [v_j-w_k]}
\right)_{1 \leq j,k \leq N}
=
\frac{\displaystyle{
\prod_{1\leq j<k \leq N} 
\langle v_j-v_k \rangle [v_j-v_k]
\langle w_k -w_j \rangle [w_k-w_j]
}}
{\displaystyle{
\prod_{j,k=1}^{N} \langle v_j-w_k \rangle [v_j-w_k]
}}
\label{Cauchy}
\end{align}

\noindent Substituting (\ref{Cauchy}) into (\ref{IK-ff1}) and performing various cancellations, the domain wall partition function has the factorized expression

\begin{align}
Z_N \Big|_{\gamma = \pi i/2}
&=
\label{IK-ff3}
\langle 0\rangle^N
\prod_{1\leq j<k \leq N} 
\langle v_j-v_k \rangle \langle w_k-w_j \rangle
\end{align}

\noindent Equation (\ref{IK-ff3}) may also be found, for example, in \cite{bpz}. The aim of this chapter is to find analogues of the formula (\ref{IK-ff3}) across several different free fermionic models. 

\subsection{Partition function and free fermions}

In this subsection we study the appearance of free fermions in the partition function at the point $\gamma = \pi i/2$. We achieve this by writing lemma 2 from chapter 1 in terms of charged fermion generating functions, as follows

\begin{align}
\Big\langle
\Psi(y_1)
\ldots
\Psi(y_N)
\Psis(1/z_N)
\ldots
\Psis(1/z_1)
\Big\rangle
=
\det\left(
\sum_{k \in \mathbb{Z}}
\sum_{l \in \mathbb{Z}}
\langle \psi_{k} \psis_{l} \rangle
y_p^k z_q^{l}
\right)_{1\leq p,q \leq N}
\end{align}

\noindent where $\Psi(y_p)$ and $\Psis(1/z_q)$ are given by (\ref{KPexpI9}). Evaluating the vacuum expectation value within this determinant, we find

\begin{align}
\Big\langle
\Psi(y_1)
\ldots
\Psi(y_N)
\Psis(1/z_N)
\ldots
\Psis(1/z_1)
\Big\rangle
&=
\det\left(
\sum_{k=1}^{\infty}
\sum_{l=1}^{\infty}
\delta_{k,l}
y_p^{-k} z_q^{-l}
\right)_{1\leq p,q \leq N}
\nonumber
\\
&=
\det\left(
\frac{1}{y_p z_q-1}
\right)_{1 \leq p,q \leq N}
\end{align}

\noindent for arbitrary $\{y_1,\ldots,y_N\}$ and $\{z_1,\ldots,z_N\}$. Using this result in the expression (\ref{z'}) for the rescaled partition function, we obtain 

\begin{align}
Z_N'\Big|_{q=-1}
&=
\frac{\displaystyle{
2^N \prod_{j,k=1}^{N} (1-y_j^2 z_k^2)
}}
{\displaystyle{
\prod_{1 \leq j < k \leq N}
(y_j-y_k)(z_k-z_j)
}}
\det\left(
\frac{1}{y_j^2 z_k^2 - 1}
\right)_{1\leq j,k \leq N}
\label{z'ferm}
\\
&=
\frac{\displaystyle{
2^N 
\prod_{j,k=1}^{N} (1-y_j^2 z_k^2)
}}
{\displaystyle{
\prod_{1 \leq j < k \leq N}
(y_j-y_k)(z_k-z_j)
}}
\Big\langle \Psi(y_1^2)\ldots \Psi(y_N^2)
\Psis(1/z_N^2) \ldots \Psis(1/z_1^2) \Big\rangle
\nonumber
\end{align}

\noindent Equation (\ref{z'ferm}) demonstrates the presence of charged fermions in the partition function when $e^{2\gamma}=q= -1$, and provides a justification for the free fermionic nomenclature which is assigned to this point.  

\subsection{Bethe scalar product in the limit $\gamma = \pi i/2$}

Consider the determinant expression (\ref{betscal}) for the Bethe scalar product of the XXZ model. Setting $\gamma = \pi i/2$, we find that 

{\small
\begin{align}
S_N\Big(\{u\}_N,\{v\}_N,\{w\}_M\Big)\Big|_{\gamma=\pi i/2}
&=
\frac{\displaystyle{
\langle 0 \rangle^N
\prod_{j=1}^{N}
\prod_{k=1}^{M}
[v_j-w_k]
\prod_{j,k=1}^{N}
\langle u_j-v_k \rangle
}}
{\displaystyle{
\prod_{1\leq j < k \leq N}
[u_k-u_j][v_j-v_k]
}}
\label{scal-ff}
\\
&
\times
\det\left(
\frac{\displaystyle{\prod_{l=1}^{M}[ u_j-w_l]+(-)^N\prod_{l=1}^{M}\langle u_j-w_l\rangle}}{\langle u_j-v_k \rangle [ u_j-v_k]}
\right)_{1 \leq j,k \leq N}
\nonumber
\end{align}
}

\noindent where we have obtained the prefactor $\prod_{j,k=1}^{N}\langle u_j-v_k\rangle$ in (\ref{scal-ff}) using the fact that $\prod_{k=1}^{N}\langle u_j-v_k \rangle = (-)^N \prod_{k=1}^{N} \langle u_j-v_k-\pi i\rangle$. Since the numerator of the determinant in (\ref{scal-ff}) is common to the entire $j^{\rm th}$ row, we extract it as another prefactor, which gives 

{\small
\begin{align}
&
S_N\Big(\{u\}_N,\{v\}_N,\{w\}_M\Big)\Big|_{\gamma = \pi i/2}
=
\frac{\displaystyle{
\langle 0 \rangle^N
\prod_{j=1}^{N}
\prod_{k=1}^{M}
[v_j-w_k]
\prod_{j,k=1}^{N}
\langle u_j-v_k\rangle
}}
{\displaystyle{
\prod_{1\leq j < k \leq N}
[u_k-u_j][v_j-v_k]
}}
\times
\label{scal-ff2}
\\
&
\prod_{j=1}^{N}
\left(
\prod_{k=1}^{M}
[ u_j-w_k]
+
(-)^N
\prod_{k=1}^{M}
\langle u_j-w_k \rangle
\right)
\frac{\displaystyle{
\prod_{1\leq j < k \leq N}
\langle u_j -u_k \rangle [u_j-u_k]
\langle v_k -v_j \rangle [v_k-v_j]
}}
{\displaystyle{
\prod_{j,k=1}^{N} 
\langle u_j -v_k \rangle [u_j-v_k]
}}
\nonumber
\end{align}
}

\noindent where we have used the factorization of the resulting Cauchy determinant to produce the final term in (\ref{scal-ff2}). Cancelling various factors within (\ref{scal-ff2}), we obtain the formula

\begin{align}
S_N\Big|_{\gamma=\pi i/2}
&=
\langle 0 \rangle^N
\prod_{1\leq j < k \leq N}
\langle u_k-u_j\rangle \langle v_j-v_k \rangle
\prod_{j=1}^{N}
\prod_{k=1}^{M}
[v_j-w_k]
\label{scal-ff3}
\\
&
\times
\prod_{j,k=1}^{N}
\frac{1}{[u_j-v_k]}
\prod_{j=1}^{N}
\left(
\prod_{k=1}^{M}
[ u_j-w_k ]
+
(-)^N
\prod_{k=1}^{M}
\langle u_j-w_k \rangle
\right)
\nonumber
\end{align}

\noindent Let us remark that when $\gamma = \pi i/2$ the Bethe equations (\ref{bethe2}) decouple into the form 

\begin{align}
\prod_{k=1}^{M}
[ v_j-w_k ]
+
(-)^N
\prod_{k=1}^{M}
\langle v_j-w_k \rangle
=
0
\label{ffbethe}
\end{align} 

\noindent for all $1\leq j \leq N$. Hence all poles present in (\ref{scal-ff3}) are removable, by virtue of the constraints (\ref{ffbethe}) on $\{v\}_N$. In the next section we will derive an analogue of (\ref{scal-ff3}) in the context of a more general free fermionic model.

\section{Trigonometric Felderhof model}
\label{ff-fel}

In this section we devote our attention to the trigonometric Felderhof model. Most of the material that we present originally appeared in \cite{cfwz}, \cite{fwz1}. 

\subsection{$R$-matrix and Yang-Baxter equation}

The $R$-matrix (\ref{Rmat-ff}) may be generalized to include extra variables, in such a way that the Yang-Baxter equation remains satisfied. This leads to the trigonometric limit of the model introduced by Felderhof in \cite{fel1},\cite{fel2},\cite{fel3}, and accordingly we call it the {\it trigonometric Felderhof model.} This model was also studied in \cite{da1}, as the first in a hierarchy of vertex models with increasing spin. Explicitly speaking, the $R$-matrix for the trigonometric Felderhof model is given by

\begin{align}
&
R_{ab}(u,p,v,q)
=
\label{Rmat-tf}
\left(
\begin{array}{cccc}
a_{+}(u,p,v,q) & 0 & 0 & 0
\\
0 & b_{+}(u,p,v,q) & c_{+}(u,p,v,q) & 0
\\
0 & c_{-}(u,p,v,q) & b_{-}(u,p,v,q) & 0
\\
0 & 0 & 0 & a_{-}(u,p,v,q)
\end{array}
\right)_{ab}
\end{align}

\noindent where we have defined the functions

\begin{align}
&
a_{\pm}(u,p,v,q) = [\pm(u-v)+p+q]
\label{a-tf}
\\
&
b_{\pm}(u,p,v,q) = [u-v\pm(q-p)]
\label{b-tf}
\\
&
c_{\pm}(u,p,v,q) = [2p]^{\frac{1}{2}} [2q]^{\frac{1}{2}}
\label{c-tf}
\end{align}

\noindent with $[u] = 2\sinh u$ as usual.\footnote{The parametrization of \cite{da1} is recovered by multiplying all weights by $e^{u-v+p+q}$ and setting $e^{2p} = \alpha, e^{2q} = \beta$.} The $R$-matrix is an element of ${\rm End}(\mathcal{V}_a \otimes \mathcal{V}_b)$, and the variables $u,v$ are rapidities associated to the respective vector spaces $\mathcal{V}_a,\mathcal{V}_b$. The new features of this $R$-matrix are the variables $p,q$. These are called {\it external fields}, and are associated to the respective vector spaces $\mathcal{V}_a,\mathcal{V}_b$. We recover the free fermion point of the six-vertex model by setting $p=q=\frac{\pi i}{4}$.

The entries of the $R$-matrix (\ref{Rmat-tf}) admit the same graphical representation as those of the $R$-matrix (\ref{Rmat1}) in chapter 5. The only difference is that each vertex line now accommodates a rapidity variable and an external field. Hence we identify the functions (\ref{a-tf})--(\ref{c-tf}) with the vertices shown below.

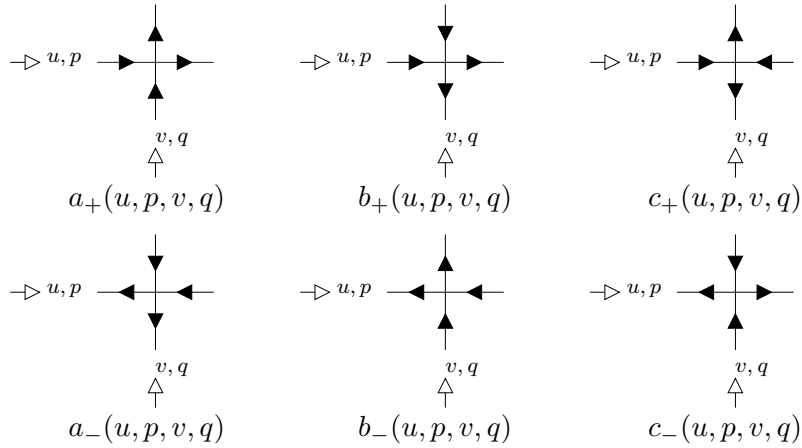
\begin{figure}[H]
\begin{center}
\begin{minipage}{4.3in}

\setlength{\unitlength}{0.00038cm}
\begin{picture}(20000,14000)(-4500,-12000)

\path(-2000,0000)(2000,0000)
\blacken\path(-1250,250)(-1250,-250)(-750,0)(-1250,250)
\blacken\path(750,250)(750,-250)(1250,0)(750,250)
\put(-3750,0){\scriptsize{$u,p$}}
\path(-5000,0)(-4000,0)
\whiten\path(-4500,250)(-4500,-250)(-4000,0)(-4500,250)
\path(0000,-2000)(0000,2000)
\blacken\path(-250,-1250)(250,-1250)(0,-750)(-250,-1250)
\blacken\path(-250,750)(250,750)(0,1250)(-250,750)
\put(-3000,-5000){$a_{+}(u,p,v,q)$}
\put(0,-2750){\scriptsize{$v,q$}}
\path(0,-4000)(0,-3000)
\whiten\path(-250,-3500)(250,-3500)(0,-3000)(-250,-3500)

\path(8000,0000)(12000,0000)
\blacken\path(8750,250)(8750,-250)(9250,0)(8750,250)
\blacken\path(10750,250)(10750,-250)(11250,0)(10750,250)
\put(6250,0){\scriptsize{$u,p$}}
\path(5000,0)(6000,0)
\whiten\path(5500,250)(5500,-250)(6000,0)(5500,250)
\path(10000,-2000)(10000,2000)
\blacken\path(9750,-750)(10250,-750)(10000,-1250)(9750,-750)
\blacken\path(9750,1250)(10250,1250)(10000,750)(9750,1250)
\put(7000,-5000){$b_{+}(u,p,v,q)$}
\put(10000,-2750){\scriptsize{$v,q$}}
\path(10000,-4000)(10000,-3000)
\whiten\path(9750,-3500)(10250,-3500)(10000,-3000)(9750,-3500)

\path(18000,0000)(22000,0000)
\blacken\path(18750,250)(18750,-250)(19250,0)(18750,250)
\blacken\path(21250,250)(21250,-250)(20750,0)(21250,250)
\put(16250,0){\scriptsize{$u,p$}}
\path(15000,0)(16000,0)
\whiten\path(15500,250)(15500,-250)(16000,0)(15500,250)
\path(20000,-2000)(20000,2000)
\blacken\path(19750,-750)(20250,-750)(20000,-1250)(19750,-750)
\blacken\path(19750,750)(20250,750)(20000,1250)(19750,750)
\put(17000,-5000){$c_{+}(u,p,v,q)$}
\put(20000,-2750){\scriptsize{$v,q$}}
\path(20000,-4000)(20000,-3000)
\whiten\path(19750,-3500)(20250,-3500)(20000,-3000)(19750,-3500)

\path(-2000,-8000)(2000,-8000)
\blacken\path(-750,-7750)(-750,-8250)(-1250,-8000)(-750,-7750)
\blacken\path(1250,-7750)(1250,-8250)(750,-8000)(1250,-7750)
\put(-3750,-8000){\scriptsize{$u,p$}}
\path(-5000,-8000)(-4000,-8000)
\whiten\path(-4500,-7750)(-4500,-8250)(-4000,-8000)(-4500,-7750)
\path(0000,-10000)(0000,-6000)
\blacken\path(-250,-8750)(250,-8750)(0,-9250)(-250,-8750)
\blacken\path(-250,-6750)(250,-6750)(0,-7250)(-250,-6750)
\put(-3000,-13000){$a_{-}(u,p,v,q)$}
\put(0,-10750){\scriptsize{$v,q$}}
\path(0,-12000)(0,-11000)
\whiten\path(-250,-11500)(250,-11500)(0,-11000)(-250,-11500)

\path(8000,-8000)(12000,-8000)
\blacken\path(9250,-7750)(9250,-8250)(8750,-8000)(9250,-7750)
\blacken\path(11250,-7750)(11250,-8250)(10750,-8000)(11250,-7750)
\put(6250,-8000){\scriptsize{$u,p$}}
\path(5000,-8000)(6000,-8000)
\whiten\path(5500,-7750)(5500,-8250)(6000,-8000)(5500,-7750)
\path(10000,-10000)(10000,-6000)
\blacken\path(9750,-9250)(10250,-9250)(10000,-8750)(9750,-9250)
\blacken\path(9750,-7250)(10250,-7250)(10000,-6750)(9750,-7250)
\put(7000,-13000){$b_{-}(u,p,v,q)$}
\put(10000,-10750){\scriptsize{$v,q$}}
\path(10000,-12000)(10000,-11000)
\whiten\path(9750,-11500)(10250,-11500)(10000,-11000)(9750,-11500)

\path(18000,-8000)(22000,-8000)
\blacken\path(19250,-7750)(19250,-8250)(18750,-8000)(19250,-7750)
\blacken\path(20750,-7750)(20750,-8250)(21250,-8000)(20750,-7750)
\put(16250,-8000){\scriptsize{$u,p$}}
\path(15000,-8000)(16000,-8000)
\whiten\path(15500,-7750)(15500,-8250)(16000,-8000)(15500,-7750)
\path(20000,-10000)(20000,-6000)
\blacken\path(19750,-9250)(20250,-9250)(20000,-8750)(19750,-9250)
\blacken\path(19750,-6750)(20250,-6750)(20000,-7250)(19750,-6750)
\put(17000,-13000){$c_{-}(u,p,v,q)$}
\put(20000,-10750){\scriptsize{$v,q$}}
\path(20000,-12000)(20000,-11000)
\whiten\path(19750,-11500)(20250,-11500)(20000,-11000)(19750,-11500)

\end{picture}

\end{minipage}
\end{center}

\caption[Six vertices of the trigonometric Felderhof model]{Six vertices of the trigonometric Felderhof model.} 
\end{figure}

\noindent With the following result we see that the Yang-Baxter equation continues to hold, even in the presence of the external fields.

\begin{lemma}
{\rm 
The $R$-matrix (\ref{Rmat-tf}) obeys the Yang-Baxter equation

\begin{align}
R_{ab}(u,p,v,q)
R_{ac}(u,p,w,r)
&
R_{bc}(v,q,w,r)
=
\label{yb-tf}
R_{bc}(v,q,w,r)
R_{ac}(u,p,w,r)
R_{ab}(u,p,v,q)
\end{align}

\noindent This is an identity in ${\rm End}(\mathcal{V}_a \otimes \mathcal{V}_b \otimes \mathcal{V}_c)$, true for all $u,v,w$ and $p,q,r$.
}
\end{lemma}

\begin{proof}
By direct computation.
\end{proof}

\begin{figure}[H]
\begin{center}
\begin{minipage}{4.3in}

\setlength{\unitlength}{0.00038cm}
\begin{picture}(20000,9500)(-2000,-3500)

\path(0,0)(5000,5000)
\put(1250,1250){\circle*{300}}
\put(1250,650){\tiny{$i_b$}}
\put(5000,5000){\circle*{300}}
\put(5000,4400){\tiny{$k_b$}}
\put(8750,5000){\circle*{300}}
\put(8750,4400){\tiny{$j_b$}}
\path(-3000,0)(-2000,0)
\whiten\path(-2250,250)(-2250,-250)(-2000,0)(-2250,250)
\put(-1750,0){\scriptsize{$v,q$}}
\path(5000,0)(0,5000)
\put(1250,3750){\circle*{300}}
\put(1250,4000){\tiny{$i_a$}}
\put(5000,0){\circle*{300}}
\put(5000,-600){\tiny{$k_a$}}
\put(8750,0){\circle*{300}}
\put(8750,-600){\tiny{$j_a$}}
\path(-3000,5000)(-2000,5000)
\whiten\path(-2250,5250)(-2250,4750)(-2000,5000)(-2250,5250)
\put(-1750,5000){\scriptsize{$u,p$}}
\path(5000,5000)(10000,5000)
\path(5000,0)(10000,0)
\path(7500,-2500)(7500,7500)
\put(7500,-1250){\circle*{300}}
\put(7750,-1250){\tiny{$i_c$}}
\put(7500,2500){\circle*{300}}
\put(7750,2500){\tiny{$k_c$}}
\put(7500,6250){\circle*{300}}
\put(7750,6250){\tiny{$j_c$}}
\put(7500,-3250){\scriptsize{$w,r$}}
\path(7500,-4500)(7500,-3500)
\whiten\path(7250,-3750)(7750,-3750)(7500,-3500)(7250,-3750)


\put(11000,2300){$=$}

\path(12000,5000)(13000,5000)
\whiten\path(12750,5250)(12750,4750)(13000,5000)(12750,5250)
\put(13250,5000){\scriptsize{$u,p$}}
\path(15000,5000)(20000,5000)
\put(16250,5000){\circle*{300}}
\put(16250,4400){\tiny{$i_a$}}
\path(20000,5000)(25000,0)
\put(20000,5000){\circle*{300}}
\put(20000,5250){\tiny{$k_a$}}
\put(23750,1250){\circle*{300}}
\put(23750,1600){\tiny{$j_a$}}
\path(12000,0)(13000,0)
\whiten\path(12750,250)(12750,-250)(13000,0)(12750,250)
\put(13250,0){\scriptsize{$v,q$}}
\path(15000,0)(20000,0)
\put(16250,0){\circle*{300}}
\put(16250,-600){\tiny{$i_b$}}
\path(20000,0)(25000,5000)
\put(20000,0){\circle*{300}}
\put(20000,-600){\tiny{$k_b$}}
\put(23750,3750){\circle*{300}}
\put(23750,3250){\tiny{$j_b$}}
\path(17500,-4500)(17500,-3500)
\whiten\path(17250,-3750)(17750,-3750)(17500,-3500)(17250,-3750)
\put(17500,-3250){\scriptsize{$w,r$}}
\path(17500,-2500)(17500,7500)
\put(17500,-1250){\circle*{300}}
\put(17750,-1250){\tiny{$i_c$}}
\put(17500,2500){\circle*{300}}
\put(17750,2500){\tiny{$k_c$}}
\put(17500,6250){\circle*{300}}
\put(17750,6250){\tiny{$j_c$}}

\end{picture}

\end{minipage}
\end{center}

\caption[Yang-Baxter equation for the trigonometric Felderhof model]{Yang-Baxter equation for the trigonometric Felderhof model.}
\end{figure}
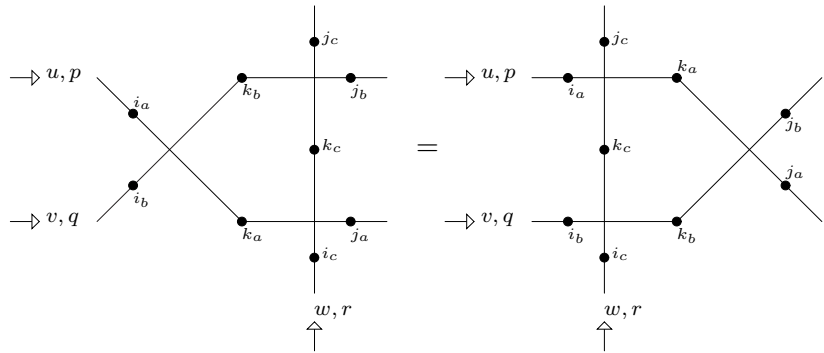

\subsection{Monodromy matrix and intertwining equation}

The monodromy matrix is an ordered product of $R$-matrices, given by

\begin{align}
T_a(u,p,\{w,r\}_M)
=
R_{a1}(u,p,w_1,r_1)
\ldots
R_{aM}(u,p,w_M,r_M)
\end{align}

\noindent with the multiplication taken in the space ${\rm End}(\mathcal{V}_a)$. We write the contribution from the space ${\rm End}(\mathcal{V}_a)$ explicitly, by defining

\begin{align}
T_a(u,p,\{w,r\}_M)
=
\left(
\begin{array}{cc}
A(u,p,\{w,r\}_M) & B(u,p,\{w,r\}_M)
\\
C(u,p,\{w,r\}_M) & D(u,p,\{w,r\}_M)
\end{array}
\right)_a
\label{tf-mon}
\end{align}

\noindent where the matrix entries are all operators acting in $\mathcal{V}_1 \otimes \cdots \otimes \mathcal{V}_M$. For a graphical representation of these operators, we refer the reader to figure \ref{6vT} in chapter 5. The correspondence is exactly the same, except that the rapidity $u$ is now accompanied by the external field $p$, and each $w_j$ by an $r_j$. Due to the Yang-Baxter equation (\ref{yb-tf}), we obtain the intertwining equation

\begin{align}
&
R_{ab}(u,p,v,q) T_a(u,p,\{w,r\}_M) T_b(v,q,\{w,r\}_M)
=
\\
&
T_b(v,q,\{w,r\}_M) T_a(u,p,\{w,r\}_M) R_{ab}(u,p,v,q)
\nonumber
\end{align}

\noindent As usual, this leads to sixteen commutation relations amongst the entries of the monodromy matrix (\ref{tf-mon}). Of these commutation relations, two have particular significance in our later calculations. They are given by

\begin{align}
& [u-v+p+q] B(u,p,\{w,r\}_M) B(v,q,\{w,r\}_M) 
=
\label{bb-tf}
\\
& [v-u+p+q] B(v,q,\{w,r\}_M)B(u,p,\{w,r\}_M)
\nonumber
\\
\nonumber
\\
& [v-u+p+q] C(u,p,\{w,r\}_M) C(v,q,\{w,r\}_M)
=
\label{cc-tf}
\\
& [u-v+p+q] C(v,q,\{w,r\}_M)C(u,p,\{w,r\}_M)
\nonumber
\end{align}

\subsection{Domain wall partition function $Z_N(\{v,q\}_N,\{w,r\}_N)$}

The domain wall partition function of the trigonometric Felderhof model has the algebraic definition

\begin{align}
Z_N\Big(
\{v,q\}_N,\{w,r\}_N
\Big)
=
\langle \Downarrow_N|
\lprod_{j=1}^{N}
B(v_j,q_j,\{w,r\}_N)
|\Uparrow_N\rangle
\label{pf-tf}
\end{align}

\noindent This naturally extends the definition of the domain wall partition function (\ref{pf-xxz}) to a model containing external fields. Notice that we must define an ordering of the $B$-operators in (\ref{pf-tf}), since by (\ref{bb-tf}) they do not commute. 

\begin{figure}[H]

\begin{center}
\begin{minipage}{4.3in}

\setlength{\unitlength}{0.00038cm}
\begin{picture}(20000,12500)(-10000,-3000)

\path(-2000,0)(10000,0)
\put(-4250,0){\tiny$v_N,q_N$}
\path(-5500,0)(-4500,0)
\whiten\path(-4750,250)(-4750,-250)(-4500,0)(-4750,250)
\blacken\path(-1250,250)(-1250,-250)(-750,0)(-1250,250)
\blacken\path(9250,250)(9250,-250)(8750,0)(9250,250)

\path(-2000,2000)(10000,2000)
\path(-5500,2000)(-4500,2000)
\whiten\path(-4750,2250)(-4750,1750)(-4500,2000)(-4750,2250)
\blacken\path(-1250,2250)(-1250,1750)(-750,2000)(-1250,2250)
\blacken\path(9250,2250)(9250,1750)(8750,2000)(9250,2250)

\path(-2000,4000)(10000,4000)
\path(-5500,4000)(-4500,4000)
\whiten\path(-4750,4250)(-4750,3750)(-4500,4000)(-4750,4250)
\blacken\path(-1250,4250)(-1250,3750)(-750,4000)(-1250,4250)
\blacken\path(9250,4250)(9250,3750)(8750,4000)(9250,4250)

\path(-2000,6000)(10000,6000)
\path(-5500,6000)(-4500,6000)
\whiten\path(-4750,6250)(-4750,5750)(-4500,6000)(-4750,6250)
\blacken\path(-1250,6250)(-1250,5750)(-750,6000)(-1250,6250)
\blacken\path(9250,6250)(9250,5750)(8750,6000)(9250,6250)

\path(-2000,8000)(10000,8000)
\put(-4250,8000){\tiny$v_1,q_1$}
\path(-5500,8000)(-4500,8000)
\whiten\path(-4750,8250)(-4750,7750)(-4500,8000)(-4750,8250)
\blacken\path(-1250,8250)(-1250,7750)(-750,8000)(-1250,8250)
\blacken\path(9250,8250)(9250,7750)(8750,8000)(9250,8250)


\path(0,-2000)(0,10000)
\put(-500,-2750){\tiny$w_1,r_1$}
\path(0,-4000)(0,-3000)
\whiten\path(-250,-3250)(250,-3250)(0,-3000)(-250,-3250)
\blacken\path(-250,-750)(250,-750)(0,-1250)(-250,-750)
\blacken\path(-250,8750)(250,8750)(0,9250)(-250,8750)

\path(2000,-2000)(2000,10000)
\path(2000,-4000)(2000,-3000)
\whiten\path(1750,-3250)(2250,-3250)(2000,-3000)(1750,-3250)
\blacken\path(1750,-750)(2250,-750)(2000,-1250)(1750,-750)
\blacken\path(1750,8750)(2250,8750)(2000,9250)(1750,8750)

\path(4000,-2000)(4000,10000)
\path(4000,-4000)(4000,-3000)
\whiten\path(3750,-3250)(4250,-3250)(4000,-3000)(3750,-3250)
\blacken\path(3750,-750)(4250,-750)(4000,-1250)(3750,-750)
\blacken\path(3750,8750)(4250,8750)(4000,9250)(3750,8750)

\path(6000,-2000)(6000,10000)
\path(6000,-4000)(6000,-3000)
\whiten\path(5750,-3250)(6250,-3250)(6000,-3000)(5750,-3250)
\blacken\path(5750,-750)(6250,-750)(6000,-1250)(5750,-750)
\blacken\path(5750,8750)(6250,8750)(6000,9250)(5750,8750)

\path(8000,-2000)(8000,10000)
\put(7500,-2750){\tiny$w_N,r_N$}
\path(8000,-4000)(8000,-3000)
\whiten\path(7750,-3250)(8250,-3250)(8000,-3000)(7750,-3250)
\blacken\path(7750,-750)(8250,-750)(8000,-1250)(7750,-750)
\blacken\path(7750,8750)(8250,8750)(8000,9250)(7750,8750)

\end{picture}

\end{minipage}
\end{center}

\caption[Domain wall partition function of the trigonometric Felderhof model]{Domain wall partition function of the trigonometric Felderhof model. The top row of arrows corresponds with the state vector $|\Uparrow_N\rangle$. The bottom row of arrows corresponds with the dual state vector $\langle \Downarrow_N|$. Each horizontal lattice line corresponds to multiplication by a $B(v_j,q_j,\{w,r\}_N)$ operator. Notice that the ordering of these lattice lines respects the ordering of $B$-operators defined in (\ref{pf-tf}).} 

\label{da-part}
\end{figure}
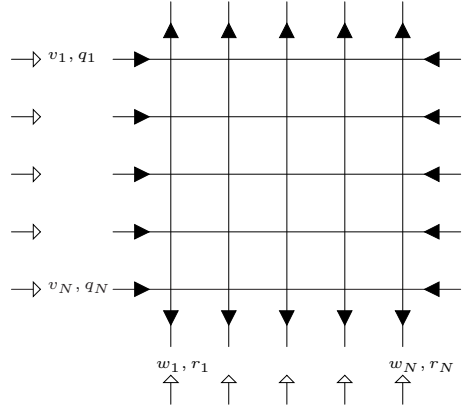

Similarly to the previous chapter, we represent the domain wall partition function by an $N \times N$ lattice, as shown in figure \ref{da-part}.

\subsection{Conditions on $Z_N( \{v,q\}_N, \{w,r\}_N )$}

We now progress towards calculating the domain wall partition function (\ref{pf-tf}). The procedure begins with the following result from \cite{fwz1}, which establishes a set of Korepin-type conditions on $Z_N(\{v,q\}_N,\{w,r\}_N)$. 

\begin{lemma}
{\rm We adopt the shorthand $Z_N = Z_N(\{v,q\}_N,\{w,r\}_N)$. For all $N \geq 2$ we claim that

\begin{property}
{\rm 
$Z_N$ is a trigonometric polynomial of degree $N-1$ in the rapidity variable $v_N$.
}
\end{property}

\begin{property}
{\rm
The zeros of $Z_N$ occur at the points $v_N = v_j+q_j+q_N$, for all $1 \leq j \leq N-1$.
}
\end{property}

\begin{property}
{\rm 
Setting $v_N=w_N+q_N+r_N$, $Z_N$ satisfies the recursion relation

{
\small
\begin{align}
&
Z_N \Big|_{v_N = w_N+q_N+r_N}
=
[2q_N]^{\frac{1}{2}} [2r_N]^{\frac{1}{2}}
\label{pfrec1-tf}
\prod_{j=1}^{N-1}
[w_N-w_j+r_j+r_N]
[v_j-w_N+q_j-r_N]
Z_{N-1}
\end{align}
}

\noindent where $Z_{N-1}$ is the domain wall partition function on a square lattice of size $N-1$.
}
\end{property}

In addition, we have the supplementary condition

\begin{property}
{\rm 
The partition function on the $1\times 1$ lattice is given by
$
Z_1
=
[2q_1]^{\frac{1}{2}}
[2r_1]^{\frac{1}{2}}
$.
}
\end{property}
}
\end{lemma}

\begin{proof}

\begin{property2}
{\rm 
By inserting the set of states $\sum_{n=1}^{N} \sigma_n^{+} |\Downarrow_N\rangle \langle \Downarrow_N| \sigma_n^{-}$ after the first $B$-operator appearing in (\ref{pf-tf}), we obtain the expansion

\begin{align}
Z_N\Big(\{v,q\}_N,\{w,r\}_N\Big)
&=
\sum_{n=1}^{N}
\langle \Downarrow_N|
B(v_N,q_N,\{w,r\}_N)
\sigma_n^{+}
|\Downarrow_N\rangle
\label{tf-peel2}
\\
&
\times
\langle \Downarrow_N|
\sigma_n^{-}
\lprod_{j=1}^{N-1}
B(v_j,q_j,\{w,r\}_N)
|\Uparrow_N\rangle
\nonumber
\end{align}

\noindent in which all dependence on $v_N$ appears in the first factor within the sum. Hence we shall calculate $\langle \Downarrow_N| B(v_N,q_N,\{w,r\}_N) \sigma_n^{+} |\Downarrow_N\rangle$ for all $1\leq n \leq N$, as shown below.

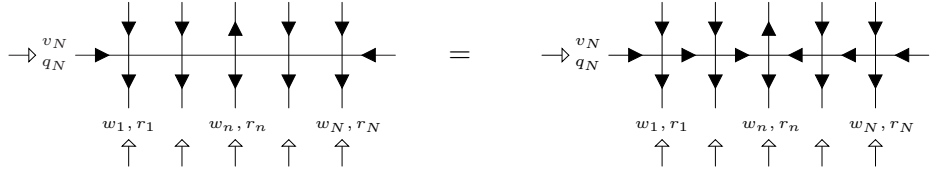
\begin{figure}[H]

\begin{center}
\begin{minipage}{4.3in}

\setlength{\unitlength}{0.00035cm}
\begin{picture}(20000,6000)(-3000,-3000)

\path(-2000,0)(10000,0)
\put(-3250,400){\tiny$v_N$}
\put(-3250,-400){\tiny$q_N$}
\path(-4500,0)(-3500,0)
\whiten\path(-3750,250)(-3750,-250)(-3500,0)(-3750,250)
\blacken\path(-1250,250)(-1250,-250)(-750,0)(-1250,250)
\blacken\path(9250,250)(9250,-250)(8750,0)(9250,250)


\path(0,-2000)(0,2000)
\put(-1000,-2750){\tiny$w_1,r_1$}
\path(0,-4200)(0,-3200)
\whiten\path(-250,-3450)(250,-3450)(0,-3200)(-250,-3450)
\blacken\path(-250,-750)(250,-750)(0,-1250)(-250,-750)
\blacken\path(-250,1250)(250,1250)(0,750)(-250,1250)

\path(2000,-2000)(2000,2000)
\path(2000,-4200)(2000,-3200)
\whiten\path(1750,-3450)(2250,-3450)(2000,-3200)(1750,-3450)
\blacken\path(1750,-750)(2250,-750)(2000,-1250)(1750,-750)
\blacken\path(1750,1250)(2250,1250)(2000,750)(1750,1250)

\path(4000,-2000)(4000,2000)
\put(3000,-2750){\tiny$w_n,r_n$}
\path(4000,-4200)(4000,-3200)
\whiten\path(3750,-3450)(4250,-3450)(4000,-3200)(3750,-3450)
\blacken\path(3750,-750)(4250,-750)(4000,-1250)(3750,-750)
\blacken\path(3750,750)(4250,750)(4000,1250)(3750,750)

\path(6000,-2000)(6000,2000)
\path(6000,-4200)(6000,-3200)
\whiten\path(5750,-3450)(6250,-3450)(6000,-3200)(5750,-3450)
\blacken\path(5750,-750)(6250,-750)(6000,-1250)(5750,-750)
\blacken\path(5750,1250)(6250,1250)(6000,750)(5750,1250)

\path(8000,-2000)(8000,2000)
\put(7000,-2750){\tiny$w_N,r_N$}
\path(8000,-4200)(8000,-3200)
\whiten\path(7750,-3450)(8250,-3450)(8000,-3200)(7750,-3450)
\blacken\path(7750,-750)(8250,-750)(8000,-1250)(7750,-750)
\blacken\path(7750,1250)(8250,1250)(8000,750)(7750,1250)


\put(12000,-250){$=$}


\path(18000,0)(30000,0)
\put(16750,400){\tiny$v_N$}
\put(16750,-400){\tiny$q_N$}
\path(15500,0)(16500,0)
\whiten\path(16250,250)(16250,-250)(16500,0)(16250,250)
\blacken\path(18750,250)(18750,-250)(19250,0)(18750,250)
\blacken\path(20750,250)(20750,-250)(21250,0)(20750,250)
\blacken\path(22750,250)(22750,-250)(23250,0)(22750,250)
\blacken\path(25250,250)(25250,-250)(24750,0)(25250,250)
\blacken\path(27250,250)(27250,-250)(26750,0)(27250,250)
\blacken\path(29250,250)(29250,-250)(28750,0)(29250,250)


\path(20000,-2000)(20000,2000)
\put(19000,-2750){\tiny$w_1,r_1$}
\path(20000,-4200)(20000,-3200)
\whiten\path(19750,-3450)(20250,-3450)(20000,-3200)(19750,-3450)
\blacken\path(19750,-750)(20250,-750)(20000,-1250)(19750,-750)
\blacken\path(19750,1250)(20250,1250)(20000,750)(19750,1250)

\path(22000,-2000)(22000,2000)
\path(22000,-4200)(22000,-3200)
\whiten\path(21750,-3450)(22250,-3450)(22000,-3200)(21750,-3450)
\blacken\path(21750,-750)(22250,-750)(22000,-1250)(21750,-750)
\blacken\path(21750,1250)(22250,1250)(22000,750)(21750,1250)

\path(24000,-2000)(24000,2000)
\put(23000,-2750){\tiny$w_n,r_n$}
\path(24000,-4200)(24000,-3200)
\whiten\path(23750,-3450)(24250,-3450)(24000,-3200)(23750,-3450)
\blacken\path(23750,-750)(24250,-750)(24000,-1250)(23750,-750)
\blacken\path(23750,750)(24250,750)(24000,1250)(23750,750)

\path(26000,-2000)(26000,2000)
\path(26000,-4200)(26000,-3200)
\whiten\path(25750,-3450)(26250,-3450)(26000,-3200)(25750,-3450)
\blacken\path(25750,-750)(26250,-750)(26000,-1250)(25750,-750)
\blacken\path(25750,1250)(26250,1250)(26000,750)(25750,1250)

\path(28000,-2000)(28000,2000)
\put(27000,-2750){\tiny$w_N,r_N$}
\path(28000,-4200)(28000,-3200)
\whiten\path(27750,-3450)(28250,-3450)(28000,-3200)(27750,-3450)
\blacken\path(27750,-750)(28250,-750)(28000,-1250)(27750,-750)
\blacken\path(27750,1250)(28250,1250)(28000,750)(27750,1250)

\end{picture}

\end{minipage}
\end{center}

\caption[Peeling away the bottom row of the trigonometric Felderhof partition function]{Peeling away the bottom row of the trigonometric Felderhof partition function. The diagram on the left represents $\langle \Downarrow_N |B(v_N,q_N,\{w,r\}_N) \sigma_n^{+} |\Downarrow_N \rangle$, with the internal black arrows being summed over all configurations. The diagram on the right represents the only surviving configuration.}

\label{tf-zequiv3}
\end{figure}

\noindent The right hand side of figure \ref{tf-zequiv3} represents a product of vertices. Replacing each vertex with its corresponding trigonometric weight, we have

\begin{align}
&
\langle \Downarrow_N| B(v_N,q_N,\{w,r\}_N)
\sigma_n^{+} |\Downarrow_N\rangle
=
\label{tf-peel}
\\
&
[2q_N]^{\frac{1}{2}} [2r_n]^{\frac{1}{2}}
\prod_{1 \leq j < n} [v_N-w_j+r_j-q_N]
\prod_{n < j \leq N} [w_j-v_N+q_N+r_j]
\nonumber
\end{align}

\noindent Substituting (\ref{tf-peel}) into the expansion (\ref{tf-peel2}) gives

{\small
\begin{align}
Z_N\Big(\{v,q\}_N,\{w,r\}_N\Big)
&=
\sum_{n=1}^{N}
[2q_N]^{\frac{1}{2}} [2r_n]^{\frac{1}{2}}
\prod_{1\leq j <n} [v_N-w_j+r_j-q_N]
\label{pfexp-tf}
\\
&
\times
\prod_{n < j \leq N} [w_j-v_N+q_N+r_j]
\langle \Downarrow_N| \sigma_n^{-}
\lprod_{j=1}^{N-1} B(v_j,q_j,\{w,r\}_N)
|\Uparrow_N\rangle
\nonumber
\end{align}
}

\noindent From (\ref{pfexp-tf}) we see that every term in $Z_N(\{v,q\}_N,\{w,r\}_N)$ contains a product of exactly $N-1$ trigonometric functions with argument $v_N$. Thus $Z_N(\{v,q\}_N,\{w,r\}_N)$ is a trigonometric polynomial of degree $N-1$ in the variable $v_N$.
}
\end{property2}

\begin{property2}
{\rm 
We multiply the partition function (\ref{pf-tf}) by $\prod_{j=1}^{N-1}[v_N-v_j+q_j+q_N]$ and repeatedly use the commutation relation

\begin{align}
&
[v_N-v_j+q_j+q_N]
B(v_N,q_N,\{w,r\}_N)
B(v_j,q_j,\{w,r\}_N)
=
\\
&
[v_j-v_N+q_j+q_N]
B(v_j,q_j,\{w,r\}_N)
B(v_N,q_N,\{w,r\}_N)
\nonumber
\end{align}

\noindent which is a rewriting of (\ref{bb-tf}), to change the order of the $B$-operators. We obtain 

\begin{align}
&
\prod_{j=1}^{N-1}
[v_N-v_j+q_j+q_N]
Z_N\Big( \{v,q\}_N,\{w,r\}_N\Big)
=
\label{reorder}
\\
&
\prod_{j=1}^{N-1}
[v_j-v_N+q_j+q_N]
\langle \Downarrow_N|
\lprod_{j=1}^{N-1} B(v_j,q_j,\{w,r\}_N)
B(v_N,q_N,\{w,r\}_N)
|\Uparrow_N\rangle
\nonumber
\end{align}

\noindent Graphically, we depict (\ref{reorder}) with the following diagrams.

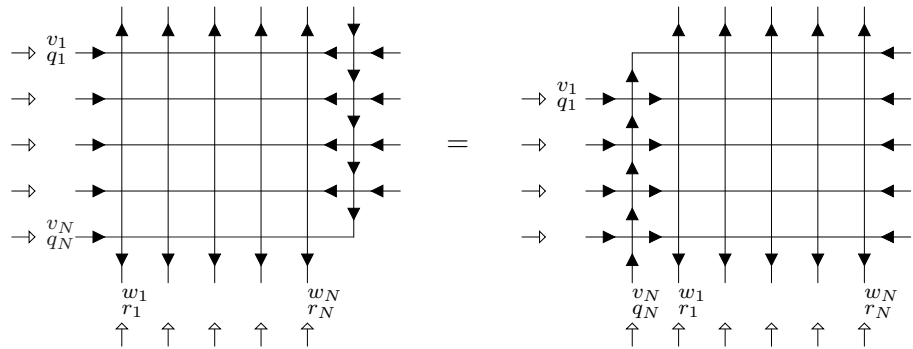
\begin{figure}[H]

\begin{center}
\begin{minipage}{4.3in}

\setlength{\unitlength}{0.00030cm}
\begin{picture}(20000,14000)(-4000,-4000)

\path(-2000,0)(10000,0)
\put(-3250,350){\scriptsize{$v_N$}}
\put(-3250,-350){\scriptsize{$q_N$}}
\path(-4750,0)(-3750,0)
\whiten\path(-4000,250)(-4000,-250)(-3750,0)(-4000,250)
\blacken\path(-1250,250)(-1250,-250)(-750,0)(-1250,250)

\path(-2000,2000)(12000,2000)
\path(-4750,2000)(-3750,2000)
\whiten\path(-4000,2250)(-4000,1750)(-3750,2000)(-4000,2250)
\blacken\path(-1250,2250)(-1250,1750)(-750,2000)(-1250,2250)
\blacken\path(9250,2250)(9250,1750)(8750,2000)(9250,2250)
\blacken\path(11250,2250)(11250,1750)(10750,2000)(11250,2250)

\path(-2000,4000)(12000,4000)
\path(-4750,4000)(-3750,4000)
\whiten\path(-4000,4250)(-4000,3750)(-3750,4000)(-4000,4250)
\blacken\path(-1250,4250)(-1250,3750)(-750,4000)(-1250,4250)
\blacken\path(9250,4250)(9250,3750)(8750,4000)(9250,4250)
\blacken\path(11250,4250)(11250,3750)(10750,4000)(11250,4250)

\path(-2000,6000)(12000,6000)
\path(-4750,6000)(-3750,6000)
\whiten\path(-4000,6250)(-4000,5750)(-3750,6000)(-4000,6250)
\blacken\path(-1250,6250)(-1250,5750)(-750,6000)(-1250,6250)
\blacken\path(9250,6250)(9250,5750)(8750,6000)(9250,6250)
\blacken\path(11250,6250)(11250,5750)(10750,6000)(11250,6250)

\path(-2000,8000)(12000,8000)
\put(-3250,8350){\scriptsize{$v_1$}}
\put(-3250,7650){\scriptsize{$q_1$}}
\path(-4750,8000)(-3750,8000)
\whiten\path(-4000,8250)(-4000,7750)(-3750,8000)(-4000,8250)
\blacken\path(-1250,8250)(-1250,7750)(-750,8000)(-1250,8250)
\blacken\path(9250,8250)(9250,7750)(8750,8000)(9250,8250)
\blacken\path(11250,8250)(11250,7750)(10750,8000)(11250,8250)


\path(0,-2000)(0,10000)
\put(0,-2650){\scriptsize{$w_1$}}
\put(0,-3350){\scriptsize{$r_1$}}
\path(0,-4750)(0,-3750)
\whiten\path(-250,-4000)(250,-4000)(0,-3750)(-250,-4000)
\blacken\path(-250,-750)(250,-750)(0,-1250)(-250,-750)
\blacken\path(-250,8750)(250,8750)(0,9250)(-250,8750)

\path(2000,-2000)(2000,10000)
\path(2000,-4750)(2000,-3750)
\whiten\path(1750,-4000)(2250,-4000)(2000,-3750)(1750,-4000)
\blacken\path(1750,-750)(2250,-750)(2000,-1250)(1750,-750)
\blacken\path(1750,8750)(2250,8750)(2000,9250)(1750,8750)

\path(4000,-2000)(4000,10000)
\path(4000,-4750)(4000,-3750)
\whiten\path(3750,-4000)(4250,-4000)(4000,-3750)(3750,-4000)
\blacken\path(3750,-750)(4250,-750)(4000,-1250)(3750,-750)
\blacken\path(3750,8750)(4250,8750)(4000,9250)(3750,8750)

\path(6000,-2000)(6000,10000)
\path(6000,-4750)(6000,-3750)
\whiten\path(5750,-4000)(6250,-4000)(6000,-3750)(5750,-4000)
\blacken\path(5750,-750)(6250,-750)(6000,-1250)(5750,-750)
\blacken\path(5750,8750)(6250,8750)(6000,9250)(5750,8750)

\path(8000,-2000)(8000,10000)
\put(8000,-2650){\scriptsize{$w_N$}}
\put(8000,-3350){\scriptsize{$r_N$}}
\path(8000,-4750)(8000,-3750)
\whiten\path(7750,-4000)(8250,-4000)(8000,-3750)(7750,-4000)
\blacken\path(7750,-750)(8250,-750)(8000,-1250)(7750,-750)
\blacken\path(7750,8750)(8250,8750)(8000,9250)(7750,8750)

\path(10000,0)(10000,10000)
\blacken\path(9750,1250)(10250,1250)(10000,750)(9750,1250)
\blacken\path(9750,3250)(10250,3250)(10000,2750)(9750,3250)
\blacken\path(9750,5250)(10250,5250)(10000,4750)(9750,5250)
\blacken\path(9750,7250)(10250,7250)(10000,6750)(9750,7250)
\blacken\path(9750,9250)(10250,9250)(10000,8750)(9750,9250)


\put(14000,3750){$=$}


\path(20000,0)(34000,0)
\path(17250,0)(18250,0)
\whiten\path(18000,250)(18000,-250)(18250,0)(18000,250)
\blacken\path(20750,250)(20750,-250)(21250,0)(20750,250)
\blacken\path(22750,250)(22750,-250)(23250,0)(22750,250)
\blacken\path(33250,250)(33250,-250)(32750,0)(33250,250)

\path(20000,2000)(34000,2000)
\path(17250,2000)(18250,2000)
\whiten\path(18000,2250)(18000,1750)(18250,2000)(18000,2250)
\blacken\path(20750,2250)(20750,1750)(21250,2000)(20750,2250)
\blacken\path(22750,2250)(22750,1750)(23250,2000)(22750,2250)
\blacken\path(33250,2250)(33250,1750)(32750,2000)(33250,2250)

\path(20000,4000)(34000,4000)
\path(17250,4000)(18250,4000)
\whiten\path(18000,4250)(18000,3750)(18250,4000)(18000,4250)
\blacken\path(20750,4250)(20750,3750)(21250,4000)(20750,4250)
\blacken\path(22750,4250)(22750,3750)(23250,4000)(22750,4250)
\blacken\path(33250,4250)(33250,3750)(32750,4000)(33250,4250)

\path(20000,6000)(34000,6000)
\put(18750,6350){\scriptsize$v_1$}
\put(18750,5650){\scriptsize$q_1$}
\path(17250,6000)(18250,6000)
\whiten\path(18000,6250)(18000,5750)(18250,6000)(18000,6250)
\blacken\path(20750,6250)(20750,5750)(21250,6000)(20750,6250)
\blacken\path(22750,6250)(22750,5750)(23250,6000)(22750,6250)
\blacken\path(33250,6250)(33250,5750)(32750,6000)(33250,6250)

\path(22000,8000)(34000,8000)
\blacken\path(33250,8250)(33250,7750)(32750,8000)(33250,8250)


\path(22000,-2000)(22000,8000)
\put(22000,-2650){\scriptsize$v_N$}
\put(22000,-3350){\scriptsize$q_N$}
\path(22000,-4750)(22000,-3750)
\whiten\path(21750,-4000)(22250,-4000)(22000,-3750)(21750,-4000)
\blacken\path(21750,-1250)(22250,-1250)(22000,-750)(21750,-1250)
\blacken\path(21750,750)(22250,750)(22000,1250)(21750,750)
\blacken\path(21750,2750)(22250,2750)(22000,3250)(21750,2750)
\blacken\path(21750,4750)(22250,4750)(22000,5250)(21750,4750)
\blacken\path(21750,6750)(22250,6750)(22000,7250)(21750,6750)

\path(24000,-2000)(24000,10000)
\put(24000,-2650){\scriptsize$w_1$}
\put(24000,-3350){\scriptsize$r_1$}
\path(24000,-4750)(24000,-3750)
\whiten\path(23750,-4000)(24250,-4000)(24000,-3750)(23750,-4000)
\blacken\path(23750,-750)(24250,-750)(24000,-1250)(23750,-750)
\blacken\path(23750,8750)(24250,8750)(24000,9250)(23750,8750)

\path(26000,-2000)(26000,10000)
\path(26000,-4750)(26000,-3750)
\whiten\path(25750,-4000)(26250,-4000)(26000,-3750)(25750,-4000)
\blacken\path(25750,-750)(26250,-750)(26000,-1250)(25750,-750)
\blacken\path(25750,8750)(26250,8750)(26000,9250)(25750,8750)

\path(28000,-2000)(28000,10000)
\path(28000,-4750)(28000,-3750)
\whiten\path(27750,-4000)(28250,-4000)(28000,-3750)(27750,-4000)
\blacken\path(27750,-750)(28250,-750)(28000,-1250)(27750,-750)
\blacken\path(27750,8750)(28250,8750)(28000,9250)(27750,8750)

\path(30000,-2000)(30000,10000)
\path(30000,-4750)(30000,-3750)
\whiten\path(29750,-4000)(30250,-4000)(30000,-3750)(29750,-4000)
\blacken\path(29750,-750)(30250,-750)(30000,-1250)(29750,-750)
\blacken\path(29750,8750)(30250,8750)(30000,9250)(29750,8750)

\path(32000,-2000)(32000,10000)
\put(32000,-2650){\scriptsize$w_N$}
\put(32000,-3350){\scriptsize$r_N$}
\path(32000,-4750)(32000,-3750)
\whiten\path(31750,-4000)(32250,-4000)(32000,-3750)(31750,-4000)
\blacken\path(31750,-750)(32250,-750)(32000,-1250)(31750,-750)
\blacken\path(31750,8750)(32250,8750)(32000,9250)(31750,8750)

\end{picture}

\end{minipage}
\end{center}

\caption[Reordering the lattice lines of the trigonometric Felderhof partition function]{Reordering the lattice lines of the trigonometric Felderhof partition function. The diagram on the left is the domain wall partition function multiplied by the string of vertices $\prod_{j=1}^{N-1} a_{-}(v_j,q_j,v_N,q_N)$, and it corresponds with the left hand side of (\ref{reorder}). Each vertex can be threaded through the lattice using the Yang-Baxter equation, which ultimately produces the diagram on the right. This diagram represents the domain wall partition function with its $N^{\rm th}$ row transferred to the top of the lattice, multiplied by the string of vertices $\prod_{j=1}^{N-1} a_{+}(v_j,q_j,v_N,q_N)$. Clearly, this corresponds with the right hand side of (\ref{reorder}).  
}
\end{figure}

\noindent The right hand side of (\ref{reorder}) is a trigonometric polynomial of degree $2N-2$ in $v_N$, with zeros at the points $v_N = v_j + q_j +q_N$ for all $1 \leq j \leq N-1$. Therefore the partition function $Z_N(\{v,q\}_N,\{w,r\}_N)$ must have zeros at the same points.

}
\end{property2}

\begin{property2}
{\rm 
We start from the expansion (\ref{pfexp-tf}) of the domain wall partition function, and set $v_N = w_N+q_N+r_N$. This causes all terms in the summation over $1\leq n \leq N$ to collapse to zero except the $n=N$ term, and we obtain

\begin{align}
&
Z_N\Big(\{v,q\}_N,\{w,r\}_N\Big)\Big|_{v_N = w_N + q_N + r_N}
=
\label{pfrec2-tf}
\\
&
[2q_N]^{\frac{1}{2}} [2r_N]^{\frac{1}{2}}
\prod_{j=1}^{N-1} [w_N-w_j+r_j+r_N]
\langle \Downarrow_N| \sigma_N^{-}
\lprod_{j=1}^{N-1} B(v_j,q_j,\{w,r\}_N)
|\Uparrow_N\rangle
\nonumber
\end{align}

\noindent This equation can be further simplified by considering the graphical representation of $\langle \Downarrow_N | \sigma_N^{-}\ \lprod_{j=1}^{N-1} B(v_j,q_j,\{w,r\}_N) |\Uparrow_N\rangle$, as shown below.

\begin{figure}[H]

\begin{center}
\begin{minipage}{4.3in}

\setlength{\unitlength}{0.000325cm}
\begin{picture}(20000,12500)(-4000,-3000)

\path(-2000,0)(10000,0)
\put(-4250,400){\tiny$v_{N-1}$}
\put(-4250,-400){\tiny$q_{N-1}$}
\path(-5500,0)(-4500,0)
\whiten\path(-4750,250)(-4750,-250)(-4500,0)(-4750,250)
\blacken\path(-1250,250)(-1250,-250)(-750,0)(-1250,250)
\blacken\path(9250,250)(9250,-250)(8750,0)(9250,250)

\path(-2000,2000)(10000,2000)
\path(-5500,2000)(-4500,2000)
\whiten\path(-4750,2250)(-4750,1750)(-4500,2000)(-4750,2250)
\blacken\path(-1250,2250)(-1250,1750)(-750,2000)(-1250,2250)
\blacken\path(9250,2250)(9250,1750)(8750,2000)(9250,2250)

\path(-2000,4000)(10000,4000)
\path(-5500,4000)(-4500,4000)
\whiten\path(-4750,4250)(-4750,3750)(-4500,4000)(-4750,4250)
\blacken\path(-1250,4250)(-1250,3750)(-750,4000)(-1250,4250)
\blacken\path(9250,4250)(9250,3750)(8750,4000)(9250,4250)

\path(-2000,6000)(10000,6000)
\put(-4250,6400){\tiny$v_1$}
\put(-4250,5600){\tiny$q_1$}
\path(-5500,6000)(-4500,6000)
\whiten\path(-4750,6250)(-4750,5750)(-4500,6000)(-4750,6250)
\blacken\path(-1250,6250)(-1250,5750)(-750,6000)(-1250,6250)
\blacken\path(9250,6250)(9250,5750)(8750,6000)(9250,6250)


\path(0,-2000)(0,8000)
\put(-1000,-2750){\tiny$w_1,r_1$}
\path(0,-4200)(0,-3200)
\whiten\path(-250,-3450)(250,-3450)(0,-3200)(-250,-3450)
\blacken\path(-250,-750)(250,-750)(0,-1250)(-250,-750)
\blacken\path(-250,6750)(250,6750)(0,7250)(-250,6750)

\path(2000,-2000)(2000,8000)
\path(2000,-4200)(2000,-3200)
\whiten\path(1750,-3450)(2250,-3450)(2000,-3200)(1750,-3450)
\blacken\path(1750,-750)(2250,-750)(2000,-1250)(1750,-750)
\blacken\path(1750,6750)(2250,6750)(2000,7250)(1750,6750)

\path(4000,-2000)(4000,8000)
\path(4000,-4200)(4000,-3200)
\whiten\path(3750,-3450)(4250,-3450)(4000,-3200)(3750,-3450)
\blacken\path(3750,-750)(4250,-750)(4000,-1250)(3750,-750)
\blacken\path(3750,6750)(4250,6750)(4000,7250)(3750,6750)

\path(6000,-2000)(6000,8000)
\path(6000,-4200)(6000,-3200)
\whiten\path(5750,-3450)(6250,-3450)(6000,-3200)(5750,-3450)
\blacken\path(5750,-750)(6250,-750)(6000,-1250)(5750,-750)
\blacken\path(5750,6750)(6250,6750)(6000,7250)(5750,6750)

\path(8000,-2000)(8000,8000)
\put(7000,-2750){\tiny$w_N,r_N$}
\path(8000,-4200)(8000,-3200)
\whiten\path(7750,-3450)(8250,-3450)(8000,-3200)(7750,-3450)
\blacken\path(7750,-1250)(8250,-1250)(8000,-750)(7750,-1250)
\blacken\path(7750,6750)(8250,6750)(8000,7250)(7750,6750)


\put(12000,2750){$=$}


\path(18000,0)(30000,0)
\put(15750,400){\tiny$v_{N-1}$}
\put(15750,-400){\tiny$q_{N-1}$}
\path(14500,0)(15500,0)
\whiten\path(15250,250)(15250,-250)(15500,0)(15250,250)
\blacken\path(18750,250)(18750,-250)(19250,0)(18750,250)
\blacken\path(27250,250)(27250,-250)(26750,0)(27250,250)
\blacken\path(29250,250)(29250,-250)(28750,0)(29250,250)

\path(18000,2000)(30000,2000)
\path(14500,2000)(15500,2000)
\whiten\path(15250,2250)(15250,1750)(15500,2000)(15250,2250)
\blacken\path(18750,2250)(18750,1750)(19250,2000)(18750,2250)
\blacken\path(27250,2250)(27250,1750)(26750,2000)(27250,2250)
\blacken\path(29250,2250)(29250,1750)(28750,2000)(29250,2250)

\path(18000,4000)(30000,4000)
\path(14500,4000)(15500,4000)
\whiten\path(15250,4250)(15250,3750)(15500,4000)(15250,4250)
\blacken\path(18750,4250)(18750,3750)(19250,4000)(18750,4250)
\blacken\path(27250,4250)(27250,3750)(26750,4000)(27250,4250)
\blacken\path(29250,4250)(29250,3750)(28750,4000)(29250,4250)

\path(18000,6000)(30000,6000)
\put(15750,6400){\tiny$v_1$}
\put(15750,5600){\tiny$q_1$}
\path(14500,6000)(15500,6000)
\whiten\path(15250,6250)(15250,5750)(15500,6000)(15250,6250)
\blacken\path(18750,6250)(18750,5750)(19250,6000)(18750,6250)
\blacken\path(27250,6250)(27250,5750)(26750,6000)(27250,6250)
\blacken\path(29250,6250)(29250,5750)(28750,6000)(29250,6250)


\path(20000,-2000)(20000,8000)
\put(19000,-2750){\tiny$w_1,r_1$}
\path(20000,-4200)(20000,-3200)
\whiten\path(19750,-3450)(20250,-3450)(20000,-3200)(19750,-3450)
\blacken\path(19750,-750)(20250,-750)(20000,-1250)(19750,-750)
\blacken\path(19750,6750)(20250,6750)(20000,7250)(19750,6750)

\path(22000,-2000)(22000,8000)
\path(22000,-4200)(22000,-3200)
\whiten\path(21750,-3450)(22250,-3450)(22000,-3200)(21750,-3450)
\blacken\path(21750,-750)(22250,-750)(22000,-1250)(21750,-750)
\blacken\path(21750,6750)(22250,6750)(22000,7250)(21750,6750)

\path(24000,-2000)(24000,8000)
\path(24000,-4200)(24000,-3200)
\whiten\path(23750,-3450)(24250,-3450)(24000,-3200)(23750,-3450)
\blacken\path(23750,-750)(24250,-750)(24000,-1250)(23750,-750)
\blacken\path(23750,6750)(24250,6750)(24000,7250)(23750,6750)

\path(26000,-2000)(26000,8000)
\path(26000,-4200)(26000,-3200)
\whiten\path(25750,-3450)(26250,-3450)(26000,-3200)(25750,-3450)
\blacken\path(25750,-750)(26250,-750)(26000,-1250)(25750,-750)
\blacken\path(25750,6750)(26250,6750)(26000,7250)(25750,6750)

\path(28000,-2000)(28000,8000)
\put(27000,-2750){\tiny$w_N,r_N$}
\path(28000,-4200)(28000,-3200)
\whiten\path(27750,-3450)(28250,-3450)(28000,-3200)(27750,-3450)
\blacken\path(27750,-1250)(28250,-1250)(28000,-750)(27750,-1250)
\blacken\path(27750,750)(28250,750)(28000,1250)(27750,750)
\blacken\path(27750,2750)(28250,2750)(28000,3250)(27750,2750)
\blacken\path(27750,4750)(28250,4750)(28000,5250)(27750,4750)
\blacken\path(27750,6750)(28250,6750)(28000,7250)(27750,6750)

\end{picture}

\end{minipage}
\end{center}

\caption[Peeling the right-most column of the trigonometric Felderhof partition function]{Peeling the right-most column of the trigonometric Felderhof partition function. The diagram on the left represents $\langle \Downarrow_N | \sigma_N^{-}\ \lprod_{j=1}^{N-1} B(v_j,q_j,\{w,r\}_N) |\Uparrow_N\rangle$, with the internal black arrows being summed over all configurations. The diagram on the right contains all surviving configurations.}

\label{tf-zequiv5}
\end{figure}

\noindent The right hand side of figure \ref{tf-zequiv5} represents the $(N-1)\times (N-1)$ partition function, multiplied by a column of vertices. Replacing these vertices with their trigonometric weights, we have

\begin{align}
&
\langle \Downarrow_N| \sigma_N^{-}
\lprod_{j=1}^{N-1} B(v_j,q_j,\{w,r\}_N)
|\Uparrow_N\rangle
=
\label{tf-peel3}
\prod_{j=1}^{N-1} [v_j-w_N+q_j-r_N]
Z_{N-1}
\end{align}

\noindent Substituting (\ref{tf-peel3}) into (\ref{pfrec2-tf}) we recover the required recursion relation (\ref{pfrec1-tf}).

}
\end{property2}

\begin{property2}
{\rm
Specializing the definition (\ref{pf-tf}) to the case $N=1$ gives

\begin{align}
Z_1(v_1,q_1,w_1,r_1)
&=
\langle \Downarrow_1|
B(v_1,q_1,\{w,r\}_1)
|\Uparrow_1\rangle
\\
&=
\uparrow_{a_1}^{*} \otimes \downarrow_{1}^{*}
R_{a_1 1}(v_1,q_1,w_1,r_1)
\uparrow_1 \otimes \downarrow_{a_1}
=
[2q_1]^{\frac{1}{2}} [2r_1]^{\frac{1}{2}}
\nonumber
\end{align}

\noindent as required. Alternatively, the $1\times 1$ partition function is the top-right vertex in figure 6.1, whose weight is equal to $[2q_1]^{\frac{1}{2}} [2r_1]^{\frac{1}{2}}$.
}
\end{property2}
\end{proof}

\subsection{Factorized expression for $Z_{N}(\{v,q\}_{N},\{w,r\}_{N})$}

The conditions {\bf 1}--{\bf 4} are strong constraints on $Z_{N}(\{v,q\}_{N},\{w,r\}_{N})$. Not only do they specify $Z_{N}(\{v,q\}_{N},\{w,r\}_{N})$ uniquely, they lead to its direct evaluation, as we demonstrate below.

\begin{lemma}
{\rm 
The domain wall partition function has the factorized expression

{\small
\begin{align}
&
Z_N\Big(\{v,q\}_N,\{w,r\}_N\Big)
=
\label{lem-tf}
\prod_{j=1}^{N}
[2q_j]^{\frac{1}{2}} [2r_j]^{\frac{1}{2}}
\prod_{1 \leq j < k \leq N}
[v_j-v_k+q_j+q_k] [w_k-w_j+r_j+r_k]
\end{align}
}

\noindent Specializing the external fields to $q_j=r_j=\frac{\pi i}{4}$ for all $1\leq j \leq N$, we recover the partition function of the six-vertex model at its free fermion point (\ref{IK-ff3}). The result (\ref{lem-tf}) was first obtained in \cite{cfwz} using a complicated recursion relation. A more straightforward proof, based on solving the conditions {\bf 1}--{\bf 4}, subsequently appeared in \cite{fwz1}. It is the latter proof which we present below. 
}
\end{lemma} 

\begin{proof}
From condition {\bf 1} and {\bf 2} on $Z_{N}(\{v,q\}_{N},\{w,r\}_{N})$, we know that it must have the form

\begin{align}
Z_{N}\Big(\{v,q\}_{N},\{w,r\}_{N}\Big)
=
\mathcal{C}\Big(\{v\}_{N-1},\{q\}_N,\{w,r\}_N\Big)
\prod_{j=1}^{N-1}
[v_j-v_N+q_j+q_N]
\label{p1-tf}
\end{align}

\noindent where $\mathcal{C}$ does not depend on $v_N$, but depends on all other variables. Evaluating (\ref{p1-tf}) at $v_N = w_N+q_N+r_N$ and comparing with condition {\bf 3} on $Z_{N}$, we obtain 

{\small
\begin{align}
Z_N\Big|_{v_N = w_N+q_N+r_N}
&=
\mathcal{C} \Big(\{v\}_{N-1},\{q\}_N,\{w,r\}_N\Big)
\prod_{j=1}^{N-1}
[v_j-w_N+q_j-r_N]
\\
&=
[2q_N]^{\frac{1}{2}} [2r_N]^{\frac{1}{2}}
\prod_{j=1}^{N-1}
[w_N-w_j+r_j+r_N]
[v_j-w_N+q_j-r_N]
Z_{N-1}
\nonumber
\end{align}
}

\noindent from which we extract the equation

\begin{align}
\mathcal{C}
=
[2q_N]^{\frac{1}{2}} [2r_N]^{\frac{1}{2}}
\prod_{j=1}^{N-1}
[w_N-w_j+r_j+r_N]
Z_{N-1}\Big(\{v,q\}_{N-1},\{w,r\}_{N-1}\Big)
\end{align}

\noindent Substituting this expression for $\mathcal{C}$ into (\ref{p1-tf}), we obtain the recurrence 

\begin{align}
&
Z_N\Big(\{v,q\}_N,\{w,r\}_N\Big)
=
[2q_N]^{\frac{1}{2}} [2r_N]^{\frac{1}{2}}
\times
\\
&
\prod_{j=1}^{N-1}
[v_j-v_N+q_j+q_N]
[w_N-w_j+r_j+r_N]
Z_{N-1}\Big(\{v,q\}_{N-1},\{w,r\}_{N-1}\Big)
\nonumber
\end{align}

\noindent whose basis is given by condition {\bf 4}. This recurrence is trivially solved to produce the formula (\ref{lem-tf}).
\end{proof}

\subsection{Scalar products $S_n(\{u,p\}_n,\{v,q\}_N,\{w,r\}_M)$}

Let $\{u\}_n = \{u_1,\ldots,u_n\},\{v\}_N = \{v_1,\ldots,v_N\},\{w\}_M = \{w_1,\ldots,w_M\}$ be sets of rapidities, and $\{p\}_n = \{p_1,\ldots,p_n\},\{q\}_N = \{q_1,\ldots,q_N\},\{r\}_M = \{r_1,\ldots,r_M\}$ the corresponding sets of external fields. The cardinalities of these sets are assumed to satisfy $0 \leq n \leq N$ and $1 \leq N \leq M$. For $n=0$ we define

\begin{align}
S_0\Big(\{v,q\}_N,\{w,r\}_M\Big)
=
\langle \Downarrow_{N/M} |
\lprod_{k=1}^{N}
B(v_k,q_k,\{w,r\}_M)
|\Uparrow_M\rangle
\label{tf-sp1}
\end{align}

\noindent Similarly to the last chapter, we will find that $S_0$ is equal to the trigonometric Felderhof partition function $Z_N$, up to an overall normalization. Next, for all $1\leq n \leq N-1$ we define

{\small
\begin{align}
&
S_n\Big(\{u,p\}_n,\{v,q\}_N,\{w,r\}_M\Big)
=
\label{tf-sp2}
\langle \Downarrow_{\widetilde{N}/M} |
\lprod_{j=1}^{n} C(u_j,p_j,\{w,r\}_M)
\lprod_{k=1}^{N} B(v_k,q_k,\{w,r\}_M)
|\Uparrow_M\rangle
\end{align}
}

\noindent with $\widetilde{N} = N-n$. Finally, in the case $n=N$ we fix

{\small
\begin{align}
&
S_N\Big(
\{u,p\}_N,\{v,q\}_N,\{w,r\}_M
\Big)
=
\label{tf-sp3}
\langle \Uparrow_M|
\lprod_{j=1}^{N}
C(u_j,p_j,\{w,r\}_M)
\lprod_{k=1}^{N}
B(v_k,q_k,\{w,r\}_M)
|\Uparrow_M \rangle
\end{align}
}

\noindent The scalar products (\ref{tf-sp1})--(\ref{tf-sp3}) are the trigonometric Felderhof analogues of those defined in subsection \ref{xxz-sp-def}. They have identical graphical representations to those described in subsection \ref{xxz-sp-graph}, except that every rapidity variable is now accompanied by an appropriate external field. In the following subsection we give a set of conditions on these scalar products, using similar techniques to those of the previous chapter.

\subsection{Conditions on $S_n(\{u,p\}_n,\{v,q\}_N,\{w,r\}_M)$}

\begin{lemma}
{\rm 
For all $1\leq n \leq N$ we claim that

\begin{property6}
{\rm 
$S_n$ is invariant under the simultaneous permutation of variables
$\{w_j,r_j\} \leftrightarrow \{w_k,r_k\}$ for all $j,k \in \{\widetilde{N}+1,\ldots,M\}$.
}
\end{property6}

\begin{property6}
{\rm
$S_n$ is a trigonometric polynomial of degree $M-1$ in $u_n$, with zeros occurring at the points $u_n =p_n+w_j+r_j$, for all $1\leq j \leq \widetilde{N}$.
}
\end{property6}

\begin{property6}
{\rm
Setting $u_n+p_n = w_{\widetilde{N}+1}+r_{\widetilde{N}+1}$, $S_n$ satisfies the recursion relation

\begin{align}
S_n
\Big|_{u_n+p_n=w_{\widetilde{N}+1}+r_{\widetilde{N}+1}}
&=
[2p_n]^{\frac{1}{2}} [2r_{\widetilde{N}+1}]^{\frac{1}{2}}
\prod_{1\leq j < \widetilde{N}+1}
[w_j-w_{\widetilde{N}+1}+r_j-r_{\widetilde{N}+1}+2p_n]
\nonumber
\\
&
\times
\prod_{\widetilde{N}+1 < j \leq M}
[w_{\widetilde{N}+1}-w_j+r_j+r_{\widetilde{N}+1}]
S_{n-1}
\label{Srec0-tf}
\end{align}

\noindent where we have abbreviated $S_{n-1} = S_{n-1}(\{u,p\}_{n-1},\{v,q\}_N,\{w,r\}_M)$.
}
\end{property6}

In addition, we have the supplementary condition

\begin{property6}
{\rm $S_0$ and $Z_N$ are related via the equation

\begin{align}
S_0\Big(\{v,q\}_N,\{w,r\}_M\Big)
=
\prod_{j=1}^{N}
\prod_{k=N+1}^{M}
[v_j-w_k+q_j-r_k]
Z_N\Big(\{v,q\}_N,\{w,r\}_N\Big)
\label{S0-tf}
\end{align}

}
\end{property6}
}
\end{lemma}

\begin{proof}
The proof of properties {\bf 1}--{\bf 4} is analogous to the proof of lemma 9 in chapter 5. There, we presented an algebraic proof of the properties. Here, we outline a less technical graphical proof.

\begin{property7}
{\rm
For any $\widetilde{N}+1 < j \leq M$, multiplying $S_n(\{u,p\}_n,\{v,q\}_N,\{w,r\}_M)$ by the function $a_{+}(w_{j},r_{j},w_{j-1},r_{j-1})$ is equivalent to attaching an $a_{+}$ vertex at the base of the lattice, as shown in figure \ref{multbya}.

\begin{figure}[H]

\begin{center}
\begin{minipage}{4.3in}

\setlength{\unitlength}{0.0003cm}
\begin{picture}(20000,22000)(-9000,-11000)

\put(-8000,-4000){\fbox{$u,p$}}

\path(-2000,-6000)(20000,-6000)
\put(-3250,-6000){\tiny{$n$}}
\path(-4750,-6000)(-3750,-6000)
\whiten\path(-4000,-5750)(-4000,-6250)(-3750,-6000)(-4000,-5750)
\blacken\path(-750,-5750)(-750,-6250)(-1250,-6000)(-750,-5750)
\blacken\path(18750,-5750)(18750,-6250)(19250,-6000)(18750,-5750)

\path(-2000,-4000)(20000,-4000)
\path(-4750,-4000)(-3750,-4000)
\whiten\path(-4000,-3750)(-4000,-4250)(-3750,-4000)(-4000,-3750)
\blacken\path(-750,-3750)(-750,-4250)(-1250,-4000)(-750,-3750)
\blacken\path(18750,-3750)(18750,-4250)(19250,-4000)(18750,-3750)

\path(-2000,-2000)(20000,-2000)
\put(-3250,-2000){\tiny{1}}
\path(-4750,-2000)(-3750,-2000)
\whiten\path(-4000,-1750)(-4000,-2250)(-3750,-2000)(-4000,-1750)
\blacken\path(-750,-1750)(-750,-2250)(-1250,-2000)(-750,-1750)
\blacken\path(18750,-1750)(18750,-2250)(19250,-2000)(18750,-1750)

\put(-8000,5000){\fbox{$v,q$}}

\path(-2000,0)(20000,0)
\put(-3250,0){\tiny{$N$}}
\path(-4750,0)(-3750,0)
\whiten\path(-4000,250)(-4000,-250)(-3750,0)(-4000,250)
\blacken\path(-1250,250)(-1250,-250)(-750,0)(-1250,250)
\blacken\path(19250,250)(19250,-250)(18750,0)(19250,250)

\path(-2000,2000)(20000,2000)
\path(-4750,2000)(-3750,2000)
\whiten\path(-4000,2250)(-4000,1750)(-3750,2000)(-4000,2250)
\blacken\path(-1250,2250)(-1250,1750)(-750,2000)(-1250,2250)
\blacken\path(19250,2250)(19250,1750)(18750,2000)(19250,2250)

\path(-2000,4000)(20000,4000)
\path(-4750,4000)(-3750,4000)
\whiten\path(-4000,4250)(-4000,3750)(-3750,4000)(-4000,4250)
\blacken\path(-1250,4250)(-1250,3750)(-750,4000)(-1250,4250)
\blacken\path(19250,4250)(19250,3750)(18750,4000)(19250,4250)

\path(-2000,6000)(20000,6000)
\path(-4750,6000)(-3750,6000)
\whiten\path(-4000,6250)(-4000,5750)(-3750,6000)(-4000,6250)
\blacken\path(-1250,6250)(-1250,5750)(-750,6000)(-1250,6250)
\blacken\path(19250,6250)(19250,5750)(18750,6000)(19250,6250)

\path(-2000,8000)(20000,8000)
\path(-4750,8000)(-3750,8000)
\whiten\path(-4000,8250)(-4000,7750)(-3750,8000)(-4000,8250)
\blacken\path(-1250,8250)(-1250,7750)(-750,8000)(-1250,8250)
\blacken\path(19250,8250)(19250,7750)(18750,8000)(19250,8250)

\path(-2000,10000)(20000,10000)
\put(-3250,10000){\tiny{$1$}}
\path(-4750,10000)(-3750,10000)
\whiten\path(-4000,10250)(-4000,9750)(-3750,10000)(-4000,10250)
\blacken\path(-1250,10250)(-1250,9750)(-750,10000)(-1250,10250)
\blacken\path(19250,10250)(19250,9750)(18750,10000)(19250,10250)


\path(0,-8000)(0,12000)
\put(-200,-10750){\tiny{1}}
\path(0,-12500)(0,-11500)
\whiten\path(-250,-11750)(250,-11750)(0,-11500)(-250,-11750)
\blacken\path(-250,-6750)(250,-6750)(0,-7250)(-250,-6750)
\blacken\path(-250,10750)(250,10750)(0,11250)(-250,10750)

\path(2000,-8000)(2000,12000)
\path(2000,-12500)(2000,-11500)
\whiten\path(1750,-11750)(2250,-11750)(2000,-11500)(1750,-11750)
\blacken\path(1750,-6750)(2250,-6750)(2000,-7250)(1750,-6750)
\blacken\path(1750,10750)(2250,10750)(2000,11250)(1750,10750)

\path(4000,-8000)(4000,12000)
\put(3800,-10750){\tiny{$\widetilde{N}$}}
\path(4000,-12500)(4000,-11500)
\whiten\path(3750,-11750)(4250,-11750)(4000,-11500)(3750,-11750)
\blacken\path(3750,-6750)(4250,-6750)(4000,-7250)(3750,-6750)
\blacken\path(3750,10750)(4250,10750)(4000,11250)(3750,10750)

\path(6000,-8000)(6000,12000)
\put(5800,-10750){\tiny{$\widetilde{N}+1$}}
\path(6000,-12500)(6000,-11500)
\whiten\path(5750,-11750)(6250,-11750)(6000,-11500)(5750,-11750)
\blacken\path(5750,-7250)(6250,-7250)(6000,-6750)(5750,-7250)
\blacken\path(5750,10750)(6250,10750)(6000,11250)(5750,10750)

\path(8000,-8000)(8000,12000)
\path(8000,-12500)(8000,-11500)
\whiten\path(7750,-11750)(8250,-11750)(8000,-11500)(7750,-11750)
\blacken\path(7750,-7250)(8250,-7250)(8000,-6750)(7750,-7250)
\blacken\path(7750,10750)(8250,10750)(8000,11250)(7750,10750)

\path(10000,-8000)(10000,12000)
\path(10000,-12500)(10000,-11500)
\whiten\path(9750,-11750)(10250,-11750)(10000,-11500)(9750,-11750)
\blacken\path(9750,-7250)(10250,-7250)(10000,-6750)(9750,-7250)
\blacken\path(9750,10750)(10250,10750)(10000,11250)(9750,10750)

\path(12000,-12500)(12000,-11500)
\whiten\path(11750,-11750)(12250,-11750)(12000,-11500)(11750,-11750)
\put(13800,-10750){\tiny{$j-1$}}
\path(12000,-7250)(12000,12000)
\path(12000,-7250)(14000,-9250)
\blacken\path(13750,-9750)(14250,-9750)(14000,-9250)(13750,-9750)
\shade\path(11750,-7250)(12250,-7250)(12000,-6750)(11750,-7250)
\blacken\path(11750,10750)(12250,10750)(12000,11250)(11750,10750)

\path(14000,-12500)(14000,-11500)
\whiten\path(13750,-11750)(14250,-11750)(14000,-11500)(13750,-11750)
\put(11800,-10750){\tiny{$j$}}
\path(14000,-7250)(14000,12000)
\path(14000,-7250)(12000,-9250)
\blacken\path(11750,-9750)(12250,-9750)(12000,-9250)(11750,-9750)
\shade\path(13750,-7250)(14250,-7250)(14000,-6750)(13750,-7250)
\blacken\path(13750,10750)(14250,10750)(14000,11250)(13750,10750)

\path(16000,-8000)(16000,12000)
\path(16000,-12500)(16000,-11500)
\whiten\path(15750,-11750)(16250,-11750)(16000,-11500)(15750,-11750)
\blacken\path(15750,-7250)(16250,-7250)(16000,-6750)(15750,-7250)
\blacken\path(15750,10750)(16250,10750)(16000,11250)(15750,10750)

\path(18000,-8000)(18000,12000)
\put(17800,-10750){\tiny{$M$}}
\path(18000,-12500)(18000,-11500)
\whiten\path(17750,-11750)(18250,-11750)(18000,-11500)(17750,-11750)
\blacken\path(17750,-7250)(18250,-7250)(18000,-6750)(17750,-7250)
\blacken\path(17750,10750)(18250,10750)(18000,11250)(17750,10750)

\put(20000,-9000){\fbox{$w,r$}}

\end{picture}

\end{minipage}
\end{center}

\caption[Attaching an $a_{+}(w_j,r_j,w_{j-1},r_{j-1})$ vertex to the $S_n$ lattice]{Attaching an $a_{+}(w_j,r_j,w_{j-1},r_{j-1})$ vertex to the $S_n$ lattice. The points marked with grey arrows are considered to be summed over all arrow configurations, but the only non-zero configuration is the one shown.}

\label{multbya}
\end{figure}
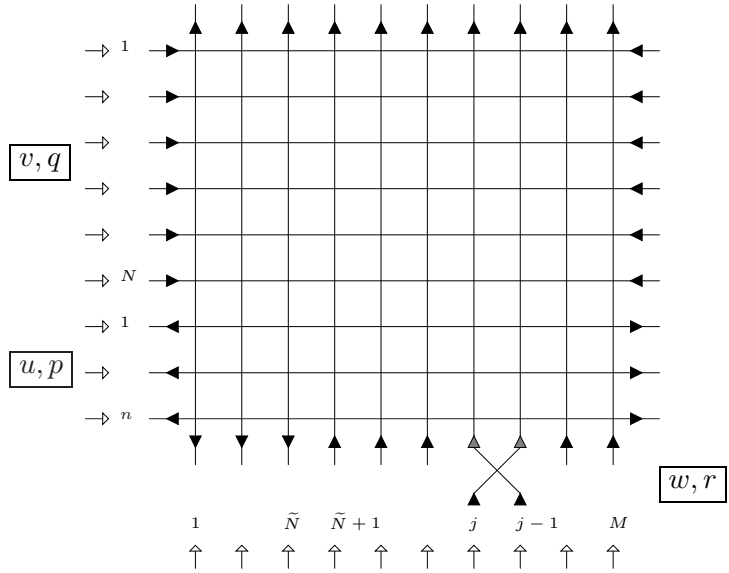

\noindent The attached vertex can be translated vertically through the lattice using the graphical version of the Yang-Baxter equation, as given by figure 6.2. It ultimately emerges from the top of the lattice, still as an $a_{+}(w_{j},r_{j},w_{j-1},r_{j-1})$ vertex, and the $(j-1)^{\rm th}$ and $j^{\rm th}$ lattice columns are swapped in the process. The result of this procedure is shown in figure \ref{extracta}.

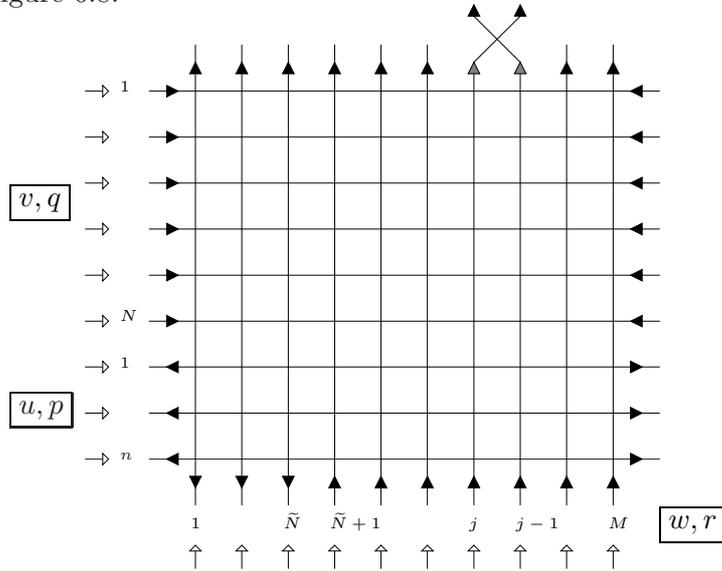
\begin{figure}[H]

\begin{center}
\begin{minipage}{4.3in}

\setlength{\unitlength}{0.00030cm}
\begin{picture}(20000,22000)(-9000,-10000)

\put(-8000,-4000){\fbox{$u,p$}}

\path(-2000,-6000)(20000,-6000)
\put(-3250,-6000){\tiny{$n$}}
\path(-4750,-6000)(-3750,-6000)
\whiten\path(-4000,-5750)(-4000,-6250)(-3750,-6000)(-4000,-5750)
\blacken\path(-750,-5750)(-750,-6250)(-1250,-6000)(-750,-5750)
\blacken\path(18750,-5750)(18750,-6250)(19250,-6000)(18750,-5750)

\path(-2000,-4000)(20000,-4000)
\path(-4750,-4000)(-3750,-4000)
\whiten\path(-4000,-3750)(-4000,-4250)(-3750,-4000)(-4000,-3750)
\blacken\path(-750,-3750)(-750,-4250)(-1250,-4000)(-750,-3750)
\blacken\path(18750,-3750)(18750,-4250)(19250,-4000)(18750,-3750)

\path(-2000,-2000)(20000,-2000)
\put(-3250,-2000){\tiny{1}}
\path(-4750,-2000)(-3750,-2000)
\whiten\path(-4000,-1750)(-4000,-2250)(-3750,-2000)(-4000,-1750)
\blacken\path(-750,-1750)(-750,-2250)(-1250,-2000)(-750,-1750)
\blacken\path(18750,-1750)(18750,-2250)(19250,-2000)(18750,-1750)

\put(-8000,5000){\fbox{$v,q$}}

\path(-2000,0)(20000,0)
\put(-3250,0){\tiny{$N$}}
\path(-4750,0)(-3750,0)
\whiten\path(-4000,250)(-4000,-250)(-3750,0)(-4000,250)
\blacken\path(-1250,250)(-1250,-250)(-750,0)(-1250,250)
\blacken\path(19250,250)(19250,-250)(18750,0)(19250,250)

\path(-2000,2000)(20000,2000)
\path(-4750,2000)(-3750,2000)
\whiten\path(-4000,2250)(-4000,1750)(-3750,2000)(-4000,2250)
\blacken\path(-1250,2250)(-1250,1750)(-750,2000)(-1250,2250)
\blacken\path(19250,2250)(19250,1750)(18750,2000)(19250,2250)

\path(-2000,4000)(20000,4000)
\path(-4750,4000)(-3750,4000)
\whiten\path(-4000,4250)(-4000,3750)(-3750,4000)(-4000,4250)
\blacken\path(-1250,4250)(-1250,3750)(-750,4000)(-1250,4250)
\blacken\path(19250,4250)(19250,3750)(18750,4000)(19250,4250)

\path(-2000,6000)(20000,6000)
\path(-4750,6000)(-3750,6000)
\whiten\path(-4000,6250)(-4000,5750)(-3750,6000)(-4000,6250)
\blacken\path(-1250,6250)(-1250,5750)(-750,6000)(-1250,6250)
\blacken\path(19250,6250)(19250,5750)(18750,6000)(19250,6250)

\path(-2000,8000)(20000,8000)
\path(-4750,8000)(-3750,8000)
\whiten\path(-4000,8250)(-4000,7750)(-3750,8000)(-4000,8250)
\blacken\path(-1250,8250)(-1250,7750)(-750,8000)(-1250,8250)
\blacken\path(19250,8250)(19250,7750)(18750,8000)(19250,8250)

\path(-2000,10000)(20000,10000)
\put(-3250,10000){\tiny{1}}
\path(-4750,10000)(-3750,10000)
\whiten\path(-4000,10250)(-4000,9750)(-3750,10000)(-4000,10250)
\blacken\path(-1250,10250)(-1250,9750)(-750,10000)(-1250,10250)
\blacken\path(19250,10250)(19250,9750)(18750,10000)(19250,10250)


\path(0,-8000)(0,12000)
\put(-200,-9000){\tiny{1}}
\path(0,-10750)(0,-9750)
\whiten\path(-250,-10000)(250,-10000)(0,-9750)(-250,-10000)
\blacken\path(-250,-6750)(250,-6750)(0,-7250)(-250,-6750)
\blacken\path(-250,10750)(250,10750)(0,11250)(-250,10750)

\path(2000,-8000)(2000,12000)
\path(2000,-10750)(2000,-9750)
\whiten\path(1750,-10000)(2250,-10000)(2000,-9750)(1750,-10000)
\blacken\path(1750,-6750)(2250,-6750)(2000,-7250)(1750,-6750)
\blacken\path(1750,10750)(2250,10750)(2000,11250)(1750,10750)

\path(4000,-8000)(4000,12000)
\put(3800,-9000){\tiny{$\widetilde{N}$}}
\path(4000,-10750)(4000,-9750)
\whiten\path(3750,-10000)(4250,-10000)(4000,-9750)(3750,-10000)
\blacken\path(3750,-6750)(4250,-6750)(4000,-7250)(3750,-6750)
\blacken\path(3750,10750)(4250,10750)(4000,11250)(3750,10750)

\path(6000,-8000)(6000,12000)
\put(5800,-9000){\tiny{$\widetilde{N}+1$}}
\path(6000,-10750)(6000,-9750)
\whiten\path(5750,-10000)(6250,-10000)(6000,-9750)(5750,-10000)
\blacken\path(5750,-7250)(6250,-7250)(6000,-6750)(5750,-7250)
\blacken\path(5750,10750)(6250,10750)(6000,11250)(5750,10750)

\path(8000,-8000)(8000,12000)
\path(8000,-10750)(8000,-9750)
\whiten\path(7750,-10000)(8250,-10000)(8000,-9750)(7750,-10000)
\blacken\path(7750,-7250)(8250,-7250)(8000,-6750)(7750,-7250)
\blacken\path(7750,10750)(8250,10750)(8000,11250)(7750,10750)

\path(10000,-8000)(10000,12000)
\path(10000,-10750)(10000,-9750)
\whiten\path(9750,-10000)(10250,-10000)(10000,-9750)(9750,-10000)
\blacken\path(9750,-7250)(10250,-7250)(10000,-6750)(9750,-7250)
\blacken\path(9750,10750)(10250,10750)(10000,11250)(9750,10750)

\path(12000,-8000)(12000,10750)
\put(11800,-9000){\tiny{$j$}}
\path(12000,-10750)(12000,-9750)
\whiten\path(11750,-10000)(12250,-10000)(12000,-9750)(11750,-10000)
\blacken\path(11750,-7250)(12250,-7250)(12000,-6750)(11750,-7250)
\shade\path(11750,10750)(12250,10750)(12000,11250)(11750,10750)
\path(12000,11250)(14000,13250)
\blacken\path(13750,13250)(14250,13250)(14000,13750)(13750,13250)

\path(14000,-8000)(14000,10750)
\put(13800,-9000){\tiny{$j-1$}}
\path(14000,-10750)(14000,-9750)
\whiten\path(13750,-10000)(14250,-10000)(14000,-9750)(13750,-10000)
\blacken\path(13750,-7250)(14250,-7250)(14000,-6750)(13750,-7250)
\shade\path(13750,10750)(14250,10750)(14000,11250)(13750,10750)
\path(14000,11250)(12000,13250)
\blacken\path(11750,13250)(12250,13250)(12000,13750)(11750,13250)

\path(16000,-8000)(16000,12000)
\path(16000,-10750)(16000,-9750)
\whiten\path(15750,-10000)(16250,-10000)(16000,-9750)(15750,-10000)
\blacken\path(15750,-7250)(16250,-7250)(16000,-6750)(15750,-7250)
\blacken\path(15750,10750)(16250,10750)(16000,11250)(15750,10750)

\path(18000,-8000)(18000,12000)
\put(17800,-9000){\tiny{$M$}}
\path(18000,-10750)(18000,-9750)
\whiten\path(17750,-10000)(18250,-10000)(18000,-9750)(17750,-10000)
\blacken\path(17750,-7250)(18250,-7250)(18000,-6750)(17750,-7250)
\blacken\path(17750,10750)(18250,10750)(18000,11250)(17750,10750)

\put(20000,-9000){\fbox{$w,r$}}

\end{picture}

\end{minipage}
\end{center}

\caption[Extracting the $a_{+}(w_j,r_j,w_{j-1},r_{j-1})$ vertex from the $S_n$ lattice]{Extracting the $a_{+}(w_j,r_j,w_{j-1},r_{j-1})$ vertex from the $S_n$ lattice. Once again, the grey arrows indicate the only surviving configuration in the summation at those points.}

\label{extracta}
\end{figure}

\noindent Cancelling the common factor $a_{+}(w_j,r_j,w_{j-1},r_{j-1})$ from figures \ref{multbya} and \ref{extracta}, we conclude that $S_n$ is invariant under swapping the $(j-1)^{\rm th}$ and $j^{\rm th}$ lattice columns, for all $\widetilde{N}+1 < j \leq M$. An arbitrary permutation of the lattice columns is just a composition of such swaps. Therefore $S_n$ is invariant under permuting its $j^{\rm th}$ and $k^{\rm th}$ columns, for all $\widetilde{N}+1 \leq j,k \leq M$. 
}
\end{property7}

\begin{property7}
{\rm
Consider the graphical representation of the scalar product $S_n$, as given below.

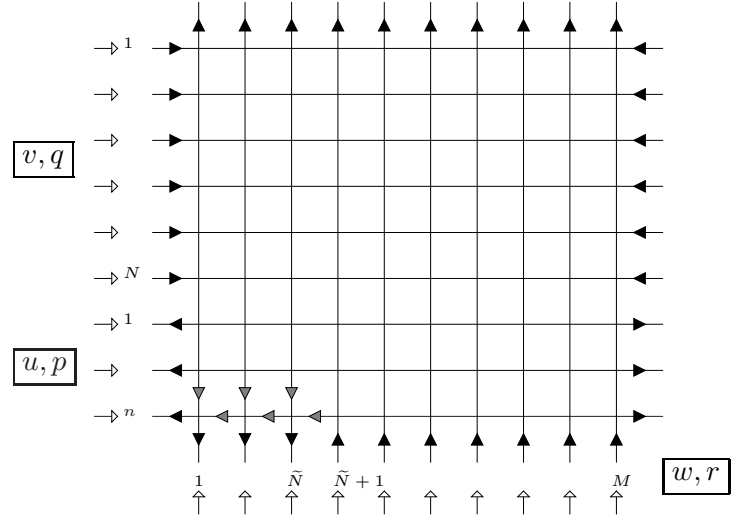
\begin{figure}[H]

\begin{center}
\begin{minipage}{4.3in}

\setlength{\unitlength}{0.0003cm}
\begin{picture}(20000,20000)(-9000,-9000)

\put(-8000,-4000){\fbox{$u,p$}}

\path(-2000,-6000)(20000,-6000)
\put(-3250,-6000){\tiny$n$}
\path(-4500,-6000)(-3500,-6000)
\whiten\path(-3750,-5750)(-3750,-6250)(-3500,-6000)(-3750,-5750)
\blacken\path(-750,-5750)(-750,-6250)(-1250,-6000)(-750,-5750)
\shade\path(1250,-5750)(1250,-6250)(750,-6000)(1250,-5750)
\shade\path(3250,-5750)(3250,-6250)(2750,-6000)(3250,-5750)
\shade\path(5250,-5750)(5250,-6250)(4750,-6000)(5250,-5750)
\blacken\path(18750,-5750)(18750,-6250)(19250,-6000)(18750,-5750)

\path(-2000,-4000)(20000,-4000)
\path(-4500,-4000)(-3500,-4000)
\whiten\path(-3750,-3750)(-3750,-4250)(-3500,-4000)(-3750,-3750)
\blacken\path(-750,-3750)(-750,-4250)(-1250,-4000)(-750,-3750)
\blacken\path(18750,-3750)(18750,-4250)(19250,-4000)(18750,-3750)

\path(-2000,-2000)(20000,-2000)
\put(-3250,-2000){\tiny$1$}
\path(-4500,-2000)(-3500,-2000)
\whiten\path(-3750,-1750)(-3750,-2250)(-3500,-2000)(-3750,-1750)
\blacken\path(-750,-1750)(-750,-2250)(-1250,-2000)(-750,-1750)
\blacken\path(18750,-1750)(18750,-2250)(19250,-2000)(18750,-1750)

\put(-8000,5000){\fbox{$v,q$}}

\path(-2000,0)(20000,0)
\put(-3250,0){\tiny$N$}
\path(-4500,0)(-3500,0)
\whiten\path(-3750,250)(-3750,-250)(-3500,0)(-3750,250)
\blacken\path(-1250,250)(-1250,-250)(-750,0)(-1250,250)
\blacken\path(19250,250)(19250,-250)(18750,0)(19250,250)

\path(-2000,2000)(20000,2000)
\path(-4500,2000)(-3500,2000)
\whiten\path(-3750,2250)(-3750,1750)(-3500,2000)(-3750,2250)
\blacken\path(-1250,2250)(-1250,1750)(-750,2000)(-1250,2250)
\blacken\path(19250,2250)(19250,1750)(18750,2000)(19250,2250)

\path(-2000,4000)(20000,4000)
\path(-4500,4000)(-3500,4000)
\whiten\path(-3750,4250)(-3750,3750)(-3500,4000)(-3750,4250)
\blacken\path(-1250,4250)(-1250,3750)(-750,4000)(-1250,4250)
\blacken\path(19250,4250)(19250,3750)(18750,4000)(19250,4250)

\path(-2000,6000)(20000,6000)
\path(-4500,6000)(-3500,6000)
\whiten\path(-3750,6250)(-3750,5750)(-3500,6000)(-3750,6250)
\blacken\path(-1250,6250)(-1250,5750)(-750,6000)(-1250,6250)
\blacken\path(19250,6250)(19250,5750)(18750,6000)(19250,6250)

\path(-2000,8000)(20000,8000)
\path(-4500,8000)(-3500,8000)
\whiten\path(-3750,8250)(-3750,7750)(-3500,8000)(-3750,8250)
\blacken\path(-1250,8250)(-1250,7750)(-750,8000)(-1250,8250)
\blacken\path(19250,8250)(19250,7750)(18750,8000)(19250,8250)

\path(-2000,10000)(20000,10000)
\put(-3250,10000){\tiny$1$}
\path(-4500,10000)(-3500,10000)
\whiten\path(-3750,10250)(-3750,9750)(-3500,10000)(-3750,10250)
\blacken\path(-1250,10250)(-1250,9750)(-750,10000)(-1250,10250)
\blacken\path(19250,10250)(19250,9750)(18750,10000)(19250,10250)


\path(0,-8000)(0,12000)
\put(-200,-9000){\tiny$1$}
\path(0,-10250)(0,-9250)
\whiten\path(-250,-9500)(250,-9500)(0,-9250)(-250,-9500)
\blacken\path(-250,-6750)(250,-6750)(0,-7250)(-250,-6750)
\shade\path(-250,-4750)(250,-4750)(0,-5250)(-250,-4750)
\blacken\path(-250,10750)(250,10750)(0,11250)(-250,10750)

\path(2000,-8000)(2000,12000)
\path(2000,-10250)(2000,-9250)
\whiten\path(1750,-9500)(2250,-9500)(2000,-9250)(1750,-9500)
\blacken\path(1750,-6750)(2250,-6750)(2000,-7250)(1750,-6750)
\shade\path(1750,-4750)(2250,-4750)(2000,-5250)(1750,-4750)
\blacken\path(1750,10750)(2250,10750)(2000,11250)(1750,10750)

\path(4000,-8000)(4000,12000)
\put(3800,-9000){\tiny$\widetilde{N}$}
\path(4000,-10250)(4000,-9250)
\whiten\path(3750,-9500)(4250,-9500)(4000,-9250)(3750,-9500)
\blacken\path(3750,-6750)(4250,-6750)(4000,-7250)(3750,-6750)
\shade\path(3750,-4750)(4250,-4750)(4000,-5250)(3750,-4750)
\blacken\path(3750,10750)(4250,10750)(4000,11250)(3750,10750)

\path(6000,-8000)(6000,12000)
\put(5800,-9000){\tiny$\widetilde{N}+1$}
\path(6000,-10250)(6000,-9250)
\whiten\path(5750,-9500)(6250,-9500)(6000,-9250)(5750,-9500)
\blacken\path(5750,-7250)(6250,-7250)(6000,-6750)(5750,-7250)
\blacken\path(5750,10750)(6250,10750)(6000,11250)(5750,10750)

\path(8000,-8000)(8000,12000)
\path(8000,-10250)(8000,-9250)
\whiten\path(7750,-9500)(8250,-9500)(8000,-9250)(7750,-9500)
\blacken\path(7750,-7250)(8250,-7250)(8000,-6750)(7750,-7250)
\blacken\path(7750,10750)(8250,10750)(8000,11250)(7750,10750)

\path(10000,-8000)(10000,12000)
\path(10000,-10250)(10000,-9250)
\whiten\path(9750,-9500)(10250,-9500)(10000,-9250)(9750,-9500)
\blacken\path(9750,-7250)(10250,-7250)(10000,-6750)(9750,-7250)
\blacken\path(9750,10750)(10250,10750)(10000,11250)(9750,10750)

\path(12000,-8000)(12000,12000)
\path(12000,-10250)(12000,-9250)
\whiten\path(11750,-9500)(12250,-9500)(12000,-9250)(11750,-9500)
\blacken\path(11750,-7250)(12250,-7250)(12000,-6750)(11750,-7250)
\blacken\path(11750,10750)(12250,10750)(12000,11250)(11750,10750)

\path(14000,-8000)(14000,12000)
\path(14000,-10250)(14000,-9250)
\whiten\path(13750,-9500)(14250,-9500)(14000,-9250)(13750,-9500)
\blacken\path(13750,-7250)(14250,-7250)(14000,-6750)(13750,-7250)
\blacken\path(13750,10750)(14250,10750)(14000,11250)(13750,10750)

\path(16000,-8000)(16000,12000)
\path(16000,-10250)(16000,-9250)
\whiten\path(15750,-9500)(16250,-9500)(16000,-9250)(15750,-9500)
\blacken\path(15750,-7250)(16250,-7250)(16000,-6750)(15750,-7250)
\blacken\path(15750,10750)(16250,10750)(16000,11250)(15750,10750)

\path(18000,-8000)(18000,12000)
\put(17800,-9000){\tiny$M$}
\path(18000,-10250)(18000,-9250)
\whiten\path(17750,-9500)(18250,-9500)(18000,-9250)(17750,-9500)
\blacken\path(17750,-7250)(18250,-7250)(18000,-6750)(17750,-7250)
\blacken\path(17750,10750)(18250,10750)(18000,11250)(17750,10750)

\put(20000,-8750){\fbox{$w,r$}}

\end{picture}

\end{minipage}
\end{center}

\caption[Lattice representation of $S_n$, with frozen vertices included]{Lattice representation of $S_n$, with frozen vertices included. The grey arrows indicate points which are summed over all configurations, but whose only non-zero configuration is the one shown.}

\label{snfroze}
\end{figure} 

\noindent We examine the final row of this lattice, through which the variable $u_n$ flows. Every non-zero configuration of this row contains a $c_{-}(u_n,p_n,w_j,r_j)$ vertex, which by (\ref{c-tf}) does not depend on $u_n$, and $M-1$ other vertices which are trigonometric polynomials of degree 1 in $u_n$. It follows that $S_n$ is a trigonometric polynomial of degree $M-1$ in $u_n$. 

Furthermore, all surviving configurations of the final row contain the $\widetilde{N}$ vertices as shown in figure \ref{snfroze}. Consequentially, $S_n$ contains the factor

\begin{align}
\prod_{j=1}^{\widetilde{N}} a_{-}(u_n,p_n,w_j,r_j)
=
\prod_{j=1}^{\widetilde{N}} [w_j-u_n+p_n+r_j]
\end{align} 

\noindent which gives rise to zeros at $u_n = p_n+w_j+r_j$ for all $1\leq j \leq \widetilde{N}$.
  
}
\end{property7}

\begin{property7}
{\rm
Consider the vertex at the intersection of the $u_n$ and $w_{\widetilde{N}+1}$ lines in figure \ref{snfroze}. In any given lattice configuration, this can be of type $b_{-}(u_n,p_n,w_{\widetilde{N}+1},r_{\widetilde{N}+1})$ or $c_{-}(u_n,p_n,w_{\widetilde{N}+1},r_{\widetilde{N}+1})$. Setting $u_n+p_n = w_{\widetilde{N}+1}+r_{\widetilde{N}+1}$ results in the cancellation of all terms containing $b_{-}(u_n,p_n,w_{\widetilde{N}+1},r_{\widetilde{N}+1})$, and freezes the entire final row of the lattice to the configuration below.

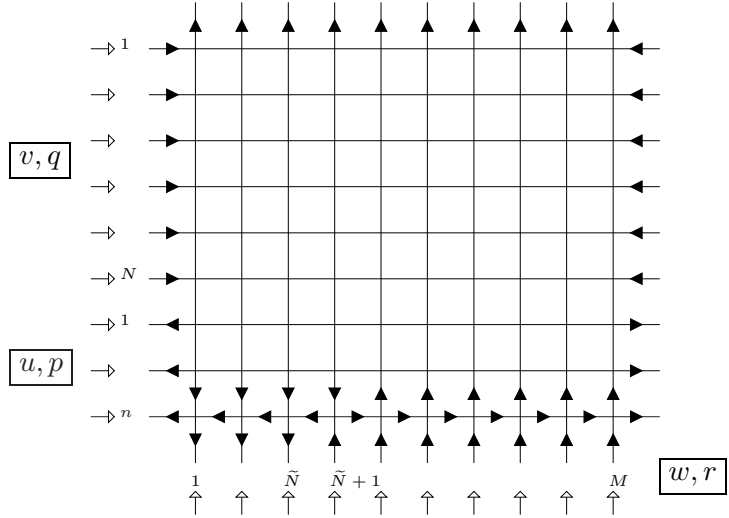
\begin{figure}[H]

\begin{center}
\begin{minipage}{4.3in}

\setlength{\unitlength}{0.0003cm}
\begin{picture}(20000,20000)(-9000,-9000)

\put(-8000,-4000){\fbox{$u,p$}}

\path(-2000,-6000)(20000,-6000)
\put(-3250,-6000){\tiny$n$}
\path(-4500,-6000)(-3500,-6000)
\whiten\path(-3750,-5750)(-3750,-6250)(-3500,-6000)(-3750,-5750)
\blacken\path(-750,-5750)(-750,-6250)(-1250,-6000)(-750,-5750)
\blacken\path(1250,-5750)(1250,-6250)(750,-6000)(1250,-5750)
\blacken\path(3250,-5750)(3250,-6250)(2750,-6000)(3250,-5750)
\blacken\path(5250,-5750)(5250,-6250)(4750,-6000)(5250,-5750)
\blacken\path(6750,-5750)(6750,-6250)(7250,-6000)(6750,-5750)
\blacken\path(8750,-5750)(8750,-6250)(9250,-6000)(8750,-5750)
\blacken\path(10750,-5750)(10750,-6250)(11250,-6000)(10750,-5750)
\blacken\path(12750,-5750)(12750,-6250)(13250,-6000)(12750,-5750)
\blacken\path(14750,-5750)(14750,-6250)(15250,-6000)(14750,-5750)
\blacken\path(16750,-5750)(16750,-6250)(17250,-6000)(16750,-5750)
\blacken\path(18750,-5750)(18750,-6250)(19250,-6000)(18750,-5750)

\path(-2000,-4000)(20000,-4000)
\path(-4500,-4000)(-3500,-4000)
\whiten\path(-3750,-3750)(-3750,-4250)(-3500,-4000)(-3750,-3750)
\blacken\path(-750,-3750)(-750,-4250)(-1250,-4000)(-750,-3750)
\blacken\path(18750,-3750)(18750,-4250)(19250,-4000)(18750,-3750)

\path(-2000,-2000)(20000,-2000)
\put(-3250,-2000){\tiny$1$}
\path(-4500,-2000)(-3500,-2000)
\whiten\path(-3750,-1750)(-3750,-2250)(-3500,-2000)(-3750,-1750)
\blacken\path(-750,-1750)(-750,-2250)(-1250,-2000)(-750,-1750)
\blacken\path(18750,-1750)(18750,-2250)(19250,-2000)(18750,-1750)

\put(-8000,5000){\fbox{$v,q$}}

\path(-2000,0)(20000,0)
\put(-3250,0){\tiny$N$}
\path(-4500,0)(-3500,0)
\whiten\path(-3750,250)(-3750,-250)(-3500,0)(-3750,250)
\blacken\path(-1250,250)(-1250,-250)(-750,0)(-1250,250)
\blacken\path(19250,250)(19250,-250)(18750,0)(19250,250)

\path(-2000,2000)(20000,2000)
\path(-4500,2000)(-3500,2000)
\whiten\path(-3750,2250)(-3750,1750)(-3500,2000)(-3750,2250)
\blacken\path(-1250,2250)(-1250,1750)(-750,2000)(-1250,2250)
\blacken\path(19250,2250)(19250,1750)(18750,2000)(19250,2250)

\path(-2000,4000)(20000,4000)
\path(-4500,4000)(-3500,4000)
\whiten\path(-3750,4250)(-3750,3750)(-3500,4000)(-3750,4250)
\blacken\path(-1250,4250)(-1250,3750)(-750,4000)(-1250,4250)
\blacken\path(19250,4250)(19250,3750)(18750,4000)(19250,4250)

\path(-2000,6000)(20000,6000)
\path(-4500,6000)(-3500,6000)
\whiten\path(-3750,6250)(-3750,5750)(-3500,6000)(-3750,6250)
\blacken\path(-1250,6250)(-1250,5750)(-750,6000)(-1250,6250)
\blacken\path(19250,6250)(19250,5750)(18750,6000)(19250,6250)

\path(-2000,8000)(20000,8000)
\path(-4500,8000)(-3500,8000)
\whiten\path(-3750,8250)(-3750,7750)(-3500,8000)(-3750,8250)
\blacken\path(-1250,8250)(-1250,7750)(-750,8000)(-1250,8250)
\blacken\path(19250,8250)(19250,7750)(18750,8000)(19250,8250)

\path(-2000,10000)(20000,10000)
\put(-3250,10000){\tiny$1$}
\path(-4500,10000)(-3500,10000)
\whiten\path(-3750,10250)(-3750,9750)(-3500,10000)(-3750,10250)
\blacken\path(-1250,10250)(-1250,9750)(-750,10000)(-1250,10250)
\blacken\path(19250,10250)(19250,9750)(18750,10000)(19250,10250)


\path(0,-8000)(0,12000)
\put(-200,-9000){\tiny$1$}
\path(0,-10250)(0,-9250)
\whiten\path(-250,-9500)(250,-9500)(0,-9250)(-250,-9500)
\blacken\path(-250,-6750)(250,-6750)(0,-7250)(-250,-6750)
\blacken\path(-250,-4750)(250,-4750)(0,-5250)(-250,-4750)
\blacken\path(-250,10750)(250,10750)(0,11250)(-250,10750)

\path(2000,-8000)(2000,12000)
\path(2000,-10250)(2000,-9250)
\whiten\path(1750,-9500)(2250,-9500)(2000,-9250)(1750,-9500)
\blacken\path(1750,-6750)(2250,-6750)(2000,-7250)(1750,-6750)
\blacken\path(1750,-4750)(2250,-4750)(2000,-5250)(1750,-4750)
\blacken\path(1750,10750)(2250,10750)(2000,11250)(1750,10750)

\path(4000,-8000)(4000,12000)
\put(3800,-9000){\tiny$\widetilde{N}$}
\path(4000,-10250)(4000,-9250)
\whiten\path(3750,-9500)(4250,-9500)(4000,-9250)(3750,-9500)
\blacken\path(3750,-6750)(4250,-6750)(4000,-7250)(3750,-6750)
\blacken\path(3750,-4750)(4250,-4750)(4000,-5250)(3750,-4750)
\blacken\path(3750,10750)(4250,10750)(4000,11250)(3750,10750)

\path(6000,-8000)(6000,12000)
\put(5800,-9000){\tiny$\widetilde{N}+1$}
\path(6000,-10250)(6000,-9250)
\whiten\path(5750,-9500)(6250,-9500)(6000,-9250)(5750,-9500)
\blacken\path(5750,-7250)(6250,-7250)(6000,-6750)(5750,-7250)
\blacken\path(5750,-4750)(6250,-4750)(6000,-5250)(5750,-4750)
\blacken\path(5750,10750)(6250,10750)(6000,11250)(5750,10750)

\path(8000,-8000)(8000,12000)
\path(8000,-10250)(8000,-9250)
\whiten\path(7750,-9500)(8250,-9500)(8000,-9250)(7750,-9500)
\blacken\path(7750,-7250)(8250,-7250)(8000,-6750)(7750,-7250)
\blacken\path(7750,-5250)(8250,-5250)(8000,-4750)(7750,-5250)
\blacken\path(7750,10750)(8250,10750)(8000,11250)(7750,10750)

\path(10000,-8000)(10000,12000)
\path(10000,-10250)(10000,-9250)
\whiten\path(9750,-9500)(10250,-9500)(10000,-9250)(9750,-9500)
\blacken\path(9750,-7250)(10250,-7250)(10000,-6750)(9750,-7250)
\blacken\path(9750,-5250)(10250,-5250)(10000,-4750)(9750,-5250)
\blacken\path(9750,10750)(10250,10750)(10000,11250)(9750,10750)

\path(12000,-8000)(12000,12000)
\path(12000,-10250)(12000,-9250)
\whiten\path(11750,-9500)(12250,-9500)(12000,-9250)(11750,-9500)
\blacken\path(11750,-7250)(12250,-7250)(12000,-6750)(11750,-7250)
\blacken\path(11750,-5250)(12250,-5250)(12000,-4750)(11750,-5250)
\blacken\path(11750,10750)(12250,10750)(12000,11250)(11750,10750)

\path(14000,-8000)(14000,12000)
\path(14000,-10250)(14000,-9250)
\whiten\path(13750,-9500)(14250,-9500)(14000,-9250)(13750,-9500)
\blacken\path(13750,-7250)(14250,-7250)(14000,-6750)(13750,-7250)
\blacken\path(13750,-5250)(14250,-5250)(14000,-4750)(13750,-5250)
\blacken\path(13750,10750)(14250,10750)(14000,11250)(13750,10750)

\path(16000,-8000)(16000,12000)
\path(16000,-10250)(16000,-9250)
\whiten\path(15750,-9500)(16250,-9500)(16000,-9250)(15750,-9500)
\blacken\path(15750,-7250)(16250,-7250)(16000,-6750)(15750,-7250)
\blacken\path(15750,-5250)(16250,-5250)(16000,-4750)(15750,-5250)
\blacken\path(15750,10750)(16250,10750)(16000,11250)(15750,10750)

\path(18000,-8000)(18000,12000)
\put(17800,-9000){\tiny$M$}
\path(18000,-10250)(18000,-9250)
\whiten\path(17750,-9500)(18250,-9500)(18000,-9250)(17750,-9500)
\blacken\path(17750,-7250)(18250,-7250)(18000,-6750)(17750,-7250)
\blacken\path(17750,-5250)(18250,-5250)(18000,-4750)(17750,-5250)
\blacken\path(17750,10750)(18250,10750)(18000,11250)(17750,10750)

\put(20000,-8750){\fbox{$w,r$}}

\end{picture}

\end{minipage}
\end{center}

\caption[Freezing the entire last row of the $S_n$ lattice]{Freezing the entire last row of the $S_n$ lattice. The last row of vertices produces the prefactor in (\ref{Srec0-tf}), while the remainder of the lattice represents $S_{n-1}$.}

\label{snfroze2}
\end{figure}

\noindent From the diagram we see that setting $u_n+p_n = w_{\widetilde{N}+1}+r_{\widetilde{N}+1}$ reduces $S_n$ to $S_{n-1}$, up to a multiplicative factor. This factor is evaluated by matching each vertex in the final row with its trigonometric weight, giving  

{\small
\begin{align}
&
S_n\Big|_{u_n+p_n = w_{\widetilde{N}+1}+r_{\widetilde{N}+1}}
=
\prod_{1\leq j < \widetilde{N}+1}
a_{-}\Big(w_{\widetilde{N}+1}+r_{\widetilde{N}+1}-p_n,p_n,w_j,r_j\Big)
\times
\label{Srec-tf}
\\
&
c_{-}\Big(w_{\widetilde{N}+1}+r_{\widetilde{N}+1}-p_n,p_n,w_{\widetilde{N}+1},r_{\widetilde{N}+1}\Big)
\prod_{\widetilde{N}+1 < j \leq M}
a_{+}\Big(w_{\widetilde{N}+1}+r_{\widetilde{N}+1}-p_n,p_n,w_j,r_j\Big)
S_{n-1}
\nonumber
\end{align}
}

\noindent Using the explicit formulae (\ref{a-tf}) and (\ref{c-tf}) for the functions appearing in (\ref{Srec-tf}), we obtain the required recursion relation (\ref{Srec0-tf}).
}
\end{property7}

\begin{property7}
{\rm
The scalar product $S_0$ is represented by the lattice below.

\begin{figure}[H]

\begin{center}
\begin{minipage}{4.3in}

\setlength{\unitlength}{0.0003cm}
\begin{picture}(20000,15000)(-9000,-3000)

\put(-8000,5000){\fbox{$v,q$}}

\path(-2000,0)(20000,0)
\put(-3250,0){\tiny$N$}
\path(-4500,0)(-3500,0)
\whiten\path(-3750,250)(-3750,-250)(-3500,0)(-3750,250)
\blacken\path(-1250,250)(-1250,-250)(-750,0)(-1250,250)
\blacken\path(11250,250)(11250,-250)(10750,0)(11250,250)
\blacken\path(13250,250)(13250,-250)(12750,0)(13250,250)
\blacken\path(15250,250)(15250,-250)(14750,0)(15250,250)
\blacken\path(17250,250)(17250,-250)(16750,0)(17250,250)
\blacken\path(19250,250)(19250,-250)(18750,0)(19250,250)

\path(-2000,2000)(20000,2000)
\path(-4500,2000)(-3500,2000)
\whiten\path(-3750,2250)(-3750,1750)(-3500,2000)(-3750,2250)
\blacken\path(-1250,2250)(-1250,1750)(-750,2000)(-1250,2250)
\blacken\path(11250,2250)(11250,1750)(10750,2000)(11250,2250)
\blacken\path(13250,2250)(13250,1750)(12750,2000)(13250,2250)
\blacken\path(15250,2250)(15250,1750)(14750,2000)(15250,2250)
\blacken\path(17250,2250)(17250,1750)(16750,2000)(17250,2250)
\blacken\path(19250,2250)(19250,1750)(18750,2000)(19250,2250)

\path(-2000,4000)(20000,4000)
\path(-4500,4000)(-3500,4000)
\whiten\path(-3750,4250)(-3750,3750)(-3500,4000)(-3750,4250)
\blacken\path(-1250,4250)(-1250,3750)(-750,4000)(-1250,4250)
\blacken\path(11250,4250)(11250,3750)(10750,4000)(11250,4250)
\blacken\path(13250,4250)(13250,3750)(12750,4000)(13250,4250)
\blacken\path(15250,4250)(15250,3750)(14750,4000)(15250,4250)
\blacken\path(17250,4250)(17250,3750)(16750,4000)(17250,4250)
\blacken\path(19250,4250)(19250,3750)(18750,4000)(19250,4250)

\path(-2000,6000)(20000,6000)
\path(-4500,6000)(-3500,6000)
\whiten\path(-3750,6250)(-3750,5750)(-3500,6000)(-3750,6250)
\blacken\path(-1250,6250)(-1250,5750)(-750,6000)(-1250,6250)
\blacken\path(11250,6250)(11250,5750)(10750,6000)(11250,6250)
\blacken\path(13250,6250)(13250,5750)(12750,6000)(13250,6250)
\blacken\path(15250,6250)(15250,5750)(14750,6000)(15250,6250)
\blacken\path(17250,6250)(17250,5750)(16750,6000)(17250,6250)
\blacken\path(19250,6250)(19250,5750)(18750,6000)(19250,6250)

\path(-2000,8000)(20000,8000)
\path(-4500,8000)(-3500,8000)
\whiten\path(-3750,8250)(-3750,7750)(-3500,8000)(-3750,8250)
\blacken\path(-1250,8250)(-1250,7750)(-750,8000)(-1250,8250)
\blacken\path(11250,8250)(11250,7750)(10750,8000)(11250,8250)
\blacken\path(13250,8250)(13250,7750)(12750,8000)(13250,8250)
\blacken\path(15250,8250)(15250,7750)(14750,8000)(15250,8250)
\blacken\path(17250,8250)(17250,7750)(16750,8000)(17250,8250)
\blacken\path(19250,8250)(19250,7750)(18750,8000)(19250,8250)

\path(-2000,10000)(20000,10000)
\put(-3250,10000){\tiny$1$}
\path(-4500,10000)(-3500,10000)
\whiten\path(-3750,10250)(-3750,9750)(-3500,10000)(-3750,10250)
\blacken\path(-1250,10250)(-1250,9750)(-750,10000)(-1250,10250)
\blacken\path(11250,10250)(11250,9750)(10750,10000)(11250,10250)
\blacken\path(13250,10250)(13250,9750)(12750,10000)(13250,10250)
\blacken\path(15250,10250)(15250,9750)(14750,10000)(15250,10250)
\blacken\path(17250,10250)(17250,9750)(16750,10000)(17250,10250)
\blacken\path(19250,10250)(19250,9750)(18750,10000)(19250,10250)


\path(0,-2000)(0,12000)
\put(-200,-2750){\tiny$1$}
\path(0,-4250)(0,-3250)
\whiten\path(-250,-3500)(250,-3500)(0,-3250)(-250,-3500)
\blacken\path(-250,-750)(250,-750)(0,-1250)(-250,-750)
\blacken\path(-250,10750)(250,10750)(0,11250)(-250,10750)

\path(2000,-2000)(2000,12000)
\path(2000,-4250)(2000,-3250)
\whiten\path(1750,-3500)(2250,-3500)(2000,-3250)(1750,-3500)
\blacken\path(1750,-750)(2250,-750)(2000,-1250)(1750,-750)
\blacken\path(1750,10750)(2250,10750)(2000,11250)(1750,10750)

\path(4000,-2000)(4000,12000)
\path(4000,-4250)(4000,-3250)
\whiten\path(3750,-3500)(4250,-3500)(4000,-3250)(3750,-3500)
\blacken\path(3750,-750)(4250,-750)(4000,-1250)(3750,-750)
\blacken\path(3750,10750)(4250,10750)(4000,11250)(3750,10750)

\path(6000,-2000)(6000,12000)
\path(6000,-4250)(6000,-3250)
\whiten\path(5750,-3500)(6250,-3500)(6000,-3250)(5750,-3500)
\blacken\path(5750,-750)(6250,-750)(6000,-1250)(5750,-750)
\blacken\path(5750,10750)(6250,10750)(6000,11250)(5750,10750)

\path(8000,-2000)(8000,12000)
\path(8000,-4250)(8000,-3250)
\whiten\path(7750,-3500)(8250,-3500)(8000,-3250)(7750,-3500)
\blacken\path(7750,-750)(8250,-750)(8000,-1250)(7750,-750)
\blacken\path(7750,10750)(8250,10750)(8000,11250)(7750,10750)

\path(10000,-2000)(10000,12000)
\put(9800,-2750){\tiny$N$}
\path(10000,-4250)(10000,-3250)
\whiten\path(9750,-3500)(10250,-3500)(10000,-3250)(9750,-3500)
\blacken\path(9750,-750)(10250,-750)(10000,-1250)(9750,-750)
\blacken\path(9750,10750)(10250,10750)(10000,11250)(9750,10750)

\path(12000,-2000)(12000,12000)
\put(11800,-2750){\tiny$N+1$}
\path(12000,-4250)(12000,-3250)
\whiten\path(11750,-3500)(12250,-3500)(12000,-3250)(11750,-3500)
\blacken\path(11750,-1250)(12250,-1250)(12000,-750)(11750,-1250)
\blacken\path(11750,750)(12250,750)(12000,1250)(11750,750)
\blacken\path(11750,2750)(12250,2750)(12000,3250)(11750,2750)
\blacken\path(11750,4750)(12250,4750)(12000,5250)(11750,4750)
\blacken\path(11750,6750)(12250,6750)(12000,7250)(11750,6750)
\blacken\path(11750,8750)(12250,8750)(12000,9250)(11750,8750)
\blacken\path(11750,10750)(12250,10750)(12000,11250)(11750,10750)

\path(14000,-2000)(14000,12000)
\path(14000,-4250)(14000,-3250)
\whiten\path(13750,-3500)(14250,-3500)(14000,-3250)(13750,-3500)
\blacken\path(13750,-1250)(14250,-1250)(14000,-750)(13750,-1250)
\blacken\path(13750,750)(14250,750)(14000,1250)(13750,750)
\blacken\path(13750,2750)(14250,2750)(14000,3250)(13750,2750)
\blacken\path(13750,4750)(14250,4750)(14000,5250)(13750,4750)
\blacken\path(13750,6750)(14250,6750)(14000,7250)(13750,6750)
\blacken\path(13750,8750)(14250,8750)(14000,9250)(13750,8750)
\blacken\path(13750,10750)(14250,10750)(14000,11250)(13750,10750)

\path(16000,-2000)(16000,12000)
\path(16000,-4250)(16000,-3250)
\whiten\path(15750,-3500)(16250,-3500)(16000,-3250)(15750,-3500)
\blacken\path(15750,-1250)(16250,-1250)(16000,-750)(15750,-1250)
\blacken\path(15750,750)(16250,750)(16000,1250)(15750,750)
\blacken\path(15750,2750)(16250,2750)(16000,3250)(15750,2750)
\blacken\path(15750,4750)(16250,4750)(16000,5250)(15750,4750)
\blacken\path(15750,6750)(16250,6750)(16000,7250)(15750,6750)
\blacken\path(15750,8750)(16250,8750)(16000,9250)(15750,8750)
\blacken\path(15750,10750)(16250,10750)(16000,11250)(15750,10750)

\path(18000,-2000)(18000,12000)
\put(17800,-2750){\tiny$M$}
\path(18000,-4250)(18000,-3250)
\whiten\path(17750,-3500)(18250,-3500)(18000,-3250)(17750,-3500)
\blacken\path(17750,-1250)(18250,-1250)(18000,-750)(17750,-1250)
\blacken\path(17750,750)(18250,750)(18000,1250)(17750,750)
\blacken\path(17750,2750)(18250,2750)(18000,3250)(17750,2750)
\blacken\path(17750,4750)(18250,4750)(18000,5250)(17750,4750)
\blacken\path(17750,6750)(18250,6750)(18000,7250)(17750,6750)
\blacken\path(17750,8750)(18250,8750)(18000,9250)(17750,8750)
\blacken\path(17750,10750)(18250,10750)(18000,11250)(17750,10750)

\put(20000,-2750){\fbox{$w,r$}}

\end{picture}

\end{minipage}
\end{center}

\caption[Frozen vertices within $S_0$]{Frozen vertices within $S_0$. The final $M-N$ columns of vertices produce the prefactor in (\ref{S0-tf}), while the remainder of the lattice represents $Z_N$.}
\end{figure}
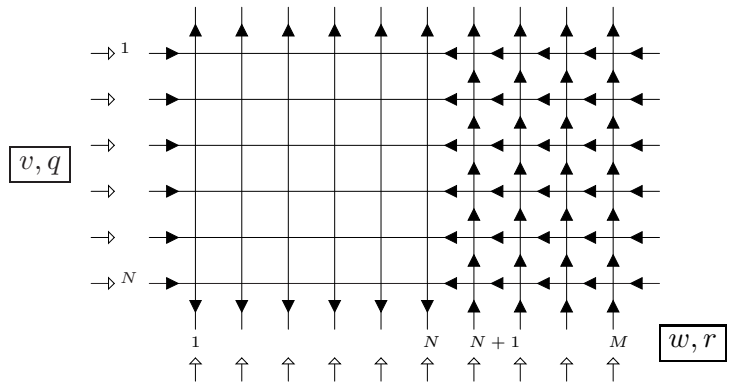

\noindent The vertices in the last $M-N$ columns of a given lattice configuration must be of the form $b_{-}(v_j,q_j,w_k,r_k)$, or else the configuration vanishes. Peeling away this block of frozen vertices, we find that $S_0$ is equal to $Z_N$ up to the overall factor $\prod_{j=1}^{N}\prod_{k=N+1}^{M} b_{-}(v_j,q_j,w_k,r_k)$.
}
\end{property7}

\end{proof}

\subsection{Factorized expression for $S_n(\{u,p\}_n,\{v,q\}_N,\{w,r\}_M)$}

\begin{lemma}
{\rm We shall assume that $(v_j+q_j)$ satisfies the equation

\begin{align}
(-)^N
\prod_{k=1}^{M}
[v_j+q_j-w_k+r_k]
+
\prod_{k=1}^{M}
[v_j+q_j-w_k-r_k]
=
0
\label{tf-bethe}
\end{align}

\noindent for all $1\leq j \leq N$.\footnote{The equations (\ref{tf-bethe}) constitute the Bethe equations for the trigonometric Felderhof model. By this, we mean that the state vector $\ \lprod_{j=1}^{N} B(v_j,q_j,\{w,r\}_M) |\Uparrow_M\rangle$ is an eigenvector of $A(u,p,\{w,r\}_M) + D(u,p,\{w,r\}_M)$ if and only if the equations (\ref{tf-bethe}) are obeyed. We will not prove this fact explicitly, but it can be derived by extending theorem 1 in chapter 2 to models with external fields.} In the presence of this constraint, the scalar product $S_n$ has the factorized expression

{\small
\begin{align}
&
S_n\Big( \{u,p\}_n, \{v,q\}_N, \{w,r\}_M \Big)
=
\prod_{j=1}^{n}
[2p_j]^{\frac{1}{2}}
\prod_{j=1}^{N}
[2q_j]^{\frac{1}{2}}
\prod_{j=1}^{\widetilde{N}}
[2r_j]^{\frac{1}{2}}
\times
\label{Scalc-tf}
\\
 &
\prod_{1 \leq j<k \leq n}
[u_k-u_j+p_j+p_k]
\prod_{1\leq j<k \leq N}
[v_j-v_k+q_j+q_k]
\prod_{1\leq j<k \leq \widetilde{N}}
[w_k-w_j+r_j+r_k]
\times
\nonumber
\\
 &
\prod_{j=1}^{n}
\prod_{k=1}^{\widetilde{N}}
[w_k-u_j+p_j+r_k]
\prod_{j=1}^{N}
\prod_{k=\widetilde{N}+1}^{M}
[v_j-w_k+q_j-r_k]
\times
\nonumber
\\
 &
\prod_{j=1}^{n}
\prod_{k=1}^{N}
\frac{1}{\displaystyle{
[u_j-v_k+p_j-q_k]
}}
\prod_{j=1}^{n}
\left(
(-)^N
\prod_{k=1}^{M}
[u_j-w_k+p_j+r_k]
+
\prod_{k=1}^{M}
[u_j-w_k+p_j-r_k]
\right)
\nonumber
\end{align}
}

}
\end{lemma}

\begin{proof}

We begin by stating that the conditions {\bf 1}--{\bf 4} are uniquely determining.\footnote{This is proved along very similar lines to lemma 6 in the previous chapter.} Hence we need only verify that (\ref{Scalc-tf}) satisfies properties {\bf 1}--{\bf 4}.

\begin{property8}
{\rm
By studying (\ref{Scalc-tf}) we see that $S_n$ has dependence on $\{w_{\widetilde{N}+1},\ldots,w_M\}$ and $\{r_{\widetilde{N}+1},\ldots,r_M\}$ only through the terms

{\footnotesize
\begin{align*}
&
\prod_{j=1}^{N}
\prod_{k=\widetilde{N}+1}^{M}
[v_j-w_k+q_j-r_k],
\quad\quad
\prod_{j=1}^{n}
\left(
(-)^N
\prod_{k=1}^{M}
[u_j-w_k+p_j+r_k]
+
\prod_{k=1}^{M}
[u_j-w_k+p_j-r_k]
\right)
\nonumber
\end{align*}
}

\noindent Both of these terms are invariant under the permutation $\{w_j,r_j\} \leftrightarrow \{w_k,r_k\}$ for all $j,k \in \{\widetilde{N}+1,\ldots,M\}$. 

}
\end{property8}

\begin{property8}
{\rm
Since $(v_j+q_j)$ is a root of the equation (\ref{tf-bethe}) for all $1\leq j \leq N$, it follows that

{\small
\begin{align*}
\prod_{j=1}^{n}
\prod_{k=1}^{N}
\frac{1}{[u_j-v_k+p_j-q_k]}
\prod_{j=1}^{n}
\left(
(-)^N
\prod_{k=1}^{M}
[u_j-w_k+p_j+r_k]
+
\prod_{k=1}^{M}
[u_j-w_k+p_j-r_k]
\right)
\end{align*}
}

\noindent is a trigonometric polynomial in $u_n$ of degree $M-N$. The remaining terms in (\ref{Scalc-tf}) comprise a trigonometric polynomial of degree $N-1$ in $u_n$. Therefore the entire expression (\ref{Scalc-tf}) is a trigonometric polynomial of degree $M-1$ in $u_n$. In addition, the required factor $\prod_{j=1}^{\widetilde{N}}[w_j-u_n+p_n+r_j]$ is present in (\ref{Scalc-tf}).

}
\end{property8}

\begin{property8}
{\rm
The recursion relation (\ref{Srec0-tf}) is proved by setting $u_n+p_n=w_{\widetilde{N}+1}+r_{\widetilde{N}+1}$ in (\ref{Scalc-tf}) and rearranging the factors in the resulting equation. Since this procedure is elementary in nature, we shall omit the details. 
}
\end{property8}

\begin{property8}
{\rm
Setting $n=0$ in (\ref{Scalc-tf}) gives

\begin{align}
&
S_0\Big(\{v,q\}_N,\{w,r\}_M\Big)
=
\prod_{j=1}^{N} \prod_{k=N+1}^{M}
[v_j-w_k+q_j-r_k]
\times
\label{S0proof-tf}
\\
&
\prod_{j=1}^{N} [2q_j]^{\frac{1}{2}} [2r_j]^{\frac{1}{2}}
\prod_{1\leq j < k \leq N} [v_j-v_k+q_j+q_k] [w_k-w_j+r_j+r_k]
\nonumber
\end{align}

\noindent Comparing equation (\ref{S0proof-tf}) with the factorized expression (\ref{lem-tf}) for the domain wall partition function, we verify (\ref{S0-tf}).
}
\end{property8}

\end{proof}

\subsection{Evaluation of $S_N(\{u,p\}_N,\{v,q\}_N,\{w,r\}_M)$}

For completeness, we write the $n=N$ case of equation (\ref{Scalc-tf}) explicitly. We have

{\small
\begin{align}
&
S_N\Big(
\{u,p\}_N,\{v,q\}_N,\{w,r\}_M
\Big)
=
\prod_{j=1}^{N}
[2p_j]^{\frac{1}{2}} [2q_j]^{\frac{1}{2}}
\times
\\
&
\prod_{1\leq j<k \leq N}
[u_k-u_j+p_j+p_k]
[v_j-v_k+q_j+q_k]
\prod_{j=1}^{N} \prod_{k=1}^{M}
[v_j-w_k+q_j-r_k]
\times
\nonumber
\\
&
\prod_{j,k=1}^{N}
\frac{1}{[u_j-v_k+p_j-q_k]}
\prod_{j=1}^{N}
\left(
(-)^N \prod_{k=1}^{M} [u_j-w_k+p_j+r_k]
+
\prod_{k=1}^{M} [u_j-w_k+p_j-r_k]
\right)
\nonumber
\end{align}
}

\noindent Specializing the external fields to $p_j = q_j = \frac{\pi i}{4}$, $r_k = \frac{\pi i}{4}$ for all $1\leq j \leq N$, $1\leq k \leq M$ gives the XXZ Bethe scalar product at the free fermion point (\ref{scal-ff3}).

\section{Free fermion point of SOS model}
\label{ff-sos}

\subsection{Jacobi theta functions}

We begin by presenting some basic theory on elliptic functions, taken from chapter 15 of \cite{bax1}. Up to overall normalization and scaling of the variable $u$, the Jacobi theta functions $H(u),H_1(u)$ are given by 

\begin{align}
H(u) 
&= 
2  \sinh  u
\prod_{n=1}^{\infty}
\left( 
1-2\nome^{2n}\cosh (2 u)+\nome^{4n}
\right) 
\left( 
1-\nome^{2n} 
\right)
\label{jac-ell}
\\
H_1(u)
&=
2 i \cosh u 
\prod_{n=1}^{\infty}
\left( 
1+2\nome^{2n}\cosh (2 u)+\nome^{4n}
\right) 
\left( 
1-\nome^{2n} 
\right)
\label{jac-ell3}
\end{align}

\noindent where $0 < \nome < 1$ is the elliptic nome. Taking the limit $\nome \rightarrow 0$, these functions collapse to their trigonometric versions, $ 2 \sinh u$ and $2 i \cosh u $ respectively. Henceforth, we will find it convenient to define

\begin{align}
[u]
=
H\left(\frac{\pi i u}{2}\right),
\quad\quad
\langle u \rangle
=
H_1\left( \frac{\pi i u}{2} \right)
=
[u+1]
\label{jac-ell2}
\end{align}

\noindent which should not be confused with the previous usage of this notation, in the first part of the chapter. The theta function $[u]$ is entire and has the quasi-periodicity properties



\begin{align}
[u+2] = -[u],
\quad 
[u-2i\log(\nome)/\pi] 
=
-\frac{1}{\nome} \exp(-\pi i u) [u] 
\label{quasip}
\end{align}

\noindent From these properties we obtain the following result, which will be used later in the chapter.

\begin{theorem}
{\rm 
If $f(u)$ is an entire function that 
satisfies the quasi-periodicity conditions

\begin{align}
f(u+2)
&=
(-)^N f(u),
\quad
f(u-2i\log(\nome)/\pi)
=
\frac{(-)^N}{\nome^N}
\exp\Big( {\pi i}(\eta- N u) \Big)
f(u)
\end{align}

\noindent for some constant $\eta$, then 

\begin{align}
f(u)
=
\mathcal{C}
\left(
\prod_{j=1}^{N-1}
[u-\zeta_j]
\right)
[u-\eta+\sum_{j=1}^{N-1}\zeta_j ]
\end{align}

\noindent where $\mathcal{C}$ and $\zeta_1,\ldots,\zeta_{N-1}$ are suitably chosen constants.
}
\end{theorem}

\begin{proof} 
Choose a period rectangle $R$ such that 
$f(u)$ has no zeros on the boundary $\partial R$, and integrate 
$\frac{f'(u)}{f(u)}$ around the anti-clockwise contour formed by 
$\partial R$. From the quasi-periodicity conditions it follows 
that

\begin{align}
\oint_{\partial R} \frac{f'(u)}{f(u)}du=2\pi i N
\end{align}

\noindent Hence the sum of residues of $\frac{f'(u)}{f(u)}$ in $R$ is equal to $N$, proving that $f(u)$ has exactly $N$ zeros in $R$, if we count a zero of order $n$ with multiplicity $n$. Write the zeros as $\zeta_1,\ldots,\zeta_N$, and define the function 
$\phi(u)=\prod_{j=1}^{N}[u-\zeta_j]$. By construction $\frac{f'(u)}{f(u)}- \frac{\phi'(u)}{\phi(u)}$ is doubly periodic and holomorphic, and therefore 

\begin{align}
\frac{f'(u)}{f(u)}-\frac{\phi'(u)}{\phi(u)}
=
\frac{d}{du} \log 
\frac{f(u)}{\phi(u)} 
= 
\kappa
\end{align}

\noindent where $\kappa$ is a constant. Integrating, we obtain 
$f(u)= \mathcal{C} e^{\kappa u}\prod_{j=1}^{N}[u-\zeta_j]$. Finally, using the quasi-periodicity properties of $f(u)$ we obtain $\kappa=0$ and $\eta=\sum_{j=1}^{N}\zeta_j$, which 
concludes the proof.
\end{proof}

\subsection{$R$-matrix and dynamical Yang-Baxter equation}

The eight-vertex solid-on-solid (SOS) model was introduced by Baxter in \cite{bax4}, using the vertex/interaction-round-a-face(IRF) correspondence. The main feature distinguishing this model from its vertex counterpart is the {\it dynamical} or {\it height parameter} appearing in its weights, as we describe below. In the parametrization of \cite{djmo}, the $R$-matrix for the SOS model is given by

\begin{align}
R_{ab}(u,v,h)
=
\left(
\begin{array}{cccc}
a_{+}(u,v) & 0 & 0 & 0
\\
0 & b_{+}(u,v,h) & c_{+}(u,v,h) & 0
\\
0 & c_{-}(u,v,h) & b_{-}(u,v,h) & 0
\\
0 & 0 & 0 & a_{-}(u,v)
\end{array}
\right)_{ab}
\label{Rmat-sos}
\end{align}

\noindent where we have defined the weights

\begin{align}
a_{\pm}(u,v) &= H(\gamma(u-v+1))
\label{a-sos}
\\
b_{\pm}(u,v,h) &= 
\pm i
\frac{H(\gamma(\xi+h\mp 1))}{H(\gamma(\xi+h))}
H(\gamma(u-v))
\label{b-sos}
\\
c_{\pm}(u,v,h) &= \frac{H(\gamma(\xi+h\pm v\mp u))}{H(\gamma(\xi+h))}H(\gamma)
\label{c-sos}
\end{align}

\noindent with $H(u)$ given by (\ref{jac-ell}), $\gamma$ denoting the crossing parameter, $h$ the height variable, and $\xi$ an arbitrary extra parameter which we shall hereafter fix to $\xi = 1$. The factors of $\pm i$ appearing in the $b_{\pm}$ weights constitute a trivial gauge transformation which does not affect the Yang-Baxter equation. We have introduced these factors to unify this model with the one discussed in the next section.

\begin{figure}[H]
\begin{center}
\begin{minipage}{4.3in}

\setlength{\unitlength}{0.0004cm}
\begin{picture}(20000,17500)(-3500,-15000)

\put(-2000,2000){\circle*{300}}
\put(-2750,2000){\scriptsize$h$}
\put(2000,2000){\circle*{300}}
\put(2250,2000){\scriptsize$h+1$}
\put(-2000,-2000){\circle*{300}}
\put(-4250,-2000){\scriptsize$h+1$}
\put(2000,-2000){\circle*{300}}
\put(2250,-2000){\scriptsize$h+2$}

\path(-2000,0000)(2000,0000)
\blacken\path(-2000,250)(-2000,-250)(-1500,0)(-2000,250)
\blacken\path(2000,250)(2000,-250)(2500,0)(2000,250)
\path(-2000,-2000)(2000,-2000)
\path(-2000,2000)(2000,2000)
\put(-3250,0){$u$}
\path(-4500,0)(-3500,0)
\whiten\path(-4000,250)(-4000,-250)(-3500,0)(-4000,250)
\path(0000,-2000)(0000,2000)
\blacken\path(-250,-2000)(250,-2000)(0,-1500)(-250,-2000)
\blacken\path(-250,2000)(250,2000)(0,2500)(-250,2000)
\path(-2000,-2000)(-2000,2000)
\path(2000,-2000)(2000,2000)
\put(-1800,-5500){$a_{+}(u,v)$}
\put(0,-3250){$v$}
\path(0,-4500)(0,-3500)
\whiten\path(-250,-4000)(250,-4000)(0,-3500)(-250,-4000)

\put(8000,2000){\circle*{300}}
\put(7250,2000){\scriptsize$h$}
\put(12000,2000){\circle*{300}}
\put(12250,2000){\scriptsize$h-1$}
\put(8000,-2000){\circle*{300}}
\put(5750,-2000){\scriptsize$h+1$}
\put(12000,-2000){\circle*{300}}
\put(12250,-2000){\scriptsize$h$}

\path(8000,0000)(12000,0000)
\blacken\path(8000,250)(8000,-250)(8500,0)(8000,250)
\blacken\path(12000,250)(12000,-250)(12500,0)(12000,250)
\path(8000,-2000)(12000,-2000)
\path(8000,2000)(12000,2000)
\put(6750,0){$u$}
\path(5500,0)(6500,0)
\whiten\path(6000,250)(6000,-250)(6500,0)(6000,250)
\path(10000,-2000)(10000,2000)
\blacken\path(9750,-2000)(10250,-2000)(10000,-2500)(9750,-2000)
\blacken\path(9750,2000)(10250,2000)(10000,1500)(9750,2000)
\path(8000,-2000)(8000,2000)
\path(12000,-2000)(12000,2000)
\put(7900,-5500){$b_{+}(u,v,h)$}
\put(10000,-3250){$v$}
\path(10000,-4500)(10000,-3500)
\whiten\path(9750,-4000)(10250,-4000)(10000,-3500)(9750,-4000)

\put(18000,2000){\circle*{300}}
\put(17250,2000){\scriptsize$h$}
\put(22000,2000){\circle*{300}}
\put(22250,2000){\scriptsize$h+1$}
\put(18000,-2000){\circle*{300}}
\put(15750,-2000){\scriptsize$h+1$}
\put(22000,-2000){\circle*{300}}
\put(22250,-2000){\scriptsize$h$}

\path(18000,0000)(22000,0000)
\blacken\path(18000,250)(18000,-250)(18500,0)(18000,250)
\blacken\path(22000,250)(22000,-250)(21500,0)(22000,250)
\path(18000,-2000)(22000,-2000)
\path(18000,2000)(22000,2000)
\put(16750,0){$u$}
\path(15500,0)(16500,0)
\whiten\path(16000,250)(16000,-250)(16500,0)(16000,250)
\path(20000,-2000)(20000,2000)
\blacken\path(19750,-2000)(20250,-2000)(20000,-2500)(19750,-2000)
\blacken\path(19750,2000)(20250,2000)(20000,2500)(19750,2000)
\path(18000,-2000)(18000,2000)
\path(22000,-2000)(22000,2000)
\put(17900,-5500){$c_{+}(u,v,h)$}
\put(20000,-3250){$v$}
\path(20000,-4500)(20000,-3500)
\whiten\path(19750,-4000)(20250,-4000)(20000,-3500)(19750,-4000)

\put(-2000,-8000){\circle*{300}}
\put(-2750,-8000){\scriptsize$h$}
\put(2000,-8000){\circle*{300}}
\put(2250,-8000){\scriptsize$h-1$}
\put(-2000,-12000){\circle*{300}}
\put(-4250,-12000){\scriptsize$h-1$}
\put(2000,-12000){\circle*{300}}
\put(2250,-12000){\scriptsize$h-2$}

\path(-2000,-10000)(2000,-10000)
\blacken\path(-2000,-9750)(-2000,-10250)(-2500,-10000)(-2000,-9750)
\blacken\path(2000,-9750)(2000,-10250)(1500,-10000)(2000,-9750)
\path(-2000,-12000)(2000,-12000)
\path(-2000,-8000)(2000,-8000)
\put(-3250,-10000){$u$}
\path(-4500,-10000)(-3500,-10000)
\whiten\path(-4000,-9750)(-4000,-10250)(-3500,-10000)(-4000,-9750)
\path(0000,-12000)(0000,-8000)
\blacken\path(-250,-12000)(250,-12000)(0,-12500)(-250,-12000)
\blacken\path(-250,-8000)(250,-8000)(0,-8500)(-250,-8000)
\path(-2000,-12000)(-2000,-8000)
\path(2000,-12000)(2000,-8000)
\put(-1800,-15500){$a_{-}(u,v)$}
\put(0,-13250){$v$}
\path(0,-14500)(0,-13500)
\whiten\path(-250,-14000)(250,-14000)(0,-13500)(-250,-14000)

\put(8000,-8000){\circle*{300}}
\put(7250,-8000){\scriptsize$h$}
\put(12000,-8000){\circle*{300}}
\put(12250,-8000){\scriptsize$h+1$}
\put(8000,-12000){\circle*{300}}
\put(5750,-12000){\scriptsize$h-1$}
\put(12000,-12000){\circle*{300}}
\put(12250,-12000){\scriptsize$h$}

\path(8000,-10000)(12000,-10000)
\blacken\path(8000,-9750)(8000,-10250)(7500,-10000)(8000,-9750)
\blacken\path(12000,-9750)(12000,-10250)(11500,-10000)(12000,-9750)
\path(8000,-12000)(12000,-12000)
\path(8000,-8000)(12000,-8000)
\put(6750,-10000){$u$}
\path(5500,-10000)(6500,-10000)
\whiten\path(6000,-9750)(6000,-10250)(6500,-10000)(6000,-9750)
\path(10000,-12000)(10000,-8000)
\blacken\path(9750,-12000)(10250,-12000)(10000,-11500)(9750,-12000)
\blacken\path(9750,-8000)(10250,-8000)(10000,-7500)(9750,-8000)
\path(8000,-12000)(8000,-8000)
\path(12000,-12000)(12000,-8000)
\put(7900,-15500){$b_{-}(u,v,h)$}
\put(10000,-13250){$v$}
\path(10000,-14500)(10000,-13500)
\whiten\path(9750,-14000)(10250,-14000)(10000,-13500)(9750,-14000)

\put(18000,-8000){\circle*{300}}
\put(17250,-8000){\scriptsize$h$}
\put(22000,-8000){\circle*{300}}
\put(22250,-8000){\scriptsize$h-1$}
\put(18000,-12000){\circle*{300}}
\put(15750,-12000){\scriptsize$h-1$}
\put(22000,-12000){\circle*{300}}
\put(22250,-12000){\scriptsize$h$}

\path(18000,-10000)(22000,-10000)
\blacken\path(18000,-9750)(18000,-10250)(17500,-10000)(18000,-9750)
\blacken\path(22000,-9750)(22000,-10250)(22500,-10000)(22000,-9750)
\path(18000,-12000)(22000,-12000)
\path(18000,-8000)(22000,-8000)
\put(16750,-10000){$u$}
\path(15500,-10000)(16500,-10000)
\whiten\path(16000,-9750)(16000,-10250)(16500,-10000)(16000,-9750)
\path(20000,-12000)(20000,-8000)
\blacken\path(19750,-12000)(20250,-12000)(20000,-11500)(19750,-12000)
\blacken\path(19750,-8000)(20250,-8000)(20000,-8500)(19750,-8000)
\path(18000,-12000)(18000,-8000)
\path(22000,-12000)(22000,-8000)
\put(17900,-15500){$c_{-}(u,v,h)$}
\put(20000,-13250){$v$}
\path(20000,-14500)(20000,-13500)
\whiten\path(19750,-14000)(20250,-14000)(20000,-13500)(19750,-14000)

\end{picture}

\end{minipage}
\end{center}

\caption[Weights of the SOS model]{Weights of the SOS model. Each entry of the $R$-matrix (\ref{Rmat-sos}) is paired with a {\it face.} The faces contain all the information normally present in a vertex, plus four surrounding points with prescribed dynamical variables. The dynamical variable in the top-left corner matches that of the $R$-matrix itself. The values at the remaining corners are obtained by circuiting the face clockwise from the top-left, adding 1 to the previous value each time a black arrow points outward, and subtracting 1 from the previous value each time a black arrow points inward.}

\end{figure}
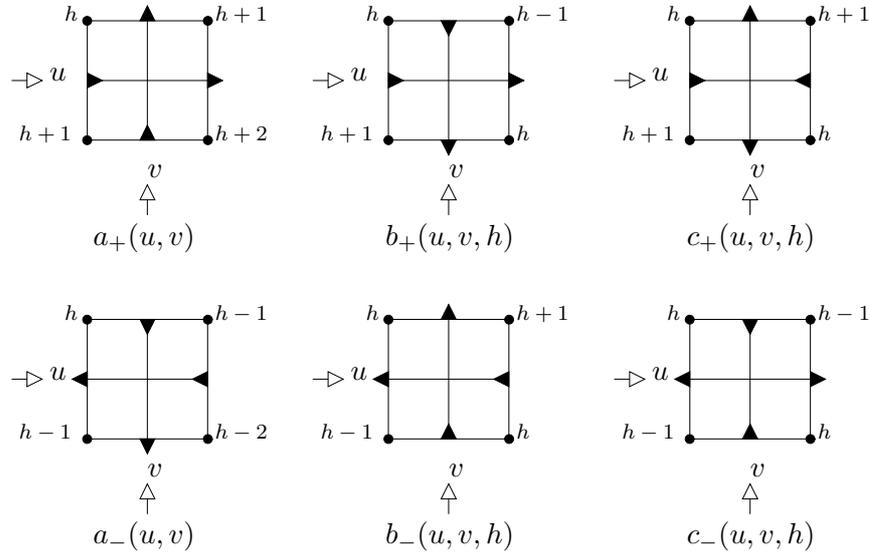

\noindent Often, when the variables of a particular $R$-matrix are clear from context, we will abbreviate $R_{ab}(u,v,h) = R_{ab}(h)$. As usual, we have placed the subscript $ab$ on the $R$-matrix to denote the fact that it is an element of ${\rm End}(\mathcal{V}_a\otimes\mathcal{V}_b)$. Let 

\begin{align}
e_a^{+} = \uparrow_a,
\quad 
e_a^{-} = \downarrow_a
\end{align} 

\noindent be the canonical basis vectors of the two dimensional vector space $\mathcal{V}_a$. The {\it dynamical} $R$-matrix $R_{ab}(h+\sigma_c^{z})$ is an element of ${\rm End}(\mathcal{V}_a\otimes\mathcal{V}_b\otimes\mathcal{V}_c)$, whose action is given by

\begin{align}
R_{ab}(h+\sigma_c^{z})
(
e_a^{i} \otimes e_b^{j} \otimes e_c^{k}
)
=
R_{ab}(h+k)
(
e_a^i \otimes e_b^j
)
\otimes
e_c^k 
\end{align}

\noindent where $\sigma_c^{z}$ denotes the third of the Pauli matrices (\ref{pauli}) acting in $\mathcal{V}_c$, and $i,j,k \in \{+1,-1\}$ are fixed indices. 

\begin{lemma}
{\rm 
Associate the variables $u,v,w$ to the respective vector spaces $\mathcal{V}_a,\mathcal{V}_b,\mathcal{V}_c$ so that, for example, $R_{ab}(h)$ is understood to equal $R_{ab}(u,v,h)$. The {\it dynamical Yang-Baxter equation} is the identity

\begin{align}
R_{ab}(h) R_{ac}(h+\sigma_b^{z}) R_{bc}(h)
=
R_{bc}(h+\sigma_a^{z}) R_{ac}(h) 
R_{ab}(h+\sigma_c^{z})
\label{yb-sos}
\end{align}

\noindent which holds in ${\rm End}(\mathcal{V}_a\otimes\mathcal{V}_b\otimes\mathcal{V}_c)$.\footnote{See \cite{kit}.} 
}
\end{lemma}

\begin{proof}
We act with the equation (\ref{yb-sos}) on the vector $e_a^{j_a}\otimes e_b^{j_b} \otimes e_c^{j_c}$, where $j_a,j_b,j_c \in \{+1,-1\}$ are fixed indices. We obtain

{\footnotesize
\begin{align*}
R_{ab}(h) R_{ac}(h+\sigma_b^{z}) R_{bc}(h)
e_a^{j_a}\otimes e_b^{j_b} \otimes e_c^{j_c}
=
&
R_{ab}(h) R_{ac}(h+\sigma_b^{z})
R^{k_b j_b}_{k_c j_c}(h)
e_a^{j_a} \otimes e_b^{k_b} \otimes e_c^{k_c}
\\
=
&
R_{ab}(h)
R^{k_a j_a}_{i_c k_c}(h+k_b)
R^{k_b j_b}_{k_c j_c}(h)
e_a^{k_a} \otimes e_b^{k_b} \otimes e_c^{i_c}
\\
=
&
R^{i_a k_a}_{i_b k_b}(h)
R^{k_a j_a}_{i_c k_c}(h+k_b)
R^{k_b j_b}_{k_c j_c}(h)
e_a^{i_a} \otimes e_b^{i_b} \otimes e_c^{i_c} 
\end{align*}
} 

\noindent for the left hand side of (\ref{yb-sos}), and 

{\footnotesize
\begin{align*}
R_{bc}(h+\sigma_a^{z}) 
R_{ac}(h) 
R_{ab}(h+\sigma_c^{z})
e_a^{j_a} \otimes e_b^{j_b} \otimes e_c^{j_c}
=
&
R_{bc}(h+\sigma_a^{z}) R_{ac}(h)
R^{k_a j_a}_{k_b j_b}(h+j_c)
e_a^{k_a} \otimes e_b^{k_b} \otimes e_c^{j_c}
\\
=
&
R_{bc}(h+\sigma_a^{z})
R^{i_a k_a}_{k_c j_c}(h)
R^{k_a j_a}_{k_b j_b}(h+j_c)
e_a^{i_a} \otimes e_b^{k_b} \otimes e_c^{k_c}
\\
=
&
R^{i_b k_b}_{i_c k_c}(h+i_a)
R^{i_a k_a}_{k_c j_c}(h)
R^{k_a j_a}_{k_b j_b}(h+j_c)
e_a^{i_a} \otimes e_b^{i_b} \otimes e_c^{i_c}
\end{align*}
}

\noindent for the right hand side of (\ref{yb-sos}), where in both cases all repeated indices are summed over $\{+1,-1\}$. Equating these two sides again, we have

\begin{align}
R^{i_a k_a}_{i_b k_b}(h)
R^{k_a j_a}_{i_c k_c}(h+k_b)
R^{k_b j_b}_{k_c j_c}(h)
=
R^{i_b k_b}_{i_c k_c}(h+i_a)
R^{i_a k_a}_{k_c j_c}(h)
R^{k_a j_a}_{k_b j_b}(h+j_c)
\label{SOS-YB}
\end{align}

\noindent for all fixed indices $\{i_a,i_b,i_c,j_a,j_b,j_c\} \in \{+1,-1\}$, where summation is implied over $k_a,k_b,k_c$. Equation (\ref{SOS-YB}) matches the Yang-Baxter equation listed in \cite{djkmo} and can be verified directly using the entries of the $R$-matrix (\ref{Rmat-sos}).
\end{proof}

\begin{figure}[H]
\begin{center}
\begin{minipage}{4.3in}

\setlength{\unitlength}{0.00031cm}
\begin{picture}(20000,12500)(-1000,-3500)

\put(-250,-750){\tiny$i_b$}

\put(-250,4250){\tiny$i_a$}

\put(3750,0){\tiny$k_a$}

\put(3750,5000){\tiny$k_b$}

\put(10350,0){\tiny$j_a$}

\put(10250,5000){\tiny$j_b$}

\put(4000,-2250){\tiny$i_c$}

\put(10000,2750){\tiny$k_c$}

\put(5000,8000){\tiny$j_c$}

\put(-2500,2500){\circle*{300}}

\put(2500,7500){\circle*{300}}
\put(2500,7750){\tiny$h$}

\put(2500,-2500){\circle*{300}}

\put(7500,-2500){\circle*{300}}

\put(7500,2500){\circle*{300}}
\put(4600,2400){\tiny$h+k_b$}

\put(7500,7500){\circle*{300}}

\put(12500,2500){\circle*{300}}

\path(2500,7500)(7500,2500)
\path(-2500,2500)(2500,7500)
\path(2500,-2500)(-2500,2500)
\path(2500,-2500)(7500,2500)

\path(2500,-2500)(7500,-2500)
\path(7500,-2500)(12500,2500)
\path(7500,2500)(12500,2500)
\path(2500,7500)(7500,7500)
\path(7500,7500)(12500,2500)


\path(0,0)(5000,5000)
\path(-2350,0)(-1350,0)
\whiten\path(-1600,250)(-1600,-250)(-1350,0)(-1600,250)
\put(-1100,0){\scriptsize$v$}
\path(5000,0)(0,5000)
\path(-2350,5000)(-1350,5000)
\whiten\path(-1600,5250)(-1600,4750)(-1350,5000)(-1600,5250)
\put(-1100,5000){\scriptsize$u$}
\path(5000,5000)(10000,5000)
\path(5000,0)(10000,0)
\path(5000,-2500)(10000,2500)
\path(10000,2500)(5000,7500)
\put(5000,-3250){\scriptsize$w$}
\path(5000,-4500)(5000,-3500)
\whiten\path(4750,-3750)(5250,-3750)(5000,-3500)(4750,-3750)


\put(16000,2250){$=$}


\put(23750,-750){\tiny$i_b$}

\put(24000,4400){\tiny$i_a$}

\put(28750,750){\tiny$k_b$}

\put(28750,4000){\tiny$k_a$}

\put(34250,-250){\tiny$j_a$}

\put(34000,3900){\tiny$j_b$}

\put(29000,-2200){\tiny$i_c$}

\put(23500,1900){\tiny$k_c$}

\put(29000,7800){\tiny$j_c$}

\put(21500,2500){\circle*{300}}
\put(18600,2500){\tiny$h+i_a$}

\put(26500,7500){\circle*{300}}
\put(26500,7750){\tiny$h$}

\put(26500,-2500){\circle*{300}}

\put(31500,-2500){\circle*{300}}

\put(31500,7500){\circle*{300}}
\put(31750,7500){\tiny$h+j_c$}

\put(36500,2500){\circle*{300}}

\put(26500,2500){\circle*{300}}

\path(21500,2500)(26500,7500)
\path(21500,2500)(26500,-2500)
\path(26500,-2500)(31500,-2500)
\path(31500,-2500)(36500,2500)
\path(36500,2500)(31500,7500)
\path(26500,7500)(31500,7500)

\path(21500,2500)(26500,2500)
\path(26500,2500)(31500,7500)
\path(26500,2500)(31500,-2500)

\path(21500,5000)(22500,5000)
\whiten\path(22250,5250)(22250,4750)(22500,5000)(22250,5250)
\put(22750,5000){\scriptsize$u$}
\path(24000,5000)(29000,5000)
\path(29000,5000)(34000,0)
\path(21500,0)(22500,0)
\whiten\path(22250,250)(22250,-250)(22500,0)(22250,250)
\put(22750,0){\scriptsize$v$}
\path(24000,0)(29000,0)
\path(29000,0)(34000,5000)


\path(29000,-2500)(24000,2500)
\path(24000,2500)(29000,7500)

\put(29000,-3250){\scriptsize$w$}
\path(29000,-4500)(29000,-3500)
\whiten\path(28750,-3750)(29250,-3750)(29000,-3500)(28750,-3750)

\end{picture}

\end{minipage}
\end{center}

\caption[Yang-Baxter equation for the SOS model]{Yang-Baxter equation for the SOS model. The external indices $\{i_a,i_b,i_c,j_a,j_b,j_c\}$ are held fixed on both sides of the equation, while $\{k_a,k_b,k_c\}$ are summed over $\{+1,-1\}$. This figure is the graphical equivalent of (\ref{SOS-YB}), as can be seen by matching each face with its corresponding $R$-matrix entry.}

\end{figure}
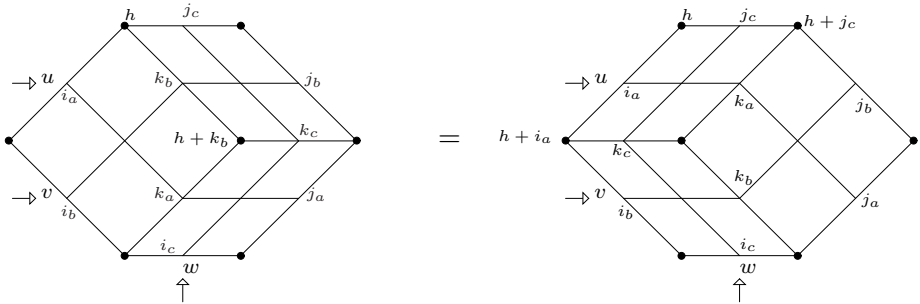

\begin{remark}
{\rm
Consider the case when the crossing parameter is set to $\gamma = \pi i /2$. From the form of the $R$-matrix weights (\ref{a-sos})--(\ref{c-sos}), the fact that $[u] = H(\pi i u/2)$ and the first of the quasi-periodicity conditions (\ref{quasip}), we see that

\begin{align}
R_{ab}(u,v,h+1)
=
R_{ab}(u,v,h-1)
\end{align}

\noindent at this special value of $\gamma$. This means that the action of the dynamical $R$-matrix trivializes, because

\begin{align}
R_{ab}(h+\sigma_c^{z})
(e_a^i\otimes e_b^j \otimes e_c^k)
=
R_{ab}(h+1) (e_a^i \otimes e_b^j)\otimes e_c^k
\label{actff-sos}
\end{align}

\noindent regardless of the value of $k \in \{+1,-1\}$. Therefore when $\gamma = \pi i /2$, the Yang-Baxter equation (\ref{yb-sos}) becomes

\begin{align}
R_{ab}(h) R_{ac}(h+1) R_{bc}(h)
=
R_{bc}(h+1) R_{ac}(h) R_{ab}(h+1)
\label{ffyb-sos}
\end{align}

\noindent which is the form more typical of a vertex model. In the next section we will consider a model whose $R$-matrix satisfies a more general version of (\ref{ffyb-sos}). 
}
\end{remark}

\subsection{Important commutation relation}

\begin{lemma}
{\rm We have the commutation relation 

\begin{align}
R_{ab}(u_a,v_b,h+\sigma_c^{z}+\sigma_d^{z})
&
R_{cd}(u_c,v_d,h)
=
R_{cd}(u_c,v_d,h)
R_{ab}(u_a,v_b,h+\sigma_c^{z}+\sigma_d^{z})
\label{com-sos}
\end{align}

\noindent in ${\rm End}(\mathcal{V}_a\otimes\mathcal{V}_b\otimes\mathcal{V}_c\otimes\mathcal{V}_d)$.
}
\end{lemma}

\begin{proof} 
We act with (\ref{com-sos}) on the vector $e_a^{j_a} \otimes e_b^{j_b} \otimes e_c^{j_c} \otimes e_d^{j_d}$, where $j_a,j_b,j_c,j_d \in \{+1,-1\}$ are fixed indices. We obtain

{\footnotesize
\begin{align*}
R_{ab}(h+\sigma_c^{z}+\sigma_d^{z})
R_{cd}(h)
e_a^{j_a} \otimes e_b^{j_b} \otimes e_c^{j_c} \otimes e_d^{j_d}
=
&
R_{ab}(h+\sigma_c^{z}+\sigma_d^{z})
R^{k_c j_c}_{k_d j_d}(h)
e_a^{j_a} \otimes e_b^{j_b} \otimes e_c^{k_c} \otimes e_d^{k_d}
\\
=
&
R^{k_a j_a}_{k_b j_b}(h+k_c+k_d)
R^{k_c j_c}_{k_d j_d}(h)
e_a^{k_a} \otimes e_b^{k_b} \otimes e_c^{k_c} \otimes e_d^{k_d}
\end{align*}
}

\noindent for the left hand side of (\ref{com-sos}), and

{\footnotesize
\begin{align*}
R_{cd}(h)
R_{ab}(h+\sigma_c^{z}+\sigma_d^{z})
e_a^{j_a} \otimes e_b^{j_b} \otimes e_c^{j_c} \otimes e_d^{j_d}
=
&
R_{cd}(h)
R^{k_a j_a}_{k_b j_b}(h+j_c+j_d)
e_a^{k_a} \otimes e_b^{k_b} \otimes e_c^{j_c} \otimes e_d^{j_d}
\\
=
&
R^{k_c j_c}_{k_d j_d}(h)
R^{k_a j_a}_{k_b j_b}(h+j_c+j_d)
e_a^{k_a} \otimes e_b^{k_b} \otimes e_c^{k_c} \otimes e_d^{k_d}
\end{align*}
}

\noindent for the right hand side of (\ref{com-sos}), with all repeated indices summed over $\{+1,-1\}$. These two sides are equal since $R^{k_c j_c}_{k_d j_d}(h)$ is zero unless $k_c+k_d = j_c+j_d$.
\end{proof}

\subsection{Dynamical monodromy matrix}

The dynamical monodromy matrix is defined as

\begin{align}
T_{a}(u,\{w\}_M,h)
&=
R_{a1}(u,w_1,h)
\ldots
R_{aM}\Big(u,w_M,h+\sum_{j=1}^{M-1} \sigma_j^{z}\Big)
\label{mon-sos}
\\
&
=
\rprod_{m=1}^{M}
R_{am}
\Big(u,w_m,h+\sum_{j=1}^{m-1} \sigma_j^{z}\Big)
\nonumber
\end{align}

\noindent with the multiplication taken in the space ${\rm End}(\mathcal{V}_a)$. We write the contribution from the space ${\rm End}(\mathcal{V}_a)$ explicitly, by defining

\begin{align}
T_a(u,\{w\}_M,h)
=
\left(
\begin{array}{cc}
A(u,\{w\}_M,h)
&
B(u,\{w\}_M,h)
\\
C(u,\{w\}_M,h)
&
D(u,\{w\}_M,h)
\end{array}
\right)_a
\end{align}

\begin{lemma}
{\rm By virtue of the dynamical Yang-Baxter equation (\ref{yb-sos}), we obtain the intertwining equation

\begin{align}
&
R_{ab}(u,v,h)
T_{a}(u,\{w\}_M,h+\sigma_b^{z})
T_{b}(v,\{w\}_M,h)
=
\label{int0-sos}
\\
&
T_{b}(v,\{w\}_M,h+\sigma_a^{z})
T_{a}(u,\{w\}_M,h)
R_{ab}\Big(u,v,h+\sum_{j=1}^{M}\sigma_j^{z}\Big)
\nonumber
\end{align}
}
\end{lemma}

\begin{proof} Starting from the definition (\ref{mon-sos}) of the dynamical monodromy matrix, we find

\begin{align}
&
R_{ab}(u,v,h)
T_{a}(u,\{w\}_M,h+\sigma_b^{z})
T_{b}(v,\{w\}_M,h)
=
\label{int1-sos}
\\
&
R_{ab}(u,v,h)
\rprod_{m=1}^{M}
R_{am}\Big(u,w_m,h+\sigma_b^{z}+\sum_{j=1}^{m-1} \sigma_j^{z}\Big)
\rprod_{n=1}^{M}
R_{bn}\Big(v,w_n,h+\sum_{j=1}^{n-1} \sigma_j^{z}\Big)
\nonumber
\end{align}

\noindent Using the commutation relation (\ref{com-sos}) we can change the order of the $R$-matrices appearing on the right hand side of (\ref{int1-sos}), giving

\begin{align}
&
R_{ab}(u,v,h)
T_{a}(u,\{w\}_M,h+\sigma_b^{z})
T_{b}(v,\{w\}_M,h)
=
\label{int2-sos}
\\
&
R_{ab}(u,v,h)
\rprod_{m=1}^{M}
\left[
R_{am}
\Big(u,w_m,h+\sigma_b^{z}+\sum_{j=1}^{m-1} \sigma_j^{z}\Big)
R_{bm}
\Big(v,w_m,h+\sum_{j=1}^{m-1} \sigma_j^{z}\Big)
\right]
\nonumber
\end{align}

\noindent Successively applying the dynamical Yang-Baxter equation (\ref{yb-sos}) to the right hand side of (\ref{int2-sos}), we obtain

\begin{align}
&
R_{ab}(u,v,h)
T_{a}(u,\{w\}_M,h+\sigma_b^{z})
T_{b}(v,\{w\}_M,h)
=
\label{int3-sos}
\\
&
\rprod_{m=1}^{M}
\left[
R_{bm}
\Big(v,w_m,h+\sigma_a^{z}+\sum_{j=1}^{m-1} \sigma_j^{z}\Big)
R_{am}
\Big(u,w_m,h+\sum_{j=1}^{m-1} \sigma_j^{z}\Big)
\right]
R_{ab}\Big(u,v,h+\sum_{j=1}^{M} \sigma_j^{z}\Big)
\nonumber
\end{align}

\noindent Using the commutation relation (\ref{com-sos}) we then restore the previous ordering of $R$-matrices on the right hand side of (\ref{int3-sos}), giving the result (\ref{int0-sos}).
\end{proof}

\begin{remark}
{\rm
By virtue of equation (\ref{actff-sos}), in the limit $\gamma = \pi i /2$ the dynamical monodromy matrix (\ref{mon-sos}) becomes

{\small 
\begin{align}
T_a(u,\{w\}_M,h)
&=
R_{a1}(u,w_1,h)
\ldots
R_{aM}(u,w_M,h+M-1)
\label{ffmon-sos}
=
\rprod_{m=1}^{M}
R_{am}(u,w_m,h+m-1)
\end{align}
}

\noindent and satisfies the intertwining equation

\begin{align}
&
R_{ab}(u,v,h)
T_a(u,\{w\}_M,h+1)
T_b(v,\{w\}_M,h)
=
\label{gen-version}
\\
&
T_b(v,\{w\}_M,h+1)
T_a(u,\{w\}_M,h)
R_{ab}(u,v,h+M)
\nonumber
\end{align}

\noindent In the next section, we will consider a model whose monodromy matrix is a generalization of (\ref{ffmon-sos}), and satisfies a more general version of (\ref{gen-version}).
}
\end{remark}

\subsection{Domain wall partition function}

The domain wall partition function of the SOS model is defined as

%

\begin{align}
Z_N\Big(\{v\}_N,\{w\}_N,h\Big)
=
\langle \Downarrow_N|
\lprod_{j=1}^{N}
B(v_j,\{w\}_N,h+j-1)
|\Uparrow_N\rangle
\label{PF-sos}
\end{align}

\noindent An explicit expression for (\ref{PF-sos}) was obtained by H~Rosengren in \cite{ros}. It was observed that $Z_N(\{v\}_N,\{w\}_N,h)$ does {\it not} admit a determinant representation for general values of the crossing parameter $\gamma$. However in the limit $\gamma = \pi i /2$, we have the following result.

\begin{lemma}
{\rm
Setting $\gamma = \pi i/2$ the domain wall partition function (\ref{PF-sos}) has the factorized form

\begin{align}
&
Z_N
\Big|_{\gamma=\pi i/2}
=
\label{PFexp-sos}
\frac{\Big[h+N+\displaystyle{\sum_{j=1}^{N}}(w_j-v_j)\Big]}{[h+N]}
\langle 0 \rangle^N
\prod_{1\leq j<k \leq N} \langle v_j-v_k\rangle \langle w_k-w_j \rangle
\end{align}

\noindent This is the height model analogue of equation (\ref{IK-ff3}). The vertex model expression (\ref{IK-ff3}) can be recovered by taking the trigonometric $\nome \rightarrow 0$ and heightless $h \rightarrow i\infty$ limits. We also remark that just as (\ref{IK-ff3}) is related to the factorization of a Cauchy-type determinant, the expression (\ref{PFexp-sos}) seems to be related to the factorization of a Frobenius-type determinant \cite{fro}. 
}
\end{lemma}

\begin{proof}
{\rm We will defer the proof to the next section. There, we will calculate the domain wall partition function of a model which generalizes the $\gamma = \pi i /2$ limit of the SOS model. Specializing that result, we will recover (\ref{PFexp-sos}) as a corollary.}
\end{proof}

\section{Elliptic Deguchi-Akutsu height model}
\label{ff-elda}

In this section we study an elliptic extension of the height model discovered by Deguchi and Akutsu in \cite{da2}. The results which we present were originally obtained in \cite{fwz2}.

\subsection{$R$-matrix and Yang-Baxter equation}

The $R$-matrix for the elliptic Deguchi-Akutsu height model is given by

{\small
\begin{align}
&
R_{ab}(u,p,v,q,h)
=
\label{Rmat-da}
\left(
\begin{array}{cccc}
a_{+}(u,p,v,q) & 0 & 0 & 0
\\
0 & b_{+}(u,p,v,q,h) & c_{+}(u,p,v,q,h) & 0
\\
0 & c_{-}(u,p,v,q,h) & b_{-}(u,p,v,q,h) & 0
\\
0 & 0 & 0 & a_{-}(u,p,v,q)
\end{array}
\right)_{ab}
\end{align}
}

\noindent where we have defined the functions

\begin{align}
a_{\pm}(u,p,v,q) &= [\pm(u-v)+p+q]
\label{a-da}
\\
b_{\pm}(u,p,v,q,h) &= 
\frac{
[h]^\frac{1}{2}
[h+2p+2q]^\frac{1}{2}
}
{
[h+2p]^\frac{1}{2}
[h+2q]^\frac{1}{2}
}
[u-v\pm(q-p)]
\label{b-da}
\\
c_{\pm}(u,p,v,q,h) &=
\frac{
[2p]^\frac{1}{2}
[2q]^\frac{1}{2}
}
{
[h+2p]^\frac{1}{2}
[h+2q]^\frac{1}{2}
}
[\pm(v-u)+p+q+h]
\label{c-da}
\end{align}

\noindent with $[u]$ given by (\ref{jac-ell2}). We have placed the subscript $ab$ on the $R$-matrix to denote the fact that it is an element of ${\rm End}(\mathcal{V}_a\otimes\mathcal{V}_b)$. When the variables of a particular $R$-matrix are clear from context, we will abbreviate $R_{ab}(u,p,v,q,h)=R_{ab}(h)$. The $\gamma=\pi i/2$ limit of the SOS $R$-matrix (\ref{Rmat-sos}) is recovered by setting $p=q=\frac{1}{2}$ in (\ref{Rmat-da}).

\begin{figure}[H]
\begin{center}
\begin{minipage}{4.3in}

\setlength{\unitlength}{0.00038cm}
\begin{picture}(20000,17500)(-3500,-15000)

\put(-4000,2000){\circle*{300}}
\put(-4750,2000){\tiny$h$}
\put(0000,2000){\circle*{300}}
\put(250,2000){\tiny$h+2q$}
\put(-4000,-2000){\circle*{300}}
\put(-6450,-2000){\tiny$h+2p$}
\put(000,-2000){\circle*{300}}
\put(250,-2000){\tiny$h+2p+2q$}

\path(-4000,0000)(000,0000)
\blacken\path(-4000,250)(-4000,-250)(-3500,0)(-4000,250)
\blacken\path(0,250)(0,-250)(500,0)(0,250)
\path(-4000,-2000)(000,-2000)
\path(-4000,2000)(000,2000)
\put(-6250,0){\scriptsize$u,p$}
\path(-7500,0)(-6500,0)
\whiten\path(-7000,250)(-7000,-250)(-6500,0)(-7000,250)
\path(-2000,-2000)(-2000,2000)
\blacken\path(-2250,-2000)(-1750,-2000)(-2000,-1500)(-2250,-2000)
\blacken\path(-2250,2000)(-1750,2000)(-2000,2500)(-2250,2000)
\path(-4000,-2000)(-4000,2000)
\path(0000,-2000)(000,2000)
\put(-4500,-5500){$a_{+}(u,p,v,q)$}
\put(-2500,-3250){\scriptsize$v,q$}
\path(-2000,-4500)(-2000,-3500)
\whiten\path(-2250,-4000)(-1750,-4000)(-2000,-3500)(-2250,-4000)

\put(8000,2000){\circle*{300}}
\put(7250,2000){\tiny$h$}
\put(12000,2000){\circle*{300}}
\put(12250,2000){\tiny$h+2q$}
\put(8000,-2000){\circle*{300}}
\put(5550,-2000){\tiny$h+2p$}
\put(12000,-2000){\circle*{300}}
\put(12250,-2000){\tiny$h+2p+2q$}

\path(8000,0000)(12000,0000)
\blacken\path(8000,250)(8000,-250)(8500,0)(8000,250)
\blacken\path(12000,250)(12000,-250)(12500,0)(12000,250)
\path(8000,-2000)(12000,-2000)
\path(8000,2000)(12000,2000)
\put(5750,0){\scriptsize$u,p$}
\path(4500,0)(5500,0)
\whiten\path(5000,250)(5000,-250)(5500,0)(5000,250)
\path(10000,-2000)(10000,2000)
\blacken\path(9750,-2000)(10250,-2000)(10000,-2500)(9750,-2000)
\blacken\path(9750,2000)(10250,2000)(10000,1500)(9750,2000)
\path(8000,-2000)(8000,2000)
\path(12000,-2000)(12000,2000)
\put(7500,-5500){$b_{+}(u,p,v,q,h)$}
\put(9500,-3250){\scriptsize$v,q$}
\path(10000,-4500)(10000,-3500)
\whiten\path(9750,-4000)(10250,-4000)(10000,-3500)(9750,-4000)

\put(20000,2000){\circle*{300}}
\put(19250,2000){\tiny$h$}
\put(24000,2000){\circle*{300}}
\put(24250,2000){\tiny$h+2q$}
\put(20000,-2000){\circle*{300}}
\put(17550,-2000){\tiny$h+2p$}
\put(24000,-2000){\circle*{300}}
\put(24250,-2000){\tiny$h+2p+2q$}

\path(20000,0000)(24000,0000)
\blacken\path(20000,250)(20000,-250)(20500,0)(20000,250)
\blacken\path(24000,250)(24000,-250)(23500,0)(24000,250)
\path(20000,-2000)(24000,-2000)
\path(20000,2000)(24000,2000)
\put(17750,0){\scriptsize$u,p$}
\path(16500,0)(17500,0)
\whiten\path(17000,250)(17000,-250)(17500,0)(17000,250)
\path(22000,-2000)(22000,2000)
\blacken\path(21750,-2000)(22250,-2000)(22000,-2500)(21750,-2000)
\blacken\path(21750,2000)(22250,2000)(22000,2500)(21750,2000)
\path(20000,-2000)(20000,2000)
\path(24000,-2000)(24000,2000)
\put(19500,-5500){$c_{+}(u,p,v,q,h)$}
\put(21500,-3250){\scriptsize$v,q$}
\path(22000,-4500)(22000,-3500)
\whiten\path(21750,-4000)(22250,-4000)(22000,-3500)(21750,-4000)

\put(-4000,-8000){\circle*{300}}
\put(-4750,-8000){\tiny$h$}
\put(000,-8000){\circle*{300}}
\put(250,-8000){\tiny$h+2q$}
\put(-4000,-12000){\circle*{300}}
\put(-6450,-12000){\tiny$h+2p$}
\put(000,-12000){\circle*{300}}
\put(250,-12000){\tiny$h+2p+2q$}

\path(-4000,-10000)(000,-10000)
\blacken\path(-4000,-9750)(-4000,-10250)(-4500,-10000)(-4000,-9750)
\blacken\path(000,-9750)(000,-10250)(-500,-10000)(000,-9750)
\path(-4000,-12000)(000,-12000)
\path(-4000,-8000)(000,-8000)
\put(-6250,-10000){\scriptsize$u,p$}
\path(-7500,-10000)(-6500,-10000)
\whiten\path(-7000,-9750)(-7000,-10250)(-6500,-10000)(-7000,-9750)
\path(-2000,-12000)(-2000,-8000)
\blacken\path(-2250,-12000)(-1750,-12000)(-2000,-12500)(-2250,-12000)
\blacken\path(-2250,-8000)(-1750,-8000)(-2000,-8500)(-2250,-8000)
\path(-4000,-12000)(-4000,-8000)
\path(000,-12000)(000,-8000)
\put(-4500,-15500){$a_{-}(u,p,v,q)$}
\put(-2500,-13250){\scriptsize$v,q$}
\path(-2000,-14500)(-2000,-13500)
\whiten\path(-2250,-14000)(-1750,-14000)(-2000,-13500)(-2250,-14000)

\put(8000,-8000){\circle*{300}}
\put(7250,-8000){\tiny$h$}
\put(12000,-8000){\circle*{300}}
\put(12250,-8000){\tiny$h+2q$}
\put(8000,-12000){\circle*{300}}
\put(5550,-12000){\tiny$h+2p$}
\put(12000,-12000){\circle*{300}}
\put(12250,-12000){\tiny$h+2p+2q$}

\path(8000,-10000)(12000,-10000)
\blacken\path(8000,-9750)(8000,-10250)(7500,-10000)(8000,-9750)
\blacken\path(12000,-9750)(12000,-10250)(11500,-10000)(12000,-9750)
\path(8000,-12000)(12000,-12000)
\path(8000,-8000)(12000,-8000)
\put(5750,-10000){\scriptsize$u,p$}
\path(4500,-10000)(5500,-10000)
\whiten\path(5000,-9750)(5000,-10250)(5500,-10000)(5000,-9750)
\path(10000,-12000)(10000,-8000)
\blacken\path(9750,-12000)(10250,-12000)(10000,-11500)(9750,-12000)
\blacken\path(9750,-8000)(10250,-8000)(10000,-7500)(9750,-8000)
\path(8000,-12000)(8000,-8000)
\path(12000,-12000)(12000,-8000)
\put(7500,-15500){$b_{-}(u,p,v,q,h)$}
\put(9500,-13250){\scriptsize$v,q$}
\path(10000,-14500)(10000,-13500)
\whiten\path(9750,-14000)(10250,-14000)(10000,-13500)(9750,-14000)

\put(20000,-8000){\circle*{300}}
\put(19250,-8000){\tiny$h$}
\put(24000,-8000){\circle*{300}}
\put(24250,-8000){\tiny$h+2q$}
\put(20000,-12000){\circle*{300}}
\put(17550,-12000){\tiny$h+2p$}
\put(24000,-12000){\circle*{300}}
\put(24250,-12000){\tiny$h+2p+2q$}

\path(20000,-10000)(24000,-10000)
\blacken\path(20000,-9750)(20000,-10250)(19500,-10000)(20000,-9750)
\blacken\path(24000,-9750)(24000,-10250)(24500,-10000)(24000,-9750)
\path(20000,-12000)(24000,-12000)
\path(20000,-8000)(24000,-8000)
\put(17750,-10000){\scriptsize$u,p$}
\path(16500,-10000)(17500,-10000)
\whiten\path(17000,-9750)(17000,-10250)(17500,-10000)(17000,-9750)
\path(22000,-12000)(22000,-8000)
\blacken\path(21750,-12000)(22250,-12000)(22000,-11500)(21750,-12000)
\blacken\path(21750,-8000)(22250,-8000)(22000,-8500)(21750,-8000)
\path(20000,-12000)(20000,-8000)
\path(24000,-12000)(24000,-8000)
\put(19500,-15500){$c_{-}(u,p,v,q,h)$}
\put(21500,-13250){\scriptsize$v,q$}
\path(22000,-14500)(22000,-13500)
\whiten\path(21750,-14000)(22250,-14000)(22000,-13500)(21750,-14000)

\end{picture}

\end{minipage}
\end{center}

\caption[Weights of the elliptic Deguchi-Akutsu height model]{Weights of the elliptic Deguchi-Akutsu height model. Each entry of the $R$-matrix (\ref{Rmat-da}) is paired with a face. Similarly to the SOS model, the dynamical variable at the top-left corner matches that of the $R$-matrix. In contrast, the dynamical variables at the remaining corners are fixed and do not depend on the black arrows.}

\end{figure}
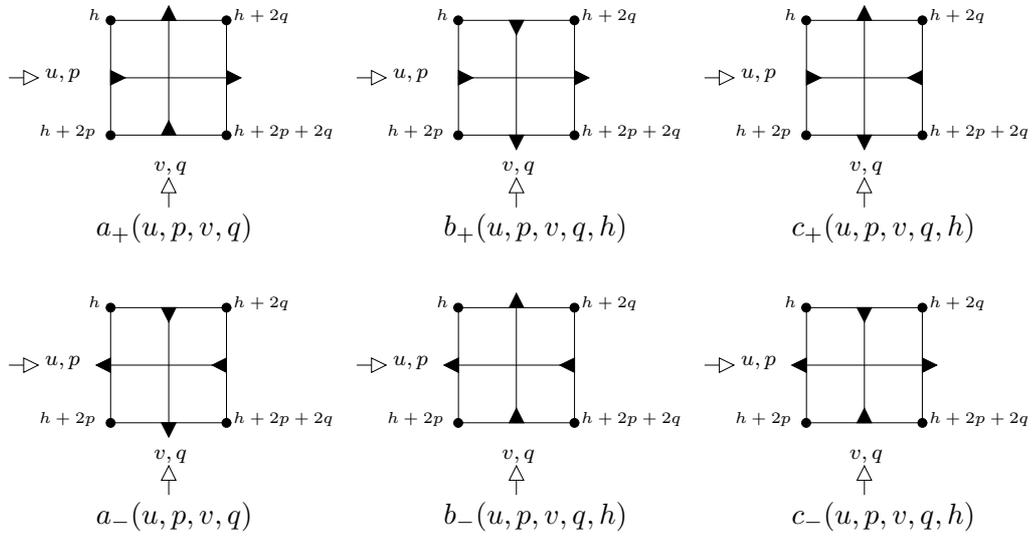

\begin{lemma}
{\rm 
Associate the variables $\{u,p\},\{v,q\},\{w,r\}$ to the respective vector spaces $\mathcal{V}_a,\mathcal{V}_b,\mathcal{V}_c$ so that, for example, $R_{ab}(h)$ is understood to equal $R_{ab}(u,p,v,q,h)$. The $R$-matrix (\ref{Rmat-da}) satisfies the Yang-Baxter equation

\begin{align}
R_{ab}(h) R_{ac}(h+2q) R_{bc}(h)
=
R_{bc}(h+2p) R_{ac}(h) R_{ab}(h+2r)
\label{yb-da}
\end{align}

\noindent which holds in ${\rm End}(\mathcal{V}_a\otimes\mathcal{V}_b\otimes\mathcal{V}_c)$. Setting $p=q=r=\frac{1}{2}$, we recover (\ref{ffyb-sos}).
}
\end{lemma}

\begin{proof}

By direct computation. To check each component of (\ref{yb-da}) one needs various theta function identities, which may be found in chapter 15 of \cite{bax1}. 

\end{proof}

\begin{figure}[H]
\begin{center}
\begin{minipage}{4.3in}

\setlength{\unitlength}{0.00031cm}
\begin{picture}(20000,12500)(-1000,-3500)

\put(-250,-750){\tiny$i_b$}

\put(-250,4250){\tiny$i_a$}

\put(3750,0){\tiny$k_a$}

\put(3750,5000){\tiny$k_b$}

\put(10350,0){\tiny$j_a$}

\put(10250,5000){\tiny$j_b$}

\put(4000,-2250){\tiny$i_c$}

\put(10000,2750){\tiny$k_c$}

\put(5000,8000){\tiny$j_c$}

\put(-2500,2500){\circle*{300}}

\put(2500,7500){\circle*{300}}
\put(2500,7750){\tiny$h$}

\put(2500,-2500){\circle*{300}}

\put(7500,-2500){\circle*{300}}

\put(7500,2500){\circle*{300}}
\put(4600,2400){\tiny$h+2q$}

\put(7500,7500){\circle*{300}}

\put(12500,2500){\circle*{300}}

\path(2500,7500)(7500,2500)
\path(-2500,2500)(2500,7500)
\path(2500,-2500)(-2500,2500)
\path(2500,-2500)(7500,2500)

\path(2500,-2500)(7500,-2500)
\path(7500,-2500)(12500,2500)
\path(7500,2500)(12500,2500)
\path(2500,7500)(7500,7500)
\path(7500,7500)(12500,2500)


\path(0,0)(5000,5000)
\path(-4350,0)(-3350,0)
\whiten\path(-3600,250)(-3600,-250)(-3350,0)(-3600,250)
\put(-3000,-100){\fbox{\scriptsize$v,q$}}
\path(5000,0)(0,5000)
\path(-4350,5000)(-3350,5000)
\whiten\path(-3600,5250)(-3600,4750)(-3350,5000)(-3600,5250)
\put(-3000,5000){\fbox{\scriptsize$u,p$}}
\path(5000,5000)(10000,5000)
\path(5000,0)(10000,0)
\path(5000,-2500)(10000,2500)
\path(10000,2500)(5000,7500)
\put(4500,-3750){\fbox{\scriptsize$w,r$}}
\path(5000,-5500)(5000,-4500)
\whiten\path(4750,-4750)(5250,-4750)(5000,-4500)(4750,-4750)


\put(16000,2250){$=$}


\put(23750,-750){\tiny$i_b$}

\put(24000,4400){\tiny$i_a$}

\put(28750,750){\tiny$k_b$}

\put(28750,4000){\tiny$k_a$}

\put(34250,-250){\tiny$j_a$}

\put(34000,3900){\tiny$j_b$}

\put(29000,-2200){\tiny$i_c$}

\put(23500,1900){\tiny$k_c$}

\put(29000,7800){\tiny$j_c$}

\put(21500,2500){\circle*{300}}
\put(18600,2500){\tiny$h+2p$}

\put(26500,7500){\circle*{300}}
\put(26500,7750){\tiny$h$}

\put(26500,-2500){\circle*{300}}

\put(31500,-2500){\circle*{300}}

\put(31500,7500){\circle*{300}}
\put(31750,7500){\tiny$h+2r$}

\put(36500,2500){\circle*{300}}

\put(26500,2500){\circle*{300}}

\path(21500,2500)(26500,7500)
\path(21500,2500)(26500,-2500)
\path(26500,-2500)(31500,-2500)
\path(31500,-2500)(36500,2500)
\path(36500,2500)(31500,7500)
\path(26500,7500)(31500,7500)

\path(21500,2500)(26500,2500)
\path(26500,2500)(31500,7500)
\path(26500,2500)(31500,-2500)

\path(19500,5000)(20500,5000)
\whiten\path(20250,5250)(20250,4750)(20500,5000)(20250,5250)
\put(20850,5000){\fbox{\scriptsize$u,p$}}
\path(24000,5000)(29000,5000)
\path(29000,5000)(34000,0)
\path(19500,0)(20500,0)
\whiten\path(20250,250)(20250,-250)(20500,0)(20250,250)
\put(20850,-100){\fbox{\scriptsize$v,q$}}
\path(24000,0)(29000,0)
\path(29000,0)(34000,5000)


\path(29000,-2500)(24000,2500)
\path(24000,2500)(29000,7500)

\put(28500,-3750){\fbox{\scriptsize$w,r$}}
\path(29000,-5500)(29000,-4500)
\whiten\path(28750,-4750)(29250,-4750)(29000,-4500)(28750,-4750)

\end{picture}

\end{minipage}
\end{center}

\caption[Yang-Baxter equation for the elliptic Deguchi-Akutsu height model]{Yang-Baxter equation for the elliptic Deguchi-Akutsu height model.}

\end{figure}
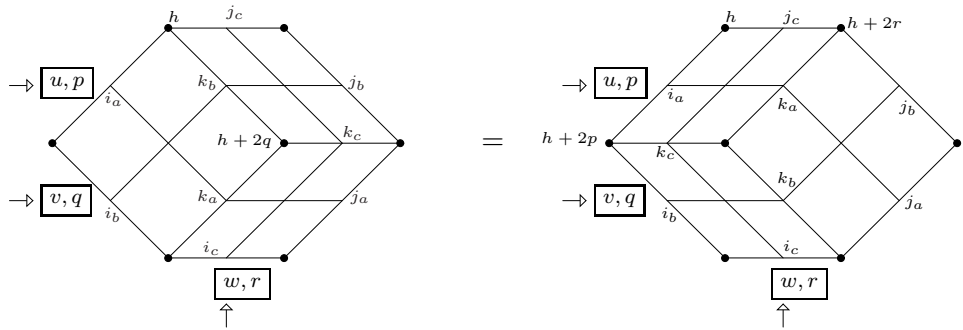 

\subsection{Monodromy matrix and intertwining equation}

For all integers $n \geq 1$, we define partial sums of the external field variables $\{p\},\{q\},\{r\}$ as follows

\begin{align}
\overline{p}_n
=
\sum_{j=1}^{n}
p_j,
\quad
\overline{q}_n
=
\sum_{j=1}^{n}
q_j,
\quad
\overline{r}_n
=
\sum_{j=1}^{n}
r_j
\end{align}

\noindent In addition, we fix $\overline{p}_0 = \overline{q}_0 = \overline{r}_0 = 0$. The monodromy matrix is an ordered product of $R$-matrices, given by

\begin{align}
T_{a}(u,p,\{w,r\}_M,h)
&=
R_{a1}(u,p,w_1,r_1,h)
\ldots
R_{aM}(u,p,w_M,r_M,h+2\overline{r}_{M-1}
)
\label{mon-da}
\\
&=
\rprod_{m=1}^{M}
R_{am}(u,p,w_m,r_m,h+2\overline{r}_{m-1}
)
\nonumber
\end{align}

\noindent with the multiplication taken in the space ${\rm End}(\mathcal{V}_a)$. We write the contribution from the space ${\rm End}(\mathcal{V}_a)$ explicitly, by defining

\begin{align}
T_a(u,p,\{w,r\}_M,h)
=
\left(
\begin{array}{cc}
A(u,p,\{w,r\}_M,h)
&
B(u,p,\{w,r\}_M,h)
\\
C(u,p,\{w,r\}_M,h)
&
D(u,p,\{w,r\}_M,h)
\end{array}
\right)_a
\label{mon2-da}
\end{align}

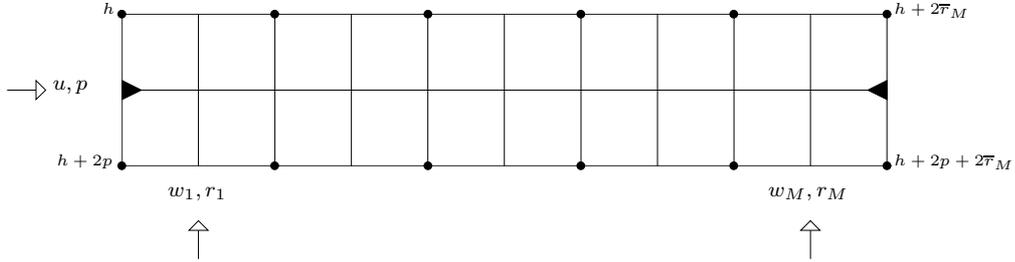
\begin{figure}[H]
\begin{center}
\begin{minipage}{4.3in}

\setlength{\unitlength}{0.0005cm}
\begin{picture}(20000,6000)(-3000,-3500)

\path(-2000,-2000)(18000,-2000)
\path(-2000,2000)(18000,2000)

\path(-2000,-2000)(-2000,2000)
\put(-2000,2000){\circle*{200}}
\put(-2500,2000){\tiny$h$}

\path(2000,-2000)(2000,2000)
\put(2000,2000){\circle*{200}}

\path(6000,-2000)(6000,2000)
\put(6000,2000){\circle*{200}}

\path(10000,-2000)(10000,2000)
\put(10000,2000){\circle*{200}}

\path(14000,-2000)(14000,2000)
\put(14000,2000){\circle*{200}}

\path(18000,-2000)(18000,2000)

\put(18000,2000){\circle*{200}}
\put(18200,2000){\tiny$h+2\overline{r}_M$}

\put(18000,-2000){\circle*{200}}
\put(18200,-2000){\tiny$h+2p+2\overline{r}_M
$}

\put(14000,-2000){\circle*{200}}
\put(10000,-2000){\circle*{200}}
\put(6000,-2000){\circle*{200}}
\put(2000,-2000){\circle*{200}}

\put(-2000,-2000){\circle*{200}}
\put(-3700,-2000){\tiny$h+2p$}


\path(-2000,0)(18000,0)
\blacken\path(-2000,-250)(-2000,250)(-1500,0)(-2000,-250)
\blacken\path(18000,-250)(18000,250)(17500,0)(18000,-250)

\path(-5000,0)(-4000,0)
\whiten\path(-4250,250)(-4250,-250)(-4000,0)(-4250,250)
\put(-3800,0){\scriptsize$u,p$}

\path(0,-2000)(0,2000)

\path(0,-4450)(0,-3450)
\whiten\path(-250,-3700)(250,-3700)(0,-3450)(-250,-3700)
\put(-800,-2800){\scriptsize$w_1,r_1$}

\path(4000,-2000)(4000,2000)

\path(8000,-2000)(8000,2000)

\path(12000,-2000)(12000,2000)

\path(16000,-2000)(16000,2000)

\path(16000,-4450)(16000,-3450)
\whiten\path(15750,-3700)(16250,-3700)(16000,-3450)(15750,-3700)
\put(14900,-2800){\scriptsize$w_M,r_M$}

\end{picture}

\end{minipage}
\end{center}

\caption[Graphical representation of the operator $B(u,p,\{w,r\}_M,h)$]{Graphical representation of the operator $B(u,p,\{w,r\}_M,h)$. This should be viewed as a string of $M$ conjoined faces, each representing an entry of the $R$-matrix (\ref{Rmat-da}). The external horizontal black arrows are frozen to values which select the $(+1,-1)$ component of the matrix (\ref{mon2-da}), while the internal horizontal black arrows are summed over all configurations.}

\end{figure}

\begin{lemma}
{\rm
By virtue of the Yang-Baxter equation (\ref{yb-da}), we obtain the intertwining equation

\begin{align}
&
R_{ab}(u,p,v,q,h)
T_{a}(u,p,\{w,r\}_M,h+2q)
T_{b}(v,q,\{w,r\}_M,h)
=
\label{int0-da}
\\
&
T_{b}(v,q,\{w,r\}_M,h+2p)
T_{a}(u,p,\{w,r\}_M,h)
R_{ab}(u,p,v,q,h+2\overline{r}_M
)
\nonumber
\end{align}

\noindent Setting $p=q=\frac{1}{2}$ and $r_j = \frac{1}{2}$ for all $1\leq j \leq M$, we recover the equation (\ref{gen-version}).
}
\end{lemma}

\begin{proof} Starting from the definition (\ref{mon-da}) of the monodromy matrix, we find 

\begin{align}
&
R_{ab}(u,p,v,q,h)
T_{a}(u,p,\{w,r\}_M,h+2q)
T_{b}(v,q,\{w,r\}_M,h)
=
\label{int1-da}
\\
&
R_{ab}(h)
\rprod_{m=1}^{M}
R_{am}(h+2q+2\overline{r}_{m-1}
)
\rprod_{n=1}^{M}
R_{bn}(h+2\overline{r}_{n-1}
)
\nonumber
\end{align}

\noindent We change the order of the $R$-matrices on the right hand side of (\ref{int1-da}), by commuting those which act in different spaces to obtain

\begin{align}
&
R_{ab}(u,p,v,q,h)
T_{a}(u,p,\{w,r\}_M,h+2q)
T_{b}(v,q,\{w,r\}_M,h)
=
\label{int2-da}
\\
&
R_{ab}(h)
\rprod_{m=1}^{M}
\Big(
R_{am}(h+2q+2\overline{r}_{m-1}
)
R_{bm}(h+2\overline{r}_{m-1}
)
\Big)
\nonumber
\end{align}

\noindent Successively applying the Yang-Baxter equation (\ref{yb-da}) to the right hand side of (\ref{int2-da}), we have

\begin{align}
&
R_{ab}(u,p,v,q,h)
T_{a}(u,p,\{w,r\}_M,h+2q)
T_{b}(v,q,\{w,r\}_M,h)
=
\label{int3-da}
\\
&
\rprod_{m=1}^{M}
\Big(
R_{bm}(h+2p+2\overline{r}_{m-1}
)
R_{am}(h+2\overline{r}_{m-1}
)
\Big)
R_{ab}(h+2\overline{r}_M
)
\nonumber
\end{align}

\noindent Restoring the previous ordering to the $R$-matrices on the right hand side of (\ref{int3-da}), we recover the result (\ref{int0-da}).

\end{proof}

As always, (\ref{int0-da}) generates sixteen commutation relations amongst the entries of the monodromy matrix (\ref{mon2-da}). For our purposes, the most important of these is

\begin{align}
&[u-v+p+q]B(u,p,\{w,r\}_M,h+2q)B(v,q,\{w,r\}_M,h)=
\label{bb-da}
\\
&[v-u+p+q]B(v,q,\{w,r\}_M,h+2p)B(u,p,\{w,r\}_M,h)
\nonumber
\end{align}

\noindent and is graphically depicted by the diagrams below. 

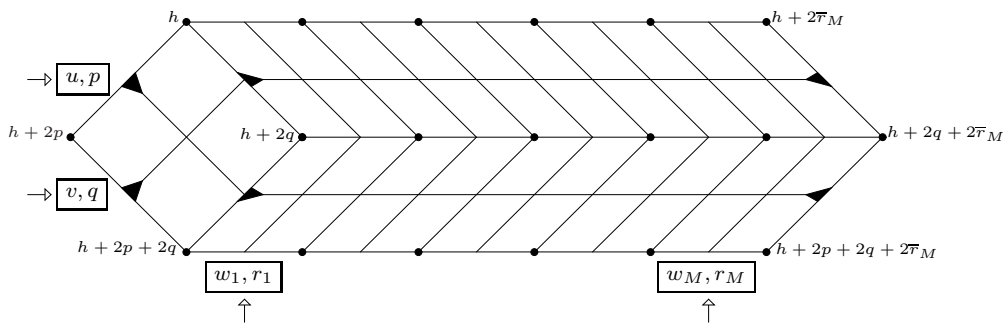
\begin{figure}[H]

\begin{center}
\begin{minipage}{4.3in}

\setlength{\unitlength}{0.0003cm}
\begin{picture}(20000,12000)(-2000,-3500)










\put(-5200,2500){\tiny$h+2p$}

\put(2500,7500){\circle*{300}}
\put(1700,7500){\tiny$h$}

\put(-2200,-2500){\tiny$h+2p+2q$}


\put(7500,2500){\circle*{300}}
\put(4800,2400){\tiny$h+2q$}



\path(2500,7500)(7500,2500)
\path(-2500,2500)(2500,7500)
\path(2500,-2500)(-2500,2500)
\path(2500,-2500)(7500,2500)

\blacken\path(-300,4700)(300,5300)(600,4400)(-300,4700)
\blacken\path(4700,5300)(5300,4700)(5800,5000)(4700,5300)
\blacken\path(29700,5300)(30300,4700)(29200,5000)(29700,5300)

\blacken\path(-300,300)(300,-300)(600,600)(-300,300)
\blacken\path(4700,-300)(5300,300)(5800,0)(4700,-300)
\blacken\path(29700,-300)(30300,300)(29200,0)(29700,-300)

\path(2500,-2500)(27500,-2500)
\path(7500,-2500)(12500,2500)
\path(7500,2500)(32500,2500)
\path(2500,7500)(27500,7500)
\path(7500,7500)(12500,2500)

\path(10000,7500)(15000,2500)
\path(12500,7500)(17500,2500)
\path(15000,7500)(20000,2500)
\path(15000,2500)(10000,-2500)
\path(17500,2500)(12500,-2500)
\path(20000,2500)(15000,-2500)

\path(17500,7500)(22500,2500)
\path(20000,7500)(25000,2500)
\path(22500,2500)(17500,-2500)
\path(25000,2500)(20000,-2500)

\path(22500,7500)(27500,2500)
\path(25000,7500)(30000,2500)
\path(27500,2500)(22500,-2500)
\path(30000,2500)(25000,-2500)

\path(27500,7500)(32500,2500)
\path(32500,2500)(27500,-2500)

\put(7500,7500){\circle*{300}}
\put(12500,7500){\circle*{300}}
\put(17500,7500){\circle*{300}}
\put(22500,7500){\circle*{300}}
\put(27500,7500){\circle*{300}}
\put(27700,7500){\tiny$h+2\overline{r}_M$}

\put(-2500,2500){\circle*{300}}
\put(7500,2500){\circle*{300}}
\put(12500,2500){\circle*{300}}
\put(17500,2500){\circle*{300}}
\put(22500,2500){\circle*{300}}
\put(27500,2500){\circle*{300}}
\put(32500,2500){\circle*{300}}
\put(32700,2500){\tiny$h+2q+2\overline{r}_M$}

\put(2500,-2500){\circle*{300}}
\put(7500,-2500){\circle*{300}}
\put(12500,-2500){\circle*{300}}
\put(17500,-2500){\circle*{300}}
\put(22500,-2500){\circle*{300}}
\put(27500,-2500){\circle*{300}}
\put(27900,-2600){\tiny$h+2p+2q+2\overline{r}_M$}







\path(0,0)(5000,5000)
\path(-4350,0)(-3350,0)
\whiten\path(-3600,250)(-3600,-250)(-3350,0)(-3600,250)
\put(-3100,-100){\fbox{\scriptsize$v,q$}}
\path(5000,0)(0,5000)
\path(-4350,5000)(-3350,5000)
\whiten\path(-3600,5250)(-3600,4750)(-3350,5000)(-3600,5250)
\put(-3100,4900){\fbox{\scriptsize$u,p$}}
\path(5000,5000)(30000,5000)
\path(5000,0)(30000,0)
\path(5000,-2500)(10000,2500)
\path(10000,2500)(5000,7500)
\put(3300,-3700){\fbox{\scriptsize$w_1,r_1$}}
\path(5000,-5550)(5000,-4550)
\whiten\path(4750,-4800)(5250,-4800)(5000,-4550)(4750,-4800)

\put(22800,-3700){\fbox{\scriptsize$w_M,r_M$}}
\path(25000,-5550)(25000,-4550)
\whiten\path(24750,-4800)(25250,-4800)(25000,-4550)(24750,-4800)

\end{picture}

\end{minipage}
\end{center}

\caption[Product of two $B$-operators before commutation]{Product of two $B$-operators before commutation.}

\end{figure}

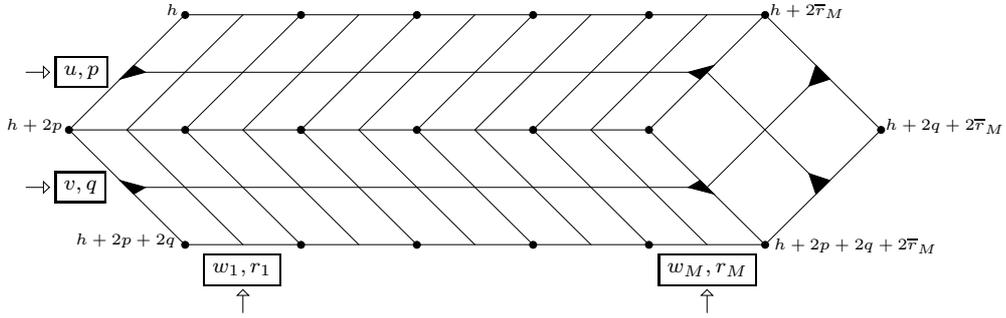
\begin{figure}[H]

\begin{center}
\begin{minipage}{4.3in}

\setlength{\unitlength}{0.0003cm}
\begin{picture}(20000,11000)(-2000,-3500)











\put(2500,7500){\circle*{300}}
\put(1700,7500){\tiny$h$}



\put(7500,2500){\circle*{300}}



\blacken\path(-300,300)(300,-300)(800,0)(-300,300)
\blacken\path(24700,300)(25300,-300)(24200,0)(24700,300)
\blacken\path(29700,-300)(30300,300)(29400,600)(29700,-300)

\blacken\path(-300,4700)(300,5300)(800,5000)(-300,4700)
\blacken\path(24700,4700)(25300,5300)(24200,5000)(24700,4700)
\blacken\path(29700,5300)(30300,4700)(29400,4400)(29700,5300)

\path(-2500,2500)(2500,7500)
\path(2500,-2500)(-2500,2500)

\path(2500,-2500)(27500,-2500)
\path(7500,-2500)(2500,2500)
\path(-2500,2500)(22500,2500)
\path(2500,7500)(27500,7500)
\path(7500,7500)(2500,2500)

\path(10000,7500)(5000,2500)
\path(12500,7500)(7500,2500)
\path(15000,7500)(10000,2500)
\path(5000,2500)(10000,-2500)
\path(7500,2500)(12500,-2500)
\path(10000,2500)(15000,-2500)

\path(17500,7500)(12500,2500)
\path(20000,7500)(15000,2500)
\path(12500,2500)(17500,-2500)
\path(15000,2500)(20000,-2500)

\path(22500,7500)(17500,2500)
\path(25000,7500)(20000,2500)
\path(17500,2500)(22500,-2500)
\path(20000,2500)(25000,-2500)

\path(27500,7500)(32500,2500)
\path(32500,2500)(27500,-2500)

\put(7500,7500){\circle*{300}}
\put(12500,7500){\circle*{300}}
\put(17500,7500){\circle*{300}}
\put(22500,7500){\circle*{300}}
\put(27500,7500){\circle*{300}}
\put(27700,7500){\tiny$h+2\overline{r}_M$}

\put(-5200,2500){\tiny$h+2p$}
\put(-2500,2500){\circle*{300}}
\put(2500,2500){\circle*{300}}
\put(7500,2500){\circle*{300}}
\put(12500,2500){\circle*{300}}
\put(17500,2500){\circle*{300}}
\put(22500,2500){\circle*{300}}
\put(32500,2500){\circle*{300}}
\put(32700,2500){\tiny$h+2q+2\overline{r}_M$}

\put(-2200,-2500){\tiny$h+2p+2q$}
\put(2500,-2500){\circle*{300}}
\put(7500,-2500){\circle*{300}}
\put(12500,-2500){\circle*{300}}
\put(17500,-2500){\circle*{300}}
\put(22500,-2500){\circle*{300}}
\put(27500,-2500){\circle*{300}}
\put(27900,-2700){\tiny$h+2p+2q+2\overline{r}_M$}







\path(25000,0)(30000,5000)
\path(-4350,0)(-3350,0)
\whiten\path(-3600,250)(-3600,-250)(-3350,0)(-3600,250)
\put(-3100,-100){\fbox{\scriptsize$v,q$}}
\path(25000,5000)(30000,0)
\path(-4350,5000)(-3350,5000)
\whiten\path(-3600,5250)(-3600,4750)(-3350,5000)(-3600,5250)
\put(-3100,4900){\fbox{\scriptsize$u,p$}}
\path(0,5000)(25000,5000)
\path(0,0)(25000,0)
\path(5000,-2500)(0000,2500)
\path(00000,2500)(5000,7500)
\put(3300,-3700){\fbox{\scriptsize$w_1,r_1$}}
\path(5000,-5450)(5000,-4450)
\whiten\path(4750,-4700)(5250,-4700)(5000,-4450)(4750,-4700)

\path(27500,-2500)(22500,2500)
\path(22500,2500)(27500,7500)
\put(22900,-3700){\fbox{\scriptsize$w_M,r_M$}}
\path(25000,-5450)(25000,-4450)
\whiten\path(24750,-4700)(25250,-4700)(25000,-4450)(24750,-4700)

\end{picture}

\end{minipage}
\end{center}

\caption[Product of two $B$-operators after commutation]{Product of two $B$-operators after commutation.}

\end{figure}

\noindent Similarly to equation (\ref{bb-tf}) in the context of the trigonometric Felderhof model, (\ref{bb-da}) plays a crucial role in the evaluation of the domain wall partition function, which we define in the next subsection.

\subsection{Domain wall partition function}

The domain wall partition function of the elliptic Deguchi-Akutsu height model is defined as

\begin{align}
Z_N\Big(\{v,q\}_N,\{w,r\}_N,h\Big)
=
\langle \Downarrow_N|
\lprod_{j=1}^{N}
B\Big(v_j,q_j,\{w,r\}_N,h+2\overline{q}_{j-1}
\Big)
|\Uparrow_N\rangle
\label{pfpfpf}
\end{align}

\begin{figure}[H]

\begin{center}
\begin{minipage}{4.3in}

\setlength{\unitlength}{0.00025cm}
\begin{picture}(30000,25000)(-10000,-2000)

\path(0,0)(24000,0)
\path(0,4000)(24000,4000)
\path(0,8000)(24000,8000)
\path(0,12000)(24000,12000)
\path(0,16000)(24000,16000)
\path(0,20000)(24000,20000)
\path(0,24000)(24000,24000)

\path(0,0)(0,24000)
\path(4000,0)(4000,24000)
\path(8000,0)(8000,24000)
\path(12000,0)(12000,24000)
\path(16000,0)(16000,24000)
\path(20000,0)(20000,24000)
\path(24000,0)(24000,24000)

\put(0,0){\circle*{300}}
\put(-4200,0){\tiny$h+2\overline{q}_N$}
\put(4000,0){\circle*{300}}
\put(8000,0){\circle*{300}}
\put(12000,0){\circle*{300}}
\put(16000,0){\circle*{300}}
\put(20000,0){\circle*{300}}
\put(24000,0){\circle*{300}}
\put(24300,0){\tiny$h+2\overline{q}_N+2\overline{r}_N$}

\put(0,4000){\circle*{300}}
\put(4000,4000){\circle*{300}}
\put(8000,4000){\circle*{300}}
\put(12000,4000){\circle*{300}}
\put(16000,4000){\circle*{300}}
\put(20000,4000){\circle*{300}}
\put(24000,4000){\circle*{300}}

\put(0,8000){\circle*{300}}
\put(4000,8000){\circle*{300}}
\put(8000,8000){\circle*{300}}
\put(12000,8000){\circle*{300}}
\put(16000,8000){\circle*{300}}
\put(20000,8000){\circle*{300}}
\put(24000,8000){\circle*{300}}

\put(0,12000){\circle*{300}}
\put(4000,12000){\circle*{300}}
\put(8000,12000){\circle*{300}}
\put(12000,12000){\circle*{300}}
\put(16000,12000){\circle*{300}}
\put(20000,12000){\circle*{300}}
\put(24000,12000){\circle*{300}}

\put(0,16000){\circle*{300}}
\put(4000,16000){\circle*{300}}
\put(8000,16000){\circle*{300}}
\put(12000,16000){\circle*{300}}
\put(16000,16000){\circle*{300}}
\put(20000,16000){\circle*{300}}
\put(24000,16000){\circle*{300}}

\put(0,20000){\circle*{300}}
\put(4000,20000){\circle*{300}}
\put(8000,20000){\circle*{300}}
\put(12000,20000){\circle*{300}}
\put(16000,20000){\circle*{300}}
\put(20000,20000){\circle*{300}}
\put(24000,20000){\circle*{300}}

\put(0,24000){\circle*{300}}
\put(-1000,24000){\scriptsize$h$}
\put(4000,24000){\circle*{300}}
\put(8000,24000){\circle*{300}}
\put(12000,24000){\circle*{300}}
\put(16000,24000){\circle*{300}}
\put(20000,24000){\circle*{300}}
\put(24000,24000){\circle*{300}}
\put(24300,24000){\tiny$h+2\overline{r}_N$}

\path(0,2000)(24000,2000)
\blacken\path(0,2400)(0,1600)(800,2000)(0,2400)
\blacken\path(24000,2400)(24000,1600)(23200,2000)(24000,2400)
\path(-6500,2000)(-5500,2000)
\whiten\path(-5900,2200)(-5900,1800)(-5500,2000)(-5900,2200)
\put(-5000,2000){\fbox{\scriptsize$v_N,q_N$}}

\path(0,6000)(24000,6000)
\blacken\path(0,6400)(0,5600)(800,6000)(0,6400)
\blacken\path(24000,6400)(24000,5600)(23200,6000)(24000,6400)
\path(-6500,6000)(-5500,6000)
\whiten\path(-5900,6200)(-5900,5800)(-5500,6000)(-5900,6200)

\path(0,10000)(24000,10000)
\blacken\path(0,10400)(0,9600)(800,10000)(0,10400)
\blacken\path(24000,10400)(24000,9600)(23200,10000)(24000,10400)
\path(-6500,10000)(-5500,10000)
\whiten\path(-5900,10200)(-5900,9800)(-5500,10000)(-5900,10200)

\path(0,14000)(24000,14000)
\blacken\path(0,14400)(0,13600)(800,14000)(0,14400)
\blacken\path(24000,14400)(24000,13600)(23200,14000)(24000,14400)
\path(-6500,14000)(-5500,14000)
\whiten\path(-5900,14200)(-5900,13800)(-5500,14000)(-5900,14200)

\path(0,18000)(24000,18000)
\blacken\path(0,18400)(0,17600)(800,18000)(0,18400)
\blacken\path(24000,18400)(24000,17600)(23200,18000)(24000,18400)
\path(-6500,18000)(-5500,18000)
\whiten\path(-5900,18200)(-5900,17800)(-5500,18000)(-5900,18200)

\path(0,22000)(24000,22000)
\blacken\path(0,22400)(0,21600)(800,22000)(0,22400)
\blacken\path(24000,22400)(24000,21600)(23200,22000)(24000,22400)
\path(-6500,22000)(-5500,22000)
\whiten\path(-5900,22200)(-5900,21800)(-5500,22000)(-5900,22200)
\put(-5000,22000){\fbox{\scriptsize$v_1,q_1$}}

\path(2000,0)(2000,24000)
\blacken\path(1600,0)(2400,0)(2000,-800)(1600,0)
\blacken\path(1600,24000)(2400,24000)(2000,24800)(1600,24000)
\path(2000,-4200)(2000,-3200)
\whiten\path(1800,-3600)(2200,-3600)(2000,-3200)(1800,-3600)
\put(0,-2200){\fbox{\scriptsize$w_1,r_1$}}

\path(6000,0)(6000,24000)
\blacken\path(5600,0)(6400,0)(6000,-800)(5600,0)
\blacken\path(5600,24000)(6400,24000)(6000,24800)(5600,24000)
\path(6000,-4200)(6000,-3200)
\whiten\path(5800,-3600)(6200,-3600)(6000,-3200)(5800,-3600)

\path(10000,0)(10000,24000)
\blacken\path(9600,0)(10400,0)(10000,-800)(9600,0)
\blacken\path(9600,24000)(10400,24000)(10000,24800)(9600,24000)
\path(10000,-4200)(10000,-3200)
\whiten\path(9800,-3600)(10200,-3600)(10000,-3200)(9800,-3600)

\path(14000,0)(14000,24000)
\blacken\path(13600,0)(14400,0)(14000,-800)(13600,0)
\blacken\path(13600,24000)(14400,24000)(14000,24800)(13600,24000)
\path(14000,-4200)(14000,-3200)
\whiten\path(13800,-3600)(14200,-3600)(14000,-3200)(13800,-3600)

\path(18000,0)(18000,24000)
\blacken\path(17600,0)(18400,0)(18000,-800)(17600,0)
\blacken\path(17600,24000)(18400,24000)(18000,24800)(17600,24000)
\path(18000,-4200)(18000,-3200)
\whiten\path(17800,-3600)(18200,-3600)(18000,-3200)(17800,-3600)

\path(22000,0)(22000,24000)
\blacken\path(21600,0)(22400,0)(22000,-800)(21600,0)
\blacken\path(21600,24000)(22400,24000)(22000,24800)(21600,24000)
\path(22000,-4200)(22000,-3200)
\whiten\path(21800,-3600)(22200,-3600)(22000,-3200)(21800,-3600)
\put(19500,-2200){\fbox{\scriptsize$w_N,r_N$}}

\end{picture}

\end{minipage}
\end{center}

\caption[Domain wall partition function of the elliptic Deguchi-Akutsu model]{Domain wall partition function of the elliptic Deguchi-Akutsu model. The top row of arrows corresponds with the state vector $|\Uparrow_N\rangle$. The bottom row of arrows corresponds with the dual state vector $\langle \Downarrow_N|$. Each horizontal lattice line represents multiplication by a $B(v_j,q_j,\{w,r\}_N,h+2\overline{q}_{j-1})$ operator. The ordering of lattice lines agrees with the ordering of $B$-operators in (\ref{pfpfpf}).}
\end{figure}
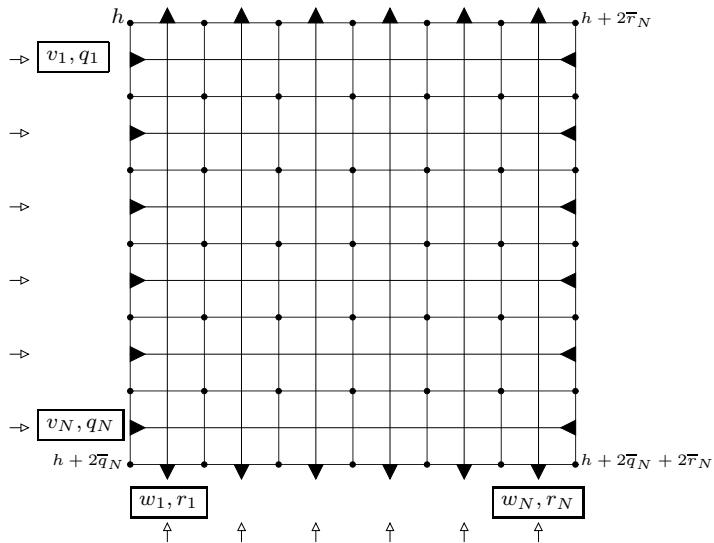

\subsection{Conditions on $Z_N(\{v,q\}_N,\{w,r\}_N,h)$}

We provide a set of Korepin-type conditions on $Z_N(\{v,q\}_N,\{w,r\}_N,h)$, which were originally obtained in \cite{fwz2}. 

\begin{lemma} 
{\rm As always, we make the abbreviation $Z_N = Z_N(\{v,q\}_N,\{w,r\}_N,h)$. For all $N \geq 2$ we claim that

\begin{property4}
{\rm 
$Z_N$ is an entire function in $v_N$ that satisfies the quasi-periodicity conditions

\begin{align}
&
Z_N\Big|_{v_N \rightarrow v_N+2}
=
(-)^N Z_N
\label{quas1-da}
\\
&
Z_N
\Big|_{v_N \rightarrow v_N - 2i\log(\nome)/\pi}
=
\label{quas2-da}
\frac{(-)^N}{\nome^N}
\exp\Big(\pi i(\eta-Nv_N)\Big)
Z_N
\end{align}

\noindent where we have defined

\begin{align}
\eta
&=
h+2\sum_{j=1}^{N-1}q_j
+
Nq_N
+
\sum_{j=1}^{N}(w_j+r_j)
\label{eta-da}
\end{align}

}
\end{property4}

\begin{property4}
{\rm
$Z_N$ has simple zeros at the points $v_N = v_j+q_j+q_N$, for all $1\leq j \leq N-1$.
}
\end{property4}

\begin{property4}
{\rm
Setting $v_N = w_N+q_N+r_N$, $Z_N$ satisfies the recursion relation

\begin{align}
Z_N\Big|_{v_N = w_N+q_N+r_N}
&=
\frac{\displaystyle{
[h+2\overline{q}_{N-1}
]^{\frac{1}{2}}
[h+2\overline{r}_{N-1}
]^{\frac{1}{2}}
}}
{\displaystyle{
[h+2\overline{q}_N
]^{\frac{1}{2}}
[h+2\overline{r}_N
]^{\frac{1}{2}}
}}
[2q_N]^{\frac{1}{2}} [2r_N]^{\frac{1}{2}}
\label{PFrec-da}
\\
&
\times
\prod_{j=1}^{N-1}
[w_N-w_j+r_j+r_N][v_j-w_N+q_j-r_N]
Z_{N-1}
\nonumber
\end{align}

\noindent where $Z_{N-1}$ is the domain wall partition function on a square lattice of size $N-1$.


}
\end{property4}

In addition, we have the supplementary condition

\begin{property4}
{\rm The partition function on the $1\times 1$ lattice is given by 

\begin{align}
Z_1(v_1,q_1,w_1,r_1,h)
=
\frac{[2q_1]^{\frac{1}{2}} [2r_1]^{\frac{1}{2}}}
{[h+2q_1]^{\frac{1}{2}} [h+2r_1]^{\frac{1}{2}}}
[w_1-v_1+q_1+r_1+h]
\end{align}
}
\end{property4}

}
\end{lemma}

\begin{proof}

\begin{property5}
{\rm
Inserting the set of states $\sum_{n=1}^{N} \sigma_n^{+} |\Downarrow_N\rangle \langle \Downarrow_N| \sigma_n^{-}$ after the first $B$-operator appearing in (\ref{pfpfpf}), the domain wall partition function may be written in the form

\begin{align}
Z_N\Big(
\{v,q\}_N,\{w,r\}_N,h
\Big)
&=
\sum_{n=1}^{N}
\langle \Downarrow_N | 
B\Big(v_N,q_N,\{w,r\}_N,h+2\overline{q}_{N-1}
\Big)
\sigma_n^{+}
|\Downarrow_N \rangle
\nonumber
\\
&
\times
\langle \Downarrow_N| \sigma_n^{-}
\lprod_{j=1}^{N-1}
B\Big(v_j,q_j,\{w,r\}_N,h+2\overline{q}_{j-1}
\Big)
|\Uparrow_N\rangle
\label{PFexp-da}
\end{align}

\noindent in which all dependence on $v_N$ appears in the first factor within the sum. We therefore proceed to calculate 
$
\langle \Downarrow_N | 
B(v_N,q_N,\{w,r\}_N,h+2\overline{q}_{N-1}
)
\sigma_n^{+}
|\Downarrow_N \rangle
$ for all $1\leq n \leq N$, as shown below.

\begin{figure}[H]

\begin{center}
\begin{minipage}{4.3in}

\setlength{\unitlength}{0.00030cm}
\begin{picture}(20000,18000)(-6000,-12000)

\path(0,0)(24000,0)
\path(0,4000)(24000,4000)

\path(0,0)(0,4000)
\path(4000,0)(4000,4000)
\path(8000,0)(8000,4000)
\path(12000,0)(12000,4000)
\path(16000,0)(16000,4000)
\path(20000,0)(20000,4000)
\path(24000,0)(24000,4000)

\put(0,0){\circle*{300}}
\put(-3500,0){\tiny$h+2\overline{q}_N$}
\put(4000,0){\circle*{300}}
\put(8000,0){\circle*{300}}
\put(12000,0){\circle*{300}}
\put(16000,0){\circle*{300}}
\put(20000,0){\circle*{300}}
\put(24000,0){\circle*{300}}
\put(24300,0){\tiny$h+2\overline{q}_N+2\overline{r}_N$}

\put(0,4000){\circle*{300}}
\put(-4500,4000){\tiny$h+2\overline{q}_{N-1}$}
\put(4000,4000){\circle*{300}}
\put(8000,4000){\circle*{300}}
\put(12000,4000){\circle*{300}}
\put(16000,4000){\circle*{300}}
\put(20000,4000){\circle*{300}}
\put(24000,4000){\circle*{300}}
\put(24300,4000){\tiny$h+2\overline{q}_{N-1}+2\overline{r}_N$}


\path(0,2000)(24000,2000)
\blacken\path(0,2400)(0,1600)(800,2000)(0,2400)
\blacken\path(24000,2400)(24000,1600)(23200,2000)(24000,2400)
\path(-6500,2000)(-5500,2000)
\whiten\path(-5900,2200)(-5900,1800)(-5500,2000)(-5900,2200)
\put(-5000,2000){\fbox{\tiny$v_N,q_N$}}


\path(2000,0)(2000,4000)
\blacken\path(1600,0)(2400,0)(2000,-800)(1600,0)
\blacken\path(1600,4000)(2400,4000)(2000,3200)(1600,4000)
\path(2000,-3500)(2000,-2500)
\whiten\path(1800,-2900)(2200,-2900)(2000,-2500)(1800,-2900)
\put(500,-1750){\fbox{\tiny$w_1,r_1$}}

\path(6000,0)(6000,4000)
\blacken\path(5600,0)(6400,0)(6000,-800)(5600,0)
\blacken\path(5600,4000)(6400,4000)(6000,3200)(5600,4000)
\path(6000,-3500)(6000,-2500)
\whiten\path(5800,-2900)(6200,-2900)(6000,-2500)(5800,-2900)

\path(10000,0)(10000,4000)
\blacken\path(9600,0)(10400,0)(10000,-800)(9600,0)
\blacken\path(9600,4000)(10400,4000)(10000,4800)(9600,4000)
\path(10000,-3500)(10000,-2500)
\whiten\path(9800,-2900)(10200,-2900)(10000,-2500)(9800,-2900)
\put(8500,-1750){\fbox{\tiny$w_n,r_n$}}

\path(14000,0)(14000,4000)
\blacken\path(13600,0)(14400,0)(14000,-800)(13600,0)
\blacken\path(13600,4000)(14400,4000)(14000,3200)(13600,4000)
\path(14000,-3500)(14000,-2500)
\whiten\path(13800,-2900)(14200,-2900)(14000,-2500)(13800,-2900)

\path(18000,0)(18000,4000)
\blacken\path(17600,0)(18400,0)(18000,-800)(17600,0)
\blacken\path(17600,4000)(18400,4000)(18000,3200)(17600,4000)
\path(18000,-3500)(18000,-2500)
\whiten\path(17800,-2900)(18200,-2900)(18000,-2500)(17800,-2900)

\path(22000,0)(22000,4000)
\blacken\path(21600,0)(22400,0)(22000,-800)(21600,0)
\blacken\path(21600,4000)(22400,4000)(22000,3200)(21600,4000)
\path(22000,-3500)(22000,-2500)
\whiten\path(21800,-2900)(22200,-2900)(22000,-2500)(21800,-2900)
\put(20000,-1750){\fbox{\tiny$w_N,r_N$}}


\put(-8000,-8250){$=$}


\path(0,-10000)(24000,-10000)
\path(0,-6000)(24000,-6000)

\path(0,-10000)(0,-6000)
\path(4000,-10000)(4000,-6000)
\path(8000,-10000)(8000,-6000)
\path(12000,-10000)(12000,-6000)
\path(16000,-10000)(16000,-6000)
\path(20000,-10000)(20000,-6000)
\path(24000,-10000)(24000,-6000)

\put(0,-10000){\circle*{300}}
\put(-3500,-10000){\tiny$h+2\overline{q}_N$}
\put(4000,-10000){\circle*{300}}
\put(8000,-10000){\circle*{300}}
\put(12000,-10000){\circle*{300}}
\put(16000,-10000){\circle*{300}}
\put(20000,-10000){\circle*{300}}
\put(24000,-10000){\circle*{300}}
\put(24300,-10000){\tiny$h+2\overline{q}_N+2\overline{r}_N$}

\put(0,-6000){\circle*{300}}
\put(-4500,-6000){\tiny$h+2\overline{q}_{N-1}$}
\put(4000,-6000){\circle*{300}}
\put(8000,-6000){\circle*{300}}
\put(12000,-6000){\circle*{300}}
\put(16000,-6000){\circle*{300}}
\put(20000,-6000){\circle*{300}}
\put(24000,-6000){\circle*{300}}
\put(24300,-6000){\tiny$h+2\overline{q}_{N-1}+2\overline{r}_N$}


\path(0,-8000)(24000,-8000)
\blacken\path(0,-7600)(0,-8400)(800,-8000)(0,-7600)
\blacken\path(4000,-7600)(4000,-8400)(4800,-8000)(4000,-7600)
\blacken\path(8000,-7600)(8000,-8400)(8800,-8000)(8000,-7600)
\blacken\path(12000,-7600)(12000,-8400)(11200,-8000)(12000,-7600)
\blacken\path(16000,-7600)(16000,-8400)(15200,-8000)(16000,-7600)
\blacken\path(20000,-7600)(20000,-8400)(19200,-8000)(20000,-7600)
\blacken\path(24000,-7600)(24000,-8400)(23200,-8000)(24000,-7600)
\path(-6500,-8000)(-5500,-8000)
\whiten\path(-5900,-7800)(-5900,-8200)(-5500,-8000)(-5900,-7800)
\put(-5000,-8000){\fbox{\tiny$v_N,q_N$}}


\path(2000,-10000)(2000,-6000)
\blacken\path(1600,-10000)(2400,-10000)(2000,-10800)(1600,-10000)
\blacken\path(1600,-6000)(2400,-6000)(2000,-6800)(1600,-6000)
\path(2000,-13500)(2000,-12500)
\whiten\path(1800,-12900)(2200,-12900)(2000,-12500)(1800,-12900)
\put(500,-11750){\fbox{\tiny$w_1,r_1$}}

\path(6000,-10000)(6000,-6000)
\blacken\path(5600,-10000)(6400,-10000)(6000,-10800)(5600,-10000)
\blacken\path(5600,-6000)(6400,-6000)(6000,-6800)(5600,-6000)
\path(6000,-13500)(6000,-12500)
\whiten\path(5800,-12900)(6200,-12900)(6000,-12500)(5800,-12900)

\path(10000,-10000)(10000,-6000)
\blacken\path(9600,-10000)(10400,-10000)(10000,-10800)(9600,-10000)
\blacken\path(9600,-6000)(10400,-6000)(10000,-5200)(9600,-6000)
\path(10000,-13500)(10000,-12500)
\whiten\path(9800,-12900)(10200,-12900)(10000,-12500)(9800,-12900)
\put(8500,-11750){\fbox{\tiny$w_n,r_n$}}

\path(14000,-10000)(14000,-6000)
\blacken\path(13600,-10000)(14400,-10000)(14000,-10800)(13600,-10000)
\blacken\path(13600,-6000)(14400,-6000)(14000,-6800)(13600,-6000)
\path(14000,-13500)(14000,-12500)
\whiten\path(13800,-12900)(14200,-12900)(14000,-12500)(13800,-12900)

\path(18000,-10000)(18000,-6000)
\blacken\path(17600,-10000)(18400,-10000)(18000,-10800)(17600,-10000)
\blacken\path(17600,-6000)(18400,-6000)(18000,-6800)(17600,-6000)
\path(18000,-13500)(18000,-12500)
\whiten\path(17800,-12900)(18200,-12900)(18000,-12500)(17800,-12900)

\path(22000,-10000)(22000,-6000)
\blacken\path(21600,-10000)(22400,-10000)(22000,-10800)(21600,-10000)
\blacken\path(21600,-6000)(22400,-6000)(22000,-6800)(21600,-6000)
\path(22000,-13500)(22000,-12500)
\whiten\path(21800,-12900)(22200,-12900)(22000,-12500)(21800,-12900)
\put(20000,-11750){\fbox{\tiny$w_N,r_N$}}

\end{picture}

\end{minipage}
\end{center}

\caption[Peeling away the bottom row of the elliptic Deguchi-Akutsu partition function]{Peeling away the bottom row of the elliptic Deguchi-Akutsu partition function. The top diagram represents $\langle \Downarrow_N | 
B(v_N,q_N,\{w,r\}_N,h+2\overline{q}_{N-1})
\sigma_n^{+}|\Downarrow_N \rangle$, with the internal black arrows being summed over all configurations. The lower diagram represents the only surviving configuration.}

\label{da-peel}
\end{figure}
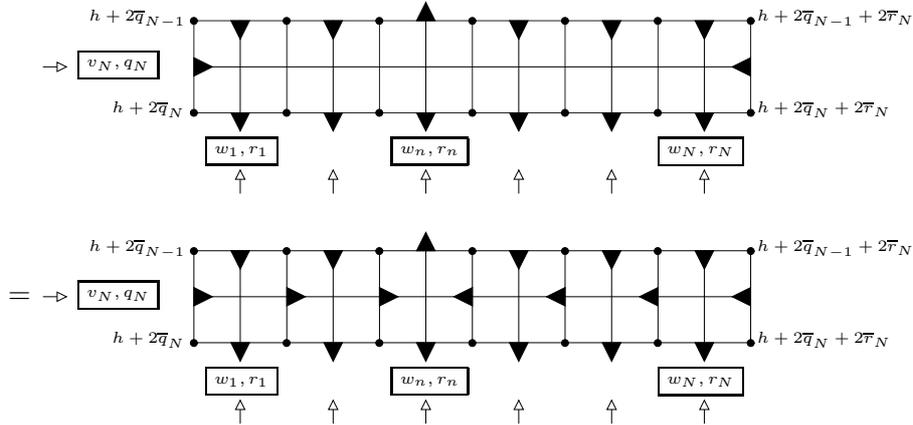

\noindent Replacing each face in figure \ref{da-peel} with its corresponding elliptic weight, we thus obtain

%
%
%
%

{\small
\begin{align}
&
\langle \Downarrow_N|
B\Big(v_N,q_N,\{w,r\}_N,h+2\overline{q}_{N-1}
\Big)
\sigma_n^{+}
|\Downarrow_N\rangle
=
c_{+}(v_N,q_N,w_n,r_n,h+2\overline{q}_{N-1}
+2\overline{r}_{n-1}
)
\times
\nonumber
\\
&
\prod_{1\leq j<n}
b_{+}(
v_N,q_N,w_j,r_j,h+2\overline{q}_{N-1}
+2\overline{r}_{j-1}
)
\prod_{n < j \leq N}
a_{-}(v_N,q_N,w_j,r_j)
\label{da-peel2}
\end{align}
}

\noindent Recalling the explicit form of these functions, as given by equations (\ref{a-da})--(\ref{c-da}), it is clear that $\langle \Downarrow_N|
B(v_N,q_N,\{w,r\}_N,h+2\overline{q}_{N-1}
)
\sigma_n^{+}
|\Downarrow_N\rangle$ is entire in $v_N$. Furthermore, we immediately see that 

\begin{align}
&
\langle \Downarrow_N|
B\Big(v_N+2,q_N,\{w,r\}_N,h+2\overline{q}_{N-1}
\Big)
\sigma_n^{+}
|\Downarrow_N\rangle
=
\\
&
(-)^N
\langle \Downarrow_N|
B\Big(v_N,q_N,\{w,r\}_N,h+2\overline{q}_{N-1}
\Big)
\sigma_n^{+}
|\Downarrow_N\rangle
\nonumber
\end{align}

\noindent which proves that the first quasi-periodicity condition (\ref{quas1-da}) holds. The proof of the second quasi-periodicity condition (\ref{quas2-da}) is established using the identities

{\small
\begin{align}
&
b_{+}(v_N-2i\log(\nome)/\pi,q_N,w_j,r_j,h+2\overline{q}_{N-1}
+2\overline{r}_{j-1}
)
=
\\
&
-\frac{1}{\nome}
\exp\Big(\pi i(w_j-r_j+q_N-v_N)\Big)
b_{+}(v_N,q_N,w_j,r_j,h+2\overline{q}_{N-1}
+2\overline{r}_{j-1}
)
\nonumber
\end{align}
}

{\footnotesize
\begin{align}
&
c_{+}(v_N-2i\log(\nome)/\pi,q_N,w_n,r_n,h+2\overline{q}_{N-1}
+2\overline{r}_{n-1}
)
=
\\
&
-\frac{1}{\nome}
\exp\Big(\pi i(w_n+q_N+r_n+h+2\overline{q}_{N-1}
+2\overline{r}_{n-1}-v_N
)\Big)
c_{+}(v_N,q_N,w_n,r_n,h+2\overline{q}_{N-1}
+2\overline{r}_{n-1}
)
\nonumber
\end{align}
}

{\small
\begin{align}
&
a_{-}(v_N-2i\log(\nome)/\pi,q_N,w_j,r_j)
=
-\frac{1}{\nome}
\exp\Big(\pi i(w_j+r_j+q_N-v_N)\Big)
a_{-}(v_N,q_N,w_j,r_j)
\end{align}
}

\noindent which, when substituted into (\ref{da-peel2}), produce the equation

\begin{align}
&
\langle \Downarrow_N|
B\Big(v_N-2i\log(q)/\pi,q_N,\{w,r\}_N,h+2\overline{q}_{N-1}
\Big)
\sigma_n^{+}
|\Downarrow_N\rangle
=
\\
&
\frac{(-)^N}{\nome^N}
\exp\Big(\pi i(\eta-Nv_N)\Big)
\langle \Downarrow_N|
B\Big(v_N,q_N,\{w,r\}_N,h+2\overline{q}_{N-1}
\Big)
\sigma_n^{+}
|\Downarrow_N\rangle
\nonumber
\end{align}

\noindent where $\eta$ is given by (\ref{eta-da}). Hence the second quasi-periodicity condition (\ref{quas2-da}) also holds.
 
}
\end{property5}

\begin{property5}
{\rm
We rearrange the expression (\ref{pfpfpf}) for the partition function by using the commutation relation

{\small
\begin{align}
&
[v_N-v_j+q_j+q_N]
B\Big(v_N,q_N,\{w,r\}_N,h+2\overline{q}_j
\Big)
B\Big(v_j,q_j,\{w,r\}_N,h+2\overline{q}_{j-1}
\Big)
=
\\
&
[v_j-v_N+q_j+q_N]
B\Big(v_j,q_j,\{w,r\}_N,h+2\overline{q}_{j-1}
+2q_N
\Big)
B\Big(v_N,q_N,\{w,r\}_N,h+2\overline{q}_{j-1}
\Big)
\nonumber
\end{align}
}

\noindent repeatedly to change the order of the $B$-operators. This effectively amounts to using figures 6.17 and 6.18 to reorder the horizontal lattice lines in figure 6.19. We obtain

{\footnotesize
\begin{align}
&
\prod_{j=1}^{N-1}
[v_N-v_j+q_j+q_N]
Z_N\Big(\{v,q\}_N,\{w,r\}_N,h\Big)
=
\label{reorder-da}
\\
&
\prod_{j=1}^{N-1}
[v_j-v_N+q_j+q_N]
\langle \Downarrow_N|
\lprod_{j=1}^{N-1}
B\Big(v_j,q_j,\{w,r\}_N,h+2\overline{q}_{j-1}
+2q_N
\Big)
B\Big(v_N,q_N,\{w,r\}_N,h\Big)
|\Uparrow_N \rangle
\nonumber
\end{align}
}


%

\noindent The right hand side of (\ref{reorder-da}) is an entire function in $v_N$, with simple zeros at the points $v_N = v_j+q_j+q_N$ for all $1\leq j \leq N-1$. Therefore the partition function $Z_N(\{v,q\}_N,\{w,r\}_N,h)$ must have simple zeros at the same points.

}
\end{property5}

\begin{property5}
{\rm
We start from the expansion (\ref{PFexp-da}) of the domain wall partition function and set $v_N = w_N + q_N + r_N$. This causes all terms in the summation over $1\leq n \leq N$ to collapse to zero except the $n=N$ term, giving

\begin{align}
Z_N
\Big|_{v_N = w_N + q_N + r_N}
&=
c_{+}(w_N+q_N+r_N,q_N,w_N,r_N,h+2\overline{q}_{N-1}+2\overline{r}_{N-1}
)
\nonumber
\\
&
\times
\prod_{j=1}^{N-1}
b_{+}(w_N+q_N+r_N,q_N,w_j,r_j,h+2\overline{q}_{N-1}
+2\overline{r}_{j-1}
)
\nonumber
\\
&
\times
\langle \Downarrow_N|
\sigma_N^{-}
\lprod_{j=1}^{N-1}
B\Big(v_j,q_j,\{w,r\}_N,h+2\overline{q}_{j-1}
\Big)
|\Uparrow_N\rangle
\label{PFexp2-da}
\end{align}

\noindent We then calculate 
$
\langle \Downarrow_N|
\sigma_N^{-}\ 
\lprod_{j=1}^{N-1}
B(v_j,q_j,\{w,r\}_N,h+2\overline{q}_{j-1})
|\Uparrow_N\rangle
$ 
using the diagram shown below.

\begin{figure}[H]

\begin{center}
\begin{minipage}{4.3in}

\setlength{\unitlength}{0.00028cm}
\begin{picture}(20000,23000)(-8000,1500)

\path(0,4000)(24000,4000)
\path(0,8000)(24000,8000)
\path(0,12000)(24000,12000)
\path(0,16000)(24000,16000)
\path(0,20000)(24000,20000)
\path(0,24000)(24000,24000)

\path(0,4000)(0,24000)
\path(4000,4000)(4000,24000)
\path(8000,4000)(8000,24000)
\path(12000,4000)(12000,24000)
\path(16000,4000)(16000,24000)
\path(20000,4000)(20000,24000)
\path(24000,4000)(24000,24000)

\put(-5200,4000){\tiny$h+2\overline{q}_{N-1}$}
\put(24300,4000){\tiny$h+2\overline{q}_{N-1}+2\overline{r}_N$}

\put(0,4000){\circle*{300}}
\put(4000,4000){\circle*{300}}
\put(8000,4000){\circle*{300}}
\put(12000,4000){\circle*{300}}
\put(16000,4000){\circle*{300}}
\put(20000,4000){\circle*{300}}
\put(24000,4000){\circle*{300}}

\put(0,8000){\circle*{300}}
\put(4000,8000){\circle*{300}}
\put(8000,8000){\circle*{300}}
\put(12000,8000){\circle*{300}}
\put(16000,8000){\circle*{300}}
\put(20000,8000){\circle*{300}}
\put(24000,8000){\circle*{300}}

\put(0,12000){\circle*{300}}
\put(4000,12000){\circle*{300}}
\put(8000,12000){\circle*{300}}
\put(12000,12000){\circle*{300}}
\put(16000,12000){\circle*{300}}
\put(20000,12000){\circle*{300}}
\put(24000,12000){\circle*{300}}

\put(0,16000){\circle*{300}}
\put(4000,16000){\circle*{300}}
\put(8000,16000){\circle*{300}}
\put(12000,16000){\circle*{300}}
\put(16000,16000){\circle*{300}}
\put(20000,16000){\circle*{300}}
\put(24000,16000){\circle*{300}}

\put(0,20000){\circle*{300}}
\put(4000,20000){\circle*{300}}
\put(8000,20000){\circle*{300}}
\put(12000,20000){\circle*{300}}
\put(16000,20000){\circle*{300}}
\put(20000,20000){\circle*{300}}
\put(24000,20000){\circle*{300}}

\put(0,24000){\circle*{300}}
\put(-1000,24000){\tiny$h$}
\put(4000,24000){\circle*{300}}
\put(8000,24000){\circle*{300}}
\put(12000,24000){\circle*{300}}
\put(16000,24000){\circle*{300}}
\put(20000,24000){\circle*{300}}
\put(24000,24000){\circle*{300}}
\put(24300,24000){\tiny$h+2\overline{r}_N$}


\path(0,6000)(24000,6000)
\blacken\path(0,6400)(0,5600)(800,6000)(0,6400)
\blacken\path(20000,6400)(20000,5600)(19200,6000)(20000,6400)
\blacken\path(24000,6400)(24000,5600)(23200,6000)(24000,6400)
\path(-9000,6000)(-8000,6000)
\whiten\path(-8400,6200)(-8400,5800)(-8000,6000)(-8400,6200)
\put(-7500,6000){\fbox{\tiny$v_{N-1},q_{N-1}$}}

\path(0,10000)(24000,10000)
\blacken\path(0,10400)(0,9600)(800,10000)(0,10400)
\blacken\path(20000,10400)(20000,9600)(19200,10000)(20000,10400)
\blacken\path(24000,10400)(24000,9600)(23200,10000)(24000,10400)
\path(-9000,10000)(-8000,10000)
\whiten\path(-8400,10200)(-8400,9800)(-8000,10000)(-8400,10200)

\path(0,14000)(24000,14000)
\blacken\path(0,14400)(0,13600)(800,14000)(0,14400)
\blacken\path(20000,14400)(20000,13600)(19200,14000)(20000,14400)
\blacken\path(24000,14400)(24000,13600)(23200,14000)(24000,14400)
\path(-9000,14000)(-8000,14000)
\whiten\path(-8400,14200)(-8400,13800)(-8000,14000)(-8400,14200)

\path(0,18000)(24000,18000)
\blacken\path(0,18400)(0,17600)(800,18000)(0,18400)
\blacken\path(20000,18400)(20000,17600)(19200,18000)(20000,18400)
\blacken\path(24000,18400)(24000,17600)(23200,18000)(24000,18400)
\path(-9000,18000)(-8000,18000)
\whiten\path(-8400,18200)(-8400,17800)(-8000,18000)(-8400,18200)

\path(0,22000)(24000,22000)
\blacken\path(0,22400)(0,21600)(800,22000)(0,22400)
\blacken\path(20000,22400)(20000,21600)(19200,22000)(20000,22400)
\blacken\path(24000,22400)(24000,21600)(23200,22000)(24000,22400)
\path(-9000,22000)(-8000,22000)
\whiten\path(-8400,22200)(-8400,21800)(-8000,22000)(-8400,22200)
\put(-7500,22000){\fbox{\tiny$v_1,q_1$}}


\path(2000,4000)(2000,24000)
\blacken\path(1600,4000)(2400,4000)(2000,3200)(1600,4000)
\blacken\path(1600,24000)(2400,24000)(2000,24800)(1600,24000)
\path(2000,0)(2000,1000)
\whiten\path(1800,600)(2200,600)(2000,1000)(1800,600)
\put(500,2000){\fbox{\tiny$w_1,r_1$}}

\path(6000,4000)(6000,24000)
\blacken\path(5600,4000)(6400,4000)(6000,3200)(5600,4000)
\blacken\path(5600,24000)(6400,24000)(6000,24800)(5600,24000)
\path(6000,0)(6000,1000)
\whiten\path(5800,600)(6200,600)(6000,1000)(5800,600)

\path(10000,4000)(10000,24000)
\blacken\path(9600,4000)(10400,4000)(10000,3200)(9600,4000)
\blacken\path(9600,24000)(10400,24000)(10000,24800)(9600,24000)
\path(10000,0)(10000,1000)
\whiten\path(9800,600)(10200,600)(10000,1000)(9800,600)

\path(14000,4000)(14000,24000)
\blacken\path(13600,4000)(14400,4000)(14000,3200)(13600,4000)
\blacken\path(13600,24000)(14400,24000)(14000,24800)(13600,24000)
\path(14000,0)(14000,1000)
\whiten\path(13800,600)(14200,600)(14000,1000)(13800,600)

\path(18000,4000)(18000,24000)
\blacken\path(17600,4000)(18400,4000)(18000,3200)(17600,4000)
\blacken\path(17600,24000)(18400,24000)(18000,24800)(17600,24000)
\path(18000,0)(18000,1000)
\whiten\path(17800,600)(18200,600)(18000,1000)(17800,600)

\path(22000,4000)(22000,24000)
\blacken\path(21600,4000)(22400,4000)(22000,4800)(21600,4000)
\blacken\path(21600,8000)(22400,8000)(22000,8800)(21600,8000)
\blacken\path(21600,12000)(22400,12000)(22000,12800)(21600,12000)
\blacken\path(21600,16000)(22400,16000)(22000,16800)(21600,16000)
\blacken\path(21600,20000)(22400,20000)(22000,20800)(21600,20000)
\blacken\path(21600,24000)(22400,24000)(22000,24800)(21600,24000)
\path(22000,0)(22000,1000)
\whiten\path(21800,600)(22200,600)(22000,1000)(21800,600)
\put(20000,2000){\fbox{\tiny$w_N,r_N$}}

\end{picture}

\end{minipage}
\end{center}

\caption[Peeling the right-most column of the elliptic Deguchi-Akutsu partition function]{Peeling the right-most column of the elliptic Deguchi-Akutsu partition function. This diagram represents $\langle \Downarrow_N|\sigma_N^{-}\ 
\lprod_{j=1}^{N-1} B(v_j,q_j,\{w,r\}_N,h+2\overline{q}_{j-1})
|\Uparrow_N\rangle$, with the internal black arrows being summed over all configurations. Each surviving configuration must contain the column of faces shown on the right of the lattice.}

\label{da-peel3}
\end{figure}
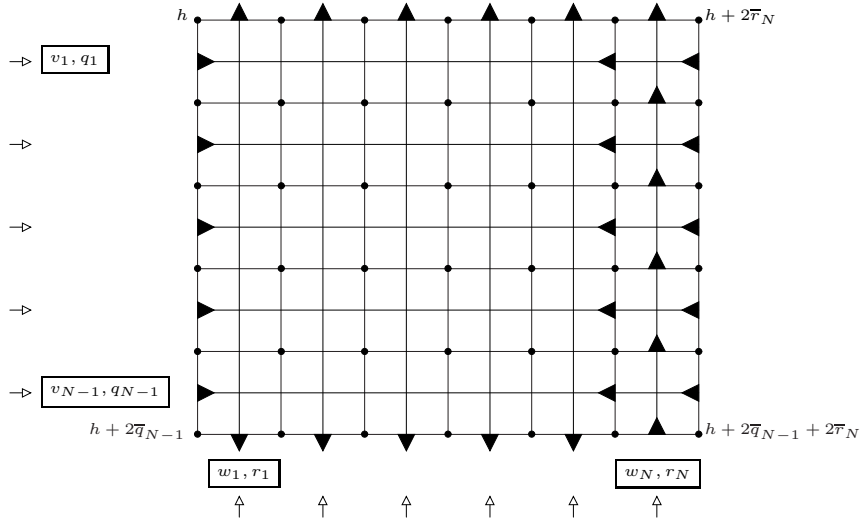

\noindent Figure \ref{da-peel3} represents the $(N-1)\times (N-1)$ domain wall partition function, multiplied by a column of faces. Replacing these faces with their elliptic weights, we obtain

\begin{align}
&
\langle \Downarrow_N|
\sigma_N^{-}
\lprod_{j=1}^{N-1}
B\Big(v_j,q_j,\{w,r\}_N,h+2\overline{q}_{j-1}
\Big)
|\Uparrow_N\rangle
=
\label{da-peel4}
\\
&
\prod_{j=1}^{N-1}
b_{-}(v_j,q_j,w_N,r_N,h+2\overline{q}_{j-1}
+2\overline{r}_{N-1}
)
Z_{N-1}\Big(\{v,q\}_{N-1},\{w,r\}_{N-1},h\Big)
\nonumber
\end{align}

\noindent Finally, substituting (\ref{da-peel4}) into (\ref{PFexp2-da}) gives

\begin{align}
Z_N\Big|_{v_N = w_N + q_N + r_N}
&=
c_{+}(w_N+q_N+r_N,q_N,w_N,r_N,h+2\overline{q}_{N-1}+2\overline{r}_{N-1}
)
\label{PFexp3-da}
\\
&
\times
\prod_{j=1}^{N-1}
b_{+}(w_N+q_N+r_N,q_N,w_j,r_j,h+2\overline{q}_{N-1}
+2\overline{r}_{j-1}
)
\nonumber
\\
&
\times
\prod_{j=1}^{N-1}
b_{-}(v_j,q_j,w_N,r_N,h+2\overline{q}_{j-1}
+2\overline{r}_{N-1}
)
Z_{N-1}
\nonumber
\end{align}

\noindent Using the explicit formulae (\ref{b-da}) and (\ref{c-da}) for the functions in (\ref{PFexp3-da}), it is straightforward to recover the recursion relation (\ref{PFrec-da}).

}
\end{property5}

\begin{property5}
{\rm
Specializing the definition (\ref{pfpfpf}) to the case $N=1$ gives

\begin{align}
Z_1(v_1,q_1,w_1,r_1,h)
&=
\langle \Downarrow_1|
B(v_1,q_1,\{w,r\}_1,h)
|\Uparrow_1\rangle
\\
&=
\uparrow_{a_1}^{*} \otimes \downarrow_1^{*}
R_{a_1 1}(v_1,q_1,w_1,r_1,h)
\uparrow_1 \otimes \downarrow_{a_1}
=
c_{+}(v_1,q_1,w_1,r_1,h)
\nonumber
\end{align}

\noindent as required. Alternatively, the $1\times 1$ partition function is the top-right face in figure 6.14, whose weight is equal to $c_{+}(v_1,q_1,w_1,r_1,h)$.

}
\end{property5}

\end{proof}

\subsection{Evaluation of $Z_N(\{v,q\}_N,\{w,r\}_N,h)$}

The conditions {\bf 1}--{\bf 4} determine $Z_N(\{v,q\}_N,\{w,r\}_N,h)$ uniquely and give a direct algorithm for its evaluation.

\begin{lemma}
{\rm
The domain wall partition function has the factorized expression

\begin{align}
&
Z_N\Big(\{v,q\}_N,\{w,r\}_N,h\Big)
=
\frac{\displaystyle{
[h+\overline{w}_N-\overline{v}_N+\overline{q}_N+\overline{r}_N
]
}}
{\displaystyle{
[h+2\overline{q}_N
]^{\frac{1}{2}}
[h+2\overline{r}_N
]^{\frac{1}{2}}
}}
\times
\label{PFans-da}
\\
&
\prod_{j=1}^{N} [2q_j]^{\frac{1}{2}}[2r_j]^{\frac{1}{2}}
\prod_{1 \leq j < k \leq N}
[v_j-v_k+q_j+q_k][w_k-w_j+r_j+r_k]
\nonumber
\end{align}

\noindent This expression was originally proved in \cite{fwz2}. Setting $q_j=r_j=\frac{1}{2}$ for all $1\leq j \leq N$, we recover the formula (\ref{PFexp-sos}) for the partition function of the SOS model at its free fermion point. In addition, taking the trigonometric $\nome \rightarrow 0$ and heightless $h\rightarrow i\infty$ limits of (\ref{PFans-da}), we recover the partition function of the trigonometric Felderhof model (\ref{lem-tf}).

}
\end{lemma}

\begin{proof}
From theorem 1 and conditions {\bf 1} and {\bf 2} on $Z_N(\{v,q\}_N,\{w,r\}_N,h)$, we know that it must have the form

\begin{align}
&
Z_N\Big(\{v,q\}_N,\{w,r\}_N,h\Big)
=
\mathcal{C}\Big(\{v\}_{N-1},\{q\}_N,\{w,r\}_N,h\Big)
\times
\label{coeff-da}
\\
&
\prod_{j=1}^{N-1} [v_j-v_N+q_j+q_N]
[h+\overline{w}_N-\overline{v}_N+\overline{q}_N+\overline{r}_N
]
\nonumber
\end{align}

\noindent where $\mathcal{C}$ does not depend on $v_N$, but depends on all other variables. Evaluating (\ref{coeff-da}) at $v_N = w_N+q_N+r_N$, we obtain

\begin{align}
&
Z_N\Big|_{v_N = w_N+q_N+r_N}
=
\mathcal{C}\Big(\{v\}_{N-1},\{q\}_N,\{w,r\}_N,h\Big)
\times
\label{coeff2-da}
\\
&
\prod_{j=1}^{N-1} [v_j-w_N+q_j-r_N]
[h+\overline{w}_{N-1}-\overline{v}_{N-1}+\overline{q}_{N-1}+\overline{r}_{N-1}
]
\nonumber
\end{align}

\noindent Comparing (\ref{coeff2-da}) with condition {\bf 3} on $Z_N(\{v,q\}_N,\{w,r\}_N,h)$, we arrive at the expression

\begin{align}
&
\mathcal{C}\Big(\{v\}_{N-1},\{q\}_N,\{w,r\}_N,h\Big)
=
\frac{\displaystyle{
[h+2\overline{q}_{N-1}
]^{\frac{1}{2}}
[h+2\overline{r}_{N-1}
]^{\frac{1}{2}}
}}
{\displaystyle{
[h+2\overline{q}_N
]^{\frac{1}{2}}
[h+2\overline{r}_N
]^{\frac{1}{2}}
}}
\times
\\
&
\frac{\displaystyle{
[2q_N]^{\frac{1}{2}}
[2r_N]^{\frac{1}{2}}
\prod_{j=1}^{N-1}
[w_N-w_j+r_j+r_N]
}}
{\displaystyle{
[h+\overline{w}_{N-1}-\overline{v}_{N-1}+\overline{q}_{N-1}+\overline{r}_{N-1}
]
}}
Z_{N-1}\Big(\{v,q\}_{N-1},\{w,r\}_{N-1},h\Big)
\nonumber
\end{align}

\noindent Finally, substituting this expression for $\mathcal{C}$ into (\ref{coeff-da}), we obtain the recurrence

\begin{align}
&
Z_N\Big(\{v,q\}_N,\{w,r\}_N,h\Big)
=
\\
&
\frac{\displaystyle{
[h+2\overline{q}_{N-1}
]^{\frac{1}{2}}
[h+2\overline{r}_{N-1}
]^{\frac{1}{2}}
}}
{\displaystyle{
[h+2\overline{q}_N
]^{\frac{1}{2}}
[h+2\overline{r}_N
]^{\frac{1}{2}}
}}
\frac{
[h+\overline{w}_N-\overline{v}_N+\overline{q}_N+\overline{r}_N
]
}{
[h+\overline{w}_{N-1}-\overline{v}_{N-1}+\overline{q}_{N-1}+\overline{r}_{N-1}
]
}
[2q_N]^{\frac{1}{2}}
[2r_N]^{\frac{1}{2}}
\times
\nonumber
\\
&
\prod_{j=1}^{N-1}
[v_j-v_N+q_j+q_N][w_N-w_j+r_j+r_N]
Z_{N-1}\Big(\{v,q\}_{N-1},\{w,r\}_{N-1},h\Big)
\nonumber
\end{align}

\noindent whose basis is given by condition {\bf 4}. This recurrence is trivially solved to produce the formula (\ref{PFans-da}).

\end{proof}

\section{Conclusion}

In this chapter we have investigated the free fermion condition in lattice models. The starting point in our studies was the free fermion point of the six-vertex model, whose partition function and Bethe scalar product both factorize into product form. The main result of the chapter is that this factorization persists when non-trivial external fields are introduced into the model. Indeed, we found that the partition function and Bethe scalar product both factorize when considering the trigonometric Felderhof model, which is an external field deformation of the free fermionic six-vertex model.

Another key result was the observation that these ideas can be extended to models with a height parameter. We claimed that at its free fermion point, the SOS model has a factorized domain wall partition function. This result indicates that the free fermion point is a powerful restriction, since for general values of the crossing parameter the partition function has a relatively complicated non-determinant expression \cite{ros}. Furthermore, we showed that this factorization persists at the level of the elliptic Deguchi-Akutsu height model, which contains external fields.

There is potential for further work in the context of these models. We list some of these problems below.  

{\bf 1.} It should be possible to extend the results obtained for the trigonometric Felderhof model to the calculation of its one and two-point correlation functions. In particular, drawing on the work of \cite{kmt}, one could express the local spin operators $\sigma_m^{\pm},\sigma_m^{z}$ in terms of the monodromy matrix operators. At the very least, the calculation of the one-point functions is then facilitated by our expression for the Bethe scalar product. 

{\bf 2.} The situation is more complicated for the elliptic Deguchi-Akutsu height model, and the evaluation of its scalar product remains unsolved. The difficulty essentially arises from the fact that the $c_{\pm}$ faces depend on the rapidities, unlike the $c_{\pm}$ vertices in the trigonometric Felderhof model. This leads to an extra zero in the height model scalar product, compared with its vertex model counterpart.


\newpage

\thispagestyle{empty}

\phantom{nothing}


\addcontentsline{toc}{chapter} 
{\protect\numberline{Bibliography\hspace{-96pt}}}


\end{document}